\documentclass[11pt,dutch,british]{book}
\usepackage{amsfonts,amsmath,amssymb} 
\usepackage{comment} 
\excludecomment{comment}  

\usepackage{color} 
\usepackage{pslatex}
\usepackage[section]{placeins} 
\usepackage{float}
\usepackage{subfigure}
\usepackage{graphicx,wrapfig,hyperref}
\usepackage[small,bf]{caption}
\usepackage{slashed}
\usepackage{cite}

\numberwithin{equation}{section}
\definecolor{Darkgreen}{RGB}{0,100,0}
\definecolor{Forestgreen}{RGB}{34,139,34}
\definecolor{Mediumblue}{RGB}{0,0,205} 

\newdimen\mylength

\begin{document} 

\begin{titlepage}
\pagestyle{empty}
\setlength{\topmargin}{0cm}

\large

\begin{figure}
\hspace{2.5cm}
\subfigure{\includegraphics[width=30mm]{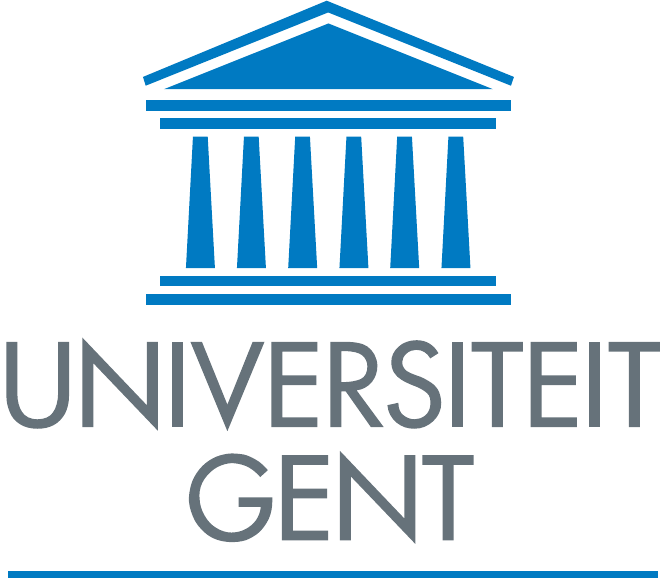}}
\hspace{3cm}
\subfigure{\includegraphics[width=60mm]{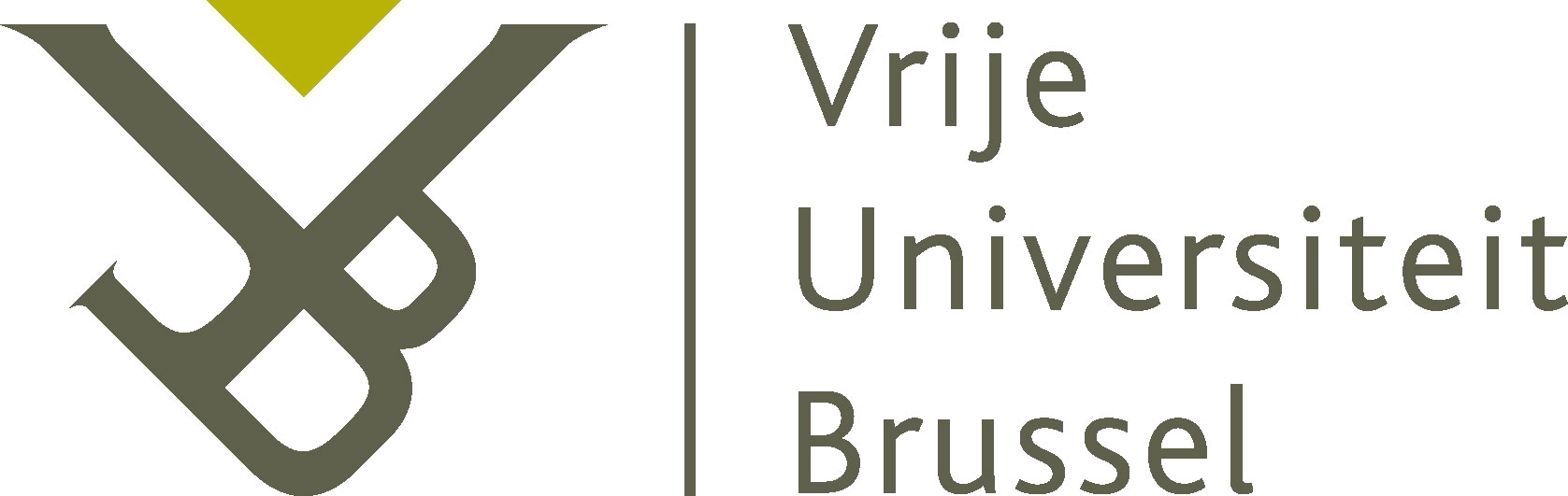}}
\end{figure}

\begin{minipage}[t]{0.45\textwidth}
\begin{center}
Faculty of Sciences \\
\end{center}
\end{minipage}
\begin{minipage}[t]{0.45\textwidth}  
\begin{center}
Faculty of Science and \\ Bio-Engineering Sciences \\
\end{center}
\end{minipage} 

\vspace{0.5cm}

\begin{minipage}[t]{0.45\textwidth}
\begin{center}
Department of Physics \\and Astronomy 
\end{center}
\end{minipage}
\begin{minipage}[t]{0.45\textwidth}  
\begin{center}
Department of Physics 
\end{center}
\end{minipage}

\begin{center}

\vspace{1cm}

Academic year 2013-2014 \\

\vspace*{2cm}

{\Huge\bf Background field and time dependence effects in holographic models} \\

\vspace*{1cm}

{\LARGE Nele \textsc{Callebaut}} 

\vspace*{3cm}

Promotors at Ghent University: Prof.\ Dr.\ David Dudal and Prof.\ Dr.\ Henri Verschelde \\ 
Promotor at Vrije Universiteit Brussel: Prof.\ Dr.\ Ben Craps  

\vspace*{2cm}

Thesis submitted in fulfillment of the requirements for the degree of  \\ 
\textsc{Doctor in Sciences: Physics} \\ 
at Ghent University, and  \\
\textsc{Doctor in Sciences} \\ 
at Vrije Universiteit Brussel 

\end{center}

\end{titlepage}

\newpage
\thispagestyle{empty}
\mbox{}

{\large Promoters} \vspace{-2mm}\\
\rule{\linewidth}{0.25mm}\vspace{2mm}
Prof.\ Dr.\ David Dudal,\\ 
Universiteit Gent \vspace{\mylength} \\  
Prof.\ Dr.\ Ben Craps,\\ 
Vrije Universiteit Brussel \vspace{\mylength}\\ 
Prof.\ Dr.\ Henri Verschelde,\\
Universiteit Gent \vspace{\mylength}\\ 

{
\large Members of the examination committee} \vspace{-2mm}\\
\rule{\linewidth}{0.25mm}\vspace{2mm}
Prof.\ Dr.\ David Dudal (promoter),\\ 
Universiteit Gent \vspace{\mylength} \\  
Prof.\ Dr.\ Ben Craps (promoter),\\ 
Vrije Universiteit Brussel  \vspace{\mylength}\\ 
Prof.\ Dr.\ Henri Verschelde (co-promoter),\\
Universiteit Gent \vspace{\mylength}\\ 
Prof.\ Dr.\ Jan Ryckebusch (UGent president), \\ 
Universiteit Gent \vspace{\mylength}\\ 
Prof.\ Dr.\ Alexander Sevrin (VUB president),\\
Vrije Universiteit Brussel  \vspace{\mylength}\\
Prof.\ Dr.\ Freya Blekman (secretary), \\ 
Vrije Universiteit Brussel \vspace{\mylength} \\
Prof.\ Dr.\ Maxim Chernodub,\\ 
Universit\'e Fran\c{c}ois-Rabelais Tours 
and 
Universiteit Gent \vspace{\mylength}\\ 
Prof.\ Dr.\ Frank Ferrari, \\ 
Universit\'e Libre de Bruxelles \vspace{\mylength}\\
Prof.\ Dr.\ Andrei Starinets,\\
University of Oxford \vspace{\mylength} \vspace{\mylength} \\

\newpage
\thispagestyle{empty}
\mbox{}

\tableofcontents
\newpage

\chapter*{A word of thanks to} 
\addcontentsline{toc}{chapter}{Word of thanks}

The FWO, for making it possible for me to carry out research through its funding.  \vspace{\mylength}  \\ 
All my jury members, for accepting to be part of my jury, reading the thesis and asking questions. \vspace{\mylength} \\ 
Henri, for the opportunity to carry out a PhD at UGent. \vspace{\mylength} \\
Alex and Ben, for the trust and the opportunity to finish my PhD at VUB, as well as for the support and help in applying for postdocs, without which my near future in physics could have been very different.  \vspace{\mylength} \\ 
David and Ben, who deserve a very special thanks for having been great advisors, teaching me so much. Of course, without you, there is no way I could have finished this PhD. Thanks also for a careful reading of first versions of the thesis and the helpful comments to it. \vspace{\mylength} \\
David, for the indispensable guidance on the first part of the thesis, and general help throughout the PhD. \\ 
Dan, Fede, Joris, Hongbao, Ben and Jan for a very pleasant collaboration on the last part of the thesis.
 \vspace{\mylength} \\ 
My physics friends from all around the world,  
from whom I've learned a lot. Let me in particular mention (although I am forgetting many) the much appreciated help and answers to all my questions from Dan, Ioannis P., Maxim, Shigeki S.\ and Umut. \\ 
But also  
all the others who I didn't mention by name here explicitly, for opening up my world. \vspace{\mylength} \\    
My local physics friends too of course, starting with all my colleagues in Ghent and Brussels.  \\ 
My office mates Thomas, Dirk and Ben B.\ in Ghent and Tassos in Brussels, for the good times. And Dirk in particular for helping me discover the Sakai-Sugimoto model in my Master's thesis. \\ 
Inge en Merel, voor alle administratieve hulp. \\ 
Joke en Loes, om vanaf dag 1 het vrouwenteam te vormen. \\  
Ben B.\ en Katleen, om samen de theoretische masters aan te gaan en Ben B.\ om samen alle hoogte- en dieptepunten die een doctoraat met zich meebrengt te beleven, \\
maar ook allen voor al het plezier daarbuiten. 
\vspace{\mylength} \\
Dat brengt mij bij de andere vrienden die de voorbije 5 jaar of langer belangrijk voor me zijn geweest en daardoor er onrechtstreeks voor hebben gezorgd dat ik niet onderweg heb opgegeven. 
\vspace{\mylength} \\
Bart, Kristof en Stefanie van de WWI, omdat we de beste zijn! \\
De rest van de OLVC/The Vicious bende, voor alle optredens die mijn gedachten altijd konden verzetten. \\
In het bijzonder Erwin en Ine, om mij tandem-gewijs te helpen met mijn armproblemen toen ik net moest beginnen schrijven, met de kinesitherapie en het typen. Voor hulp bij het typen, ook dank aan mam, Vin, Saar en Bert. \vspace{\mylength} \\
Stefaan, om mij misschien wel bijna 15 jaar lang viool te leren met het grootste geduld. \\
De Klezmerelband, Robin, Annelien en Femke voor het samenspelen. \\
De Umani's voor het samen zingen. \vspace{\mylength} \\
Tante Truus en Pierre, oma, meme en pepe, tante Nadine, Gautier en Wout, om mijn kleine maar sterke 
 familie te zijn. 
\vspace{\mylength} \\  
Bert, voor de knuffel wanneer ik hem nodig had en om Saar's grote geluk te zijn. \\
Fien en Lore, omdat ik mijn schoonzussen zelf niet beter had kunnen kiezen. 
\vspace{\mylength} \\
Saar, om al bijna dertig jaar mijn beste vriendin te zijn en Vin, mijn beste vriend. \\
Pap, om mij al die jaren te hebben gesteund. En tenslotte mam,  
mijn grootste voorbeeld, voor alles. \\  
Duizendmaal bedankt om er altijd voor mij te zijn.   
\vspace{\mylength} \\

\hfill \textit{With love,} 

\hfill \textit{Nele   } \\

\hfill \textit{Gent, 2014}

\chapter*{Dutch summary \\ Nederlandse samenvatting} 
\addcontentsline{toc}{chapter}{Dutch summary -  Nederlandse samenvatting}

In het grootste gedeelte van deze thesis  
worden twee effecten besproken die mogelijks optreden in de theorie van de sterke interactie als gevolg van de aanwezigheid van sterke magnetische velden. Hiervoor wordt gebruik gemaakt van  
een `holografisch QCD-model'. 
Er volgt hieronder een uitleg voor de woorden `holografisch' en `QCD', maar kort gezegd komt het hierop neer: bepaalde aspecten van de sterke kernkracht, die verantwoordelijk is voor het bij elkaar houden 
van protonen en neutronen in de kernen van atomen, kunnen  
bestudeerd worden via een snaartheorie. Dit is het geval dankzij de ontdekking van een dualiteit tussen kwantumveldentheorie\"en enerzijds en snaartheorie anderzijds: deze op zich compleet verschillende theorie\"en blijken dezelfde fysica te beschrijven. Er is tot op vandaag geen strikt bewijs voor de dualiteit, maar waar het mogelijk is om een berekening zowel in de veldentheorie als in de snaartheorie uit te voeren, lijken ze altijd hetzelfde resultaat op te leveren (tenminste in de originele formulering van de dualiteit, de AdS/CFT correspondentie).

Quantum chromodynamica (QCD) is de kwantumveldentheorie die de sterke interactie 
tussen quarks (de constituenten van hadronen) en gluonen (de `glue' die de quarks bindt) beschrijft. 
Voordat QCD ten tonele verscheen, eind 1960, probeerde men de sterke kernkracht te beschrijven met behulp van snaren. Deze aanpak was gebaseerd op de experimenteel waargenomen `Regge'-relatie tussen spin en massa van een meson, die verkregen kan worden in een vortex-model voor het meson: een smalle fluxtube van veldlijnen, de `QCD-snaar', verbindt quark en antiquark.   
Snaartheorie is echter geen consistente theorie in vier dimensies, en het enthousiasme om snaartheorie te gebruiken ter beschrijving van de sterke kernkracht verdween met de komst van QCD. 

Sindsdien werd snaartheorie, met als fundamentele objecten snaren in plaats van puntdeeltjes, verder uitgebreid tot supersnaartheorie (met inbegrip van supersymmetrie) en op zichzelf bestudeerd als mo\-ge\-lij\-ke kandidaat voor een kwantumgravitatietheorie (nodig voor een correcte beschrijving van bijvoorbeeld zwarte gaten en de Big Bang),  aangezien snaartheorie op een natuurlijke manier gravitatie bleek te omvatten. 
Er ontstond een kloof tussen de QCD-gemeenschap en de snaar-gemeenschap. 

In 1997 kwam daar weer verandering in dankzij de `AdS/CFT correspondentie', voorgesteld door Maldacena, die een dualiteit formuleert tussen een kwantumveldentheorie enerzijds (de `CFT'-kant) en een snaartheorie gedefinieerd op een welbepaalde achtergrond anderzijds (de `AdS'-kant). Het wordt een ``holografische dualiteit'' genoemd, omdat beide kanten in een verschillend aantal dimensies `leven'. 
Snaartheorie is een consistente theorie in 10 ruimtetijdsdimensies. De achtergrond waarop de snaartheorie in de context van de AdS/CFT correspondentie is gedefinieerd, is het product van een 5-sfeer en een 5-dimensionale Anti de Sitter (AdS$_5$)  ruimte. AdS$_5$ is een ruimte met de eigenschap dat haar 4-dimensionale rand een Minkowski ruimtetijd is. Het is op deze rand dat de duale kwantumveldentheorie, die in het bijzonder een conforme theorie (CFT) is, invariant onder de conforme groep van onder andere schaaltransformaties, gedefinieerd is. In analogie met een hologram, waarbij  
een driedimensionaal beeld volledig ge\"encodeerd wordt op een tweedimensionaal oppervlak, ontstond de term `holografische dualiteit'. 
Het is een `dualiteit' omdat de CFT- en de snaartheoriebeschrijving, die radicaal verschillende beschrijvingen van hetzelfde fysisch systeem leveren, perturbatief geldig zijn in 
tegengestelde regimes (respectievelijk $\lambda \ll 1$ en $\lambda \gg 1$) in de ruimte van de CFT-koppelingsconstante $\lambda$. 

De AdS/CFT correspondentie wordt besproken in hoofdstuk \ref{stringchapter}, na 
een overzicht te hebben gegeven van QCD in hoofdstuk \ref{QCDchapter} en van snaartheorie in hoofdstuk \ref{stringchapter}. 

De kwantumveldentheorie voor de beschrijving van de elektromagnetische kracht (quantum elektrodynamica of QED), die de elektronen in atomen in een baan rond de kern houdt, is zeer succesvol in het voorspellen van elektrodynamische grootheden met zeer hoge precisie. In QCD echter is het veel moei\-lij\-ker om voorspellingen te doen. De reden is dat de sterke nucleaire kracht te sterk is om de wiskundige techniek te gebruiken die zo goed werkt in QED, namelijk perturbatietheorie, waarin berekeningen kunnen worden gedaan in een reeksontwikkeling in de koppelingsconstante op voorwaarde dat die klein genoeg is.  
De AdS/CFT correspondentie zou een doorbraak kunnen betekenen in de niet-perturbatieve studie van het sterk gekoppelde QCD, als het mogelijk zou zijn om een veralgemening van de dualiteit te vinden naar een (non)AdS/QCD dualiteit. We bespreken in hoofdstuk \ref{nonAdSchapter} strategie\"en om de AdS/CFT dualiteit uit te breiden zodat de `CFT'-kant veralgemeend wordt tot een veel minder symmetrische theorie, zoals QCD. Dit blijkt niet zo makkelijk te zijn en verschillende voorstellen voor dualiteiten bestaan, onder de noemer `holografische QCD-modellen'.  
Ze delen enkele eigenschappen met QCD, maar niet alle. Het is niet duidelijk of d\'e gezochte snaar-duale theorie van QCD zelfs bestaat (alhoewel er wel aanwijzingen zijn, zoals de structuur van QCD in de limiet van een oneindig aantal kleuren in plaats van drie). 
\'E\'en van de meest succesvolle holografische QCD-modellen is het Sakai-Sugimoto model, besproken in hoofdstuk \ref{SSMchapter}. Het slaagt erin om veel van de lage-energie fysica van 
QCD te reproduceren: het verschaft niet enkel een mooie geometrische interpretatie van confinement en spontane chirale symmetriebreking, maar reproduceert ook effectieve lage-energie QCD-modellen in termen van mesonen (in plaats van quarks en gluonen), die geconstrueerd werden v\'o\'or de ontwikkeling van QCD op basis van fenomenologische observaties van meson-dynamica.

Om het effect te introduceren dat in hoofdstuk \ref{rhochapter} besproken wordt, keren 
we eventjes terug naar nog een andere tak in de theoretische fysica, die van de studie van materialen. 
In 2011 werd het 100-jarige bestaan gevierd van de ontdekking van supergeleidbaarheid, i.e.\ geleidbaarheid van elektrische lading zonder weerstand. Dit is een kwantummechanisch fenomeen op macroscopische schaal dat optreedt in bepaalde materialen wanneer ze tot onder een zeer lage kritische temperatuur gekoeld worden. Het wordt verklaard door de microscopische Bardeen-Cooper-Schrieffer (BCS) theorie: elektronen die bewegen in het kristalrooster van het materiaal vormen bosonische `Cooper'-paren van elektronen, die vervolgens `condenseren'. Dit wil zeggen dat ze zich allen  in dezelfde kwantummechanische grondtoestand bevinden, en het resulterende `condensaat' beweegt als \'e\'en geheel, zonder weerstand.  
In 2010 wees Maxim Chernodub op de mogelijkheid van een exotische vorm van supergeleiding, namelijk dat het vacu\"um zelf (in plaats van een specifiek materiaal) supergeleidend kan worden. De voorwaarde is dat er een zeer sterk magnetisch veld aanwezig is. Bij een kritische waarde van het magnetisch veld $B_c$ wordt er spontaan (omdat het geen energie kost) een condensaat gevormd van rho mesonen, of juister, van combinaties van quarks en antiquarks tot toestanden met de kwantumgetallen van geladen rho mesonen met spin in de richting van het magnetisch veld. Deze voorspelling van een `rho meson condensatie' werd gemaakt op basis van berekeningen in een effectief QCD-model in termen van mesonen.  

We bestuderen in hoofdstuk \ref{rhochapter} of het rho meson condensatie effect ook verschijnt in een holografisch QCD-model, namelijk het Sakai-Sugimoto model. 
We vinden inderdaad dezelfde instabiliteit van het vacu\"um terug. 
Aan de ene kant vormt dit onderzoek een test van het holografische model in de zin dat het erin slaagt om een QCD-effect te beschrijven, aan de andere kant verschaft het een onafhankelijke aanwijzing voor rho meson condensatie  gebaseerd op een totaal andere aanpak.  
Als we bijkomende effecten van het magnetisch veld op de constituentenquarks van de mesonen in rekening brengen, vinden we een aanzienlijke verhoging van de kritische waarde van het magnetisch veld $B_c$ ten opzichte van de eerste schattingen in het effectief QCD-model in termen van mesonen.

De studie van magnetisch ge\"induceerde effecten in QCD is relevant omwille van de experimentele omstandigheden waarin QCD-materie wordt gecre/-\"eerd in bijvoorbeeld de Large Hadron Collider (LHC) in Cern. Daar worden bundels van zware ionen (bijvoorbeeld goud-kernen) versneld tot relativistische snelheden (bijna de snelheid van het licht), waardoor een enorme elektrische stroom ontstaat. Als de zware ionen botsen met een eindige impactparameter dan zullen die stromen via een klassiek elektrodynamisch proces een maximaal magnetisch veld cre\"eren in het centrum van de botsing, i.e.\ precies daar waar een soep van quarks en gluonen genaamd `quark-gluon plasma' (QGP) ontstaat. Het QGP wordt bestudeerd als nabootsing van de primordiale vorm van materie die bestond in het universum kort na de Big Bang, en in het algemeen als bron van informatie over de natuur van QCD-materie. 
De gevormde magnetische velden zijn gigantisch, namelijk van de orde $10^{15}$ Tesla. Ter vergelijking, het voordien als hoogste magnetisch veld in de natuur beschouwde veld van een magnetar (i.e.\ een sterk gemagnetizeerde neutron ster) is ``slechts'' van de orde $10^9$ Tesla.  Een gewone dagdagelijkse magneet is $10^{-2}$ Tesla sterk, en de sterkste magnetisch velden gecre\"eerd in laboratoria bereiken (voor korte tijd) 10$^{3}$ Tesla. 

Bij zulke hoge magnetische velden als in het QGP kunnen verschillende kritische fasetransitie-temperaturen in het fasediagram van QCD, die samenvallen bij de afwezigheid van magnetische velden, splitsen. Dit geeft  aanleiding tot bijkomende fases in het $(T,B)$ fasediagram van QCD. We bespreken dit in de context van het Sakai-Sugimoto model in hoofdstuk \ref{SStempchapter}, met de benodigde review van 
holografische modellen bij eindige temperatuur $T$ gegeven in hoofdstuk \ref{finiteThol}.

Een tweede aspect aan de vorming van het QGP in het LHC dat nog slecht begrepen is, naast de aanwezigheid en invloed van sterke magnetische velden, is de thermalizatie van de sterk ge\"exciteerde toestand net na de botsing van de nucleonen naar de quark-gluon plasma evenwichtstoestand. Om dit proces in een ver-van-equilibrium sterk gekoppeld kwantumsysteem te beschrijven zijn er weinig beschikbare technieken. Holografie is er \'e\'en van, en we bespreken in hoofdstuk \ref{vaidyachapter} een holografisch model dat hiervoor gebruikt kan worden. Vervolgens gebruiken we dat model om een notie van een tijds\-af\-han\-ke\-lij\-ke spectrale functie te introduceren (waar het begrip `spectrale functie' normaal gedefinieerd is in een evenwichtssituatie), en enkele zulke spectrale functies te berekenen. Het gebruikte model laat interpretaties buiten QCD toe, bijvoorbeeld in de context van materiaalfysica. 
Mijn bijdrage tot dit werk was voornamelijk het implementeren van numerieke code, en dat is ook waar hoofdstuk \ref{vaidyachapter} verder op focust. 

We eindigen met een conclusie in hoofdstuk \ref{conclchapter}.


\chapter{Introduction}

This dissertation deals with the general subject of holographic dualities, and more specifically applications of it to the study of strongly coupled quantum systems, in the presence of a background field   \cite{Callebaut:2011ab,Callebaut:2013wba,Callebaut:2013ria},  and in non-equilibrium  \cite{preliminary}. 

The main part is based on the following published papers \cite{Callebaut:2011ab,Callebaut:2013wba,Callebaut:2013ria} (and proceedings \cite{Callebaut:2011uc,Callebaut:2011zz} on the same subject) on my work at Ghent University in collaboration with David Dudal, and also Henri Verschelde on \cite{Callebaut:2011ab}. A last chapter concerns 
unpublished preliminary results \cite{preliminary} obtained during the last year of my PhD, which I spent  
at the Vrije Universiteit Brussel. This part is based on work in collaboration with Ben Craps, Federico Galli, Dan Thompson, Joris Vanhoof, Hongbao Zhang and Jan Zaanen. 

In \cite{Callebaut:2011ab,Callebaut:2013wba} we used a top-down holographic QCD model known as the Sakai-Sugimoto model to study a conjectured instability of the QCD vacuum which goes by the name of `rho meson condensation'. The instability is induced by the presence of a very strong magnetic field. A motivation for the study of magnetically induced effects in QCD is the production of extremely intense magnetic fields in non-central heavy ion collisions at RHIC or LHC,  induced 
precisely where the quark-gluon plasma forms. 

In \cite{Callebaut:2013ria} we analyzed the temperature-dependent version of the Sakai-Sugimoto model in the presence of a strong magnetic field. We examined the $(T,B)$ phase diagram in this set-up, and its dependence on a holographic parameter $L$ (whose physical interpretation will be discussed). We showed that the magnetic field induces a split between the chiral transition temperatures per flavour, as well as a possible split between the chiral transition temperatures and the deconfinement temperature, dependent on the value of $L$.  

In \cite{preliminary} we use a bottom-up AdS-Vaidya model to study time-dependent notions of spectral functions for  scalar and spinor operators in a far from equilibrium strongly coupled quantum theory.  
The AdS-Vaidya model is used as a holographic model for thermalization processes, for example in the formation of the thermal equilibrium quark-gluon plasma state starting from the highly non-equilibrium state right after the heavy ion collision, in which case it serves as a holographic QCD model. In \cite{preliminary} however, we 
interpret the results in a condensed matter physics context (because of the bottom-up nature of the model there is more freedom in the field theory interpretation). 
My contribution to this research was mainly focused on the implementation of numerical code.

\section{General introduction}

Quantum chromodynamics (QCD) is the quantum field theory that describes the strong interactions between quarks (the constituents of hadrons) and gluons (the `glue' between the quarks). Before QCD emerged as the correct theory, there were attempts, end 1960s, to describe the strong nuclear force u\-sing 
 strings. This approach was based on the experimentally observed linear `Regge' relation between spin and mass of a meson. The same relation can be obtained in a vortex model for the meson, where a small flux tube of field lines referred to as the `QCD string' connects quark and antiquark. String theory however is not a consistent theory in 4 dimensions, and the enthusiasm to use string theory for the description of the strong nuclear force disappeared with the advent of QCD.

String theory itself, with 1-dimensional strings instead of point particles as fundamental objects, was further generalized to superstring theory 
and studied as a possible candidate for a quantum gravity theory when it appeared that string theory naturally seems to incorporate gravity. The QCD and string community grew apart.  

In 1997, the conjecture of an AdS/CFT correspondence \cite{Maldacena:1997re} would pave the way for a reconciliation.  
This correspondence formulates a duality between a quantum field theory on the one hand (the `CFT' side) and a string theory on a particular background on the other (the `AdS' side). It is called a ``holographic duality'' because both sides `live' in a different amount of dimensions. 
Superstring theory is a consistent theory in 10 spacetime dimensions. The particular background on which the superstring theory in the context of the AdS/CFT correspondence is defined, is the product of a 5-sphere and a 5-dimensional Anti de Sitter (AdS$_5$) space. AdS$_5$  has the property that its (conformal) boundary is a 4-dimensional Minkowski spacetime. It is on this boundary that the dual quantum field theory, which is in particular a conformal field theory (CFT), is defined. In analogy with a hologram, where a 3-dimensional image is fully encoded on a 2-dimensional surface, the term `holographic' duality arose. It is a `duality' because the CFT and the string theory description, which are believed to provide radically different descriptions of the same physical system, are perturbatively valid at opposite limits 
in the space of the CFT coupling constant $\lambda$. 
This means that the strongly coupled regime of the CFT ($\lambda \gg 1$) becomes accessible through a perturbative analysis in a dual string theory. 

While the quantum field theory for the description of the electromagnetic force (quantum electrodynamics or QED) is highly successful in predicting various electrodynamic quantities with unprecedented precision,  it is a lot harder 
to make predictions in QCD. The reason is that the strong nuclear force is basically too strong to be able to use the mathematical techniques (perturbation theory) that are used to perform calculations in QED. Further development of non-perturbative techniques is ongoing in QCD research. 
The promise of the AdS/CFT correspondence for QCD lies in the hope that a generalization of the duality to a (non)AdS/QCD duality could be found which would provide an analytical setting for studying non-perturbative QCD effects. 
It is not clear that such a dual string theory for QCD would exist (although there are some clues, such as the structure of QCD in the limit of infinite number of colours instead of three). 
A lot of effort has gone into constructing generalized dualities known as `holographic QCD models'. They share some features with QCD, but it is safe to say that none of them is without problems.  The `QCD string' mentioned in the first paragraph corresponds to a fundamental string in a higher-dimensional background in these models. 

One of the most successful holographic QCD models  
is the Sakai-Sugimoto model.
It manages to reproduce much of the low-energy physics of quenched QCD in the chiral limit: 
not only does it provide a nice geometrical interpretation of confinement and spontaneous chiral symmetry breaking at low temperatures, it also succeeds in reproducing low-energy effective models in terms of mesons that were built based on phenomenological observations of the behaviour of mesons in the pre-QCD era. 

The inspiration for the original work presented in this thesis is to 
be found in the heavy-ion collision experiments at RHIC and LHC. 
There, the set-up is such that extremely strong magnetic fields are produced in non-central collisions. Beams of heavy ions are accelerated to velocities close to the speed of light, thereby creating an enormous electric current. If the heavy ions collide with a non-zero impact parameter, these currents induce (via a basic classical electrodynamics process) a maximum magnetic field at the center of the collision, i.e.\ exactly where the resulting quark-gluon plasma medium forms. Aside from the magnetic field, the thermalization process of forming the quark-gluon plasma state from the initial highly excited state right after the heavy ion collision, is a poorly understood process that might teach us a lot about QCD.  
Both the problem of QCD in the presence of strong magnetic fields and the problem of far from equilibrium strongly coupled physics 
can be studied using a holographic approach (in simplifying set-ups of course). 
They are moreover complementary in the way of viewing the use of 
holographic methods. The first problem can be handled in a number of varying phenomenological QCD models, but 
also in lattice QCD. This is to be contrasted with the problem of finite density QCD, where the presence of a chemical potential $\mu$ is the source of the so-called sign problem on the lattice. 
The availability of results on a particular $QCD + B$ problem from various models including lattice QCD, is very interesting from the point of view of testing holographic models. It is important for these type of models (which can describe some but certainly not all features of QCD) to be able to check by comparison with different approaches what it can and cannot describe well. 
The second problem of time evolution of strongly coupled systems is much harder to tackle with conventional methods, 
and the holographic method naturally is of interest to gain some insight in the situation. 

We used the Sakai-Sugimoto model to study two conjectured effects in QCD in the presence of strong magnetic fields, where we explicitly compare to other methods by fixing the holographic parameters to GeV units.  
In the last part of the thesis we employ an AdS-Vaidya model to study aspects of thermalization in a strongly coupled system (but not necessarily in the context of the quark-gluon plasma, which I mentioned here rather as motivation for the two subjects handled in the thesis than the exact setting).

\section{Outline of the thesis} 

We begin by reviewing the basics of quantum chromodynamics in chapter \ref{QCDchapter} and of string theory in chapter \ref{stringchapter}. The necessary ingredients are covered to be able to discuss the AdS/CFT correspondence in section \ref{sectionAdSCFT}. In chapter \ref{nonAdSchapter} we explain how one can modify the AdS/CFT duality to generalize it to (non)AdS/QCD models. Demanding the CFT side of the duality to transform into a non-supersymmetric and non-conformal theory that looks like QCD imposes conditions on the AdS side of the duality: we discuss the desired features of the gravitational background in section \ref{desiredf}. The D4-brane background introduced in section \ref{D4section} appears to be a good candidate and forms the basis of the Sakai-Sugimoto model. String theory in this background is argued to be dual to a pure QCD-like theory. 
The approach to holographic QCD dualities followed in chapter \ref{nonAdSchapter} is the 
top-down one 
(i.e.\ deriving holographic models from string theory in the same spirit as the AdS/CFT correspondence, as opposed to the bottom-up approach explained in section \ref{limitationsbottomup}). 
Sakai and Sugimoto added $N_f$ flavour degrees of freedom to the D4-brane background by adding $N_f$ pairs of D8-$\overline{\text{D8}}$ probe branes, resulting in the Sakai-Sugimoto model reviewed in chapter \ref{SSMchapter}: string theory in the D4/D8/$\overline{\text{D8}}$ background is dual to a 4-dimensional quenched and massless QCD-like theory. The shape of the embedding of the D8-branes in interpreted as modeling the spontaneous chiral symmetry breaking in the dual field 
theory (see section \ref{interpretationD8}). 
Mesons manifest themselves as fluctuations of the flavour gauge field living on the D8-branes ($\pi$, $\rho$, $a_1$ ... are unified in the 5-dimensional gauge field) (see section \ref{sectionrelation}). The QCD-like dual contains a large number of redundant degrees of freedom (in the sense that they are not present in real QCD), including an infinite tower of (in principle non-decoupling) Kaluza-Klein modes.  Despite these limitations, the Sakai-Sugimoto model has succeeded in rediscovering previously known effective QCD models. We discuss in particular how the 4-dimensional Proca action for vector mesons (in (\ref{procaaction})) and the 4-dimensional Skyrme action for pions (in (\ref{skyrme}) and (\ref{skyrmeintro})) emerge from the model. (We leave out the discussion of the hidden local symmetry (HLS) like formalism for the coupling between rho mesons and pions, vector meson dominance, baryons as skyrmions, ... and other interesting accomplishments of the model, for which we refer to \cite{Sakai:2004cn,Sakai:2005yt}.) The effective 4-dimensional meson theory is distilled from the 5-dimensional DBI-CS action ((\ref{nonabelian}) and (\ref{CSSSM})),  
describing the dynamics of the 5-dimensional flavour gauge field on the D8-branes, by integrating out the extra holographic radial dimension.

In chapter \ref{rhochapter} we finally come to our own work, based on  \cite{Callebaut:2011ab,Callebaut:2013wba}. 
We are able to show that the non-Abelian ($N_f=2$) Sakai-Sugimoto model develops a DBI-induced tachyonic instability in its vector meson sector in the presence of a strong, constant background magnetic field. The presence of the instability is not influenced by the CS-part of the action or,  in dual field theory language,  by the WZW-term describing the chiral anomaly (\ref{CSchiralanomaly}). This delivers a holographic description of the `rho meson condensation effect' discussed first in \cite{Chernodub:2010qx}. The conjecture is that, at a critical value of the magnetic field $B_c$,  the QCD vacuum might be unstable towards forming a superconducting state where the condensed particles are the combinations of charged rho mesons that have their spin aligned with the magnetic field. 

This effect has been studied in several different phenomenological models as well as in lattice QCD (and later also in complementary holographic bottom-up models \cite{Ammon:2011je,Bu:2012mq,Cai:2013pda}). 
We are able to compare directly to the results in those models by fixing the holographic parameters of the Sakai-Sugimoto model. In particular, we use the most general embedding of the flavour branes $u_0>u_K$ (whereas $u_0=u_K$ corresponds to the original `antipodal' Sakai-Sugimoto model), which allows to model non-zero constituent quark masses. We extend the analysis of \cite{Sakai:2004cn} on fixing the holographic parameters to this more general embedding, by matching to certain  QCD input parameters (at zero magnetic field)  in section \ref{B}. By using the obtained values in that section, it should be possible 
to perform explicit comparisons of se\-veral effects described in the (generalized) Sakai-Sugimoto model with other approaches. Though the Sakai-Sugimoto model does not claim to provide quantitative predictions, it is interesting to perform this exercise, and often turns out to be surprisingly `accurate' (cfr.\ comparisons of meson masses in \cite{Sakai:2004cn} and of the glueball spectrum in figure \ref{glueballspectrumfig2}). 
In the limit of the simplest embedding  ($u_0=u_K$, which comes closest to the bottom-up holographic configurations  \cite{Ammon:2011je,Bu:2012mq,Cai:2013pda}) we recover exactly the Landau levels for an effective 4-dimensional rho meson in the background of the magnetic field, which through the argument of \cite{Chernodub:2010qx} also means the prediction for the critical magnetic field $B_c$ is the same as the one obtained in \cite{Chernodub:2010qx}, namely $eB_c = m_\rho^2 \approx$ 0.6 GeV$^2$.  
This estimate is ``naive'' in the sense that it is based on ignoring any substructure of the rho mesons even in the presence of such strong fields. 
We improve on it by taking into account the constituent masses of the rho mesons through the generalized embedding. The effect of the magnetic field on the constituent quarks is that they gain more mass through the effect of `chiral magnetic catalysis'. Its holographic realization is explained in section \ref{D} in the discussion of the $B$-dependent embedding of the flavour branes, and it results in an increased value of $B_c$, the calculation of which involves numerically solving for $m_\rho(B)$ (see also appendix \ref{appendixnum}). The geometric stability of this embedding is proven in section \ref{4.2}.  

We believe we are the first to apply a more realistic magnetic field $B$ in the context of the non-Abelian Sakai-Sugimoto model, in the sense that it couples to up- and down-charges with different electromagnetic charges ($2e/3$ and $-e/3$). Usually, the charges are averaged and as a result the flavour branes remain coincident even in the presence of $B$. In our set-up, this is no longer the case and the general $B$-dependent embedding corresponds to separated branes, see figure \ref{changedembedding}. This severely complicates the mathematical analysis, mainly because of the non-triviality of the STr-prescription in the DBI-action in that situation (with STr a gauge trace supplemented with an ordering prescription of the traced matrices). We calculate the critical magnetic field in chapter \ref{rhochapter} in increasingly more complicated (i.e.\ general) set-ups, with the various results summarized in figure \ref{summaryfig}. Rather than repeating the detailed outline here, let us refer to the outline of chapter \ref{rhochapter} in section \ref{goalrho}. 
The final set-up is that of the most general embedding ($u_0 > u_K$ and $B$ coupling twice as strong to the up-brane compared to the down-brane) and using the full (non-linear in the field strength) DBI-action. 
An exact evaluation of the STr is possible to second order in the fluctuations (see appendix \ref{appendix}).  
Our final value for the critical magnetic field is $eB_c \approx 0.85$ GeV$^2$, obtained from our end result (\ref{omegasquared}) (combined with (\ref{defomega})) for the generalized Landau levels, as we were able to compute the complete energy spectrum for the generalized Proca equations of motion exactly. The energy eigenstates are no longer spin eigenstates, except for the condensing one. 
Apart from this result in the context of rho meson condensation, the investigation of the general non-Abelian Sakai-Sugimoto model is interesting in its own right, where we clarified several difficulties regarding the handling of separated branes (such as the STr-evaluation and the gauge fixing necessary to disentangle scalar and vector fluctuations in section \ref{gaugefixing}),  
and our stability analysis is complementary to other investigations of the stability of the model (see the references in section \ref{goalrho}, most of which focussed on chemical potential instead of magnetic fields).

In chapter \ref{finiteThol} we review how to turn on a finite temperature in respectively a quantum field theory, the boundary field theory in the AdS/CFT correspondence and the Sakai-Sugimoto model. 
This provides us with the necessary background to also investigate the finite-temperature $N_f=2$ Sakai-Sugimoto model in the presence of an external magnetic field, in chapter \ref{SStempchapter} which is based on \cite{Callebaut:2013ria}. The splitting of the flavour branes leads us to the conclusion that the chiral restoration temperatures will split per flavour, creating an intermediate phase where some of the chiral symmetry is restored, as depicted in figure \ref{TchiralSS}. 
The flavour-dependent $(T,L,eB)$ phase diagram, with variable asymptotic brane-antibrane separation $L$, is presented 
in figure \ref{Tchifig} and cross sections of it in figure \ref{Tchicrosssections}. A split between the chiral restoration temperatures and the deconfinement temperature can emerge only for small enough values of $L$, which is in  correspondence with phenomenological and lattice data on the subject. 

Next, in chapter \ref{vaidyachapter} based on 
\cite{preliminary}, we turn to a different type of holographic model: 
the AdS-Vaidya background (introduced in section \ref{vaidyasection}). This background allows to study a thermalization process holographically as the dual of a black hole formation process in the bulk. Time dependent spectral functions of the boundary theory can be defined using a Wigner transform in eq.\ (\ref{spectralf}) \cite{Balasubramanian:2012tu}. We focus on the used numerical  techniques (pseudospectral method) in obtaining time-dependent spectral functions of scalar and spinor bulk fields in thin-shell AdS$_3$-Vaidya in section \ref{vaidyaresults}. In figure \ref{finalrhofigs} an example of a numerically obtained spectral function $\rho(\omega)$ for a spinor operator of fixed scaling dimension is shown for increasing values of average time. 
In section \ref{latestpaper} the calculation of time-dependent spectral functions in Reissner-Nordstr\"om-AdS$_4$-Vaidya is presented as a very first step towards extracting in principle measurable quantities 
in time-resolved photoemission spectroscopy (ARPES) experiments. 

Finally, we end with a conclusion and outlook in chapter \ref{conclchapter}.

\section{Comment on notations}

We use the mostly minus convention for the Minkowski metric $\eta_{\mu\nu}$ and natural units $\hbar = c= 1$. 
Notations can be different per chapter, but this is always signaled in the text. For example, $\tau$ is used to denote  Euclidean time in chapter \ref{vaidyachapter}, while in the chapters on the  Sakai-Sugimoto model $\tau$ refers to a compactified spatial dimension.

\chapter{Quantum chromodynamics}  \label{QCDchapter}

In this chapter you will find an attempt at a short overview of the theory of quantum chromodynamics (QCD), focused on the concepts that will appear further on in the thesis. Used references include 
\cite{QCDcursus,PS,banks,scrucca,schwartz,Manohar:1996cq,bhaduri}
and references therein.

\section{Classical}

The strong interaction, indirectly responsible for the attraction between protons and neutrons in nuclei, is written down as the Yang-Mills theory\footnote{It will become clear in section \ref{sectionquantum} why we already refer to this Lagrangian (\ref{QCDlagr}) as the one for \textit{quantum} chromodynamics (QCD).} with gauge group $SU(N_c)$ -- where $N_c$ is the number of colours equal to 3 -- describing the interactions between fermionic quarks $\psi$ and bosonic gluons (in the form of gauge bosons) $A_\mu$: 
\begin{equation}
\mathcal L_{QCD} =  -\frac{1}{2} \text{Tr} (F_{\mu\nu}  F^{\mu\nu}) + \bar \psi (i D_\mu \gamma^\mu - m) \psi.    \label{QCDlagr} 
\end{equation}
Here $F_{\mu\nu} = \partial_\mu A_\nu - \partial_\nu A_\mu + i g [A_\mu, A_\nu]$ 
is the field strength with $g$ the coupling constant (which is dimensionless in 4 dimensions) and $D_\mu = \partial_\mu + i g A_\mu$ 
the covariant derivative such that the above Lagrangian is indeed invariant under local $SU(N_c)$ transformations of the quarks $\psi'=U \psi$ (in the fundamental representation) and the gluons $A'_\mu = U A_\mu U^\dagger + \frac{i}{g} (\partial_\mu U) U^\dagger$ 
 (in the adjoint representation), 
with symmetry element $U=\exp(-i\theta_a X_a)$ 
 in terms of the Hermitian traceless generators $X_a$, which fulfill the Lie algebra $[X_a,X_b] = i f_{abc} X_c$ 
as well as $\text{Tr}[X_a X_b] = \frac{\delta_{ab}}{2}$, 
and form a basis for objects in the adjoint representation of the group: $A_\mu = A_\mu^a X^a$, $F_{\mu\nu}^a X^a$. 
$\gamma_\mu$ are the $4 \times 4$ Dirac matrices. 

The first term of the Lagrangian (\ref{QCDlagr}) is referred to as \textit{pure QCD}, describing the interactions between gluonic degrees of freedom only (which are present due to the non-Abelian character of the theory as $N_c>1$). 
When included, the quark spinor $\psi$ is an object $\psi_{j \alpha A}$ that carries 3 silent indices, indicating its transformation properties under the symmetry groups of the Lagrangian: a Dirac spinor index $\alpha$ w.r.t. the Lorentz group (under which the gluon $A_\mu$ transforms as a vector), a colour index $A$ in the fundamental representation of $SU(N_c)$ (under which the gluon transforms in the adjoint representation) and a flavour index $j$ w.r.t. the global chiral symmetry (under which the gluon is invariant), which will be discussed in detail next. 
The mass $m$ in (\ref{QCDlagr}) is a diagonal $N_f \times N_f$ mass matrix in flavour space, where the number of flavours $N_f$ will usually be set to 
2 in this work, including only the light up and down flavours. The corresponding masses are of the order of 5 MeV and thus negligible in comparison to the proton mass $M_p \approx 1$ GeV. They are therefore often omitted from the action, in which case we will refer to it as \textit{massless QCD} or \textit{QCD in the chiral limit}. 

How can (massless) QCD explain the emergence of the heavy proton? How can it explain the experimental fact that only colourless 
composite objects such as protons are observed and the isolation 
of a single coloured quark seems to be impossible - a phenomenon known as confinement? And how can it explain the effect of spontaneous chiral symmetry breaking - which will be covered in the section on global chiral symmetry? 
These three observations are examples of non-perturbative effects, inaccessible to standard perturbative quantum field techniques. To date, they are confirmed by numeric lattice QCD techniques, but not yet fully understood from first principles.   

The local $SU(N_c)$ gauge symmetry of (\ref{QCDlagr}), inherent to Yang-Mills theories, is redundant in the sense that physical observables are gauge invariant. Global symmetries on the other hand do correspond to true physical symmetries and are therefore interesting for effective model building, on which more later. We discuss the global dilatation and chiral symmetry. 

\subsection{Dilatation symmetry} 

The classical action of massless QCD, $S = \int d^4 x \mathcal L_{QCD,m=0}$, is invariant under scale transformations 
\begin{equation}
x^\mu \rightarrow \lambda x^\mu
\end{equation}
under which $\psi \rightarrow \lambda^{-3/2} \psi$ and $A_\mu \rightarrow \lambda^{-1} A_\mu$. 
Indeed, in the absence of quark masses and with dimensionless coupling constant $g$, there is no fundamental scale present in the theory. 
Scale invariance can be expressed as the tracelessness $T_\mu^\mu=0$ of a manifestly symmetric energy-momentum tensor $T_{\mu\nu}$, defined alternatively\footnote{Based on rewriting invariance of the action $\delta S = \int d^d x T^{\mu\nu} \partial_\mu \epsilon_\nu = 0$ 
under  general coordinate transformations 
 $x^\mu \rightarrow x^\mu+ \epsilon^\mu(x)$ as $\delta S =  \frac{1}{2} \int d^d x T^{\mu\nu} (\partial_\mu \epsilon_\nu + \partial_\nu \epsilon_\mu)$ and observing that $g'_{\mu\nu} = \frac{\partial x^\alpha}{\partial x'^{\mu}} \frac{\partial x^\beta}{\partial x'^{\nu}} g_{\alpha \beta} = g_{\mu\nu} - (\partial_\mu \epsilon_\nu + \partial_\nu \epsilon_\mu)$. 
} as the functional derivative of the action w.r.t. the metric, evaluated in flat space: 

\begin{equation}
\delta S = -\frac{1}{2} \int d^d x \sqrt{\det g_{\mu\nu}}  T^{\mu\nu} \delta g_{\mu\nu}. \label{gTcoupling}
\end{equation}

A scale transformation gives rise to a variation of the metric 
\begin{equation}
\delta g_{\alpha \beta} = \epsilon \, g_{\alpha \beta}
\end{equation}
and a variation of the action 
\begin{equation}
\delta S = - \frac{1}{2} \int d^d x \,T^{\mu\nu} \, \epsilon \, g_{\mu\nu} = - \frac{1}{2} \int d^d x \, \epsilon \,  T_\mu^\mu 
\end{equation}
which disappears if 
\begin{equation}
 T_\mu^\mu=0. 
\end{equation}
The corresponding conserved current is the dilatation current $D^\mu = T^{\mu\nu} x_\nu$ with $\partial_\mu T^{\mu\nu} = 0$  (translational invariance): 
\begin{equation}
\partial_\mu D^\mu = (\partial_\mu T^{\mu\nu}) x_\nu + T^{\mu\nu} \eta_{\mu\nu} = T_\mu^\mu = 0. 
\end{equation}
Scale transformations are a special case of conformal transformations $x \rightarrow x'$ 
which leave the metric invariant up to a scale  
\begin{equation}
g'_{\mu\nu}(x') = \Lambda(x) g_{\mu\nu}(x) 
\end{equation} 
and form the \emph{conformal group}, with the Poincar\'e group as subgroup $\Lambda(x)=1$. Next to dilatations $x'^{\mu} = \alpha \, x^\mu$, 
the conformal group contains translations $x'^\mu = x^\mu + a^\mu$, rigid rotations $x'^\mu = M^\mu_{\; \;\nu} x^\nu$ and special conformal transformations $x'^\mu = \frac{x^\mu - b^\mu x^2}{1 - 2 b_\nu x^\nu + b^2 x^2}$.

This dilatation 
or scale invariance is only slightly broken by the addition of small quark masses. To be able to explain a 1 GeV proton mass, a dynamically generated scale is 
necessary, and we will indeed encounter it at the quantum level.

\subsection{Global (approximate) chiral symmetry and its spontaneous breaking} \label{sectionchiralsymm}

Massless QCD 
is invariant under global transformations $\Lambda_V$ in flavour space 
\begin{align*}
\Lambda_V: &\psi \rightarrow e^{-i \vec T \cdot \vec \theta} \psi \simeq (1-i \vec T \cdot \vec \theta) \psi \\
					 &\bar \psi \rightarrow e^{i \vec T \cdot \vec \theta} \bar \psi \simeq (1+i \vec T \cdot \vec \theta) \bar \psi,
\end{align*}
with $T^a$ ($a=1, ..., N_f^2-1$) the Hermitian traceless generators of $SU(N_f)$ transformations, satisfying $\text{Tr} (T^a T^b) = \delta^{ab}/2$, as well as for $T^0 = \textbf{1}_2$ (the $2\times2$ unity matrix) generating a $U(1)_V$ transformation. 
This $U(N_f)_V = SU(N_f)_V \times U(1)_V$ flavour symmetry is manifestly present in the 
multiplets observed in hadronic spectra, and persists when a (bare) flavour independent mass ($m_u = m_d = m_s = \cdots$) is assigned to the quarks in the Lagrangian. The associated conserved currents are the baryon number and isospin\footnote{Using this terminology we implicitly focus on the $N_f=2$ case, referring to the isospin symmetry that transforms neutron and proton into each other by interchanging up and down quarks.} vector currents $j_V^\mu = \bar \psi \gamma^\mu \psi$ and $j_V^{\mu a} = \bar \psi \gamma^\mu T^a \psi$.

The Lagrangian for massless QCD is also invariant under global axial transformations $\Lambda_A$ in flavour space 
\begin{align*}
\Lambda_A: &\psi \rightarrow e^{-i \gamma_5 \vec T \cdot \vec \theta} \psi \simeq (1-i \gamma_5 \vec T \cdot \vec \theta) \psi \\
           &\bar \psi \rightarrow \bar \psi e^{-i \gamma_5 \vec T \cdot \vec \theta} \simeq \bar \psi (1-i \gamma_5 \vec T \cdot \vec \theta), 
\end{align*}
forming the $U(N_f)_A = SU(N_f)_A \times U(1)_A$ group with associated conserved axial vector currents $j_5^{\mu} = \bar \psi \gamma^\mu \gamma_5 \psi$ and $j_5^{\mu a} = \bar \psi \gamma^\mu \gamma_5 T^a \psi$ (with $\gamma_5 = \gamma^5 = \gamma^0 \gamma^1 \gamma^2 \gamma^3$ and $\{\gamma_5,\gamma_\mu \}=0$). Quark masses (of the order 5 to 10 MeV) 
slightly break the $U(N_f)_A$ symmetry: the symmetry is imperfect and the associated current is only partially conserved. 
Pion decay (via the weak force) and pion nucleon scattering (via the strong force) appear to be consistent with the `Partially Conserved Axial Current' (PCAC) hypothesis \cite{Koch:1995vp}. 

The QCD Lagrangian (\ref{QCDlagr}) is invariant under $U(N_f)_V$ transformations when quark masses are assumed to be flavour independent, and approximately invariant under $U(N_f)_A$ transformations when the quarks are approximately massless.
The approximate $U(N_f)_V \times U(N_f)_A$ symmetry is called the chiral symmetry  (in practice however, the term `chiral symmetry' sometimes only refers to the $U(N_f)_A$ symmetry). 
The generators of $SU(N_f)_V$ and $SU(N_f)_A$ are respectively given by
\begin{equation}
Q^a = \int j_V^{0a}(\vec x, t) d^3x  \quad \text{and} \quad 
Q^a_5 = \int j_5^{0a}(\vec x, t) d^3x, 
\end{equation}
and obey the commutation relations 
\begin{equation}
[Q^a, Q^b] = i f^{abc} Q^c, \quad  [Q^a, Q^b_5] = i f^{abc} Q^c_5, \quad [Q^a_5, Q^b_5] = i f^{abc} Q^c,
\end{equation}
with $f^{abc}$ the structure constants of the group ($f^{abc} = \epsilon^{abc}$ for $N_f=2$).
The charges $Q_5^a$ do not form a closed algebra. 
One therefore defines the combinations 
\begin{equation}
Q_R^a = \frac{1}{2}(Q^a + Q_5^a), \quad Q_L^a = \frac{1}{2}(Q^a- Q_5^a) 
\end{equation}
with
\begin{equation}
[Q^a_R, Q^b_R] = i f^{abc} Q^c_R, \quad  [Q^a_L, Q^b_L] = i f^{abc} Q^c_L, \quad 
[Q^a_R, Q^b_L] = 0 
\end{equation}
to obtain two decoupled $SU(N_f)$ algebras. The $SU(N_f)_V \times SU(N_f)_A$ symmetry from now on is denoted as $SU(N_f)_L \times SU(N_f)_R$: $Q^a_L$ generates a $SU(N_f)$ transformation of $\psi_L = \frac{1}{2}(1-\gamma_5) \psi$ and $Q^a_R$ generates a $SU(N_f)$ transformation of $\psi_R = \frac{1}{2}(1+\gamma_5) \psi$, that leaves the massless QCD Lagrangian, with $\psi = \psi_L+ \psi_R$, invariant.

We set $N_f=2$, such that $T^a = \tau^a/2$ ($\tau^a$ the Pauli matrices, $a=1,2,3$)
, for which:
\begin{align}
[\tau^a, \tau^b] &= 2 i \epsilon^{abc} \tau^c, \label{commutatie}\\
\{\tau^a, \tau^b\} &= 2 \delta_{ab} \textbf{1}_2. \label{anticommutatie}
\end{align}
To investigate how mesons transform under the symmetry transformations $\Lambda_V$ and $\Lambda_A$, we consider the following mesonic combinations of the quark fields $\psi$:
\begin{align*}
 \vec \pi &\equiv i \bar \psi \vec \tau \gamma_5 \psi \quad &\text{pseudoscalar combination (pion state)} \\ 
 \sigma &\equiv  \bar \psi \psi \quad &\text{scalar combination (sigma state)} \\ 
 \vec \rho_\mu &\equiv \bar \psi \vec \tau \gamma_\mu \psi \quad &\text{vector combination (rho state)} \\
 \vec a_{1\mu} &\equiv \bar \psi \vec \tau \gamma_\mu \gamma_5 \psi \quad &\text{axial vector combination ($a_1$ state)}. 
\end{align*}  
Using (\ref{commutatie}) and (\ref{anticommutatie}) it is possible to show that $\Lambda_V$ transformations correspond to isospin rotations of pions and $\rho$ mesons: 
\begin{equation} \label{gammaV}
\Lambda_V: \vec \pi \rightarrow \vec \pi + \vec \theta \times \vec \pi \quad \text{and} \quad 
					 \vec \rho_\mu \rightarrow \vec \rho_\mu + \vec \theta \times \vec \rho_\mu, 
\end{equation}
and that pions and $\sigma$ mesons on one hand and $\rho$ and $a_1$ on the other are rotated into each other under $\Lambda_A$:
\begin{equation} \label{pion}
\Lambda_A: \vec \pi \rightarrow \vec \pi + \vec \theta \sigma \quad \text{and} \quad 
					 \vec \rho_\mu \rightarrow \vec \rho_\mu + \vec \theta \times \vec a_{1\mu}. 
\end{equation}					 
We expect that states that are rotated into each other under a symmetry of the QCD Hamiltonian, would have equal eigenvalues, i.e.\ masses. This however is not the case: $m_\rho = 770$ MeV and $m_{a_1} = 1260$ MeV, and the difference in mass is too large (namely of the order of $m_\rho$) to be a consequence of the light symmetry breaking caused by the small quark masses. 
The axial symmetry is thus not reflected in the mass spectrum of mesons, but there are experimental indications (mentioned earlier)  that the axial vector current is partially conserved. An explanation for these apparently contradictory observations is a spontaneous breaking of the axial symmetry: in the Nambu-Goldstone realization of chiral symmetry, the symmetry of the Hamiltonian is invisible in the ground state or $Q_5^a|\Omega \rangle \neq 0$ (while still $Q^a|\Omega \rangle =0$) -- in contrast to $Q^a|\Omega \rangle = Q_5^a|\Omega \rangle = 0$ in the Wigner-Weyl realization.

\begin{figure}[h!]
  \centering
  \scalebox{0.4}{
  \includegraphics{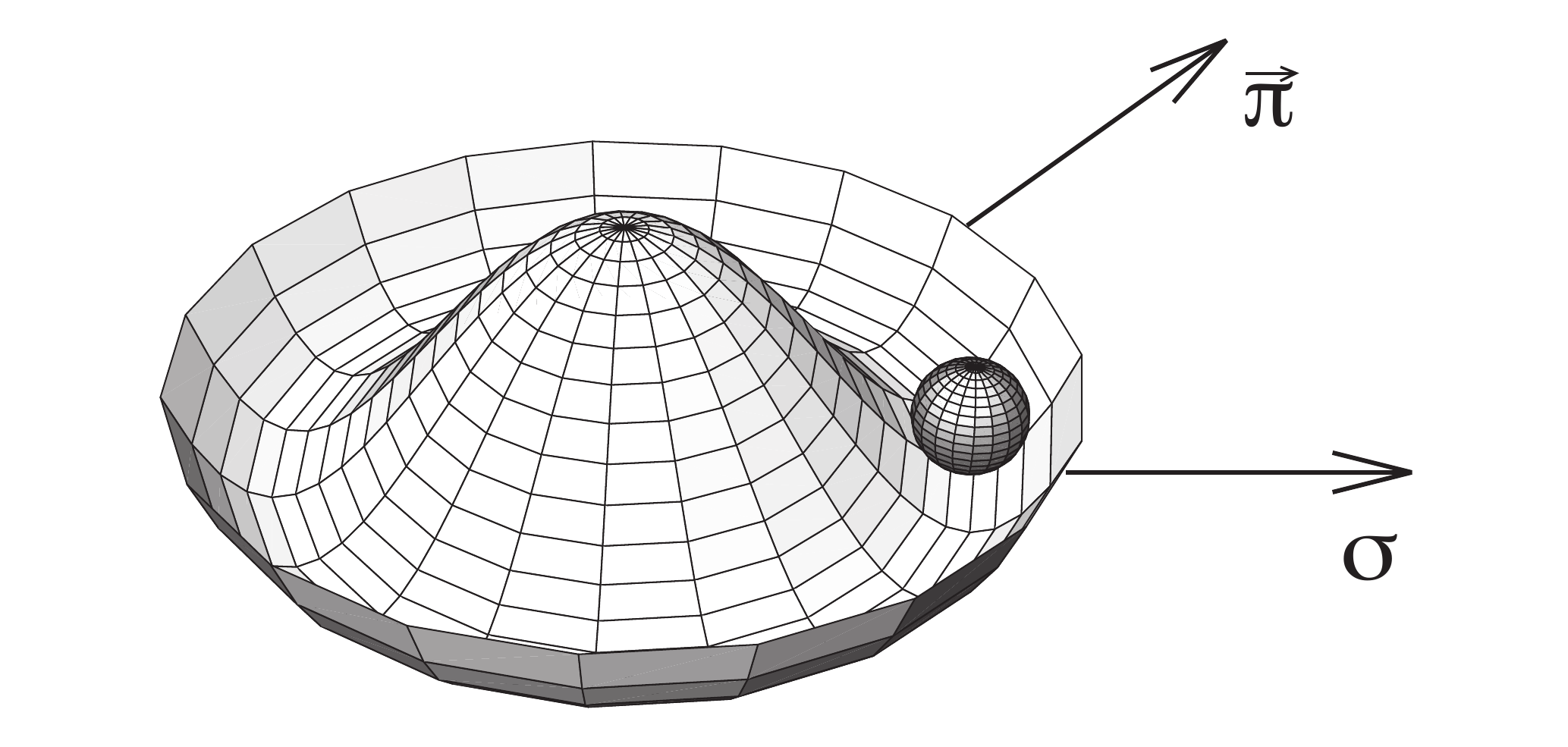}}
  \caption{Spontaneous chiral symmetry breaking \cite{Koch:1995vp}.}\label{kochfig}
\end{figure}

To get an intuitive picture of spontaneous symmetry breaking, we assume that the effective QCD Hamiltonian at zero temperature takes the form of a Mexican hat potential, with the $(x,y)$ coordinates replaced by  $(\sigma, \vec \pi)$ fields (see figure \ref{kochfig}). Spatial rotations in the valley of the hat serve as a mechanic analogon for the axial rotation $\Lambda_A$, rotating $\vec \pi$ in $\sigma$ (\ref{pion}). 
To end up in one of the ground states in the valley, a direction has to be chosen:the axial symmetry is dynamically broken $SU(N_f)_A \rightarrow \textbf{1}$, and thus $SU(N_f)_V \times SU(N_f)_A \rightarrow SU(N_f)_V$, or\begin{equation} \label{sXSB}
U(N_f)_L \times U(N_f)_R \rightarrow U(N_f)_V \quad \text{(\emph{spontaneous chiral symmetry breaking})}.
\end{equation}
As the ground state is located at a finite distance from the origin, the scalar $\sigma$ field, carrying the quantum numbers of the vacuum,  acquires a finite expectation value. 
In quark language this means that a finite scalar quark condensate arises, $\langle \bar \psi \psi \rangle \neq 0$, as order parameter of the spontaneously broken chiral symmetry. 
Pion excitations correspond to small, massless rotations along the valley (away from the ground state): \emph{the massless pion}, carrying the quantum numbers of broken generators $Q_5^a$, \emph{serves as the  Gold$\-$stone bo$\-$son of spontaneous axial symmetry breaking}. Excitations in the $\sigma$ direction correspond to massive, radial excitations. (The mass difference between $\rho$ and $a_1$ is also explained within the idea of spontaneous axial symmetry breaking, see for instance \cite{sakurai,Weinberg:1967kj}.) 
The small explicit axial symmetry breaking caused by small quark masses corresponds in this picture to a slight tilt of the hat. 
While the toy picture with the Mexican hat focuses on the $SU(N_f)$ part of the symmetry, it is in fact the full $U(N_f)_L \times U(N_f)_R$ which is broken down to the subgroup of $U(N_f)_V$ vector symmetries, as stated in (\ref{sXSB}), consistent with the vacuum expectation value for the $\bar \psi \psi$ operator. With the four spontaneously broken continuous symmetries, associated with four axial vector currents, there only seem to correspond three Goldstone bosons, in the form of the isospin triplet of pions. The reason for the absence of a fourth will become clear in section \ref{sectionchiralanomaly}.  

Assuming that the pion, as Goldstone boson, dominates the spectrum of pseudoscalar isovector excitations, in combination with  the PCAC hypothesis leads to the Gell-Mann Oakes Renner (GMOR) relation 
\begin{equation} \label{GMOR}
m_\pi^2 f_\pi^2 = - \frac{1}{2} (m_u + m_d) \langle \bar \psi \psi \rangle 
\end{equation} 
relating bare quark $m_f$ and pion masses $m_\pi$, with $f_\pi$ the pion decay constant (mass dimension 1) from $\langle 0| j_A^{\mu a}(x)|\pi^b(p)\rangle = i \delta_{ab} f_\pi p^\mu e^{-i p\cdot x}$.

\section{Quantum} \label{sectionquantum}

At the quantum level, $\mathcal L_{QCD}$ in (\ref{QCDlagr}) is supplemented with a gauge-fixing term $\mathcal L_{gauge fixing}$ (and a resulting $\mathcal L_{ghost}$ containing unphysical ghost particles), 
to make sure that in the path integral formulation of the quantized field theory (in the Faddeev-Popov quantization) one is integrating out gauge inequivalent fields.  
This is a complicated story involving residual overcounting after gauge fixing (``Gribov copies") that will not be discussed here further. 
In the operator formulation of quantum field theory, the fields are promoted to operators, so with this interpretation understood we can use $\mathcal L_{QCD}$ in (\ref{QCDlagr}) as the \textit{quantum} chromodynamics Lagrangian. 
The basic objects considered at the quantum level are correlation functions, 
defined by the vacuum expectation value of time-ordered products of field operators -- in a general notation for general fields $\phi$ and where we will write $|0 \rangle$ for the perturbative and $| \Omega \rangle$ for the non-perturbative vacuum: 
\begin{equation} 
\langle \phi(x_1) \cdots \phi(x_n) \rangle = \langle \Omega | T  \phi(x_1) \cdots \phi(x_n) | \Omega \rangle   
\end{equation} 
or in the path integral formalism 
\begin{equation} 
\langle \phi(x_1) \cdots \phi(x_n) \rangle = \frac{\int \mathcal D \phi \, \phi(x_1) \cdots \phi(x_n) e^{i S}}{\int \mathcal D \phi \, e^{i S}}  = \left. \mathcal Z[J]^{-1} \left( \frac{1}{i} \right)^n \frac{\delta^n  \mathcal Z[J]}{\delta J(x_1) \cdots \delta J(x_n) } \right|_{J=0}
\end{equation} 
with the path integral or partition function 
\begin{equation} 
\mathcal Z[J] = \int \mathcal D \phi e^{i \left( S + \int d^4 x \, J \phi \right)} = e^{-i W[J]}
\end{equation} 
the generating functional of correlation functions, and $W[J]$ the generating functional of connected correlation functions 
\begin{equation} 
\langle \phi(x_1) \cdots \phi(x_n) \rangle_{conn} = \left. \left( \frac{1}{i} \right)^{n+1} \frac{\delta^n W[J] }{\delta J(x_1) \cdots \delta J(x_n) } \right|_{J=0}. 
\end{equation} 

The chiral condensate from the previous section is defined in this quantum language as 
\begin{equation} 
\langle \bar \psi \psi  \rangle = - \text{Tr} \lim_{y \rightarrow x+} \langle \Omega | N \psi(x) \bar \psi(y) | \Omega \rangle 
\end{equation} 
with the minus sign and trace over a product of Dirac matrices 
from the closed fermion loop, and $N$ the notation for the normal ordered product of fields, related to the time-ordered product through Wick's theorem  $ T\phi(x_1) \cdots \phi(x_n) =  N \{ \phi(x_1) \cdots \phi(x_n) + \text{all possible contractions} \}$. 
From this definition we can appreciate that the chiral condensate is perturbatively zero, since 
 $\langle 0| N \psi(x) \bar \psi(y) |0 \rangle = 0$, but $\langle \Omega| N \psi(x) \bar \psi(y) |\Omega \rangle \neq 0$. From dimensional analysis it is moreover clear that $\langle \bar \psi  \psi \rangle$ has mass dimension 3, signaling the need for a dynamically generated non-perturbative mass scale in (approximately) massless QCD. The same is true for confinement, which can be formulated in terms of infinitely heavy quarks separated by a distance $L$, between which the field lines are squeezed into a flux tube. They feel a confining linear potential $V \sim \sigma L$ with $\sigma$ the ``QCD string tension"  with mass dimension 2. The mass scale will appear from the process of renormalization.

\subsection{$\beta$-function and trace anomaly}  \label{sectionbeta}

If the interaction can be treated as a perturbation of the free theory, the correlation functions can be expanded perturbatively in powers of the coupling constant, giving rise to Feynman diagrams
\begin{equation} 
\langle \Omega | T  \phi(x_1) \cdots \phi(x_n) | \Omega \rangle = \left( \text{sum of all connected diagrams with $n$ external points} \right). 
\end{equation} 
These graphs contain more loops of virtual particles at each order, leading to integrals over all loop momenta that are typically UV-divergent\footnote{In the operational formalism the singular behaviour typically comes from products of operators at the same point, in the path integral formalism from the formal definition of the measure as an infinite-dimensional product of fields.}.  
This can be demonstrated in the simpler case of $\phi^n$ scalar field theory,  
 where a 1-particle irreducible Feynman graph with $E$ external legs and $V$ vertices has $I=(n V-E)/2$ internal lines and the number of loop integrals is $L=I-V+1$. Focusing on the region of integration where all loop momenta are large, the evaluation of the graphs in momentum space gives rise to $\int \frac{d^{4L} p}{p^{2I}}$, which diverges for $4L \geq 2I$ or $4 \geq (4-n)V+E$. For $n>4$ or negative mass dimension of the $\phi^n$ coupling (which is equal to $4-n$) the divergence increases with each order $V$ in the perturbation, and the interaction is non-renormalizable.

For $n = (<) 4$ 
or zero (positive) mass dimension of the coupling, the interaction is (super-)re\-nor\-ma\-li\-za\-ble: it is possible to add counterterms to the action (order by order) and absorb them in a redefinition of the previously divergent bare quantities (such as charge, mass, ...) 
into finite, renormalized ones - as a function of which the new action takes the same form as before. 
In order to define the infinite counterterms, the infinities of the bare quantities have to be parameterized in terms of a cut-off in a process called regularization (for example $\epsilon \rightarrow 0$ in dimensional regularization, $d=4-\epsilon$). 
The renormalized quantities will depend on the cut-off or related renormalization energy 
scale $\mu$. In particular, the renormalized coupling constant $g_R$ in QCD will `run' with the scale $\mu$ (for example $g_R \sim g \,  \mu^{-\epsilon/2}$ in dimensional regularization); 
 this is captured by the beta function defined as 
\begin{equation}
\beta(g_R) = \mu \frac{d}{d \mu} g_R(\mu) 
\end{equation}
and given in QCD at one loop by 
\begin{align}  
\beta(g_R) &= -\frac{g_R^3}{16 \pi^2} \left( \frac{11}{3} N_c - \frac{2}{3} N_f \right) + \mathcal O(g_R^5) \quad \\ \label{beta} 
\text{or} \quad  \beta(g_R^2) &= - \beta_0 \, g_R^4  + \mathcal O(g_R^6)  \quad \text{with} \quad  \beta_0 = \frac{1}{8 \pi^2} \left( \frac{11}{3} N_c - \frac{2}{3} N_f \right). 
\end{align}
QCD can be shown to be renormalizable at all orders in perturbation theory, e.g.\ by using a technique called algebraic renormalization.

The introduction of the energy scale $\mu$ in the renormalization process, is responsible for breaking the scale invariance of classical massless QCD. This is an example of an anomaly or the breaking of a classical symmetry at quantum level. Anomalies of global symmetries are consistent and correspond to true restrictions on the theory, anomalies of local symmetries are not allowed (because of e.g.\ unitarity issues). 
A scale transformation $x \rightarrow e^{-\sigma} x$ transforming the energy scale $\mu \rightarrow e^\sigma \mu = (1+\sigma)\mu + \cdots$, now induces an infinitesimal variation of the coupling constant $g_R \rightarrow g_R + \sigma \, \beta(g_R)$ 
and corresponding variation of the Lagrangian $\delta \mathcal L = \sigma \, \beta(g_R) \frac{\partial}{\partial g_R} \mathcal L$, such that the quantum corrected trace of the symmetric energy-momentum tensor reads 
\begin{equation}
\partial_\mu D^\mu = T_\mu^\mu = \frac{\delta \mathcal L}{\delta \sigma} = \beta(g_R) \frac{\partial}{\partial g_R} \mathcal L. 
\end{equation}
The coupling constant can be removed from the covariant derivative by a rescaling of the gauge field $g_R A_\mu^a \rightarrow A_\mu^a$ after which the only $g_R$-dependence of the action is in the kinetic term $-\frac{1}{4 g_R^2} F_{\mu\nu}^a F^{\mu\nu \, a}$, resulting in 
\begin{equation} 
\partial_\mu D^\mu = T_\mu^\mu = \frac{\beta(g_R)}{2 g_R^3} F_{\mu\nu}^a F^{\mu\nu \, a} \neq 0.  
\label{betaTrF2}
\end{equation}

\begin{figure}[h!]
  \centering
  \scalebox{0.4}{
  \includegraphics{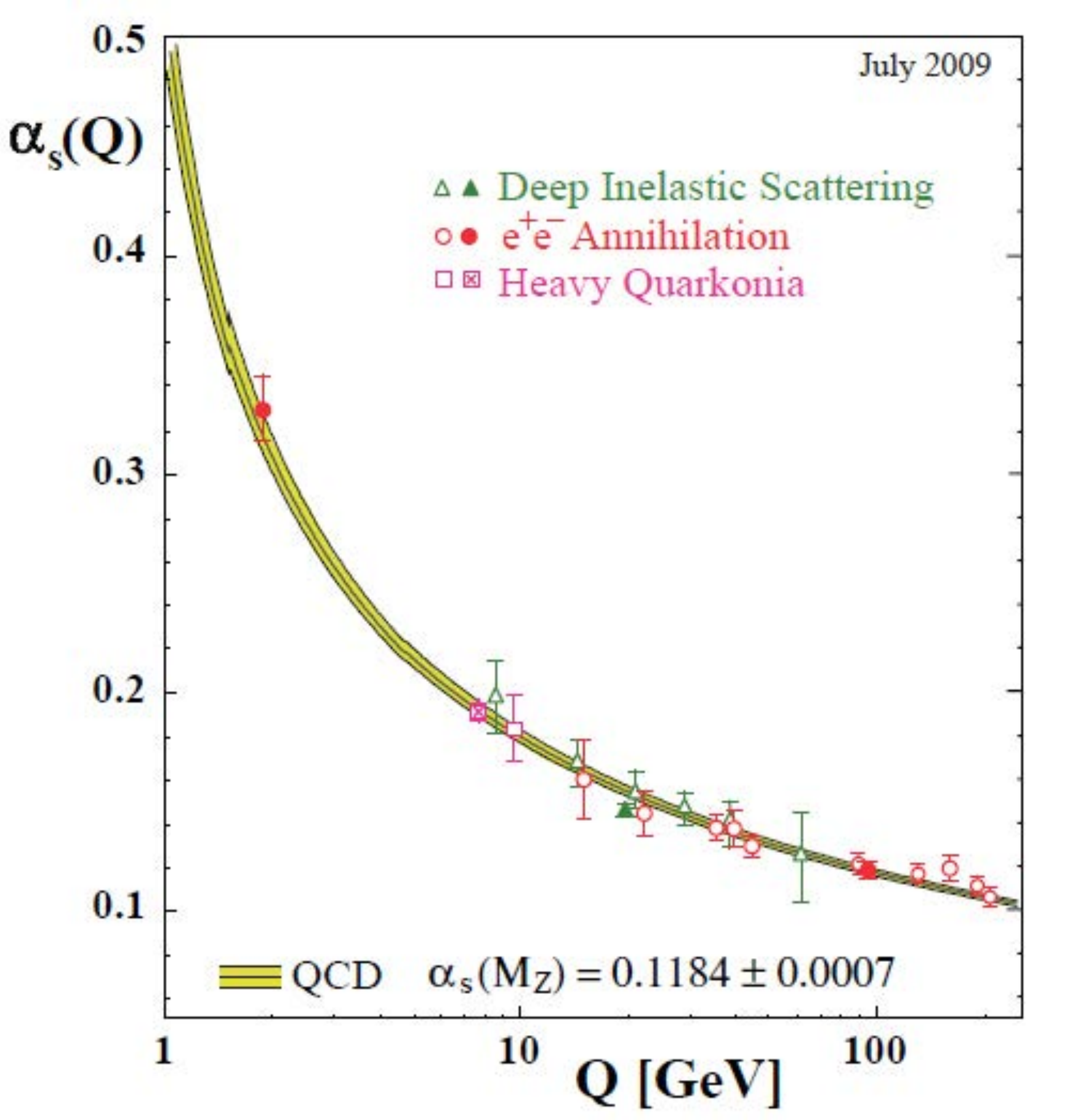}}
  \caption{Summary of measurements of $g^2_{YM} \sim \alpha_s$ as a function of energy scale $\mu$ or $Q$  \cite{Bethke:2009jm}.}\label{alphasfig} 
\end{figure}

We interpret the physical implications of the running of the QCD coupling with energy scale $\mu$, obtained by integrating (\ref{beta}): 
\begin{equation}
g_R^2(\mu) = \frac{g_R^2(\mu_0)}{1 + \beta_0 g_R^2(\mu_0) \ln \frac{\mu}{\mu_0}} \quad \text{or} \quad g_R^2(\mu) = \frac{g^2}{1 + \beta_0 g^2 \ln \frac{\mu}{\Lambda}} \label{gR} 
\end{equation}
with integration constant $\mu_0$ and bare coupling $g = g_R(\Lambda)$. 
The sign of $\beta_0 \sim \left( \frac{11}{3} N_c - \frac{2}{3} N_F \right)$ is positive, since $N_c=3$ and $N_f=6$. This means that $g_R$ decreases as the scale $\mu$ increases, and in the limit of $\mu \rightarrow \infty$
\begin{equation}
g_R^2(\mu) \sim \frac{1}{(11 N_c-2 N_f) \ln \frac{\mu}{\mu_0}}  \rightarrow 0,   
\end{equation}
QCD becomes \emph{asymptotically free}: quarks and gluons interact increasingly weaker at short distances.   
Conversely, the coupling grows at large distances, signaling \emph{confinement} in the IR. In particular the perturbative coupling will diverge at the Landau pole 
\begin{equation}
 \mu = \mu_0 \exp{\left(-\frac{1}{\beta_0 g_R^2(\mu_0)}\right)}. 
\end{equation}
How the coupling actually will grow at low energies can only be asked in a non-perturbative QCD treatment. A figure combining different experimental results is given in figure \ref{alphasfig}.
The expression (\ref{gR}) can be further rewritten, without reference to the coupling constant $g_R^2(\mu_0)$ and $\mu_0$ on the right hand side, but instead as a function of a dynamic mass scale $\Lambda_{QCD}$: 
\begin{equation}
g_R^2(\mu) = \frac{1}{\beta_0 \ln(\mu/\Lambda_{QCD})}, \qquad 
\label{couplingLambda}
\end{equation}
with $\Lambda_{QCD}$ defined in lowest order as 
\begin{equation}
\Lambda_{QCD} = \mu \exp{\left(- \frac{1}{\beta_0 g_R^2(\mu)}\right)} \label{LambdaQCD}. 
\end{equation}
This scale can be introduced by solving 
\begin{equation}
\mu \frac{d}{d\mu} M_h \left( \mu,g_R^2(\mu) \right) = 0,  
\end{equation}
expressing the independence of a hadron mass $M_h = \mu f \left( g_R^2(\mu) \right)$ 
on the renormalization scale $\mu$. The equation for the dimensionless $f$, given by $f + \beta f' = 0$, can be solved in lowest order (with $\beta$ given in (\ref{beta})), 
to obtain $f(g_R^2) = C_h \exp{\left( -1/\{\beta_0 g_R^2(\mu)\}\right)}$, with $C_h$ an integration constant, such that the result for the hadron mass can be written as 
\begin{equation} 
M_h = C_h \Lambda_{QCD}.  
\end{equation}
From the expression (\ref{LambdaQCD}) it follows that this scale is zero at each order in the expansion around $g_R = 0$ and is hence a purely non-perturbative object.  
The three examples of non-perturbative effects mentioned earlier, can be expressed in terms of this intrinsic QCD scale 
$\Lambda_{QCD}$: 
the hadron mass $M_h \sim \Lambda_{QCD}$, the QCD string tension $\sigma \sim \Lambda_{QCD}^2$ associated with confinement, and the chiral condensate $\langle \psi \bar \psi \rangle \sim \Lambda_{QCD}^3$ associated with spontaneous chiral symmetry breaking.  
$\Lambda_{QCD}$ is the momentum scale at which the coupling (\ref{couplingLambda}) becomes strong (technically diverges) as the energy scale $\mu$ is decreased; its estimated value from experimental measurements (and lattice studies) is $\Lambda_{QCD} \approx 200$ MeV.

\subsection{Chiral anomaly} \label{sectionchiralanomaly}

\begin{figure}[h!]
  \centering
  \scalebox{0.25}{
  \includegraphics{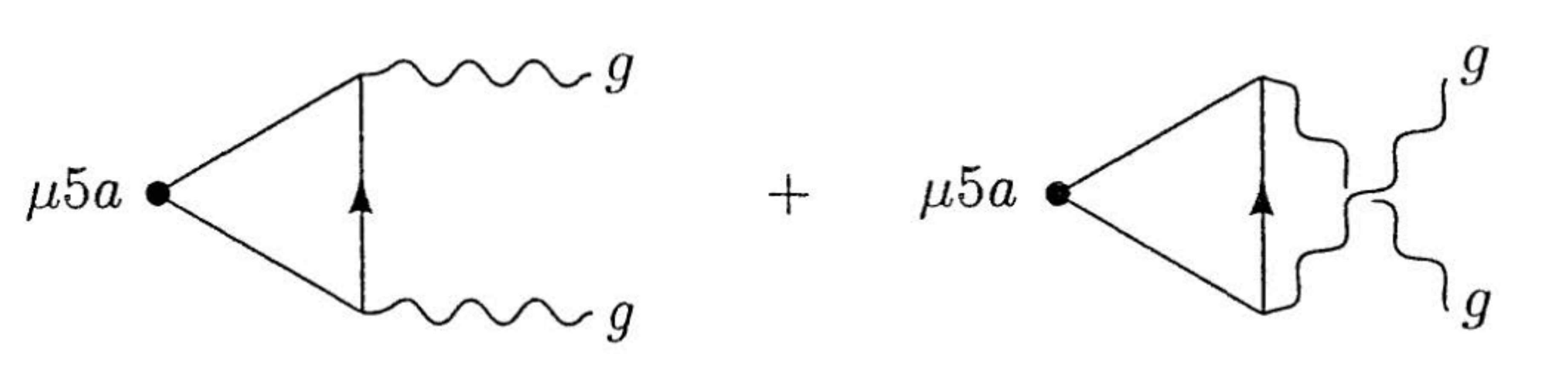}}
  \caption{Diagrams that lead to an axial vector anomaly for a chiral current in QCD \cite{PS}. 
}\label{anomalyfig} 
\end{figure}

The 
axial vector current $j_5^{\mu} = \bar \psi \gamma^\mu \gamma_5  
\psi$ of section \ref{sectionchiralsymm} is no longer conserved at the quantum level: 
\begin{equation}
\partial_\mu j_5^{\mu} \sim N_f \epsilon^{\mu\nu\rho\sigma} \text{Tr}( F_{\mu\nu} F_{\rho\sigma}) \neq 0 
\end{equation}
This is known as the $U(1)_A$ anomaly. 
An indication of its origin is the need for a redefinition of $\gamma_5$ (defined previously to anticommute with all other $\gamma_\mu$ in $d=4$ dimensions) in the process of dimensional regularization $d=4-\epsilon$. 
It explains why there does not seem to be a light isosinglet pseudoscalar with mass comparable to that of the pions, which would be necessary as Goldstone boson of the spontaneous breaking of $U(1)_A$, if it were not explicitly broken by quantum corrections. Typical triangle Feynman diagrams that contribute to the anomaly are shown in figure \ref{anomalyfig}. 
In the limit of infinite number of colours, $N_c \rightarrow \infty$, which is the relevant limit in holographic QCD models, these diagrams are suppressed because of their 
quark loops (see section \ref{sectionlargeN}). This leads to the restoration of $U(1)_A$ and the $\eta'$ meson is identified as the corresponding `missing' Goldstone boson. 
To summarize, due to quantum effects (the $U(1)_A$ anomaly), the classical expression (\ref{sXSB}) for spontaneous chiral symmetry breaking ($\chi SB$) becomes 
\begin{equation}
U(N_f)_V \times SU(N_f)_A \rightarrow U(N_f)_V \quad \text{(\emph{spontaneous $\chi SB$ incl.\ quantum effects})},  \label{chiSBquantum}
\end{equation}
but in the large $N_c$ limit ($U(1)_A$ restored), the classical symmetry is restored and again we find 
\begin{equation}
U(N_f)_V \times U(N_f)_A \rightarrow U(N_f)_V \quad \text{(\emph{spontaneous $\chi SB$ at large $N_c$})}.  \label{chiSBlargeN}
\end{equation}

Although the axial isospin currents have no axial anomaly from QCD interactions, they do have an anomaly associated with the coupling of quarks to electromagnetism (with electromagnetic coupling constant $e$ and electromagnetic field strength $F_{\mu\nu}$): 
\begin{equation}
\partial_\mu j_5^{\mu a} \sim \delta_{a3} e^2 \epsilon^{\mu\nu\rho\sigma} F_{\mu\nu} F_{\rho\sigma} \label{CSchiralanomaly}
\end{equation}
which gives the leading contribution to the amplitude of $\pi^0 \rightarrow 2 \gamma$ decay. 
In low-energy effective QCD models, the effect of this chiral anomaly is modeled by the Wess-Zumino-Witten (WZW) action (see section \ref{sectioneffQCD}). In the Sakai-Sugimoto model it will correspond to the Chern-Simons action on the flavour branes.

\section{More inspiration from statistical mechanics} \label{statmech}

We already encountered an example of a concept in field theory that was borrowed from statistical mechanics, namely the Mexican hat potential associated with spontaneous breaking of a global symmetry. Here, we will mention some more examples, most notably the concept of the renormalization group: although it is not directly used in the work discussed in this thesis, it 
has a close link 
with the AdS/CFT correspondence and is thus not unimportant to mention. 

The renormalization group (RG) method was first developed in the context of spin systems, to deal with a class of problems, including critical phenomena, 
which are characterized by having very many degrees of freedom in a region the size of the correlation length\footnote{Any relativistic quantum field theory falls into this class, as the field $\phi$ at each point $x$ is a separate degree of freedom, so any region of finite size contains an infinite number of degrees of freedom.} (which could be defined as the minimum size of a system, e.g.\ a gas, that one can reach before it qualitatively changes properties).     
In this type of systems it turns out to be the cooperative behaviour between the many degrees of freedom that is a determining factor, rather than the precise form of the initial interactions, giving rise to the concepts of universal `critical' behaviour and \emph{universality classes} of theories which lead to the same long-distance behaviour.  
The idea of RG is to successively perform scale transformations on the system, and while zooming out from the UV to the IR (``RG flow") define effective degrees of freedom with effective local interactions (the assumption being that the interactions only couple directly to nearby degrees of freedom). 
The flow is expected to be characterized by scale invariant `fixed points' where the beta function vanishes. Asymptotic freedom corresponds to a free 
fixed point in this language. The scale invariance at the fixed point usually 
belongs to the larger conformal symmetry group, and conformal field theories are used to describe statistical systems at criticality (non-trivial fixed point).

In its application to quantum field theory (QFT), RG comes down to integrating out high energy degrees of freedom, with momentum $p$ say between $\mu < p < \Lambda$: $e^{ -S_{eff}(\phi;\mu)} = \int \mathcal D \phi_{\mu < p < \Lambda} e^{- S_{eff}(\phi;\Lambda)}$, with $S_{eff}(\phi;\Lambda)$ the Wilsonian effective action 
$e^{-S_{eff}(\phi;\Lambda)} = \int \mathcal D \phi_{|p|>\Lambda} e^{-S_E(\phi)}$ 
such that $\mathcal Z[J] = \int \mathcal D \phi \, e^{-S_E - \int J \phi} = \int \mathcal D \phi_{|p|<\Lambda} e^{-S_{eff}(\phi;\Lambda) - \int J \phi}$, where we performed the typical Wick rotation to Euclidean time $\tau = i t$, $-S_E = i S$, for the sake of convergence properties of the path integrals. 
The renormalization scale $\mu = \frac{\Lambda}{s}$, related to $\Lambda$ with a scale factor $s$, is interpreted as the `detector scale' beyond which we are not yet able to probe the theory further. $\Lambda$ is the true UV cutoff of the QFT, beyond which the field theory breaks down and the correct but unknown high-energy theory, termed \emph{UV-completion}, has to take over. In this vision on renormalization, the infinities encountered in QFT signal a physical reality, namely that it is wrong to extrapolate a theory that works at $E= 1$ TeV to $E \rightarrow \infty$, as one implicitly does in a \emph{local} QFT. The fact that strong interaction physics can be described by a renormalizable gauge theory (it is possible to take the limit $\Lambda \rightarrow \infty$) is believed to be the unavoidable consequence of two `theorems'. The first one is unproven but believed to be true for lack of counterexamples: ``any sensible physical theory reduces at low enough energy to a renormalizable quantum field theory". The second one says that ``a unitary quantum field theory which contains vector particles must be a local gauge theory" (which can be proven).  

The RG leads to a low-energy effective theory with an in principle infinite number of possible interaction terms $\sum_i c^i  \hat O^i$, which do not necessarily appear in the Lagrangian of the fundamental microscopic theory but are compatible with its symmetries. They can be classified according to the mass dimension $d_i$ of the coupling $c^i$ (or equivalently in terms of the mass dimension $d_{\mathcal O^i}$ and the spacetime dimension $d$), 
which determines not only renormalizability (as discussed in section \ref{sectionbeta}) but also scaling behaviour of the interaction term ($\sim s^{d_i}$) under scaling transformations: 
the super-renormalizable ($d_i>0$ or $d_{\mathcal O^i}<d$), renormalizable  ($d_i=0$ or $d_{\mathcal O^i}=d$) and non-renormalizable ($d_i<0$ or $d_{\mathcal O^i}>d$) interactions correspond respectively to \emph{relevant}, \emph{marginal} and \emph{irrelevant} interactions in the sense that they amplify, remain the same, or die out under the flow. 
In this argument it is the quantum dimension, which differs from the classical mass dimension value in an anomalous term (anomalous dimension), that has to be used for the classification. The beta functions parameterizing RG flow associated with different operators can mix  
and are diagonalized near a fixed point of the flow, with the eigenvalues determining the anomalous dimensions, and the eigenvectors the directions which can be classified into relevant (flowing away from the fixed point), marginal or irrelevant (flowing towards the fixed point). 
The effective field theory Lagrangian, of the form 
\begin{equation}
\mathcal L_{eft} = \mathcal L_{\leq d} + \mathcal L_{d+1} + \mathcal L_{d+2} + \cdots, 
\end{equation}
with subscripts referring to operator dimensions $d_{\mathcal O^i}$, is valid at energy scale $\mu = \Lambda/s$. Retaining only the renormalizable $\mathcal L_{\leq d}$ terms means we are dealing with an error of $1/s$. Higher order corrections in $1/s$ can be systematically included by retaining irrelevant terms up to $\mathcal L_{d+r}$ - to compute with an error of $1/s^{r+1}$: the effective field theory is predictive but has a finite accuracy (so non-renormalizable theories are not without physical meaning). 
The renormalizable QFT result is obtained by assuming $\Lambda \rightarrow \infty$ and thus $s=\Lambda/\mu \rightarrow \infty$, in which case the renormalizable $\mathcal L_{\leq d}$ gives the `exact' result (where `exact' means that the QFT is erroneously assumed to be the correct theory for all energy scales, and in fact the calculations are approximate with powers of $1/s$ neglected).

The advantage of the renormalization group view by Wilson, is that it provides a non-perturbative definition of the theory. Imposing a spacetime lattice (in Euclidean time $\tau$ after a Wick rotation $\tau = i t$) to provide the UV cutoff, lead to the numeric non-perturbative approach of \emph{lattice QCD}, which is for example able to demonstrate confinement and succeeds in reproducing the low-lying spectrum of hadrons \cite{latticebook}. 
Its main limitations are dealing with non-zero baryon densities (because of the numerical sign problem) and real-time dynamics (because of Euclidean time). 

\begin{figure}[h!]
  \centering
  \scalebox{0.35}{
  \includegraphics{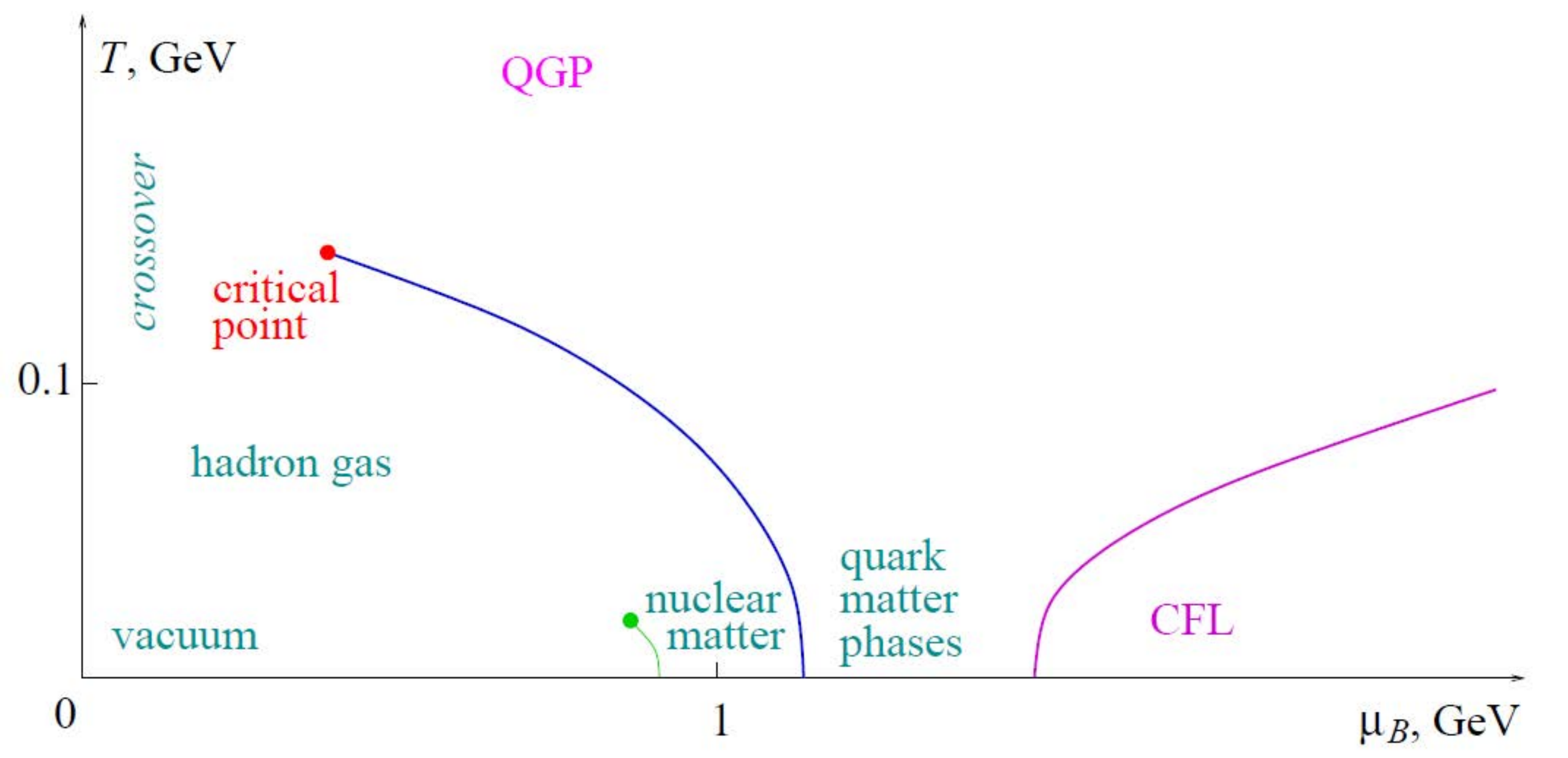}}
  \caption{Sketch of the suspected QCD phase diagram \cite{Stephanov:2007fk}. 
}\label{QCDphasediagramfig} 
\end{figure}

The analogy between field theory and statistical mechanics, which lies at the basis of RG methods in field theory, furthermore suggests that spaces of field theoretic interactions also divide into phase domains. 
A cartoon of the suspected phase diagram of QCD, as a function of external thermodynamic parameters temperature $T$ and baryon chemical potential $\mu_B = \frac{1}{N_f} \sum_f \mu_f$, 
is presented in figure \ref{QCDphasediagramfig}. Each point on the diagram corresponds to a stable thermodynamic state. 

At sufficiently high temperatures $T \gg \Lambda_{QCD}$ one expects to find a phase of asymptotically `free' quarks and gluons, termed \emph{quark-gluon plasma} (QGP), where chiral symmetry is restored. Indeed, data from the Relativistic Heavy Ion Collider (RHIC) \cite{Adcox:2004mh} and LHC mark the production of such a phase 
(see e.g.\ \cite{Martinez:2013xka} and references therein). 
The low-energy effective degrees of freedom of QCD on the other hand are hadronic in nature. Lowering the temperature one should encounter a phase transition for confinement as well as for the spontaneous breaking of chiral symmetry. The nature of this phase transition depends on the parameters. For large $N_c$ it is believed to be strongly first order, 
while for $N_c=3$ in the chiral limit it is second order, and for $N_c=3$ with $N_f=2$ light quarks, a crossover occurs. 
Near the crossover the phase is referred to as strongly coupled QGP (sQGP). It is in the strongly coupled regime that one can hope for viable input from holographic QCD models. For example, the shear viscosity over entropy density $\eta/s$ is expected to reach a minimum near the crossover (and extend to infinity far from it, for $T \rightarrow 0$ (dilute hadron gas) as well as $T \rightarrow \infty$) \cite{Stephanov:2007fk, Csernai:2006zz}.  
Experimental data from heavy ion collisions indicate that it is indeed small, by comparing to hydrodynamic calculations, and plausibly saturating the lower bound of $(4 \pi)^{-1}$ conjectured in holography \cite{Kovtun:2004de,Buchel:2003tz}. 

Where perturbative techniques are applicable for very high $T$ and/or $\mu_B$, and lattice QCD is restricted primarily to low $\mu_B$, the largest part of the pictured phase diagram is at this point only accessible through phenomenological or effective QCD models. Holographic QCD can be placed in this category of models. 
In the bulk of this thesis we will focus on two effects that possibly occur in the $(T,B)$ plane of the QCD phase diagram, considering the effects of the presence of an external magnetic field $B$ (at $\mu_B=0$).

\section{Effective low-energy QCD models} \label{sectioneffQCD}

In the IR, the relevant QCD degrees of freedom are not quarks and gluons as in the UV, but mesons and nucleons (and glueballs). Effective models in terms of these observable asymptotic states, that respect the global QCD symmetries, are useful to describe low-energy phenomenology. 

In order to construct a Lagrangian that is invariant under chiral symmetry, the building block to use is $\vec \pi^2 + \sigma^2$ (cfr. (\ref{gammaV}) and (\ref{pion})). Including the Mexican hat potential of figure \ref{kochfig}, this leads to the \emph{Linear Sigma model} (terms in the nucleonic degrees of freedom $\mathcal L_{\pi N}, \mathcal L_N$ left out) 
\begin{equation} \label{sigmamodel}
\mathcal L_{LS} =  \frac{1}{2}(\partial_\mu \vec\pi)^2 + \frac{1}{2}(\partial_\mu \sigma)^2 - C^2 \left( (\vec \pi^2+\sigma^2) - f_\pi^2\right)^2 
\end{equation}
from which the $\sigma$-field (which is unidentifiable with a physical particle) can be eliminated by restricting the dynamics to pionic excitations in the valley of the hat $\vec \pi^2 + \sigma^2 = f_\pi^2$ (send $C \rightarrow \infty$), giving rise to the Non-Linear Sigma model 
\begin{align}
\mathcal L_{NLS} 
                 &= \frac{1}{2} (\partial_\mu \vec \pi)^2 + \frac{1}{2f_\pi^2}(\vec \pi \cdot \partial_\mu \vec \pi)^2 
 =  \frac{f_\pi^2}{4} \text{Tr}(\partial_\mu U \partial^\mu U^+) =  -\frac{f_\pi^2}{4} \text{Tr}(U^+ \partial^\mu U)^2   \label{NLS} 
\end{align} 
where in the last line we introduced 
\begin{equation} \label{U}
U(x) = e^{2i\frac{\vec T \cdot \vec \pi(x)}{f_\pi}} \quad \in SU(N_f) 
\end{equation}
for the `pion field' $U(x) \in SU(N_f)$, transforming under chiral transformations $(h_L,h_R) \in SU(N)_L \times SU(N)_R$ as $U'(x) = h_L U(x) h_R^{-1}$. 
The three pions $\pi^a$ in (\ref{U}) (for $N_f=2$) are the three Goldstone bosons associated with the spontaneous breaking of $SU(N_f)_A$ in (\ref{chiSBquantum}). 
In the \emph{Skyrme Lagrangian}, non-linear $\pi \pi$ interactions are added 
\begin{align} 
\mathcal L &= \frac{f_\pi^2}{4} \text{Tr}(\partial_\mu U \partial^\mu U^+) - \frac{1}{32 g^2} \text{Tr}[U^+\partial_\mu U, U^+ \partial_\nu U]^2 
\label{skyrme}
\end{align}  
describing an effective theory of weakly coupled mesons, valid in the limit $f_\pi^2 \rightarrow \infty$ (the $\pi \pi$ coupling can be seen to be proportional to $1/f_\pi^2$ in (\ref{NLS})), and with dimensionless parameter $g \sim 1/\sqrt{N_c}$ identifiable with the coupling $f_{\rho \pi\pi}$. This limit corresponds to the large $N_c$ limit discussed in the next section, where we will find that $f_\pi$ scales as $\sqrt{N_c}$. 
Going further and further in order, one constructs what is called `the \emph{chiral Lagrangian}' in chiral perturbation theory for mesons.

(\ref{skyrme}) contains a redundant symmetry that is not respected by nature: it is invariant under  $U(x) \rightarrow U^+(x)$, which forbids processes transforming an even number of mesons into an odd number of mesons, such as  $K^+ K^- \rightarrow \pi^+ \pi^- \pi^0$ and $\pi^0 \eta \rightarrow \pi^+ \pi^- \pi^0$. These processes are however observed experimentally, and they go through in QCD thanks to the chiral anomaly. To include the chiral anomaly into the Skyrme model, the redundant
 $U(x) \rightarrow U^+(x)$ symmetry has to be removed. This effect is achieved by adding the Wess-Zumino term  $n \Gamma_{WZ}$ ($n \in \mathbb{Z}$), consisting of an odd number of factors $U^+ \partial_\mu U$, to the Skyrme action. This corresponds to adding terms with an odd number of Goldstone bosons, referred to as ``odd intrinsic parity terms".  
In the presence of an electromagnetic field, extra terms have to be added to the action in order for it to be gauge invariant, resulting in the 
\emph{Wess-Zumino-Witten} (WZW) action $\tilde \Gamma_{WZ}$ (and with $n=N_c$), the precise expression of which can be found in  \cite{Witten:1983tw}. 
The resulting theory contains soliton solutions (`skyrmions') that can be identified with strongly interacting nucleons \cite{Witten:1979kh}, 
and  succeeds in predicting the correct decay rates of electromagnetic  (e.g. $\pi^0 \rightarrow 2 \gamma$) as well as hadronic (e.g. $\omega \rightarrow 3 \pi$) decay processes in the meson sector. 

Another, much used low-energy effective QCD model is the \emph{Nambu-Jona-Lasinio (NJL) model} \cite{Nambu:1961tp,Buballa:2003qv}. 
It is defined in terms of quark degrees of freedom, the gluon-mediated interactions between which are integrated out and replaced by effective four-fermion interactions, such that effective nucleonic and mesonic degrees of freedom are obtained through a kind of Cooper pairing mechanism as in the BCS theory of superconductivity. The Lagrangian 
\begin{equation}
\mathcal L_{NJL} = \bar \psi (i \slashed \partial - m) \psi  + G \left\{ (\bar\psi \psi)^2 + (\bar \psi i \gamma_5 \vec \tau \psi)^2 \right\},  
\end{equation}
with $G$ a dimensionful coupling constant, respects the global symmetries of QCD.

It is perhaps important to note that the above models are not effective QCD theories in the `Wilson effective field theory' sense: there are no derivations of these models from integrating out degrees of freedom starting from QCD, in fact they stem from the pre-QCD era.  QCD could be regarded as a UV-completion of the chiral model, but - as they do not share the same degrees of freedom - a very different kind of UV-completion than for example the electroweak theory is to the Fermi theory (where the fermions are present in both theories but with different interactions); one cannot ask about pion scattering at high energy in QCD (but lattice QCD can show the existence of pions as poles in correlation functions).

\section{Large $N_c$ QCD: link to string theory}  \label{sectionlargeN}

There is no obvious small expansion parameter present in (approximately massless) QCD, whereas a definite need for approximation schemes presents itself. An unobvious small parameter was suggested 
by 't Hooft \cite{'tHooft:1973jz} in the form of the inverse of the number of colours, for large $N_c$.   
In the limit of an infinite number of colours $N_c \rightarrow \infty$, QCD Feynman diagrams can be associated with 2-dimensional surfaces and in this way linked to world surfaces of strings in string theory. 
The dualities between field and string theories presented further on, such as the AdS/CFT-correspondence and the Sakai-Sugimoto model, are precisely valid in the $N_c \rightarrow \infty$ limit. 

\begin{figure}[h!]
  \centering
  \scalebox{0.5}{
  \includegraphics[width= 27 cm]{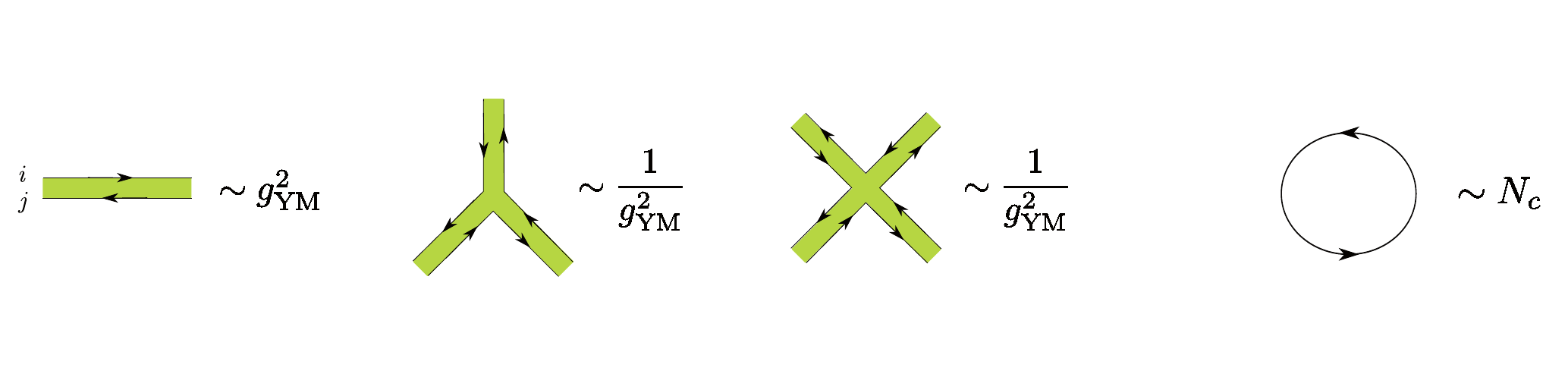}}
  \caption{Feynman  rules for a pure $SU(N_c)$ gauge theory in the 'double line notation'   
\cite{Peeters:2007ab}. } 
 \label{fig1peeters}
\end{figure}

The Feynman rules for pure QCD\footnote{In the $N_c \rightarrow \infty$ limit, quark loops are suppressed compared to gluon loops, as the latter contain an extra colour loop and thus an extra factor $N_c$. Quarks will therefore be ignored here (until the discussion of mesons).} (in the convention scheme used in \cite{Peeters:2007ab}) are given in figure \ref{fig1peeters} in the `double line notation' introduced by 't Hooft. The colour lines represent the indices $i$ and $j$ of $(A_\mu)_{ij}$, and Feynman diagrams correspond to surfaces of polygons glued to each other along the gluons: the propagators are the edges of polygons and every colour index loop represents a polygon face. 

\begin{figure}[h!]
  \centering
  \scalebox{0.5}{
  \includegraphics[width=27 cm]{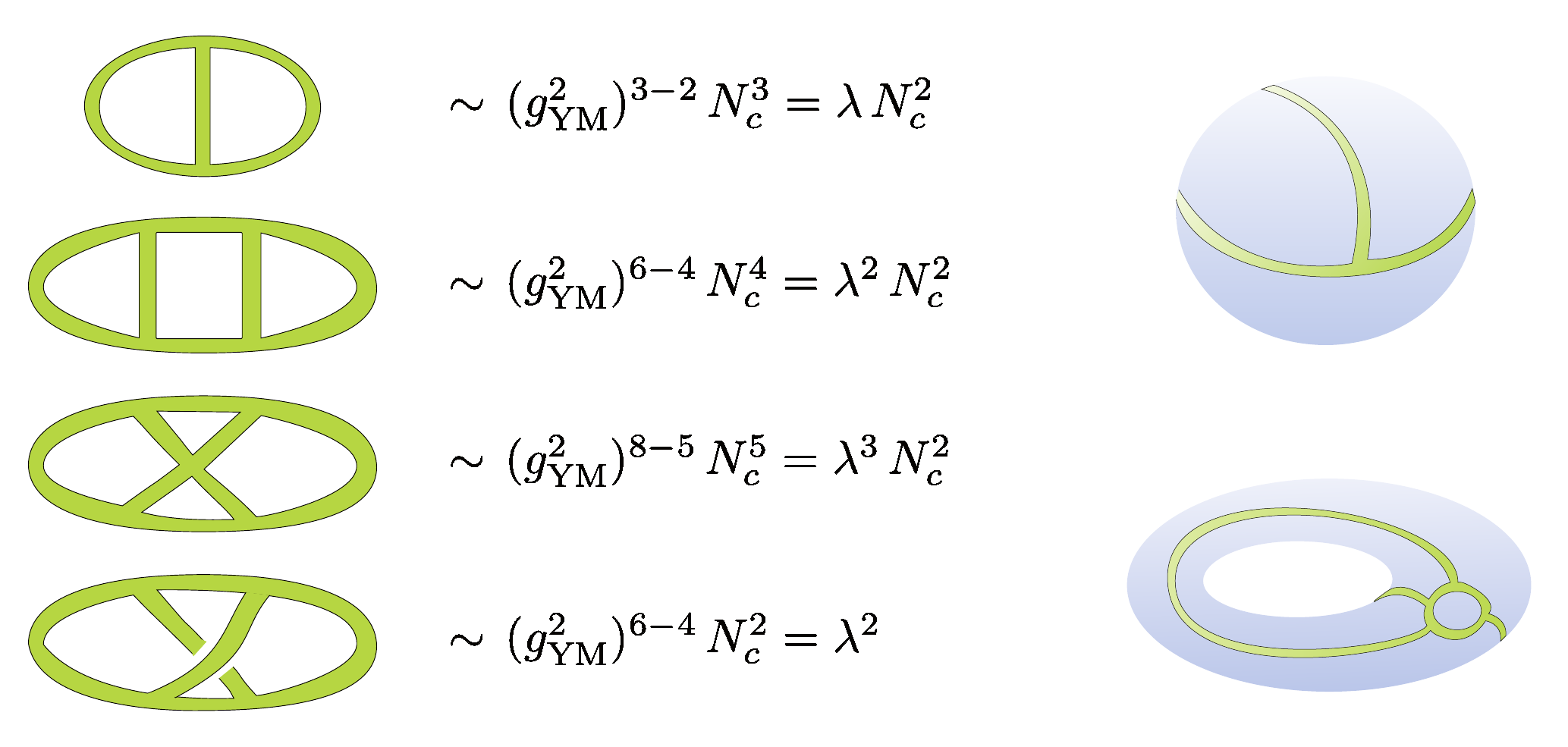}}
  \caption{By grouping the coupling constant $g_{YM}$ and the number of colours $N_c$ into the 't Hooft coupling constant $\lambda = g^2_{YM}N_c$, one can see that $N_c^{-2}$ counts the degree of non-planarity. Planar diagrams can be mapped on a sphere, while non-planar diagrams have to be mapped to surfaces of higher genus \cite{Peeters:2007ab}.}  
 \label{fig2peeters} 
\end{figure}

It will 
appear that Feynman diagrams will reorganize themselves in an expansion in degree of non-planarity in the perturbation series in the limit $N_c\rightarrow \infty$ and  $g^2_{YM}N_c$ finite and constant. In order to see this, we look at some simple diagrams in figure \ref{fig2peeters} and count the powers of $g^2_{YM}$ and $N_c$ as follows:  
\begin{equation} \label{eq:FD}
\text{a Feynman diagram} \sim (g^2_{YM})^{\text{number of propagators} \hspace{1 mm} - \hspace{1 mm} \text{number of vertices}} \hspace{1 mm} (N_c)^{\text{number colour loops}}  
\end{equation}
(every closed loop corresponds to a sum over $N_c$ colours and therefore gives a factor $N_c$).
The defining quantity that appears is  
\begin{eqnarray*}
\text{the number of vertices (V)} & - & \\ 
\text{the number of propagators  (= the number of edges E)} & + & \\ 
\text{the number of loops (= the number of faces F)},
\end{eqnarray*}
better known as the topologically invariant Euler number $\chi \equiv V-E+F$. 
In this notation we rewrite (\ref{eq:FD}) as
\begin{eqnarray} 
\text{a Feynman diagram} & \sim & (g^2_{YM})^{E - V} \hspace{1 mm} (N_c)^{F} \nonumber\\
 & \sim & (g^2_{YM} N_c)^{E - V} \hspace{1 mm} (N_c)^{V-E+F} \nonumber\\
 & \sim & \lambda^{E - V} \hspace{1 mm} (N_c)^{\chi}, \label{Feynmancounting}
\end{eqnarray} 
where we grouped the coupling constant $g_{YM}$ and the number of colours $N_c$ in the 't Hooft coupling constant $\lambda = g^2_{YM}N_c$.

For a connected orientable surface 
we have $\chi = 2-2g-b$, with $b$ the number of boundaries, 
and $g$ the genus (= the number of handles). 
For a sphere $g=0$, $b=0$, $\chi=2$; for a torus $g=1$, $b=0$, $\chi=0$. 
The first three planar diagrams in figure \ref{fig2peeters} have $\chi=2$, and can therefore be mapped onto a sphere. The last, non-planar diagram in figure \ref{fig2peeters} has $\chi=0$ and can be mapped onto a torus. 
When the 't Hooft coupling constant $\lambda$ is kept constant\footnote{To demonstrate why $\lambda$ constant or $g_{YM} \sim 1/\sqrt{N_c}$ is the only sensible large $N_c$ behaviour, consider the beta function (\ref{beta}). 
After rescaling $g_{YM} = \bar g_{YM} N_c^{-1/2+\epsilon}$, with $\bar g$ finite as $N_c \rightarrow \infty$, it reads 
\begin{equation} 
\mu \frac{d \bar g_{YM}(\mu)}{d \mu} = - \left( \frac{11}{3} - \frac{2}{3} \frac{N_f}{N_c} \right) \frac{\bar g_{YM}^3}{16 \pi^2} N_c^{2 \epsilon} + \mathcal O(\bar g_R^5). 
\end{equation} 
For $\epsilon < 0$, $\bar g$ is constant and $g_{YM}$ runs to zero, leading to a trivial theory, while for $\epsilon >0$, $\bar g$ runs infinitely fast. 
A non-trivial and sensible theory is only obtained when $g_{YM} \sim 1/\sqrt{N_c}$. 
Moreover, the RG equation for $\lambda$ is $N_c$-independent (up to an $N_f/N_c$-suppressed term), 
\begin{equation} 
\mu \frac{d \lambda}{d \mu} = - \frac{11}{3} \frac{\lambda^2}{8 \pi^2} + \frac{2}{3} \frac{N_f}{N_c} \frac{\lambda^2}{8 \pi^2}, 
\end{equation} 
such that $\lambda$ is the natural large $N_c$ coupling constant \cite{utexas}. 
\label{largeNfootnote}}, the non-planar diagram in figure \ref{fig2peeters} is suppressed by a factor $N_c^{-2}$. 
In general, the expansion in $1/N_c$ of gauge theories in the limit $N_c\rightarrow \infty$ with $\lambda$ constant, is an expansion in degree of non-planarity, which is given by  
the  minimal genus of the surface onto which the diagram can be mapped without crossings of gluon propagators.
QCD diagrams in the large $N_c$ limit then look like 2-dimensional surfaces  which can be interpreted as world surfaces of closed strings. This suggests a possible relation between large $N_c$ QCD and string theory, which however is yet to be made explicit. One of the obstacles is understanding the meaning of the 't Hooft coupling $\lambda$ in terms of string theory.

There are further clues for stringy behaviour in the strong coupling regime of QCD: spectra of mesons and baryons display a linear relation between the spin $J$ and the mass squared $M^2$ (`Regge trajectories'), which can be explained within a stringy model where mesons are represented as massive quarks connected by a relativistic string.

\begin{figure}[h!]
  \centering
  \scalebox{0.5}{
  \includegraphics[width= 27 cm]{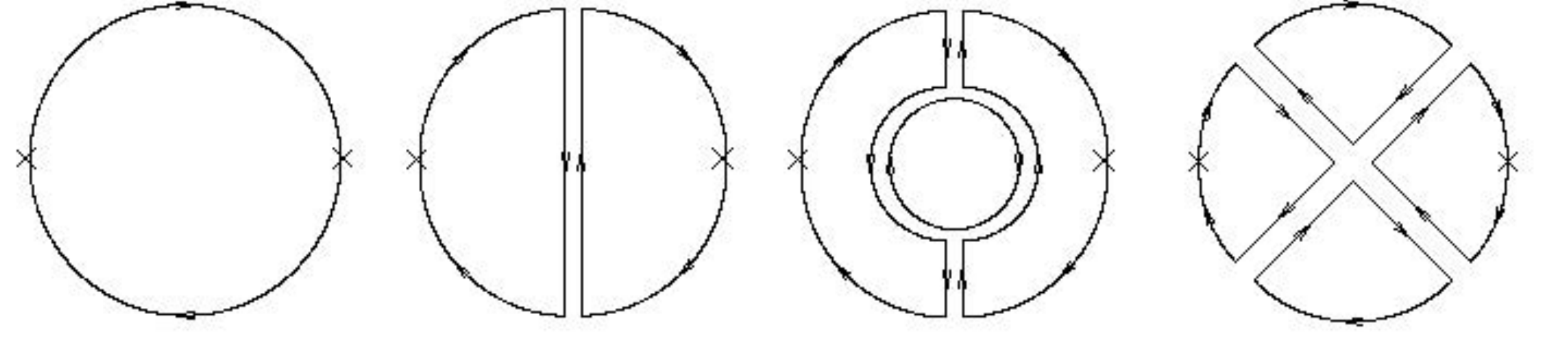}}
  \caption{Example planar two-point diagrams. From left to right, large $N_c$ counting for the Feynman diagrams gives $(g_{YM}^2)^0 N_c^1 = \lambda^0 N_c$, $(g_{YM}^2)^1 N_c^2 = \lambda N_c$, $(g_{YM}^2)^2 N_c^3 = \lambda^2 N_c$, $(g_{YM}^2)^3 N_c^4 = \lambda^3 N_c$, all scaling with $N_c^1$.} 
\label{mesonlargeNfig} 
\end{figure}

Let us finish this section by discussing the large-$N_c$ limit in terms of low-energy meson degrees of freedom \cite{'tHooft:1974hx,Lebed:1998st}. 
Consider a local quark bilinear operator $J(x)$ that creates a $\psi \bar \psi$ pair or scalar meson state at position $x$, under the extra assumption that the sum of planar diagrams leads to a confining theory. In figure \ref{mesonlargeNfig} a couple of diagrams are drawn that represent the creation of a meson by $J$, 
followed by the propagation and interaction of the quarks for some interval, and then its annihilation by the current $J^\dagger$.  
The sum of planar two-point diagrams with momentum transfer $k$ is equal to a sum of meson propagators \cite{'tHooft:1974hx}:  
\begin{equation} 
\langle J(k) J(-k) \rangle = \text{ planar diagram sum} = \sum_n \frac{f_n^2}{k^2-m_n^2} \sim N_c^1
\end{equation} 
with $m_n$ the mass of the $n$-th meson and $f_n=\langle 0|J|n \rangle$ the $n$-th meson decay constant. 
The order in $N_c$ is demonstrated in figure \ref{mesonlargeNfig} to be one. 
Demanding the equality to hold for any $N_c$ and arbitrary momentum $k$, gives 
\begin{equation} 
m_n^2 \sim N_c^0 \quad \text{ and} \quad f_n \sim \sqrt{N_c},  \label{fpibehaviour}
\end{equation}  
as alluded to in section \ref{sectioneffQCD}. 
From the requirement that the diagram sum approaches its perturbative QCD form for very large momentum, one deduces that the number of meson states is infinite \cite{'tHooft:1974hx}. 
It can further be shown \cite{Veneziano:1976wm} that 
the $m$-meson vertex 
scales as $N_c^{1-m/2}$. 
Because each additional meson in the vertex gives a $1/\sqrt{N_c}$ suppression, the large $N_c$ limit thus implies a weakly coupled 
meson theory, 
such as the phenomenological chiral Lagrangian (\ref{skyrme}) \cite{Manohar:1998xv}. 
Despite $N_c=3$, the large $N_c$ limit succeeds in making contact to several observational facts. It explains for example why mesons occur in nonets for three light quark flavours, that is the $SU(3)$ flavour octet ($\bar \psi T^a \psi$) and singlet ($\bar \psi \psi$) mesons tend to mix and have comparable masses. Usually they 
are not related, because the flavour singlet meson can mix with gluonic operators. This mixing is suppressed in the large $N_c$ limit, and the singlet and octet mesons combine in a $U(N_f)$ multiplet (for example $m_\rho = m_\omega$ and $f_\pi = f_{\eta'}$ for $N_c \rightarrow \infty$). 

Baryons (containing $N_c$ quarks) have masses that scale as $(N_c)^1$ with sizes and shapes that have an $N_c$-independent limit. As we will not be concerned with baryons in this thesis, we will not further discuss them but refer the interested reader to \cite{Witten:1979kh}.

\chapter{String theory and AdS/CFT} \label{stringchapter}

In this chapter we review the basic ingredients of string theory which are needed to discuss the AdS/CFT correspondence \cite{Maldacena:1997re}. This correspondence is a concrete realization of a duality between string theory and a large $N_c$ gauge theory, more precisely a string theory on a 5-dimensional Anti de Sitter space and a 4-dimensional conformal field theory. General references on string theory include \cite{polchinskiboek, urangalect, johnson}.

\section{Quantum gravity and string theory}  \label{introstring}

The standard model for the fundamental forces is formulated in terms of quantum field theories, but it does not contain gravity. Is it possible to reconcile quantum mechanics and gravity in one theory?  
The general relativity theory of gravity can be derived as the low-energy limit of the only consistent (read gauge invariant upon gauging Lorentz invariance) 
field theory of a massless 
spin-2 field in flat space, serving as gauge field of general coordinate transformations $x ^\mu \rightarrow x^\mu + \xi^\mu(x)$,  
with action (see \cite{feynmangravity} and references therein)  
\begin{equation} 
S = \int d^4 x \sqrt{|\det g_{\mu\nu}|} \left\{ \Lambda + \frac{2}{\kappa^2} \mathcal R + c_1 \mathcal R^2 + c_2 \mathcal R_{\mu\nu} \mathcal R^{\mu\nu} + \cdots + \mathcal L_{matter} \right\}.   \label{gravityEFT}
\end{equation}  
This action contains, in addition to the Einstein-Hilbert action of general relativity, an infinite number of terms which are higher order in curvatures $\mathcal R, \mathcal R_{\mu\nu}$, ... or derivatives, with $\mathcal R (\sim \partial \partial g)$ the Ricci-scalar associated with the metric $g_{\mu\nu} = \eta_{\mu\nu} + \kappa \, h_{\mu\nu}$ in the weak field limit ($\eta_{\mu\nu}$ the Minkowski metric), and with $\kappa^2 = 32 \pi G$ proportional to Newton's constant $G$, and $\Lambda$ the cosmological constant. 
It can be quantized using covariant quantization with the background field method ($g_{\mu\nu} = \bar g_{\mu\nu} + \kappa \,  h_{\mu\nu}$) \cite{Feynman:1963ax,DeWitt:1967ub,'tHooft:1974bx,Donoghue:1995cz}. 
However, because of the dimensionful coupling $\kappa$ and the nonlinear interactions to all orders in $h_{\mu\nu}$, it is a non-renormalizable theory.
This resulted in statements that the above covariant perturbation approach to formulating a quantum theory of 
gravity cannot make meaningful physical predictions, but in the modern viewpoint \cite{Donoghue:1995cz,feynmangravity} 
(\ref{gravityEFT}) is interpreted as a (non-renormalizable) effective field theory (as discussed in section \ref{sectioneffQCD}) for quantum gravity at low energies compared to the Planck scale: 
predictions (including quantum predictions \cite{'tHooft:1974bx}) 
can be made, but with a finite accuracy at a given order, where the higher order curvature terms are negligible only when spacetime curvature is weak.  		
It describes the low-energy phenomenology of a more fundamental theory, which might involve new degrees of freedom 
at length scales of order the Planck length $l_P = (G \hbar/c^3)^{1/2} \simeq 10^{-33}$ cm, where a quantum description of black holes would force itself. 
A candidate for this UV-completion is \emph{string theory} (another is loop quantum gravity). 
The fundamental degrees of freedom are, instead of point particles, 1-dimensional strings, in terms of which there exists 
a perturbative formulation of the theory. Although non-perturbative string theoretic objects can be found (see section \ref{sugrasection}), there is no non-perturbative formulation, i.e.\ no (simple) spacetime action for the full string theory such as $S = \int d^4 x \mathcal L_{QCD}$ 
for QCD (there is an effective one at low energies, which is the supergravity action (\ref{sugraaction})). In fact, there are different types of string theories, which are connected through dualities and believed to possibly unify in a further, non-perturbative UV-completion often called 
\emph{M-theory}, where degrees of freedom are branes (``M-branes''), not strings.

\section{String theory basics}

\subsection{Strings and D-branes} \label{Dbranesection}

String theory\footnote{To be correct, not all types of 
string theory contain D-branes, we will only be concerned with the ones that do, such as type IIA and IIB. 
} contains two types of fundamental objects. The first are 1-dimensional 
strings - of which the different vibrational modes are associated with elementary particles in nature - tracing out a world surface or `worldsheet' in spacetime that is minimized by the variational law for the classical Nambu-Goto action 
\begin{equation} \label{NGACTIEannex}
S_{NG} = -T \int dA = -\frac{1}{2\pi\alpha'} \int d\tau d\sigma \sqrt{-\det_{ab} g_{ab}},    
\end{equation} 
with string tension $T=1/(2 \pi \alpha')$ a function of the fundamental string length $l_s = \sqrt{\alpha'}$, and $g_{ab} = \partial_a X^\mu \partial_b X^\nu G_{\mu\nu} (a,b=\tau,\sigma)$ the induced metric on the world surface spanned by $\tau,\sigma$ coordinates, as it is embedded in spacetime with metric $G_{\mu\nu}$. 
Classical string theory dynamics is described by a  2-dimensional conformal field theory  
for the (bosonic) embedding fields $X^\mu(\tau,\sigma)$. Upon addition of worldsheet fermion fields, one obtains superstring theory. Bosonic string theory is a consistent theory only when defined in 26-dimensional spacetimes, superstring theory in 10-dimensional spacetimes. We will be concerned with superstring theory. 
The quantization of strings leads to a tower of modes.  

The second type of objects are $p$-dimensional D$p$-branes 
- defined in perturbative string theory as 
hypersurfaces in spacetime in which endpoints of attaching strings with Dirichlet boundary conditions are restricted to move - of which the low-energy dynamics is governed by the 
Dirichlet-Born-Infeld (DBI) action
\begin{equation} \label{DBIstringannex}
S_{DBI}=-T_p \int d^{p+1}\xi \hspace{1mm} e^{-\phi} \sqrt{-\det_{mn}(G_{mn}+B_{mn}+ 2\pi\alpha' F_{mn})}, 
\end{equation} 
with D$p$-brane tension \cite{johnson} 
\begin{equation} \label{Dbranetensionstringannex}
T_p = \frac{1}{\sqrt{\alpha'}}\frac{1}{(2\pi\sqrt{\alpha'})^p} = (2\pi)^{-p} l_s^{-(p+1)}.
\end{equation} 
For $B_{mn} = F_{mn} = \phi = 0$ 
the variational principle 
for the classical DBI-action just describes minimization of the $(p+1)$-dimensional worldvolume of the brane.  
The fields appearing in this action come from the consideration of massless 
 modes of the attaching open strings (determining the dynamics of the brane) and of closed strings (in order to describe the dynamics of the brane in the background of, and coupling to, the modes of curved spacetime). 
The open string modes contain the massless photon field excitation $A_\mu(x)$ $(\mu=0,...,9)$, which can be decomposed in a Maxwell field $A_m(x)$ ($m=0,...,p$) living on the brane and $9-p$ scalars $\phi^m$ ($m=p+1,...,9)$ (w.r.t. Lorentz symmetry of the brane) corresponding to fluctuations of the brane in its $9-p$ transversal directions. $F_{mn}$ is the field strength of $A_m$.   
Here we assumed the static gauge in which worldvolume coordinates $\xi^m (m=0,...,p)$ coincide with spacetime coordinates $X^m$. 
The closed string modes contain a massless two-index tensor field, which can be decomposed in its symmetric (traceless) part, its antisymmetric part, and its trace. They correspond respectively to the graviton $G_{\mu\nu}$, the Kalb-Ramond field $B_{\mu\nu}$ and the dilaton $\phi$. $G_{mn}$ and $B_{mn}$ are the pullbacks of these background fields: $(B,G)_{mn}(\xi) = (B,G)_{\mu\nu}(X(\xi)) \partial_m X^\mu \partial_n X^\nu$.  
There is another massless bosonic mode in the closed string spectrum that does not appear in the DBI-action, namely the $n$-form Ramond-Ramond\footnote{The worldsheet fermion fields added to bosonic string theory in superstring theory can obey Ramond (R) or Neveu-Schwarz (NS) boundary conditions, characterizing the modes. (The graviton, Kalb-Ramond field and dilaton belong to the NS-NS sector.) \label{footnoteRNS}} (RR) field $F^{(n)}$. 
The D$p$-brane couples to the associated RR potentials $C^{(n-1)}$ with a charge equal to its tension $T_p$, 
described by the topological Chern-Simons (CS) action: 
\begin{equation} 
S_{CS} = T_p \int_{\mathcal M_{p+1}} e^{ 2 \pi \alpha' F-B} \sum_n C^{(n)} 
\label{SWZ}, 
\end{equation} 
where the integral is over the worldvolume $\mathcal M_{p+1}$ of the brane, with $C^{(n)}$ the pullback of the background RR potential, $F$ the two-form field strength of the gauge field $A_m$ on the brane, and the sum over $n$ running over odd or even values for respectively type IIA or IIB superstring theory. 
The first term in the expansion of the exponential, 
\begin{equation} 
S_{CS} = T_p \int_{\mathcal M_{p+1}} C^{(p+1)} + \mathcal O(F-B), \label{CSzero}
\end{equation} 
describes the `electric' coupling of the D$p$-brane to the RR gauge field $C^{(p+1)}$.

Expanding the flat space ($G_{\mu\nu}=\eta_{\mu\nu}$, $B_{\mu\nu}=0$) and constant dilaton
($\phi=\phi_0$)  DBI-action for slowly varying fields to second order in the fields, it can be seen to reduce to the action of a\footnote{More precisely the dimensional reduction of 10-dimensional $U(1)$ Yang-Mills to $p+1$ dimensions, for gauge fields that are only dependent on the brane coordinates.} 
$U(1)$ gauge theory in $p+1$ dimensions with $(9-p)$ real scalar fields (all in the adjoint representation), that has a Yang-Mills coupling constant 
\begin{equation} 
g_{YM}^2 = g_s T_p^{-1} (2\pi \alpha')^{-2} = \frac{g_s}{\sqrt{\alpha'}} (2\pi\sqrt{\alpha'})^{p-2},  \label{YMcoupling}
\end{equation} 
with the string coupling constant $g_s$ defined as $g_s=e^{\phi_0}$ (on which more in the following section \ref{sectionstringexp} and eq. (\ref{dilatonpbranesol})).

When $N$ D$p$-branes coincide, the $U(1)^N$ gauge group of separated branes becomes enhanced to a $U(N)$ gauge group. The full non-Abelian generalization of the non-linear DBI-action, including all stringy $\alpha'$-corrections, is not yet known. To lowest order however, the non-Abelian DBI-action for a system of $N$ coincident 
D$p$-branes in a flat supergravity background is just the non-Abelian generalization of the $U(1)$ Yang-Mills theory of the previous paragraph to the  
$U(N)$ one, namely the dimensional reduction of 10-dimensional $U(N)$ $\mathcal N=1$ (with $\mathcal N$ the number of spinor supercharges $Q^i_\alpha, i=1,...,\mathcal N,\alpha=1,...16$) supersymmetric Yang-Mills theory (SYM) to $p+1$ dimensions \cite{Witten:1995im}.  
To make sense of the non-Abelianization of the higher order terms in the DBI-action, Tseytlin \cite{Tseytlin:1997csa} suggested that, at least for the bosonic terms, the gauge trace should be replaced by a symmetrized trace STr (discussed in detail later) to resolve ordering ambiguities. We refer also to \cite{Taylor:1999pr,Myers:1999ps,Myers:2003bw,Wulff:2007vj}. 
The STr-prescription has not been derived from more fundamental principles, but gives results that are compatible to direct string computations at low order in the field strength. We will come back to this discussion in section \ref{subsecnonab}. 
Similarly, for the CS-action, a symmetrized trace is added to the definition (\ref{SWZ}) \cite{Myers:1999ps,Myers:2003bw,Wulff:2007vj}.

\subsection{Perturbative expansions in string theory and relation to large $N_c$ gauge theory} \label{sectionstringexp}

\begin{figure}[h!]
  \centering
  \scalebox{0.7}{
  \includegraphics{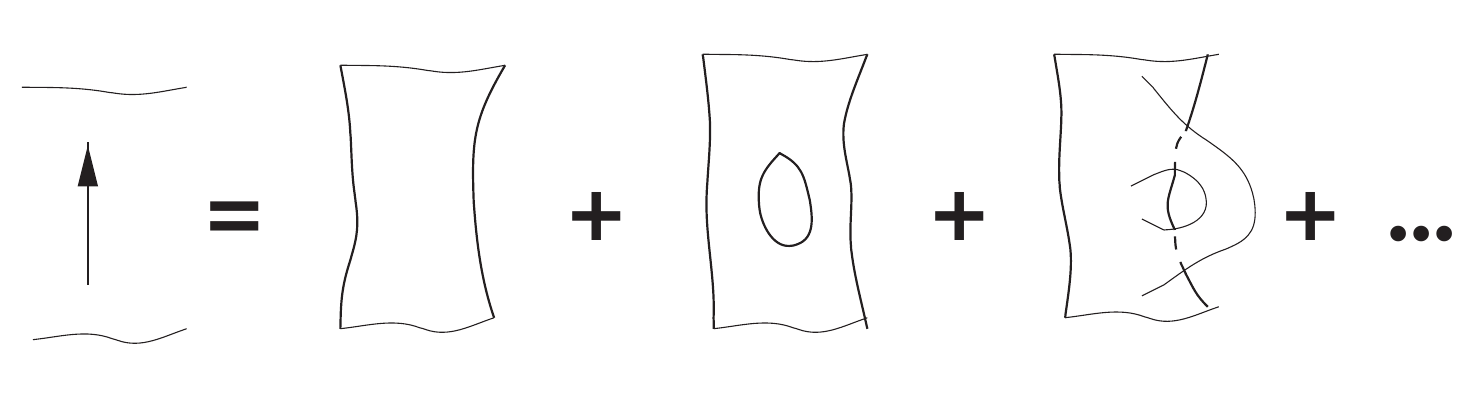}}
  \caption{The genus expansion for open string theories \cite{urangalect}.}\label{uranga1}
\end{figure}

\begin{figure}[h!]
  \centering
  \scalebox{0.8}{
  \includegraphics{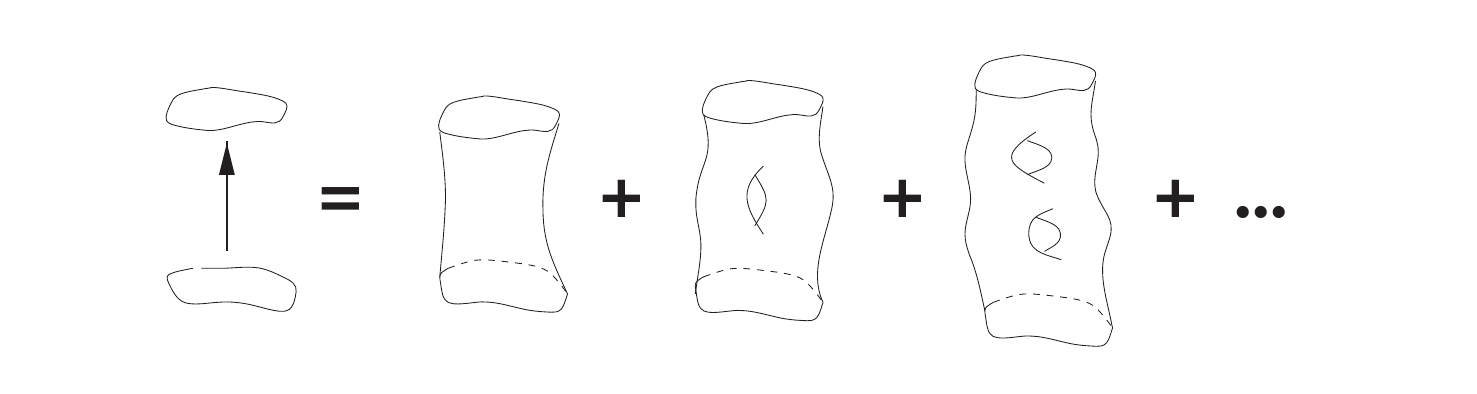}}
  \caption{The genus expansion for closed string theories \cite{urangalect}.}\label{uranga2}
\end{figure}

String theory in an arbitrary curved background is characterized by two expansions. Firstly there is the quantum loop or genus expansion in the string coupling constant $g_s$ for scattering amplitudes in spacetime. It corresponds to a sum over all possible worldsheet 
topologies with increasing number of handles or genus $g$, counting the number of splittings and rejoinings of the string (see figure \ref{uranga1},\ref{uranga2}): $\mathcal A \sim \sum_{g=0}^\infty g_s^{2g} \mathcal A_g$. Or, slightly more general, a worldsheet with $g$ handles and $b$ boundaries is weighted by $g_s^{-\chi}$, with the topological invariant $\chi$ defined in section \ref{sectionlargeN}. 
The string coupling constant depends on one of the dynamic modes of the string, namely (the background value or vacuum expectation value of) the dilaton field\footnote{This follows from considering the generalization of the string worldsheet action (\ref{NGACTIEannex}) for a string in a curved background ($G_{MN}, B_{MN}, \phi$), with the coupling term to the dilaton $S_\phi= \frac{1}{4 \pi \alpha'} \int d \sigma d \tau \sqrt{|\det g_{MN}|} \alpha' R^{(2)} \phi$ (with $R^{(2)}$ the 2-dimensional Ricci scalar of the worldsheet) reducing to $\phi \chi$ 
for the case of a constant background dilaton, such that each worldsheet diagram in the Euclidean path integral is weighted by a factor $e^{-\phi \chi}$.}: 
$g_s = e^{\langle \phi(x) \rangle}$. This is consistent with the general property of string theory that there are no arbitrary external parameters. Secondly there is the $\alpha'$-expansion in $\alpha'/R^2$ (with $R$ a curvature radius of the background) for interactions in the two-dimensional field theory on the world surface, for any given world surface topology.  

At this point it is interesting to couple back to the discussion of large $N_c$ QCD in section \ref{sectionlargeN}, where from (\ref{Feynmancounting}), we can write the expansion of a large $N_c$ gauge theory\footnote{The expression (\ref{largeNampl}) is quite generally valid for a $U(N_c)$ gauge theory coupled to adjoint matter fields, such as for example $\mathcal N=4$ Super-Yang-Mills.} amplitude in (connected) Feynman diagrams as 
\begin{equation} 
\mathcal A = \sum_{g=0}^\infty N_c^\chi \sum_{n=0}^\infty c_{g,n} \lambda^n \quad \text{($c_{g,n}$ constants)}  \label{largeNampl}
\end{equation}
with Euler constant $\chi = 2-2 g - b$. The expansion in non-planarity $\sim 1/N_c^2$ in the first sum of (\ref{largeNampl}) can now be recognized as the genus expansion of a string theory, with expansion parameter $g_s^2 \sim 1/N_c^2$. 
The boundaries $b$ - in the field theory obtained by replacing a gluon by a quark loop and thus $N_f/N_c$ suppressed - correspond to inclusion of open strings in the expansion, with a coupling constant $1/N_c \sim \sqrt{g_s}$ (consistent with being able to make one closed string vertex diagram out of two open string vertex diagrams).  
The interpretation of the expansion in 't Hooft coupling $\lambda$ in the second sum of (\ref{largeNampl}) depends on the string theory identification of $\lambda$. In the AdS/CFT-correspondence (see section \ref{sectionAdSCFT}) it will turn out to be given by $\lambda \sim (\alpha'/R^2)^{-2}$, in which case the second sum can be associated with the $\alpha'$-expansion, or more precisely: the $\alpha'$-expansion on the string theory side of the duality corresponds to a strong-coupling expansion in $1/\sqrt{\lambda}$ on the field theory side.

\subsection{Supergravity} \label{sugrasection}

The spectrum of closed string theory contains the graviton, and superstring theory reduces in the low-energy limit to $(D=10)$-dimensional supergravity, with the effective low-energy type II superstring action in the Einstein frame given by 
\begin{equation} \label{sugraaction}
S = \frac{1}{2\kappa_{D}^2} \int d^{D} x \sqrt{|\det G_{\mu\nu}|} \left( \mathcal R-\frac{1}{2}G_{\mu\nu}\partial_\mu \phi \partial_\nu \phi - \frac{1}{2} \sum_n \frac{1}{n!} e^{a_n \phi} \left(F^{(n)}\right)^2 + (\cdots) \right).
\end{equation} 
with $2 \kappa_D^2 = 16 \pi G_D$ the Newton constant in $D=10$ dimensions, $a_n = \frac{1}{2}(n-5)$ and $n$ taking on even (odd) values for type IIA (IIB) strings; the $(\cdots)$ stand for fermionic terms and a term in the NS-NS 3-form field strength. Supergravity theories are locally supersymmetric and can exist in a number of dimensions less than or equal to 11.  

When D$p$-branes are added to a flat background, they will -- as massive and charged objects\footnote{Being charged  under RR potentials $C^{(p+1)}$ (cfr.\ eq.\ (\ref{CSzero})), D$p$-branes generate a $F^{(p+2)}$-flux, which in its turn will contribute to the energy-momentum tensor.} -- 
cause a curvature of the geometry. Solutions of the supergravity equations derived from (\ref{sugraaction}) that carry the corresponding fluxes and have an event horizon, were originally termed $p$-branes -- `black holes' extended in $p$ spatial dimensions. Extremal $p$-branes saturate the condition that the RR charge $Q$ of the brane has to be smaller or equal than its mass (per unit of volume) $M$, a condition that watches over the fact that singularities are hidden behind the horizon. 
They 
are solitonic 
and thus non-perturbative in nature (with an energy per unit volume of order $1/g_s$). Initially, there was no reason to assume that these solutions of the truncated string action to supergravity would extrapolate to string theoretic objects, but they do thanks to their invariance under half of the supersymmetry transformations of the vacuum theory (BPS states), preserving the maximum amount of supersymmetry. 
The dynamics of the soliton is described by zero mode fluctuations (massless modes that do not change the energy of the soliton, which can often 
be interpreted as Goldstone bosons of the symmetries broken in the soliton background). A stringy description of the spectrum of fluctuations of the theory around the $p$-brane state, is given in terms of oscillation modes of open strings with ends on the D$p$-brane worldvolume. 
The insight that D$p$-branes 
establish the full string theoretic description of extremal $p$-brane supergravity solutions\footnote{This is not true for all $p$-branes, for example for NS5-branes there is no simple stringy description of their spectrum of fluctuations. 
} is due to Polchinski \cite{Polchinski:1995mt}.  One can compute scattering amplitudes of massless modes in perturbative string theory and build an effective action that reproduces them, which results in the 
DBI-action (\ref{DBIstringannex}) \cite{urangalect}. 

The general $p$-brane solution of (\ref{sugraaction}) can for example be found in \cite{Aharony:1999ti} 
but let us here mention the dilaton part of the solution: 
\begin{equation}
e^{-2 \phi} = g_s^{-2} f_-(\rho)^{-\frac{p-3}{2}} \quad \text{with} \quad f_-(\rho) = 1 - \left( \frac{r_-}{\rho} \right)^{7-p}. \label{dilatonpbranesol} 
\end{equation}
The metric solution in the Einstein frame has a horizon at $\rho=r_+$ and a curvature singularity at $\rho=r_-$ for $p \leq 6$. At $\rho \rightarrow \infty$ the metric is asymptotically flat 
and the dilaton takes on its constant asymptotic value $\phi_0$ with $g_s = e^{\phi_0}$ the asymptotic string coupling constant.  
When referring to (\ref{YMcoupling}) in the context of the AdS/CFT-correspondence, $g_s$ is to be interpreted in this asymptotic sense.

\section{The AdS/CFT correspondence} \label{sectionAdSCFT}

\begin{figure}[h!]
  \hfill
  \begin{minipage}[t]{.45\textwidth}
    \begin{center}
      \scalebox{0.2}{
  \includegraphics{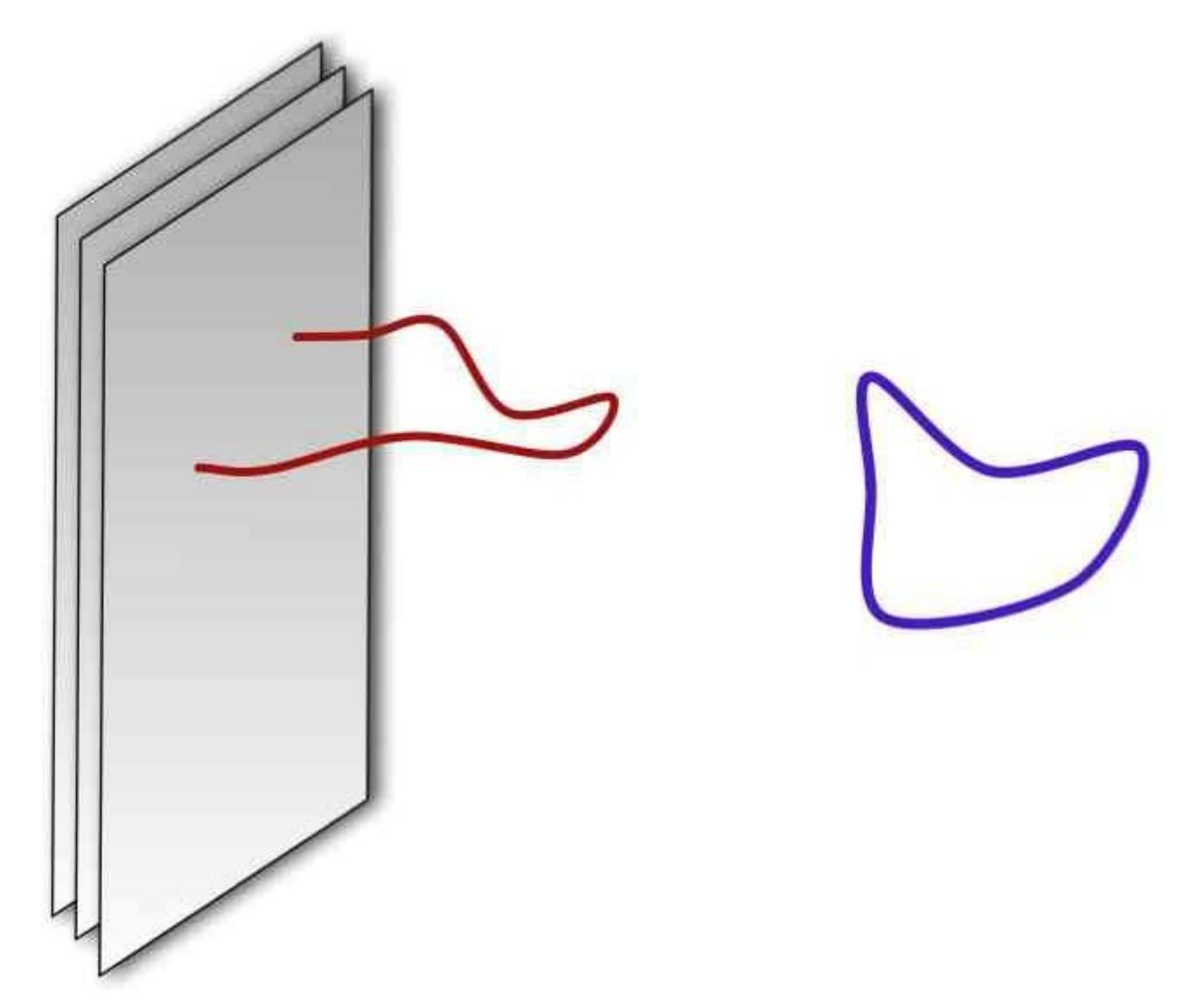}}
    \end{center}
  \end{minipage}
  \hfill
  \begin{minipage}[t]{.45\textwidth}
    \begin{center}
      \scalebox{0.16}{
  \includegraphics{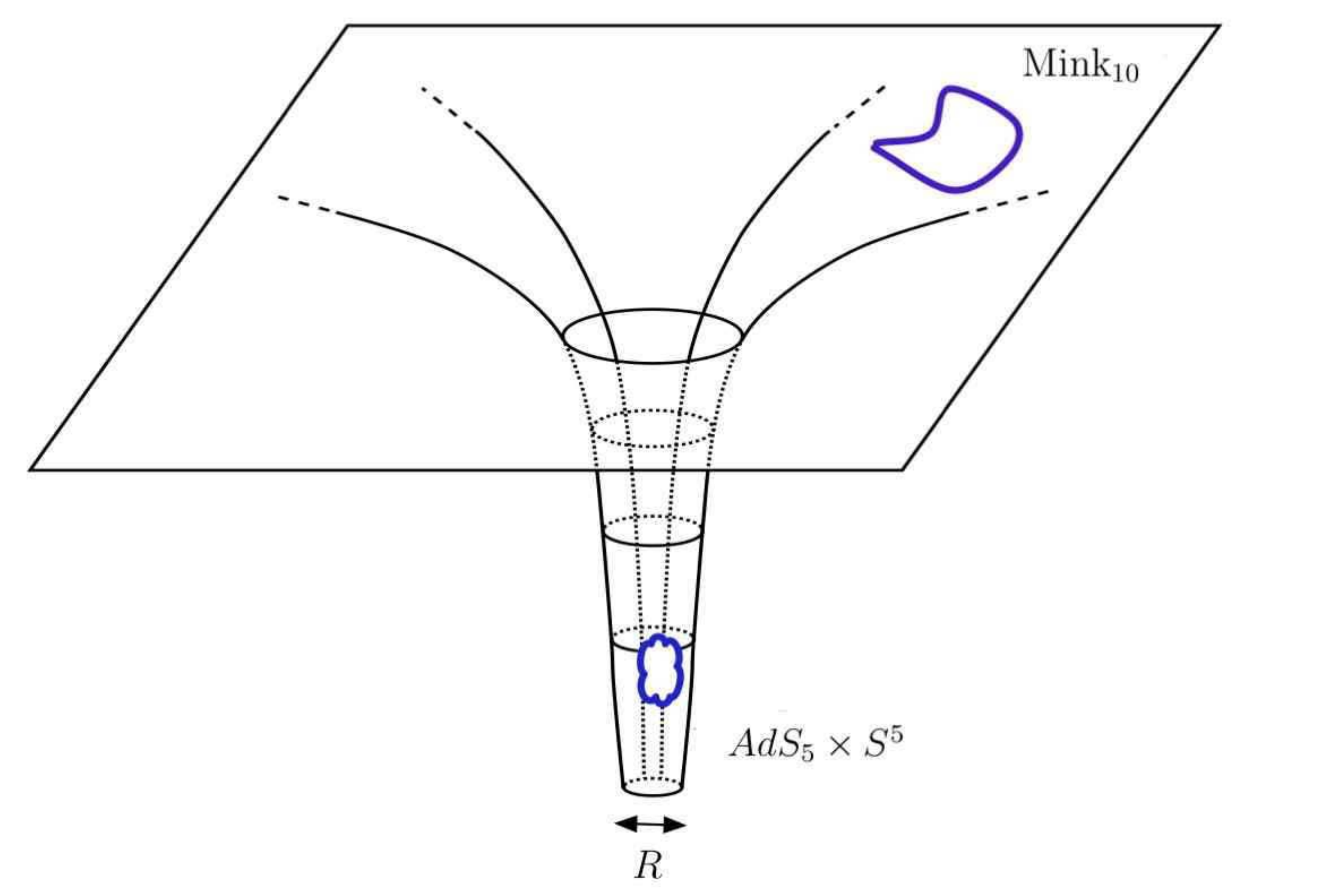}}
    \end{center}
  \end{minipage}
      \caption{ A sketch of the two descriptions of the D3-brane system, resp.\ for small and large coupling. The right figure represents the $ds^2$ of (\ref{3branesol}) \cite{Mateos:2007ay}.}
	\label{mateosfig}
  \hfill
\end{figure}

Consider a system of $N$ coinciding D3-branes. The 3-brane supergravity solution is given by 
\begin{align} 
ds^2 &= f(r)^{-1/2} (-dt^2 + dx_1^2 + dx_2^2 + dx_3^3) + f(r)^{1/2} (dr^2 + r^2 d \Omega_5^2) \nonumber\\
  F_5  &\sim d(\text{Vol})_{S_5}, \qquad 
 f(r)= 1 + \frac{R^4}{r^4}, \qquad R^4 = 4 \pi g_s N \alpha'^2.  \label{3branesol}
\end{align}
The dilaton is constant and 
$F_5$ is such that there are $N$ units of RR 5-form flux on the angular 5-sphere ($S_5$ with line element $d \Omega_5^2$) in the transverse 6-dimensional space (with radial coordinate $r$): $\int_{S_5} F_5 = N$. 
$R$ gives the characteristic size of the developed `throat' (see right figure of \ref{mateosfig}). The standard argument for the AdS/CFT duality comes from considering the low-energy limit of the D3-brane system in its two possible interpretations (see figure \ref{mateosfig}), which are perturbatively valid at opposite regimes in the space of coupling constants. 

On the one hand the stack represents an extremal black brane, characterized by the flat Minkowski metric at radial position $r\rightarrow \infty$, and a horizon that coincides with the singularity at $r=0$. The region near 
 the horizon and the asymptotic region, i.e. the bulk of spacetime at large distances from the $p$-brane, form two decoupled systems of low energy excitations (we are 
in the supergravity approximation 
of closed string theory). The energy $E_p$ of an excitation, as it is measured by an observer at position $r$, is observed as a redshifted energy $E = f(r)^{-1/4} E_p$ 
by a Minkowski observer at infinity. From the point of view of the asymptotic observer there are two types of low-energy excitations: massless particles with very large wavelengths propagating in the bulk, 
as well as all the excitations close to $r=0$ that are consequently infinitely redshifted. On top of this, both types of excitations are decoupled: the near-horizon excitations cannot escape the gravitational well, and conversely asymptotic excitations (with wavelengths much larger then the typical size of the black brane) cannot resolve the region close to the horizon. The two decoupled low-energy systems are: gravity near $r\rightarrow 0$ in the 3-brane background and linearized gravity in the asymptotic bulk region (gravity becomes a free theory at large distances (low energies $E\rightarrow 0$) where the effective dimensionless gravitational coupling $G_D E^8\rightarrow 0$). 

The classical supergravity approximation is valid only when quantum fluctuations of the strings are negligible, $g_s\rightarrow 0$, and if the curvature radius $R$ of the background is large compared to the fundamental string length $\sqrt{\alpha'}$ (in the low energy limit $\alpha' \rightarrow 0$ the infinite tower of massive string excitations decouples and only low-energy degrees of freedom $E \ll 1/\sqrt \alpha'$ remain), which appears to be satisfied when $g_s N \gg 1$ (since $R \sim \sqrt{\alpha'} (g_s N)^{1/4}$). $g_s N$ gives a measure for the total gravitational deformation caused by $N$ D3-branes that each couple to the gravitational degrees of freedom with a strength $g_s$. We conclude that classical supergravity is justified for 
\begin{equation} 
g_s\rightarrow 0, \quad  N\rightarrow \infty  \quad \text{and} \quad g_s N\gg 1. 
\end{equation}

On the other hand, for $g_s N \ll 1$, spacetime will be almost flat, and the low-energy dynamics of the $N$ D3-branes is described by the to $D=4$ dimensionally reduced version of $\mathcal N=1$ SYM in $D=10$, given by $\mathcal N=4$ SYM with 16 Majorana supercharges $Q_\alpha^i$, $i=1,...,4$, $\alpha =1,...,4$ and coupling constant $g_{YM}^2$ in (\ref{YMcoupling}). The massless closed string states in the bulk of spacetime describe linearized supergravity, and are decoupled from the open string modes on the brane because the string coupling $g_s$ is weak. We thus again find two decoupled low-energy systems: an $\mathcal N=4$ SYM theory on the branes and linearized gravity in the asymptotic bulk region.

The asymptotic region is described by decoupled linearized gravity in both interpretations - in the $g_s N\ll 1$ as well as the $g_s N\gg 1$ regime. The Maldacena conjecture states that the remaining decoupled systems can be identified: the $\mathcal N=4$ SYM theory with gauge group $U(N)$ on the one hand and supergravity near the horizon ($r\rightarrow 0$) in the 3-brane background on the other, give two different descriptions of the same physical system in the low energy limit $\alpha' \rightarrow 0$.

Near the horizon $r\rightarrow 0$, the geometry of the 3-brane looks like the product AdS$_5 \times S^5$ of a 5-dimensional Anti de Sitter space (i.e.\ the maximally symmetric solution of the Einstein equations with negative cosmological constant, see appendix \ref{adsspaceapp}) and a 5-sphere: 
\begin{equation}
ds^2 = \frac{r^2}{R^2} (-dt^2 + dx_1^2+ dx_2^2 + dx_3^2) + \frac{R^2}{r^2} dr^2 + R^2 d \Omega_5^2  \label{adsmetric}
\end{equation}
of which the AdS-part (with boundary at $r \rightarrow \infty$) can be recast into the form (\ref{adspoincare}) (with boundary at $x_0 \rightarrow 0$) after a change in coordinates $r/R = R/x_0$. To be more precise, the near-horizon limit $r \rightarrow 0$ is taken simultaneously with the low-energy limit $\alpha' \rightarrow 0$ such that $U \equiv r/\alpha'$ is fixed and can be considered the new radial variable in (\ref{adsmetric}) after an overall rescaling of the metric with $1/l_s^2$ (we consider the region very close to $r=0$ and subsequently scale this region up in a singular way). 
This ensures that energies of objects in the throat are fixed in string units ($\sqrt{\alpha'} E_p \sim$ fixed) 
and energies in the field theory, measured from infinity, are fixed as well ($E \sim E_p r/\sqrt{\alpha'} \sim$ fixed).   
The SYM theory lives 
at the Minkowski conformal boundary $r \rightarrow \infty$ of the AdS$_5$ space (\ref{adsmetric}), which we will refer to as `the boundary' (strictly speaking it is the conformally equivalent metric $d\tilde s^2 = R^2 ds^2 / r^2$ which has a boundary $\mathbb R^{1,3}$ at $r \rightarrow \infty$).  

\begin{figure}[h!]
  \centering
  \scalebox{0.35}{
  \includegraphics{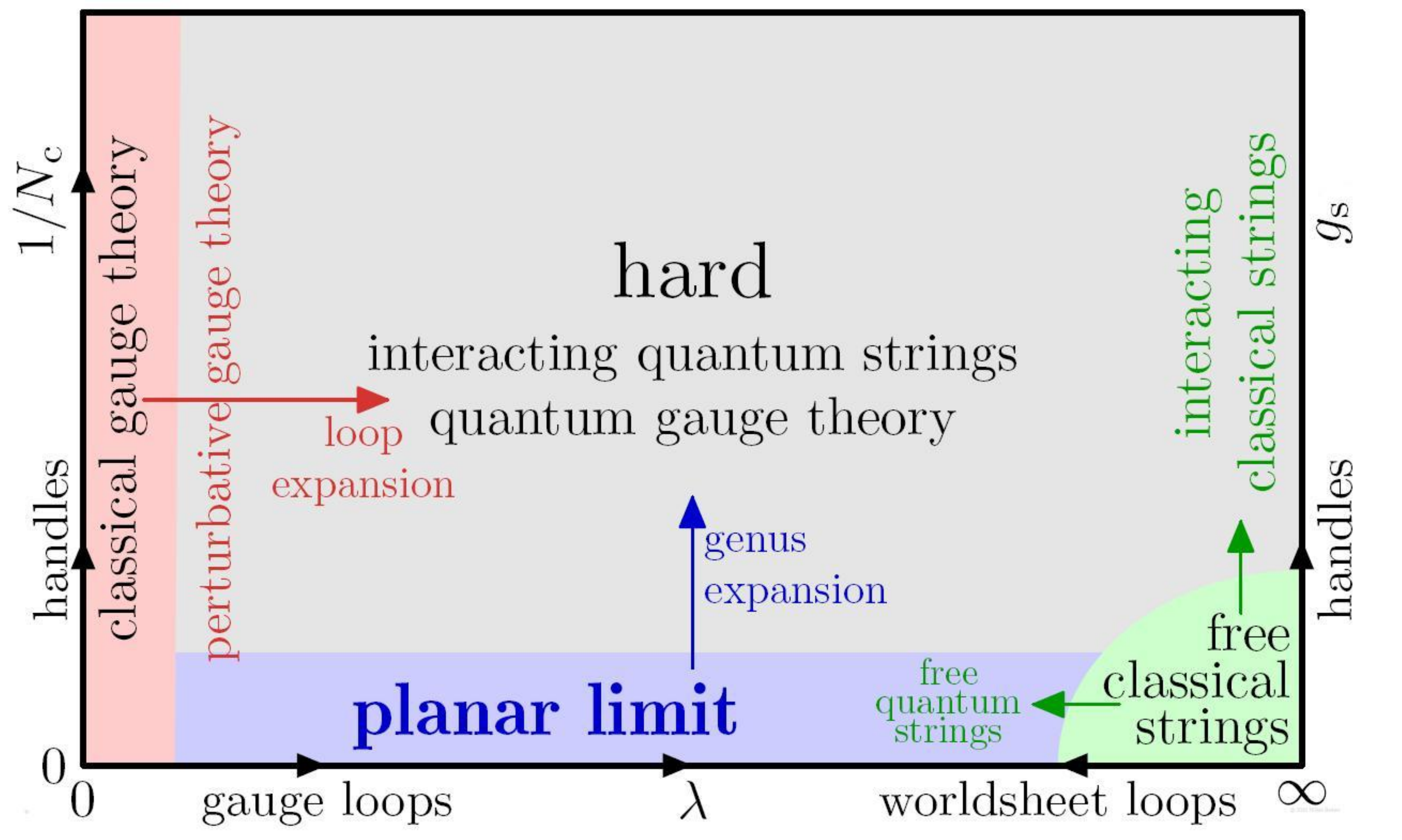}}
  \caption{Map of the parameter space of $\mathcal N = 4$ SYM or strings on AdS$_5 \times S_5$ \cite{Beisert:2010jr}.} 
\label{adspar}
\end{figure}

We can reformulate the conjecture 
as follows. \emph{The AdS/CFT correspondence is the duality between type IIB superstring theory in AdS$_5 \times S^5$ on the one hand and the 4-dimensional $\mathcal N=4$ SYM theory with gauge group $U(N)$, living on the (3+1)-dimensional Minkowski 
boundary of AdS$_5$ space, on the other}. 
The 't Hooft coupling $\lambda$ of the gauge theory can be identified on the gravity side as (using (\ref{YMcoupling}) with $p=3$ for the relation between $g_{YM}$ and $g_s$, and (\ref{3branesol})) 
\begin{equation}
\lambda = g^2_{YM} N = 2 \pi g_s N \simeq \frac{R^4}{l_s^4}. 
\label{2.2.8}
\end{equation}
The classical limit $g_s \rightarrow 0$ can be seen to correspond to the relevant limit in large $N$ gauge theory of $N \rightarrow \infty$ while keeping $\lambda$ fixed, as discussed in section \ref{sectionlargeN}. Taking $\lambda$ large on top of this, gives the supergravity limit (see also discussion in section \ref{sectionstringexp}). 
The correspondence is called a duality because the two different descriptions are perturbatively valid in opposite regimes ($\lambda \gg 1$ and $\lambda \ll 1$), see figure \ref{adspar}. This makes the duality hard to prove but also useful: 
the strongly coupled, non-perturbative regime of a gauge theory becomes accessible through perturbative calculations in the string theory dual, valid in the limits $g_s\rightarrow 0$, $N\rightarrow \infty$ and $g_s N$ constant and large. In its strongest form the conjecture is to hold for all values of $g_s$ and $N$. 
(The possibility exists that there is a phase transition between the gravity regime $\lambda \gg 1$ and the gauge regime $\lambda \ll 1$, in which case there would not be a correspondence between both regimes.) 
A proof of the duality would require a full non-perturbative treatment of string theory, whereas the above argument is not even non-perturbative in $\alpha'$.

\FloatBarrier

\subsection{Symmetries and interpretation of extra dimension} \label{sectionsymm}

$\mathcal{N}=4$ SYM is invariant under the conformal group $SO(4,2)$ in 4 dimensions and has a \emph{global} $SU(4) \simeq SO(6)$ R-symmetry mixing the 4 supercharges. These symmetries are represented in the dual string theory by the $SO(4,2)$ isometry of AdS$_5$ and the $SO(6)$ isometry of $S_5$ respectively. 
More specifically, the isometries on the gravity side 
 that leave the asymptotic form of the metric invariant. In this sense there is a \emph{correspondence between global symmetries in the gauge theory and gauge symmetries in the dual string theory}, which is a general feature of known gauge-gravity dualities \cite{Mateos:2007ay}. 

The conformal group in particular contains scaling transformations $D: x^\mu \rightarrow \Lambda x^\mu$. Indeed, as string theory contains a scale, set by the string tension, the only way that a string (with the usual Nambu-Goto action) can be symmetric under the scaling, is when it corresponds to an isometry of the metric: (\ref{adsmetric}) is invariant under $D$ if $r \rightarrow r/\Lambda$. This means that short-distance (UV) physics in the gauge theory is associated to physics near the AdS boundary (large scale $r \rightarrow \infty$ or IR), and long-distance (IR) physics to the bulk physics near the horizon (short scale $r \rightarrow 0$ or UV). This is the so-called \emph{UV/IR correspondence} and it is natural to \emph{identify the holographic radial direction $r$ with the renormalization group scale}. For a conformal theory the RG flow is trivial and this is reflected in the exact isometry of the AdS background. 
It can be shown that invariance under the 5-dimensional general coordinate transformations implies the Callan-Symanzik
renormalization group equations in the dual field theory \cite{deBoer:2002da, deBoer:1999xf, Fukuma:2002sb}. 
This leads to the statement that from the 5-dimensional point of view, the renormalization group is on equal footing with Poincar\'e invariance.

\subsection{The dictionary of AdS/CFT} \label{adsEucdictionary} 

We briefly discuss the map between observables on (Euclidean versions of) the AdS and the CFT side \cite{Gubser:1998bc, Witten:1998qj}. 
As motivation for the main idea, consider a change in the value of the coupling constant $g_{YM}$ corresponding to a deformation of the $\mathcal N=4$ SYM theory by 
a marginal operator (such that none of the symmetries of the theory are broken). 
Through $g^2_{YM} \sim g_s$ from (\ref{YMcoupling}), this changes the value of the string coupling constant $g_s$ and thus of the background dilaton field in the 3-brane supergravity solution, given by the value of the dilaton at the AdS boundary. This suggests, more generally, that the asymptotic value of a string bulk field $\phi$ at the AdS boundary, $\phi_0(x) = \phi(x,r)|_{\partial \text{AdS}}$, acts as a source $\phi_0(x)$ for a gauge-invariant, local operator $\mathcal O(x)$ in the field theory, deforming the field theory action to $S \rightarrow S + \int d^4 x \, \phi_0(x) \mathcal O(x)$.  This is the \emph{field-operator correspondence}. 
The AdS/CFT prescription then naturally consists of identifying the (Euclidean) partition functions on both sides: 
\begin{equation}
\mathcal Z_{CFT}[\phi_0] = \mathcal Z_{string}[\phi ; \phi|_{\partial \text{AdS}}=\phi_0] 
\end{equation}
with $\mathcal Z_{CFT}[\phi_0] = \left \langle e^{ \int d^4 x \phi_0(x) \mathcal O(x)} \right \rangle_{CFT}$ 
the generating function of correlation functions of operators by multiple functional differentiation, 
 and $W_{CFT}[\phi_0] = -\log \mathcal Z_{CFT}[\phi_0]$ the generating function of connected correlation functions. The string partition function simplifies drastically in the classical (or saddle point) supergravity approximation where $\mathcal Z_{string} \approx e^{-S_{sugra}[\phi_{cl}]}$, in which case the prescription can be reformulated as 
\begin{equation}
W_{CFT}[\phi_0] \approx S_{sugra}[\phi_{cl}; \phi_{cl}|_{\partial \text{AdS}} = \phi_0] 
\end{equation}
with $ S_{sugra}$ the gravitational on-shell action as a functional of the boundary conditions on the classical bulk solution $\phi_{cl}$. 
This action is typically divergent because of the infinite bulk volume. These IR divergences on the gravity side correspond to UV divergences of correlators on the field theory side. 
In the process of \emph{holographic renormalization}, local covariant counterterms are added to the on-shell action to remove the volume divergences. From the renormalized action one can then extract renormalized correlators for the quantum field theory. This is equivalent to renormalizing the correlators in the field theory directly. 
The scaling dimension $\Delta$ ($=$ what we referred to as $d_{\mathcal O}$ in section \ref{statmech}) of a scalar operator $\mathcal O$ and the mass $m$ of the associated supergravity bulk field $\phi$ are related through $m^2 = \Delta(\Delta-d)$, from which it follows that massless, massive and tachyonic fields on the supergravity side correspond to marginal, irrelevant and relevant operators, respectively, on the field theory side.

\subsubsection{Some examples}

A conserved current $J^\mu(x)$ in the field theory, associated with a global symmetry, couples to a gauge field $A_\mu(x,r)$ in the bulk: this makes more explicit the 
 correspondence between local symmetries in the bulk and global symmetries on the boundary, discussed in section \ref{sectionsymm}. Indeed, the coupling $\int d^4 x A_\mu(x) J^\mu(x)$ is invariant under gauge transformations $\delta A_\mu = \partial_\mu f$ by virtue of $\partial_\mu J^\mu=0$. 
The conserved energy-momentum tensor $T_{\mu\nu}$ in a translationally invariant field theory for example, couples to or ``is dual to" the metric field $g_{\mu\nu}(x,r)$ (cfr.\ form of the coupling term (\ref{gTcoupling})).  
Returning to the original motivating example of this section, the dilaton field is dual to (roughly) the operator Tr $F^2$ because of its relation to $g_{YM}$ and the form of the coupling term $\frac{1}{g^2_{YM}} \text{Tr } F^2$. In a conformal setting such as AdS/CFT, the dilaton is constant and the supergravity solutions with deformed boundary conditions, i.e. changed value of $g_{YM}$, are just the AdS$_5 \times S_5$ solution with any value of the string coupling.  
A running dilaton however, i.e.\ a dilaton that depends on the holographic radius $r$ identified with the field theory energy scale, will be encountered later and can be associated with a non-zero beta function $\beta(g_{YM})$ ($\langle T_\mu^\mu \rangle$ of eq.\ (\ref{betaTrF2}) can be calculated holographically \cite{Kanitscheider:2008kd}).

\chapter{Towards a (non)AdS/QCD correspondence}  \label{nonAdSchapter}

Since the discovery of the AdS/CFT-correspondence, a lot of effort has gone into finding ge\-ne\-ra\-li\-za\-tions of the duality, with applications to QCD and condensed matter physics (as well as fluid mechanics, cosmology, etc.)\ in mind. In this more general context, one speaks of \emph{gauge-gravity dualities}. In contrast to $\mathcal N=4$ SYM, QCD is neither conformal nor supersymmetric. We discuss in this section how the gravitational background should be deformed to be able to give a dual description of these broken symmetries compared to the AdS/CFT case. The required features can be summarized as: asymptotically AdS to model conformal invariance in the UV, a cut-off deep in the bulk to model confinement and a mass gap in the IR, and a compact dimension to break supersymmetry via dimensional reduction. 
We will see that the D4-brane background is a good candidate, forming the basis of the Sakai-Sugimoto model.

\section{Desired features of the gravitational background}   \label{desiredf} 

Consider a general background 
\begin{equation}
ds^2 = w(u) ^2 (-dt^2 + d\vec x^2) + \cdots du^2 + \cdots,
\end{equation}
with $w(u)$ a warpfactor, i.e. a redshift factor (multiplying $dt$) that multiplies the field theory space coordinates as well, insuring 4-dimensional Poincar\'e invariance. 
The warpfactor is assumed to be dependent on the holographic radial dimension $u$ only, and more precisely to be a monotonically increasing function with typically $w(\infty) \rightarrow \infty$ such that the boundary of the background at $u \rightarrow \infty$ is Minkowski space. 
Distances and time intervals measured in the field theory with coordinates $\vec x$ and $t$ are related to distances and time intervals in the bulk through the warp factor. An object in the bulk (with energy $E_{bulk}$ and size $d_{bulk}$)  will correspond to a field theoretic configuration with different size ($d_{bdy}$) and energy ($E_{bdy}$):
\begin{eqnarray}
E_{bdy} & = & w(u) E_{bulk} \\
d_{bdy} & = & \frac{1}{w(u)} d_{bulk}.
\end{eqnarray}
Taking $u \rightarrow \infty$ $(0)$ in the bulk, corresponds to probing the dual field theory in the UV region (IR region), making it natural to interpret the holographic direction $u$ as an energy scale in the boundary field theory.  
We thus again find a UV/IR correspondence (as in section \ref{sectionsymm}), this time in a more general background without referring to conformal symmetry. 
\begin{figure}[h!]
  \centering
  \scalebox{0.5}{
  \includegraphics{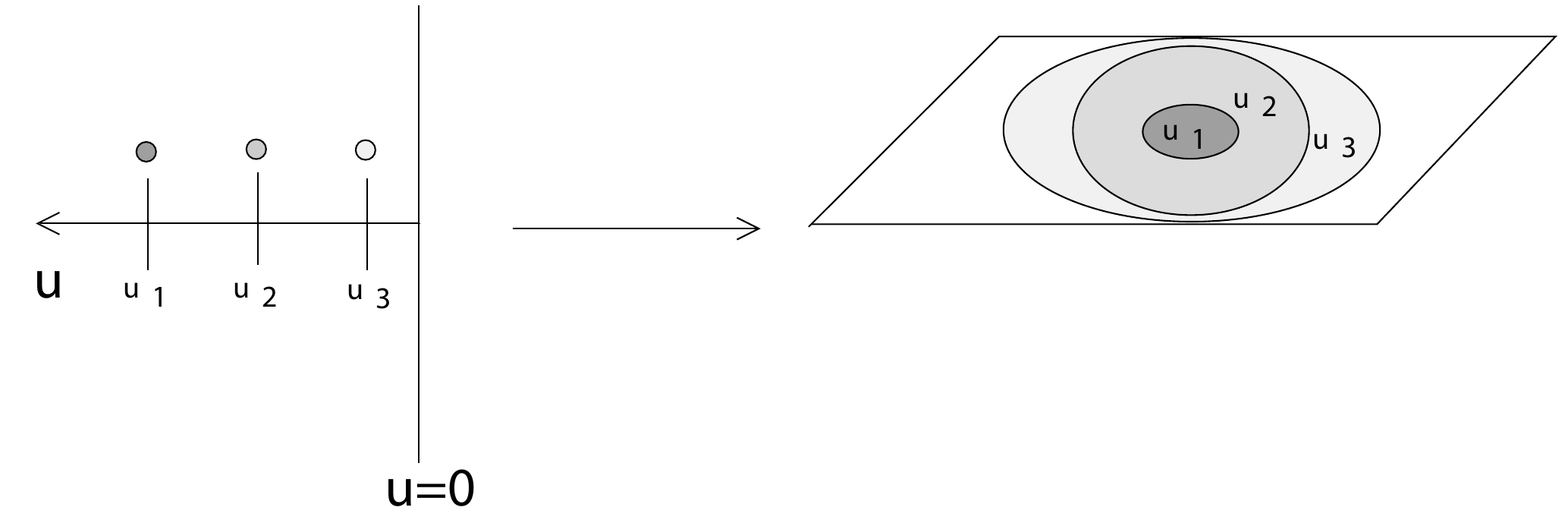}}
  \caption{The UV/IR correspondence: the same bulk object at different radial positions $u_1, u_2, u_3$ in the gravitational background  corresponds to an object in the dual field theory with different sizes    
  \cite{Maldacena:2003nj}.}\label{maldacena}
\end{figure}

\subsubsection{Asymptotically AdS at high energy} 

QCD looks conformal in the UV: at energies $E \gg \Lambda_{QCD}$ there is effectively no scale in the theory. 
One therefore can impose that the gravitational background asymptotes to an AdS space (times a compact space). 
In this way, the deviation of the original AdS/CFT duality, with dictionary defined at the AdS boundary, is limited.  
To model asymptotic freedom however, the duality should be applicable at weak coupling, i.e.\ it would be necessary to go beyond the supergravity approximation (the correspondence between the supergravity approximation and strong coupling of section \ref{sectionstringexp} and the paragraph below eq.\ (\ref{2.2.8}) extends to more general gauge-gravity dualities). For this reason, asymptotic freedom is always problematic in holographic QCD theories.

\subsubsection{Cut-off at low energy} 

QCD exhibits confinement in the IR, modeled in the field theory as follows: an infinitely heavy quark $q$ and antiquark $\bar q$, separated by a distance $L$, are trapped in a confining linear potential $V(L) = \sigma L$. The colour field lines between $q$ and $\bar q$ form a stringy flux tube with constant string tension $\sigma$. 
In order to determine the conditions on the gravitational background for it to allow the formation of a flux tube in the dual field theory, first we discuss how to introduce external, non-dynamic quarks in the holographic theory.  

Consider a stack of $N+1$ coinciding D-branes, on which a $U(N+1)$ gauge theory resides. Separating one of the D-branes from the stack, corresponds to giving a vacuum expectation value to one of the scalars on the brane. The $U(N+1)$ symmetry is broken to $U(N) \times U(1)$. Open strings with one end on the separated brane and the other on the stack, have become massive (with mass equal to the string tension times the length of the string), 
and the endpoint of the string coupling to the $U(N)$ gauge theory is interpreted as a quark in the fundamental representation of $U(N)$. A static quark or external probe quark can then be modeled by an infinitely heavy string attached to the probe brane at large $u \rightarrow \infty$. 

\begin{figure}[h!]
  \centering
  \scalebox{0.4}{
  \includegraphics{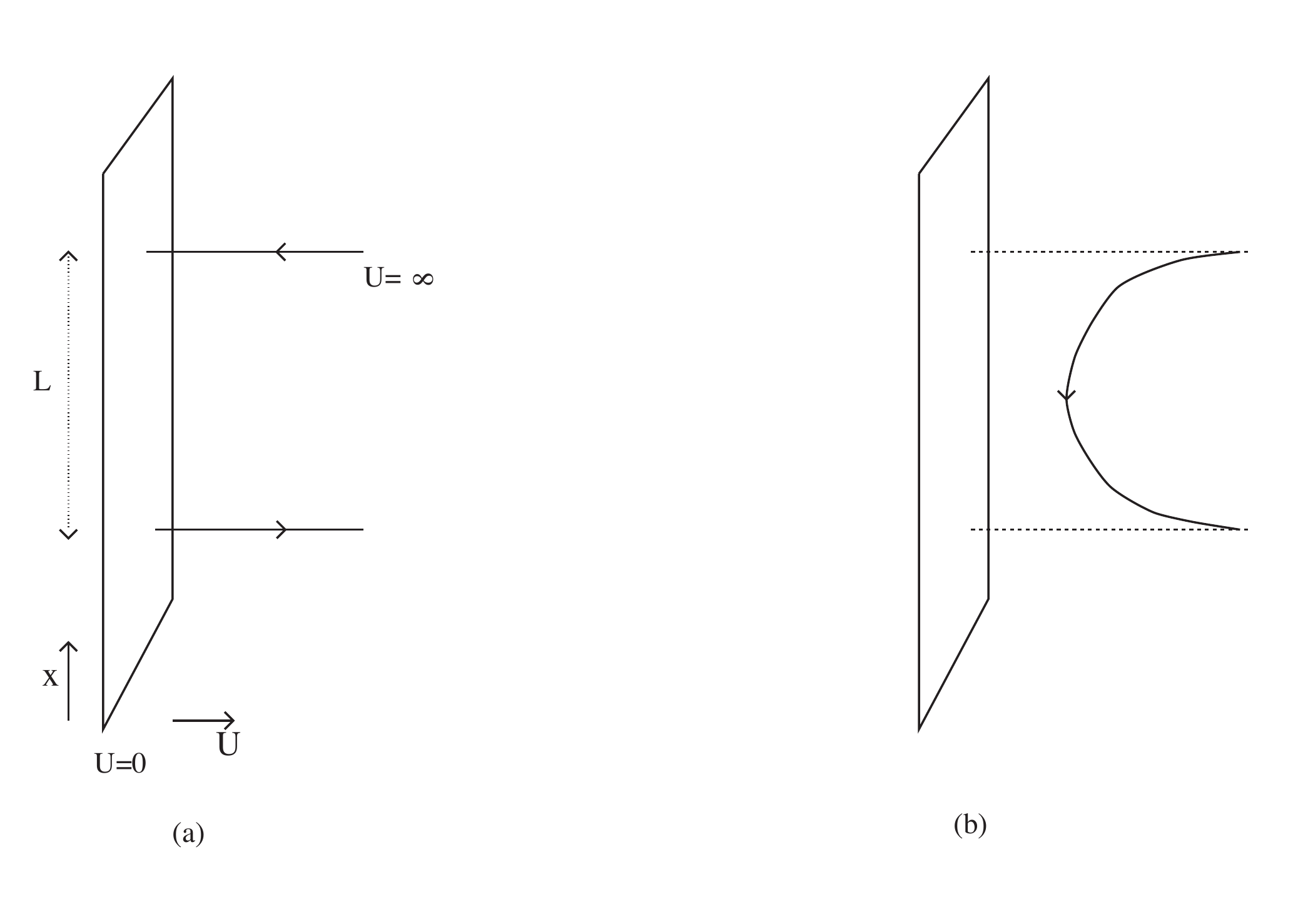}}
  \caption{Configuration of two probe quarks (a) before and (b) after turning on their coupling to the $U(N)$ gauge theory. Configuration (b) minimizes the Nambu-Goto action  
  \cite{Maldacena:1998im}.}\label{wilson}
\end{figure}

\begin{figure}[h!]
  \centering
  \scalebox{0.4}{
  \includegraphics{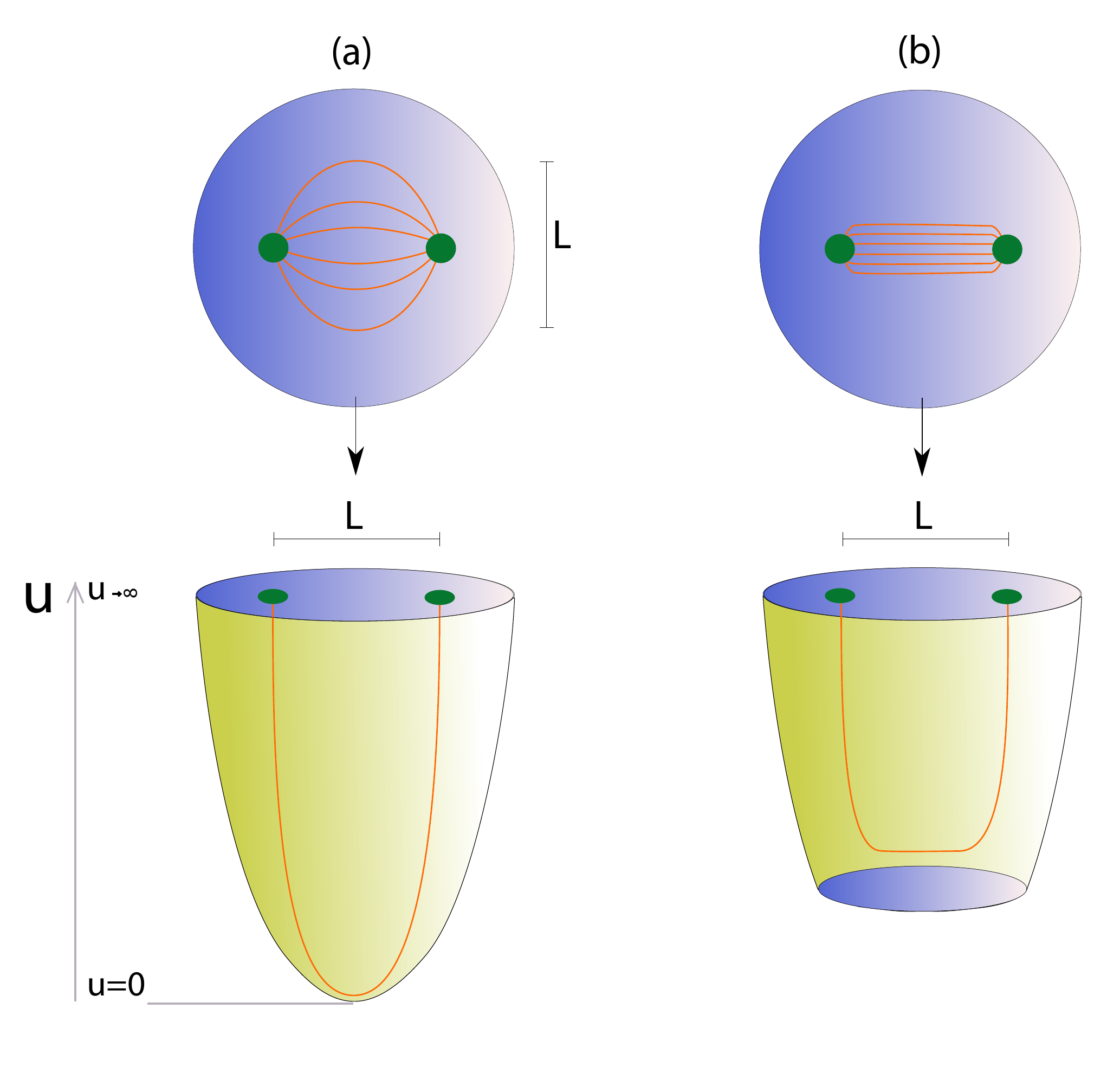}}
  \caption{A $q \bar q$-pair in an AdS space (b) with and (a) without cut-off, and the associated field theoretic interpretation. The orange $q \bar q$-string represents the colour flux in the dual field theory. To model confinement, a cut-off has to be present in the gravitational background.}\label{UVIR}
\end{figure}

We do not expect to see confinement when considering a static $q \bar q$-pair in the AdS$_5 \times S_5$ background. The free $q$ and $\bar q$, at positions $x=-L/2$ and $L/2$  in the field theory, are represented holographically by two strings hanging from the AdS boundary at $u \rightarrow \infty$, and stretching to $u=0$. The interaction between the quarks is turned on by considering their coupling to the $U(N)$ gauge theory, or, in the dual gravitational picture, by turning on gravity in the AdS$_5\times S^5$ background. The strings interact by merging their end points into one long string (see figure \ref{wilson}). This string will minimize its energy: on the one hand it wants to be short because of its finite string tension, on the other it is drawn to the bulk of AdS space, where the warpfactor is small (the string seeks the minimum of the gravitational potential). This competition resolves in a $\cup$-shaped string that reaches its minimum at a finite radial value $u=u_0$. The $\cup$-shaped string represents the colour field lines, that in the field theory describe the interaction between $q$ and $\bar q$. We call the projection of this string on the AdS boundary the `QCD string'. The UV/IR correspondence tells us that the QCD string will `thicken' in the center (see figure \ref{UVIR}(a)). Since we know that the field theory on the AdS boundary is conformal, the thickening can only be of order $L$. For the same reason the interaction can only be given by a static Coulomb potential $V_{q\bar q} \simeq \frac{1}{L}$. The full 
calculation 
of the potential comes down to calculating the energy of the $\cup$-shaped Nambu-Goto string and can be found in \cite{Maldacena:1998im}. 
The above reasoning based on the UV/IR correspondence leads to the insight that a confining flux tube with a constant thickening can be modeled holographically by a rectangular (instead of $\cup$-shaped) string with a long horizontal segment (see figure \ref{UVIR}(b)). This creates an effective string tension $\sigma$ of the QCD string, leading to a linear potential $V=\sigma L$ ($+$ correction terms). For a background to allow such a rectangular string configuration, it has to have  
a cut-off at a finite value of the radial coordinate $u=u_\textit{cut-off}$. The warpfactor then has a finite lower bound and the string will rest at the artificial wall. 
One can conclude that confinement in a holographic model can only be described by introducing a cut-off scale $u_\textit{cut-off}$ in the gravitational background, which explicitly breaks conformal invariance and is related to the confinement scale $\Lambda_{QCD}$, introduced in eq.\ (\ref{LambdaQCD}). 


The effective QCD string tension associated with a long $q\bar q$-string stretching in the $u$- and $x$-direction of a general background 
\begin{equation} \label{metriek}
ds^2 = g_{MN} dX^M dX^N = -G_{tt}(u) dt^2 + G_{xx}(u)d\vec x^2 + G_{uu}(u) du^2 + \cdots
\end{equation}
can be calculated \cite{Kinar:1998vq} 
to be 
\begin{equation}
\sigma =  \frac{1}{2\pi\alpha'}\sqrt{G_{tt}(u_\textit{cut-off})G_{xx}(u_\textit{cut-off})}.  \label{Teff}
\end{equation} 
A confining background is then a metric (\ref{metriek}) for which $\sigma$ in (\ref{Teff}) is non-zero. 

The introduction of a cut-off scale in AdS space also leads to a mass gap in the dual gauge theory, i.e.\ all excitations have a positive mass. To see this, consider the metric of an AdS$_d$ space in Poincar\'e coordinates (\ref{adspoincare}). In these coordinates the boundary (center) of AdS is at $x_0=0$ ($x_0=\infty$). A radial photon ($dt^2=dx_0^2$) can reach the boundary in a finite amount of time:  $\int dt = |\int_{finite}^{0} dx_0| = \text{ finite}$, but needs an infinite amount of time to reach the center: $\int dt = |\int_{finite}^{\infty} dx_0| = \infty$. If however space ends at $x_0=x_\textit{cut-off} < \infty$, the photon experiences the deformed AdS space as a finite box with consequently a discrete spectrum and a mass gap. 

Introducing an artificial cut-off of the AdS radius $u$ (this is the so-called hard wall approximation), does not fulfill the supergravity equations of motion, hence leading to the risk that the gauge theory dynamics are not correctly encoded. Another way to obtain a cut-off is to look for a supergravity solution that naturally ends at a finite value $u=u_\textit{cut-off}$ of the radial coordinate.

\subsubsection{Compact dimension} 

In order to break the gauge theory supersymmetry, consider a geometry with a compact dimension, such that the boundary at $u \rightarrow \infty$ where the, a priori supersymmetric, $(D+1)$-dimensional field theory lives, is Minkowski$_D \times S^1$. The theory can be compactified on the circle $S^1$ via the standard procedure of dimensional reduction: the fields are expanded in Fourier modes around the circle, 
\begin{equation}
\phi(x^\mu, y) = \sum_{k \in \mathbb{Z}} e^{\frac{iky}{L}} \phi_k(x^\mu), \qquad \mu=0, ..., D-1; \,\, 0 \leq y \leq 2 \pi L,   \label{fourierexp}
\end{equation}
and only the lowest modes\footnote{The reduction is consistent if the light modes do not source the heavy modes. 
} of the resulting infinite tower of massive Kaluza-Klein modes, $m_k^2 = \left(\frac{k}{L}\right)^2 + m^2$, are kept, assuming the radius $L$ of $S^1$ is small enough. For bosons, which are necessarily periodic along the circle, this means that massless bosons remain massless (at a classical level) upon keeping only the zero mode after dimensional reduction. For fermions however, a non-trivial spin structure can be chosen: imposing anti-periodic instead of periodic boundary conditions, gives $k \rightarrow k + \frac{1}{2}$ in the exponential in (\ref{fourierexp}) and the lack of a zero mode in this case results in decoupling, massive (of the order $1/L$) fermions. 
Supersymmetry hence is broken and the scalars (those present in the $(D+1)$-dimensional field theory plus extra scalars from the $S^1$-component of the gauge fields) obtain a mass of order $\lambda/L$ at one loop in the $D$-dimensional gauge theory with 't Hooft coupling $\lambda$. Only the gauge bosons, protected by gauge invariance, remain massless under dimensional reduction, such that the end result is a non-supersymmetric, pure gauge theory in $D$ dimensions.

\section{D4-brane background}   \label{D4section}

We introduce the D4-brane background \cite{Petersen:1999zh,Aharony:1999ti,Kruczenski:2003uq}, proposed by Witten \cite{Witten:1998zw} as dual description of pure 4-dimensional QCD. It forms the main ingredient of the Sakai-Sugimoto model discussed in the next chapter. 
 
To begin with, consider a system of $N_c$ coinciding M5-branes wrapped around the 11th-dimensional circle in M-theory (we mentioned M-theory in the introductory section \ref{introstring} of the chapter on string theory). The low-energy limit of M-theory is 11-dimensional supergravity and the effective action, at least for static solutions corresponding to flat translationally invariant $p$-branes that are isotropic in transverse directions, 
 is given by (\ref{sugraaction}) with $D=11$, $n=4$, $a_n=0$ 
and in particular no dilaton $\phi \equiv 0$ (so $g_s$ is absent and only the Einstein frame metric is relevant). The 5-brane supergravity solution is 
\begin{align} 
ds^2 &= f(r)^{-1/3} \left( -dt^2 + \sum_{i=1}^5 dx_i^2 \right) + f(r)^{2/3} (dr^2 + r^2 d \Omega_4^2)   \\
f(r) &= 1 + \frac{\pi N l_p^3}{r^3} \nonumber
\end{align} 
(in the convention that the tension of 2-branes, in terms of which excitations of the 5-brane are understood instead of open strings, is $T_2 = 1/((2 \pi)^2 l_p^3$). There is also a 4-form flux of $N_c$ units on the $S^4$. 
Near the horizon, $r \rightarrow 0$, the geometry is of the form AdS$_7 \times S_4$. Similar arguments as for the AdS/CFT correspondence lead to the conjecture \cite{Maldacena:1997re} that in he large $N_c$ limit, M-theory on AdS$_7 \times S_4$ is dual to the low-energy field theory living on the M5-branes' worldvolume, namely a 6-dimensional $\mathcal N=(2,0)$ superconformal theory (SCFT). 
Upon a first compactification on 
a circle $C^1$ with radius $R_1$ and supersymmetry preserving boundary conditions for the fermions, one obtains at low energies a (4+1)-dimensional $SU(N_c)$ SYM theory on the worldvolume of extremal D4-branes, with correspon\-ding dual the near-horizon geometry of the (type IIA supergravity) extremal 4-brane solution. 
A second compactification on a circle $C^2$ with radius\footnote{Because the 5-dimensional coupling constant $g_5^2 \sim g_s l_s \sim  
R_1$ (from the standard identification between M-theory and IIA string theory upon compactification on a circle of 10-dimensional radius $R_1$) 
and the 4-dimensional one $g_{YM}^2 = \frac{g_5^2}{2 \pi R_2} \sim \frac{R_1}{R_2}$ (from comparing coefficients of $F^2$ in the expanded DBI action: $\int d^5 x \frac{1}{g_5^2} \rightarrow \int d^4 x \frac{2 \pi R_2}{g_5^2} \Rightarrow \frac{1}{g_{YM}^2} = \frac{2 \pi R_2}{g_5^2}$), the relation between the 't Hooft coupling $\lambda = g^2_{YM} N_c$ and the radii is $R_1 \sim \frac{\lambda R_2}{N_c}$, such that the condition $R_1 \ll R_2$ corresponds to the limit $N_c \rightarrow \infty$ with $\lambda$ fixed. 
\label{D4par}
} 
$R_2 \gg R_1$ and anti-periodic boundary conditions for the fermions is necessary to break supersymmetry, and one ends up with a $(3+1)$-dimensional pure QCD-like theory. 

The pure QCD-like boundary theory is interpreted to be dual to supergravity in the geometry which results from performing the two compactifications 
in the bulk. This turns out to be the near-horizon geometry of a non-extremal 4-brane solution. We try to provide some intuition for this result. 
On the supergravity side it is more convenient to do the second compactification first \cite{Witten:1998zw}, upon which the original near-horizon AdS$_7 \times S_4$ metric is replaced by a Euclidean\footnote{This is the first Wick rotation, to Euclidean ``time" $\tau$.} 
AdS$_7$-Schwarzschild $\times S_4$ geometry 
with a ``temperature" scale $\sim 1/R_2 =  M_K$ 
giving the mass scale $M_K$ of the Kaluza-Klein modes. (This should be more clear after reading sections \ref{QFTfiniteT} and \ref{AdSfiniteTsection}, where it is explained that compactification of Euclidean time in the boundary field theory corresponds to turning on temperature in the field theory, and to the addition of a Schwarzschild black hole in the bulk.) 
The Euclidean AdS$_7$-Schwarzschild $\times S_4$ metric can be recognized as the near-horizon region of non-extremal M5-branes. 
Compactifying further on $C^1$, gives the near-horizon geometry of the non-extremal 4-brane solution, where we Wick-rotate back to Lorentzian signature, hereby taking one of the $x_i$ coordinates as time $t$. The result is the \emph{D4-brane background}
\begin{align} 
ds^2 &= \left(\frac{u}{R}\right)^{3/2} (\eta_{\mu\nu}dx^\mu dx^\nu + f(u)d\tau^2) + \left(\frac{R}{u}\right)^{3/2} \left( \frac{du^2}{f(u)} + u^2 d\Omega_4^2 \right), \label{D4} \\
e^\phi &= g_s \left(\frac{u}{R}\right)^{3/4} \hspace{2mm}, \quad   \label{ephiD4}\\
F_4 &= \frac{2\pi N_c}{V_4}\epsilon_4 \hspace{2mm}, \quad f(u) = 1-\frac{u_K^3}{u^3}, \quad R^3 = \pi g_s N_c l_s^3, \label{F4D4}
\end{align}
with $u_K$ the degree of non-extremality. 
It is the doubly Wick-rotated near-horizon geometry of a near-extremal  (horizon $\rightarrow$ singularity) type IIA 4-brane solution, or the deformation of spacetime at $g_s N_c \gg 1$ of $N_c$ D4-branes located at $u=0$, with $u$ the radial coordinate in the 56789-directions tranverse to the brane, and stretching out in 4 non-compact directions $x^\mu (\mu=0, ..., 3)$ and 1 compact direction $\tau$.  
$d\Omega_4^2$ and $\epsilon_4$ respectively are the line element and volume form of a  $SO(5)$ invariant unit 4-sphere and  $V_4 = 8\pi^2/3$ its volume. The 6-form RR field strength $F_6$ generates a flux 
$\frac{1}{2\pi} \int_{S^4} F_4 = N_c$ 
through the 4-sphere, with $F_4  = * F_6$ the Hodge dual 4-form field strength.

\begin{figure}[h!]
  \centering
  \scalebox{0.4}{
  \includegraphics{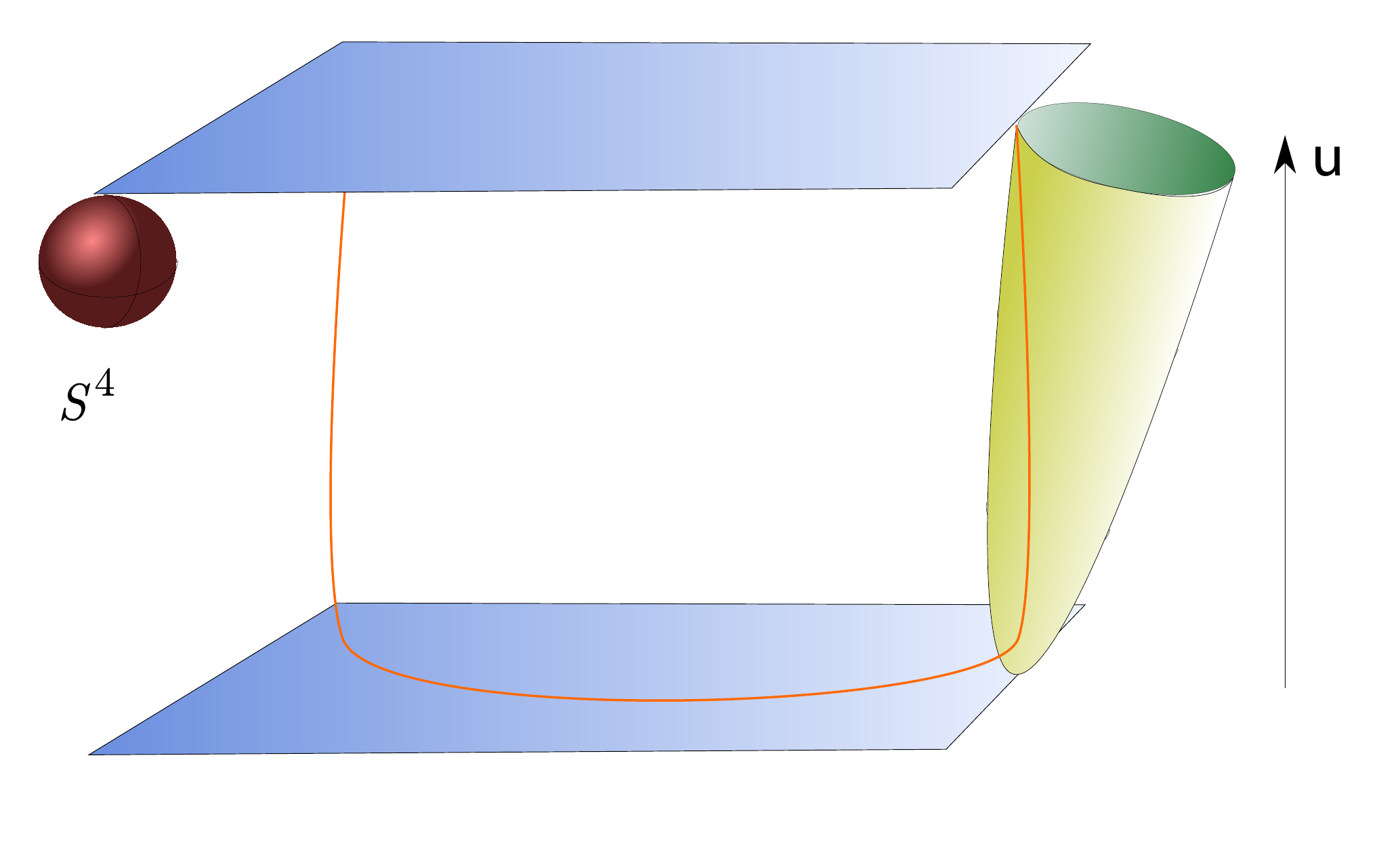}}
  \caption{The geometry of the D4-brane background (\ref{D4}): spacetime ends smoothly at the tip of the cigar ($u=u_K$). This natural cutoff leads to the formation of a $q \bar q$ flux tube in the dual field theory, represented by   the orange string. 
The cigar-shaped subspace provides both a compact dimension for supersymmetry breaking and a low-energy cut-off scale to model confinement \cite{Peeters:2007ab}.} \label{peetersfig7}
\end{figure}

The most important characteristic of the D4-brane background, of which an impression is shown in figure  \ref{peetersfig7}, 
is the cigar-shaped subspace formed by the $\tau$-circle shrinking as $u$ decreases, till it disappears at $u=u_K$. 
To avoid a conical singularity at $u=u_K$, the metric $ds^2_{(u,\tau)}$ should take the form of a 2-dimensional plane there, which uniquely determines the period $\delta \tau$ of $\tau$ to be\footnote{
From the demand 
\begin{equation}
 ds^2_{(u,\tau)} = F(u) d\tau^2 + F(u)^{-1} du^2
 \stackrel{\text{at }u=u_K}{=} r^2 d\theta^2 + dr^2,
\end{equation}
with $\theta$ periodic with period $2\pi$, the conditions 
(at $u=u_K$) $dr = F^{-1/2} du$ and $r d\theta = F^{1/2} d\tau$ can be combined to $d(F^{1/2}) \frac{d\tau}{d\theta} = F^{-1/2} du$,  uniquely determining $\delta \tau$ of $\tau$: 
\begin{equation} \label{omtrekalgemeen}
\delta \tau = \frac{4\pi}{\left.F'(u)\right|_{u=u_K}}, 
\end{equation} 
with $F(u) = \left(\frac{u}{R}\right)^{3/2} f(u)$. 
}
\begin{equation} \label{omtrek}
\delta \tau 
= \frac{4\pi}{3} \frac{R^{3/2}}{u_K^{1/2}}. 
\end{equation} 
The geometry has a natural cutoff at $u=u_K$, providing a scale that breaks conformal invariance and modeling a mass gap and confinement in the dual field theory. 
The condition (\ref{Teff}) for a non-zero QCD string tension is satisfied:
\begin{equation} \label{tension}
\sigma 
 = \frac{1}{2\pi\alpha'} \sqrt{G_{tt}(u_{K}) G_{xx}(u_{K})}  
          = \frac{1}{2\pi\alpha'} \left(\frac{u_K}{R}\right)^{3/2} 
          \neq 0.
\end{equation}
At $u \rightarrow u_K$ the $(u, \tau)$-metric is smooth and locally $\sim \mathbb{R}^2$. 
Because $\mathbb{R}^2$ is `simply connected', every loop in the neighbourhood of $u_K$ can be continuously contracted to a point (there are no non-trivial loops). This means that a fermion cannot pick up a non-trivial phase around a circle, it 
must have anti-periodic boundary conditions along the contractible $\tau$-circle.  
The anti-periodic fermion field (from a 4-4 superstring\footnote{A string with both endpoints attached to the D4-branes is denoted a 4-4 string.}) 
thus breaks supersymmetry via dimensional reduction over the $\tau$-circle with radius $M_K^{-1}$ at $u \rightarrow \infty$. 
At energy scales lower than $M_K$, i.e.\ the mass scale of the infinite tower of Kaluza-Klein modes, one obtains an effective, 4-dimensional $U(N_c)$ Yang-Mills theory on the compactified worldvolume of the D4-branes, describing the dynamics of massless gluons. 

\subsubsection{Relations between string theory and gauge theory parameters}

The compactification scale $M_K$ is given by 
\begin{equation} \label{M}
M_K = \frac{2\pi}{\delta \tau} = \frac{3}{2}\frac{u_K^{1/2}}{R^{3/2}} 
    = \frac{3}{2\sqrt \pi} \frac{u_K^{1/2}}{(g_s N_c)^{1/2} l_s^{3/2}}. 
\end{equation}
The 't Hooft coupling constant $\lambda$ of the effective 4-dimensional field theory is 
\begin{equation} \label{verbandgYMg5}
\lambda = g^2_{YM} N_c = \frac{g_5^2 N_c}{\delta \tau}  \stackrel{(\ref{YMcoupling})}{=} \frac{(2\pi)^2 g_s l_s N_c}{\delta \tau} = 3\sqrt{\pi} \left(\frac{g_s u_K N_c}{l_s}\right)^{1/2}, 
\end{equation}
with $g_5$ the 5-dimensional coupling of the supersymmetric field theory on the D4-branes before compactification, and the relation between $g_5$ and $g_{YM}$ explained in footnote \ref{D4par} on page \pageref{D4par}. 
Inverting the above relations 
gives the expressions for the string theory parameters as a function of the gauge theory parameters :
\begin{equation} \label{relaties}
R^3 = \frac{1}{2} \frac{\lambda l_s^2}{M_K}, \quad g_s = \frac{1}{2\pi}\frac{g^2_{YM}}{M_K l_s}, 
      \quad u_K = \frac{2}{9} \lambda M_K l_s^2, 
\end{equation}      
which can be further combined to the handy relation 
$M_K^2 = \frac{9}{4} \frac{u_K}{R^3}$. 
The fundamental string length $l_s$ will disappear in every calculation of physical observables in the field theory. The QCD string tension (\ref{tension}) can for example be rewritten as  
$\sigma 
= \frac{2}{27\pi} \lambda M_K^2.$
The independence of all physical results on the choice of $\lambda l_s^2$ allows us to set $\frac{2}{9} M_K^2 l_s^2 = \frac{1}{\lambda}$ \cite{Sakai:2005yt} without loss of generality.  
It is the same as saying
\begin{equation} \label{SS2(2.4)a}
u_K = \frac{1}{M_K} 
\end{equation}
and consequently 
\begin{eqnarray} R^3 = \frac{9}{4} \frac{1}{M_K^3} \quad \mbox{and}\quad \frac{1}{g_s l_s^3} = \frac{4\pi}{9} N_c M_K^3. 
\end{eqnarray}

\subsubsection{Validity regime of the supergravity approximation} \label{Geldigheidsregimevandesupergravitatiebenadering}

For the supergravity approximation to be valid, the curvature radius of the background has to be much larger than the fundamental string length, $\alpha'/R^2 \ll 1$. 
The maximal curvatures in the D4-brane background are reached near  the tip of the cigar ($u\approx u_K$): the Ricci scalar associated with the ($u,\tau$)-metric becomes of the order $(u_K R^{3})^{-1/2}$ there, just as the curvature ($\sim$ 1/radius$^2$) of the 4-sphere $S^4$ with radius $(R^{3/2}u^{1/2})^{1/2}$. 
We therefore demand that
\begin{equation} \label{geldigheidseissugra}
\frac{u_K^{1/2}R^{3/2}}{l_s^2} \stackrel{(\ref{relaties})}{\simeq} g^2_{YM} N_c \gg 1.
\end{equation}
The restriction to weak curvature of the geometry thus corresponds to large 't Hooft coupling in the effective 4-dimensional gauge theory, just as in the AdS/CFT-correspondence. 
The classical supergravity approximation is moreover only justified when the local string coupling $e^\phi$ is small enough to suppress quantum loop effects. Using (\ref{ephiD4}) and (\ref{relaties}) the demand $e^\phi \ll 1$ leads to
\begin{equation}
u \ll u_{crit} \simeq \frac{N_c^{1/3}M_K l_s^2}{g^2_{YM}},
\end{equation}
with $u_{crit} \gg u_K$ satisfied when $\frac{1}{g^2_{YM} N_c} \gg g^4_{YM}$. 
In conclusion the classical supergravity analysis of the D4-brane background is reliable when
\begin{equation}
 g^4_{YM} \ll \frac{1}{g^2_{YM} N_c} \ll 1.
\end{equation}
This is precisely the strong coupling regime of the 4-dimensional gauge theory in the 't Hooft limit 
\begin{equation}
g_{YM} \rightarrow 0, \quad N_c \rightarrow \infty, \quad g^2_{YM} N_c \gg 1, 
\end{equation} 
(which was also obtained from the reasoning in footnote \ref{D4par} on page \pageref{D4par}). 
From the expressions (\ref{relaties}), with the choice $l_s^2 \sim \frac{1}{\lambda}$, we can write 
\begin{equation}
g_s = \frac{1}{2\pi} \frac{g^2_{YM}}{M_K l_s} \propto \frac{\lambda^{3/2}}{N_c}  \quad \text{and} \quad \frac{\alpha'}{R^2} \propto \frac{1}{\lambda}
\end{equation} 
such that the quantum loop expansion and the $\alpha'$-expansion of perturbative string theory (discussed in section \ref{sectionstringexp}) correspond respectively to the expansion in $\lambda^{3/2}/N_c$ and $1/\lambda$ in the dual gauge theory. 

In the AdS/CFT correspondence, there is no ambiguity concerning the identification of the constant 't Hooft coupling in (\ref{2.2.8}) in the scale-invariant boundary field theory. Here, while the pure QCD-like coupling runs, the coupling 
\begin{equation}
\lambda \simeq \frac{M_K R^3}{l_s^2}
\end{equation}
(from the first equation in (\ref{relaties})) is again constant. 
It is interpreted as the bare coupling at the UV cutoff scale $\Lambda = M_K$ (beyond which extra scalars and fermions appear) in the dual field theory \cite{Gross:1998gk}: 
\begin{equation}
\lambda = \lambda_R(\Lambda=M_K) = \frac{1}{\beta_0 \ln(M_K/\Lambda_{QCD})}, 
\end{equation}
where the last equality, following from eq.\ (\ref{couplingLambda}) with the relevant coupling at large $N_c$ the 't Hooft coupling (see footnote \ref{largeNfootnote} on page \pageref{largeNfootnote}), is only perturbatively valid for small $\lambda$ or $M_K \gg \Lambda_{QCD}$. 
To make contact with (continuum) QCD we should be looking at the limit of weak  
bare coupling $\lambda \ll 1$ (at $\Lambda \rightarrow \infty$), which is the opposite regime of the classical supergravity dual valid at strong coupling $\lambda \gg 1$. This implies $M_K  \gg \Lambda_{QCD}$ is not true in the supergravity limit. 
It is useful to compare the situation with 
lattice QCD, to see that indeed $M_K$ is actually of the order of $\Lambda_{QCD}$. 
The UV cut-off $M_K$ is analogous to the UV cut-off $\Lambda=1/a$ in lattice QCD (with $a$ the lattice spacing). There, the bare coupling has to be tuned when changing $\Lambda$ to make sure that IR physical observables, such as for example the rho meson mass $m_\rho$, remain the same. The dependence of the bare coupling on the UV cut-off, $\lambda(M_K)$, is thus determined by demanding that (e.g.)\ $d m_\rho/d M_K = 0$ for $m_\rho = M_K f(\lambda)$. The result for $f(\lambda) = c + \frac{c'}{\lambda} + \cdots$ is only known to be constant at leading order in the large coupling expansion. Subleading terms would result from taking $\alpha'$-corrections in string theory into account.  Dimensional analysis imposes $\Lambda_{QCD} = M_K  f(\lambda)$. From  $\Lambda_{QCD} \approx m_\rho$, to leading order in the supergravity approximation it follows that $\Lambda_{QCD}$ is of the order of $M_K$. 

\begin{figure}[h!]
  \centering
  \scalebox{0.5}{
  \includegraphics{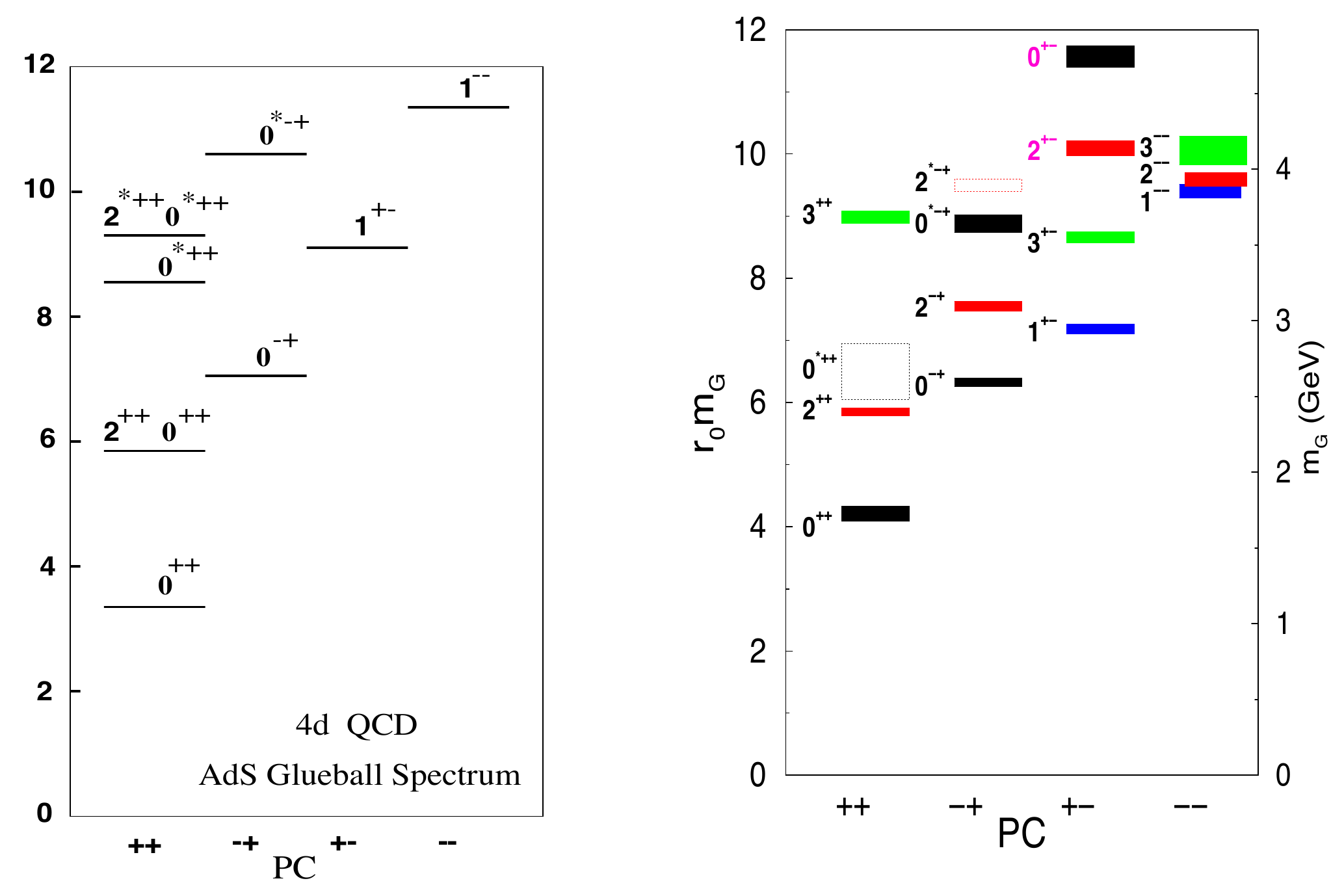}}
  \caption{The holographically obtained glueball spectrum for 4-dimensional QCD in the strong coupling regime (left) in comparison to the lattice spectrum for pure $SU(3)$ QCD (right) (in units $1/r_0 = 410$ MeV)   
  \cite{Brower:2000rp}.}\label{glueballspectrumfig2}
\end{figure}

Indeed, unfortunately the mass scale of QCD states (such as mesons and glueballs) turns out to be of the order $M_K$ (the lowest glueball masses correspond to the zero modes of closed strings and their mass is proportional to $1/R_2=M_K$) 
such that the Kaluza-Klein modes don't actually decouple. They should in the small bare coupling limit because they reflect physics of higher dimensions, but this is technically indescribable in the dual string regime. 
In \cite{Hashimoto:1998if} 
the lowest glueball masses, calculated in the D4-brane background in units of $M_K$ at large $\lambda$, are compared to lattice QCD results at a `correspondence point' in parameter space, $\lambda$ finite and $M_K \approx \Lambda_{QCD} \approx 200$ MeV, where neither the supergravity calculation nor the lattice one can be trusted, and the results are surprisingly good. The comparison for the full glueball spectrum \cite{Brower:2000rp} 
 is shown in figure \ref{glueballspectrumfig2}. The glueball masses are obtained in QCD by computing correlation functions of gauge invariant local operators, such as Tr $F^2$ for the lightest glueball, 
and looking for particle poles. On the supergravity side, you have to identify the supergravity fluctuations that couple to these operators, which is roughly the dilaton $\phi$ for Tr $F^2$ (there is a subtlety in disentangling the mixing between the dilaton and volume factor fluctuations, in general the diagonalization of the fluctuation equations can be very involved). 
 
The fact that the dilaton $\phi(u)$ runs as a function of the RG scale $u$ refers to the running of the coupling in the dual field theory, which is not conformal. It is not clear how to actually identify the beta function $\beta(g_{YM})$, but what can be calculated holographically is $\langle T_\mu^\mu \rangle$ in eq.\ (\ref{betaTrF2}). For the D4-brane background it is shown in \cite{Kanitscheider:2008kd} to be non-zero (no scale invariance) 
and constant. 
Where it usually relies 
on the asymptotically AdS nature of the geometry, the holographic renormalization performed in \cite{Kanitscheider:2008kd} to obtain $\langle T_\mu^\mu \rangle$ relies on the underlying `generalized'  
conformal structure of the -- non-renormalizable because of its dimensionful coupling $g_5^2 \sim g_s l_s$ -- D4-brane background, inherited from its UV-completion. 
The non-conformal D4-brane background is not asymptotically AdS, but asymptotically AdS up to a conformal factor\footnote{ 
To see this, switching to the coordinate $\rho = \sqrt u$ gives 
\begin{equation}
ds^2 = \frac{\rho}{R^{3/2}} \left[ \rho^2\left(\eta_{\mu\nu}dx^\mu dx^\nu + f(\rho)d\tau^2\right) + 4R^3 \frac{d\rho^2}{f(\rho) \rho^2} 
+ R^3 d\Omega_4^2 \right].
\end{equation}
At $u \rightarrow \infty$, $f(\rho) \rightarrow 1$ and one finds the metric of  $AdS_6 \times S^4$ (in coordinates (\ref{adspetersen})) up to a conformal factor $\frac{\rho}{R^{3/2}}$: 
\begin{equation}
ds^2_{AdS_6 \times S^4} = b^2 \frac{d\rho^2}{\rho^2} + \rho^2 dx_\mu^2 + b_4^2 d\Omega_4^2, 
\end{equation}
with AdS radius $b=2 R^{3/2}$ and $S^4$-radius $b_4 = R^{3/2}$. 
}, i.e.\ the dual field theory does not flow to a (5-dimensional) UV fixed point. However, in the far UV (at energy scales high enough to resolve $R_1$) an M-theory dimension opens up and the 11-dimensional M-theory defined on an  AdS$_7 \times S_4$ background appears. The corresponding UV-completion of the dual field theory is a 6-dimensional superconformal field theory, i.e.\ the theory flows to a 6-dimensional UV fixed point.

\subsection{Limitations of the model and other approaches}  \label{limitationsbottomup} 

The statement is that in the limits $N_c \rightarrow \infty$ and $g^2_{YM} N_c$ fixed and large, classical supergravity in the D4-brane background is dual to a 4-dimensional $U(N_c)$ gauge theory which is pure QCD at energies lower than the compactification scale $M_K$, but diverges from QCD at the mass scale $M_K$ of the extra massive modes. 
Because the mass scale of typical QCD states is also of order $M_K$, the holographic  
theory does contain an infinite tower of Kaluza-Klein modes and is therefore ``QCD-like" rather than QCD. 
Due to the limitation of the supergravity approximation, all holographic QCD models suffer from the fact that they substantially differ from QCD in the UV, where they are strongly coupled. 
The IR behavior can however be very similar to that of QCD. 

The followed approach to holographic QCD, where the duality is derived from string theory, like the AdS/CFT-correspondence (with dual backgrounds (\ref{3branesol}) and (\ref{D4}) solutions of the supergravity action (\ref{sugraaction})), is referred to as \emph{top-down.} The advantage is that it includes all gauge theory operators. (QCD has an infinite number of operators with any set of quantum numbers. In general these operators mix and, in principle, an arbitrarily large subset of them can contribute to any given process \cite{Cohen:2008hd}.) The disadvantage is that it also includes non-QCD operators associated with the higher-dimension physics encoded in the Kaluza-Klein modes.  

In the \emph{bottom-up} approach, to avoid Kaluza-Klein modes, one starts by guessing a 5-dimensional background. Next, one chooses the field content (for example in the simplest models one keeps the bulk fields that correspond to boundary operators of UV dimension 3, hoping the higher-dimensional ones dual to more massive fields can be safely neglected). With these ingredients an action is built, that typically uses 5-dimensional masses obtained from the AdS/CFT dictionary (relating the masses to the engineering dimensions of the QCD operators -- which is not a very well-justified assumption when it comes to QCD operators that receive anomalous dimension corrections  \cite{Cohen:2008hd,Alvares:2011wb}). 
Often, parameters of the model are fixed by matching to QCD in the UV (where the dual is usually asymptotically AdS), based on matching to QCD in its weak coupling regime -- hereby extending the validity of the AdS/CFT duality to weak coupling, though it is actually restricted to the strong coupling regime. The difficulty here is that there is no region where QCD is both conformally invariant and strongly coupled. 
The simplest bottom-up holographic QCD models are hard wall and soft wall AdS, with respectively a hard cut-off and a smooth cut-off by tuning a dilaton field. An advantage is that calculations are 
 less involved than in top-down, because of the inherent freedom in the approach. More advanced bottom-up models can for example be found in \cite{Gursoy:2007cb,Gursoy:2010fj,Gubser:2008ny}.

\section{Incorporating flavour degrees of freedom} 

Karch and Katz \cite{Karch:2002sh} proposed to incorporate flavour degrees of freedom in holographic models by introducing $N_f$ `flavour branes'. Strings that are attached to the stack of flavour branes with one end, and to the stack of colour branes (at $u=0$ in the background geometry) with the other, represent fermions in the fundamental representation of $U(N_c)$, appearing in $N_f$ flavours. Strings with both endpoints on the flavour brane correspond to mesonic degrees of freedom in the field theory. 

For the backreaction of these new branes on the gravitational background to be negligible, the number of flavour branes has to be much smaller than the number of colour branes $N_f \ll N_c$. This is the so-called `probe approximation' and the flavour branes serve as probe branes in the background. In lattice QCD this limit is known as the \emph{quenched approximation}: the full dynamics of the gluons and their effect on the fermions is taken into account, but the reaction of the fermions on the gluons (and themselves) is neglected.

\chapter{The Sakai-Sugimoto model}  \label{SSMchapter}

\section*{D4/D8/$\overline{\text{D8}}$-configuration}

The holographic QCD-model proposed by Sakai and Sugimoto \cite{Sakai:2004cn,Sakai:2005yt}
consists of the D4-brane background (\ref{D4}), 
\begin{align} 
ds^2 &= g_{mn} dx^m dx^n \quad (m,n= 0 \cdots 9) \nonumber\\
&= \left(\frac{u}{R}\right)^{3/2} (\eta_{\mu\nu}dx^\mu dx^\nu + f(u)d\tau^2) + \left(\frac{R}{u}\right)^{3/2} 
\left( \frac{du^2}{f(u)} + u^2 d\Omega_4^2 \right), \label{D4SS}
\end{align}
\begin{equation} \label{gsSS}
e^\phi = g_s \left(\frac{u}{R}\right)^{3/4} \hspace{2mm}, \quad 
F_4 = \frac{N_c}{V_4}\epsilon_4 \hspace{2mm}, \quad f(u) = 1-\frac{u_K^3}{u^3},
\end{equation}
to which $N_f$ probe D8-branes and $N_f$ $\overline{\text{D8}}$-branes (i.e.\ D8-branes with opposite orientation) are added in order to model  quarks and antiquarks in the probe approximation $N_f \ll N_c$.
The brane configuration consists of $N_c$ D4-branes compactified on a supersymmetry breaking $\tau$-circle $S^1$ and $N_f$ D8-$\overline{\text{D8}}$ pairs perpendicular to this circle: 
\begin{equation} \label{config}
\begin{array}{ccccccccccc}
 D4: \quad & 0 & 1 & 2 & 3 & (4) & - & - & - & - & -   \\
 D8\text{-}\overline{D8}: \quad & 0 & 1 & 2 & 3 & (-) & 5 & 6 & 7 & 8 & 9 
\end{array} 
\end{equation}
with coordinates $x^M$ ($M=0, ..., 9$) $= x^\mu$ ($\mu=0, ..., 3$), $\tau$, $u$, $\Omega_4$.

\section{Weak coupling regime: open string spectrum of  D4/$\-$D8/$\-$$\overline{\text{D8}}$ 
system} \label{subs:spectrum}

\begin{figure}[h!]
  \centering
  \scalebox{0.5}{
  \includegraphics{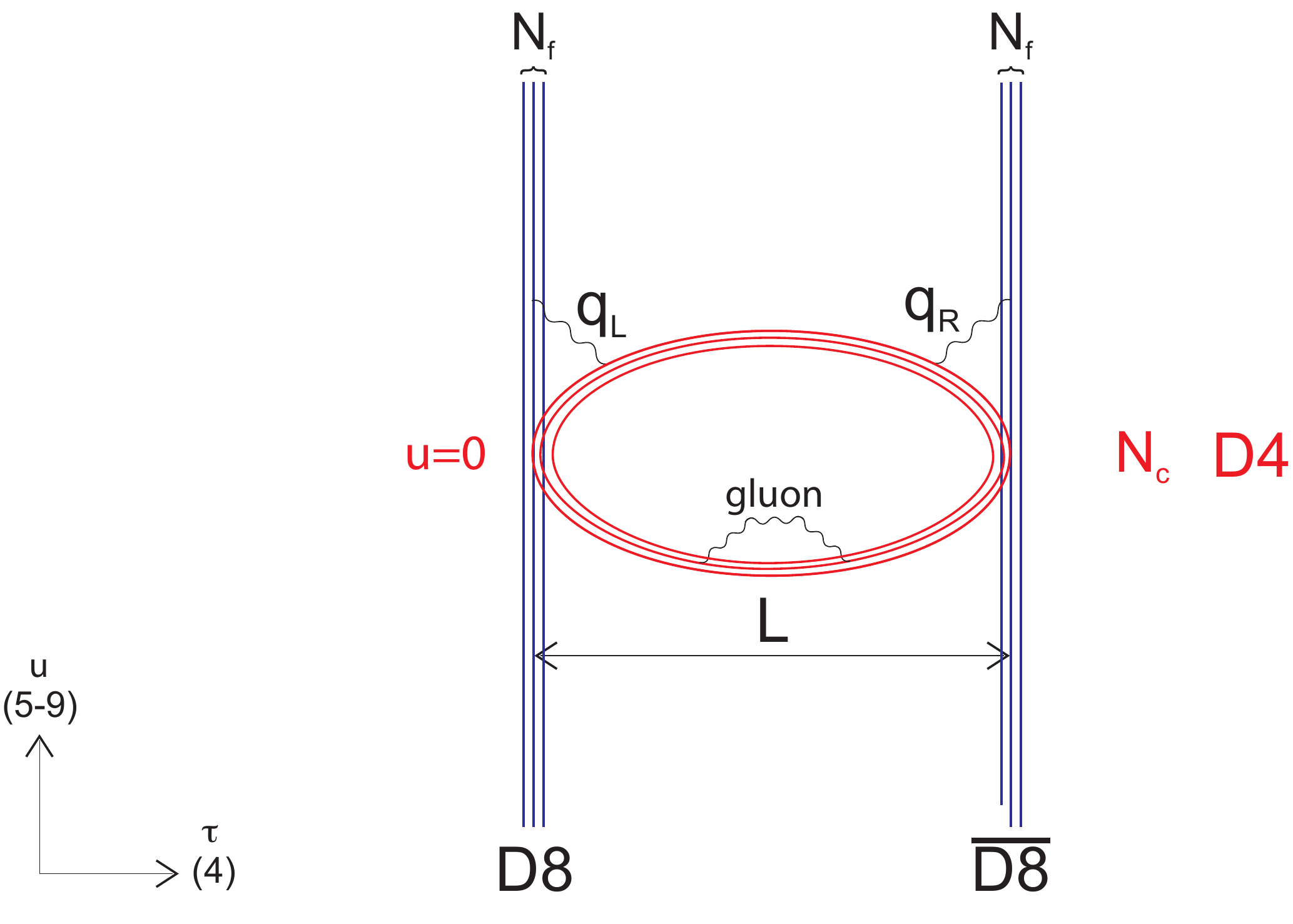}}
  \caption{The Sakai-Sugimoto D4/D8/$\overline{\text{D8}}$ configuration in the weak coupling regime, consisting of $N_f$ D8- and $N_f$ $\overline{\text{D8}}$-branes which cross the $N_c$ D4-branes, wrapped around the $\tau$-circle $S^1$.  4-4, 4-8 and 4-$\bar 8$ strings respectively represent gluons, left-handed (massless) quarks $q_L$ and right-handed (massless) quarks $q_R$   
  \cite{Antonyan:2006vw}.}\label{NJLfig1}
\end{figure}

We discuss the open string spectrum of the D4/D8/$\overline{\text{D8}}$ configuration in the weak coupling regime $g_sN_c \ll 1$.
Strings that stretch between a D$p$-brane and a D$q$-brane are referred to as $p$-$q$ strings. 

The massless fields generated by 4-4 superstrings belong to the multiplet of the 5-dimensional $U(N_c)$ $\mathcal{N}=4$ SYM with 16 symplectic supercharges (the dimensional reduced version of 10-dimensional $U(N_c)$ $\mathcal{N}=1$ SYM): the gauge field $A_M$ ($M=0, ..., 4$), scalar fields $\phi^i$ ($i=5, ..., 9$) and fermions $\psi$, all in the adjoint representation of the gauge group $U(N_c)$.
After compactification on $S^1$, with antiperiodic boundary conditions for the fermions on the circle,
the only remaining massless modes of the 4-4 strings are the gauge field $A_\mu^{(D4)}$ ($\mu=0, ..., 3$), and the fields $a_4 = \text{Tr}[A_4^{(D4)}]$ and $\Phi^i = \text{Tr}(\phi^i)$ (with Tr the trace over $U(N_c)$ indices). The $U(N_c)$ singlets $a_4$ and $\Phi^i$ cannot obtain mass through  
one loop diagrams in the $U(N_c)$ field theory (in contrast to  $A_4^{(D4)}$ and $\phi^i$).

The mass of the ground state of a 8-$\bar 8$ superstring depends on the separation $L$ between D8- and $\overline{D8}$-branes along the circle \cite{polchinskiboek}: 
\begin{equation}
m^2 = \left(\frac{L}{2\pi\alpha'}\right)^2 - \frac{1}{2\alpha'}.
\end{equation}
The corresponding field can be made massive by choosing $L > \sqrt 2 \pi l_s$. In this way one no longer has to take this mode into account in the analysis of the D8-brane action in the low energy limit.  
On the D8($\overline{D8}$)-branes lives a $U(N_f)_{D8}$ $(U(N_f)_{\overline{D8}})$ gauge field originating from the 8-8 
($\bar 8$-$\bar 8$) strings.

The zero point energy 
$a$ of the 4-8 and 4-$\bar 8$ superstrings is given by \cite{johnson} 
\begin{equation}
a^R = 0, \quad a^{NS} = -\frac{1}{2}+\frac{\nu}{8} = \frac{1}{4} \neq 0
\end{equation}
for respectively the R-sector and the NS-sector, 
with $\nu$ the number of ND plus DN coordinates (in the notation D $=$ Dirichlet and N = Neumann boundary condition for begin- and endpoint of the string) equal to 6 in the configuration (\ref{config}) at hand.
Given that $a^{NS}\neq 0$ there are no massless string states in the NS-sector so no massless bosons. 
There are however massless string states in the R-sector ($a^R = 0$). They correspond to massless fermions in spacetime, belonging to spinor representations of the Lorentz group $SO(3,1)$ on the compactified worldvolume of the D4-branes. The GSO-projection (necessary to obtain supersymmetry in spacetime) reduces the number of Dirac spinor components by demanding that they are chiral. 
The physical states created by the 4-8 (4-$\bar 8$) strings represent spinors with positive (negative) chirality. These $N_f$ massless fermions are interpreted as quarks $q_L^f$ $\left(q_R^{\bar f}\right)$ in the fundamental representation of the $U(N_c)$ gauge group.
Since the 4-8 and 4-$\bar 8$ fermions have opposite chirality, the $U(N_f)_{D8} \times U(N_f)_{\overline{D8}}$ gauge symmetry of the 
$N_f$ D8-$\overline{D8}$ pairs can be interpreted as the  $U(N_f)_{L} \times U(N_f)_{R}$
chiral symmetry of QCD. 

\begin{table}[h!]
\begin{center}
\begin{tabular}{c|cccc}
	\hline
	\hline
	veld & $U(N_c)$ & $SO(3,1)$ & $SO(5)$ & $U(N_f)_{L} \times U(N_f)_{R}$ \\
	\hline
	$A_\mu^{(D4)}$ & adj. & 4 & 1 & (1,1) \\
	$a_4$ & 1 & 1 & 1 & (1,1) \\
	$\phi^i$ & 1 & 1 & 5 & (1,1) \\
	\hline
	$q_L^{f}$ & fund. & $2_{+}$ & 1 & (fund.,1) \\
	$q_R^{\bar f}$ & fund. & $2_{-}$ & 1 & (1,fund.) \\
	\hline
\end{tabular}
\end{center} \caption{The massless fields in D4/D8/$\overline{\text{D8}}$; with adj. and fund. respectively short for adjoint and fundamental representation, and $2_{+}$ and $2_{-}$ for the positive and negative chiral spinor representation of  $SO(3,1)$. } \label{veldenTABEL}
\end{table}

The massless fields in the D4/D8/$\overline{\text{D8}}$ configuration are summarized in table \ref{veldenTABEL}, where the transformations of the fields under the gauge group $U(N_c)$, the Lorentz group $SO(3,1)$ on the D4-brane, the $U(N_f)_{L} \times U(N_f)_{R}$ chiral symmetry, and the $SO(5)$-sym$\-$me$\-$try rotating the $x^5$ to $x^9$ coordinates are specified. Kaluza-Klein modes with $S^4$ quantum numbers have no equivalent in QCD, which does not contain a $SO(5)$-sym$\-$me$\-$try, and should therefore decouple. In the analysis of the meson spectrum we will only consider $SO(5)$-sin$\-$glet modes. 

We are left with 4-dimensional $U(N_c)$ QCD with $N_f$ flavours and the $U(N_f)_{L} \times U(N_f)_{R}$ chiral symmetry manifest. 
Indeed, because the D4-branes cross the D8- and $\overline{\text{D8}}$-branes in the weak coupling regime, bare quark masses are zero and the Sakai-Sugimoto model is dual to  
QCD \emph{in the chiral limit}. 
A non-zero bare quark mass would break the chiral symmetry explicitly: left-handed quarks living at the intersection point of the $N_c$ D4-branes and the $N_f$ D8-branes (see figure \ref{NJLfig1}) would have to mix with their right-handed counterparts at the intersection of the $N_c$ D4-branes and the $N_f$ $\overline{\text{D8}}$-branes. To introduce a bare quark mass in the model, a connection between the D8- and $\overline{\text{D8}}$-branes would have to be realized, and this 
in the D-brane configuration in the flat spacetime background. The first attempts at incorporating massive bare quarks in the Sakai-Sugimoto model were based on `throat'-configurations, 
where D8- and $\overline{\text{D8}}$-branes are connected by a tube. The D4-branes are placed in the tube such that they no longer cross the D8- and $\overline{\text{D8}}$-branes in the flat geometry. Other possible mechanisms to introduce bare quark masses are discussed in \cite{Hashimoto:2007fa,Hashimoto:2008sr,Bergman:2007pm}. 
For example, in \cite{Bergman:2007pm} the bifundamental `tachyon'-field connecting D8- and $\overline{\text{D8}}$-branes is taken into account. 
We will not consider these possible extensions of the Sakai-Sugimoto model  
in this work, for reasons of simplicity.

\section{The duality}

Pure 4-dimensional $U(N_c)$ QCD, living on the compactified worldvolume of $N_c$ D4-branes at low e\-ner\-gies, is assumed to be dual to supergravity in the D4-brane background, valid in the limits $g_{YM} \rightarrow 0$, $N_c \rightarrow \infty$  and $g_{YM}^2 N_c \gg 1$ constant.
To model chiral (massless) quarks one adds $N_f$ probe D8- and $N_f$ $\overline{\text{D8}}$-branes to the background ($N_f \ll N_c)$, such that the duality in the Sakai-Sugimoto model can be formulated as follows: 
\emph{4-dimensional massless $U(N_c)$ `QCD' is dual to supergravity in the D4-brane background in which $N_f$ probe D8- and $N_f$ $\overline{\text{D8}}$-branes are embedded (with $g_{YM} \rightarrow 0$, $N_c \rightarrow \infty$, $g_{YM}^2 N_c \gg 1$, $N_f \ll N_c$)}.   
The `QCD' in the last sentence should be replaced by `QCD-like theory', as the field theory dual to supergravity in the Sakai-Sugimoto background contains extra degrees of freedom on top of the gluons and chiral quarks: the tower of Kaluza-Klein modes resulting from compactification on $S^1$, $SO(5)$-states and extra fermionic partners\footnote{The fermionic `mesons' on the D8-branes, which have no QCD counterpart, are discussed in the appendix of  \cite{Sakai:2004cn}.} of the mesons (8-8 strings) on the D8-branes (the scale of supersymmetry breaking is of the same order as the meson mass). The hope is that the dual gauge theory belongs to the same universality class as 4-dimensional massless large-$N_c$ QCD.

\section{Strong coupling regime: probe D8-branes in D4-brane background} \label{SS4.1.2}

\subsection{Embedding of the D8-branes in the D4-brane background}

\begin{figure}[h!]
  \hfill
  \begin{minipage}[t]{.45\textwidth}
    \begin{center}
      \scalebox{0.9}{
  \includegraphics{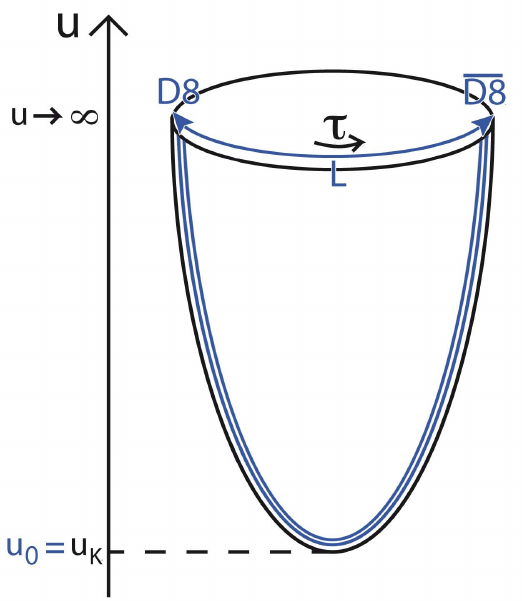}}
    \end{center}
  \end{minipage}
  \hfill
  \begin{minipage}[t]{.45\textwidth}
    \begin{center}
      \scalebox{0.9}{
  \includegraphics{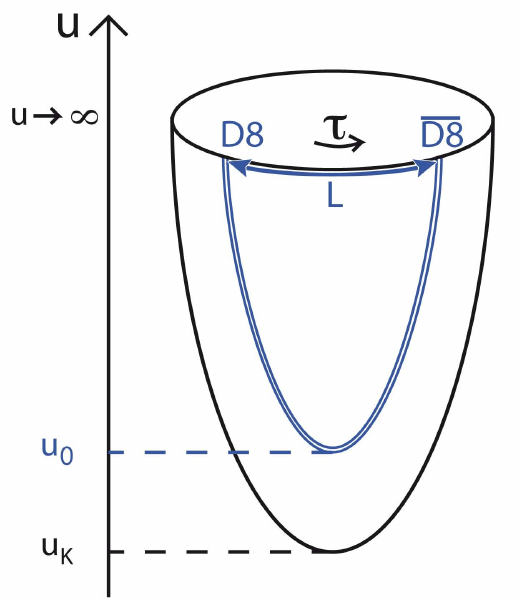}}
    \end{center}
  \end{minipage}
      \caption{The Sakai-Sugimoto model: antipodal ($u_0=u_K$) and non-antipodal ($u_0 > u_K$) embedding of the $N_f$ D8- and $N_f$ $\overline{\text{D8}}$ probe branes in the D4-brane background.}
	\label{SS}
  \hfill
\end{figure}

The embedding of the (8+1)-dimensional D8-branes in the (9+1)-dimensional D4-brane background is specified by one function only: $u=u(\tau)$. 
Close to the boundary of the  
background at $u\rightarrow \infty$, where the open string interpretation of the D-branes lives, the separation between the $N_f$ D8-branes and the $N_f$ $\overline{\text{D8}}$-branes has to be $L$. 
It is however impossible for the probe branes to end anywhere in the D4-brane background. 
We therefore expect to find a $\cup$-shaped embedding function $u(\tau)$:
the D8-branes and the  $\overline{\text{D8}}$-branes will merge smoothly ($u'(0)=0$ at $u(0)=u_0$). 
In what follows we derive the equation of motion for the D8-branes in the D4-brane background and give a dual interpretation to the solution.

The induced metric on the D8-branes is given by
\begin{equation} \label{D8inducedmetric}
ds^2_{D8} = \left(\frac{u}{R}\right)^{3/2} \eta_{\mu\nu}dx^\mu dx^\nu  +  \left( \left(\frac{u}{R}\right)^{3/2} f(u) + \left(\frac{R}{u}\right)^{3/2} \frac{u'^2}{f(u)} \right)d\tau^2  + \left(\frac{R}{u}\right)^{3/2} u^2 d\Omega_4^2,
\end{equation}
with $u' = du/d\tau$. 
The DBI-action (\ref{DBIstringannex}) for the D8-branes (with vanishing gauge field) is then proportional to
\begin{align} 
S_{DBI} &\propto  \int d^4 x d\tau \hspace{1mm} \epsilon_4 e^{-\phi}\sqrt{-\det(g_{D8})} \nonumber\\
       &\propto  \int d^4 x d\tau \hspace{1mm} u^4 \sqrt{f(u)+\left(\frac{R}{u}\right)^{3} \frac{u'^2}{f(u)}}. \label{SDBInodiginveralgemening}
\end{align}       
As the integrand $\mathcal{L}$ does not explicitly depend on $\tau$, the `Hamiltonian' $u'\frac{\partial \mathcal{L}}{\partial u'} - \mathcal{L}$, associated with translations in the $\tau$-direction, is conserved:  
\begin{equation} \label{beweg}
\frac{d}{d\tau}\left(\frac{u^4 f(u)}{\sqrt{f(u)+\left(\frac{R}{u}\right)^3 \frac{u'^2}{f(u)}}} \right)= 0 \quad 
\text{or} \quad \frac{u^4 f(u)}{\sqrt{f(u)+\left(\frac{R}{u}\right)^3 \frac{u'^2}{f(u)}}}= u_0^4 \sqrt{f(u_0)},
\end{equation}
where we assumed the initial conditions $u(0) = u_0$ and $u'(0)=0$. 
The solution of the equation of motion (\ref{beweg}) is  
given by:
\begin{equation} \label{dtaudu}
\tau(u) = \int_{u_0}^{u} \frac{du}{u'}  = \int_{u_0}^u du  \left(\frac{R}{u}\right)^{3/2} f(u)^{-1} \sqrt{\frac{u_0^8 f(u_0)}{u^8 f(u) - u_0^8 f(u_0)}}. 
\end{equation}
and the asymptotic separation $L$ (at $u\rightarrow \infty$) between D8- and $\overline{\mbox{D8}}$-branes, indicated in Figure \ref{SS}, is related to $u_0$ as
\begin{align}
L &= 2 \int_{u_0}^{\infty} \frac{du}{u'} 
\nonumber\\
 &= 2 \int_{u_0}^\infty du  \left(\frac{R}{u}\right)^{3/2} f(u)^{-1} \sqrt{\frac{u_0^8 f(u_0)}{u^8 f(u) - u_0^8 f(u_0)}}.
   \label{Lconf} 
\end{align}
In the limit $u_K \ll u_0$, the integral can be solved to $L \propto \sqrt{R^3/u_0}$.  
Large values of $u_0$ correspond to small values of $L$. In the limit of $u_0 \rightarrow \infty$ the D8-$\overline{\text{D8}}$ pairs are sent to infinity and disappear ($L \rightarrow 0$).
(It can be shown that $L(u_0)$ is a monotonically increasing function of $u_0$.)
The solution is only regular when the integral is smaller or equal to $\delta \tau/2$, for every value of $L$ there is thus only one regular configuration. A special solution is the case where the D8- and $\overline{\text{D8}}$-branes are at antipodal points on the circle, i.e. at a distance $L=\delta \tau/2$ from each other, and with minimal value $u_0=u_K$. 
It is not so clear how to interpret the strong $L$-dependence of the D8-brane configuration in the dual gauge theory, as one could expect $L$ to be irrelevant for the low-energy effective theory on the D4-brane, whose massless spectrum does not contain 8-$\bar 8$ strings. We will come back to a possible interpretation of $L$ in section \ref{D}.

The stability of the flavour brane embedding was checked in \cite{Sakai:2004cn} for $u_0=u_K$ and in \cite{Ghoroku:2009iv,Mintakevich:2008mm} for $u_0>u_K$. We will encounter them as special cases in our stability analysis of the brane embedding dependent on 
an external magnetic field in section \ref{4.2}.

\subsection{Interpretation of the D8-brane solution}   \label{interpretationD8}

For large $u$, i.e.\ at large energies in the dual field theory (where we have chiral quark degrees of freedom), the D8- and $\overline{\text{D8}}$-branes are separated, corresponding to the chiral symmetry $U(N_f)_{L} \times U(N_f)_{R}$. At low $u$, i.e.\ low energies in the dual field theory (where spectra of mesonic degrees of freedom are inconsistent with chiral symmetry in the Wigner-Weyl realization), both stacks merge continuously to one stack of  $N_f$ D8-branes, corresponding to one remaining $U(N_f)$ symmetry.
We conclude that the smooth interpolation of the D8-branes and $\overline{\text{D8}}$-branes at  $u=u_0$ (forced by the topology of the gravitational background) gives a geometrical representation of the spontaneous chiral symmetry breaking ($\chi$SB) of $U(N_f)_{L} \times U(N_f)_{R}$ to the diagonal subgroup $U(N_f)_V$ in eq.\  (\ref{chiSBlargeN}). 
For the precise identification of the elements of the global chiral symmetry, $(h_L,h_R) \in U(N_f)_L \times U(N_f)_R$, we refer to section \ref{C} further on.

\begin{figure}[h!]
  \centering
  \scalebox{0.55}{
  \includegraphics{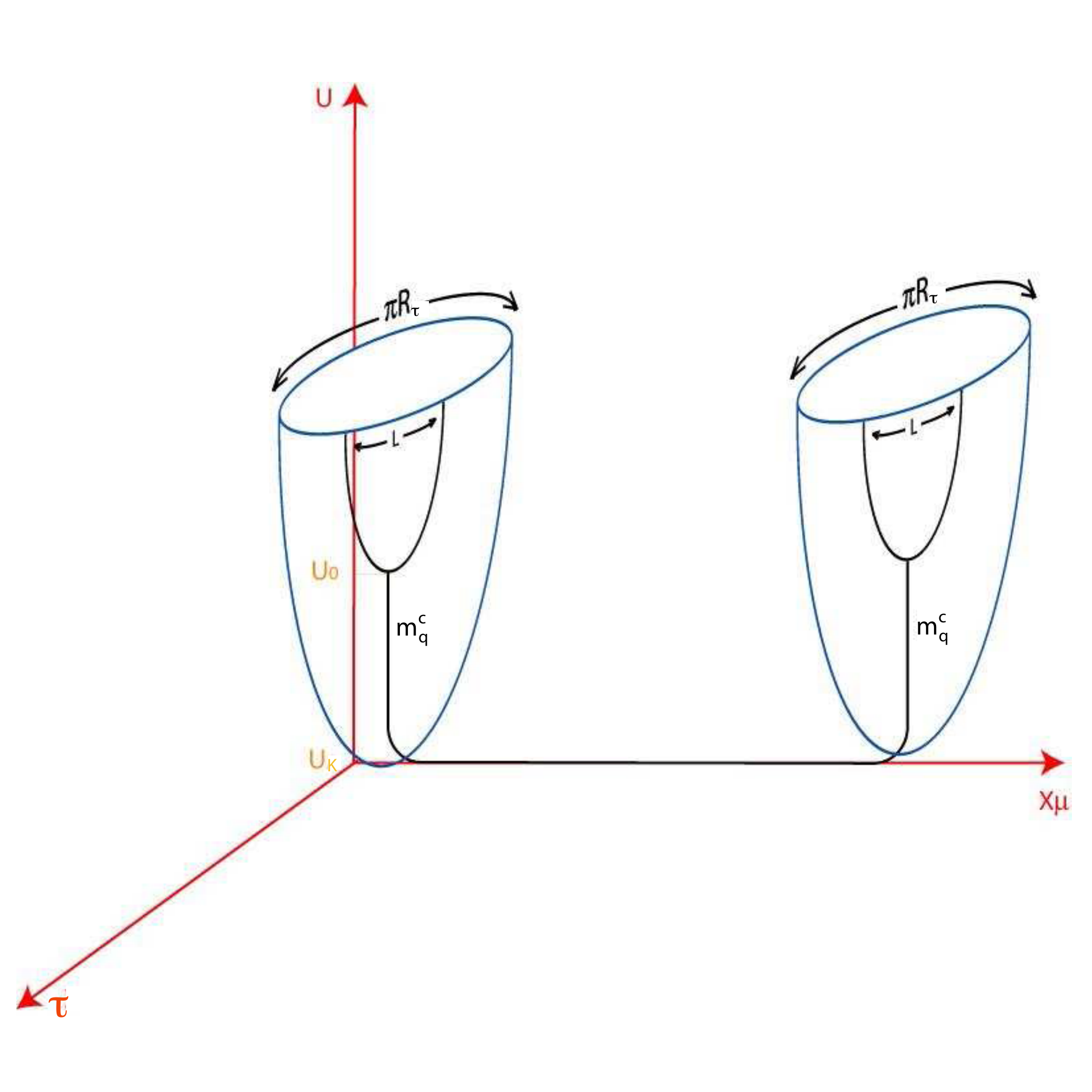}}
  \caption{A high-spin meson is represented by a macroscopic $q\bar q$ string configuration in the Sakai-Sugimoto background  
  \cite{Aharony:2006da}.}\label{finiteTfig2}
\end{figure}

There is no direct holographic description of the chiral condensate in the Sakai-Sugimoto model, related to the absence of bare quark masses as discussed in section \ref{subs:spectrum} (since $\langle \bar \psi \psi \rangle = (\delta e^{-S_{quark}}/\delta m)|_{m=0}$, the mass operator is exactly the chiral order parameter). 
In the non-antipodal embedding however, we take the approach that 
there is a quantity that can be used as an indicator for the chiral symmetry breaking order parameter\footnote{
Possible alternatives to define chiral order parameters can be found in, for example, \cite{Aharony:2008an} or \cite{Bergman:2007pm,Hashimoto:2008sr,Dhar:2007bz}.
}, namely the energy stored in a string stretching from $u_0$ to $u_K$ (the higher $u_0$, the further you move away from the chiral invariant situation of straight branes): 
\begin{equation}\label{constmass}
    m_q=\frac{1}{2\pi \alpha'}\int_{u_K}^{u_0}\frac{du}{\sqrt{f(u)}}\,\,. 
\end{equation}
This expression is obtained by considering the Nambu-Goto action (\ref{NGACTIEannex}) for a static macroscopic string, stretching in the $u$-direction of the background (\ref{D4SS}) only, in the static gauge $\tau=t$, $\sigma = x$ (with $a,b=\tau,\sigma$ and $\mathcal T$ the lifetime of the string): 
\begin{equation}
S_{NG} = -\frac{1}{2 \pi \alpha'} \mathcal T \int dx \sqrt{-\det_{ab} (\partial_a x^m \partial_b x^n g_{mn})} =   
-\frac{1}{2 \pi \alpha'} \mathcal T \int dx \sqrt{-g_{tt} g_{uu} (\partial_x u)^2} = \mathcal T \int dx \mathcal L_{NG}.  
\end{equation}
Then the corresponding energy is given by 
\begin{equation}
m_q = \int dx \left( \frac{\delta \mathcal L_{NG}}{\delta \, \partial_t u} \partial_t u - \mathcal L_{NG} \right) = -\int dx \mathcal L_{NG}
= \frac{1}{2 \pi \alpha'} \int du \sqrt{-g_{tt} g_{uu}} = (\ref{constmass}).  
\end{equation}
This energy can be understood to be related to the \emph{constituent quark mass} in the dual gauge theory by considering a high-spin meson. In contrast to low-spin mesons which are described by gauge field fluctuations on the flavour branes (with the DBI-action limited to supergravity modes), one needs to go beyond the supergravity approximation to describe high-spin mesons: they are modeled by a macroscopic string with endpoints attached to the D8-branes, as depicted in figure \ref{finiteTfig2}. This configuration is (classically) 
equivalent to a string with two massive endpoints (with mass (\ref{constmass}) in flat 4-dimensional spacetime)  \cite{Kruczenski:2004me}. 
Because part of the energy of the 4-dimensional meson comes from the horizontal part of the 5-dimensional string, the `string endpoint masses' (i.e.\ the energies of the vertical parts) can be roughly identified with the constituent quark masses for heavy quarks, but are somewhat smaller 
in the case of light quarks. 
In the following chapters we will interpret $m_q$ in (\ref{constmass}) as the (approximate) constituent quark mass, also for the light flavours. Pions remain massless for all values of $u_0$ in the Sakai-Sugimoto model, consistent with the absence of bare quark masses and the GMOR-relation (\ref{GMOR}). 
The above picture leads to the characteristic Regge relation between angular momentum $J$ and energy $M$ for rotating high-spin mesons in the Sakai-Sugimoto model \cite{Kruczenski:2004me}.

\subsection{Relation to dual low-energy effective degrees of freedom} \label{sectionrelation} 

The Sakai-Sugimoto model can be treated as an effective QCD model for the low-energy dynamics of hadrons and baryons\footnote{We will not be concerned with baryons in this thesis. Suffice it to say a baryon is represented in the Sakai-Sugimoto model by a D4-brane wrapped around the 4-sphere in the D4-brane background.}.  
The effective 4-dimensional action for mesons is obtained by integrating out the $u$-dependence (and the $S^4$-dependence, but usually assuming the D8-brane worldvolume fields to be independent of those coordinates) of the flavour branes' action, which is given by a DBI-part and a CS-part, cfr. section \ref{Dbranesection}. 

The fields living on the D8-branes are the $U(N_f)$ gauge field $A_m$ ($m=0, ..., 8$), which can be decomposed in $A_\mu$ ($\mu=0, ..., 3$), $A_u$ and $A_\alpha$ ($\alpha=5,6,7,8$ the coordinates on $S_4$), and the scalar field $\tau$. Only being interested in $SO(5)$-singlet states, we put $A_\alpha=0$ and assume $A_\mu$ and $A_u$ independent of the 4-sphere coordinates; further we assume that the remaining gauge field components can be expanded in complete sets 
$\left\{\psi_n(u)\right\}_{n\geq 1}$ and $\left\{\phi_n(u)\right\}_{n\geq 0}$ as follows \cite{Sakai:2004cn}
\begin{align}
A_\mu(x^\mu,u) &= \sum_{n \geq 1} B_\mu^{(n)}(x^\mu) \psi_n(u) = \rho_\mu(x^\mu) \psi(u) + \cdots  \label{Amuexpansion}\\
A_u(x^\mu,u) &= \sum_{n \geq 0} \phi^{(n)}(x^\mu) \phi_n(u) = \pi(x^\mu) \phi_0(u) + \cdots.  \label{Auexpansion}
\end{align} 
Then the modes $B_\mu^{(n)}(x^\mu)$ and $\phi^{(n)}(x^\mu)$ describe infinite towers of respectively (pseudo)vector and (pseu\-do)\-sca\-lar mesons in the dual field theory. The identifications of the lowest modes with mesons are given in table \ref{mesonen}. 

\begin{table}[h!]
\begin{center}
\begin{tabular}{c|cc|c}
	\hline
	\hline
	field & $P$ & $C$ & associated meson  \\
	\hline
	\hline
	$B_\mu^{(1)}$ & - & - & $\rho$  \\
	$B_\mu^{(2)}$ & + & + & $a_1(1260)$ \\
	$B_\mu^{(3)}$ & - & - & $\rho(1450)$ \\
	\hline
	$\phi^{(0)}$ & - & + & $\pi$ \\
	\hline
\end{tabular}
\end{center} \caption{Interpretation of massless fluctuations on the D8-brane as mesons in the dual QCD-like theory, with corresponding sign under parity (P) and charge conjugation (C).  
} \label{mesonen}
\end{table}

Integrating out the $u$-dependence in the DBI-action indeed gives rise to the effective 4-dimensional \emph{Proca action for vector mesons and Skyrme action for pions} \cite{Sakai:2004cn}, (this will become more clear in section \ref{B}).  
The (lowest) modes of the scalar $\tau$ were originally interpreted as scalar mesons as well \cite{Sakai:2004cn}, but this interpretation was revisited in \cite{Imoto:2010ef} and they are now cataloged as redundant (non-QCD) modes of the theory. We will come back to this in section \ref{4.2}.

\subsection{Chern-Simons action for the D8-branes}

Let us briefly discuss the CS-part of the flavour brane action and its interpretation. 
It describes the coupling of the D8-branes' gauge field to the 3-form RR potential $C_3$ associated with the 4-form RR field strength $F_4 = dC_3$ that is present in the D4-brane background (\ref{F4D4}), 
 and is given by (see eq.\ (\ref{SWZ}) and the remark in section \ref{Dbranesection} on coincident branes)  
\begin{equation} \label{SCSC3niet}
S_{CS}^{D8} = \mu \int_{\mathcal M_{D8}} C_3 \, \text{STr}(F^3), 
\end{equation} 
with $\mu = 1/(48 \pi^3)$ a normalization constant (in the normalization conventions of \cite{Sakai:2004cn}). The notation $F^3$ is short for $F \wedge F \wedge F$ with $F = dA + A^2 = dA + A\wedge A$ the 2-form field strength associated with the 1-form flavour gauge field $A = A_m dx^m$ in form notation\footnote{
A differential  $p$-form in a $D$-dimensional space is a completely antisymmetric tensor $A^{(p)}(x^\mu)$ with  $p$ indices. 
The wedge-product $\wedge$ of a $p$-form $A^{(p)}$ with a $q$-form $B^{(q)}$ is defined as  
$(A^{(p)}\wedge B^{(q)})_{\mu_1 ... \mu_{p+q}} = \frac{(p+q)!}{p!q!} A_{[\mu_1 ... \mu_p} B_{\mu_{p+1}...\mu_{p+q}]}$, 
where the square brackets denote full antisymmetrization of the indices.
The exterior derivative $dA^{(p)}$ of a  $p$-form, defined as  
$(dA^{(p)})_{\mu_1 ... \mu_{p+1}} = (p+1) \partial_{[\mu_1} A_{\mu_2... \mu_{p+1}]}$, 
$(dA^{(0)})_{\mu} = \partial_\mu f$,
creates a $p+1$-form out of a  $p$-form: $dA^{(p)} = F^{(p+1)}$.
}. 
An equivalent form of the action is 
\begin{equation} \label{SCSNc}
 S_{CS}^{D8} = \mu \int_{\mathcal M_{D8}} F_4 \omega_5(A) = \frac{N_c}{24\pi^2} \int_{\mathcal M^4 \times \mathbb{R}} \omega_5(A)  
\end{equation}
in terms of the CS 5-form 
\begin{equation}
\omega_5(A) = \text{STr} \left(A F^2 - \frac{1}{2}A^3 F + \frac{1}{10} A^5\right) 
\end{equation}
(with all wedge products implicit) constructed to satisfy  $d \omega_5 = \text{STr} \,(F^3)$,
and where $\mathcal M^4 \times \mathbb{R}$ is the 5-dimensional space parameterized by the coordinates $x^0, ..., x^3$ and $u$. In the last equality of (\ref{SCSNc}) we made use of $\frac{1}{2\pi} \int_{S_4} F_4 = N_c$ from (\ref{F4D4}). 
The variation of the CS-action $\delta S_{CS}^{D8}$ 
under an infinitesimal gauge transformation $\delta A = d\Lambda + [A,\Lambda]$ 
on the D8-branes, takes the form of the chiral anomaly $\delta \Gamma[A]$ of QCD \cite{Bilal:2008qx},   
which is calculated from triangle Feynman diagrams. The CS-term cancels the chiral anomaly on the 4-dimensional boundary $\mathcal M_4$ 
of the  5-dimensional $\mathcal M_4 \times \mathbb{R}$ space via the so-called `anomaly cancellation by inflow' mechanism \cite{Bilal:2008qx}: 
the required variation $\delta S_{CS}^{D8} = - \delta \Gamma[A]$ flows from the 5-dimensional bulk into the 4-dimensional boundary.

Sakai and Sugimoto have shown in \cite{Sakai:2004cn} that the CS action $S_{CS}^{D8}$ provides the \emph{WZW-term} in the chiral Lagrangian.

\chapter{Rho meson condensation in the Sakai-Sugimoto model} 
\chaptermark{$\rho$ meson condensation} \label{rhochapter}

At the LHC 
 very strong magnetic fields are expected to arise in non-central heavy ion collisions.  
The interest in magnetically induced QCD effects has therefore grown considerably in recent years. 
One conjectured effect of this kind is the possible instability of the QCD vacuum towards formation of an electromagnetically superconducting phase where rho mesons are condensed, in the presence of a strong (Abelian) magnetic field.  This so-called  `rho meson condensation' is also present in the holographic Sakai-Sugimoto model (SSM), as we were able to show in \cite{Callebaut:2011ab,Callebaut:2013wba} on which this chapter is based. That is, in \cite{Callebaut:2011ab} the appearance of a tachyonic instability in the SSM was observed in the simplest set-up as well as in the more involved non-antipodal embedding, used to include effects of the rho meson constituents, but in an approximated way. In \cite{Callebaut:2013wba} the remaining approximations in \cite{Callebaut:2011ab} were uplifted, resulting in a higher value for the critical magnetic field.  
Many other magnetic effects have been investigated in the Sakai-Sugimoto model, so, to avoid incompleteness, let us refer here to the review paper \cite{Bergman:2012na} for a nice overview.

\section{Rho meson condensation in QCD}

We first review the magnetically induced effect of rho meson condensation in QCD. This story begins with some motivation for the study of QCD in the presence of strong magnetic fields. 

\subsection{QCD and strong magnetic fields}   \label{QGPB}

\begin{figure}[h!]
  \hfill
  \begin{minipage}[t]{.45\textwidth}
    \begin{center}
      \scalebox{0.25}{
  \includegraphics{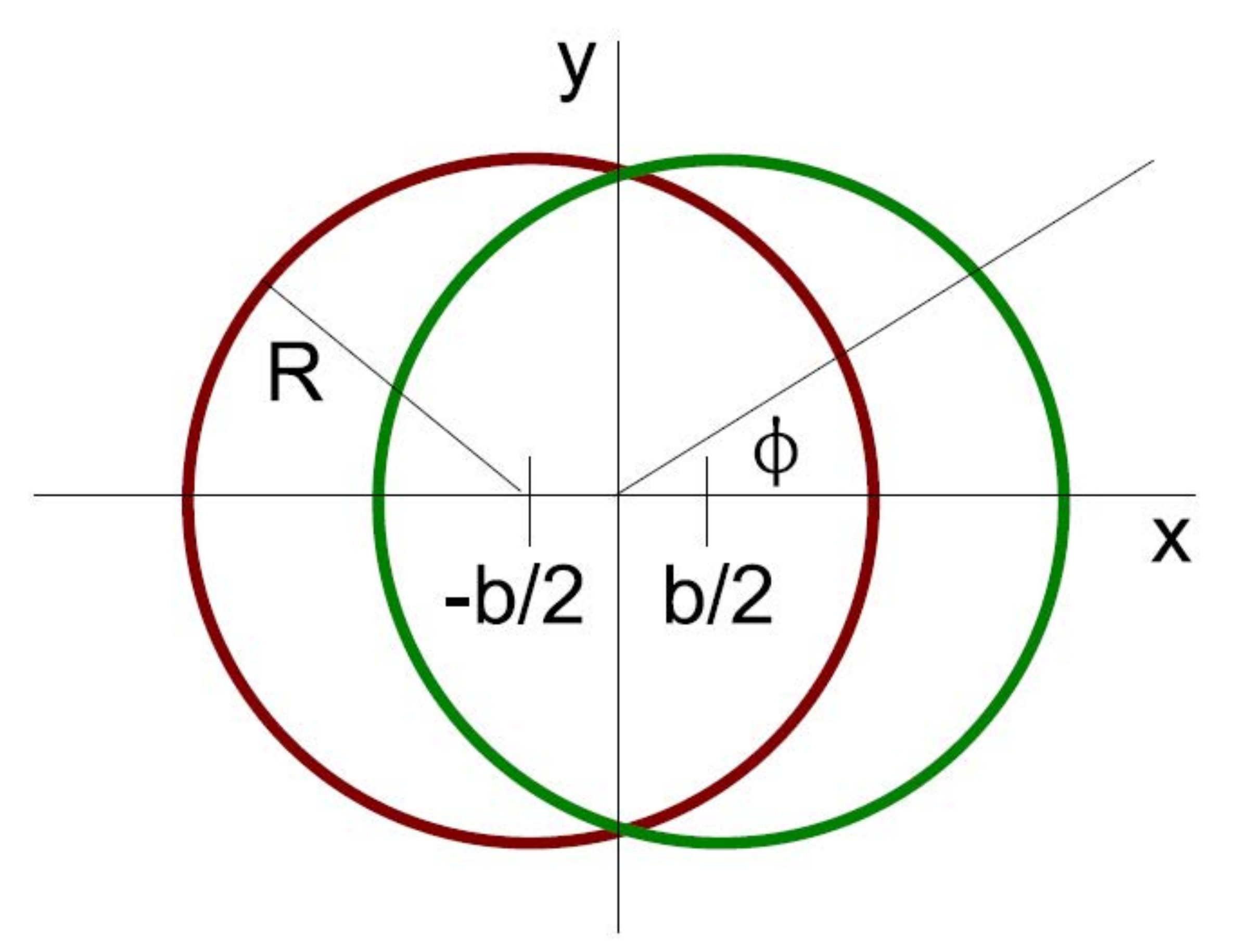}}
    \end{center}
  \end{minipage}
  \hfill
  \begin{minipage}[t]{.45\textwidth}
    \begin{center}
      \scalebox{0.35}{
  \includegraphics{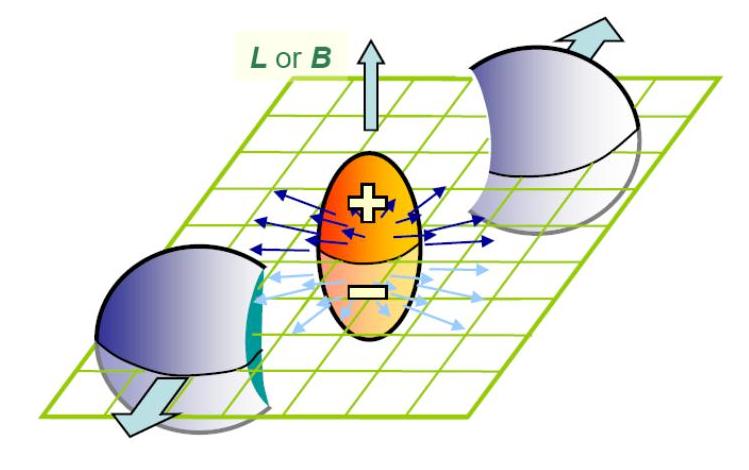}}
    \end{center}
  \end{minipage}
      \caption{Creation of magnetic fields in non-central heavy ion collisions \cite{Kharzeev:2007jp}. 
}
	\label{QGPBfig}
  \hfill
\end{figure}

Consider a collision of two heavy ions with radius $R$ and electric charge $Z e$ at non-zero impact parameter $b$, as represented in figure \ref{QGPBfig} \cite{Kharzeev:2007jp}. 
The beam direction lies along the $z$-axis, with the $(x,z)$-plane referred to as reaction plane and $(x,y)$ as transverse plane. In the region where the nuclei overlap, the QGP (discussed (briefly) in section \ref{sectioneffQCD}) will be formed. It are the charged nucleons that fly past each other rather than colliding which produce a magnetic field in the transverse plane due to their current along $z$.  
Through the Biot-Savart law $\vec B = \frac{\mu_0 q \vec v}{4\pi} \times \frac{\vec e_r}{r^2}$, 
the magnitude in the center-of-mass frame is estimated as $B \sim \gamma Z e b/R^3$ \cite{Tuchin:2013ie}, 
 with $\gamma$ the Lorentz contraction factor and $e$ the electromagnetic coupling constant. For typical RHIC gold-gold 
collisions (at center of mass energies of 200 GeV per nucleon pair), 
$\gamma \approx 100$ such that the nuclei are Lorentz contracted in the $z$-direction to $1 \%$ of their size and can therefore be approximated as `pancakes'.   
The direction of the magnetic field at position $\vec x = (\vec x_\perp,z)$ caused by the moving charged particle at position $\vec x'_\perp$ at time $t=0$ is along $\vec e_v \times (\vec x_\perp - \vec x'_\perp) = (\pm \vec e_z) \times (\vec x_\perp - \vec x'_\perp)$ and contributions from the nuclei moving in opposite directions add up. 
The magnetic field at the origin 
is pointing in the $y$-direction, and because the magnetic field in the overlap region is to a good degree homogeneous in the transverse plane, it is a good estimate for the magnetic field at the surface of the interacting region, especially for large impact parameters. (Moreover, when averaged over the transverse plane, only the $y$-component of the magnetic field survives, see for example \cite{Tuchin:2013apa}.) 
For typical LHC-values of the heavy-ion collision parameters, the magnetic field is estimated to be as large as $eB_y \sim 15 m_\pi^2 \approx 0.3$ GeV$^2$ right after the collision \cite{Skokov:2009qp}.  
This is probably the strongest encountered magnetic field in nature (in other units: $e B \sim 10^5$ MeV$^2$ $\sim 10^{19}$ Gauss $\sim 10^{15}$ Tesla); a magnetar or highly magnetized neutron star reaches ``merely" $10^9$ Tesla. Another context in which gigantic magnetic fields occurred, was during the cosmological electroweak phase transition \cite{Vachaspati:1991nm}.

The above is clearly an oversimplified picture to illustrate the origin of the magnetic field (to name the most obvious simplification: the nuclei collide and travel on, the current is not static as the Biot-Savart law assumes). 
More careful analyses can be found e.g.\ in \cite{Skokov:2009qp,Deng:2012pc,Bzdak:2011yy,Tuchin:2013ie,Tuchin:2013apa,McLerran:2013hla}, 
where more attention is given to the time and space dependence of $B$ \cite{Tuchin:2013apa,Deng:2012pc}, 
the effect of the electric conductivity of the QGP on the evolution and lifetime of $B$ \cite{Deng:2012pc,Tuchin:2013ie,McLerran:2013hla}, 
the presence of electric fields (argued in \cite{Bzdak:2011yy} to be comparable in strength to the magnetic fields), etc. 
It is argued though in \cite{Tuchin:2013ie} that, after dropping by about two orders of magnitude during the first fm/c of plasma expansion, the magnetic field freezes out and lasts for as long as the QGP lives (as a consequence of the finite electrical conductivity of the plasma). 
It is concluded in \cite{Tuchin:2013ie} that the $B$-field must affect all the processes in the QGP.

Studying the possible effects that the presence of strong magnetic fields can have on the QCD phase diagram, might lead to new insights in QCD 
(e.g.~the chiral magnetic effect, which is related to CP-violating processes \cite{Abelev:2009ac}, chiral magnetic spiral/wave, ...), where the QGP created at LHC seems to offer a possible experimental setting.  
This has resulted in a lot of activity in this relatively young research field. We refer to 
the review paper \cite{Kharzeev:2012ph} 
and references therein. 
In our discussion of rho meson condensation, we will ignore any time or space dependence of the magnetic field. The appearance of strong magnetic fields in the QGP is used as motivation for the study of magnetically induced QCD effects, but does not serve as the actual setting of the discussed problem at zero temperature and constant background magnetic field.

\FloatBarrier

\subsection{Rho meson condensation} \label{sectionrholattice}

\begin{figure}[h!]
  \hfill
  \begin{minipage}[t]{\textwidth}
    \begin{center}
    \scalebox{0.38}{
  \includegraphics{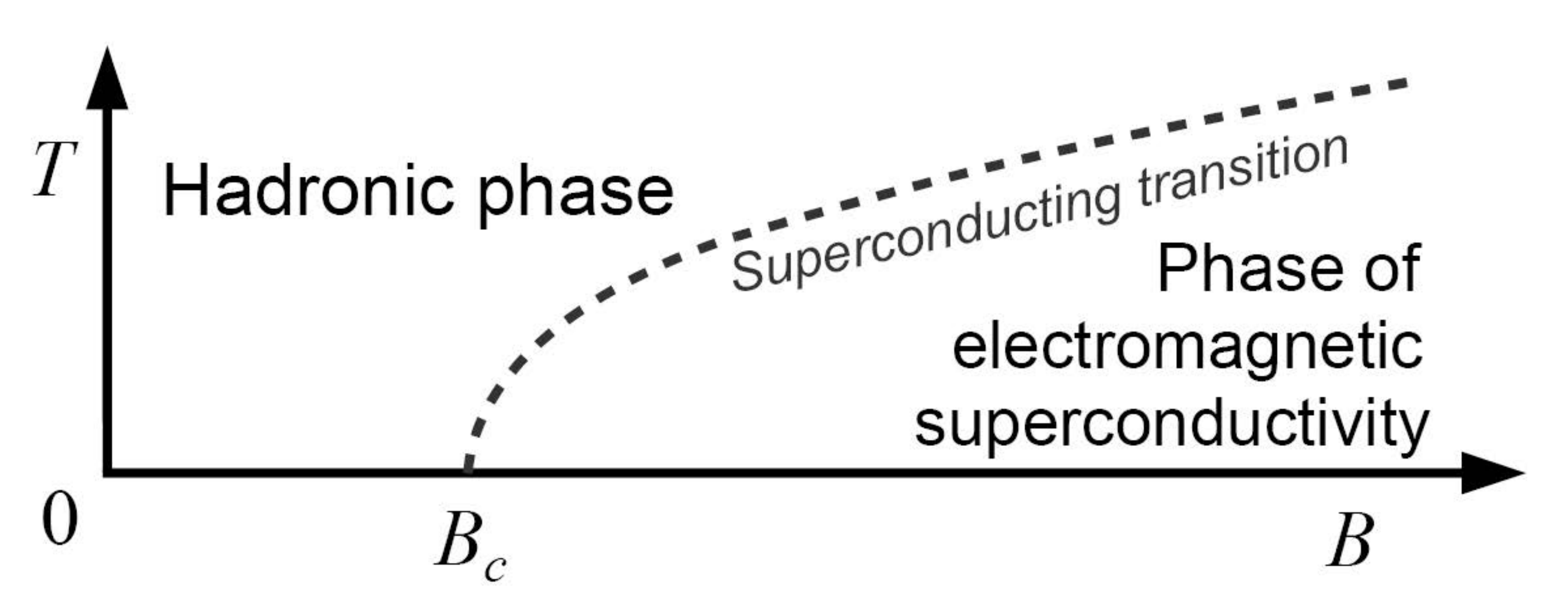}}
    \end{center}
  \end{minipage}
  \hfill
  \begin{minipage}[t]{\textwidth}
  \hspace{2cm}  \begin{center}
      \scalebox{0.7}{
  \includegraphics{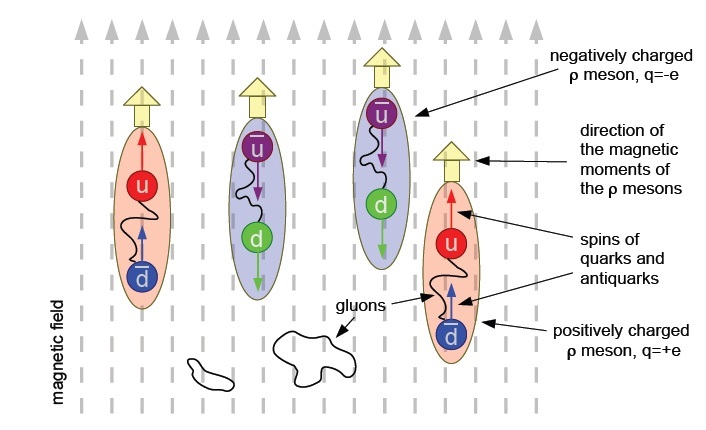}}
    \end{center}
  \end{minipage}
      \caption{Rho meson condensation effect: suggested phase transition in the $(T,B)$ plane of the QCD phase diagram (top) corresponding to an instability of the QCD vacuum towards forming a superconducting state of condensed charged rho mesons (bottom) at critical magnetic field $B_c$ \cite{Chernodub:2011mc,Chernodub:2011tv}. 
}
	\label{rho1}
  \hfill
\end{figure}

\begin{figure}[h!]
  \hfill
  \begin{minipage}[t]{\textwidth}
    \begin{center}
    \scalebox{1.3}{
  \includegraphics{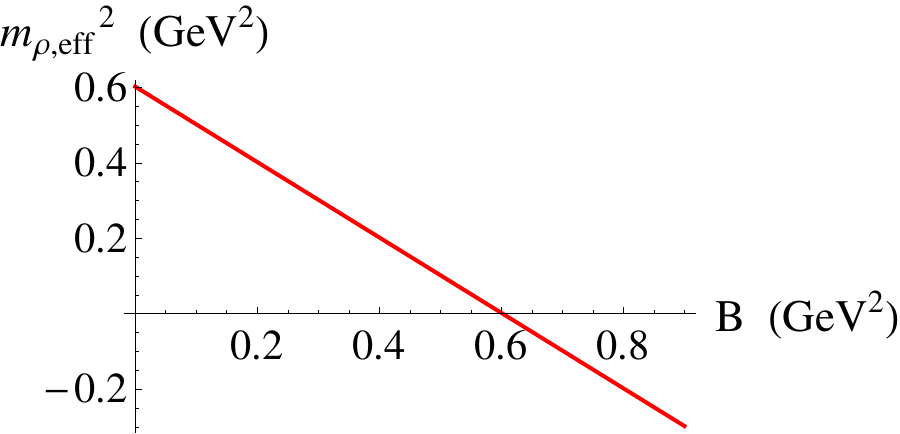}}
    \end{center}
  \end{minipage}
  \hfill
  \begin{minipage}[t]{\textwidth}
  \hspace{2cm}  \begin{center}
      \scalebox{0.33}{ 
  \includegraphics{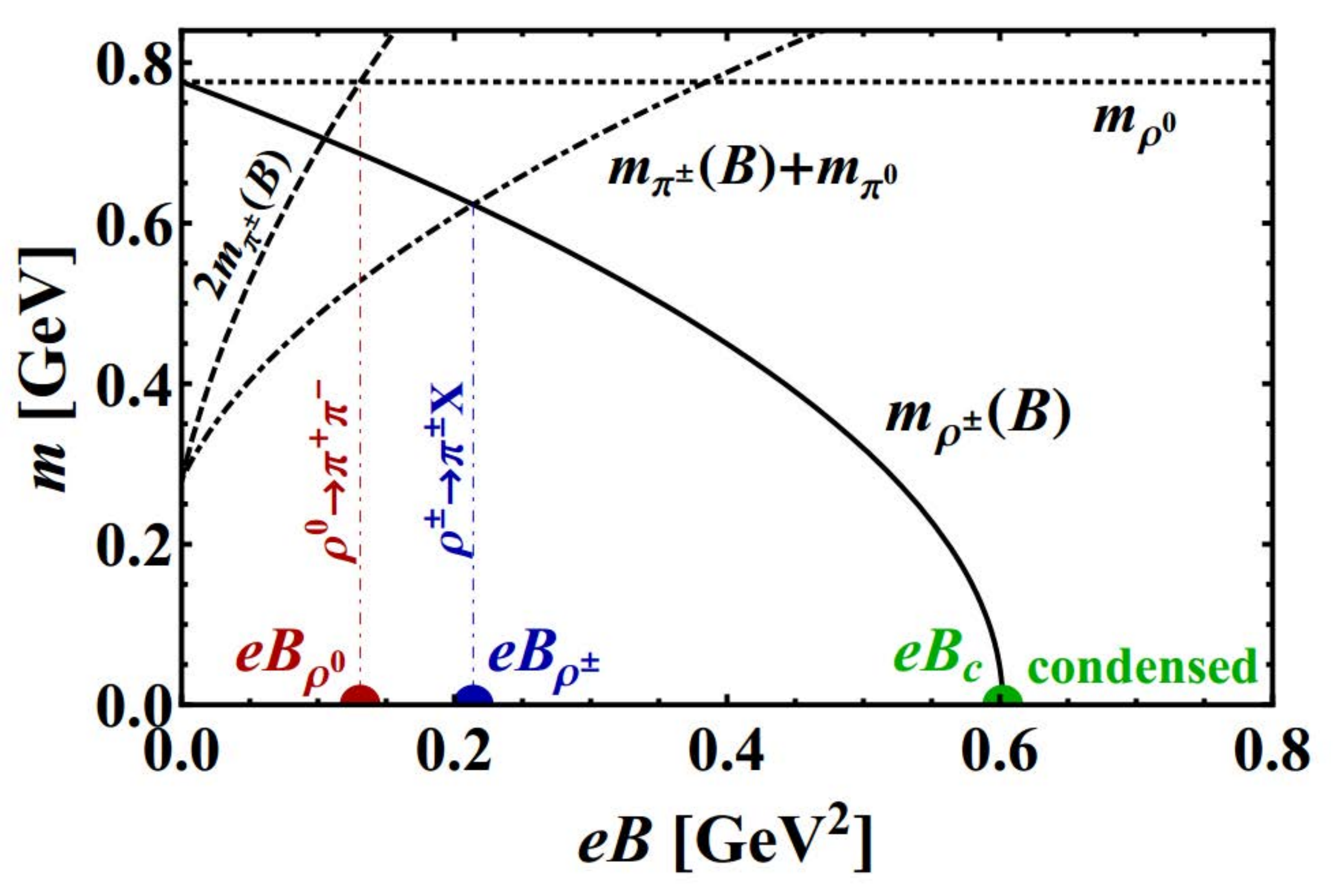}}
    \end{center}
  \end{minipage}
      \caption{The effective rho meson mass squared as a function of $eB$ (from eq.\ (\ref{meff})) becomes negative at $eB_c$,  signaling a tachyonic instability (where in the top figure the electric charge $e$ is absorbed into the notation of $B$).  The lower figure of $\mathit{m_{\rho,eff}}(eB)$, taken from \cite{Chernodub:2010qx}, includes the $B$-dependences for the effective masses of pions (from eq.\ (\ref{Elevels}), with spin zero) showing that the decay processes of the rho meson are kinematically suppressed at high values of $eB$. 
}
	\label{rho2}
  \hfill
\end{figure}

\begin{figure}[h!]
  \hfill
  \begin{minipage}[t]{\textwidth}
    \begin{center}
    \scalebox{0.3}{
  \includegraphics{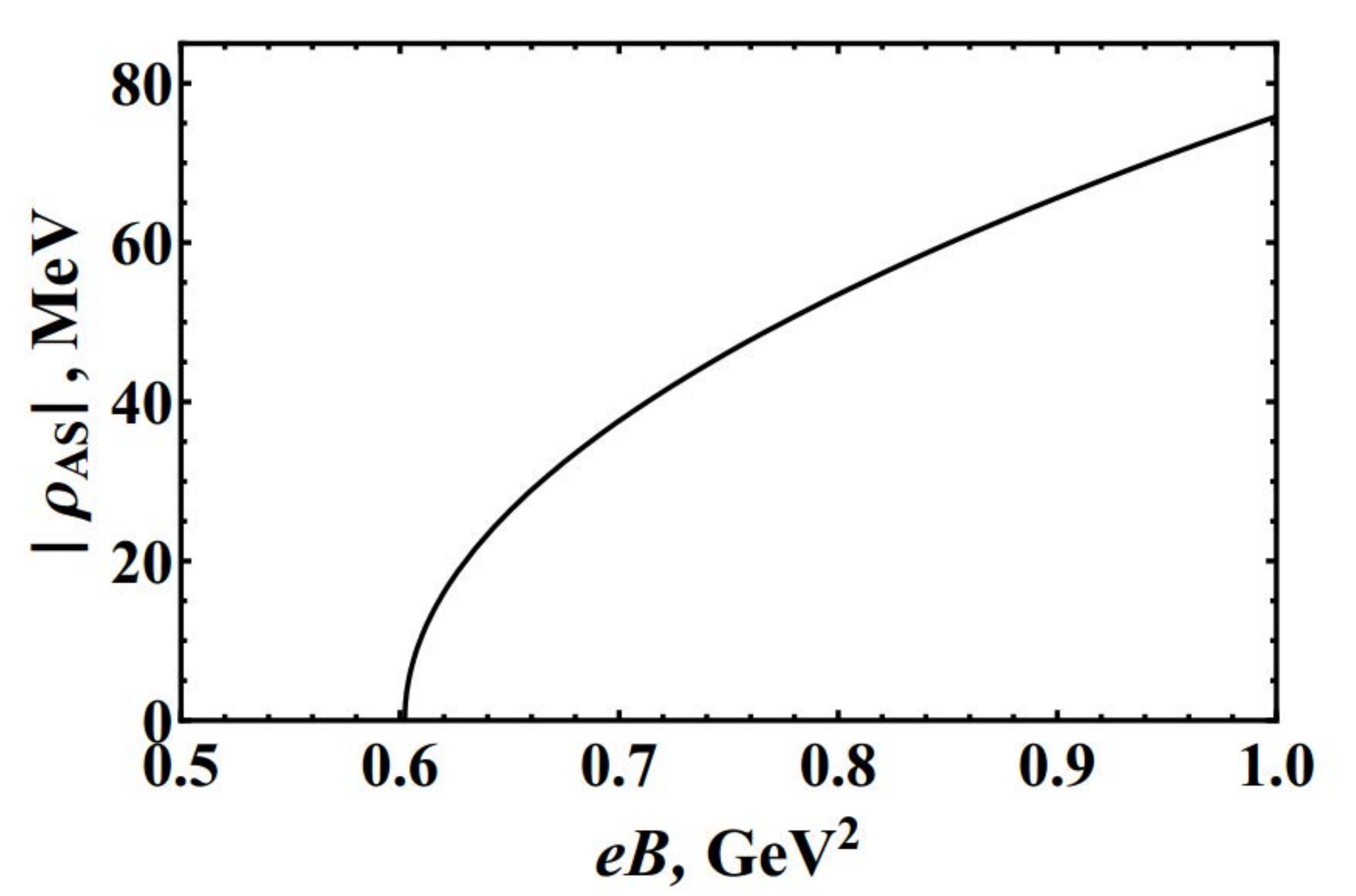}}
    \end{center}
  \end{minipage}
  \hfill
  \begin{minipage}[t]{\textwidth}
  \hspace{2cm}  \begin{center}
      \scalebox{0.4}{
  \includegraphics{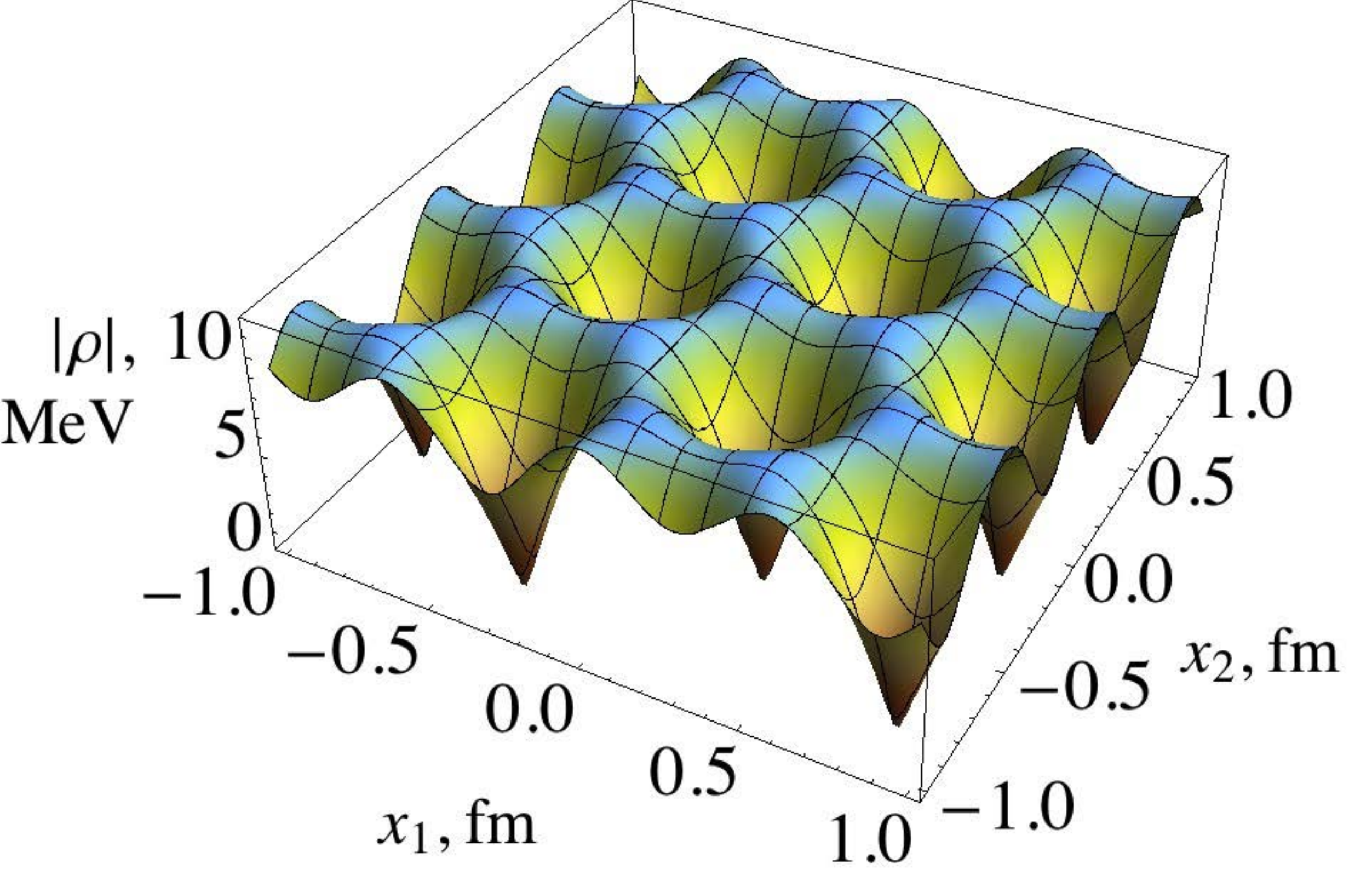}}
    \end{center}
  \end{minipage}
      \caption{Top figure: The superconducting condensate $|\rho_{AS}|$ 
(with subscript $AS$ short for `anisotropic superconductor') as a function of $eB$. The single curve describes both the uniform condensate in the homogeneous approximation of \cite{Chernodub:2010qx} and the mean-cell value of the inhomogeneous condensate in the weak-amplitude approximation \cite{Chernodub:2010qx}. 
Bottom figure: Absolute value of the superconducting condensate $\rho$ at $B = 1.01 B_c$ in the transversal $(x_1,x_2)$-plane, forming an Abrikosov lattice structure \cite{Chernodub:2011gs}.  
}
	\label{rhocondfig}
  \hfill
\end{figure}

A possible magnetically induced tachyonic instability of the QCD vacuum towards a phase where charged rho mesons -- or better quark-antiquark combinations 
with the quantum numbers of rho mesons -- are condensed (figure \ref{rho1}), was first discussed in \cite{Chernodub:2010qx}. The basic argument for this rho meson condensation at some critical value of the magnetic field $B_c$, is that the charged rho meson combinations which have their spin aligned with the magnetic field $B$, have an effective mass squared
\begin{equation} \label{meff}
\mathit{m_{\rho,eff}^2}(B) = m_\rho^2 - eB
\end{equation}
which vanishes at
\begin{equation} \label{BcLandau}
eB_c = m_\rho^2 = 0.602 \text{ GeV}^2,
\end{equation}
based on the fact that the $n$-th energy level of a free, structureless spin-$s$ particle with mass $m$ in the presence of a background magnetic field $\vec B=B \vec e_3$ is given by the well-known Landau level quantization formula
\begin{align}\label{Elevels}
E^2 = m^2 + p_3^2 + (2n - 2 s_3 + 1) eB, 
\end{align}
with $p_3$ the particle's momentum in the direction of the magnetic field, and $s_3$ its spin projection on the same direction.  This leads to (\ref{meff}) for the lowest-energy rho meson $p_3=0$, $n=0$ with spin $s_3=1$, plotted in  figure \ref{rho2}. The mechanism is similar 
to a possible $W^\pm$-boson condensation in the electroweak model \cite{Ambjorn:1989sz,VanDoorsselaere:2012zb,Chernodub:2012fi}. 
It is further shown in \cite{Chernodub:2010qx} that the electrically charged condensate almost inevitably implies electromagnetic superconductivity (along the magnetic field axis) of the new vacuum ground state. 
The QCD vacuum at zero temperature and zero density becoming superconducting at sufficiently large magnetic field  would thus result in 
a quite exotic phase: 
it would not only be anisotropic (exhibiting superconductivity only along the direction of $\vec B$)  
but also spatially inhomogeneous \cite{Chernodub:2010qx,Chernodub:2011mc,Chernodub:2011gs,Chernodub:2012tf}, as the vacuum structure in the transverse directions turns out to be formed of repelling rho meson vortices, resembling  an Abrikosov lattice of a type II superconductor (see figure \ref{rhocondfig}). 
Such a lattice was constructed in the DSGS-model in \cite{Chernodub:2011gs} and in a bottom-up holographic model in \cite{Bu:2012mq} using similar methods. 
It has the interesting property that the magnetic field creates the superconducting state instead of destroying it (cfr.\ Meissner effect). In the bottom-up holographic study of \cite{Cai:2013pda}, the real part of the optical conductivity in the condensed phase is shown to contain a delta peak at the origin, consistent with a superconducting condensed state.

The above argument based on Landau levels holds in the context of the bosonic effective DSGS-model \cite{Djukanovic:2005ag} for rho meson quantum electrodynamics, used in \cite{Chernodub:2010qx}. Somewhat later, the rho meson condensation effect was also shown to emerge in the NJL-model \cite{Chernodub:2011mc}, which was introduced in section \ref{sectioneffQCD}.  
Lattice evidence in favour of the effect appeared in \cite{Braguta:2011hq}.
It should be clear however that rho meson condensation is merely conjectured to occur in QCD based on these descriptions in effective QCD-models, not proven nor experimentally observed.
To date, the effect of rho meson condensation has been discussed in  \cite{Chernodub:2010qx,Chernodub:2011mc,Braguta:2011hq,Chernodub:2011gs,Chernodub:2012tf} using phenomenological and lattice approaches, in our work \cite{Callebaut:2011ab} using 
the Sakai-Sugimoto model, and in \cite{Ammon:2011je,Bu:2012mq,Cai:2013pda} using a bottom-up holographic approach. Its possible occurrence has been argued against in \cite{Hidaka:2012mz} -- followed by a rebuttal in \cite{Chernodub:2012zx,Chernodub:2013uja} showing that the counterarguments of \cite{Hidaka:2012mz} should not apply.

The value for the critical magnetic field $B_c$ marking the onset of the condensation differs from approach to approach. It is minimal in the DSGS-model, which predicts $eB_c = m_\rho^2 \approx 0.6$ GeV$^2$. Both the lattice simulation ($eB_c \approx 0.9$ GeV$^2$) and the NJL calculation ($eB_c \approx 1$ GeV$^2$) show an increase in $eB_c$, which is considerable in the used GeV units.

\section{Goal and strategy}  \label{goalrho}

We set out to give further evidence for a magnetically induced $\rho$ meson condensation using a (top-down) holographic approach. 
We will study 
first of all \textit{if} the rho meson condensation effect can be modeled using holography, and second, whether the top-down holographic QCD approach can deliver new insights, in particular when it comes to taking into account effects from the constituents of the rho mesons. 
We will be able to show that a condensation should occur as we encounter a tachyonic instability in the $\rho$ meson sector, $\mathit{m^2_{\rho,eff}}(B_c)=0$, relying on the much studied Sakai-Sugimoto model, which was reviewed in chapter \ref{SSMchapter}.

In holographic language: we investigate the stability of the two-flavour Sakai-Sugimoto model in the presence of a magnetic field, and this in the confinement phase, finding 
stability in the scalar and an instability in the charged vector sector.
Previous stability analyses of the Sakai-Sugimoto model (SSM) have mainly focused on the case of a background chemical potential. In particular Chern-Simons-induced instabilities to spatially modulated phases have received quite some attention recently
\cite{Ooguri:2010xs,Fukushima:2013zga,deBoer:2012ij,Bayona:2011ab,BallonBayona:2012wx,Domokos:2013kha}. Earlier works in this context include \cite{Chuang:2010ku}, and \cite{Bergman:2007wp} on the $(T,\mu,B)$ phase diagram in the Sakai-Sugimoto model.
More relevant for our current purposes is the DBI-induced instability in the presence of an isospin  chemical potential studied in \cite{Aharony:2007uu},
where a tachyonic instability of the rho meson and ensuing rho meson condensation was described.
We will encounter a somewhat similar phenomenon here, but as a result of the presence of a background magnetic field $B$ at zero chemical potential. 
Most of the technical complications we will encounter, as well as the occurrence of the condensation effect itself, stem directly from the non-Abelian nature of the problem ($N_f=2$). 
The complications can be avoided by resorting to bottom-up holographic models \cite{Ammon:2011je,Bu:2012mq,Cai:2013pda}, as it turns out without having to give up on being able to reproduce the general physical picture, but hereby moving away further from a QCD-like dual. The model of   \cite{Ammon:2011je,Bu:2012mq} involves an $SU(2)$ Einstein-Yang-Mils action for an $SU(2)$ bulk gauge field in a (4+1)-dimensional AdS-Schwarzschild background (we will discuss this type of background in section \ref{AdS5-SS}).  
\cite{Cai:2013pda} considers a (3+1)-dimensional DSGS-model generalized to AdS. 
In both cases, the analysis remains much closer to the phenomenological picture of \cite{Chernodub:2010qx}. 
Other examples of magnetic instabilities of bulk charged vectors in an AdS/CFT context can be found in \cite{Donos:2011pn,Almuhairi:2011ws}.

\paragraph{Outline}

We start with the discussion of the holographic set-up in section \ref{setup}. 
We fix the number of colours $N_c=3$, number of flavours $N_f=2$ and the rest of the holographic parameters to numerical GeV units, in order to obtain results for $\mathit{m^2_{\rho,eff}}$ and $B_c$ in physical units, comparable to other -- phenomenological and lattice -- approaches. That is, we extend the work of Sakai and Sugimoto on the numerical fixing of the holographic parameters for the $u_0 = u_K$ case to the non-antipodal case where $u_0 > u_K$. 
In the same section, the effect of the magnetic field on the probe branes' embedding is studied.

In section \ref{stabilityfluctuations} we discuss the stability in the presence of fluctuations. 
For that purpose we plug a flavour gauge field ansatz containing a background ($\sim B$) and a fluctuation part ($\sim$ mesons) into the non-Abelian DBI-action governing the dynamics of the flavour gauge field living on the probe branes, and expand the action to second order in the fluctuations.
The eventual goal is to extract the effective rho meson mass (as a function of $B$) from the 4-dimensional mass equation for the vector meson, the effective 4-dimensional action to be obtained from the DBI-action by integrating out the extra dimensions. 

Before tackling the full complications of the stability analysis, we focus on two simplifying cases for the vector meson sector in section \ref{simpleholorho1}. Within the approximation of the action to second order in the field strength, we consider the effect of $B$ on the rho meson mass in the antipodal embedding (this comes closest to the analyses in bottom-up approaches to the problem) and in the non-antipodal embedding with equal electric charges assigned to up and down quarks (i.e. flavour branes). The antipodal case exactly reproduces the Landau levels and consequential prediction for the value of $B_c$ 
obtained in the DSGS-model in \cite{Chernodub:2010qx}. The non-antipodal embedding is used to include at least some effects of the constituent quarks of the rho meson. In particular, already in the approximation used in this section, the so-called  `chiral magnetic catalysis effect' translates in an increase of $B_c$.  

The rest of section \ref{stabilityfluctuations} is devoted to the general non-antipodal case with up and down D8-branes coupling to the magnetic field with different electric charges ($2 e/3$ and $-e/3$). 
In this technically more involved scenario, 
we have to choose a particular gauge to disentangle the scalar and vector fluctuations in the action, this is done in section \ref{gaugefixing}.
Then we
discuss the stability with respect to scalar fluctuations, corresponding to the positions of the probe branes.
Next, we consider the vector fluctuations. 
For the now magnetically separated branes, we repeat the analysis of $\mathit{m^2_{\rho,eff}}(B)$, both in the case of using the action expanded to second order in $F$ (section \ref{F2approx}) and the full non-linear DBI-action in $F$ (section \ref{4.4}).
Because the field strength $F$ in the DBI-action is accompanied with a factor proportional to the inverse of the 't Hooft coupling $\lambda$, which is large in the validity range of the gauge-gravity duality, the expansion to second order in $F$ is commonly used.  However, in the presence of large background fields, the higher order terms may become important (see section \ref{ambiguities}). We therefore compare the outcome of using the $F^2$-approximated action versus the full DBI-action, from which we can conclude that the difference in $B_c$ is very small and the $F^2$-expansion was justified in our case after all. At that point, we also revisit the antipodal embedding, where the value of $B_c$ seems to be more sensitive to taking into account all higher order terms in the DBI-action, see section \ref{commentantipodal}. 

In section \ref{F2approx} the focus is on handling the magnetically separated branes. For non-coinciding branes, the symmetrized trace (STr) over flavour indices in the DBI-action no longer simplifies to a standard Tr. Instead, evaluating the STr (which can be done exactly to second order in the fluctuations) gives rise to complicated functions in the action (defined via integrals), 
which depend on the background fields and are discontinuous in the holographic radius $u$. We pay some attention
to solving the eigenvalue equation for the rho meson eigenfunction with these functions present.
The evaluation of the STr is discussed in section \ref{4.1.1}, with the used -- exact -- prescriptions outlined in  appendix \ref{appendix}, including a sketch of their derivation. 
In section \ref{4.3.2}, for completeness, we briefly discuss the pions in the DBI-action. The section ends with a comment on the validity of the use of the non-Abelian DBI-action for non-coincident branes.

In section \ref{4.4} the focus is on handling the extra dependences on the magnetic field from considering the full DBI-action. The resulting effective 4-dimensional equation of motion (EOM) (to second order in the rho meson fields) has extra terms compared to the standard Proca EOM used in phenomenological descriptions of the rho meson in a background magnetic field, making it
harder to analyze. We solve the EOMs for the complete energy spectrum exactly
in section \ref{4.4.3}, with the main result for the generalized Landau levels given in eq.\ (\ref{omegasquared}). The energy eigenstates are no longer spin eigenstates (as opposed to the Proca energy eigenstates), except for the condensing state.

A summary of this chapter can be found in section \ref{summary}. 
A small remark regarding notation is due: we have on occasion absorbed the electron charge $e$ into the notation for the magnetic field, $eB \rightarrow B$, in particular in the figures of the effective rho meson masses as a function of $(e)B$ and most other figures. We try to signal this in the text, but may have switched between one or the other notation (ironically for reasons of clarity) without explicit warning. It should hopefully be clear from the context where this has been done.

\section{Holographic set-up} \label{setup}

\subsection{Basic set-up}

We will work in the general non-antipodal Sakai-Sugimoto model ($u_0 > u_K$) for quenched QCD in the chiral limit, reviewed in the previous chapter, where in particular non-zero constituent quark masses (\ref{constmass}) are modeled. The special case $u_0=u_K$ will also be discussed. 
For the purpose of presenting the end results in physical GeV units, we will fix the number of colours to three, $N_c=3$. This means we will be comparing our results from a classical supergravity approximation to phenomenological and lattice results in a parameter regime where the used approximation is actually expected to break down, see also the discussion of the glueball spectrum in figure \ref{glueballspectrumfig2} regarding this issue. 
We choose the number of flavours to be two, $N_f=2$, in order to be able to model electromagnetically charged mesons consisting of up- and down quarks. This means we are stretching the validity of the probe approximation $N_c \gg N_f$, but we will nonetheless
ignore the backreaction. Unquenching the Sakai-Sugimoto model is a very difficult task, see the work of \cite{Burrington:2007qd}.   
 For meson phenomenology, the $N_c \gg N_f$ limit does often provide a fairly good description of the physical world (see section \ref{sectionlargeN}).

\subsection{Non-Abelian probe brane action}   \label{subsecnonab} 

While the low energy effective action for a single brane is known to be the Dirac-Born-Infeld action (\ref{DBIstringannex})  \cite{Leigh:1989jq},
valid in the static gauge (i.e.~alignment of the world volume with space-time coordinates) and for slowly varying field strengths, the full non-Abelian generalization of it
for the description of a stack of coinciding branes is not.
Tseytlin proposed in \cite{Tseytlin:1997csa} to non-Abelianize the Dirac-Born-Infeld action by introducing a symmetrized trace STr. The action is still restricted to static gauge and the (in the non-Abelian case slightly ambiguous) slowly-varying field strengths approximation, ignoring derivative terms including $[F,F] \sim [D,D]F$ terms.
This action was shown to be valid up to fourth order in the field strength, with deviations starting to appear at order $F^6$ \cite{Hashimoto:1997gm,Sevrin:2001ha}.
For the probe flavour branes we are dealing with, it is given by the following, which we will further refer to as `the' (non-Abelian) DBI-action \cite{Tseytlin:1997csa,Myers:2003bw,Howe:2006rv}:
\begin{equation} \label{nonabelian}
S_{DBI} = -T_8 \int d^4x \hspace{1mm} 2 \int_{u_0}^{\infty}  du \int \epsilon_4 \hspace{1mm} e^{-\phi}\hspace{1mm} \text{STr}
\sqrt{-\det \left[g_{mn}^{D8} + (2\pi\alpha') iF_{mn} \right]},
\end{equation}
where $T_8 = 1/((2\pi)^8 l_s^9)$ is the D8-brane tension, the factor 2 in front of the $u$-integration makes sure that we integrate over both halves of the $\cup$-shaped D8-branes,   $\text{STr}$ is the
symmetrized trace which is defined as
\begin{eqnarray} \label{STrdef}
\text{STr}(F_1 \cdots F_n) = \frac{1}{n!} \text{Tr}(F_1 \cdots F_n + \mbox{all permutations}), \end{eqnarray}
$g_{mn}^{D8}$ is the induced metric on the D8-branes,
\[
g_{mn}^{D8} = g_{mn} + g_{\tau\tau} (D_m \tau)(D_n \tau),
\]
with covariant derivative $D_m\tau = \partial_m \tau + [A_m,\tau]$,  and
\[
F_{mn} = \partial_m A_n - \partial_n A_m + [A_m, A_n] = F_{mn}^a t^a
\]
the field strength with anti-Hermitian generators 
\begin{eqnarray} t^a  =  -\frac{i}{2} (\mathbf{1}, \sigma_1, \sigma_2, \sigma_3), \quad
\text{Tr}(t^a t^b) = - \frac{\delta_{ab}}{2}   \qquad (a,b=0,1,2,3) \end{eqnarray} 
obeying 
\begin{equation}
[t^a,t^b] = \epsilon_{abc}t^c \qquad (a,b=1,2,3). 
\end{equation}

\subsection{Numerical fixing of the holographic parameters } \label{B} 

Following \cite{Sakai:2005yt}, we define
\begin{equation} \label{kappa}
\kappa = \frac{\lambda N_c}{216 \pi^3} 
\end{equation}
for notational convenience.  

In the model with antipodal embedding there are six parameters, $R$, $\kappa$, $l_s$, $M_K$, $g_s$ and $L$, related to each other through the four relations (\ref{relaties})-(\ref{SS2(2.4)a}). 
In \cite{Sakai:2005yt} the remaining independent parameters $M_K$ and $\kappa$ were fixed to GeV units by matching to the QCD input values
\begin{eqnarray} \quad f_\pi = 0.093  \mbox{ GeV} \quad \mbox{and} \quad m_\rho = 0.776 \mbox{ GeV}, \end{eqnarray}
resulting in
\begin{equation} \label{antipodalvalues}
M_K \approx 0.949 \mbox{  GeV} \quad
\mbox{and}\quad \kappa  = \frac{\lambda N_c}{216 \pi^3} \approx 0.00745.
\end{equation}

We extend this to the model with non-antipodal embedding where there are seven parameters, $R$, $\kappa$, $l_s$, $M_K$, $u_0$, $g_s$ and $L$.
These are completely determined by the four relations (\ref{relaties})-(\ref{SS2(2.4)a}), 
and the three extra requirements that the computable numerical values for the constituent quark mass $m_q$, the pion decay constant $f_\pi$ and the $\rho$ meson mass $m_\rho$, in absence of magnetic field, match to the phenomenologically or experimentally obtained QCD input values
\begin{eqnarray} m_q = 0.310 \mbox{ GeV}, \quad f_\pi = 0.093  \mbox{ GeV} \quad \mbox{and} \quad m_\rho = 0.776 \mbox{ GeV}. \end{eqnarray}
Adapting the analysis of Sakai and Sugimoto \cite{Sakai:2004cn} to the more general case $u_0 > u_K$, see also \cite{Peeters:2006iu,Peeters:2007ab}, we derive the mass eigenvalue equation for the vector meson sector and the expression for $f_\pi$ as  functions of
$M_K$, $u_0$ and $\kappa$, and thus indirectly (through the relation (\ref{Lconf})) as  functions of the three unknown parameters 
$M_K$, $L$ and $\kappa$.\\

In the confinement phase, the non-Abelian DBI-action (\ref{nonabelian})
becomes, to second order in the field strength,
\begin{equation} \label{secondorder}
S_{DBI} = \widetilde V  \int d^4x \hspace{1mm}2 \int_{u_0}^\infty du  \hspace{1mm}\text{Tr} \left\{ u^{-1/2} \gamma^{1/2} R^3 F_{\mu\nu}F^{\mu\nu} + 2u^{5/2} \gamma^{-1/2} \eta^{\mu\nu}F_{\mu u}F_{\nu u} \right\} + \mathcal O(F^4),
\end{equation}
with \begin{eqnarray} \widetilde V = T_8 V_4 g_s^{-1} R^{3/2} \frac{1}{4} (2\pi\alpha')^2 \end{eqnarray}
and \begin{eqnarray} \gamma(u) = \frac{u^8}{f(u) u^8 - f(u_0) u_0^8}, \end{eqnarray} \label{gamma}
and where the $\text{STr}$ is replaced by $\text{Tr}$ as
\begin{eqnarray} \text{STr}(t^a t^b) = \text{Tr} (t^a t^b).\end{eqnarray}

Assuming the flavour gauge field components $A_\mu(x^\mu,u)$ can be expanded in complete sets $\left\{\psi_n(u)\right\}_{n\geq 1}$,
\begin{eqnarray}
A_\mu(x,u) &=& \sum_{n \geq 1} B_\mu^{(n)}(x) \psi_n(u)\\
\Rightarrow F_{\mu\nu}F^{\mu\nu} &=& F_{\mu\nu}^{(m)}F^{(n)\mu\nu} \psi_m \psi_n \nonumber \\\text{with}~  F_{\mu\nu}^{(m)} &=& \partial_\mu B_\nu^{(m)} - \partial_\nu B_\mu^{(m)},  \quad
F_{\mu u}F_{\nu u} ~=~ B_\mu^{(m)}B_\nu^{(n)}\partial_u \psi_m \partial_u \psi_n + \cdots,\nonumber
\end{eqnarray}
the part of the action (\ref{secondorder}) in $B_\mu^{(n)}(x)$ reduces to the effective 4-dimensional action 
\begin{eqnarray}
S = - \int d^4 x \left\{ \frac{1}{4} F_{\mu\nu}^{(n)a}F^{(n)\mu\nu a} + \frac{1}{2} m_n^2 B_\mu^{(n)a} B^{(n)\mu a} \right\}, \label{procaaction}
\end{eqnarray}
describing vector mesons $B_\mu^{(n)}$ with masses $m_n$, if the $\psi_n(u)$ are subject to:
\begin{equation} \label{orthonormconditiepsi}
\widetilde V \int_{u_0}^{\infty} du \hspace{1mm} u^{-1/2} \gamma^{1/2} R^3 \psi_m \psi_n = \frac{1}{4} \delta_{mn},
\end{equation}
and
\begin{equation} \label{blablaalg}
2 \widetilde V \int_{u_0}^{\infty} du \hspace{1mm} u^{5/2} \gamma^{-1/2} (\partial_u \psi_m)(\partial_u \psi_n) = \frac{1}{2} m_n^2 \delta_{mn} \quad (m,n \geq 1).
\end{equation}
The effective action (\ref{procaaction}) corresponds to the standard Proca action for a massive vector particle.  
The conditions (\ref{orthonormconditiepsi}) and (\ref{blablaalg}) combine, using partial integration (and hereby implicitly assuming the $\psi_n$ to be normalizable functions), to the eigenvalue equation
\begin{equation} \label{finiteTeigwvgl}
-u^{1/2} \gamma^{-1/2} \partial_u (u^{5/2} \gamma^{-1/2} \partial_u \psi_n) = R^3 m_n^2 \psi_n.
\end{equation}
To include pions, following the original discussion in \cite{Sakai:2004cn}, it is convenient to introduce the coordinate $z$ that is related to $u$ through
\begin{eqnarray} u^3 = u_0^3 + u_0 z^2, \label{uzdef} \end{eqnarray}
going from $-\infty$ to $+\infty$ along the flavour branes and thus allowing the description of both boundaries at $u\rightarrow \infty$ of the $\cup$-shaped flavour branes.
In this new coordinate $z$, the action (\ref{secondorder}) reads (denoting $u(z)$ as $u_z$ for readability)
\begin{equation}
S_{DBI} = \widetilde V \int d^4x  \int_{-\infty}^\infty dz \hspace{1mm}\text{Tr} \left\{
\frac{2}{3} u_0 u_z^{3/2} \gamma' R^3 F_{\mu\nu}F^{\mu\nu} +
\frac{3}{u_0}  \frac{u_z^{1/2}}{\gamma'} \eta^{\mu\nu}F_{\mu z}F_{\nu z}  \right\} + \mathcal O(F^4),
\end{equation}
with
\begin{eqnarray} \gamma'(z) = \frac{|z| \sqrt \gamma}{u_z^4} = \sqrt{\frac{z^2}{u_z^{5}(u_z^3 - u_K^3) - (u_0^8 - u_0^5 u_K^3)}}, \label{gammaaccentdef} \end{eqnarray}
and the condition (\ref{blablaalg})
\begin{equation} \label{blablaalgz}
\frac{\widetilde V}{2} \int_{-\infty}^{\infty} dz \hspace{1mm} \frac{3}{u_0} \frac{u_z^{1/2}}{\gamma'} (\partial_z \psi_m)(\partial_z \psi_n) = \frac{1}{2} m_n^2 \delta_{mn} \quad (m,n \geq 1).
\end{equation}
The flavour gauge field component\footnote{The gauge field components along the four-sphere are assumed to vanish, $A_\alpha$ = 0, and $A_\mu$ and $A_z$ are assumed to be independent of the four-sphere coordinates.} $A_z$  is expanded in the complete set $\left\{\phi_n(z)\right\}_{n\geq 0}$:
\begin{eqnarray} \label{gaugefieldexpansion}
A_\mu(x,z) &=& \sum_{n \geq 1} B_\mu^{(n)}(x) \psi_n(z), \quad A_z(x,z) = \sum_{n \geq 0} \phi^{(n)}(x) \phi_n(z)
\\
\Rightarrow F_{\mu\nu}F^{\mu\nu} &=& F_{\mu\nu}^{(m)}F^{(n)\mu\nu} \psi_m \psi_n\,,\quad F_{\mu z}F_{\nu z} = (\partial_\mu \phi^{(m)} \phi_m - B_\mu^{(m)}\partial_z \psi_m)(\partial_\nu \phi^{(n)} \phi_n - B_\nu^{(n)}\partial_z \psi_n).\nonumber
\end{eqnarray}
Demanding a canonical normalization of the kinetic term for the $\phi^{(n)}(x^\mu)$ fields in the effective 4-dimensional action then leads to the orthonormality condition
\begin{equation} \label{orthonormconditiephi0z}
\frac{\widetilde V}{2} \int_{-\infty}^{\infty} dz \hspace{1mm} \frac{3}{u_0} u_z^{1/2}\gamma'^{-1} \phi_m \phi_n = \frac{1}{2} \delta_{mn} \quad (m,n \geq 0).
\end{equation}
From the last condition (\ref{orthonormconditiephi0z}) and (\ref{blablaalgz}) it follows that
\begin{eqnarray} \phi_n = m_n^{-1} \partial_z \psi_n \quad \mbox{for $n\geq 1$}\end{eqnarray}
and since $\phi_0 \perp \partial_z \psi_n$ for all $n \geq 1$,
\begin{equation}
\frac{\widetilde V}{2} \int_{-\infty}^{\infty} dz \hspace{1mm}  \frac{3}{u_0} u_z^{1/2}\gamma'^{-1} \phi_0 \partial_z \psi_n = 0,
\end{equation}
we can set
\begin{eqnarray}
\phi_0 = c \gamma' u_z^{-1/2} = c \frac{\gamma'}{(u_0^3+u_0 z^2)^{1/6}}
\end{eqnarray}
with the normalization constant $c$ determined by
\begin{eqnarray} \frac{\widetilde V}{2} c^2 \int_{-\infty}^{\infty} dz \hspace{1mm} \frac{3}{u_0} u_z^{-1/2}\gamma' = \frac{1}{2}. \end{eqnarray}
Now $\psi_0$ is defined through $\phi_0 = \partial_z \psi_0$, and $\hat \psi_0$ as a multiple of $\psi_0$,
\begin{eqnarray} \hat \psi_0 = c' \int_0^z dz\frac{\gamma'}{(u_0^3+u_0 z^2)^{1/6}}, \end{eqnarray}
that fulfills
\begin{eqnarray} \hat \psi_0(\pm \infty) = \pm \frac{1}{2}. \end{eqnarray}
We can then rewrite the expansion for the gauge field as
\begin{eqnarray} A_\mu = \xi_+ \partial_\mu \xi_+^{-1} \psi_+ + \xi_- \partial_\mu \xi_-^{-1} \psi_- + \sum_n B_\mu^{(n)} \psi_n \,\, , \qquad A_z=0 \label{Aziszerogauge}
\end{eqnarray}
with
\begin{eqnarray} \psi_\pm (z) = \frac{1}{2} \pm \hat \psi_0 \quad \mbox{ such that } \psi_+(\infty) = \psi_-(-\infty) = 1 \mbox{ and } \psi_+(-\infty)= \psi_-(\infty)= 0\end{eqnarray}
and
\begin{eqnarray} \xi_\pm^{-1}(x^\mu) = \mathcal P \exp\left\{- \int_0^{\pm \infty} dz' A_z(x^\mu, z')\right\}.\end{eqnarray}
One stays in the $A_z = 0$ gauge under residual gauge transformations $g(x^\mu, z = 0) = h(x^\mu)$. Fixing the gauge to $\xi_-=1$, we have the following gauge field expansion \cite{Sakai:2004cn}
\begin{eqnarray} A_\mu(x,z) = U^{-1}(x)\partial_\mu U(x) \psi_+(z) + \sum_{n\geq1} B_\mu^{(n)}(x) \psi_n(z), \qquad A_z = 0. \end{eqnarray}
The pion field is defined as the path ordered exponential 
\begin{eqnarray} U(x^\mu) = \mathcal P \exp{ \left\{-\int_{-\infty}^{\infty} dz' A_z(x^\mu, z') \right\}}.  \end{eqnarray}
This has the correct transformation behaviour $U(x) \rightarrow h_L U(x) h_R^{-1}$ under a global chiral symmetry transformation\footnote{The holographic interpretation of a global chiral symmetry transformation $(h_L,h_R)$ will be defined in section \ref{C}.} (\ref{SSMchiralel}) 
to be interpreted as   
the pion field (\ref{U}) used in the sigma-model for low-energy effective QCD, 
\begin{equation}
U(x) = e^{2i\frac{\pi(x^\mu)}{f_\pi}} \quad \in U(N_f), \qquad \pi(x^\mu) = \pi_a t^a 
\end{equation}
(with the difference compared to (\ref{U}) that it is a $U(N_f)$ field, in line with the discussion in section \ref{sectionchiralanomaly}).   
Its action in 
(\ref{NLS}) provides the first 
part of the meson action (\ref{skyrme}): 
\begin{eqnarray} \int d^4 x \frac{f_\pi^2}{4} \hspace{1mm} \text{Tr} (U^\dagger \partial_\mu U)^2.  \label{skyrmeintro} \end{eqnarray} 
Equating this action with the corresponding term in the Sakai-Sugimoto action after plugging in the expansion for the gauge field
\begin{eqnarray} \hspace{-4mm} \widetilde V \int d^4x  \int_{-\infty}^\infty dz \hspace{1mm}\text{Tr} \left\{
\frac{3}{u_0}  \frac{(u_0^3+u_0 z^2)^{1/6}}{\gamma'} \eta^{\mu\nu}F_{\mu z}F_{\nu z}  \right\} =
\widetilde V \int d^4x  \int_{-\infty}^\infty dz \hspace{1mm} \frac{3}{u_0} \frac{c}{\phi_0} (\partial_z \psi^+)^2  \hspace{1mm} \text{Tr}(U^\dagger \partial_\mu U)^2 \end{eqnarray}
leads to the identification
\begin{eqnarray} \frac{f_\pi^2}{4} = \widetilde V \frac{3}{u_0} c' = \widetilde V \frac{3}{u_0} \left( 2 \int_{0}^\infty dz \frac{\gamma'}{(u_0^3+u_0 z^2)^{1/6}} \right)^{-1}, \end{eqnarray}
or
\begin{eqnarray} 
f_\pi^2(M_K,u_0,\kappa) = \frac{4}{3} \kappa M_K^{7/2} \frac{3}{u_0}\left( 2 \int_{0}^\infty dz \frac{\gamma'}{(u_0^3+u_0 z^2)^{1/6}} \right)^{-1}, \label{fpiSSM}
\end{eqnarray}
where we have used the relations (\ref{relaties}), (\ref{SS2(2.4)a}) 
and the definition (\ref{kappa}) to determine the volume factor $\widetilde V$ in the action:
\begin{eqnarray} \widetilde V = \frac{1}{3} \kappa M_K^{7/2}.  \end{eqnarray} 
Because $\widetilde V \sim N_c$ for fixed $\lambda$ (from (\ref{kappa})), it follows from (\ref{fpiSSM}) that $f_\pi^2  \sim N_c$ has the correct large-$N_c$ scaling behaviour (\ref{fpibehaviour}).   
Similarly, the higher order terms in $U$ can be identified with the second term in (\ref{skyrme}) with $g^2 \sim 1/N_c$. The Sakai-Sugimoto model thus succeeds in reproducing the phenomenological Skyrme Lagrangian for pions  (\ref{skyrme}). 

We now have all the ingredients to numerically fix the remaining parameters $M_K$, $u_0$ (thus $L$) and $\kappa$.
First, we determine $\kappa(M_K,u_0)$ by demanding the constituent quark mass to be 0.310 GeV,
\begin{equation}  \label{mq}
m_q(M_K,u_0,\kappa) =8\pi^2 M_K^2 \kappa \int_{1/M_K}^{u_0} du \frac{1}{\sqrt{1 - \frac{1}{(M_Ku)^3}}} = 0.310 \mbox{ GeV} \Longrightarrow \kappa(M_K,u_0).
\end{equation}
Then we use the experimental value for the pion decay constant to find $u_0(M_K)$,
\begin{eqnarray}
f_\pi(M_K,u_0,\kappa(M_K,u_0))  =  f_\pi(M_K,u_0) = 0.093 \mbox{ GeV} \Longrightarrow u_0(M_K).
\end{eqnarray}
Finally, we solve the eigenvalue equation (\ref{finiteTeigwvgl}), which is now a function of  $M_K$ only, for $m_{n=1}$. We refer the reader to appendix \ref{appendixnum} for more details.
The value of $M_K$ is then determined such that $m_{n=1} = m_\rho = 0.776$ GeV,
the $\rho$ meson being the lightest meson in the vector meson tower. One identifies  $B_\mu^{(n=1)a}$ with $\rho_\mu^{a}$ ($a=1,2$), which can be recombined into the charged $\rho_{\mu}^{\pm}$ mesons, $B_\mu^{(n=1)3}$ with the neutral $\rho_\mu^{0}$ meson, and $B_\mu^{(n=1)0}$ with the $\omega_\mu$ meson. From these identifications it follows that $m_\rho = m_\omega = m_{n=1}$ in the Sakai-Sugimoto model.\\

The results of our numerical analysis are
\begin{equation}
M_K \approx 0.7209 \mbox{  GeV}, \quad  \frac{u_0}{u_K} \approx 1.38 \quad
\mbox{and}\quad \kappa  = \frac{\lambda N_c}{216 \pi^3} \approx 0.006778,
\end{equation}
or
\begin{equation} \label{values}
M_K \approx 0.7209 \mbox{  GeV}, \quad L \approx  1.574 \mbox{  GeV}^{-1} \quad
\mbox{and}\quad \kappa  = \frac{\lambda N_c}{216 \pi^3} \approx 0.006778,
\end{equation}
where we used the formula (\ref{Lconf}) describing the one-to-one relation between $L$ and $u_0$.
The value found for $L$ is
approximately 2.8 times smaller than the maximum value of $L$, given by
\begin{eqnarray}
L_{max} = \frac{\delta \tau}{2} = \frac{\pi}{M_K} \approx 4.358 \mbox{  GeV}^{-1}. \label{Lmax}
\end{eqnarray}

From the values (\ref{values}) we do extract a relatively large 't Hooft coupling, $\lambda\approx 15$. This allows us to ignore the Chern-Simons part of the action, which is a factor $\lambda$ smaller than the DBI-part. Nevertheless we will 
comment on the contributions to the $\rho$ meson mass equation originating from the Chern-Simons action in section \ref{CSpion}. We also remark that for these values of the holographic parameters the numerical value for the effective string tension between a quark and an antiquark in this background is given by (see eq.\ (\ref{Teff})) 
\begin{equation}
\sigma = \frac{1}{2\pi\alpha'} \sqrt{-g_{00}(u_K) g_{11}(u_K)} \approx 0.19 \text{ GeV}^2. 
\end{equation}
This value is in excellent agreement with the value calculated on the lattice for pure SU(3) QCD, $\sigma \approx 0.18$-$0.19 \text{ GeV}^2$, as reported in \cite{Bali:1992ru,Sommer:1993ce}. This is a nice illustration that the fixed values do have a reasonable QCD resemblance. In fact, it means we could equally well have used the QCD string tension as input parameter instead of the constituent quark mass. 
(Neither of them are QCD observables, but in contrast with the constituent mass, $\sigma$ can be fixed from ab initio lattice computations.)

\subsection{Turning on a uniform magnetic field }\label{C}
Under gauge transformations $g\in U(N_f)$, the flavour gauge field transforms as
\begin{eqnarray} \label{gaugetransf}
A_m(x^m) \rightarrow g A_m(x^m) g^{-1} + g \partial_m g^{-1} \quad (m=\mu,z).
\end{eqnarray}
Since we have assumed the eigenfunctions $\psi_n (n\geq 1)$ to be normalizable ($\psi_n(z\rightarrow \pm \infty) = 0$), the expansion (\ref{gaugefieldexpansion}) implicitly assumes we are working in the gauge $A_\mu(z\rightarrow \pm \infty) = 0$. The gauge potential can be made to vanish asymptotically by applying a gauge transformation $g(x ^\mu,z) = U(x^\mu,z)$, that cancels the asymptotic pure gauge configuration
\begin{equation} \label{puregauge}
 A_m(x^\mu,z\rightarrow \pm \infty) = U_{\pm}^{-1}(x^\mu,z) \partial_m U_\pm(x^\mu,z)
\end{equation}
which ensures a finite effective four-dimensional action. For arbitrary $N_f>2$, the homotopy group for the functions $U:\mathbb R_4 \cup \infty \simeq S_4  \rightarrow U(N_f)$, $x^\mu \rightarrow U(x^\mu,z)$ is trivial, $\pi_4(U(N_f))=0$, so a
continuously interpolating $U(N_f)$-valued function $U(x^\mu,z)$ that fulfills $U(x^\mu,z\rightarrow \pm \infty)=U_\pm(x^\mu,z)$ can always be found. The case $N_f=2$, which we consider, is an exception since $\pi_4(U(2)) = \mathbb{Z}_2$. In the seminal paper of Sakai and Sugimoto \cite{Sakai:2004cn}, it was assumed that $N_f\neq2$. Yet it appears to be still possible to consider the gauge $A_m(z\rightarrow \pm \infty) = 0$. For $N_f=2$, there will exist a 2 by 2 matrix function $U(x^\mu,z)$ interpolating between $U_+$ and $U_-$ (if they are homotopic) \'or between $U_+$ and $\widetilde U_-$ (if $U_+$ and $U_-$ are not homotopic), with $\widetilde U_-$ defined as $-U_-$, so that  $\widetilde U_-$ is homotopic to the $\mathbb{Z}_2$ element $\mp 1$ if $U_-$ is homotopic to the $\mathbb{Z}_2$ element $\pm 1$. 
Since the asymptotic pure gauge configuration (\ref{puregauge}) also equals $\widetilde U_{\pm}^{-1}(x^\mu,z) \partial_m \widetilde U_\pm(x^\mu,z)$, 
the gauge transformation $g(x ^\mu,z) = U(x^\mu,z)$ will again cancel the asymptotic gauge potential. This argument\footnote{We thank J.~Van Doorsselaere for discussion on this point.} extends the validity of the original Sakai-Sugimoto reasoning to the $N_f=2$ case.  

One does not leave the gauge $A_m(z\rightarrow \pm \infty) = 0$  under gauge transformations $h\in U(N_f)$ that adopt $x^\mu$-independent boundary values 
\begin{equation}
(h_+, h_-) = (\lim_{z \rightarrow + \infty} h, \lim_{z \rightarrow - \infty} h). 
\end{equation}
These boundary values of the residual gauge symmetry transformation $h$ are interpreted as a global chiral symmetry transformation 
\begin{equation}
(h_+, h_-) = (h_L,h_R) \in U(N_f)_L \times U(N_f)_R   \label{SSMchiralel}
\end{equation} 
in the dual QCD-like theory. By  ``lightly'' gauging this chiral symmetry, i.e.~making $h_L=h_R=h(z\rightarrow \pm \infty)=h$ dependent on $x^\mu$, 
one leaves the $A_m(z\rightarrow \pm \infty) = 0$ gauge, the boundary value of the gauge field $A_m(z\rightarrow \pm \infty)$ to be interpreted as an external background vector field $\overline A_\mu$ in the boundary field theory coupling to the quarks through a covariant derivative $\mathcal D_\mu = \partial_\mu + \overline A_\mu$ such that the Dirac action $\overline{\psi} i \gamma_\mu \mathcal D_\mu \psi$ remains invariant under local $U(N_f)$ transformations. 

To turn on an electromagnetic background field $A_\mu^{em}$ in the boundary
field theory we put ($e$ being the electromagnetic coupling constant and $Q_{em}$ the electric charge matrix)
\begin{equation} \label{Aachtergrond}
\overline A_\mu = -i e Q_{em} A_\mu^{em} = -i e  \left(
\begin{array}{cc} 2/3 & 0 \\ 0 & -1/3 \end{array} \right) A_\mu^{em} = -i e
\left( \frac{1}{6} \textbf{1}_2 + \frac{1}{2} \sigma_3 \right)  A_\mu^{em},
\end{equation}
which assigns the appropriate charge to the up- and down-quark. For the case of a constant external magnetic field along the $x_3$-direction in
the boundary field theory ($F_{12}^{em} = \partial_1 A^{em}_2 = B$), this amounts
to setting
\begin{equation}
\overline A_\mu = -i e Q_{em} x_1 B \delta_{\mu2}
=  -i x_1  \left(
\begin{array}{cc}  \frac{2}{3}eB & 0 \\ 0 &  -\frac{1}{3}eB \end{array} \right) \delta_{\mu2}
= \frac{ x_1 e B
\delta_{\mu2}}{3} \left(  -\frac{i\textbf{1}_2}{2} \right)   +  x_1 e B
\delta_{\mu2} \left(-\frac{i\sigma_3}{2}  \right),
\end{equation}
or
\begin{equation}
\overline A_2^3 = x_1 e B  \quad \mbox{   and   } \quad \overline A_2^0 =
\overline A_2^3 / 3
\end{equation}
and
\begin{eqnarray} 
\overline F_{12}= \partial_1 \overline A_2 =  -i \left(\begin{array}{cc} \frac{2}{3}eB  & 0 \\ 0 & -\frac{1}{3}eB \end{array} \right) = -i \left(\begin{array}{cc} \overline F_u  & 0 \\ 0 & \overline F_d \end{array} \right),  \label{Fbardef}
\end{eqnarray}
where in the last line we defined the up- and down-components of the background field strength, $\overline F_u$ and $\overline F_d$.

\subsection{Effect of the magnetic field on the embedding of the probe branes} \label{D}
We determine in this section the $eB$-dependence of the embedding of the flavour D8-branes in the confi\-ning (we are working at zero temperature) D4-brane background \eqref{D4SS}. On each of the D8-branes lives an induced metric
\begin{eqnarray}
\hspace{-1cm}
ds^2_{D8} &=& g_{mn}^{D8} dx^m dx^n \quad(m,n = 0 \cdots 8) \nonumber \\
&=& \left(\frac{u}{R}\right)^{3/2} \eta_{\mu\nu}dx^\mu dx^\nu + \left( \left(\frac{R}{u}\right)^{3/2} \frac{1}{f(u)} +  \left(\frac{u}{R}\right)^{3/2} \frac{f(u)}{u'^2} \right) du^2 + \left(\frac{R}{u}\right)^{3/2} u^2 d\Omega_4^2
\end{eqnarray}
or
\begin{eqnarray}
\left( g_{00}^{D8}, g_{ii}^{D8}, g_{uu}^{D8} \right) &=& \left( -\left(\frac{u}{R}\right)^{3/2}, \left(\frac{u}{R}\right)^{3/2}, \left(\frac{u}{R}\right)^{3/2} \left[ \frac{1}{f} \left(\frac{u}{R}\right)^{-3} + \frac{f}{u'^2}\right] \right) \\
&=& \left( g_{00}, g_{ii}, G_{uu} \right) \quad \text{with $G_{uu} = g_{uu} + g_{\tau\tau}(\partial_u \tau)^2$}
 \end{eqnarray}
and a gauge field $A_\mu$, for which we assume the background gauge field ansatz
\begin{equation} \label{background}
A_\mu = \overline A_\mu =  -i e Q_{em} x_1 B \delta_{\mu 2}
\quad \mbox{(all other gauge field components zero)},
\end{equation}
modeling an external magnetic field $\vec B=B \vec e_3$ in the dual field theory. 

We plug the gauge field ansatz into the non-Abelian DBI-action
\begin{equation}
S_{DBI} = -T_8 \int d^4x \hspace{1mm}2 \int du \int \epsilon_4 \hspace{1mm} e^{-\phi}\hspace{1mm} \text{STr}
\sqrt{-\det \left[g_{mn}^{D8} + (2\pi\alpha') iF_{mn} \right]}
\end{equation}
and solve for the embedding $u'=du/d\tau$ as a function of $eB$. To allow for the possibility that each of the two flavour D8-branes responds differently to the external magnetic field, we assume the following form of the metric in flavour space
\begin{align}
g^{D8}
&= \left( \begin{array}{cc} g^{D8}_u & 0 \\ 0 & g^{D8}_d \end{array} \right),
\end{align}
the only difference between $g^{D8}_u$ and $g^{D8}_d$ being that the $u$-coordinate appearing in $g^{D8}_u$ follows the up-brane, varying from $u_{0,u}$ to infinity, whereas the the $u$-coordinate appearing in $g^{D8}_d$ follows the down-brane, varying from $u_{0,d}$ to infinity. This will turn out to generate a different embedding $u'(eB)$ for up and down. We write
\begin{eqnarray} g^{D8} = g^{D8}  \mathbf 1', \label{metricnoncoincident}\end{eqnarray}
where we introduced the notation  
\begin{eqnarray}\mathbf 1' = \left( \begin{array}{cc} \theta(u-u_{0,u}) & 0 \\ 0 & \theta(u-u_{0,d}) \end{array} \right)\end{eqnarray}
for our ``generalized unity matrix'' that indicates that everything multiplied by the Heaviside step function $\theta(u-u_{0,u})$ (respectively $\theta(u-u_{0,d})$) will have to be integrated over $u$ varying from $u_{0,u}$ (respectively $u_{0,d})$ to infinity. 

The determinant in the action is
\begin{align} \label{detmatrix}
\det(g^{D8}\mathbf 1' + i (2\pi\alpha')F) &= \det \left(\begin{array}{ccccc} g_{00}\mathbf 1' &0&0&0&0 \\
0 & g_{11}\mathbf 1' & i (2\pi\alpha')\overline F_{12} &0&0\\
0 & -i (2\pi\alpha')\overline F_{12} & g_{22}\mathbf 1' &0&0\\
0&0&0& g_{33}\mathbf 1' &0\\
0&0&0&0& G_{uu} \mathbf 1'
\end{array} \right)_{\text{Lorentz space}} \hspace{-1.5cm} \times \det S_4 \nonumber \\
&=  \underbrace{\det S_4 \times g_{00}g_{11}g_{22}g_{33}G_{uu}}_{\det g} \underbrace{\left(1 - (2\pi\alpha')^2 g_{11}^{-1} g_{22}^{-1} \overline F_{12}^2\right) }_{A} \mathbf 1'  \nonumber \\
&\hspace{-2cm} = \left(\begin{array}{cc}
 \det g \times A_u \times \theta(u-u_{0,u})   & 0 \\
0 &  \det g \times A_d \times \theta(u-u_{0,d}) \end{array} \right)_{\text{flavour space}},
\end{align}
where we defined a new matrix $A$ that collects all the $B$-dependence: 
\begin{eqnarray} \label{A}
A=\left( \begin{array}{cc} A_u & 0 \\ 0 &  A_d \end{array} \right) = 1 - (2\pi\alpha')^2 \overline F_{12}^2  \left(\frac{R}{u}\right)^{3}, \quad
A_{l} = 1 + (2\pi\alpha')^2 \overline F_{l}^2  \left(\frac{R}{u}\right)^{3}, \quad (l=u,d).
\end{eqnarray}
Since the matrix (\ref{detmatrix}) is diagonal, the square root of it is equal to the matrix of the square roots of the diagonal elements and the STr reduces to an ordinary Tr, leading to the action
\begin{equation}
S_{DBI}
= -T_8 V_{\mathbb{R}^{3+1}}  V_4 g_s^{-1} \left[ 2\int_{u_{0,u}}^\infty du \hspace{1mm} u^4
\sqrt{ \frac{1}{f} \left(\frac{u}{R}\right)^{-3} + \frac{f}{u'^2} } \sqrt{A_u} +
2\int_{u_{0,d}}^\infty du \hspace{1mm} u^4
\sqrt{ \frac{1}{f} \left(\frac{u}{R}\right)^{-3} + \frac{f}{u'^2} } \sqrt{A_d}  \right],
\end{equation}
with $V_{\mathbb{R}^{3+1}}=\int d^4x$ and $u'$ as a function of $eB$ to be determined for both the up- and down-brane. 

Omitting all the up- and down-indices for clearness,
\begin{align}
S_{DBI} &= S_{up} + S_{down} \nonumber\\
S &=  -T_8 V_{\mathbb{R}^{3+1}} V_4 g_s^{-1}  \hspace{1mm} 2\int_{u_{0}}^\infty du \hspace{1mm} u^4 \sqrt{ \frac{1}{f} \left(\frac{u}{R}\right)^{-3} + \frac{f}{u'^2} } \sqrt{A}   \label{SDBIconf}
\end{align}
and using the short-hand
\begin{eqnarray}
\mathcal L^\tau
= u^4  \sqrt{\frac{u'^2}{f} \left(\frac{u}{R}\right)^{-3} + f} \hspace{1mm} \sqrt{A}\,\,, \end{eqnarray}
we determine $u'$ for each flavour from the conserved ``Hamiltonian''
\begin{eqnarray} H = u' \frac{\delta \mathcal L^\tau}{\delta u'} - \mathcal L^\tau =  \frac{-u^4 f \sqrt{A}}{\sqrt{\frac{u'^2}{f} \left(\frac{u}{R}\right)^{-3} + f}}\,\,, \quad \partial_\tau H = 0. \end{eqnarray}
Expressing that this $H$ is conserved and assuming a $\cup$-shaped embedding $u' = 0$ at $u=u_{0}$ (with $A(u_0)$ and $f(u_0)$ denoted as $A_0$ and $f_0$):
\begin{eqnarray}
\frac{-u^4 f \sqrt{A}}{\sqrt{\frac{u'^2}{f} \left(\frac{u}{R}\right)^{-3} + f}}
=
\frac{-u_0^4 f_0 \sqrt{A_0}}{\sqrt{f_0}}
\end{eqnarray}
we find
\begin{equation}
u'^2 = \left(\frac{u}{R}\right)^{3} f^2 \frac{u^8 f A - u_0^8 f_0 A_0}{u_0^8 f_0 A_0},
\end{equation}
reducing to the known $\cup$-shaped embedding (\ref{dtaudu}) for $eB\rightarrow 0$ whereby $A\rightarrow 1$.

\subsubsection[Antipodal embedding]{Antipodal embedding ($u_0=u_K$): no dependence on $eB$}

In the case $u_0=u_K$, we have $f_0=0$ so
\begin{eqnarray}
(\partial_u \tau)^2 = \left(\frac{R}{u}\right)^{3} \frac{1}{f^2} \frac{u_0^8 f_0 A_0}{u^8 f A - u_0^8 f_0 A_0} = 0
\end{eqnarray}
and the embedding function is constant,
\begin{eqnarray}
\tau(u) = \overline \tau \sim \mathbf 1,
\end{eqnarray}
($\overline \tau = 0$ for the D8-branes and $\overline \tau=\pi/M_K \mathbf 1$ for the $\overline{\text{D8}}$-branes, see the l.h.s.~of figure  \ref{SS}), independent of the value of the magnetic field. In this case, there is thus no response of the chiral symmetry breaking to the magnetic field, a somewhat unphysical feature of the extremal Sakai-Sugimoto embedding, which is a direct consequence of the absence of a constituent quark mass in this setting.

\subsubsection[Non-antipodal embedding]{Non-antipodal embedding ($u_0>u_K$): magnetic catalysis of chiral symmetry breaking} \label{testchiralmagncat}

\begin{figure}[h!]
  \centering
  \scalebox{1}{
  \includegraphics{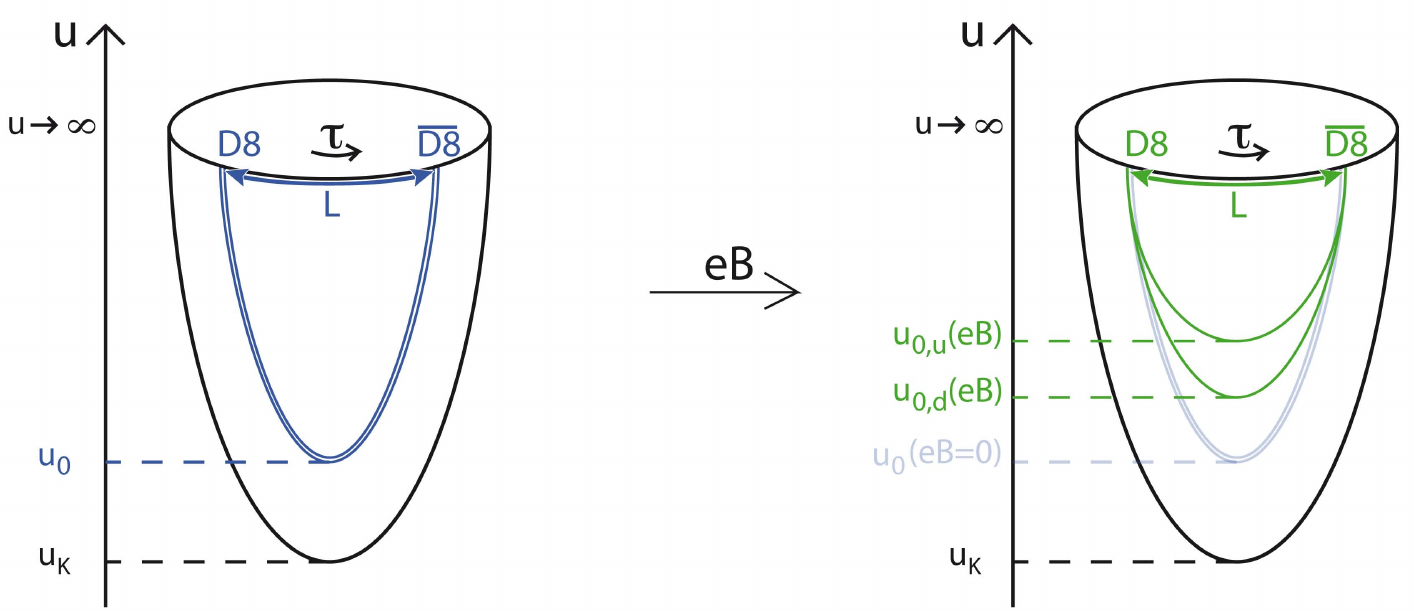}}
  \caption{The change in embedding of the flavour branes caused by the magnetic field $eB$ models the chiral magnetic catalysis effect. The up-brane reacts the strongest to $eB$, corresponding to a stronger chiral magnetic catalysis for the up-quarks than for the down-quarks.}\label{changedembedding}
\end{figure}

In the case $u_0 > u_K$ the embedding function for the flavour branes in the background is given by
\begin{eqnarray}
\tau(u) = \overline \tau =\left( \begin{array}{cc} \overline \tau_u & 0 \\ 0 &  \overline \tau_d \end{array} \right)
\end{eqnarray}
with
\begin{equation} \label{tauembedding} 
\partial_u \overline \tau_l = \sqrt{ \left(\frac{R}{u}\right)^{3} \frac{1}{f^2} \frac{u_{0,l}^8 f_{0,l} A_{0,l}}{u^8 f A_l - u_{0,l}^8 f_{0,l} A_{0,l}}}  \times \theta(u - u_{0,l}), \quad (l=u,d)
\end{equation}
The up-brane and down-brane are thus no longer coincident in the presence of a magnetic field, as sketched in figure \ref{changedembedding}.   

The $eB$-dependent induced metric on the up- and down-brane is given by
\begin{eqnarray} \left( g_{00}^{D8}, g_{ii}^{D8}, g_{uu}^{D8} \right) = \left( -\left(\frac{u}{R}\right)^{3/2}, \left(\frac{u}{R}\right)^{3/2}, \left(\frac{R}{u}\right)^{3/2} \gamma_B(u) \right)  \end{eqnarray}
and the action by
\begin{align}
S_{DBI} &= S_{up} + S_{down}, \nonumber\\
S &=  -T_8 V_{\mathbb{R}^{3+1}} V_4 g_s^{-1} \hspace{1mm} 2 \int_{u_0}^\infty du \hspace{1mm}  R^{3/2} u^{5/2} \sqrt{A} \sqrt{\gamma_B}  \label{actionmagneticfield}
\end{align}
with
\begin{eqnarray} \label{gammaB}
\gamma_B(u) = \frac{u^8 A}{u^8 f A - u_0^8 f_0 A_0}.
\end{eqnarray}
We see that the non-Abelian DBI-action for the two D8-branes in the presence of a background magnetic field reduces to the sum of two Abelian actions. This represents the explicit breaking of the global chiral symmetry,
\begin{eqnarray}
U(2)_L \times U(2)_R \stackrel{eB}{\rightarrow} (U(1)_L \times U(1)_R)^u \times (U(1)_L \times U(1)_R)^d,
\end{eqnarray}
caused by the up- and down-quarks' different coupling to the magnetic field.

The asymptotic separation $L$ between D8- and $\overline{\mbox{D8}}$-branes
as a function of the magnetic field is
\begin{align}
L     &= 2 \int_{u_0}^\infty du  \left(\frac{R}{u}\right)^{3/2} f^{-1} \sqrt{\frac{u_0^8 f_0  A_0}{u^8 f A - u_0^8 f_0  A_0}} \nonumber \\
   &= \frac{2}{3} \frac{R^{3/2}}{\sqrt {u_0}} \sqrt{f_0 A_0} \int_0^1 d\zeta \frac{f^{-1} \zeta^{1/2}}{\sqrt{fA - f_0 A_0 \zeta^{8/3}}} \label{LconfifvB}
\end{align}
where we changed the integration variable to $\zeta = (u/u_0)^{-3}$ \cite{Johnson:2008vna},
with $y_K = u_K/u_0$, $y=u/u_0$ and $f = 1 - y_K^3 \zeta$. 
In section \ref{B} the value of the geometric parameter $L$ in zero magnetic field was determined at $L(eB=0) = 1.547$ GeV$^{-1}$.
We keep $L = L(eB=0)$ fixed
while varying $eB$ to determine $u_{0,u/d}(eB)$ and consequently, via
\begin{equation}\label{masscont}
m_q(M_K,u_0,\kappa) =8\pi^2 M_K^2 \kappa \int_{1/M_K}^{u_0} du \frac{1}{\sqrt{1 - \frac{1}{(M_Ku)^3}}},
\end{equation}
the constituent quark masses $m_{u}(eB)$ and $m_d(eB)$ of up- and down-quarks. 
$L$ serves as the boundary condition on the branes' embedding (see section \ref{subs:spectrum}). 
From the perspective of the asymptotic dual field theory, the flavour branes are infinitely extended, massive objects in the bulk, requiring an infinite amount of energy to move them. In this sense it is natural to keep $L$ fixed as a boundary condition to probe the effects of the bulk dynamics in the presence of the external field. Moreover, the value of $L$ determines how much of the gluonic bulk dynamics is probed, ranging from all ($u_0=u_K$) for maximal $L$ to none ($u_0\rightarrow \infty$) for minimal $L$. In this interpretation, the choice of $L$ (which has no direct physical meaning in the dual field theory) corresponds to the choice of type of dual field theory, ranging from QCD-like to NJL-like in the limit of $L\rightarrow 0$ or $\tau$ non-compact \cite{Antonyan:2006vw}. 
Keeping $L$ fixed at $L(eB=0)$ is a choice also made in for example the work of Preis et~al.~\cite{Bergman:2008sg}.

\begin{figure}[h!]
  \hfill
    \begin{center}
      \scalebox{1.2}{
  \includegraphics{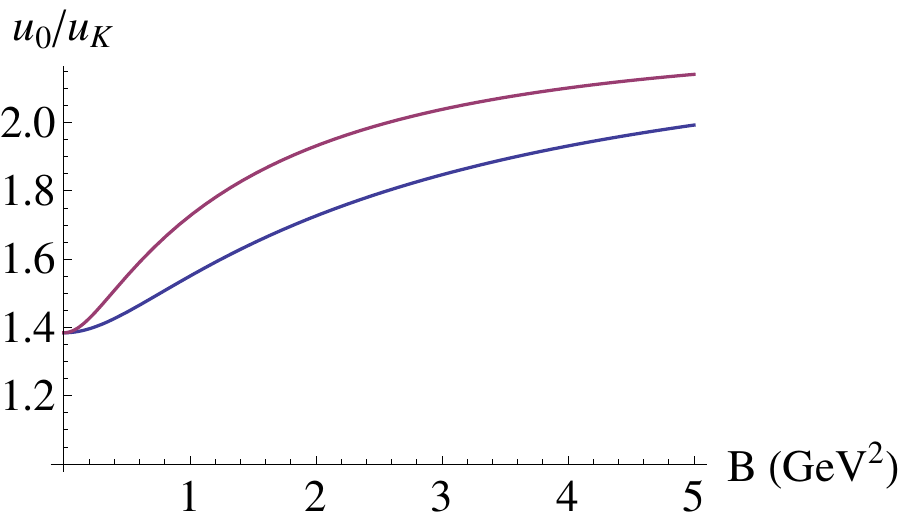}}
     \end{center}
      \caption{$\frac{u_0}{u_K}$ as a function of the magnetic field for the D8-brane corresponding to the up-quark, and the one corresponding to the down-quark.}
	\label{u0}
  \hfill
\end{figure}

\begin{figure}[h!]
  \hfill
  \begin{minipage}[t]{.45\textwidth}
    \begin{center}
      \scalebox{0.7}{
  \includegraphics{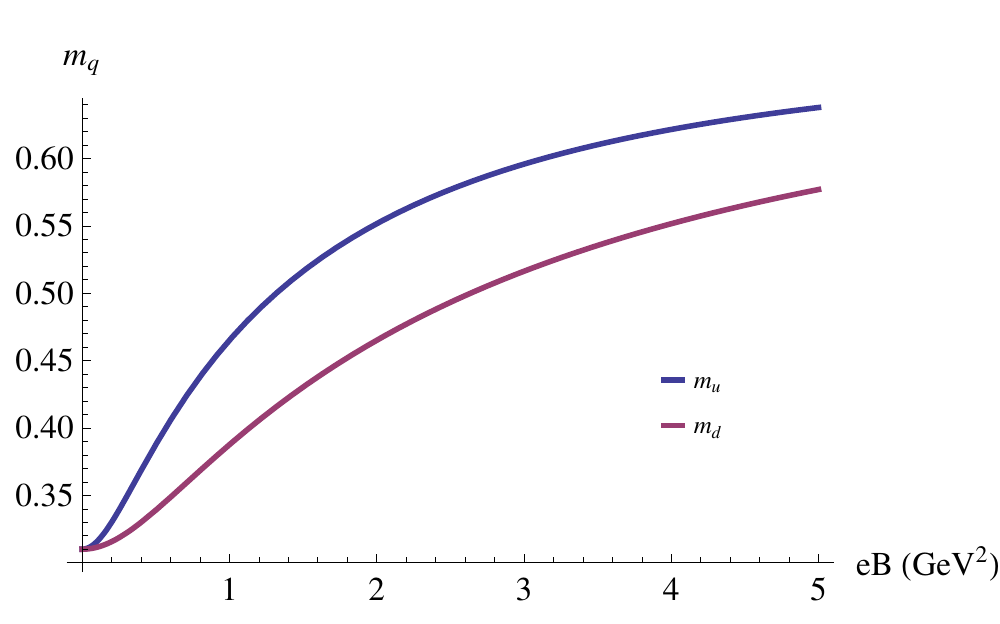}}
    \end{center}
  \end{minipage}
  \hfill
  \begin{minipage}[t]{.45\textwidth}
    \begin{center}
      \scalebox{0.7}{
  \includegraphics{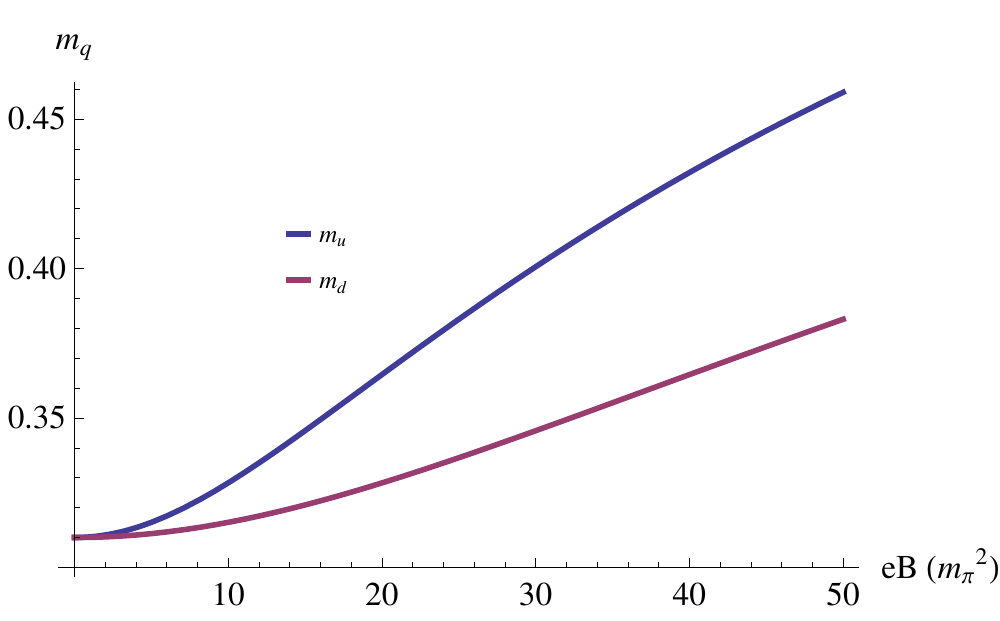}}
    \end{center}
  \end{minipage}
      \caption{The constituent masses of the up-quark and the down-quark as a function of the magnetic field (in units GeV$^{2}$ and $m_\pi^2 = (0.138$ GeV)$^2$).}
	\label{mqfig}
  \hfill
\end{figure}

In figures \ref{u0} and \ref{mqfig} the numerically obtained dependence on $eB$ of $u_{0,u/d}$, $m_{u}$ and $m_d$ are depicted. As $u_0$ rises with $eB$, the probe branes in the presence of the external magnetic field get more and more bent
towards each other, driving them further away from the chirally invariant situation of straight branes, see figure \ref{changedembedding}. This feature corresponds to a holographic modeling of the magnetic catalysis of chiral symmetry breaking \cite{Miransky:2002rp,Shushpanov:1997sf}: a magnetic field boosts the chiral symmetry breaking and hence the constituent quark masses. More precisely, the authors of \cite{Shushpanov:1997sf} discuss a low-energy theorem in the context of chiral perturbation theory, thereby finding that the chiral condensate grows (linearly) in terms of an increasing magnetic field, with the coefficient a function of the pion decay constant $f_\pi$. 
This ``\emph{chiral magnetic catalysis effect}'' was already discussed for the Sakai-Sugimoto model in \cite{Johnson:2008vna}, 
for a single flavour and without matching the free parameters onto QCD values. The constituent quark masses $m_q(eB)$, which are related to the difference 
$u_0(eB) - u_K$, accordingly increase (see figure \ref{mqfig}), leading us to expect that taking this chiral magnetic catalysis into account will translate into a corresponding increase in the $\rho$ meson mass and thus $B_c$.

\begin{figure}[h!]
  \hfill
  \begin{minipage}[t]{\textwidth} 
    \begin{center}
      \scalebox{1}{
  \includegraphics{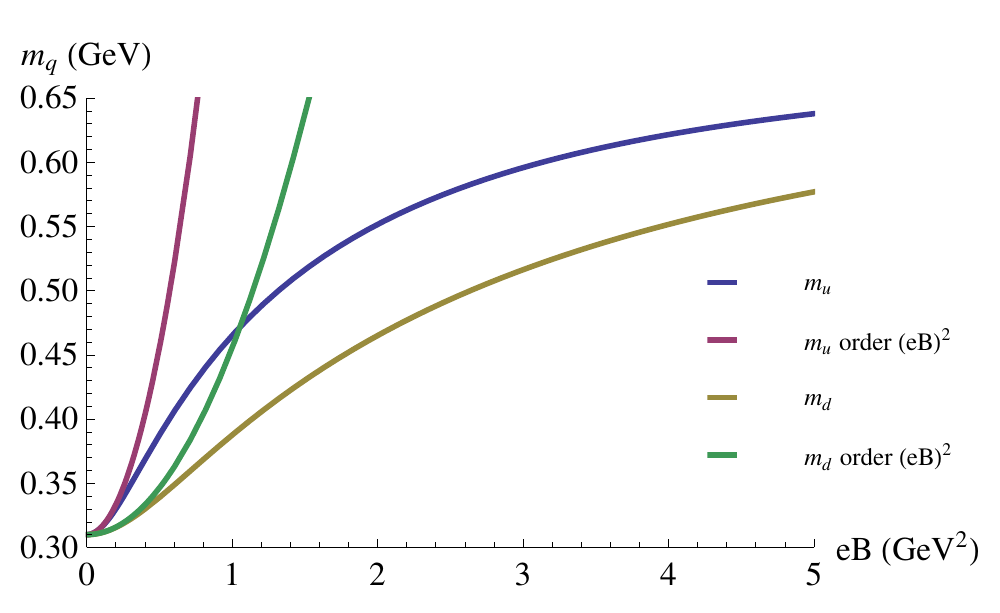}}
    \end{center}
  \end{minipage}
  \hfill
  \begin{minipage}[t]{\textwidth}
    \begin{center}
      \scalebox{1}{
  \includegraphics{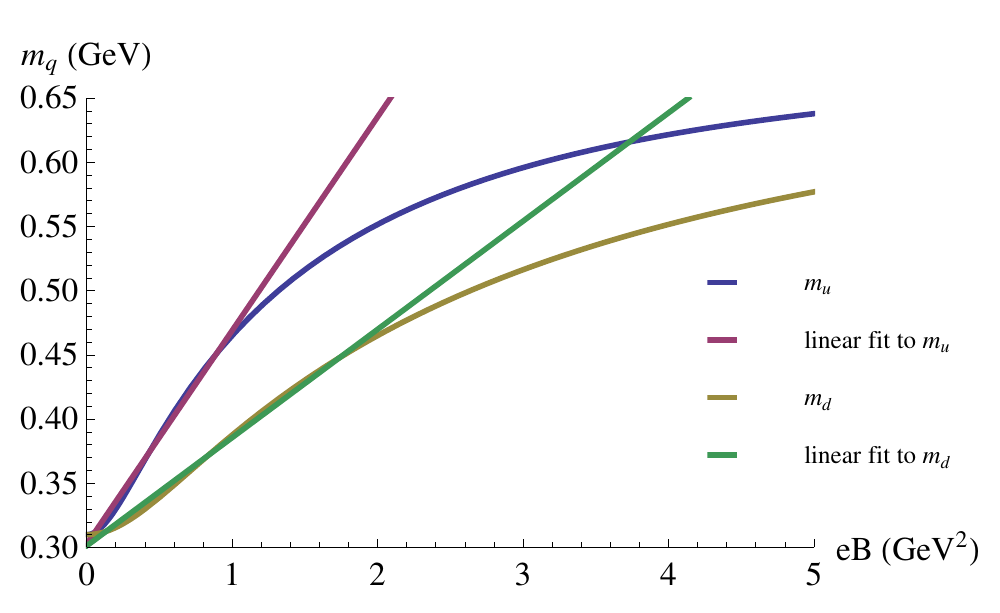}}
    \end{center}
  \end{minipage}
      \caption{Left: the constituent up-quark and down-quark masses and their quadratic approximation \eqref{kwad}. Right: a linear fit \eqref{fit} to the same quantities.}
	\label{mqkwadr}
  \hfill
\end{figure}

For small magnetic fields, we can compute the lowest order correction to $m_q(eB=0)$ analytically, confirming the typical holographic $(eB)^2$ dependence \cite{Zayakin:2008cy}.
To this end we approximate $L$ by $L = l_0(u) + (eB)^2  \, l_1(u)$ + higher order terms in $B$. The condition $l_0(u) + (eB)^2 \, l_1(u)  = L(eB=0)$ then has a solution of the form $u = u_0 + (eB)^2 \, u_1$ with $u_0 =
1.38 \, u_K = 1.92 \text{ GeV}^{-1}$ and $u_1 = -l_1(u_0) \left(\frac{dl_0}{du}(u_0)\right)^{-1}$. The corresponding constituent quark masses for small magnetic fields are
\begin{align}
 m_q(eB) &= 0.310 \mbox{  GeV} + 8\pi^2 M_K^2 \kappa \int_{u_0}^{u_0 + (eB)^2 \, u_1} du \frac{1}{\sqrt{1 - \frac{1}{(M_Ku)^3}}} \nonumber\\
&= 0.310 \mbox{  GeV} + (8\pi^2 M_K^2 \kappa) (eB)^2 \, u_1 \frac{1}{\sqrt{1 - \frac{1}{(M_K u_0)^3}}}.
\end{align}
We find
\begin{equation}\label{kwad}
m_u(eB) =  0.310 \mbox{  GeV} + 0.582 ~ (eB)^2 + \mathcal{O}(eB)^3 \quad \mbox{ and } \quad m_d(eB) =  0.310 \mbox{  GeV} + 0.145 ~ (eB)^2+ \mathcal{O}(eB)^3,
\end{equation}
depicted in figure \ref{mqkwadr}. This quadratic dependence at small magnetic field is also encountered in other effective descriptions of the constituent quark mass, as in the $\text{PLSM}_q$ model of \cite{Mizher:2011wd}, and although not explicitly mentioned, also an instanton based computation seems to give a quadratic-like power \cite{Nam:2011vn}. Also the numerical data of \cite{D'Elia:2011zu} for the up- and down-quark chiral condensates are in accordance with a quadratic behaviour at small $eB$. We must however mention that the latter lattice computations were done at nonvanishing current quark mass in an unquenched setting. Chiral perturbation theory predicts a linear behaviour \cite{Shushpanov:1997sf} (see also the comments in \cite{Zayakin:2008cy}). Quenched lattice simulations of \cite{Buividovich:2008wf} confirmed this, although we notice that the small $eB$-behaviour does not seem to be precisely caught by the proposed linear fit. In fact, we are able to fit our result quite well with a linear fit if $eB$ is not too large, see figure \ref{mqkwadr}. We used
\begin{equation}\label{fit}
m_u(eB)^{linear} \approx 0.303~\text{GeV}+0.166 ~ eB \quad \mbox{ and } \quad m_d(eB)^{linear} \approx  0.301~\text{GeV} + 0.084 ~ eB.
\end{equation}
It is also instructive to see what happens at large magnetic field. As already pointed out in \cite{Johnson:2008vna}, we observe a saturation in figure \ref{mqfig}. Such a saturation was also seen for the first time using lattice simulations in \cite{D'Elia:2011zu} for the up- and down-quark chiral condensates, carefully taken into account some subtleties related to an unphysical periodicity in the results, which is a typical lattice artefact. The results of \cite{Nam:2011vn,D'Elia:2011zu} anyhow confirm the different response of the constituent up- and down-quark masses or chiral condensates to the magnetic field, a feature which we also reproduced here for the first time in the holographic Sakai-Sugimoto setting.  It thus appears that our holographic results reproduce quite well the phenomenology of independent quenched QCD calculations. The unquenched top-of-the-bill simulations of \cite{Bali:2012zg} show a similar behaviour for the chiral condensate, at least at vanishing temperature. Similarly shaped curves for the chiral condensate extrapolated to the chiral limit can be found in \cite{Ilgenfritz:2012fw}, which would be most relevant for comparison with our analysis, nevertheless keeping in mind that the lattice study \cite{Ilgenfritz:2012fw} is for 2 rather than 3 colours.

\section{Stability analysis} \label{stabilityfluctuations}

To investigate the stability of the set-up with respect to gauge and scalar field fluctuations, let us first derive the form of the action to second order in the fluctuations by plugging the total gauge field ansatz
\begin{equation} \label{fieldansatz}
\left\{\begin{array}{ll} A_r = \overline A_r + \tilde A_r  \quad (r=\mu,u) \\ \tau = \overline \tau + \tilde \tau  \end{array} \right.
\end{equation}
with (see (\ref{background}) and (\ref{tauembedding}))
\begin{equation}
\left\{\begin{array}{ll}  \overline A_\mu = -i e Q_{em} x_1 B \delta_{\mu2}
\\ \partial_u\overline \tau =  \sqrt{\left(\frac{R}{u}\right)^{3} \frac{1}{f^2} \frac{u_0^8 f_0 A_0}{u^8 f A - u_0^8 f_0 A_0}} \times \theta(u-u_0)
\end{array} \right.,
\end{equation}
into the DBI-action (\ref{nonabelian}).
The background components of the field ansatz (\ref{fieldansatz}) describe the background magnetic field (in $\overline A_\mu$) and the ($B$-dependent) embedding of the branes (in $\partial_u \overline \tau$). The fluctuation components will be related to resp.\ vector and scalar mesons in the dual field theory.

We have to evaluate (for notational brevity we temporarily absorb the factor $(2\pi\alpha')$ into the field strength)
\begin{equation} \label{tensionabsorbed}
2 \int du \hspace{2mm} \text{STr} \sqrt{-\det (a_{rs})} = 2 \int du \hspace{2mm}  \text{STr} \sqrt{-\det (g_{rs}^{D8} + i F_{rs})},
\end{equation}
with
\begin{align}
 g_{rs}^{D8} &= g_{rs} + g_{\tau\tau} D_r \tau D_s \tau, \qquad \mbox{with $D_r \cdot = \partial_r + [\overline A_r,\cdot]$}
 \end{align}
and
\begin{align}
F_{rs} &= \partial_r A_s - \partial_s A_r + [A_r, A_s].
\end{align}
If the argument $a$ of the determinant (which runs over the Lorentz-indices) is written as
\[ a = \overline a + a^{(1)} + a^{(2)} + \cdots \]
with  $a^{(n)}$ being $n$-th order in the fluctuations $\tilde A$,
the determinant can be expanded to second order in the fluctuations as follows
\begin{align}
\sqrt{-\det a}|_{\tilde A^2} &= \sqrt{-\det \overline a}  \left\{ 1+\frac{1}{2} \text{tr} (\overline a^{-1} a^{(1)}) + \frac{1}{8} \left( \text{tr}(\overline a^{-1} a^{(1)}) \right)^2 - \frac{1}{4} \text{tr}\left( (\overline a^{-1} a^{(1)})^2 \right) + \frac{1}{2} \text{tr} (\overline a^{-1} a^{(2)}) \right\}.  \label{expansion}
\end{align}
We denote the trace in Lorentz-space with a small $\text{tr}$, and the trace in flavour space with a capital (S)Tr.
Splitting each component of $a$ in its symmetric and antisymmetric parts
\begin{equation} \label{a}
\left\{ \begin{array}{ll} \overline a^{-1} = \mathcal G + \mathcal B \\
 					    a^{(1)} = \overline a^{(1)} + \delta_1 F \\
					    a^{(2)} = \overline a^{(2)} + \delta_2 F \\
\end{array} \right.
\end{equation}
the expansion of the determinant (\ref{expansion}) to second order in the fluctuations becomes
\begin{align}
\sqrt{-\det a} |_{\tilde A^2}
&= \sqrt{-\det \overline a} + \sqrt{-\det \overline a} \times \nonumber\\
 & \hspace{-1cm}\left\{ \frac{1}{2}  \text{tr} ( \mathcal G \overline a^{(1)})  + \frac{1}{8}  \left(\text{tr} ( \mathcal G \overline a^{(1)})\right)^2  - \frac{1}{4}  \text{tr} ( \mathcal G \overline a^{(1)}\mathcal G \overline a^{(1)} + \mathcal B \overline a^{(1)}\mathcal B \overline a^{(1)}) + \frac{1}{2}  \text{tr} ( \mathcal G \overline a^{(2)}) \right.  \nonumber\\
& \hspace{-1cm} +\frac{1}{2}  \text{tr} ( \mathcal B \delta_1 F)  + \frac{1}{8}  \left(\text{tr} (\mathcal B \delta_1 F)\right)^2  - \frac{1}{4}  \text{tr} (\mathcal G \delta_1 F\mathcal G \delta_1 F + \mathcal B \delta_1 F\mathcal B \delta_1 F) + \frac{1}{2}  \text{tr} (\mathcal B \delta_2 F)   \nonumber \\
 &\hspace{-1cm} \left. +\frac{1}{4} \text{tr} ( \mathcal G \overline a^{(1)})\text{tr} ( \mathcal B \delta_1 F) - \frac{1}{2} \text{tr}(\mathcal G \overline a^{(1)} \mathcal B \delta_1 F) - \frac{1}{2} \text{tr}(
 \mathcal G \delta_1 F \mathcal B \overline a^{(1)}) \right\}. \label{derdelijnscalarplusvector}
\end{align}
For our field ansatz we have
\begin{align}
\overline a_{rs} &= g_{rs} + g_{\tau\tau} \partial_r \overline \tau \partial_s \overline \tau +  i \overline F_{rs}, \\
\overline a^{(1)}_{rs} &= g_{\tau\tau} \left(\partial_r \overline\tau \left([\tilde A_s,\overline\tau] + D_s \tilde \tau \right) + \left([\tilde A_r,\overline\tau] + D_r \tilde \tau\right) \partial_s \overline\tau \right), \\
\delta_1 F_{rs} &= i (D_r \tilde A_s - D_s \tilde A_r) \stackrel{notation}{=} i \tilde F_{rs} \label{notF1}\\
\overline a^{(2)}_{rs} &= g_{\tau\tau}\left([\tilde A_r,\overline\tau]+D_r \tilde \tau\right) \left([\tilde A_s,\overline\tau]+D_s \tilde \tau\right) + g_{\tau\tau} \left([\tilde A_r,\tilde \tau] \partial_s \overline \tau+  \partial_r \overline \tau [\tilde A_r,\tilde \tau] \right), \\
\delta_2 F_{rs} &= i [\tilde A_r, \tilde A_s].
\end{align}
The symmetric part $\mathcal G$ of $\overline a^{-1}$ is diagonal,
\begin{equation}
\mathcal G = \left( \begin{array}{ccccc}g_{00}^{-1} & & & & \\ & g_{11}^{-1} A^{-1} & & & \\ & & g_{22}^{-1} A^{-1} & & \\ & & & g_{33}^{-1} & \\ & & & & G_{uu}^{-1} \end{array} \right), \quad \mbox{ with } G_{uu} = g_{uu}+g_{\tau\tau} (\partial_u \overline \tau)^2
\end{equation}
and the antisymmetric part $\mathcal B$ has non-zero components
\begin{equation}
\mathcal B_{12}=-\mathcal B_{21} = i \overline F_{12} g_{11}^{-1} g_{22}^{-1} A^{-1}.
\end{equation}
As a check, the first order terms in (\ref{derdelijnscalarplusvector}) do vanish on-shell, that is upon using the embedding function (\ref{tauembedding}). The DBI-Lagrangian to second order in the fluctuations then reads
\begin{align}
&\text{STr} \hspace{1mm} e^{-\phi} \sqrt{-\det a} |_{\tilde A^2,\tilde \tau^2,\tilde A \tilde \tau}
= \mathcal L_1 + \mathcal L_2  + \mathcal L_3 + \mathcal L_4     \label{STrDBI}
\end{align}
with
\begin{align}
& \mathcal{L}_1 = \text{Tr} \hspace{1mm} e^{-\phi}  \sqrt{-\det \overline a} \nonumber \\
& \mathcal{L}_2 = \text{STr} \hspace{1mm} \overline x  \left\{
\frac{1}{2} \left( [\tilde A_u,\overline \tau] +D_u\tilde\tau \right)^2 G_{uu}^{-2}
+ \overline y [\tilde A_u,\tilde \tau] + \frac{1}{2} \left( [\tilde A_\mu,\overline \tau] +D_\mu \tilde\tau \right)^2 g_{\mu\mu}^{-1} A^{-1}|_{\mu=1,2} G_{uu}^{-1} \right\} \nonumber \\
& \mathcal{L}_3 = \text{STr} \hspace{1mm} \overline x  \left\{ - \overline F_{12} g_{11}^{-1} g_{22}^{-1} A^{-1} [\tilde A_1,\tilde A_2]
- \frac{1}{4} g_{\mu\mu}^{-1} g_{\nu\nu}^{-1} A^{-2}|_{\mu,\nu=1,2}\tilde  F_{\mu\nu}^2 - \frac{1}{2} g_{\mu\mu}^{-1} A^{-1}|_{\mu=1,2} G_{uu}^{-1}\tilde F_{\mu u}^2\right\}  \nonumber \\
&\mathcal{L}_4=  \text{STr} \hspace{1mm} \overline x  \left\{
-\overline z \left( \left( [\tilde A_u,\overline \tau]+ D_u \tilde\tau \right) \tilde F_{12} + \left( [\tilde A_1,\overline \tau]+ D_1 \tilde\tau \right) \tilde F_{2u} - \left( [\tilde A_2,\overline \tau]+ D_2 \tilde\tau \right) \tilde F_{1u} \right)
\right\}, \label{STR}
\end{align}
where
\begin{equation}  \label{xbardef}
\overline x =  e^{-\phi} \sqrt{- \det \overline a} =  e^{-\phi} g_{11}^2 \sqrt{G_{uu}} g_{S_4}^2 \sqrt{A}, \qquad \overline y = G_{uu}^{-1}g_{\tau\tau}\partial_u \overline \tau, \qquad \overline z =  \overline y \overline F_{12}g_{11}^{-1} g_{22}^{-1} A^{-1}
\end{equation}
are functions of the background fields $\partial_u \overline \tau$ and $\overline F_{12}$, so functions of $u$ only, and diagonal in flavour space.
The notation for the factors $g_{\mu\mu}^{-1} A^{-1}|_{\mu=1,2}$ coming from $\mathcal G$ means that $g_{\mu\mu}^{-1}$ is accompanied with a factor $A^{-1} = \frac{1}{1 - (2\pi\alpha')^2 \overline F_{12}^2  R^3/u^3}$ only for $\mu=1,2$.

\subsection{Vector sector in simplified cases} \label{simpleholorho1} 

We will begin the discussion of the effective 4-dimensional rho meson mass, obtained from the mass equation for the 5-dimensional flavour gauge field, by considering some simplifying situations. That way we hope to convey the basic method (before it gets obscured by technical details) and the main physics, which is already captured in these results. 
The first assumption is to approximate the action to second order in the field strength, based on the argument that this is an expansion in $(2 \pi \alpha') \sim \frac{1}{\lambda}$ which is small for large 't Hooft coupling $\lambda$. This leads to a typical Tr $F^2$ action as is commonly used in bottom-up holographic models.  

The vector sector in the $(2 \pi \alpha')^2 F^2$-approximation will be discussed in full detail (for the general non-antipodal embedding) in a later section \ref{F2approx}. 
Here we only consider two simplifying set-ups within this approximation. First, we study the antipodal embedding $u_0=u_K$, in which case the embedding is $B$-independent and no constituent quark mass effects are present. Indeed we will find that in this case we obtain the standard 4-dimensional Proca action for a rho meson coupling to a magnetic field, which is the one used in \cite{Chernodub:2010qx} to detect the rho meson condensation effect. We recover the Landau levels and obtain the same picture in figure \ref{meffantipodal} as the one in figure \ref{rho2} from the DSGS-prediction. 
Next, we consider the non-antipodal embedding $u_0 > u_K$ but in the approximation of coinciding branes. This comes down to taking the chiral magnetic catalysis effect on the constituent quarks into account (in an averaged way), which will cause the critical value for the magnetic field at the onset of rho meson condensation to increase, as can be seen in figure \ref{meffnonantipodal}.

\subsubsection{Antipodal embedding: Landau levels}

In the case $u_0=u_K$ we have
\begin{align}
 \partial_u \overline \tau &= 0 \Rightarrow  G_{uu} = g_{uu} \\
 \overline \tau &\sim \mathbf{1} \Rightarrow [A_r,\overline \tau] = 0,
\end{align}
simplifying (\ref{STrDBI}) drastically to a part in the scalar fluctuations and a part quadratic in the gauge fluctuations, given by
\begin{align}
\text{STr} & \hspace{1mm} e^{-\phi} \sqrt{-\det a} |_{\tilde A^2,(2\pi \alpha')^2} 
= \text{Tr}  \hspace{1mm} e^{-\phi} g_{11}^2 \sqrt{g_{uu}} g_{S_4}^2 (2\pi \alpha')^2  \times  \nonumber\\ &
\left\{
\overline F_{12} g_{11}^{-2} \tilde F_{21}
- \overline F_{12} g_{11}^{-2} [\tilde A_1,\tilde A_2]
- \frac{1}{4} g_{\mu\mu}^{-1} g_{\nu\nu}^{-1} \tilde F_{\mu\nu}^2 - \frac{1}{2} g_{\mu\mu}^{-1} g_{uu}^{-1} \tilde F_{\mu u}^2 \right\},  
\end{align}
where we chose the gauge 
\begin{equation}
A_u=0.   
\end{equation} 
In this gauge, see also (\ref{Aziszerogauge}), we have the following expansion of the gauge field \cite{Sakai:2005yt} (with our $\overline A_\mu$ equal to $A_L = A_R = \mathcal V$ in the notation of \cite{Sakai:2005yt}) 
\begin{equation}
A_\mu = (\xi_+ \partial_\mu \xi_+^{-1} + \xi_+ \overline A_\mu \xi_+^{-1}) \psi_+ + (\xi_- \partial_\mu \xi_-^{-1} + \xi_- \overline A_\mu \xi_-^{-1}) \psi_- + \sum_n B_\mu^{(n)} \psi_n, 
\end{equation}
which after fixing the residual gauge symmetry within this gauge to $\xi_+^{-1}(x^\mu) = \xi_-(x^\mu)$, 
becomes
\begin{eqnarray} \label{expansionvectorpion}
A_\mu = \overline A_\mu + \frac{1}{2 f_\pi^2} [\pi, \partial_\mu \pi] + \frac{i}{f_\pi} \left(\partial_\mu \pi - [\pi,\overline A_\mu] \right) \psi_0 +  \sum_n B_\mu^{(n)} \psi_n. \end{eqnarray}
Being mainly interested in the $\rho$ mesons here (identified with $B_\mu^{(1)}$), essentially we will thus use the gauge field expansion 
$A_\mu = \overline A_\mu +  \sum_n B_\mu^{(n)} \psi_n$.

We will only retain the first meson of the vector meson tower in the fluctuation expansion (\ref{expansionvectorpion}), 
\begin{equation}
\tilde A_\mu^a = \rho_\mu^a(x^\mu) \psi(u) + \text{pions} \qquad (a=1,2), 
\end{equation}
because it is the most likely to condense first, being the lightest spin 1 particle. 
The DBI-action for the $A_\mu^{a}$ ($a=1,2$) components can then be written as\footnote{Note that from here on contraction over Minkowski indices is assumed implicitly in the notation of squares, e.g. $(\tilde F_{\mu\nu}^a)^2 =  F_{\mu\nu}^a  F_{\mu\nu}^a \eta^{\mu\mu} \eta^{\nu\nu}$.} 
\begin{align}
S_{DBI} &= \int d^4 x \int du \sum_{a=1}^2 \left\{ -\frac{1}{4} f_1 (\tilde F_{\mu\nu}^a)^2 - \frac{1}{2} f_2 (\tilde F_{\mu u}^a)^2 - \frac{1}{2} f_3 \sum_{\mu,\nu=1}^2 \overline F_{\mu\nu}^3 \epsilon_{3ab} \tilde A^{\mu a} \tilde A^{\nu b} \right\} \nonumber\\
&=  \int d^4x \int du \sum_{a=1}^2 \left\{ -\frac{1}{4} f_1(\mathcal F_{\mu\nu}^a)^2 \psi^2 - \frac{1}{2} f_2(\rho_\mu^a)^2 (\partial_u \psi)^2 - \frac{1}{2} f_3 \sum_{\mu,\nu=1}^2 \overline F_{\mu\nu}^3 \epsilon_{3ab} \rho^{\mu a} \rho^{\nu b} \psi^2 \right\} + \text{pion action}
\end{align}
with $f_i$ ($i=1..3$) the following functions of $u$:
\begin{eqnarray}
f_1(u) = f_3(u)  &=& T_8 V_4 (2\pi\alpha')^2 e^{-\phi} g^2_{S_4} \sqrt{g_{uu}} \\
f_2(u) &=& T_8 V_4 (2\pi\alpha')^2 e^{-\phi} g^2_{S_4} g_{11} g_{uu}^{-1/2},
\end{eqnarray}
and 
\begin{equation}
\mathcal F_{\mu\nu}^a = D_\mu \rho_\nu^a - D_\nu \rho_\mu^a.
\end{equation}
There are no coupling terms between $\rho$ mesons and pions at second order in the fluctuations in the DBI-action, which can be traced back to the different parity of $\psi_0(z) \equiv \psi_0(u(z))$ and $\psi(z) \equiv \psi(u(z))$ (with $u(z) = u_K^3 + u_K z^2$), $\psi_0(z)$ being odd and $\psi(z)$ even.
In order to obtain a canonical kinetic term and mass term for the $\rho$ mesons in the effective four-dimensional action, we impose that the $\psi(u)$ fulfill the standard conditions
\begin{align}
\int_{u_K}^\infty du \hspace{1mm} f_1 \psi^2 &= 1\,, \\
\int_{u_K}^\infty du \hspace{1mm} f_2 (\partial_u \psi)^2 &= m_\rho^2.
\end{align}
These conditions combine to the eigenvalue equation (\ref{finiteTeigwvgl}) with $\gamma = f^{-1}$ for $u_0=u_K$. Per construction the lowest eigenvalue of this equation is $m_\rho^2 = 0.776^2$ GeV$^2$. The corresponding eigenfunction $\psi$ fulfilling the boundary conditions $\psi'(z=0) = 0$ and $\psi(z\rightarrow\pm\infty)=0$ can be used to evaluate the last integral over $u$ in the above action, which determines the 
magnetic moment coupling $k$ of the $\rho$ mesons to the background magnetic field,
\begin{equation}
\int_{u_K}^\infty du \hspace{1mm} f_3 \psi^2 = k,
\end{equation}
related to the magnetic moment $\mu$ as $\mu = (1+k) e /(2m)$ so to the gyromagnetic ratio $g$ as $g=1+k$ \cite{Obukhov:1984xb}. Because in this simple embedding we have $f_1 = f_3$, we immediately see from the normalization condition that $k=1$ and thus $g=2$, describing a non-minimal coupling of the $\rho$ mesons to the background magnetic field. 

The effective four-dimensional action thus takes the form of the standard four-dimensional action used to describe the coupling of charged vector mesons to an external magnetic field (i.e.\ the Proca action \cite{Obukhov:1984xb,Tsai:1972iq} or DSGS action for self-consistent $\rho$ meson quantum electrodynamics to second order in the fields \cite{Djukanovic:2005ag}):
\begin{equation}
S_{eff} = \int d^4x \sum_{a=1}^2  \left\{ -\frac{1}{4}(\mathcal F_{\mu\nu}^a)^2 - \frac{1}{2} m_\rho^2 (\rho_\mu^a)^2  - \frac{1}{2} \sum_{\mu,\nu=1}^2 \overline F_{\mu\nu}^3 \epsilon_{3ab} \rho^{\mu a} \rho^{\nu b} \right\}.   \label{procafirst}
\end{equation}
This means that the Sakai-Sugimoto model with $u_0=u_K$ automatically describes Landau levels for $\rho$ mesons moving in an external magnetic field.
Let us quickly repeat how to derive this from the effective action, following for example \cite{Obukhov:1984xb} (up to conventions). 

The equations of motion for $\rho^{\nu a}$,
\begin{equation}
D^\mu \mathcal F_{\mu\nu}^a - \epsilon_{a3b} \overline F_{\mu\nu}^3 \rho^{\mu b} - m^2 \rho_\nu^a = 0,
\end{equation}
combine to
\begin{equation}
\mathbf{D}^\mu(\mathbf D_\mu \rho_\nu - \mathbf D_\nu \rho_\mu) - i \overline F_{\mu\nu}^3 \rho^\mu - m^2 \rho_\nu = 0
\end{equation}
with $\mathbf D_\mu = \partial_\mu + i \overline A_\mu^3 = \partial_\mu + i e \overline A_\mu^{em}$ for the charged combination $\rho_\mu^- = (\rho_\mu^1 + i \rho_\mu^2)/\sqrt 2$, and the complex conjugate of this equation for the other charged combination $\rho_\mu^+ = (\rho_\mu^1 - i \rho_\mu^2)/\sqrt 2$. Acting with $\mathbf D^\nu$ on this equation of motion, and using the fact that $[\mathbf D_\mu,\mathbf D_\nu]=i \overline F_{\mu\nu}^3$, leads to the subsidiary condition
\begin{eqnarray}\label{subs}
\mathbf D^\nu \rho_\nu = 0,
\end{eqnarray}
which allows us to rewrite the equation as
\begin{equation}
\mathbf{D}_\mu^2 \rho_\nu  - 2 i \overline F_{\mu\nu}^3 \rho^\mu - m^2 \rho_\nu = 0.
\end{equation}
The condition \eqref{subs} and its conjugate are nothing else than the covariant (w.r.t.~the electromagnetic background) generalizations of the usual Proca subsidiary conditions $\partial^\nu \rho_\nu^\pm=0$. 

\begin{figure}[t]
  \centering
  \scalebox{1.1}{
  \includegraphics{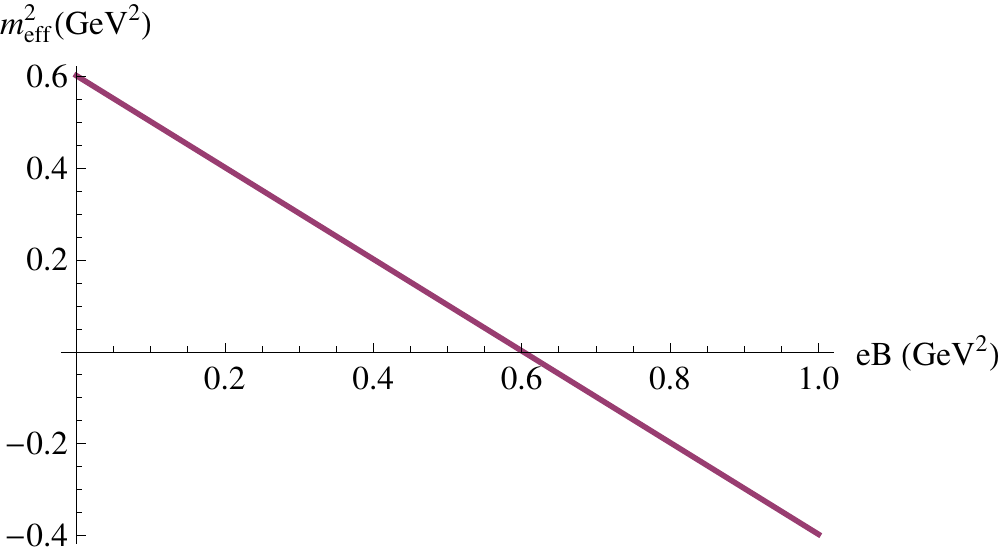}} 
  \caption{The effective mass squared $\mathit{m_{\rho,eff}^2} = m_\rho^2 - eB$ of the field combinations $\rho$ and $\rho^\dagger$ as a function of $eB$, for the case of antipodal embedding of the flavour branes. $\mathit{m_{\rho,eff}^2}$ goes through zero at $eB_c$.}
\label{meffantipodal}
\end{figure}

Fourier transforming $\rho_\nu \rightarrow e^{i (\vec k \cdot \vec x - E t )} \rho_\nu$, we find
that the transverse combinations\footnote{Because $\overline F_{\mu\nu}^{3}$ is only non-zero for $\mu,\nu \neq 0,3$ there is no magnetic moment coupling term for the longitudinal components of the $\rho$ fields, resulting in only the transverse components condensing according to the Landau equations of motion.} $\rho^-_1 \pm i \rho^-_2$ respectively get a negative or positive contribution $\mp eB$ to their effective mass squared as a consequence of their magnetic moment coupling to the magnetic field:
\begin{align}
E^2 (\rho^-_1 \pm i \rho^-_2)
&=  \left[ (k_2 + x_1 eB)^2 + k_3^2 - \partial_1^2 + m_\rho^2 \mp 2 eB \right] (\rho^-_1 \pm i \rho^-_2)  \nonumber\\
&=  \left[ - \partial_1^2 + (eB)^2 \left( x_1 + \frac{k_2}{eB}\right)^2 + k_3^2 + m_\rho^2 \mp 2 eB \right] (\rho^-_1 \pm i \rho^-_2)  \nonumber\\
&=  \left[ eB(2n+1) + k_3^2 + m_\rho^2 \mp 2 eB \right] (\rho^-_1 \pm i \rho^-_2).
\end{align}
The following combinations of the transverse (w.r.t.~$\vec{B}$) polarizations of the charged spin-1 $\rho$ meson fields,
\begin{eqnarray}\label{fieldcom}
\rho = \rho^-_1 + i \rho^-_2 \quad \mbox{ and }\quad  \rho^\dagger =  \rho_1^+ - i \rho_2^+,
\end{eqnarray}
have their spin aligned with the background magnetic field, $s_3 = 1$, which decreases their energy with an amount $eB$. As a consequence, the energy or effective mass of the fields $\rho$ and $\rho^\dagger$ in the lowest energy state, i.e.~in the lowest Landau level $n=0$ and with zero momentum $k_3$ along the direction of the magnetic field,
\begin{eqnarray} E =  \mathit{m_{\rho,eff}} = \sqrt{m_\rho^2 - eB},  \end{eqnarray} 
becomes imaginary when the magnetic field reaches the critical value of $m_\rho^2$,
\begin{eqnarray}\label{crit} eB_c = m_\rho^2 \approx 0.60 \mbox{ GeV}^2,  \end{eqnarray}
see figure \ref{meffantipodal}.  
We conclude from this that the field combinations  \eqref{fieldcom}
should experience a condensation, in accordance with \cite{Chernodub:2010qx,Chernodub:2011mc}. 

Before concluding this section, we mention that in principle there are also equations of motion for $\tilde A_u^a$, which, upon still ignoring the $F^4$ terms in the action, read (in the gauge $A_u=0$)
\begin{eqnarray}
\partial_u(\partial_\mu \omega^\mu) = \partial_u(\partial_\mu \rho^{\mu 0}) = \partial_u(\mathbf D_\mu \rho^{\mu -}) = \partial_u(\mathbf D_\mu^\dagger \rho^{\mu +}) = 0.
 \end{eqnarray}
The latter equations are automatically fulfilled due to the Proca subsidiary conditions for the neutral mesons, the subsidiary condition \eqref{subs} and its complex conjugate.

\FloatBarrier

\subsubsection{Non-antipodal embedding in coincident branes approximation: chiral magnetic catalysis}

\begin{figure}[t]
  \centering
  \scalebox{1}{
  \includegraphics{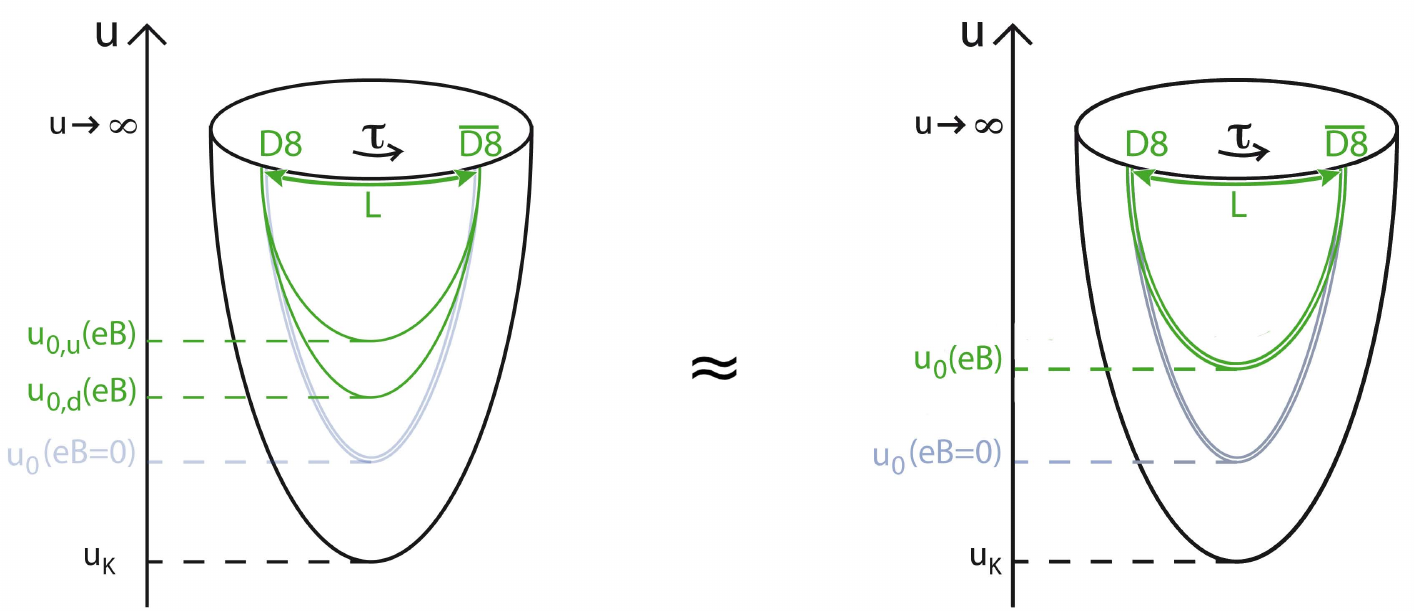}}
  \caption{The approximated $eB$-dependent embedding of the flavour branes, taking into account chiral magnetic catalysis but postulating coincident branes.}
\label{approxEmbedding}
\end{figure}

\begin{figure}[h!]
  \centering
  \scalebox{1.18}{
  \includegraphics{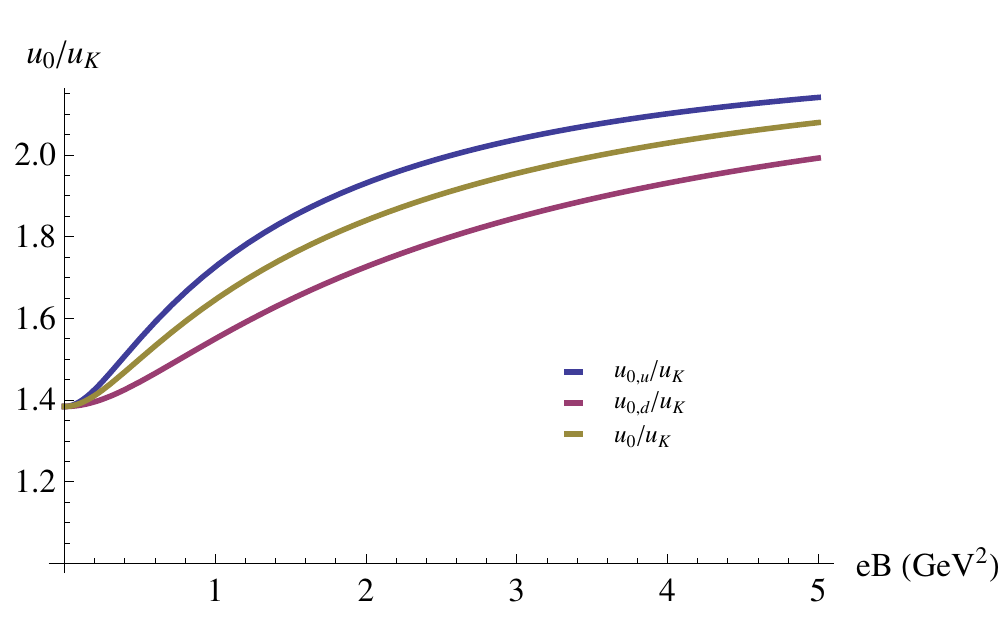}}
  \caption{The average $u_0(eB)$ (yellow) in the coincident branes approximation compared to $u_{0,u}(eB)$ (blue) and $u_{0,d}(eB)$ (red).}
\label{approxEmbeddingu0}
\end{figure}

Taking into account the splitting of the branes in figure \ref{changedembedding}, $\overline \tau \not \sim \mathbf 1$, severely complicates the analysis, mainly because the evaluation of the STr in the action (\ref{STrDBI}) becomes quite technical in the case of non-coincident branes (the STr no longer reduces to a standard Tr). We begin by 
investigating the effect of the chiral magnetic catalysis on the $\rho$ meson mass and the critical magnetic field in the approximation of  
replacing the embedding in figure \ref{changedembedding} by an embedding where the branes are still coincident in the presence of the magnetic field ($\partial_u \overline \tau \neq 0$ but $\overline \tau \sim \mathbf 1$), joining at an averaged value $u_0$ of the holographic radius, determined from
\begin{equation} \label{u0Baverage}
L_{average}(u_0,eB) = \left\{ (\ref{LconfifvB}) \text{ with every $A$ replaced by $A_{average} = \frac{(\sqrt{A_u} + \sqrt{A_d})^2}{2}$} \right\}  = L(eB=0)  \Rightarrow u_0(eB),
\end{equation}
see figure \ref{approxEmbedding} and \ref{approxEmbeddingu0}.

This form of $L_{average}(u_0,eB)$ is the one you obtain when you postulate that the branes remain coincident in the presence of the magnetic field by using $g^{D8} = g^{D8} \mathbf 1$ instead of \eqref{metricnoncoincident}.
The same approximation, i.e.~not taking into account the splitting, is implicitly done when applying the magnetic field holographically by assigning a non-zero value only to $\overline A_\mu^3$ and not $\overline A_\mu^0$ (instead of (\ref{Aachtergrond})), as is quite often done in the literature, e.g.~in \cite{Bergman:2008sg,Ammon:2011je,Bu:2012mq}. 

In the case $u_0>u_K$ and the current approximation of coincident branes we have
\begin{align}
\partial_u \overline \tau &= \sqrt{ \left(\frac{R}{u}\right)^{3} \frac{1}{f^2} \frac{u_0^8 f_0 A_0}{u^8 f A - u_0^8 f_0 A_0}}  \times \theta(u - u_0) \\
 \overline \tau &\sim \mathbf{1} \Rightarrow [A_r,\overline \tau] = 0  \text{ and $G_{uu}(\partial_u \overline \tau)$ drops out of the STr argument},
 \end{align}
simplifying (\ref{STrDBI}) to a part in the scalar fluctuations and a part quadratic in the gauge fluctuations, given by
\begin{align}
\text{STr} & \hspace{1mm} e^{-\phi} \sqrt{-\det a} |_{\tilde A^2,(2\pi \alpha')^2}
= \text{Tr}  \hspace{1mm} e^{-\phi} g_{11}^2 \sqrt{G_{uu}} g_{S_4}^2 (2\pi \alpha')^2  \times  \nonumber\\ &
\left\{
\overline F_{12} g_{11}^{-2} \tilde F_{21}
- \overline F_{12} g_{11}^{-2} [\tilde A_1,\tilde A_2]
- \frac{1}{4} g_{\mu\mu}^{-1} g_{\nu\nu}^{-1} \tilde F_{\mu\nu}^2 - \frac{1}{2} g_{\mu\mu}^{-1} G_{uu}^{-1} \tilde F_{\mu u}^2 \right\},
\end{align}
where we again chose the gauge $A_u=0$ and only retained the lowest mesons of the meson towers in the fluctuation expansions (\ref{expansionvectorpion}) for the gauge field and $\tilde \tau(x^\mu,u) = \sum_n U^{(n)}(x^\mu) \phi_n(u)$ for the scalar field:
\begin{equation}
\tilde A_\mu^a = \rho_\mu^a(x^\mu) \psi(u) \text{ and } \tilde \tau = a_0(x^\mu) \phi(u) \quad (a=1,2) .  \label{6.4.47}
\end{equation}
Because both $\psi(z) \equiv \psi(u(z))$ and $\phi(z) \equiv \phi(u(z))$ (with $u(z) = u_0^3 + u_0 z^2$)
are even functions, the term $\mathcal L_4$ in (\ref{STR})  
consists of integrals of the form $\int_{-\infty}^{\infty} dz  \{ \text{odd function of $z$} \} = 0$ and hence disappears.

\begin{figure}[h!]
  \hfill
  \begin{minipage}[t]{\textwidth}
    \begin{center}
      \scalebox{1.3}{
  \includegraphics{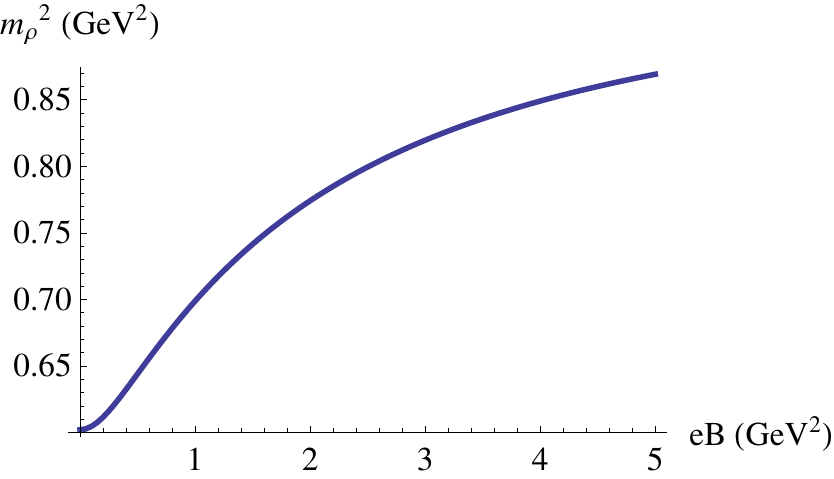}}
    \end{center}
  \end{minipage}
  \hfill
  \begin{minipage}[t]{\textwidth}
    \begin{center}
      \scalebox{1.3}{
  \includegraphics{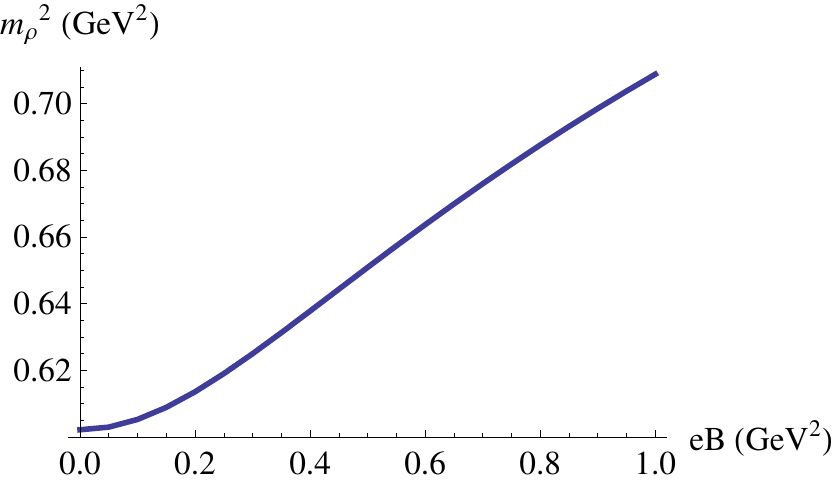}}
    \end{center}
  \end{minipage}
      \caption{The $\rho$ meson mass eigenvalue as a function of the magnetic field.}
	\label{mrhoB}
  \hfill
\end{figure}

\begin{figure}[t]
  \centering
  \scalebox{1.4}{ 
  \includegraphics{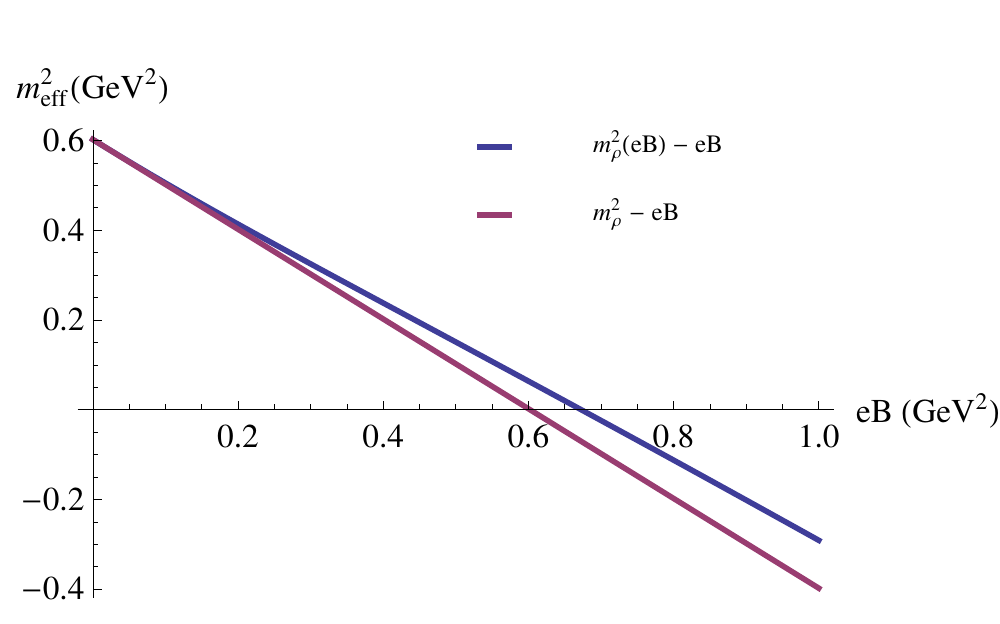}}
  \caption{The effective mass squared $\mathit{m_{\rho,eff}^2} = m_\rho^2(eB) - eB$ of the field combinations $\rho$ and $\rho^\dagger$ as a function of $eB$, for the case of non-antipodal coincident flavour branes (blue), i.e. taking into account the effect of an averaged chiral magnetic catalysis through the $eB$-dependenve of $m_\rho$, compared to the case of antipodal embedding (red). $\mathit{m_{\rho,eff}^2}$ goes through zero at $eB_c$.
}
\label{meffnonantipodal}
\end{figure}

The DBI-action to second order in the charged gauge field fluctuations can then again be written as
\begin{align}
S_{DBI} &= \int d^4 x \int du \sum_{a=1}^2  \left\{ -\frac{1}{4} f_1 (\tilde F_{\mu\nu}^a)^2 - \frac{1}{2} f_2 (\tilde F_{\mu u}^a)^2 - \frac{1}{2} f_3 \sum_{\mu,\nu=1}^2 \overline F_{\mu\nu}^3 \epsilon_{3ab} \tilde A^{\mu a} \tilde A^{\nu b} \right\} \nonumber\\
&=  \int d^4x \int du \sum_{a=1}^2 \left\{ -\frac{1}{4} f_1(\mathcal F_{\mu\nu}^a)^2 \psi^2 - \frac{1}{2} f_2(\rho_\mu^a)^2 (\partial_u \psi)^2 - \frac{1}{2} f_3 \sum_{\mu,\nu=1}^2 \overline F_{\mu\nu}^3 \epsilon_{3ab} \rho^{\mu a} \rho^{\nu b} \psi^2 \right\} + \text{pion action}
\end{align}
with the functions $f_i$ ($i=1..3$) dependent on both $u$ \'and $eB$ this time (through the $eB$-dependence of the embedding function $\partial_u \overline \tau$ in $G_{uu}$):
\begin{eqnarray}
f_1(u,eB) = f_3(u,eB)  &=& T_8 V_4 (2\pi\alpha')^2 e^{-\phi} g^2_{S_4} \sqrt{G_{uu}}\,, \\
f_2(u,eB) &=& T_8 V_4 (2\pi\alpha')^2 e^{-\phi} g^2_{S_4} g_{11} G_{uu}^{-1/2}.
\end{eqnarray}
In order to obtain a canonical kinetic term and mass term for the $\rho$ mesons in the effective four-dimensional action, we demand the $\psi(u)$ to fulfill the standard conditions
\begin{align}
\int_{u_0(eB)}^\infty du \hspace{1mm} f_1(u,eB) \psi^2 &= 1\,, \label{Bnormalization}\\
\int_{u_0(eB)}^\infty du \hspace{1mm} f_2(u,eB) (\partial_u \psi)^2 &= m_\rho^2(eB).
\end{align}
These conditions combine to the eigenvalue equation (\ref{finiteTeigwvgl}) with $\gamma =\gamma_{\langle eB \rangle} 
= (\ref{gammaB})$ with every $A$ replaced by $A_{average} = \frac{(\sqrt{A_u} + \sqrt{A_d})^2}{2}$:
\begin{equation} \label{Beigvproblem}
-u^{1/2} \gamma_{\langle eB \rangle}^{-1/2} \partial_u (u^{5/2} \gamma_{\langle eB \rangle}^{-1/2} \partial_u \psi) = R^3 m_\rho^2(eB) \psi.
\end{equation}
Numerically solving this eigenvalue equation for $m_\rho^2(eB)$ (see appendix \ref{appendixnum} for details) boils down to taking the averaged magnetic catalysis of chiral symmetry breaking into account, as we include the effect of $eB$ on the embedding of the flavour branes, both through the induced metric on the branes and the changed value of $u_0$. The resulting $\rho$ meson mass eigenvalue $m_\rho^2(eB)$ as a function of $eB$ is depicted in figure \ref{mrhoB}. As expected on grounds of figure \ref{mqfig}, it is an increasing function of magnetic field (where it should be noted that working in the approximation of coincident branes means that we consider the effect of an averaged increase for $m_q(eB)$, equal for up and down, rather than the exact $eB$-dependences in figure \ref{mqfig}).
The corresponding eigenfunction $\psi$ fulfilling the boundary conditions $\psi'(z=0) = 0$ and $\psi(z\rightarrow\pm\infty)=0$ can be used to evaluate the last integral over $u$ in the above action, which determines the gyromagnetic coupling constant $k$ to be one,
\begin{equation}
\int_{u_0(eB)}^\infty du \hspace{1mm} f_3(u,eB) \psi^2 = k = 1,
\end{equation}
or $g=k+1=2$, again because of $f_1 = f_3$. The effective four-dimensional action thus again takes the form of the standard four-dimensional action (\ref{procafirst}) used to describe the coupling of charged vector mesons to an external magnetic field,  
but this time with the influence of the constituent quarks reflected in the magnetic field dependence of $m_\rho$: 
\begin{equation}
S_{eff} = \int d^4x \sum_{a=1}^2 \left\{ -\frac{1}{4}(\mathcal F_{\mu\nu}^a)^2 - \frac{1}{2} m_\rho^2(eB) (\rho_\mu^a)^2  - \frac{1}{2} \sum_{\mu,\nu=1}^2 \overline F_{\mu\nu}^3 \epsilon_{3ab} \rho^{\mu a} \rho^{\nu b} \right\}.
\end{equation}
Completely analogous to the derivation in the preceding subsection, this action describes that the $\rho$ and $\rho^\dagger$ fields in the lowest Landau energy level will get an effective mass
\begin{eqnarray} E = \mathit{m_{\rho,eff}} = \sqrt{m_\rho^2(eB) - eB},  \end{eqnarray}
which becomes imaginary when $eB_c = m_\rho^2(eB_c)$, at 
\begin{eqnarray}\label{crit} eB_c = m_\rho^2(eB_c) \approx 0.67 \mbox{ GeV}^2 \approx 1.1 \, m_ \rho^2,   \end{eqnarray}
see figure \ref{meffnonantipodal}. 
The averaged chiral magnetic catalysis pushes the critical magnetic field for condensation to a higher value, as expected. The increase is of the order of 10 percent. In following sections, 
we will probe the further increase in $eB_c$ due to the 
flavour brane separation, to investigate to what extent we can get closer to the lattice estimate for the critical magnetic field, i.e. $eB_c$ of the order of 1 GeV$^2$ \cite{Braguta:2011hq}.

\FloatBarrier

\subsection{Gauge fixing} \label{gaugefixing}

\subsubsection{STr-evaluation} \label{4.1.1}

The action (\ref{STR}) contains mixing terms between the scalar and gauge fluctuations in $\mathcal{L}_2$ and $\mathcal{L}_4$. We will disentangle these couplings here by choosing a particular gauge.
First we work out $\mathcal{L}_2$ a bit further by evaluating the STr (\ref{STrdef}).
According to its definition in \cite{Myers:2003bw} the STr takes a symmetric average over all orderings of $F_{mn}$, $D_m \tau$ and $\tau$ appearing in the non-Abelian Taylor expansions of the fields in the action. In particular, commutators,  such as $[A_m,A_n]$ in $F_{mn}$ or $[A_m,\tau]$ in $D_m \tau$, are handled as one matrix.
The
STr-expressions we encounter in (\ref{STR}) can be classified into two types: expressions of the form STr$(\mathcal H(\partial_u \overline \tau) \mathcal G(\overline F_{12}) \tilde X)$ and STr$(\mathcal H(\partial_u \overline \tau) \mathcal G(\overline F_{12}) \tilde X^2)$. Here $\mathcal H$, resp.~$\mathcal G$ are even functions of the diagonal background field
\[
\overline \tau = \overline \tau^0 \sigma^0 + \overline \tau^3 \sigma^3,
\]
resp.~
\[
\overline F_{12}=F^0 \sigma^0 + F^3 \sigma^3 = -\frac{i}{2} \frac{B}{3} \sigma^0 -\frac{i}{2} B \, \sigma^3,
\]
and $\tilde X=\tilde X^a t^a$ is some fluctuation -- in the present case
fully general fluctuations $D_m \tilde\tau$ and off-diagonal fluctuations $[\tilde A_m,\overline \tau]$.
For expressions of these types the STr can be evaluated exactly \cite{Hashimoto:1997gm,Denef:2000rj} as elaborated on in appendix \ref{appendix}. Using the prescriptions presented and rederived there, we arrive at the following form for $\mathcal{L}_2$:
\begin{align}
\mathcal L_{2} = &\sum_{a=1}^2 \left\{ \gamma(u) \frac{1}{2} \left( [\tilde A_u,\overline \tau]^a +\partial_u\tilde\tau^a \right)^2 + \alpha(u) \frac{1}{2} \left( [\tilde A_\mu,\overline \tau]^a +D_\mu \tilde\tau^a \right)^2 + \beta(u) \sum_{\mu=1}^2 \frac{1}{2} \left( [\tilde A_\mu,\overline \tau]^a +D_\mu \tilde\tau^a \right)^2 \right\} \nonumber\\
&+  \text{Tr} \left( \overline x \overline y  [\tilde A_u,\tilde \tau]  \right) + \sum_{l=u,d} \left\{  \gamma_l(u) \frac{1}{2} \left(\partial_u\tilde\tau^l \right)^2 +  \alpha_l(u) \frac{1}{2} \left(D_\mu \tilde\tau^l \right)^2  +  \beta_l(u) \sum_{\mu=1}^2 \frac{1}{2} \left(D_\mu \tilde\tau^l \right)^2 \right\} \label{74}
\end{align}
with
\begin{align}
\gamma(u) &= -\frac{1}{2} I(\overline x G_{uu}^{-2}), \quad
\alpha(u) = -\frac{1}{2} I(\overline x g_{11}^{-1} G_{uu}^{-1}), \quad
\beta(u) = -\frac{1}{2} I(\overline x g_{11}^{-1} G_{uu}^{-1} \frac{1-A}{A}),  \label{gab} \\
\gamma_l(u) &= -\frac{1}{2} I_l(\overline x G_{uu}^{-2}), \quad
\alpha_l(u) = -\frac{1}{2} I_l(\overline x g_{11}^{-1} G_{uu}^{-1}), \quad
\beta_l(u) = -\frac{1}{2} I_l(\overline x g_{11}^{-1} G_{uu}^{-1} \frac{1-A}{A})
\end{align}
containing what we will refer to as `$I$-functions' and `$I_l$-functions', defined in (\ref{IfctionsGeneral}) and (\ref{IlfctionsGeneral}), e.g.
\begin{align*}
I(\overline x G_{uu}^{-2}) &=  e^{-\phi} g_{11}^2 g_{S_4}^2 I\left(G_{uu}^{-3/2}(\partial \overline \tau) A^{1/2}(\overline F_{12})\right) \nonumber \\
&= \frac{e^{-\phi} g_{11}^2 g_{S_4}^2}{2} \int_0^1 d\alpha \left\{G_{uu}^{-3/2}(\partial \overline \tau^0 + \alpha \partial \overline \tau^3) A^{1/2}(F^0 + \alpha F^3) + G_{uu}^{-3/2}(\partial \overline \tau^0 - \alpha \partial \overline \tau^3) A^{1/2}(F^0 - \alpha F^3)\right\},  \\
I_u(\overline x G_{uu}^{-2}) &= e^{-\phi} g_{11}^2 g_{S_4}^2 G_{uu}^{-3/2}(\partial \overline \tau^0 +  \partial \overline \tau^3) A^{1/2}(F^0 + F^3),
\end{align*}
with $\partial \overline \tau$ short for $\partial_u \overline \tau$
and (with $\tilde \tau = \tilde \tau^a t^a$)
\[
\tilde \tau^l = \frac{\tilde \tau^0 \pm \tilde \tau^3}{\sqrt 2}.
\]
Having used $g_{\mu\nu} = g_{11} \eta_{\mu\nu}$
and absorbing $\eta_{\mu\nu}$ in the notation of the squares, $(\partial_\mu \tilde\tau^a)^2 = \partial_\mu \tilde\tau^a \partial_\nu \tilde\tau^a \eta^{\mu\nu} =  \partial_\mu \tilde\tau^a \partial^\mu \tilde\tau^a$, all the products over $\mu$ in the above Lagrangian (and in all expressions following unless stated otherwise) are contracted Minkowski products.

The difficulty in evaluating the STr, although we restrict to second order in the fluctuations, comes from the presence of the background fields $\partial \overline \tau$ (appearing in the induced metric on the flavour branes through $G_{uu} = g_{uu} + g_{\tau\tau} (\partial_u \overline \tau)^2$) and $\overline F_{12}$ (appearing in $A$ as defined in (\ref{A})), which have to be ordered\footnote{There is some ambiguity here in the sense that the background scalar field $\partial_u \overline \tau$ itself depends on the background gauge field $\overline F_{12}$, so there is also the option to order the matrices $\overline F_{12}$ within $\partial_u \overline \tau$, as opposed to ordering $\partial_u \overline \tau$ as independent. We however opted for the latter, which seems more logical to us.} within the STr. The functions containing the background fields have to be Taylor expanded before the ordering and subsequently resummed. This gives rise to complicated $I$-functions as
 in (\ref
 {gab}), which in general have to be calculated numerically.

\subsubsection{Choosing a 't Hooft gauge}

We consider a 't Hooft gauge-fixing function \cite{'tHooft:1971rn} in the non-Abelian directions -- assuming the Einstein convention that double $SU(2)$-indices $b,c=1,2,3$  are summed over --
\begin{equation} \label{thooftgauge}
G^a = \frac{1}{\sqrt \xi} \left( \alpha(u) D_\mu \tilde A_\mu^a + \gamma(u) D_u \tilde A_u^a+ \sum_{i=1,2} \beta(u) D_i \tilde A_i^a \right)+ 2 i \sqrt \xi \epsilon_{abc} \tilde \tau^b \overline \tau^c \quad (a=1,2)
\end{equation}
such that the gauge-fixed Lagrangian
\begin{align}
&\mathcal L_{2} -\frac{1}{2} (G^a)^2 = \sum_{a=1}^2 \left\{ \gamma(u) \frac{1}{2}\left[ \left( [\tilde A_u,\overline \tau]^a \right)^2 +\left(\partial_u\tilde\tau^a \right)^2 \right] + \alpha(u) \frac{1}{2} \left[ \left( [\tilde A_\mu,\overline \tau]^a \right)^2 + \left(D_\mu \tilde\tau^a \right)^2 \right] \right. \nonumber\\
& \left.\quad + \beta(u) \sum_{\mu=1}^2 \frac{1}{2} \left[ \left( [\tilde A_\mu,\overline \tau]^a\right)^2 + \left(D_\mu \tilde\tau^a \right)^2 \right] -\frac{1}{2 \xi} \left[ \text{$(D\tilde A)^2$ terms} \right]
+\frac{1}{2} (\sqrt \xi \tilde \tau^a)^2 (2\overline \tau^3)^2
 +2 i \tilde A_u^a \epsilon_{abc} \tilde \tau^b \partial_u(\gamma(u)  \overline \tau^c) \right\} \nonumber\\
& \quad +  \text{Tr} \left( \overline x \overline y  [\tilde A_u,\tilde \tau]  \right) + \sum_{l=u,d} \left\{ \cdots \right\}
\label{82}
\end{align}
will be free of mixing terms for a sensible choice of the gauge parameter $\xi$. The Lagrangian
$\mathcal L$ is replaced by $\mathcal L -\frac{1}{2} (G^a)^2$ by virtue of the Faddeev-Popov trick: the partition function of a system with action $S = \int dx \mathcal L$ fulfilling the gauge-fixing constraints $G^a(A,\tau)=0$ is written
as
\begin{align}
Z = \int \mathcal D A \mathcal D \tau \hspace{1mm} e^{i \int dx \mathcal L(A,\tau)}
 \sim \int \mathcal D A \mathcal D \tau  \hspace{1mm} e^{i \int dx \mathcal L(A,\tau)} \delta\left[G(A,\tau)\right] \Delta_{G(A,\tau)}
\end{align}
with proportionality constant the volume of the
gauge group, $\delta \left[G(A,\tau)\right] = \Pi_{x,a}\left( \delta \left[G^a(A(x),\tau(x))\right] \right)$
and $\Delta_{G(A,\tau)}$ the associated Jacobian. 
Alternatively, through introducing the gauge-fixing as $\delta(G^a(A(x),\tau(x)) - \omega^a(x))$ and integrating over $\omega^a$ having a Gaussian distribution around zero, the partition function can be written as 
\[
Z \sim \int \mathcal D A \mathcal D \tau e^{i \int dx \left[ \mathcal L(A,\tau) - \frac{1}{2} \left(G^a(A,\tau)\right)^2 \right]} \Delta_{G(A,\tau)}.
\]
Now we rescale the charged scalar fluctuations $\tilde \tau^{a=1,2} \rightarrow \frac{\tilde \tau^{a=1,2}}{\sqrt{\xi}}$
and choose the so-called `unitary' gauge
\begin{equation} \label{xiunitary}
\xi\rightarrow \infty. \end{equation}
This boils down to deleting all dynamical terms for the fluctuations  $\tilde \tau^{a=1,2}$ and we are left with
\begin{align}
\mathcal L_{2} -\frac{1}{2} (G^a)^2  &= \sum_{a=1}^2 \left\{ \gamma(u) \frac{1}{2}\left( [\tilde A_u,\overline \tau]^a \right)^2 + \alpha(u) \frac{1}{2} \left( [\tilde A_\mu,\overline \tau]^a \right)^2 + \beta(u) \sum_{\mu=1}^2 \frac{1}{2} \left( [\tilde A_\mu,\overline \tau]^a\right)^2
+\frac{1}{2} (\tilde \tau^a)^2 (2\overline \tau^3)^2 \right\} \nonumber \\ &+ \sum_{l=u,d} \left\{ \cdots \right\}.
\end{align}
With the above gauge choice we can see the Higgs mechanism at work that is associated with the magnetic field pulling the up- and down-brane apart:  the charged scalar fluctuations $\tilde \tau^{1,2}$ now serve as Goldstone bosons that are eaten by the gauge bosons $\tilde A_m^{1,2}$, acquiring a mass $\sim (\overline \tau^3)^2$, where $\overline \tau^3$ is essentially the vacuum expectation value of
the diagonal component $\tau^{3}$ of the $\tau$-field. The remaining fluctuations $\tilde \tau^{0,3}$ are the Higgs bosons.

\subsubsection{Fixing the remaining gauge freedom}

In the unitary gauge, $\mathcal L_4$, containing the only remaining mixing terms between gauge and scalar fluctuations, reads
\begin{align}
\mathcal L_4 &= \frac{1}{2} \left\{ I(\overline x \overline z) \sum_{a=1}^2 \left[ [\tilde A_u,\overline \tau]^a \tilde F_{12}^a + [\tilde A_1,\overline \tau]^a \tilde F_{2u}^a -  [\tilde A_2,\overline \tau]^a \tilde F_{1u}^a \right]  + \sum_{l=u,d} I_l(\overline x \overline z) \left[ D_u \tilde\tau^l \tilde F_{12}^l + D_1 \tilde\tau^l \tilde F_{2u}^l - D_2 \tilde\tau^l \tilde F_{1u}^l \right] \right\} \nonumber\\
&=  \frac{1}{2} I(\overline x \overline z) \sum_{a=1}^2 \left( -[\tilde A_1,\overline \tau]^a \partial_u \tilde A_2^a + [\tilde A_2,\overline \tau]^a \partial_u \tilde A_1^a \right)
\end{align}
where we used partial integration. The neutral part vanishes\footnote{
This is not entirely correct. What remains after the $A_u^{a=0,3}$ gauge choice is $-(\partial_u I_l) \tilde \tau^l (\partial_1 \tilde A_2^l - \partial_2 \tilde A_1^l)$, which will however disappear under the $u$-integration, by virtue of $\int dz \{ \text{odd function of $z$} \} = 0$ for $\tilde \tau = a_0(x) \phi(u)$ as in (\ref{6.4.47}).      
} due to the gauge choice
\begin{equation} \label{a03gauge}
A_u^3 = A_u^0 = 0,
\end{equation}
hereby using the remaining gauge freedom in the $a=0,3$ directions, as the 't Hooft gauge (\ref{thooftgauge}) only fixes the gauge for $a=1,2$.

In the chosen gauge (\ref{thooftgauge}), (\ref{xiunitary}), (\ref{a03gauge}), the Lagrangian is free of $\tilde A_m \tilde \tau$ couplings:
\begin{align}
& \text{STr} \hspace{1mm} e^{-\phi} \sqrt{-\det a} |_{\tilde A^2,\tilde \tau^2}  = \overline{\mathcal L} + \mathcal L_{Higgs}  + \mathcal L_{scalar} + \mathcal L_{vector} + \mathcal L_{vector-mixing} \end{align}
with
\begin{align}
& \overline{\mathcal{L}} = \text{Tr} \hspace{1mm} e^{-\phi}  \sqrt{-\det \overline a} \nonumber \\
&\mathcal{L}_{Higgs}= \sum_{a=1}^2 \left\{ \gamma(u) \frac{1}{2}\left( [\tilde A_u,\overline \tau]^a \right)^2 + \alpha(u) \frac{1}{2} \left( [\tilde A_\mu,\overline \tau]^a \right)^2 + \beta(u) \sum_{\mu=1}^2 \frac{1}{2} \left( [\tilde A_\mu,\overline \tau]^a\right)^2
-\frac{1}{2} (\tilde \tau^a)^2 (\overline \tau^3)^2 \right\} \nonumber\\
&\mathcal{L}_{scalar}= \sum_{l=u,d} \left\{  \gamma_l(u) \frac{1}{2} \left(\partial_u\tilde\tau^l \right)^2 +  \alpha_l(u) \frac{1}{2} \left(D_\mu \tilde\tau^l \right)^2  +  \beta_l(u) \sum_{\mu=1}^2 \frac{1}{2} \left(D_\mu \tilde\tau^l \right)^2 \right\} \nonumber\\
&\mathcal{L}_{vector}= \text{STr} \hspace{1mm} \overline x \left\{
- \overline F_{12} g_{11}^{-2} A^{-1} [\tilde A_1,\tilde A_2]
- \frac{1}{4} g_{11}^{-2} \tilde F_{\mu\nu}^2 \hspace{1mm} A^{-2}|_{\mu,\nu=1,2}  - \frac{1}{2} g_{11}^{-1} G_{uu}^{-1} \tilde F_{\mu u}^2 \hspace{1mm} A^{-1}|_{\mu=1,2}  \right\} \nonumber \\
&\mathcal{L}_{vector-mixing} = \frac{1}{2} \left\{ I(\overline x \overline z) \sum_{a=1}^2 \left( -[\tilde A_1,\overline \tau]^a \partial_u \tilde A_2^a + [\tilde A_2,\overline \tau]^a \partial_u \tilde A_1^a \right) \right\}. \label{STRgauged}
\end{align}

\subsection{Stability in scalar sector} \label{4.2} 

The stability of the embedding of the flavour branes has been checked in \cite{Sakai:2004cn} for the antipodal case, and in \cite{Ghoroku:2009iv} for the non-antipodal case. We extend this analysis to the non-antipodal, $B$-dependent embedding,  finding what we 
refer to as `stability in the scalar sector'.

In this section we discuss the scalar part of the DBI-Lagrangian (\ref{STRgauged}),
\begin{align}
\mathcal L_{scalar} &=  \text{STr} \hspace{1mm} e^{-\phi} \sqrt{-\det a} |_{\tilde \tau^2} \nonumber\\
&= \sum_{l=u,d} \left\{  \gamma_l(u) \frac{1}{2} \left(\partial_u\tilde\tau^l \right)^2 +  \alpha_l(u) \frac{1}{2} \left(D_\mu \tilde\tau^l \right)^2  +  \beta_l(u) \sum_{\mu=1}^2 \frac{1}{2} \left(D_\mu \tilde\tau^l \right)^2 \right\}.
\end{align}
With the purpose of checking the stability of the $B$-dependent configuration with respect to scalar fluctuations, it is important
to keep track of the correct signs in the action.
First of all, we therefore replace $(\tilde \tau^l)^2 \rightarrow -4(\tilde \tau^l)^2$ such that the
fluctuations $\tilde \tau^l = \frac{\tilde \tau^0 \pm \tilde \tau^3}{\sqrt 2}$ are now written in terms of the real components of the scalar fluctuation $\tilde \tau = \tilde \tau^a \sigma^a$
(where in (\ref{74}) it was implicitly assumed in evaluating the STr that $\tilde \tau = \tilde \tau^a t^a = -i \tilde \tau^a \sigma^a /2$ with imaginary components $\tilde \tau^a$).
Slightly redefining $\mathcal L_{scalar}$ to incorporate the sign of the full action,
\begin{equation*}
S_{DBI} |_{\tilde \tau^2} = -T_8 \int d^4x \hspace{1mm} 2 \int_{u_0}^{\infty} du \int \epsilon_4 \hspace{1mm} e^{-\phi}\hspace{1mm} \textrm{STr}
\sqrt{-\det a} |_{\tilde \tau^2} = T_8 \int d^4x \hspace{1mm} 2 \int_{u_0}^{\infty} du \int \epsilon_4 \mathcal L_{scalar}
\end{equation*}
we then end up with
\begin{align}
& \mathcal L_{scalar}
&= - \sum_{l=u,d} \left\{  I_l(\overline x G_{uu}^{-2}) \left(\partial_u\tilde\tau^l \right)^2 +  I_l(\overline x g_{\mu\mu}^{-1} G_{uu}^{-1}) \left(D_\mu \tilde\tau^l \right)^2  +  I_l(\overline x g_{\mu\mu}^{-1} G_{uu}^{-1} \frac{1-A}{A})  \sum_{\mu=1}^2 \left(D_\mu \tilde\tau^l \right)^2 \right\}
\end{align}
with the convention $(\partial_\mu \tilde\tau^l)^2 = \partial_\mu \tilde\tau^l \partial_\nu \tilde\tau^l \eta^{\mu\nu}$.

The Hamiltonian associated with the Lagrangian is given by
\begin{align}
\mathcal H &= \frac{\delta \mathcal L_{scalar}}{\delta \partial_0 \tau^l} \partial_0 \tau^l - \mathcal L_{scalar} \nonumber\\
&= \sum_{l=u,d} \left\{  I_l(\overline x G_{uu}^{-2}) \left(\partial_u\tilde\tau^l \right)^2 +  I_l(\overline x g_{\mu\mu}^{-1} G_{uu}^{-1}) \left(\left(\partial_0 \tilde\tau^l \right)^2 + \left(\partial_3 \tilde\tau^l \right)^2\right) +  I_l(\overline x g_{\mu\mu}^{-1} G_{uu}^{-1} A^{-1})  \sum_{i=1}^2 \left(D_i \tilde\tau^l \right)^2 \right\}
\end{align}
where we switched notation again to normal squares $(\partial_\mu\tau^l)^2 = \partial_\mu\tau^l \partial_\mu\tau^l$.
For the embedding to be stable towards scalar $\tilde \tau^l$-fluctuations, the associated energy density has to obey
\begin{equation}
\mathcal E = \int_{u_{0,d}}^\infty \hspace{1mm} \mathcal H \geq 0,
\end{equation}
which will be the case if each of the $I_l$-functions is positive.

\begin{figure}[h!]
  \centering
  \scalebox{1.6}{
  \includegraphics{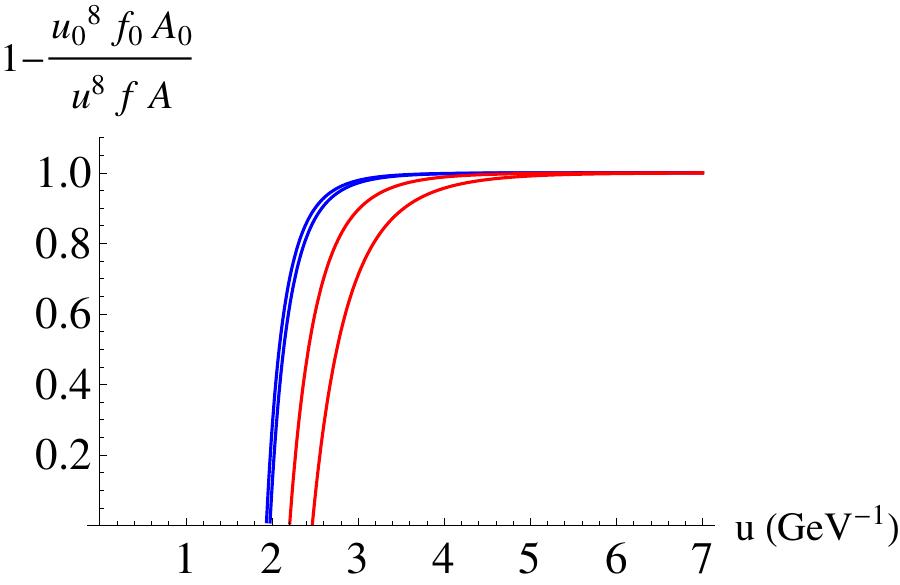}}
  \caption{The function $1 - \frac{u_{0,l}^8 f_{0,l} A_{0,l}}{u^8 f A_l}$ (for $l=u,d$) as a function of $u$ for $B=0.2$ GeV$^2$ in blue and $B=1.2$ GeV$^2$ in red. Up distinguishable from down through $u_{0,u} > u_{0,d}$.}\label{function}
\end{figure}

Let us discuss the two background functions that appear in the $I_l$-functions, $A(\overline F_{12})$ and $G_{uu}(\partial \overline \tau)$.
Using (\ref{tauembedding}), the $uu$-component of the induced metric on the D8-branes as a function of the embedding $\partial \overline \tau$ reads
\begin{align}
G_{uu}(\partial \overline \tau^0 \pm \partial \overline \tau^3) = G_{uu}(\partial \overline \tau_l)  &= g_{uu} + g_{\tau\tau} (\partial_u \overline \tau_l)^2
=  \left(\frac{R}{u}\right)^{3/2} \frac{1}{f} \frac{1}{1 - \frac{u_{0,l}^8 f_{0,l} A_{0,l}}{u^8 f A_l}}, \quad (l=u,d)
\end{align}
with $u \geq u_{0,l}$ implicitly understood,
and, from (\ref{A}),
\begin{align}
A(F^0 \pm F^3) = A_l = 1 + (2\pi\alpha')^2 \overline F_{l}^2  \left(\frac{R}{u}\right)^{3}, \quad (l=u,d)
\end{align}
with the plus (minus) sign corresponding to $l=u$ ($l=d$).
$A_l$ is an increasing function of $B$, equal to 1 for $B=0$, and a decreasing function of $u$, equal to 1 for $u=\infty$ so
\[ A_l \geq 1 \quad (\text{for all $B$ and $u$}). \]
The function $1 - \frac{u_{0,l}^8 f_{0,l} A_{0,l}}{u^8 f A_l}$
is a monotonically increasing function of $u$ going from 0 at $u_{0,l}$ to 1 at $u\rightarrow \infty$ for any fixed value of $B$, see figure \ref{function}.
Then,
\begin{align}
I_l(\overline x G_{uu}^{-2}) &= e^{-\phi} g_{11}^2 g^2_{S_4} \underbrace{I_l(G_{uu}^{-3/2} A^{1/2})}_{G_{uu}^{-3/2}(\partial \overline \tau_l) A_l^{1/2}(\overline F_l)} \nonumber\\
&\sim \underbrace{\vphantom{\left(1 - \frac{u_{0,l}^8 f_{0,l} A_{0,}l}{u^8 f A_l}\right)^{3/2}}\left(\frac{u}{R}\right)^{3/2} u^4}_{\left(\frac{u_0}{R}\right)^{3/2} u_0^4 \cdots \infty} \quad \underbrace{\vphantom{\left(1 - \frac{u_{0,l}^8 f_{0,l} A_{0,}l}{u^8 f A_l}\right)^{3/2}}\hspace{1cm}f^{3/2}\hspace{1cm}}_{\left( 1 - \frac{u_K}{u_0}^3\right)^{3/2} \cdots 1} \quad \underbrace{\hspace{0.2cm}\left(1 - \frac{u_{0,l}^8 f_{0,l} A_{0,}l}{u^8 f A_l}\right)^{3/2}\hspace{0.2cm}}_{0 \cdots 1 \hspace{1mm} \text{for any fixed value of $B$}}~~~\underbrace{\vphantom{\left(1 - \frac{u_{0,l}^8 f_{0,l} A_{0,}l}{u^8 f A_l}\right)^{3/2}}A_l^{1/2}}_{\geq 1} \nonumber\\
& \geq 0,
\end{align}
\begin{align}
I_l(\overline x g_{11}^{-1} G_{uu}^{-1}) &= e^{-\phi} g_{11}^2 g^2_{S_4} g_{11}^{-1} \underbrace{I_l( G_{uu}^{-1/2} A^{1/2})}_{G_{uu}^{-1/2}(\partial \overline \tau_l) A_l^{1/2}(\overline F_l)}  \nonumber\\
&\sim \left(\frac{u}{R}\right)^{-3/4} u^4 \left(\frac{u}{R}\right)^{-3/2} \left(\frac{R}{u}\right)^{-3/4} f^{1/2} \left(1 - \frac{u_{0,l}^8 f_{0,l} A_{0,l}}{u^8 f A_l}\right)^{1/2} A_l^{1/2} \nonumber\\
&\sim \underbrace{\vphantom{\left(1 - \frac{u_{0,l}^8 f_{0,l} A_{0,l}}{u^8 f A_l}\right)^{1/2}}\hspace{0.5cm} u^{5/2}f^{1/2}\hspace{0.5cm}}_{\sqrt{u_0^5 - u_K^3 u_0^2} \cdots \infty} \quad \underbrace{\vphantom{\left(1 - \frac{u_{0,l}^8 f_{0,l} A_{0,l}}{u^8 f A_l}\right)^{1/2}}\left(1 - \frac{u_{0,l}^8 f_{0,l} A_{0,l}}{u^8 f A_l}\right)^{1/2}}_{0 \cdots 1 \hspace{1mm}  \text{for any fixed value of $B$}}~~~\underbrace{\vphantom{\left(1 - \frac{u_{0,l}^8 f_{0,l} A_{0,l}}{u^8 f A_l}\right)^{1/2}}A_l^{1/2}}_{\geq 1}\nonumber \\
& \geq 0,
\end{align}
and for the same reasons
\begin{align}
I_l(\overline x g_{11}^{-1} G_{uu}^{-1} A^{-1}) &= e^{-\phi} g_{11}^2 g^2_{S_4} g_{11}^{-1} \underbrace{I_l( G_{uu}^{-1/2} A^{-1/2})}_{G_{uu}^{-1/2}(\partial \overline \tau_l) A_l^{-1/2}(\overline F_l)}  \geq 0.
\end{align}
This concludes the proof of stability of the flavour branes' embedding as depicted in figure \ref{changedembedding} with respect to diagonal $\tilde \tau$-fluctuations. Note that the off-diagonal $\tilde \tau$-components have disappeared through the gauge fixing in section \ref{gaugefixing} -- except for an irrelevant mass term for the undynamical $\tilde \tau^{1,2}$ in $\mathcal L_{Higgs}$. A similar mechanism in the context of the holographic description of heavy-light mesons can be found in \cite{Erdmenger:2007vj}.

Let us briefly expand on the physical interpretation of the discussion of stability in the scalar sector. While in the seminal work of \cite{Sakai:2004cn} (the $x^\mu$-dependent parts of) the scalar modes $\tilde \tau$ were identified with scalar mesons in the dual field theory, this interpretation was revisited in \cite{Imoto:2010ef}, where it is argued that the $\tilde \tau$-fluctuations are to be regarded as artifacts of the SSM\footnote{We would like to thank S.~Sugimoto for private communication about this.}. The reason is that they transform under a $\mathbb Z_2$-symmetry of the geometric configuration (strictly speaking in the antipodal set-up), which is redundant in the sense that it is not shared with QCD. This is similar to the gauge field components $A_{\Omega_4}$ not having a counterpart in the dual QCD-like field theory, as they transform under the $SO(5)$ isometry of the four-sphere in the background (\ref{D4SS}).   Any concern about the interpretation of the off-diagonal $\tilde \tau$-components disappearing in the holographic Higgs mechanism coupled to the gauge fixing, is hence resolved: the `eaten' fluctuations do not correspond to physical QCD-particles. The above discussion of the stability is not to be interpreted in terms of mesons in the dual field theory, but rather establishes that the geometrical configuration we will employ further is stable against small perturbations.

\FloatBarrier

\subsection{Vector sector in \texorpdfstring{$(2\pi\alpha')^2 F^2$-approximation}{F-squared approximation}} \label{F2approx} 

Consider the vector part of the DBI-Lagrangian (\ref{STRgauged}),
\begin{align}
\mathcal L &= \mathcal L_{Higgs} + \mathcal L_{vector} = \text{STr} \hspace{1mm} e^{-\phi} \sqrt{-\det a} |_{\tilde A^2} \nonumber\\
&=  \sum_{a=1}^2 \left\{ \gamma(u) \frac{1}{2}\left( [\tilde A_u,\overline \tau]^a \right)^2 + \alpha(u) \frac{1}{2} \left( [\tilde A_\mu,\overline \tau]^a \right)^2 + \beta(u) \sum_{\mu=1}^2 \frac{1}{2} \left( [\tilde A_\mu,\overline \tau]^a\right)^2  \right\} \nonumber\\
&+ \text{STr} \hspace{1mm} \overline x \left\{
- \overline F_{12} g_{11}^{-2} A^{-1} [\tilde A_1,\tilde A_2]
- \frac{1}{4} g_{11}^{-2} \tilde F_{\mu\nu}^2 \hspace{1mm} A^{-2}|_{\mu,\nu=1,2}  - \frac{1}{2} g_{11}^{-1} G_{uu}^{-1} \tilde F_{\mu u}^2 \hspace{1mm} A^{-1}|_{\mu=1,2}  \right\}.
\end{align}
We have anticipated the vanishing of $\mathcal L_{vector-mixing}$ upon filling in the gauge field expansion in terms of vector mesons, which we will come back to shortly.
Let us reinstate the factors $(2\pi\alpha')$ that we absorbed into the field strengths for notational convenience,  and further approximate\footnote{We assume here that the expansion in $1/\lambda$ is justified because $\lambda \approx 15$ is still large for the parameters that we fixed in section \ref{B}. We will elaborate on the validity of this expansion in the next section.} the action to second order in $(2\pi\alpha')^2 \sim 1/\lambda^2$:
\begin{align}
\hspace{-1.2cm}
\mathcal L
&\sim u^{1/4}  (2\pi\alpha')^2 \sum_{a,b=1}^2 \left\{- \frac{1}{4} f_1 (\tilde F_{\mu\nu}^a)^2- \frac{1}{2} g_{11} f_2 (\tilde F_{\mu u}^a)^2  - \frac{1}{2} \frac{g_{11}}{ (2\pi\alpha')^2} \left(f_2 -  \frac{1}{2} g_{11}^{-2}  (2\pi\alpha')^2  f_3 \right)(\tilde A_\mu^a)^2 (2\overline \tau^3)^2 \right.  \nonumber \\
& \left. \hspace{3cm}  + \sum_{\mu=1}^2 \left( -\frac{1}{2} g_{11}^{-1} f_3 (\tilde A_\mu^a)^2 (2\overline \tau^3)^2  - \frac{1}{2}(\sqrt{G_{uu}}\overline F_{\mu\nu})^3 \epsilon_{3ab}\tilde A_\mu^a \tilde A_\nu^b \right) \right. \nonumber\\
& \qquad \qquad \quad \left.  - \frac{1}{2} \frac{g_{11}^2}{ (2\pi\alpha')^2} \left(f_4 -  \frac{1}{2} g_{11}^{-2}  (2\pi\alpha')^2  f_5 \right)(\tilde A_u^a)^2 (2\overline \tau^3)^2
\right\} + \sum_{l=u,d} \left\{ \sim (\tilde F^l)^2 \right\}
\label{rhopi}
\end{align}
with proportionality factor $-\frac{1}{2} g_s^{-1} R^{\frac{3}{4}+3}$ and
\begin{align}
f_1 =  I(G_{uu}^{1/2}), \quad f_2 =  I(G_{uu}^{-1/2}), \quad f_3 = I(G_{uu}^{-1/2} \overline F_{12}^2), \quad f_4 = I(G_{uu}^{-3/2}) \quad \text{and} \quad f_5 = I(G_{uu}^{-3/2}\overline F_{12}^2) \label{f1f2f3f4f5}
\end{align}
similar $I$-functions as encountered in section \ref{4.1.1}, again arising from the evaluation of the STr using the prescriptions in appendix \ref{appendix}. Being interested in the response of charged mesons to the magnetic field, we will leave out the neutral part of the action $\sim \sum_{l=u,d}$ further on. 

Effective 4-dimensional meson fields are introduced via
the assumption that the flavour gauge field can be expanded in complete sets
$\left\{\psi_n(u)\right\}_{n\geq 1}$ and $\left\{\phi_n(u)\right\}_{n\geq 0}$ as follows \cite{Sakai:2004cn}
\begin{align}
\tilde A_\mu^a(x^\mu,u) &= \sum_{n \geq 1} B_\mu^{(n) a}(x^\mu) \psi_n(u) = \rho_\mu^a(x^\mu) \psi(u) + \cdots  \label{Amuexpansion}\\
\tilde A_u^a(x^\mu,u) &= \sum_{n \geq 0} \phi^{(n) a}(x^\mu) \phi_n(u) = \pi^a(x^\mu) \phi_0(u) + \cdots   \label{Auexpansion}
\end{align}
for $a=1,2$, consistent with the use of partial integration (and hence implicit assumption of vanishing asymptotic gauge fields) in obtaining (\ref{82}) from (\ref{thooftgauge}). The expansion of the non-Abelian components is as in (\ref{expansionvectorpion}) for $\tilde A_\mu^{a}$ and $\tilde A_u^a = 0$ for $a=0,3$ (due to the gauge choice (\ref{a03gauge})).   
The rho meson appears as the lowest mode of the infinite vector meson tower $B_\mu^{(n)}$, and the charged pion as the lowest mode of the infinite (pseudo)scalar meson tower $ \phi^{(n)}$.
We will only retain these lowest-lying mesons in the fluctuation towers, as -- with the purpose of discussing a possible tachyonic vector instability -- it makes sense that the least massive vector meson will likely be the first to condense.

One obtains an effective 4-dimensional action for the mesons by plugging the above fluctuation expansion for the gauge field into the 5-dimensional DBI-action governing the dynamics of the flavour gauge field, and subsequently integrating out the $u$-dependence.
Some terms can already be understood to vanish during the integration over the extra radial dimension $u$ by looking at the parity of $\psi(z) \equiv \psi(u(z))$ and $\phi_0(z) \equiv \phi_0(u(z))$, with $u(z) = u_0^3 + u_0 z^2$ the  commonly used coordinate transformation to the coordinate $z=-\infty \cdots \infty$ following the brane from one asymptotic endpoint to the other. Both $\psi(z)$ and $\phi_0(z)$ are even functions \cite{Sakai:2004cn}, hence coupling terms between rho mesons and pions of the form $\sim D_\mu \tilde A_u^a \partial_u \tilde A_\mu^a \sim D_\mu \pi^a \rho_\mu^a \phi_0(u) \partial_u \psi$ originating from $(\tilde F_{\mu u}^a)^2$ will give rise to vanishing integrals $\int_{-\infty}^{\infty} dz  \{ \text{odd function of $z$} \} = 0$.  This means we can discuss the rho meson and the pion terms separately.
For the same reason the terms $\sim \tilde A_i \partial_u \tilde A_j$ (with $i,j=1,2$) coming from $\mathcal L_{vector-mixing}$ will not survive the $u$-integration.
Note that this simplification is a consequence of cutting the meson towers down to their lowest states.

\subsubsection{Rho meson mass and rho meson condensation} \label{rhomesonc}

\paragraph{Background dependent functions in the action}

Before continuing with the strategy outlined above to extract the 4-dimensional effective action for the rho mesons, we take a closer look at
the relevant functions $f_1$, $f_2$ and $f_3$ as defined in (\ref{f1f2f3f4f5}),
as well as
the definitions for $\overline \tau^3$ and $(G_{uu}^{1/2}\overline F_{12})^3$ in terms of up- and down-components of the background fields. In the rest of this chapter we will absorb $e$ into the notation $B$ for the magnetic field $eB$, because the formulas will begin to take on larger proportions. 

\begin{figure}[h!]
  \hfill
  \begin{minipage}[t]{0.45\textwidth} 
    \begin{center}
      \scalebox{0.83}{
  \includegraphics{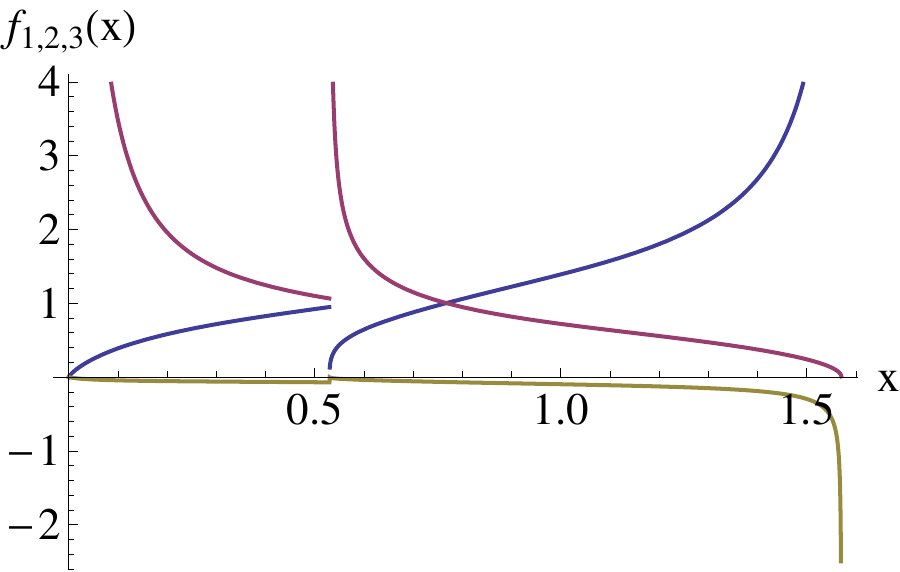}}
    \end{center}
  \end{minipage}
  \hfill
  \begin{minipage}[t]{0.45\textwidth}
    \begin{center}
      \scalebox{0.83}{
  \includegraphics{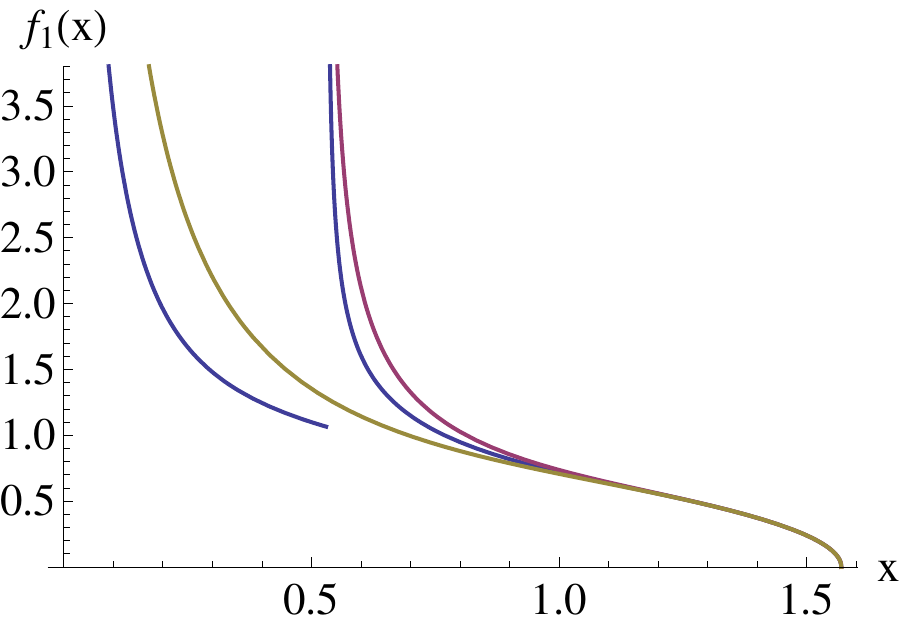}}
    \end{center}
  \end{minipage}
 \hfill
  \begin{minipage}[t]{\textwidth}
    \begin{center}
      \scalebox{0.83}{
  \includegraphics{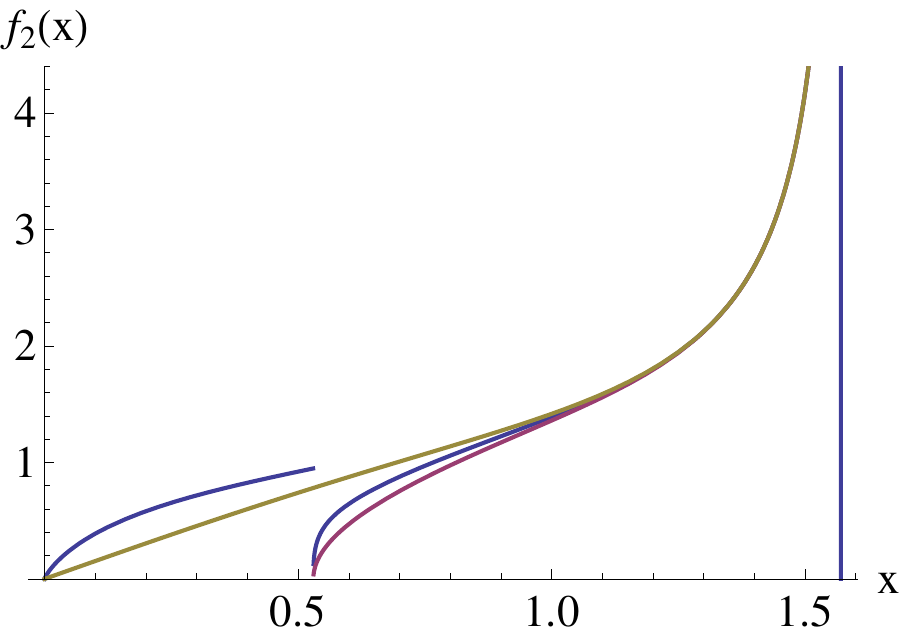}}
    \end{center}
  \end{minipage}
      \caption{(a) $f_1$ (red), $f_2$ (blue) and $f_3$ (yellow) plotted as functions of  $x$, related to $u$ through $u=u_{0,d}\cos^{-3/2}x$ mapping the infinite $u$-range to $x=0\cdots \pi/2$. (b) $f_1 = I(G_{uu}^{1/2})$ (blue) compared to $(G_{uu}^d)^{1/2}$ (yellow) and $(G_{uu}^u)^{1/2}$ (red), i.e.~the functions which would replace $f_1$ if there were a Tr instead of a STr in the action, reducing the non-Abelian to a sum of two Abelian actions. As required, $f_1 \rightarrow G_{uu}^{1/2}$ in the limit of coinciding branes at $u\rightarrow \infty$. (c) $f_2 = I(G_{uu}^{-1/2})$ (blue) compared to $(G_{uu}^d)^{-1/2}$ (yellow) and $(G_{uu}^u)^{-1/2}$ (red). All plots for $B=0.8$ GeV$^2$. }
	\label{f123fig}
  \hfill
\end{figure}

\begin{figure}[h!]
  \hfill
  \begin{minipage}[t]{\textwidth}
    \begin{center}
      \scalebox{1}{
  \includegraphics{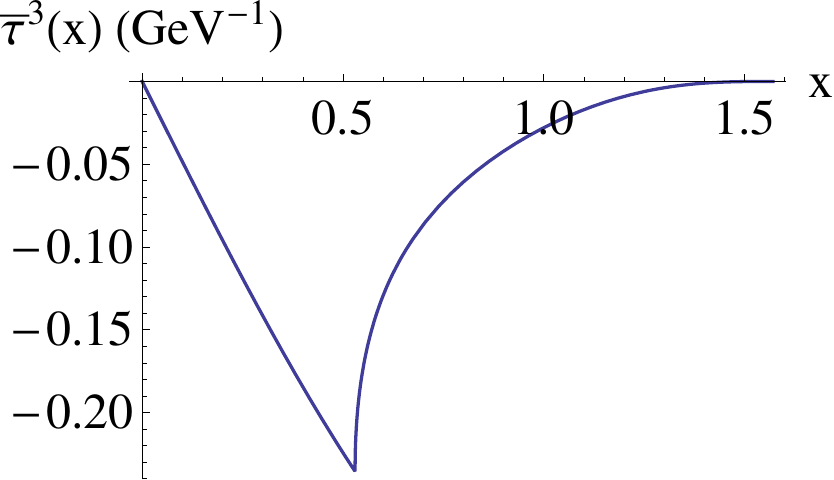}}
    \end{center}
  \end{minipage}
  \hfill
  \begin{minipage}[t]{\textwidth}
    \begin{center}
      \scalebox{1}{
  \includegraphics{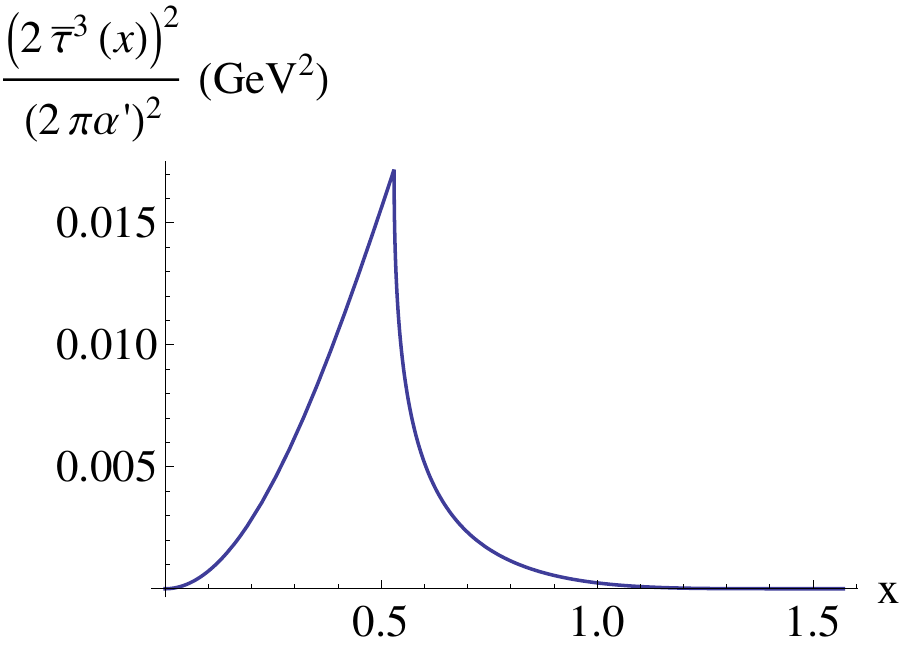}}
    \end{center}
  \end{minipage}
      \caption{The measure $\overline \tau^3(x)$ for the splitting of the branes and the resulting estimated contribution to the mass term for the flavour gauge field and indirectly the rho meson. The range $x=0\cdots \pi/2$ maps to $u=u_{0,d} \cdots \infty$ and we chose $B=0.8$ GeV$^2$. }
	\label{tau3fig}
  \hfill
\end{figure}

Using (\ref{Ifctions}) and (\ref{IfctionsGeneral}), we have
\begin{align}
f_{1} &=  I(G_{uu}^{1/2}) = \frac{1}{2(\partial \overline \tau_u-\partial \overline \tau_d)} \left( \sqrt{G_{uu}^u}\partial \overline \tau_u - \sqrt{G_{uu}^d}\partial \overline \tau_d + \frac{g_{uu}}{\sqrt{g_{\tau\tau}}} \ln \left[ \frac{\partial \overline \tau_u g_{\tau\tau} + \sqrt{g_{\tau\tau} G_{uu}^u}}{\partial \overline \tau_d g_{\tau\tau} + \sqrt{g_{\tau\tau} G_{uu}^d}} \right]\right) \\
f_{2} &= I(G_{uu}^{-1/2}) = \frac{1}{(\partial \overline \tau_u-\partial \overline \tau_d) \sqrt{g_{\tau\tau}}} \ln \left[ \frac{\partial \overline \tau_u g_{\tau\tau} + \sqrt{g_{\tau\tau} G_{uu}^u}}{\partial \overline \tau_d g_{\tau\tau} + \sqrt{g_{\tau\tau} G_{uu}^d}} \right]
\end{align}
\begin{align}
&f_3 = I(G_{uu}^{-1/2} \overline F_{12}^2) =
\frac{1}{2 (\partial \overline \tau_u - \partial \overline \tau_d)^3 g_{\tau\tau}^{3/2}} \left\{ (\overline F_d-\overline F_u)
\left[ \sqrt{g_{\tau\tau} G_{uu}^d} (\partial \overline \tau_d \overline F_d + 3 \partial \overline \tau_d \overline F_u - 4 \partial \overline \tau_u \overline F_d) \right. \right. \nonumber\\
& \qquad \qquad \qquad \qquad \qquad \qquad \qquad \qquad \left. + \sqrt{g_{\tau\tau} G_{uu}^u} (\partial \overline \tau_u \overline F_u + 3 \partial \overline \tau_u \overline F_d - 4 \partial \overline \tau_d \overline F_u) \right] \nonumber\\
& \left. - \left( 2(\partial \overline \tau_u \overline F_d - \partial \overline \tau_d \overline F_u)^2 g_{\tau\tau} - (\overline F_d-\overline F_u)^2 g_{uu} \right) \left[ \ln g_{\tau\tau} g_{uu} + \ln\left( \frac{\partial \overline \tau_u g_{\tau\tau} + \sqrt{g_{\tau\tau} G_{uu}^u}}{\partial \overline \tau_d g_{\tau\tau} - \sqrt{g_{\tau\tau} G_{uu}^d}} \right) \right] \right\},
\end{align}
with $\partial \overline \tau$ short for $\partial_u \overline \tau =$ (\ref{tauembedding}), $G_{uu}^l = G_{uu}(\partial_u \overline \tau_l)$ and $\overline F_u = \frac{2B}{3}$, $\overline F_d =- \frac{B}{3}$, as defined in (\ref{Fbardef}).
Because of the theta-functions $\theta(u-u_{0,l})$ contained in $\partial_u \overline \tau_l$, the contribution of $\partial_u \overline \tau_u$ only kicks in at $u>u_{0,u}$. Therefore these functions will all be discontinuous at $u=u_{0,u}$, as can be seen in the illustrative plot in figure \ref{f123fig} for $B=0.8$ GeV$^2$. The dependence on $B$ is implicit through the embedding,
except for $f_3$ which also depends explicitly on $B$.
Further,
$\overline \tau^3$ gives a measure for the distance between up- and down-brane, defined as
\begin{align}\label{tau3math}
\overline \tau^3(u) = \int_\infty^u  \partial_u \overline \tau^3 du  =
\int_\infty^u  \frac{\partial_u \overline \tau_u - \partial_u \overline \tau_d}{2} du  = \int_\infty^{u_{0,u}}  \frac{\partial_u \overline \tau_u - \partial_u \overline \tau_d}{2} du + \int_{u_{0,u}}^{u}  \frac{- \partial_u \overline \tau_d}{2} du \nonumber
\end{align}
such that $\overline \tau^3$ fulfills the boundary condition that the flavour branes coincide at $u \rightarrow \infty$: $\overline \tau \sim \mathbf{1} \Rightarrow \overline \tau^3(\infty)=0$. In figure \ref{tau3fig} the resulting discontinuous $\overline \tau^3$  is plotted for $B=0.8$ GeV$^2$, along with $(2\overline \tau^3)^2/(2\pi \alpha')^2$ which  contributes to the `$u$-dependent mass' of the 5-dimensional gauge field. The contribution is small -- although it is $(2\pi \alpha')^{-2}$-enhanced -- since the splitting itself is a small effect.
The last relevant background function in the action (\ref{rhopi}) for the discussion of the rho mesons is
\begin{align}
(G_{uu}^{1/2} \overline F_{12})^3 = \sqrt{G_{uu}^u} \overline F_u - \sqrt{G_{uu}^d} \overline F_d.
\end{align}

\paragraph{Eigenvalue problem} \label{paragraaf}

Upon substitution of the gauge field expansions (\ref{Amuexpansion}) and (\ref{Auexpansion}) into (\ref{rhopi}), the 5-dimensional DBI-Lagrangian to second order in the rho meson fluctuations (and second order in $(2\pi\alpha')$) reads
\begin{align}
\hspace{-1.2cm}
\mathcal L
&\sim u^{1/4} (2\pi\alpha')^2 \sum_{a,b=1}^2 \left\{- \frac{1}{4} f_1 (\mathcal F_{\mu\nu}^{a})^2 \psi^2 - \frac{1}{2} g_{11} f_2 (\rho_\mu^a)^2 (\partial_u \psi)^2  - \frac{1}{2} \frac{g_{11}}{(2\pi\alpha')^2} \tilde f_2 (\rho_\mu^a)^2 \psi^2 (2 \overline \tau^3)^2  \right.  \nonumber \\
& \left. \hspace{3cm}  + \sum_{\mu=1}^2 \left( -\frac{1}{2} g_{11}^{-1} f_3 (\rho_\mu^a)^2 \psi^2 (2 \overline \tau^3)^2     - \frac{1}{2}(\sqrt{G_{uu}}\overline F_{\mu\nu})^3 \epsilon_{3ab} \rho_\mu^a \rho_\nu^b \psi^2 \right) \right\} + \text{pions}, \label{70}
\end{align}
with $\mathcal F_{\mu\nu}^a = D_\mu \rho_\nu^a - D_\nu \rho_\mu^a$ and $\tilde f_2 = f_2 - \frac{1}{2}g_{11}^{-2}(2\pi\alpha')^2 f_3$.

Demanding that the first line of this Lagrangian reduces to the standard 4-dimensional form
\begin{align}
 \sum_{a=1,2} \left(  - \frac{1}{4} (\mathcal F_{\mu\nu}^a)^2 - \frac{1}{2} m_\rho^2 (\rho_\mu^a)^2  \right) \label{71}
\end{align}
after integrating out the $u$-dependences, leads to a normalization condition
\begin{equation} \label{normcondition}
\int_{u_{0,d}}^\infty du \hspace{1mm} u^{1/4} f_1 \psi^2 = 1
\end{equation}
and a mass term condition
\begin{equation} \label{masscondition}
\int_{u_{0,d}}^\infty du \hspace{1mm} \left\{ u^{1/4} g_{11} f_2 (\partial_u \psi)^2 +  u^{1/4} \frac{g_{11}}{(2\pi\alpha')^2} \tilde f_2  (2\overline \tau^3)^2\psi^2  \right\} = m_\rho^2
\end{equation}
on the $\psi(u)$ functions\footnote{We absorbed the total prefactor $\sqrt{V_4 T_8 g_s^{-1} R^{\frac{3}{4}+3} (2\pi\alpha')^2}$ into $\psi$ such that $\psi$ has a total mass dimension of $5/8$ instead of $2$ (without the prefactor).},
which
combine through partial integration
to an eigenvalue equation for $\psi(u)$:
\begin{equation} \label{eigeq}
u^{-1/4} f_1^{-1} \partial_u\left(u^{1/4} g_{11} f_2 \partial_u \psi \right) -  \frac{g_{11}}{(2\pi\alpha')^2} f_1^{-1}\tilde f_2 (2\overline \tau^3)^2 \psi  = -\Lambda \psi,
\end{equation}
with the eigenvalue $\Lambda = m_\rho^2$ the sought for rho meson mass squared.
We can separate the Higgs contribution to $m_\rho^2$ by defining
\begin{align} \label{Higgscontr}
\tilde m_\rho^2 &= \int_{u_{0,d}}^\infty du \hspace{1mm} u^{1/4} g_{11} f_2 (\partial_u \psi)^2 \quad \text{ and } \quad  m_{\rho,Higgs}^2 =  \int_{u_{0,d}}^\infty du \hspace{1mm}  u^{1/4} \frac{g_{11}}{(2\pi\alpha')^2} \tilde f_2  (2\overline \tau^3)^2\psi^2
\end{align}
such that
\begin{equation}
m_\rho^2 = \tilde m_\rho^2 + m_{\rho,Higgs}^2.
\end{equation}
Let us also mention that from (\ref{masscondition}) one can see that $m_\rho^2 >0$, since the integrand is positive. 

To solve the eigenvalue equation (\ref{eigeq}) numerically on a compact interval, we change to the coordinate $x=0\cdots x_{up} \cdots \frac{\pi}{2}$
related to $u=u_{0,d}\cdots u_{0,u} \cdots \infty$ by
\begin{equation}
u^3 = u_{0,d}^3 \cos^{-2} x.
\end{equation}
Rewritten as a function of $x$, the eigenvalue equation is invariant under $x\rightarrow-x$, so we can split up the eigenfunction set in even/odd $\psi_n(x)$'s, which correspond to odd/even parity mesons:
\begin{equation}
    \psi_n(0)=0\quad\text{or}\quad\partial_x\psi_n(0)=0.
\end{equation}

Asymptotically, the eigenvalue equation (\ref{eigeq}) reduces to $\partial_u \left( u^{5/2} \partial_u  \psi \right) = 0$, with the asymptotic solution $\psi(\infty) = c \frac{u^{-3/2}}{-3/2} + d$ only normalizable through (\ref{normcondition})  if $d=0$, i.e.~if
\begin{equation} \label{bdy}
\psi(u\rightarrow \infty) = 0 \quad \text{or} \quad \psi(x\rightarrow \pm \pi/2) = 0.
\end{equation}

The eigenvalue problem (\ref{eigeq}) for the (odd parity) rho meson with the appropriate boundary condition (\ref{bdy}) in the $x$-coordinate  is thus of the form
\begin{equation} \label{problem}
\cdots \partial_x^2\psi + \cdots \partial_x\psi + \cdots \psi = -\Lambda \psi \quad \text{with}\qquad \psi(\pm \pi/2)=0\,, \quad \partial_x\psi(0)=0.
\end{equation}
To solve it we employ a shooting method, which consists of temporarily replacing (\ref{problem}) with the well-defined initial value problem
\begin{equation} \label{diffeq}
    \cdots \partial_x^2\psi + \cdots \partial_x\psi + \cdots \psi = -\Lambda \psi \quad \text{with}\qquad \psi(0)=1\,, \quad \partial_x\psi(0)=0
\end{equation}
where $\Lambda$ is treated as a `shooting' parameter. We used the scaling freedom $\psi(x)\to h\psi(x)$ to impose that $\psi(0)=1$ (the value of $h$ will be fixed by the normalization condition in the end). For each value of $\Lambda$, (\ref{diffeq}) can be solved numerically for $\psi_\Lambda(x)$.
Next, solving the equation $\psi_\Lambda(\pi/2)=0$ finally determines the eigenvalue $\Lambda = m_\rho^2$.

\begin{figure}[h!]
  \hfill
  \begin{minipage}[t]{\textwidth}
    \begin{center}
      \scalebox{0.75}{
  \includegraphics{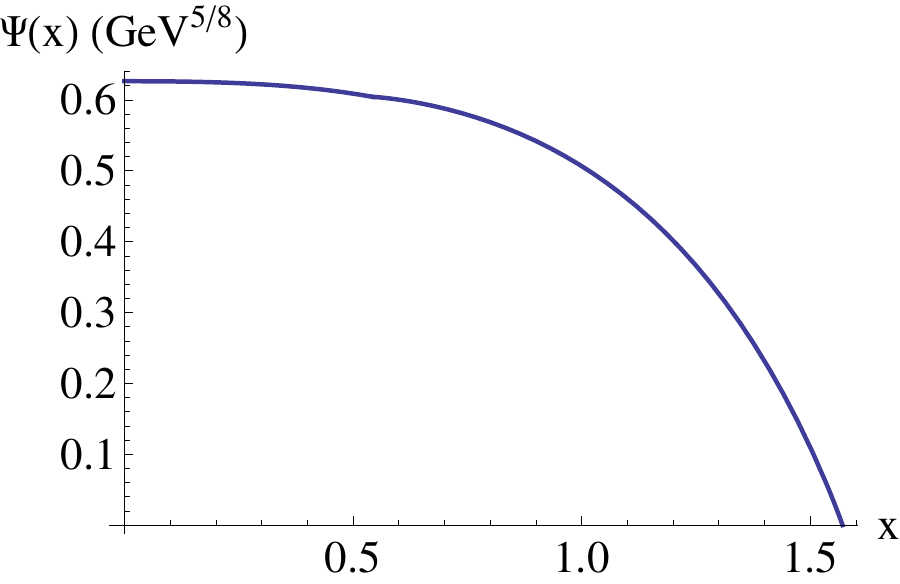}}
    \end{center}
  \end{minipage}
  \hfill
  \begin{minipage}[t]{\textwidth}
    \begin{center}
      \scalebox{0.75}{
  \includegraphics{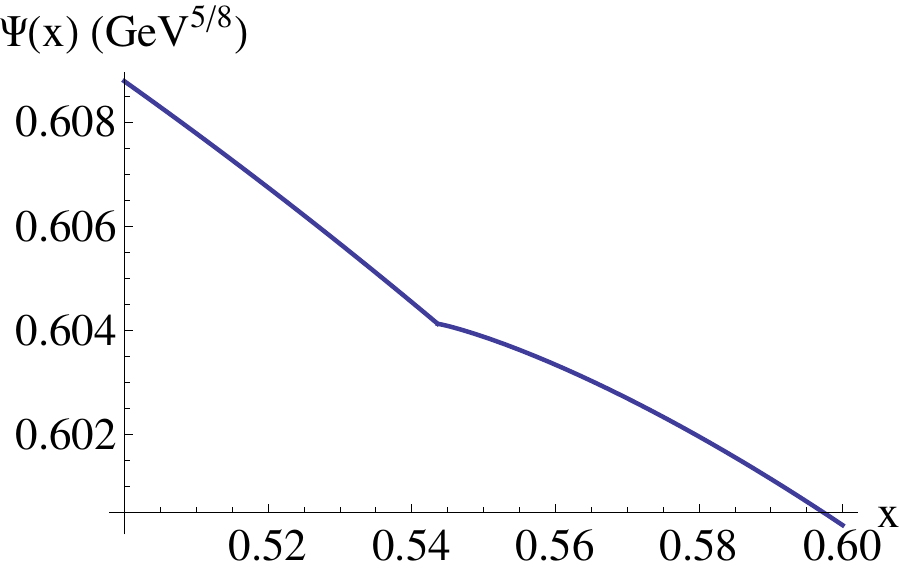}}
    \end{center}
  \end{minipage}
 \hfill
  \begin{minipage}[t]{\textwidth}
    \begin{center}
      \scalebox{0.75}{
  \includegraphics{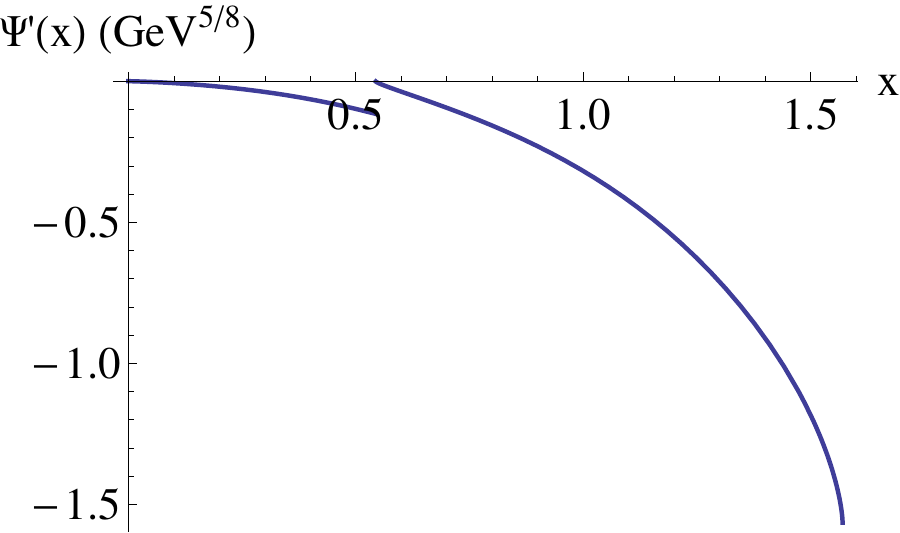}}
    \end{center}
  \end{minipage}
      \caption{Plots of the rho meson eigenfunction $\psi(x)$ and its derivative $\partial_x \psi(x)$, discontinuous at $x=x_{up} \approx 0.54$, for $B=0.9$ GeV$^2$, obtained numerically with a shooting method.}
	\label{psifigs}
  \hfill
\end{figure}

For completeness we add a few comments about the numerical method we used to solve the eigenvalue problem at hand (\ref{problem}), which in detail reads
\begin{align}
&\frac{9}{4}R^{-3/2}u_{0,d}^{-1/2} \frac{\cos^{11/6} x}{\sin x} f_{1}^{-1} \partial_x \left( f_{2}\frac{\cos^{1/2} x}{\sin x}\partial_x \psi \right)
- R^{-3/2} \frac{u_{0,d}^{3/2}}{(2\pi\alpha')^2} (\cos^{-1} x) \tilde f_{2} f_{1}^{-1} (2\overline \tau^3)^2 \psi = -m_\rho^2 \psi,
\end{align}
with $\psi(\pm \pi/2)=0$ and $\partial_x \psi(0)=0$. Near the origin $x \rightarrow 0$ the equation takes the form
\begin{align}
m_\rho^2 \psi + \partial_x^2 \psi - \ln x \hspace{1mm} \partial_x^2 \psi -\frac{1}{x}\partial_x \psi  &= 0,  \label{originbehaviour}
\end{align}
so we have to provide Mathematica with an ansatz for $\psi(x)$ at small $x$ to prevent the equation from blowing up there.
Demanding that $\partial_x \psi \sim x$ to avoid the last term in (\ref{originbehaviour}) from diverging, would still give $ \ln x \hspace{1mm} \partial_x^2 \psi \rightarrow -\infty$. Instead we demand that  $\partial_x^2 \psi \sim \frac{1}{\ln x}$ or $\psi(x\rightarrow 0) = 1 + x^2 \sum_{i=1}^n \frac{a_i}{\ln^i x}$ (in practice we have set $n=13$).
With this ansatz for $\psi \Rightarrow \partial_x \psi \sim \text{LogIntegral}(x) + c$, the term $\frac{1}{x}\partial_x \psi$ will only be finite if the integration constant $c=\partial_x \psi(0) = 0$ \footnote{This is consistent with vector mesons, but not with the initial condition on axial mesons (which we have not considered). We have not looked into it further to see if there is a way around this, in order to still be able to describe axial mesons in the presence of a magnetic field in this setting.}. Near $x = x_{up}$, or $y\rightarrow 0$ in the useful coordinate $y$ defined through $u^3 = u_{0,u}^3 \cos^{-2} y$, the differential equation's form
\begin{align}
m_\rho^2 \psi + \partial_y^2 \psi - \ln y \hspace{1mm} \partial_y^2 \psi -\frac{1}{y}\partial_y \psi  &= 0  \label{xupbehaviour}
\end{align}
again needs to be fed with an ansatz for $\psi$ that keeps the equation finite, i.e.~$\psi(y\rightarrow 0) = \psi(x=x_{up}) + y^2 \sum_{i=1}^n \frac{a_i}{\ln^i y}$ with $\partial_y \psi(0) = 0$.
This means we can demand continuity of $\psi$ at $x=x_{up}$ but not of its derivative\footnote{It is known that the Schr\"odinger wave function can display kinks (thus jumps in its derivative), depending on the potential (singularities), see e.g.~\cite{levi}. This corresponds to the singular behaviour of some of the coefficient functions for $y\to0$ in \eqref{xupbehaviour}.}.
An example result of $\psi(x)$ and its derivative is shown in figure \ref{psifigs}.

\paragraph{Effective 4-dimensional EOM and result for total eigenvalue}  \label{paragraaf2}

The effective 4-dimensional action becomes
\begin{align}
S_{4D}
&= \int d^4 x \sum_{a,b=1}^2 \left\{- \frac{1}{4}(\mathcal F_{\mu\nu}^{a})^2 -\frac{1}{2} m_\rho^2(B) (\rho_\mu^a)^2 + \sum_{\mu=1}^2 \left( -\frac{1}{2} m_{+}^2(B)  (\rho_\mu^a)^2 -\frac{1}{2} \epsilon_{3ab}\rho_\mu^a \rho_\nu^b \hspace{1mm} k(B) \overline F_{\mu\nu}^3 \right)\right\} \label{S4D}
\end{align}
with the normalized $\psi$, as determined in the previous paragraph, satisfying the normalization and mass conditions (\ref{normcondition}) and (\ref{masscondition}),
and the newly defined $m_+$ and $k$ to be calculated from
\begin{equation} \label{massplus}
\int_{u_{0,d}}^\infty du \hspace{1mm} u^{1/4} g_{11}^{-1} f_3 (2\overline \tau^3)^2 \psi^2 = m_{+}^2
\end{equation}
and
\begin{equation}
\int_{u_{0,d}}^\infty du \hspace{1mm} u^{1/4} (\sqrt{G_{uu}}\overline F_{12})^3 \psi^2 = k \hspace{1mm} \overline F_{12}^3
\end{equation}
with $\overline F_{12}^3=B$.
Here $m_+$ is an extra contribution to the mass of the transverse (w.r.t.\ the magnetic field $\vec B = B \vec e_3$) components of the charged rho meson, $\rho_{\mu=1,2}^{a=1,2}$, as a consequence of $B$ breaking Lorentz invariance. We repeat that the parameter $k$ describes a non-minimal coupling of the charged rho meson to the magnetic field, related to the magnetic moment $\mu$ via $\mu = (1+k)e/(2m)$ so to the gyromagnetic ratio $g$ via $g=1+k$.

\begin{figure}[h!]
  \hfill
  \begin{minipage}[t]{\textwidth}
    \begin{center}
      \scalebox{1.2}{
  \includegraphics{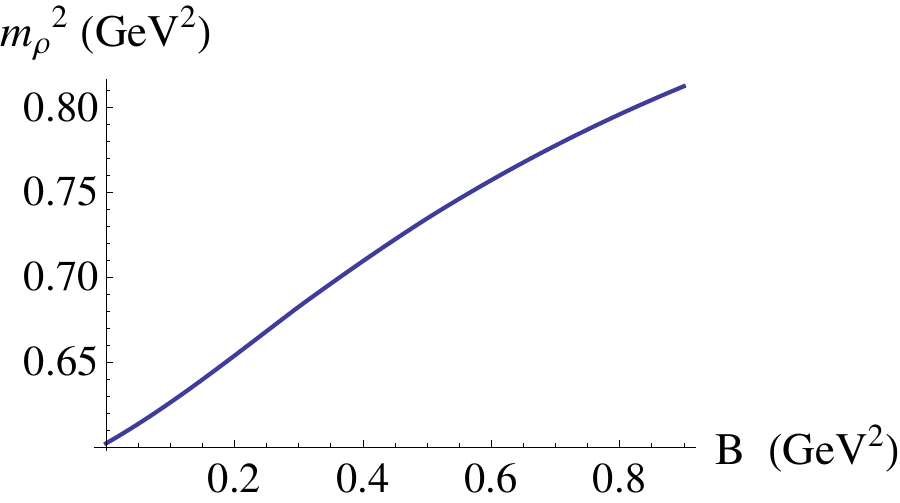}}
    \end{center}
  \end{minipage}
  \hfill
  \begin{minipage}[t]{\textwidth}
    \begin{center}
      \scalebox{1.2}{
  \includegraphics{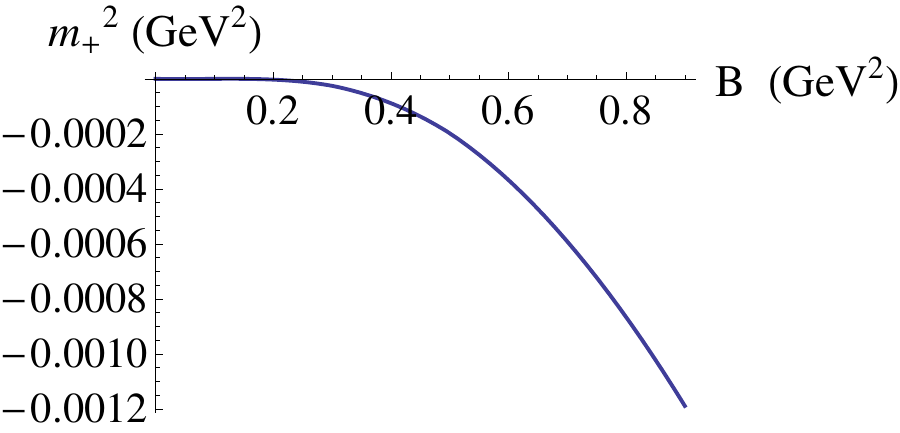}}
    \end{center}
  \end{minipage}
\hfill
  \begin{minipage}[t]{\textwidth}
    \begin{center}
      \scalebox{1.2}{
  \includegraphics{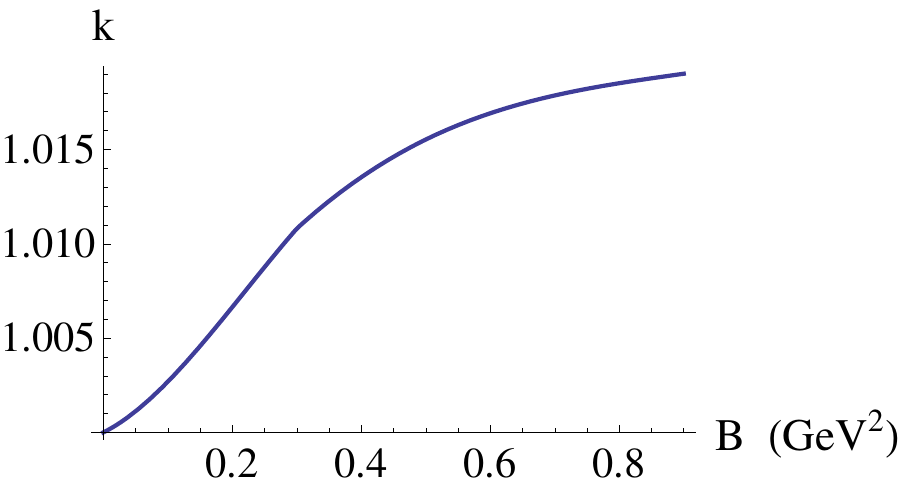}}
    \end{center}
  \end{minipage}
      \caption{Numerical results for $m_\rho^2(B)$, $m_+^2(B)$ and $k(B)$ in the $(2\pi\alpha')^2 F^2$-approximation of the DBI-action.
}
	\label{resultsfig}
  \hfill
\end{figure}

\begin{figure}[h!]
  \centering
  \scalebox{1.2}{ 
  \includegraphics{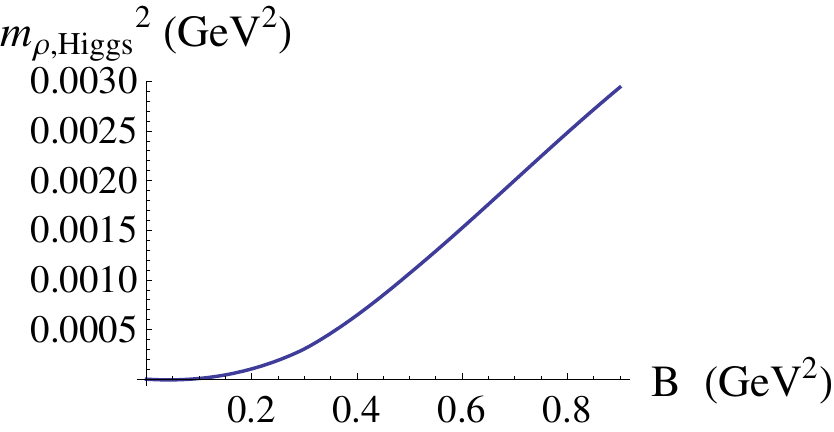}}
  \caption{The Higgs contribution $m_{\rho,Higgs}^2(B)$ to the rho meson mass squared $\mathit{m_{\rho}^2}(B)$, as defined in (\ref{Higgscontr}), in the $(2\pi\alpha')^2 F^2$-approximation of the DBI-action.
} \label{mHiggs2fig}
\end{figure}

The standard 4-dimensional action used to describe the coupling of charged rho mesons to an external magnetic field is given by the Proca action \cite{Obukhov:1984xb}. 
The Proca action is equal to (\ref{S4D}) with $m_+ = 0$ and $m_\rho$ and $k(=1)$ independent of $B$: there is only explicit dependence of the action on $B$, which is to be traced back to the treatment of the rho mesons as point-like structureless particles.
Instead, in our current approach, the effect of $B$ on the constituent quarks
is taken into account via the effect of $B$ on the embedding of the flavour probe branes, 
leading to an implicit dependence on $B$ of both the mass $m_\rho^2(B)$ and the magnetic coupling $k(B)$. The effect of $B$ on the embedding is two-fold (see section \ref{D} and in particular figure \ref{changedembedding}): the branes move upwards in the holographic direction, corresponding to chiral magnetic catalysis, and the up- and down-brane get separated, corresponding to a stronger chiral magnetic catalysis 
for the up-quark than for the down-quark.
Both effects translate into a mass generating effect for the rho meson, $m_\rho^2(B) \nearrow$, as can be seen in figure \ref{resultsfig}. The chiral magnetic catalysis causes the rho meson to get heavier as its constituents do.
The split between the branes adds to the mass of the rho meson via a holographic Higgs mechanism: as the branes separate, the flavour gauge field strings between up and down branes (i.e.~representing charged quark-antiquark combinations $u\overline d$, $\overline u d$) get stretched. Because of their string tension this results in an extra Higgs mass term in the action for $\tilde A_\mu^{a=1,2}$, and thus for $\rho_\mu^{a=1,2}$. It is of the form $(A_\mu^a)^2 (\overline \tau^3)^2$, with $\overline \tau^3 \sim  \overline \tau_u - \overline \tau_d$, originating from $(D_\mu \tau)^2 \leadsto ([\tilde A_\mu,\overline \tau])^2$ in the start action.
Where in the absence of splitted branes, $\overline \tau^3=0$, the 4-dimensional mass $m_\rho$ as defined in going from (\ref{70}) to (\ref{71}) is purely effective, i.e.~only present after integrating out the fifth dimension $u$, the Higgs contributions to the mass stem from the stringy mass of the 5-dimensional gauge field itself.
We cannot offer a direct interpretation of the stringy mass contribution in effective QCD-terms. 
Since the splitting of the branes is small though, the induced mass contribution is almost negligible, see figure \ref{mHiggs2fig}.
Further, as can be seen in figure \ref{resultsfig}, $m_+(B) \searrow$ as $f_3$ in (\ref{massplus}) is negative, so the mass of the transversal components of the charged rho mesons will already be slightly smaller than that of the longitudinal ones,
and $k(B) \nearrow$ is approximately equal to one, but not exactly, corresponding to a gyromagnetic ratio $g\approx 2$.


The 4-dimensional EOMs for the charged rho mesons $\rho_\mu^{a=1,2}$ are given by
\begin{align}\label{EOMeq}
D_\mu \mathcal F_{\mu\nu}^a - \epsilon_{a3b} \hspace{1mm} k(B) \hspace{1mm} \overline F_{\mu\nu}^3 \rho_\mu^b - M^2(B) \rho_\nu^a = 0,  \\
M^2(B) = m_\rho^2(B) + (\delta_{\nu 1}+\delta_{\nu 2}) m_{+}^2(B)
\end{align}
with $D_\mu = \partial_\mu +[\overline A_\mu,\cdot]$ and $\mathcal F_{\mu\nu}^a = D_\mu \rho_\nu^a - D_\nu \rho_\mu^a$.
They combine into the EOM
\begin{equation}  \label{rhoeq}
\text{D}_\mu(\text D_\mu \rho_\nu - \text D_\nu \rho_\mu) - i  \hspace{1mm}k(B) \hspace{1mm} \overline F_{\mu\nu}^3 \rho_\mu - M^2(B) \rho_\nu = 0
\end{equation}
with $\text D_\mu = \partial_\mu + i \overline A_\mu^3$ for the charged  combination $\rho_\mu = (\rho_\mu^1 + i \rho_\mu^2)/\sqrt 2$, and the complex conjugate of this equation for the other charged combination $\rho_\mu^* = (\rho_\mu^1 - i \rho_\mu^2)/\sqrt 2$. 
Here, we adopted a different notation for the charged fields ($\rho_\mu^- \rightarrow \rho_\mu$ and $\rho_\mu^+ \rightarrow \rho_\mu^*$) compared to section \ref{simpleholorho1}.  

Solving (\ref{rhoeq}) with $\rho_\nu \rightarrow e^{i(\vec p \cdot \vec x - E t)} \rho_\nu$ for the eigenvalues of the energy we find `modified Landau levels' that we will discuss in more detail in the next section. There they will show up as a special case of the most general form of modified Landau levels, which we encounter as  solutions of the 4-dimensional EOMs that come from 
using the full DBI-action. 
Only in the case that $k=1$, $m_+=0$ and $m_\rho(B)=m_\rho(0)$ one retrieves the standard Landau levels for a free relativistic spin-$s$ particle  moving in the background of a constant magnetic field $\vec B=B \vec e_3$ (assuming $B>0$):
\begin{align}
E^2 = m_\rho^2 + p_3^2 + (2n - 2 s_3 + 1) B
\end{align}
with $n$ the Landau level number and $s_3$ the eigenvalue of the spin operator
\begin{equation}  \label{S3spin}
S_3 = \frac{1}{2} \left( \begin{array}{cc} 0 & \sigma_2-i \sigma_1 \\
\sigma_2+i \sigma_1 & 0 \end{array}\right)
\end{equation}
giving the projection of the spin of the particle onto the direction of the magnetic field.

\begin{figure}[h!]
  \centering
  \scalebox{1.35}{
  \includegraphics{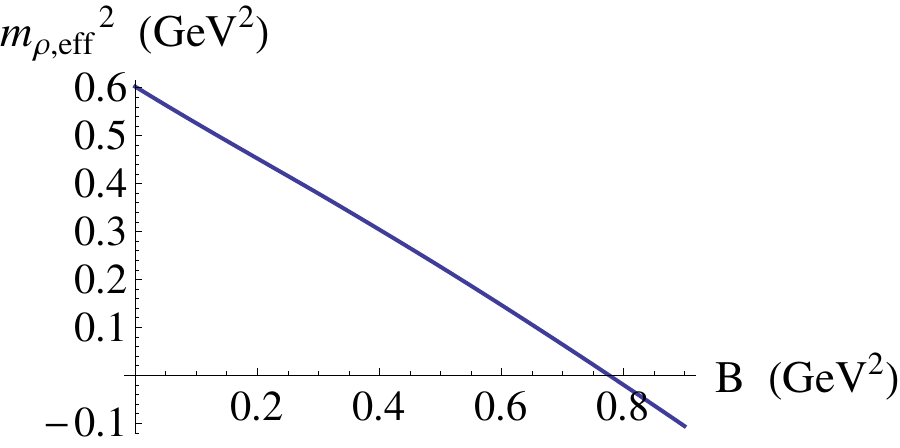}}
  \caption{The effective rho meson mass squared $\mathit{m_{\rho,eff}^2}(B)$  in the $(2\pi\alpha')^2 F^2$-approximation of the DBI-action.
} \label{mrhoeff2fig}
\end{figure}

While the modifications due to $k\neq1$, $m_+(B)\neq0$ and $m_\rho(B)$ are a bit subtle for higher levels, the energy of $s_3=1, p_3=0$ particles in the lowest Landau level $n=0$
is given by a straightforward generalization of $E^2 = m_\rho^2 - B$, namely 
\begin{equation} \label{}
E^2 = M^2(B) -  B  \hspace{1mm} k(B). 
\end{equation}
We conclude that the combinations of charged rho mesons that have their spin aligned with the magnetic field ($s_3 = 1$),  i.e.
\begin{align}
\rho = \rho_1 + i \rho_2 \quad \mbox{ and }\quad  \rho^* =  \rho_1^* - i \rho_2^*, \label{fieldcom}
\end{align}
will have an effective mass squared
\begin{equation} \label{mrhoeff2}
\mathit{m_{\rho,eff}^2} = M^2(B) - B \hspace{1mm} k(B)
\end{equation}
going through zero at a critical magnetic field
\begin{equation} \label{Bc}
B_c \approx 0.78 \text{ GeV}^2,
\end{equation}
which marks the onset of rho meson condensation. Our result for $\mathit{m_{\rho,eff}^2}$ is shown in figure \ref{mrhoeff2fig}.

The total action includes, next to the DBI-part, a Chern-Simons term. In general,  contributions from the Chern-Simons action are suppressed in the large
$\lambda$ expansion,
but in the presence of large background fields Chern-Simons effects can become important, similar to the higher order terms in the $(2\pi\alpha' \sim \frac{1}{\lambda})$-expansion of the DBI-action (see comments in the upcoming section \ref{ambiguities}).
The intrinsic-parity-odd nature of the Chern-Simons action ensures that it will not contribute $\rho^2$-terms to the effective 4-dimensional action to second order in the fluctuations, but it will describe $\rho \pi B$ coupling terms between rho mesons and pions. However, as discussed in more detail in section \ref{CSpion}, 
the antisymmetrization over spacetime indices in the Chern-Simons action (see eq.\ (\ref{SCSNc}))
\begin{align}
S_{CS}
&\sim  \int \text{STr} \left( \epsilon^{mnpqr} A_m F_{np} F_{qr} + \mathcal O(\tilde A^3) \right)
\end{align}
will make sure that the magnetic field $B = \overline F_{12}^3$ only induces couplings between longitudinal fluctuations ($\mu=0,3$), hence not affecting the
dynamics of transversal rho mesons (\ref{fieldcom}) and their condensation.

\FloatBarrier

\subsubsection{Pion mass} \label{4.3.2}

We briefly discuss the charged pion part of the DBI-Lagrangian (\ref{rhopi}), which upon substitution of the gauge field expansion (\ref{Auexpansion}) and further approximation to second order in the pion fields reads
\begin{align}
\hspace{-1.2cm}
\mathcal L \sim u^{1/4}  (2\pi\alpha')^2 \sum_{a,b=1}^2 \left\{- \frac{1}{2} g_{11} f_2 (D_\mu \pi^a)^2 \phi_0^2  - \frac{1}{2} \frac{g_{11}^2}{ (2\pi\alpha')^2} \tilde f_4 (2\overline \tau^3)^2  (\pi^a)^2 \phi_0^2
\right\}
\end{align}
with $\tilde f_4 = f_4 -  \frac{1}{2} g_{11}^{-2}  (2\pi\alpha')^2  f_5$.
Ignoring in this section the $1/\lambda$-suppressed $\rho \pi B$-contributions from the Chern-Simons action,
the effective 4-dimensional action for the charged pions becomes
\begin{align}
S_{4D} = \int d^4 x \sum_{a,b=1}^2 \left\{- \frac{1}{2}(D_\mu \pi^{a})^2 -\frac{1}{2} m_\pi^2(B) (\pi^a)^2 \right\}
\end{align}
with $\phi_0$ satisfying the normalization condition
\begin{equation} \label{normpion}
\int_{u_{0,d}}^\infty du \hspace{1mm} u^{1/4} g_{11} f_2 \phi_0^2 = 1
\end{equation}
and the pions no longer massless:
\begin{equation} \label{mpion}
\int_{u_{0,d}}^\infty du \hspace{1mm} u^{1/4} \frac{g_{11}^2}{ (2\pi\alpha')^2} \tilde f_4 (2\overline \tau^3)^2  \phi_0^2  = m_{\pi}^2.
\end{equation}
We can understand the emergence of this mass again as a consequence of the holographic Higgs mechanism. The magnetic field breaks chiral symmetry explicitly (albeit only slightly) by pulling the up- and down-brane apart. The previously massless pions, serving as Goldstone bosons associated with the spontaneous breaking of chiral symmetry, hence get a small mass, related to the distance $\overline \tau^3 \sim \overline \tau_u - \overline \tau_d$ between the branes.
Solving the effective 4-dimensional EOM for the charged pions with $\pi \rightarrow e^{i(\vec p \cdot \vec x - E t)} \pi$ for the eigenvalues of the energy, one finds `almost Landau levels' for a spinless particle
\begin{align}
E^2 = m_\pi^2(B)+ p_3^2 + (2n + 1) B
\end{align}
or an effective mass squared in the lowest Landau level
\begin{equation} \label{}
\mathit{m_{\pi,eff}^2} = m_\pi^2(B)  + B.
\end{equation}
The pion thus gets a mass in the presence of a magnetic field, although we are working in a model in the chiral limit (zero bare quark masses) and with no chiral condensate (at least not in the setting we used, without incorporating a tachyon field as was done in \cite{Bergman:2007pm}). This violates the GMOR-relation (\ref{GMOR}) relating the  bare quark masses times chiral condensate to the mass of the pion. It was however already discussed in e.g.~\cite{Shushpanov:1997sf,Orlovsky:2013gha} that the GMOR-relation is no longer valid for charged pions in the presence of a magnetic field.

\begin{figure}[h!]
  \hfill
  \begin{minipage}[t]{\textwidth}
    \begin{center}
      \scalebox{0.75}{
  \includegraphics{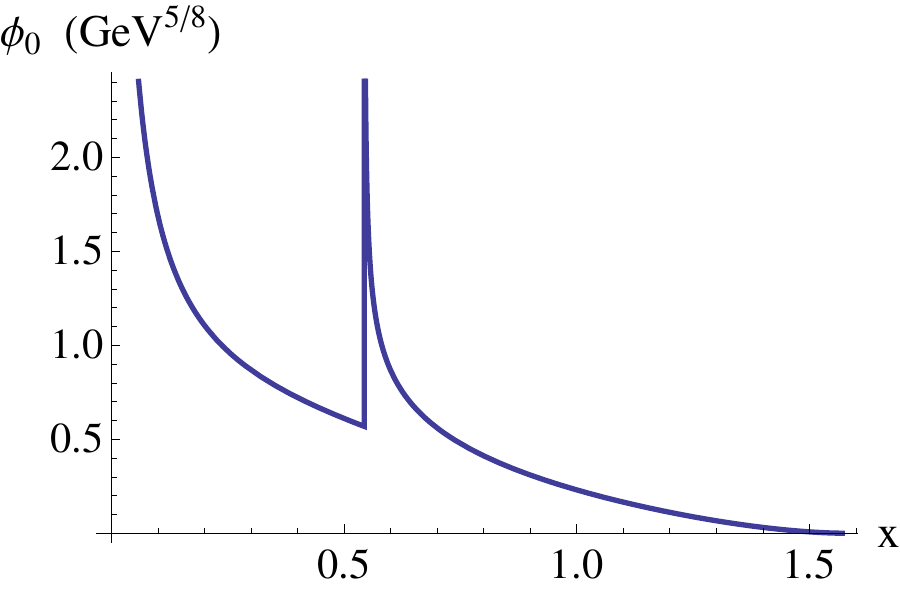}}
    \end{center}
  \end{minipage}
  \hfill
  \begin{minipage}[t]{.45\textwidth}
    \begin{center}
      \scalebox{0.9}{
  \includegraphics{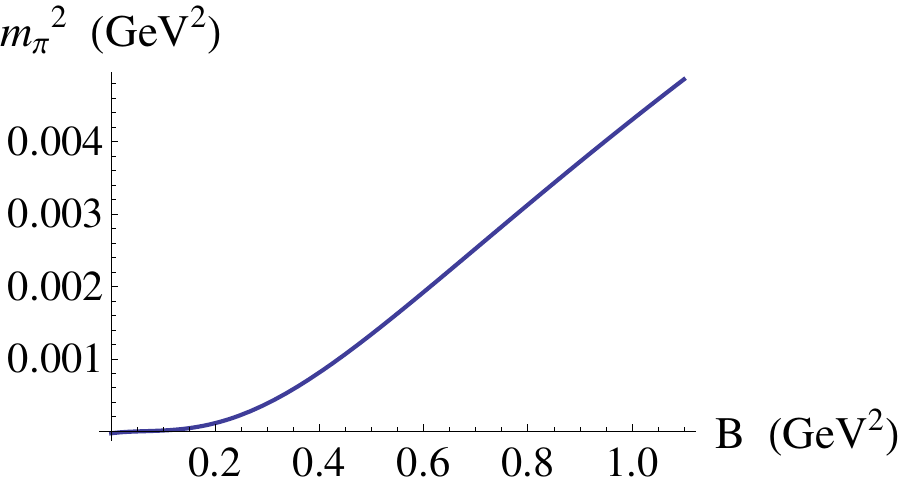}}
    \end{center}
  \end{minipage}
\hfill
  \begin{minipage}[t]{0.45\textwidth}
    \begin{center}
      \scalebox{0.9}{
  \includegraphics{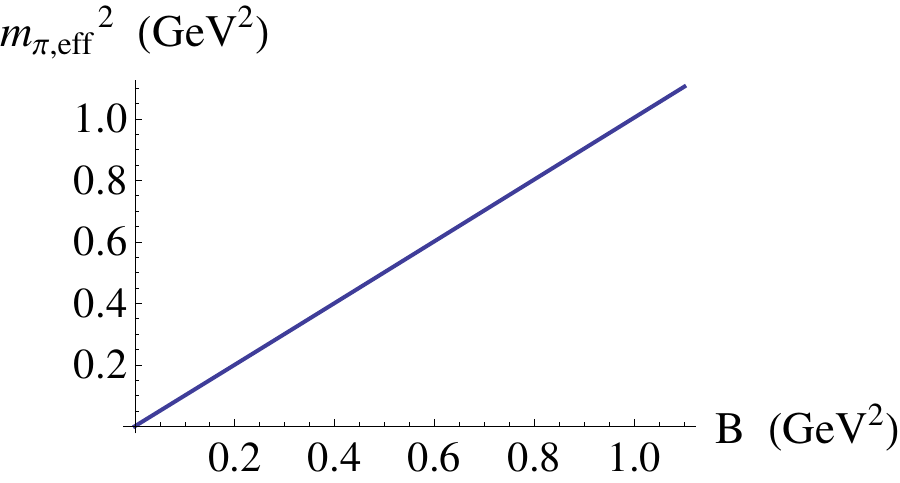}}
    \end{center}
  \end{minipage}
      \caption{Pion eigenfunction $\phi_0(x)$ (with $u=u_{0,d} \cos^{-3/2}x$) for $B=1.1$ GeV$^2$. Numerical result for $m_\pi^2(B)$ and the effective pion mass squared $\mathit{m_{\pi,eff}^2(B)}$ in the $(2\pi\alpha')^2 F^2$-approximation of the DBI-action.
}
	\label{pions}
  \hfill
\end{figure}

To calculate the mass $m_\pi$ in (\ref{mpion}),
we determine the form of the eigenfunction $\phi_0(u)$ analogously as in \cite{Sakai:2004cn}.
$\phi_0$ has to be orthogonal to all other $\phi_{n\geq 1}$ (the higher eigenfunctions that we left out in the expansion (\ref{Auexpansion})). The eigenfunctions $\phi_{n\geq 1}$ obey the same normalization condition (\ref{normpion}) as $\phi_0$, which upon comparison with the mass condition (\ref{Higgscontr}) for $\psi_{n \geq 1}$,
\begin{align}
&\int_{u_{0,d}}^\infty du \hspace{1mm} u^{1/4} g_{11} f_2 \phi_{n\geq 1}^2 = 1
\quad \text{and} \quad \int_{u_{0,d}}^\infty du \hspace{1mm} u^{1/4} g_{11} f_2 (\partial_u \psi_{n\geq1})^2  = \tilde m_\rho^2,
\end{align}
leads to
\begin{equation}
\phi_{n\geq 1} = \frac{\partial_u \psi_{n\geq 1}}{\sqrt{\tilde m_\rho^2}}.
\end{equation}
Then, orthogonality of $\phi_0$ and $\phi_{n\geq1} \sim \partial_u \psi_{n\geq1}$ is ensured
by proposing
\begin{equation} \label{phi0}
\phi_0 \sim u^{-1/4} g_{11}^{-1} f_2^{-1}
\end{equation}
(with normalization constant determined by the normalization condition (\ref{normpion})):
\begin{equation} \label{}
\int_{u_{0,d}}^\infty du \hspace{1mm} \phi_0 (u^{1/4} g_{11} f_2 \phi_{n\geq1})  \sim \int du  \hspace{1mm}\partial_u \psi_{n\geq1} = 0
\end{equation}
by virtue of the vanishing of $\psi_{n\geq1}$ at the boundary $u\rightarrow \infty$.
With $\phi_0$ given in (\ref{phi0}) we can determine the Higgs contribution to the mass $m_\pi$. In figure \ref{pions} we plot the eigenfunction $\phi_0(u)$ (which is discontinuous due to the discontinuous nature of $f_2$), the mass $m_\pi$ and the total effective 4-dimensional mass $\mathit{m_{\pi,eff}}$.

We end this section with a comment on the validity of the use of the non-Abelian DBI-action for non-coincident branes\footnote{We would like to thank  K.~Jensen for a private discussion about this.}.

In the context of heavy-light mesons, which we encounter here as magnetically induced through the splitting of the flavour branes, one often
studies the separated branes system by the use of
two (Abelian) DBI-actions plus a Nambu-Goto action for the classical, i.e.~macroscopic,  heavy-light meson string (e.g.~\cite{Erdmenger:2006bg}).
In \cite{Erdmenger:2007vj} however, one uses the non-Abelian DBI action for the description of heavy-light mesons, as we also did in this paper. They do remark
that as soon as the distance between the separated branes is larger than the fundamental string length $l_s$, the non-Abelian DBI-description is actually expected to break down.
So let us show here that in our case the separation between up- and down-brane and hence the length of the charged rho meson strings is not larger than $l_s$.

\begin{figure}[h!]
  \hfill
  \begin{minipage}[t]{.45\textwidth}
    \begin{center}
      \scalebox{0.8}{
  \includegraphics{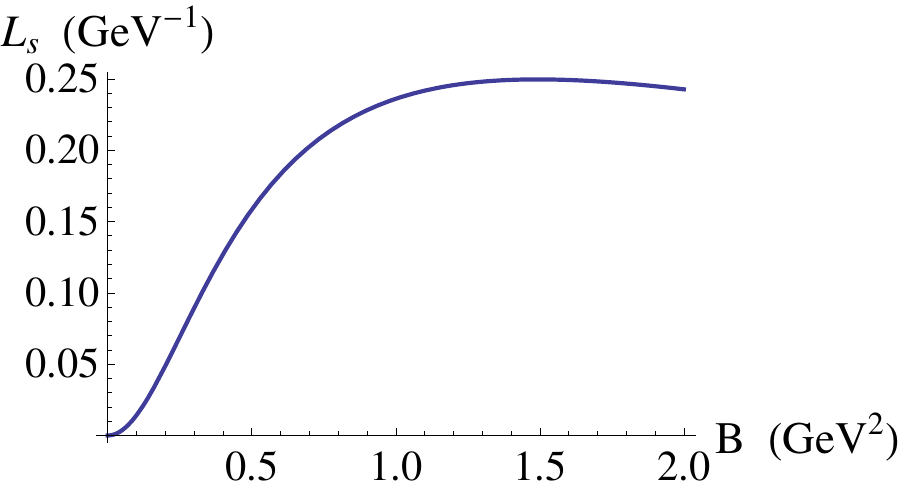}}
    \end{center}
  \end{minipage}
  \hfill
  \begin{minipage}[t]{.45\textwidth}
    \begin{center}
      \scalebox{0.8}{
  \includegraphics{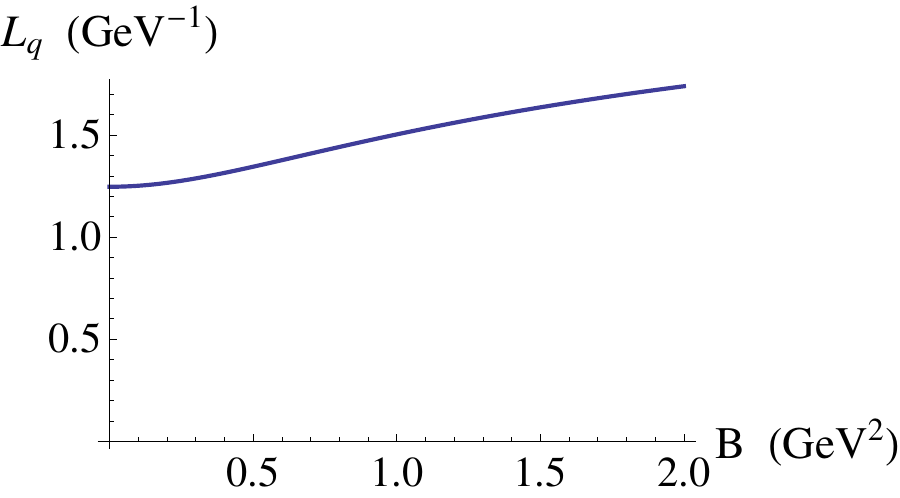}}
    \end{center}
  \end{minipage}
      \caption{The length $L_s(B)$ of a string at $\tau=0$ stretching between $u_{0,d}(B)$ and $u_{0,u}(B)$, and the length $L_q(B)$ of a down-quark string stretching between $u_K$ and $u_{0,d}(B)$.
}
	\label{Lsq}
  \hfill
\end{figure}

The total length of a string stretching in the $u$- and $\tau$-direction is given by
\[
L_s = \int ds = \int \sqrt{g_{\tau\tau} d\tau^2 + g_{uu}du^2}.
\]
Consider for example a string at $\tau=0$
 stretching from $u_{0,d}(B)$ to $u_{0,u}(B)$. It has a length
\begin{align*}
L_s &= \int ds = \int_{u_{0,d}(B)}^{u_{0,u}(B)} \sqrt{g_{uu}} du \nonumber\\
&= \int_{u_{0,d}(B)}^{u_{0,u}(B)} \left(\frac{R}{u}\right)^{3/4} f(u)^{-1/2} du  \nonumber\\
&= -\frac{R^{3/4}}{11 u_{0,d}^2 u_{0,u}^2 \sqrt{u_{0,d}^3-u_K^3} \sqrt{u_{0,u}^3-u_K^3}}4 (u_{0,d} u_{0,u})^{3/4} \nonumber\\
&\qquad \times
 \left\{ 11 u_{0,d}^3 u_{0,u}^{5/4} \sqrt{u_{0,u}^3-u_K^3}-6 u_{0,u}^{5/4} u_K^3 \sqrt{u_{0,u}^3-u_K^3}+u_{0,d}^{5/4} \sqrt{u_{0,d}^3-u_K^3} \left(-11 u_{0,u}^3+6 u_K^3\right) \right. \nonumber\\
&\qquad  \left.+6 u_K^3 \left(u_{0,u}^{5/4} \sqrt{u_{0,u}^3-u_K^3} \; {}_2 F_1\left[-\frac{11}{12},1,\frac{7}{12},\frac{u_{0,d}^3}{u_K^3}\right]-u_{0,d}^{5/4} \sqrt{u_{0,d}^3-u_K^3} \; {}_2 F_1\left[-\frac{11}{12},1,\frac{7}{12},\frac{u_{0,u}^3}{u_K^3}\right]\right)\right\},
\end{align*}
with the $B$-dependence of $u_{0,u}$ and $u_{0,d}$ implicit in the last line.
Similarly, the same string stretching between $u_0$ and $u_K$, corresponding to a constituent quark (i.e.~this one \emph{is} a macroscopic string, cfr. the use of the Nambu-Goto action to obtain the expression for the constituent quark mass (\ref{constmass})) has a length
\begin{align*}
L_q &= \int ds = \int_{u_K}^{u_0} \sqrt{g_{uu}} du \nonumber\\
&= \int_{u_K}^{u_0} \left(\frac{R}{u}\right)^{3/4} f(u)^{-1/2} du \nonumber\\
&= R^{3/4}\left(-\frac{4 \sqrt{\pi } u_K^{1/4} \Gamma \left[\frac{11}{12}\right]}{\Gamma \left[\frac{5}{12}\right]}+4 u_0^{1/4} \; {}_2 F_1 \left[-\frac{1}{12},\frac{1}{2},\frac{11}{12},\frac{u_K^3}{u_0^3}\right]\right).
\end{align*}
With our fixed holographic parameters, we have a numerical value for $l_s$ to compare these lengths to:
\begin{align*}
l_s=\sqrt{\alpha'}\approx 0.76  \hspace{1mm} \text{GeV$^{-1}$}.
\end{align*}
From the plots in figure \ref{Lsq} of $L_s$ and $L_q$ as functions of $B$ up to 2 GeV$^2$, we read of estimations of the maximal $L_s \approx 0.25$ GeV$^{-1}$ and minimal $L_q \approx 1.25$ GeV$^{-1}$, from which we can conclude that
\[
L_s < l_s \quad \text{and} \quad L_q > l_s,
\]
consistent with using the classical Nambu-Goto action for the constituent quark string, but using the non-Abelian DBI-description for the charged rho meson string.

\subsection{Vector sector for full DBI-action} \label{4.4}

\subsubsection{Comments on the validity of the \texorpdfstring{$(2\pi\alpha')$-expansion}{inverse string tension expansion}}  \label{ambiguities}

In the previous section \ref{F2approx} 
we approximated the DBI-action to second order in $(2\pi\alpha') F$. The justification that we used for this expansion is roughly that $\alpha' \sim 1/\lambda$ with $\lambda \approx 15$ `large' in our fixed units.
The reader might worry that there is some ambiguity in the proportionality factor $\alpha' \sim 1/\lambda$ since the parameter $X=\lambda l_s^2$ can be chosen freely, as we did in (\ref{SS2(2.4)a}). The ambiguity should however disappear from all physical quantities and indeed will no longer be present in the full expansion parameter.
Let us take a closer look.

Expanding $\det (g_{mn}^{D8}+(2\pi\alpha') iF_{mn}) = \det g_{mk}^{D8} \times \det(\delta_{nk}+(g_{kr}^{D8})^{-1}(2\pi\alpha') i (\overline F_{rn}+ \delta_1 F_{rn}+\delta_2 F_{rn}))$ in the action (\ref{nonabelian}), the expansion parameter $(g_{11}^{D8})^{-1}(2\pi\alpha') i \overline F_{12}$ is supposed to be small compared to 1, with (\ref{Fbardef}):
\[
\left(\frac{u}{R}\right)^{-3/2} (2\pi \alpha') \left| \left(\begin{array}{cc} \frac{2}{3}eB  & 0 \\ 0 & -\frac{1}{3}eB \end{array} \right) \right| \ll 1.
\]
The same expansion parameter can be read off from the form of the matrix A as defined in (\ref{A}).
The most strict condition would then be
\[
\left(\frac{u_{0,d}(B=0)}{R}\right)^{-3/2} (2\pi \alpha') \frac{2}{3}eB \ll 1,
\]
or, in our fixed units,
\begin{align}
eB &\ll \frac{3}{2} \left(\frac{u_{0,d}(B=0)}{R}\right)^{3/2} (2\pi \alpha')^{-1}  \equiv  0.45 \text{ GeV}^2,  \label{Bmaxestimate}
\end{align}
with the appearing combination independent of our choice of $X$ since $u_0 \sim X$, $R^3 \sim X$ and $(2\pi \alpha') \sim X$.
The instability we found in the $F^2$-approximation sets in at $B_c \approx 0.8$ GeV$^2$ (see (\ref{Bc})), where the used approximation is thus not necessarily valid anymore.
On the other hand,
the above is the most strict condition we can impose, it is not so clear what the impact of the $u$-dependence (which is integrated over) is on this argument.
We will therefore use the full STr-action
 and compare with the $F^2$-approximation results to provide a conclusive answer to the question of the validity of the $(2\pi\alpha')$-expansion in our set-up.
It will turn out that using the full STr-action the instability is still present and the value of $B_c$ is only slightly higher.

In \cite{Bolognesi:2012gr} it is argued
that $\alpha'$-corrections can cause magnetically induced tachyonic instabilities of $W$-boson strings, stretching between separated D3-branes, to disappear when the inter-brane distance becomes larger than $2\pi l_s$. The Landau level spectrum for the $W$-boson is said to receive large $\alpha'$-corrections in general \cite{Bolognesi:2012gr,Ferrara:1993sq}.
The paper \cite{Lee:2010ay} also gives an example where consideration of the full non-Abelian DBI-action in all orders of $\alpha'$ -- be it using an adapted STr-prescription -- can change the physics, that is, the order of the there discussed phase transitions changes.

\subsubsection{Deriving the effective 4-dimensional equations of motion} \label{4.4.2}

Reconsider the vector part of the DBI-Lagrangian in unitary gauge (\ref{STRgauged}),
\begin{align}
\mathcal L &= \mathcal L_{Higgs} + \mathcal L_{vector} =   \text{STr} \hspace{1mm} e^{-\phi} \sqrt{-\det a} |_{\tilde A^2} \nonumber\\
&=  \sum_{a=1}^2 \left\{ \gamma(u) \frac{1}{2}\left( [\tilde A_u,\overline \tau]^a \right)^2 + \alpha(u) \frac{1}{2} \left( [\tilde A_\mu,\overline \tau]^a \right)^2 + \beta(u) \sum_{\mu=1}^2 \frac{1}{2} \left( [\tilde A_\mu,\overline \tau]^a\right)^2  \right\} \nonumber\\
&+ \text{STr} \hspace{1mm} \overline x \left\{
- \overline F_{12} g_{11}^{-2} A^{-1} [\tilde A_1,\tilde A_2]
- \frac{1}{4} g_{11}^{-2}\tilde F_{\mu\nu}^2 \hspace{1mm} A^{-2}|_{\mu,\nu=1,2}  - \frac{1}{2} g_{11}^{-1} G_{uu}^{-1} \tilde F_{\mu u}^2 \hspace{1mm} A^{-1}|_{\mu=1,2}  \right\}
\end{align}
where the notation $|_{\mu=1,2}$ as introduced in (\ref{STR}) can be written out as
\begin{align}
\tilde F_{\mu\nu}^2 A^{-2}|_{\mu,\nu=1,2} &= 2 A^{-1} (\tilde F_{i3}^2+\tilde F_{i0}^2) + 2 \tilde F_{03}^2 + 2 A^{-2} \tilde F_{12}^2 \qquad (i=1,2) \nonumber\\
&= \tilde F_{\mu\nu}^2 + 2 \frac{1-A}{A}(\tilde F_{i3}^2 + \tilde F_{i0}^2) + 2 \frac{1-A^2}{A^2} \tilde F_{12}^2 \nonumber\\
 \text{and} \quad \tilde F_{\mu u}^2 A^{-1}|_{\mu,\nu=1,2} &= \tilde F_{\mu u}^2 + \tilde F_{iu}^2 \frac{1-A}{A} \qquad (i=1,2).
\end{align}
Instead of approximating this action further to $(2\pi\alpha')^2 F^2$, we now keep all factors of $A =  1 - (2\pi\alpha')^2 \overline F_{12}^2  \frac{R^3}{u^3}$. Upon evaluating the STr we then obtain
\begin{align}
&
\mathcal L
\sim u^{1/4} (2\pi\alpha')^2 \sum_{a,b=1}^2 \left\{-(\sqrt{G_{uu}} \overline F_{12} A^{-1/2})^3 \epsilon_{3ab} \tilde A_1^a \tilde A_2^b - \frac{1}{4} f_1 (\tilde F_{\mu\nu}^a)^2  - \frac{1}{2} \sum_{i=1}^2 f_{1A} ((\tilde F_{i 3}^a)^2 + (\tilde F_{0 i}^a)^2) \right. \nonumber\\
& \left. - \frac{1}{2} f_{1B} (\tilde F_{12}^a)^2  - \frac{1}{2}  g_{11} f_2 (\tilde F_{\mu u}^a)^2 - \frac{1}{2} g_{11} \sum_{i=1}^2 f_{2A} (\tilde F_{i u}^a)^2
- \frac{1}{2}  g_{11} \frac{1}{T^2} f_2 (\tilde A_\mu^a)^2 (2\overline \tau^3)^2
- \frac{1}{2}  g_{11} \frac{1}{T^2} f_{2A} \sum_{i=1}^2 (\tilde A_i^a)^2 (2\overline \tau^3)^2 \right\},  \label{lagra}
\end{align}
where we defined the new $I$-functions
\begin{align}
f_1 &=  I(G_{uu}^{1/2}A^{1/2}), \quad f_{1A} = I(\sqrt{G_{uu}} \frac{1-A}{\sqrt{A}}), \quad f_{1B} = I(\sqrt{G_{uu}} \sqrt{A}\frac{1-A^2}{A^2}) \label{IfunctionsfullDBIa}  \\
f_2 &=  I(G_{uu}^{-1/2} A^{1/2}), \quad f_{2A} = I(G_{uu}^{-1/2} \frac{1-A}{\sqrt{A}}),  \label{IfunctionsfullDBI}
\end{align}
with $f_1$ and $f_2$ approaching their previous definition in (\ref{f1f2f3f4f5}) and $f_{1A}$, $f_{1B}$ and $f_{2A} \rightarrow 0$ for $A \rightarrow 1$ in the $(2\pi\alpha')^2$-approximation, as they should.

Extracting the effective 4-dimensional action from (\ref{lagra}) is completely analogous to the procedure described in section \ref{F2approx}, so we will give a somewhat more schematic and short explanation here and refer to section \ref{F2approx} for more details.

After plugging in the gauge field expansions (\ref{Amuexpansion})-(\ref{Auexpansion}) into the action in the approximation of only retaining the lowest modes of the meson towers, one can already notice the vanishing of $\int du \mathcal L_{vector-mixing}$ $= 0$ and of mixing terms between pions and rho mesons.
We will focus on the instability in the rho meson sector.

\paragraph{Background dependent functions in the action}

\begin{figure}[h!]
  \hfill
  \begin{minipage}[t]{\textwidth}
    \begin{center}
      \scalebox{0.8}{
  \includegraphics{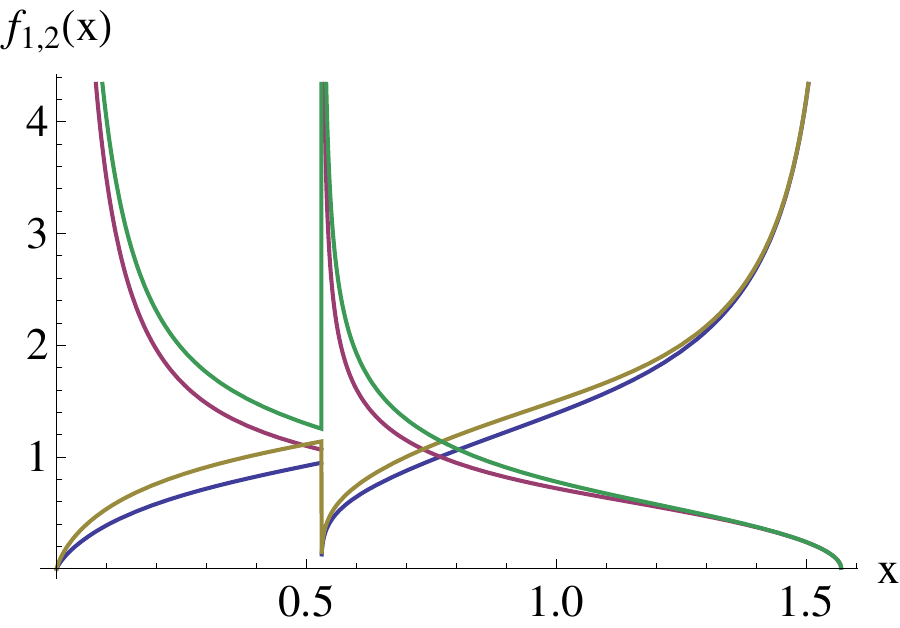}}
    \end{center}
  \end{minipage}
  \hfill
  \begin{minipage}[t]{\textwidth}
    \begin{center}
      \scalebox{0.8}{
  \includegraphics{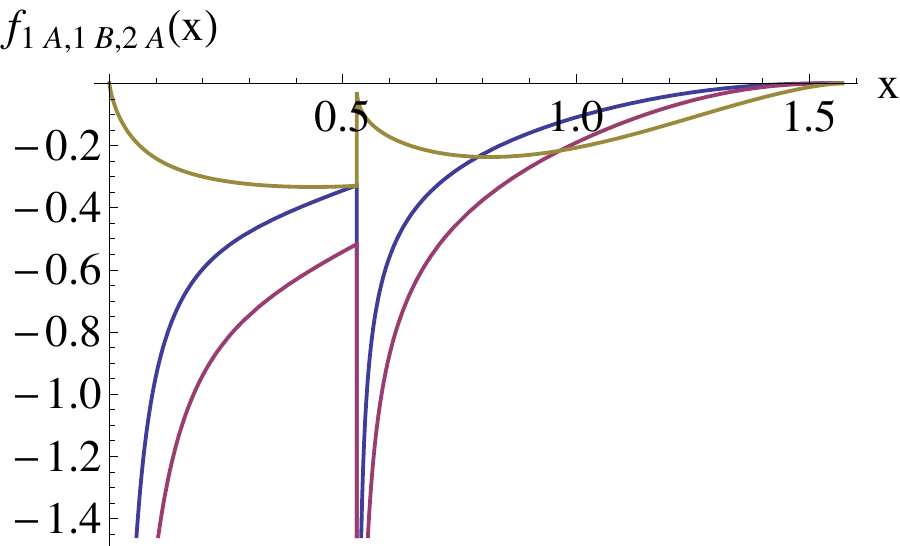}}
    \end{center}
  \end{minipage}
      \caption{(a) $f_1$ (green) and $f_2$ (yellow) compared to their $F^2$-approximated counterparts in red and blue resp. (b) $f_{1A}$ (blue), $f_{1B}$ (red) and $f_{2A}$ (yellow). For $B=0.8$ GeV$^2$ and $u=u_{0,d}\cos^{-3/2}x$.
}
	\label{fullDBIbackgroundfunctions}
  \hfill
\end{figure}

The generalized $I$-functions in (\ref{IfunctionsfullDBI}) have to be calculated numerically.
In figure \ref{fullDBIbackgroundfunctions} we compare them to their approximated counterparts for some fixed values of the magnetic field.
The measure for the distance between up- and down-brane $\overline \tau^3(u)$ is still as defined in (\ref{tau3math}), and finally
\begin{align}
(G_{uu}^{1/2} \overline F_{12} A^{-1/2})^3 = \sqrt{G_{uu}^u} \overline F_u A_u^{-1/2} - \sqrt{G_{uu}^d} \overline F_d A_d^{-1/2}
\end{align}
with $G_{uu}^l = G_{uu}(\partial_u \overline \tau^l)$ (with flavour index $l=u,d)$, $\overline F_u = \frac{2B}{3}$ and $\overline F_d =- \frac{B}{3}$ (see (\ref{Fbardef})), and $A_l$ defined in (\ref{A}).

\paragraph{Eigenvalue problem}

The rho meson part of the DBI-Lagrangian to second order in fluctuations  (\ref{lagra})  after substituting (\ref{Amuexpansion}) reads
\begin{align}
\hspace{-1.2cm}
\mathcal L
&\sim u^{1/4} (2\pi\alpha')^2 \sum_{a,b=1}^2 \left\{- \frac{1}{4} f_1 (\mathcal F_{\mu\nu}^{a})^2 \psi^2 - \frac{1}{2} g_{11} f_2(\rho_\mu^a)^2 (\partial_u \psi)^2  - \frac{1}{2} \frac{g_{11}}{(2\pi\alpha')^2} f_2 (\rho_\mu^a)^2 \psi^2 (2\overline \tau^3)^2 \right.  \nonumber \\
& - \frac{1}{2} f_{1B} (F_{12}^a)^2
 + \sum_{\mu,\nu=1}^2 \left( -\frac{1}{2} \frac{g_{11}}{ (2\pi\alpha')^2} f_{2A} (\rho_\mu^a)^2 \psi^2 (2\overline \tau^3)^2 - \frac{1}{2}(\sqrt{G_{uu}}\overline F_{\mu\nu} A^{-1/2})^3 \epsilon_{3ab} \rho_\mu^a \rho_\nu^b \psi^2
\right. \nonumber\\
& \left. \left. \qquad \qquad
- \frac{1}{2} f_{1A} ((\mathcal F_{\mu 3}^a)^2 + (\mathcal F_{\mu 0}^a)^2) \psi^2
- \frac{1}{2} g_{11} f_{2A} (\rho_\mu^a)^2 (\partial_u \psi)^2 \right)
 \right\},  \label{}
\end{align}
which results in
the following effective 4-dimensional action
\begin{align}
\hspace{-1.2cm}
S_{4D}
&= \int d^4 x \sum_{a,b=1}^2 \left\{- \frac{1}{4}(\mathcal F_{\mu\nu}^{a})^2 -\frac{1}{2} m_\rho^2(B) (\rho_\mu^a)^2 - \frac{1}{2} b(B) (\mathcal F_{12}^a)^2 \right. \nonumber\\
& \left.+ \sum_{\mu,\nu=1}^2 \left(-\frac{1}{2} a(B) ((\mathcal F_{\mu 3}^a)^2 + (\mathcal F_{\mu 0}^a)^2)  -\frac{1}{2} m_{+}^2(B) (\rho_\mu^a)^2 -\frac{1}{2} \epsilon_{3ab}\rho_\mu^a \rho_\nu^b \hspace{1mm} k(B) \overline F_{\mu\nu}^3 \right)\right\}. \label{S4Dfull}
\end{align}
The function $\psi$ (rescaled to absorb all constant prefactors in the action) satisfies the normalization condition
\begin{equation} \label{normc}
\int_{u_{0,d}}^\infty du \hspace{1mm} u^{1/4} f_1 \psi^2 = 1
\end{equation}
and
\begin{equation} \label{massc}
\int_{u_{0,d}}^\infty du \hspace{1mm} \left\{ u^{1/4} g_{11} f_2 \partial_u \psi^2 + u^{1/4} \frac{g_{11}}{(2\pi\alpha')^2} f_2   (2\overline \tau^3)^2\psi^2  \right\} = m_\rho^2,
\end{equation}
combining into the eigenvalue equation
\begin{equation} \label{eigveq}
u^{-1/4} f_1^{-1} \partial_u\left(u^{1/4} g_{11} f_2 \partial_u \psi \right) -  \frac{g_{11}}{(2\pi\alpha')^2} f_1^{-1} f_2 (2\overline \tau^3)^2 \psi  = -m_\rho^2 \psi
\end{equation}
to be solved for its $B$-dependent eigenvalue $m_\rho^2$ and eigenfunction $\psi$. The $B$-dependent numbers $m_{+}, k, a$ and $b$ can subsequently be calculated with the obtained eigenfunctions from
\begin{equation} \label{}
\int_{u_{0,d}}^\infty du \hspace{1mm} \left\{ u^{1/4} g_{11} f_{2A} \partial_u \psi^2 +  u^{1/4} \frac{g_{11}}{(2\pi\alpha')^2} f_{2A} (2\overline \tau^3)^2\psi^2  \right\} = m_{+}^2,
\end{equation}
\begin{equation}
\int_{u_{0,d}}^\infty du \hspace{1mm} u^{1/4} (\sqrt{G_{uu}}\overline F_{12} A^{-1/2})^3 \psi^2 = k  \hspace{1mm}  \overline F_{12}^3
\end{equation}
and
\begin{equation} \label{aandb}
\int_{u_{0,d}}^\infty du \hspace{1mm} u^{1/4} f_{1A} \psi^2 = a, \quad \int_{u_{0,d}}^\infty du \hspace{1mm} u^{1/4} f_{1B} \psi^2 = b.
\end{equation}

\begin{figure}[h!]
  \hfill
\hspace{-0.5cm}
  \begin{minipage}[t]{.45\textwidth}
    \begin{center}
      \scalebox{0.75}{
  \includegraphics{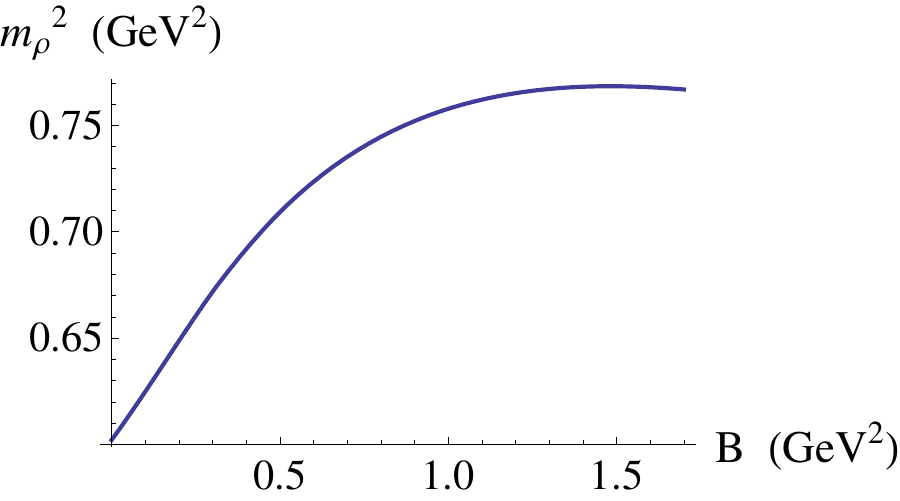}}
    \end{center}
  \end{minipage}
  \hfill
  \begin{minipage}[t]{.45\textwidth}
    \begin{center}
      \scalebox{0.75}{
  \includegraphics{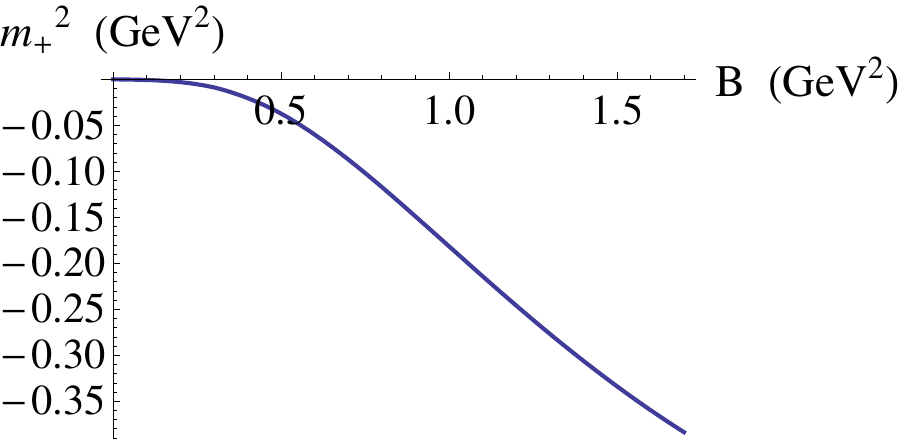}}
    \end{center}
  \end{minipage}
\hfill
  \begin{minipage}[t]{\textwidth}
    \begin{center}
      \scalebox{0.75}{
  \includegraphics{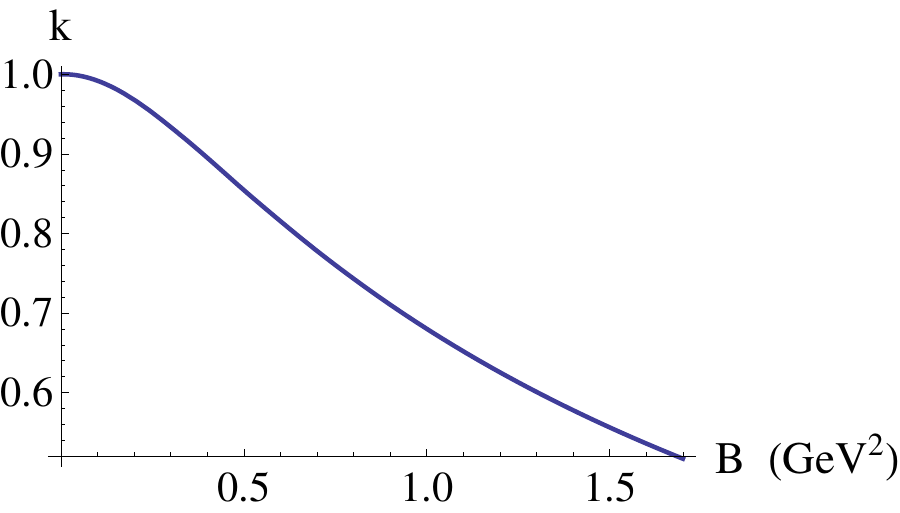}}
    \end{center}
  \end{minipage}
      \caption{Numerical results for $m_\rho^2(B)$, $m_+^2(B)$ and $k(B)$ from the full DBI-action. }
	\label{resultsfigDBI}
  \hfill
\end{figure}

\begin{figure}[h!]
  \hfill
  \begin{minipage}[t]{.5\textwidth}
    \begin{center}
      \scalebox{0.75}{
  \includegraphics{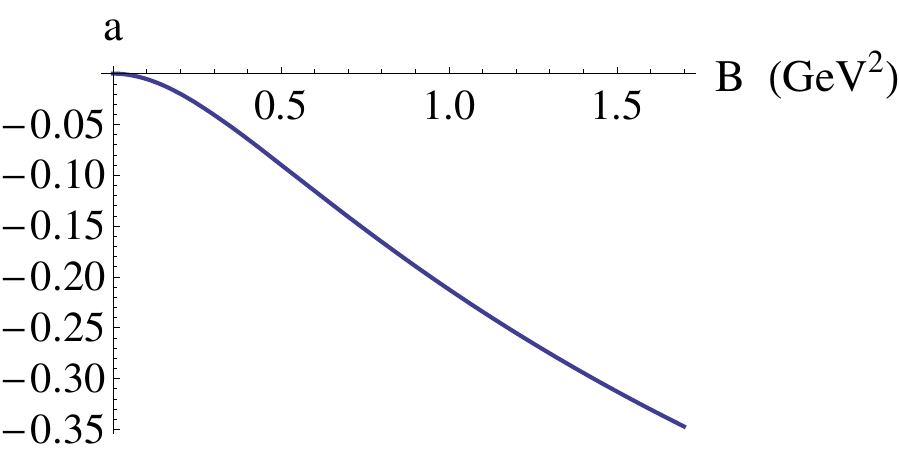}}
    \end{center}
  \end{minipage}
  \hfill
  \begin{minipage}[t]{.45\textwidth}
    \begin{center}
      \scalebox{0.75}{
  \includegraphics{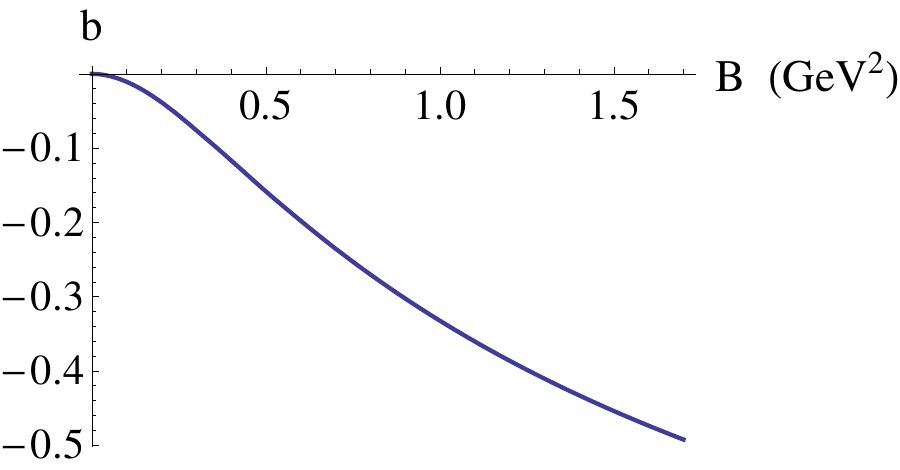}}
    \end{center}
  \end{minipage}
      \caption{Numerical results for $a(B)$ and $b(B)$.
}
	\label{resultsfigDBIpart2}
\end{figure}

The numerical results for $m_\rho^2$, $m_{+}^2$, $k$, $a$ and $b$ as functions of $B$, after having solved the eigenvalue problem with the techniques described in the second paragraph of \ref{paragraaf}, are shown in figure \ref{resultsfigDBI}-\ref{resultsfigDBIpart2}. The discussion of the behaviour of $m_\rho^2(B)$ in the third paragraph of \ref{paragraaf2} is still applicable. The parameter $k$ specifying
the strength of the coupling to the magnetic field is again approximately equal to one, but now decreasing as a function of $B$ as opposed to
increasing in the $(2\pi\alpha')^2$-approximation.

\subsubsection{Solving the 4-dimensional equations of motion} \label{4.4.3}

The 4-dimensional EOMs for $\rho_\nu^a$ derived from the effective action (\ref{S4Dfull}) are given by
\begin{align}
D_\mu \mathcal F_{\mu\nu}^a - \epsilon_{a3b} k \overline F_{\mu\nu}^3 \rho_\mu^b - m_\rho^2 \rho_\nu^a &- \delta_{\nu i} (m_{+}^2 \rho_i^a + a (D_3 \mathcal F_{i3}^a - D_0 \mathcal F_{i0}^a) + b D_j \mathcal F_{ij}^a) + \delta_{\nu 3} a D_i \mathcal F_{i3}^a - \delta_{\nu 0} a D_i \mathcal F_{i0}^a = 0  \label{EOM}
\end{align}
with $D_\mu = \partial_\mu +[\overline A_\mu,\cdot]$
and $\mathcal F_{\mu\nu}^a = D_\mu \rho_\nu^a - D_\nu \rho_\mu^a$,
and where from now on we will not only keep assuming the Einstein convention that double $\mu,\nu$ indices are Minkowski sums over $\mu,\nu=0..3$ but also that double $i,j$ indices are sums over spatial indices $i,j=1,2$.
For notational clarity we will not explicitly write out the $B$-dependence of the parameters $m_\rho$, $m_+$, $k$, $a$ and $b$ in this section, but assume it understood.

The equations $(a=1) \pm i (a=2)$ combine into the EOM for the charged rho meson $\rho_\mu = (\rho_\mu^1 + i \rho_\mu^2)/\sqrt 2$,
\begin{align}
& \text{D}_\mu \text F_{\mu\nu} - i k \overline F_{\mu\nu}^3 \rho_\mu - m_\rho^2 \rho_\nu - \delta_{\nu i} \left[ b \text D_j\text F_{ij} + a \left(\text D_3\text F_{i3}- \text D_0\text F_{i0} \right) + m_{+}^2 \rho_i \right] + \delta_{\nu 3} a \text{D}_j\text F_{j3} - \delta_{\nu 0} a \text{D}_j\text F_{j0} = 0,
\label{EOM4D}
\end{align}
with $\text D_\mu = \partial_\mu + i \overline A_\mu^3$ and $\text F_{\mu\nu} = \text D_\mu \rho_\nu - \text D_\nu \rho_\mu$ , and the complex conjugate of this equation for the other charged combination $\rho_\mu^* = (\rho_\mu^1 - i \rho_\mu^2)/\sqrt 2$.
Using $[\text D_\mu,\text D_\nu]=i \overline F_{\mu\nu}^3$, (\ref{EOM4D}) can be rewritten to the following EOMs for resp. $\nu=i$ and $\nu=3$:
\begin{itemize}
\item [\underline{$\nu=i$}]
\begin{align} \label{eom1}
& \hspace{-1cm}  (1+a) \text D_\mu^2 \rho_\nu
 - i (1+b+k) \overline F_{\mu\nu}^3 \rho_\mu - (1+a) \text D_\nu \text D_\mu \rho_\mu   - (m_\rho^2+m_{+}^2) \rho_\nu + (b-a) ( \text D_j^2 \rho_\nu - \text D_\nu \text D_j \rho_j)  = 0,\nonumber\\ & \end{align}
\item [\underline{$\nu=3$}]
\begin{align}
& \text D_\mu^2 \rho_\nu - \text D_\nu \text D_\mu \rho_\mu  - m_\rho^2 \rho_\nu + a  (\text D_j^2 \rho_\nu  - \text D_\nu \text D_j \rho_j )  = 0. \label{eom2}
\end{align}
\end{itemize}
These equations have to be complemented with a subsidiary condition, obtained
by acting with $\text D_\nu$  on the EOM (\ref{EOM4D}) and again using $[\text D_\mu,\text D_\nu]=i \overline F_{\mu\nu}^3$.
We find the generalized subsidiary condition (where by generalized we mean w.r.t. the Proca subsidiary condition $\text D_\nu \rho_\nu = 0$)
\begin{equation} \label{subsconditiongen}
\text D_\nu \rho_\nu = \frac{i}{m_\rho^2} (1+b-k)  \overline F_{\mu\nu}^3 \text D_\nu \rho_\mu  - \frac{m^2_{+}}{m_\rho^2}
\text D_i \rho_i,
\end{equation}
still relating $\text D_\nu \rho_\nu$ ($\nu=0..3)$ to transversal components $\rho_i$ $(i=1,2)$ only, such that the EOMs for the transverse rho mesons can be rewritten as independent from any longitudinal components.
Before doing so, let us remark that the above system of EOMs combined with the subsidiary condition reduces to its standard Proca form for $a$, $b$, $m_+ \rightarrow 0$, $k \rightarrow 1$ and no $B$-dependence in $m_\rho$ (or any of the previous parameters).
The non-zero and $B$-dependent $a$ and $b$ are present due to taking into account all powers in the field strength in the non-linear non-Abelian DBI-action, which is also partly the reason for
the $B$-dependence of $m_\rho$, $k$ and $m_+$, in addition to their implicit description of the response of the quark constituents to the magnetic field (cfr. the chiral magnetic catalysis and holographic Higgs mechanism for heavy-light mesons discussed earlier).

To determine the solutions of the EOMs we follow and generalize the procedure used in \cite{Mathews:1974ax}. In order to make comparisons with the original expressions in \cite{Mathews:1974ax} more clear, we temporarily change notation to
\begin{equation} \label{defphi}
\phi_\mu = \rho_\mu^* = (\rho_\mu^1 - i \rho_\mu^2)/\sqrt 2
\end{equation}
and
\begin{equation}
 i \pi_\mu = \text D_\mu^*  = \partial_\mu - i \overline A_\mu^3
\end{equation}
such that $\pi_\mu$ becomes $p_\mu - \overline A_\mu^3$ when substituting a plane wave ansatz $\phi_\mu \rightarrow e^{i \vec p \cdot \vec x - i E t} \phi_\mu$ into (\ref{eom1})-(\ref{eom2}), and in particular we can write $\pi_\nu^2 = -E^2 + \vec \pi^2$.
In this new notation the EOMs (\ref{eom1})-(\ref{eom2}) combined with (\ref{subsconditiongen}) can be recast in the form
\begin{align}
E^2 \phi_{\pm} =&  \left(\frac{m_\rho^2 + m_{+}^2}{1+a} + \mathcal B \vec \pi^2 \right) \phi_\pm + \frac{B}{2m^2}(1+b-k) \pi_\pm  (\pi_+ \phi_- - \pi_- \phi_+) \pm B \mathcal K  \phi_\pm - \frac{1}{2} \mathcal M \pi_\pm (\pi_+ \phi_- + \pi_- \phi_+)  \label{eom3}
\end{align}
with
\begin{equation}
\pi_\pm = \pi_1 \pm i \pi_2, \quad \phi_\pm = \phi_1 \pm i \phi_2,
\end{equation}
and
\begin{align}
E^2 \phi_3 =&  \left( m_\rho^2 + (1+a) \vec \pi^2 \right) \phi_3 + \frac{B}{2m^2}(1+b-k) \pi_3 (\pi_+ \phi_- - \pi_- \phi_+)  - \frac{1}{2} \left( a - \frac{m_{+}^2}{m_\rho^2} \right) \pi_3 (\pi_+ \phi_- + \pi_- \phi_+),  \label{eom4}
\end{align}
where we defined
\begin{align}
\mathcal B &= \frac{1+b}{1+a}, \quad \mathcal K = \frac{1+b+k}{1+a} \quad \text{and} \quad \mathcal M = \frac{b-a}{1+a}-\frac{m_{+}^2}{m_\rho^2}.
\end{align}
The main trick for solving the system is to notice that the operators $\pi_{\pm}$ obey the algebra of a simple harmonic oscillator, if one defines annihilation and creation operators $\hat a$ and $\hat a^\dagger$ as
\begin{align}
& \hat a = (2B)^{-1/2} \pi_+ \quad \text{and} \quad \hat a^\dagger = (2B)^{-1/2} \pi_-,
\end{align}
which obey
\begin{align}
& [\hat a,\hat a^\dagger]=1  \quad \text{and} \quad  [\hat a,\pi_3]=[\hat a^\dagger,\pi_3]=0.
\end{align}
The `number operator' $\hat N$ is then defined as
\begin{align}
 \hat N =\hat a^\dagger \hat a,
\end{align}
allowing us to rewrite the system (\ref{eom3})-(\ref{eom4}), using $\vec \pi^2 = p_3^2 + B(2 \hat N+1)$ and $\pi_+ \pi_- = 2B(1+\hat N)$,
to
\begin{align}
\hspace{-0.2cm}
\left\{ \begin{array}{ll}
(\omega^2 - \hat X_+) \phi_+ = A_\xi  \hat a^2 \phi_- \\
(\omega^2 - \hat X_-) \phi_- = -B_\xi (\hat a^\dagger)^2 \phi_+  \\
\left[ \omega_3^2 - (1+a)(2\hat N +1) \xi \right] \phi_3 = \xi^2 (1+b-k) a_3 (\hat  a \phi_- - \hat a^\dagger \phi_+) - \left( a - \frac{m_{+}^2}{m_\rho^2} \right) \xi a_3 (\hat  a \phi_- + \hat a^\dagger \phi_+),
\end{array} \right. \label{mathEOM}
\end{align}
with $\xi = \frac{B}{m_\rho^2}$ and
\begin{align}
\omega^2 &= \frac{E^2 - (m_\rho^2+m_+^2)/(1+a) - \mathcal B p_3^2}{m_\rho^2} \label{defomega}\\
\omega_3^2 &= \frac{E^2 - m_\rho^2 - (1+a) p_3^2}{m_\rho^2} \label{omega3} \\
\hat X_+ &= (2\hat N+1) \mathcal B \hspace{1mm} \xi - \frac{B_\xi}{2} + \mathcal K \hspace{1mm} \xi - (2 \hat N+1)\frac{B_\xi}{2} \\
\hat X_- &= (2\hat N+1) \mathcal B \hspace{1mm} \xi - \frac{A_\xi}{2} - \mathcal K \hspace{1mm} \xi + (2\hat N+1)\frac{A_\xi}{2} \\
A_\xi &= (1+b-k) \hspace{1mm} \xi^2 - \mathcal M \hspace{1mm} \xi \quad \text{and} \quad B_\xi = (1+b-k) \hspace{1mm} \xi^2 + \mathcal M \hspace{1mm} \xi,
\end{align}
and with $\pi_3$ replaced by its eigenvalue $p_3$ since it commutes with everything, or where convenient for the notation by the number $a_3 = (2B)^{-1/2} \pi_3$.
The system (\ref{mathEOM}) decouples completely in the special case where  $A_\xi = B_\xi = 0$ as well as $1+b-k = a - \frac{m_{+}^2}{m_\rho^2}=0$, which is for example the case for standard Proca parameters $a=b=m_+=0$ and $k=1$. In the latter situation the independent solutions for any $n$ are given by
\begin{align}
\phi_+ = |n-2\rangle, \quad \phi_-=\phi_3=0 \quad (n=2,3,\cdots) \nonumber\\
\phi_- = |n\rangle, \quad \phi_+=\phi_3=0 \quad (n=0,1,\cdots) \nonumber\\
\phi_3 = |n-1\rangle, \quad \phi_-=\phi_+=0 \quad (n=1,2,\cdots) \label{Proca}
\end{align}
with eigenvalue $\omega^2 = \omega_3^2 = (2n-1) \xi$.
Here we formally defined the `number eigenstates' $|n\rangle$ as
\begin{equation}
\hat N |n\rangle = n |n\rangle, \quad \hat a |0\rangle = 0, \quad |n\rangle = (n!)^{-1/2} (\hat a^\dagger)^n |0\rangle.
\end{equation}
In the rest of the discussion of possible solutions
below, we consider $A_\xi$ and $B_\xi$ different from zero.

\paragraph{Condensing solution}

Before decoupling the first two equations of (\ref{mathEOM}) to discuss the general form of the solution, let us first look at the one we are most interested in, the condensing solution:
\begin{equation} \label{solcondensing}
\phi_3 = \phi_+ = 0, \quad \phi_- = |0\rangle \quad (\Rightarrow \hat a \phi_- = 0),
\end{equation}
for which the EOM reduces to
\[
(\omega^2 - \hat X_-) |0\rangle = 0 \Rightarrow \omega^2 = \hat X_- (\hat N \rightarrow 0) = (\mathcal B \hspace{1mm} - \mathcal K) \xi = - \frac{k}{1+a} \xi
\]
with total eigenvalue
\begin{equation}
E^2 = \frac{m_\rho^2+m_{+}^2}{1+a} + \left(\frac{1+b}{1+a}\right) p_3^2 - \frac{k}{1+a} m_\rho^2 \xi,
\end{equation}
or, in the lowest state $p_3=0$ (and $\frac{1+b}{1+a}>0$ in the considered range of $B$):
\begin{equation}   \label{eigvcondensing}
\mathit{m_{\rho,eff}^2} = \frac{m_\rho^2+m_{+}^2}{1+a} - \frac{k}{1+a} m_\rho^2 \xi.
\end{equation}
This indeed reduces to its $(2\pi\alpha')^2$-approximated equivalent (\ref{mrhoeff2}),
$\mathit{m_{\rho,eff}^2} = m_\rho^2 + m_{+}^2  - k \xi m_\rho^2$,
for $a\rightarrow 0$.

\paragraph{Family of solutions}
We present the general discussion of the family of solutions of (\ref{mathEOM}).
One family of solutions is
\begin{equation} \label{sol1}
\phi_+ = \phi_- = 0, \quad \phi_3 = |n\rangle, \quad \omega_3^2 = (1+a)(2n+1)\xi, \quad n=0,1,2,\cdots,
\end{equation}
the other one
\begin{equation} \label{gensol}
\phi_- = |n+1\rangle, \quad \phi_+ = c_n |n-1\rangle, \quad \phi_3 = c_n' |n\rangle, \quad n=1,2,3,\cdots.
\end{equation}
The corresponding eigenvalue $\omega$ can be determined from decoupling the first two equations of  (\ref{mathEOM}) to
\begin{align}
\left\{ (\omega^2 - \hat X_-)(\omega^2 - \hat X_+) + (\hat N^2+3\hat N+2) A_\xi B_\xi - 2 (2 \mathcal B \hspace{1mm} \xi + A_\xi) (\omega^2 - \hat X_+) \right\} \phi_+ = 0   \label{16a}\\
\left\{ (\omega^2 - \hat X_-)(\omega^2 - \hat X_+) + (\hat N^2-\hat N) A_\xi B_\xi + 2 (2 \mathcal B \hspace{1mm} \xi - B_\xi) (\omega^2 - \hat X_-) \right\} \phi_- = 0. \label{16b}
\end{align}

Substitution of (\ref{gensol}) has the effect of replacing $\hat N$ in (\ref{16a}) by $(n-1)$ and in (\ref{16b}) by $(n+1)$. With these replacements, the curly-bracketed expressions in the two equations become identical, and either of them can be solved,
with the result for our \emph{generalized Landau levels} finally given by
\begin{align}
& \omega^2 = (2n+1) \xi (\mathcal B -\frac{\mathcal M}{2}) + \frac{(1+b-k)}{2} \xi^2 \nonumber\\
& \pm \xi \sqrt{\mathcal M \left(\frac{(2n+1)^2}{4} + \mathcal K - 2 \mathcal B \right) + (\mathcal K - 2 \mathcal B)^2 - (1+b-k) (2n+1) \xi (\mathcal K - 2 \mathcal B + \frac{\mathcal M}{2}) + \frac{(1+b-k)^2}{4} \xi^2}.  \label{omegasquared}
\end{align}
This reduces to Mathews' solution for general $k\neq1$, eq. (19) in \cite{Mathews:1974ax}, for $a, b,  m_{+} \rightarrow 0$, i.e.~$\mathcal B \rightarrow 1,  \mathcal M \rightarrow 0, \mathcal K \rightarrow 1+k$:
\begin{align}
\omega^2(a, b,  m_{+} \rightarrow 0) = (2n+1) \xi  + \frac{1}{2}(1-k) \xi^2 \pm (1-k) \xi \sqrt{1 + (2n+1) \xi + \frac{1}{4} \xi^2},
\end{align}
and the modified Landau levels mentioned in section \ref{rhomesonc} are given by (\ref{omegasquared}) with $a,b \rightarrow 0$. Given the value of $E^2$ from (\ref{omegasquared})
and the ansatz (\ref{gensol}) for $\phi_3$, the equation (\ref{omega3}) can be solved for $c_n'$. The constant $c_n$ can be determined from substituting the solution  (\ref{gensol}) and (\ref{omegasquared}) into either one of the first two equations of (\ref{mathEOM}).  \\
For completeness, we mention the last remaining possible solution
\begin{equation*}
\phi_- = |1\rangle, \quad \phi_+ = 0, \quad \phi_3 = c_0' |0\rangle
\end{equation*}
with $\omega^2 = \hat X_-(\hat N \rightarrow 1) = (3 \mathcal B - \mathcal K - \mathcal M) \xi + (1+b-k) \hspace{1mm} \xi^2$ and $c_0'$ to be determined from
$\omega_3^2 - (1+a) \xi c_0' =  (\xi^2 (1+b-k) - a \xi) a_3$.

In this whole discussion of the solutions of the EOMs for the rho meson, the key observation is that the energy eigenstates are so-called `number eigenstates', labeled by the Landau level number $n$. They are not necessarily spin eigenstates, as we will discuss next.

\paragraph{Discussion of the spin of the solutions}
Consider the eigenstates of the spin operator $S_3$ as defined in (\ref{S3spin}),
\begin{align*}
\phi_+ &= \phi_-= 0 \quad (s_3=0) \\
\phi_+ &= \phi_3=0 \quad (s_3=+1) \\
\phi_- &= \phi_3=0 \quad (s_3 = -1).
\end{align*}
It is clear that only the branch of solutions (\ref{sol1}) and the condensing solution (\ref{solcondensing}) are spin eigenstates, resp. with eigenvalues $s_3 = 0$ and $s_3 = +1$; the other branches of solutions for general $k\neq 1$  case
are not.  This is in contrast with the special $k=1$ Proca case (\ref{Proca}) where all Landau levels, including the excited states, are also spin eigenstates.

\begin{figure}[h!]
  \centering
  \scalebox{1.3}{
  \includegraphics{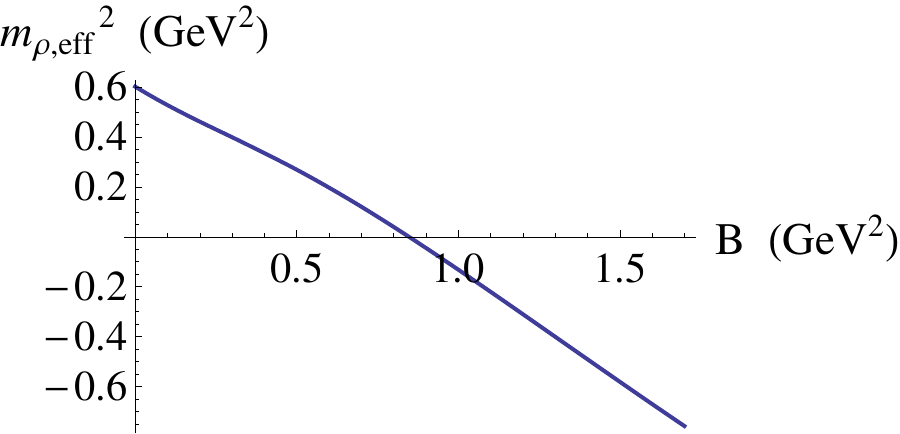}}
  \caption{The effective rho meson mass squared $\mathit{m_{\rho,eff}^2}(B)$ from the full DBI-action.
} \label{mrhoeff2figDBI}
\end{figure}

We conclude by summarizing that the condensing states are given by (\ref{solcondensing}) and its conjugate,
\[
\rho^*=\phi_- = \rho^*_1- i \rho^*_2
\quad \mbox{ and }\quad
\rho = \phi_-^* =  \rho_1 + i \rho_2
\]
--where we translated back to the previously used notation--
with energy eigenvalue $\mathit{m_{\rho,eff}^2}=$ (\ref{eigvcondensing}) and spin eigenvalue $s_3=+1$ corresponding to the spins being aligned with the magnetic field.
Our result for the effective rho meson mass squared $\mathit{m_{\rho,eff}^2}$, as  shown in figure \ref{mrhoeff2figDBI}, again demonstrates the tachyonic instability, with the critical magnetic field for rho meson condensation given this time by
\[
B_c \approx 0.85 \text{ GeV}^2.
\]
The increase compared to the estimate for $B_c$  in (\ref{Bc}) using the $(2\pi\alpha')^2$-approximation is pretty small.
This indicates that the expansion to second order in $(2\pi\alpha')F$ was a valid approximation, despite the ambiguities mentioned in section \ref{ambiguities}.

\subsection{Comment on the antipodal case} \label{commentantipodal}

For completeness, we consider the effect in the antipodal SSM, $u_0=u_K$, of including all higher order terms in the total field strength in the DBI-action. As mentioned before, the embedding of the flavour branes is independent of $B$ in this case, resulting in standard Landau levels and thus $\mathit{m_{\rho,eff}^2}(B) = m_\rho^2 - B$ if the action is approximated to second order in $(2\pi\alpha') F$.
In this set-up there is no constituent quark mass (\ref{constmass}) and no chiral magnetic catalysis.

To reproduce $m_\rho^2 = 0.602$ GeV$^2$ at zero magnetic field, along with $f_\pi = 0.093$ GeV for the pion decay constant, we have to use the holographic parameters fixed in \cite{Sakai:2005yt} to
\begin{equation} \label{antipodalvalues}
M_K \approx 0.949 \mbox{  GeV} \quad
\mbox{and}\quad \kappa  = \frac{\lambda N_c}{216 \pi^3} \approx 0.00745,
\end{equation}
instead of the values (\ref{values}) for $u_0>u_K$.
With these fixed parameters the estimate for the maximum value of the magnetic field for the  $(2\pi\alpha')$-expansion of the action to be valid, as discussed in Section \ref{ambiguities}, changes to
\begin{align}
eB \ll \frac{3}{2} \left(\frac{u_K}{R}\right)^{3/2} (2\pi \alpha')^{-1}  \equiv 0.31 \text{ GeV}^2,
\end{align}
which is even lower than the value 0.45 $\text{GeV}^2$ obtained for the non-antipodal case.

\begin{figure}[h!]
  \centering
  \scalebox{1.3}{
  \includegraphics{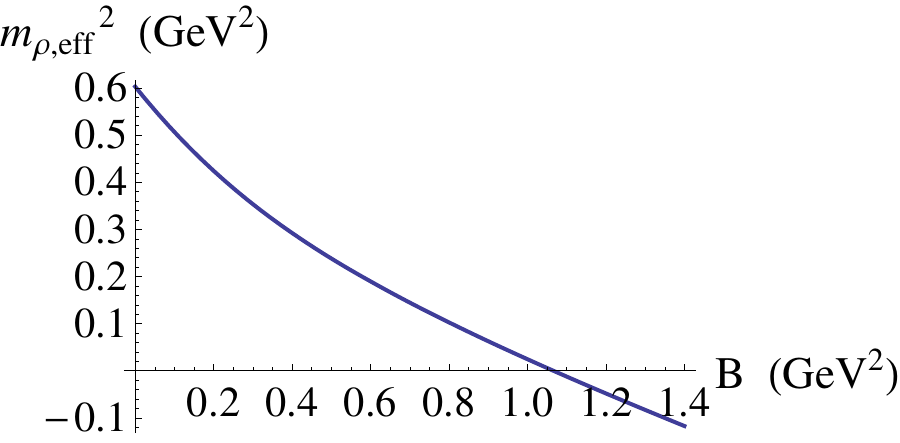}}
  \caption{
The effective rho meson mass squared $\mathit{m_{\rho,eff}^2}(B)$ from the full DBI-action for the antipodal embedded flavour branes.
} \label{antipodalfullDBI}
\end{figure}

As the flavour branes now remain coincident for any value of $B$, that is $\overline \tau \sim \mathbf{1} \Rightarrow \overline \tau^3 = 0$ and
$\partial_u \overline \tau = 0 \Rightarrow  G_{uu} = g_{uu}$,
we again obtain the effective 4-dimensional action (\ref{S4Dfull}), but with the integrals and equations (\ref{normc})-(\ref{aandb}) changed in the sense that
$u_{0,d} \rightarrow u_K$, $\overline \tau^3 \rightarrow 0$ and every $G_{uu} \rightarrow g_{uu}$, in particular in the $I$-functions $f_{1(A,B)}, f_{2(A)}$ defined in (\ref{IfunctionsfullDBIa})-(\ref{IfunctionsfullDBI}).
The eigenvalue equation can be recast in the form
\begin{equation*}
\frac{9}{4} \frac{u_K}{R^3} \cos^{4/3}x \left[ \partial_x^2 \psi + I(A^{1/2})^{-1} \partial_x  I(A^{1/2}) \partial_x \psi \right] = -m_\rho^2 \psi
\end{equation*}
with $u = u_K \cos^{-2/3}x$ this time and $I(A^{1/2})$ reducing to 1 for $B=0$.
With the numerical result for the eigenfunction $\psi$ and eigenvalue $m_\rho^2$, the total effective rho meson squared can be obtained using (\ref{eigvcondensing}),
\begin{equation*}
\mathit{m_{\rho,eff}^2} = \frac{m_\rho^2+m_{+}^2}{1+a} - \frac{k}{1+a} B.
\end{equation*}
The result is shown in figure \ref{antipodalfullDBI}, where the corresponding critical magnetic field can be read off to be $B_c \approx 1.07$ GeV$^2$.

\subsection{The Chern-Simons action and mixing with pions}\label{CSpion} 
In principle, the DBI action \eqref{nonabelian} needs to be complemented with a Chern-Simons piece (\ref{SCSNc})  which serves as the chiral anomaly in the QCD-like boundary theory:
\begin{eqnarray}
S_{CS} = \frac{N_c}{24 \pi^2}  \int \text{STr} \left(A F^2 - \frac{1}{2} A^3 F + \frac{1}{10} A^5 \right),  \label{CSSSM} 
\end{eqnarray}
where the notation implies wedge products of differential forms. Since it is a factor $\lambda$ smaller than the DBI-action and $\lambda \approx 15 \gg 1$, we have ignored the CS-action in our above analysis. 

When a magnetic field is applied, one might intuitively guess that a quark and anti-quark forming a pion bound state might try to align their spins, thereby transforming into a rho 
meson. In field theoretic terms, one might therefore expect a mixing between pions and rho 
mesons due to a magnetic field. Given the intrinsic parity odd nature of the mixing term (an odd number of pions violates the redundant ``intrinsic parity'' symmetry $U \rightarrow U^+$ discussed in section \ref{sectioneffQCD}), this would necessarily correspond to an anomaly driven process, i.e.~related to the Chern-Simons piece of the complete Sakai-Sugimoto action. It therefore looks interesting to investigate the CS-action to order meson gauge field squared in some more detail. We find that, after renormalizing by subtracting appropriate boundary terms \cite{Bergman:2008qv}, the action
\begin{align*}
S_{CS} &\sim  \int \text{STr} \left( \epsilon^{mnpqr} A_m F_{np} F_{qr} + \mathcal O(\tilde A^3) \right) \\
	     &\sim  \int 4 \, \text{STr} \left( \tilde A_3 \overline{F}_{21} \tilde F_{04} + \tilde A_0 \overline{F}_{21} \tilde F_{43} - \tilde A_3 \tilde F_{40}\overline{F}_{21} - \tilde A_0 \tilde F_{43} \overline{F}_{12} \right.\\
 & \qquad \qquad \qquad \left. + \overline{A}_2 [\tilde F_{43}\tilde F_{10} + \tilde F_{10}\tilde F_{43} + \tilde F_{41}\tilde F_{03} +\tilde F_{03}\tilde F_{41} +\tilde F_{31}\tilde F_{40} +\tilde F_{40}\tilde F_{31}] \right),
\end{align*}
does contribute interesting terms of the form $\rho \pi B$ (see also \cite{Sakai:2005yt}),
\begin{eqnarray}  \sim B \int \left\{
\partial_{[0} \pi^0 \rho_{3]}^0 + \frac{1}{2} \left(\partial_{[0} \pi^+ \rho_{3]}^-  + \partial_{[0} \pi^- \rho_{3]}^+\right) \right\},
\end{eqnarray}
indeed describing couplings between rho mesons and pions (and other axial mesons that we have not taken into account). However, since there are no direct couplings between the transverse components ($\mu = 1,2$) of the rho fields and pions, we can conclude that, at the level of a $\mathcal O(\tilde A^2)$ analysis, taking the CS-action into account will not affect the presence of the rho meson condensate which is related to the transverse field components, see eq.~\eqref{fieldcom}.

\section{Summary}\label{summary} 

\begin{figure}[h!]
  \hfill
  \begin{minipage}[t]{\textwidth}
    \begin{center}
      \scalebox{1.4}{
  \includegraphics{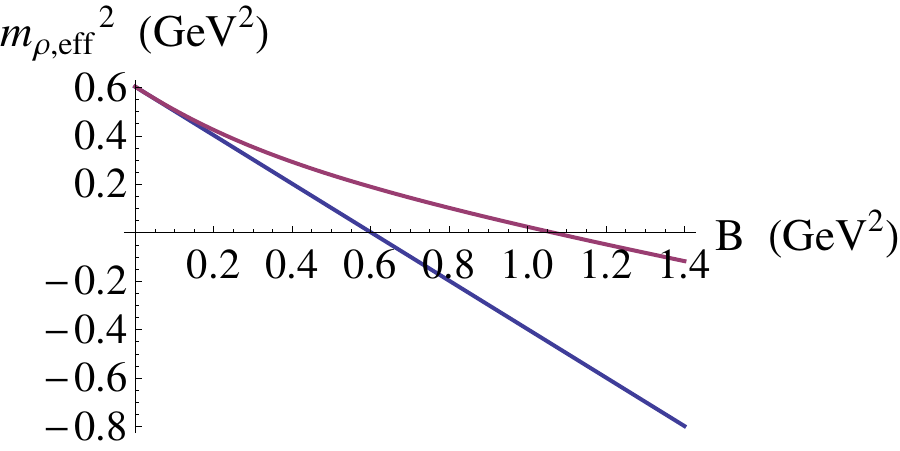}}
    \end{center}
  \end{minipage}
  \hfill
  \begin{minipage}[t]{\textwidth}
    \begin{center}
      \scalebox{1.4}{
  \includegraphics{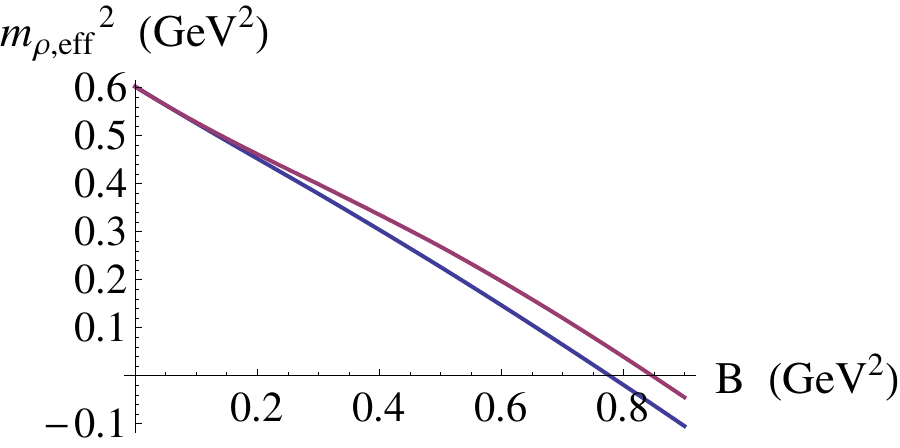}}
    \end{center}
  \end{minipage}
      \caption{The effective rho meson mass squared $\mathit{m_{\rho,eff}^2}(B)$ in the antipodal embedding (left) and the non-antipodal embedding (right), comparing the $(2\pi\alpha')^2 F^2$-approximated result in blue to the full DBI-result in red.
}\label{summaryfig}
\end{figure}

We studied a magnetically induced tachyonic instability in the charged rho meson sector, arising from the DBI-part of the two-flavour Sakai-Sugimoto model. We examined both the case of the antipodal and the more general non-antipodal embedding, each in the $(2\pi\alpha')^2 F^2$-approximation of the action versus the full DBI-action, non-linear in the total field strength $F$.
The results for the effective rho meson mass squared $\mathit{m_{\rho,eff}^2}(B)$,   vanishing at the critical magnetic field $B_c$ and thereby signaling the onset of the tachyonic instability, are shown in figure \ref{summaryfig} for each of the four set-ups.

The antipodal SSM reproduces exactly the standard 4-dimensional Proca picture and Landau levels of the effective QCD-model used in \cite{Chernodub:2010qx}, with $B_c = m_\rho^2 \approx 0.602$ GeV$^2$. The same picture was obtained in a holographic toy model involving an $SU(2)$ Einstein-Yang-Mills action for an $SU(2)$ bulk gauge field in a (4+1)-dimensional AdS-Schwarzschild black hole background \cite{Ammon:2011je},
and more recently for a 3-dimensional field theory in a (3+1)-dimensional DSGS-model generalized to AdS  \cite{Cai:2013pda}. 

The non-antipodal SSM predicts a larger value of $B_c \approx 0.78 \text{ GeV}^2$  as a result of taking two mass-generating effects for the charged rho meson into account, i.e.~chiral magnetic catalysis for the rho meson constituents on one hand, and a stringy Higgs-contribution to the mass from stretching the rho meson string between the magnetically separated up- and down-brane. Both effects are a direct result from the $B$-dependence of the non-antipodal flavour branes' embedding, and hence absent in the antipodal set-up. 

Considering the full DBI-action instead of approximating it to second order in the total field strength further increases the value of the magnetic field $B_c$ at the onset of rho meson condensation, more precisely to $B_c \approx 0.85$ GeV$^2$ in the non-antipodal case. The effect of taking the non-linear contributions in   $\overline F_{12}$
  into account seems to be stronger for the antipodal set of parameters compared to the non-antipodal one -- in both cases parameters are fixed to reproduce QCD parameters at zero magnetic field.
This leads us to conclude that the $F^2$-approximation is better justified for the considered problem in the non-antipodal embedding than in the antipodal one.
We are however very well aware of the fact that the full DBI-action is not the complete non-Abelian action for a system of $N_f$ branes -- a closed form of which is still to be found --, starting to show deviations at order $F^6$ \cite{Hashimoto:1997gm,Sevrin:2001ha}. We do not claim the DBI-result is necessarily  more correct than the $F^2$-result, yet we wanted to examine
the extent of the difference.
In conclusion, the SSM-predictions for $B_c$ are close to order 1 GeV$^2$, as  obtained in the NJL-model in \cite{Chernodub:2011mc} and on the lattice in \cite{Braguta:2011hq}.

A main motivation for these comparisons within the SSM was to investigate
what holography can add to the QCD-phenomenological picture of rho meson condensation, purposely
working in a top-down approach -- the downside of which are the technical complications. We for example elaborated on evaluating the STr exactly (to second order in fluctuations in the presence of an Abelian background field), the gauge fixing necessary to disentangle scalar and vector fluctuations, the contribution of the Chern-Simons action, the pion sector in the $F^2$-approximated DBI-part of the action, the Higgs mechanism associated with the magnetically induced heavy-light character of the charged rho mesons, numerically solving the eigenvalue equation for $m_\rho^2$ with a shooting method, and analytically solving the generalized effective 4-dimensional EOMs.
For the above reasons of complexity we have not yet been able to construct the new ground state in which the rho mesons are condensed.  This ground state is expected to be an Abrikosov lattice of rho meson vortices, as discussed in section \ref{sectionrholattice}.

We have been able to show that the SSM has a magnetically induced instability towards rho meson condensation,
consistent with the studies of this phenomenon
in phenomenological \cite{Chernodub:2010qx,Chernodub:2011mc}, lattice \cite{Braguta:2011hq} and bottom-up holographic \cite{Ammon:2011je,Cai:2013pda} approaches.
To come closer to the real-life quark-gluon plasma conditions where the presence of  magnetic fields of the order of $\sim 1$ GeV$^2$ might eventually be obtained,
it should be taken into account that there are also very high temperatures/densities present, and that the magnetic field is very localized both in space and time  (see the discussion in section \ref{QGPB} referring to  \cite{Skokov:2009qp,Deng:2012pc,Bzdak:2011yy,Tuchin:2013ie,Tuchin:2013apa,McLerran:2013hla}). 
These features may in the end seriously influence the possible occurrence of
rho meson condensation.

\chapter{Finite temperature holography}  \label{finiteThol}

We discuss in this chapter how to turn on a finite temperature in respectively a quantum field theory, the boundary field theory in the AdS/CFT correspondence and the Sakai-Sugimoto model.

\section{Quantum field theory at finite temperature} \label{QFTfiniteT}

A thermodynamic system in thermal contact with its environment at temperature $T$, and with 
volume $V$ and number of particles $N$ 
kept constant, is referred to as a canonical ensemble; its possible discrete quantum states as microstates. 
Denoting the energy of the system in microstate $s$ as $E_s$, the canonical partition function is given by  
\begin{equation} \label{Z}
Z = \sum_s e^{-\beta E_s}, \quad \beta = \frac{1}{k_B T},
\end{equation}
with $k_B$ the Boltzmann constant which we will set equal to 1 from here on. 

Consider a system at temperature $T$ consisting of a field $\phi(\vec x, t)$ that is in the state $|\phi_a\rangle$ at time $t$. 
The partition function (\ref{Z}) can be rewritten for this system as 
\begin{align}
Z &= \int d\phi_a \langle \phi_a | e^{-\beta \hat H} |  \phi_a \rangle = \text{Tr}\left(e^{-\beta \hat H}\right), \quad \beta = \frac{1}{T}. \label{partitiefunctie}
\end{align}
In the $T=0$ quantum field theory for $\phi(\vec x, t)$, the transition amplitude for a transition from an initial state $\phi_a(\vec x)$ at time $t$ to an end state $\phi_b(\vec x)$ at time $t'$ is given by the path integral 
\begin{equation}
\langle \phi_b | e^{-i \hat H (t'-t)} | \phi_a \rangle  =  \int_{\phi_a(\vec x) = \phi(\vec x, t)}^{\phi_b(\vec x) = \phi(\vec x, t')} \mathcal D \phi \hspace{1mm} e^{i S[\phi]}.
\end{equation}
After performing a Wick rotation ($t \rightarrow -it_E$, $t'-t \rightarrow -i\beta$, $iS \rightarrow -S_E$) and identifying initial and end state, we find for the Euclidian path integral along a closed path: 
\begin{equation}
\int d \phi_a \langle \phi_a | e^{-\beta \hat H } | \phi_a \rangle  =  \int_{\phi(\vec x, t_E+ \beta) = \phi(\vec x, t_E)} \mathcal D \phi \hspace{1mm} e^{- S_E[\phi]}.
\end{equation}
From comparison with (\ref{partitiefunctie}) we conclude that the Euclidian path integral along a closed path $\phi(\vec x, t_E+ \beta) = \phi(\vec x, t_E)$ is equal to the canonical partition function at temperature $T=\frac{1}{\beta}$:
\begin{equation}
Z = \int_{\phi(\vec x, t_E+ \beta) = \phi(\vec x, t_E)} \mathcal D \phi \hspace{1mm} e^{- S_E[\phi]}.
\end{equation}
In other words: turning on temperature in the field theory corresponds to compactification of Euclidean time.

\section{AdS/CFT at finite temperature} \label{AdSfiniteTsection}

\subsection[Schwarzschild geometry]{Schwarzschild geometry -- Hawking temperature of a $      $ black hole}

\begin{figure}[h!]
  \centering
  \scalebox{0.7}{
  \includegraphics{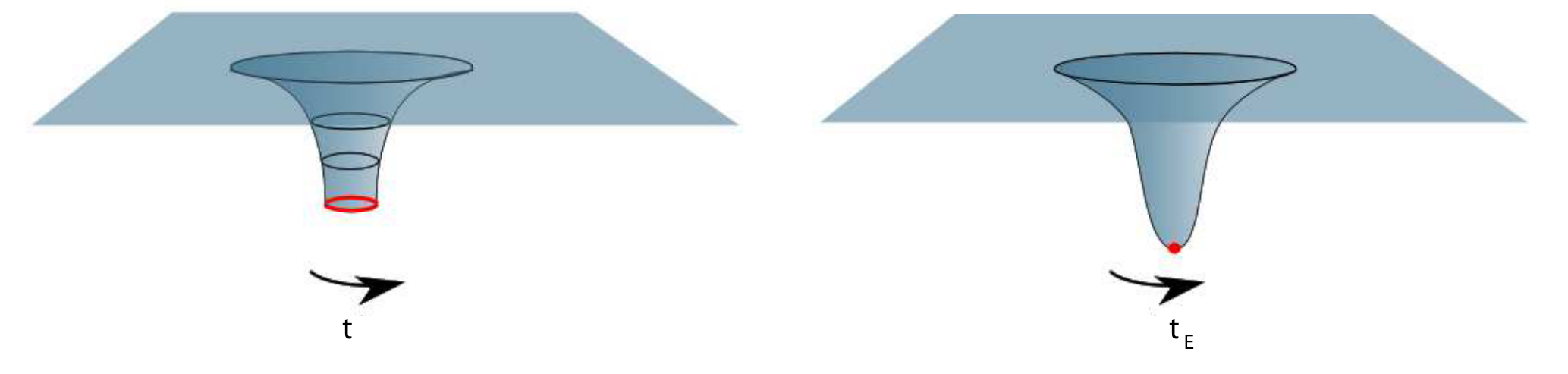}}
  \caption{Lorentzian versus Euclidean Schwarzschild geometry  
  \cite{Peeters:2007ab}.}\label{peetersfig13}
\end{figure}

To investigate how black holes are related to temperature, consider the Wick rotated  
Schwarzschild geometry
\begin{equation}
ds^2 = \left(1-\frac{2m}{r}\right) dt^2 + \left(1-\frac{2m}{r}\right)^{-1} dr^2 + r^2 d\Omega^2. 
\end{equation}
Near the horizon $r=2m$ the metric can be rewritten as a function of the new coordinate $\tilde r = r - 2m$
\begin{equation}
ds^2 = \frac{\tilde r}{2m} dt^2 + \frac{2m}{\tilde r} d\tilde r^2 + 4m^2 d\Omega^2,
\end{equation}
or 
\begin{equation}
ds^2 = \left(\frac{\rho}{4m}\right)^2 dt^2 + d\rho^2 + 4m^2 d\Omega^2
\end{equation}
with $\rho^2 = 8m\tilde r$.
In order to avoid a conical singularity at $\rho = 0$ ($r=2m$) it is required that $\frac{t}{4m}$ has a period of $2\pi$. 
In this way one naturally obtains a compactified Euclidean time $t \sim t + 2\pi \cdot 4m$ in the Euclidean Schwarzschild geometry (see figure \ref{peetersfig13}). 
A quantum field theory in the Euclidean Schwarzschild background is consequently at a temperature
\begin{equation} \label{T}
T = \frac{1}{8\pi m}.
\end{equation}
Based on its effect on the field theory, one can say that the black hole is radiating thermally as if it were a black body at a temperature $T$. This is known as `Hawking radiation' of the black hole, and (\ref{T}) is indeed precisely the temperature derived by Hawking in \cite{Hawking:1974sw}. However, the heat capacity of a black hole with mass $M$ and horizon $r=2m=2\frac{G M}{c^2}$ is negative:
\begin{equation}
C = \frac{\partial M}{\partial T} \propto \frac{\partial m}{\partial T} = -\frac{1}{8\pi T^2} < 0.
\end{equation}
This means that the Euclidean Schwarzschild geometry is thermodynamically unstable, and thus one cannot  turn on temperature in a field theory by simply adding a black hole to flat space. 
The situation is different in an AdS background.

\subsection[AdS$_5$-Schwarzschild geometry]{AdS$_5$-Schwarzschild geometry -- `Witten-prescription' for 
AdS/CFT at finite temperature} \label{AdS5-SS}

The metric of an $AdS_5$ space in global coordinates (\ref{adsglobal}) 
\begin{equation}
ds^2_{AdS_5} = R^2 \left(-\cosh^2 \rho \hspace{1mm} d\tau^2 + d\rho^2 + \sinh^2  \rho \hspace{1mm} d\Omega^2_3\right)
\end{equation}
can be rewritten via a coordinate transformation ($r=R \sinh \rho$ and $t=R \tau$) as 
\begin{equation}
ds^2_{AdS_5} = -\left(1+\frac{r^2}{R^2}\right) dt^2 + \left(1+\frac{r^2}{R^2}\right)^{-1} dr^2 + r^2 d\Omega^2_3.
\end{equation}
The Wick rotated metric  
\begin{equation} \label{metriek1}
ds^2_{AdS_5} = \left(1+\frac{r^2}{R^2}\right) dt^2 + \left(1+\frac{r^2}{R^2}\right)^{-1} dr^2 + r^2 d\Omega^2_3 
\end{equation}
with asymptotic geometry 
\begin{equation}
ds^2_{AdS_5} = \left(\frac{r}{R}dt\right)^2 + \left(\frac{R}{r} dr\right)^2 + r^2 d\Omega^2_3 \qquad (r\rightarrow \infty)
\end{equation}
is regular everywhere, for every radius of the time circle $S^1$ (the topology at infinity is 
 $S^1 \times S^3$), i.e.\ $t \sim t+\beta$ for arbitrary $\beta$.
There 
exists another solution of the supergravity equations of motion with the same asymptotic geometry $S^1 \times S^3$, whose  Euclidean continuation is 
\begin{equation} \label{dsadsSS}
ds^2_{AdS_5-Schwarzschild} = \left(1+\frac{r^2}{R^2} - \frac{\mu}{r^2}\right) dt^2 + \left(1+\frac{r^2}{R^2}- \frac{\mu}{r^2}\right)^{-1} dr^2 + r^2 d\Omega^2_3.
\end{equation}
This is the so-called $AdS_5$ black hole with a horizon 
at 
\begin{equation} \label{horizon?}
1 + \frac{r^2}{R^2} - \frac{\mu}{r^2} = 0,
\end{equation}
with $\mu$ proportional to the mass $M$ of the black hole.
The largest solution $r=r_+$ of (\ref{horizon?}),
\begin{equation}
r_+^2 = -\frac{R^2}{2} + \frac{R}{2} \sqrt{R^2 + 4 \mu},
\end{equation}
is called the outer horizon.
To avoid a conical singularity at $r=r_+$, time has to be periodic with period (see (\ref{omtrekalgemeen}))
\begin{equation}
\beta =\frac{4\pi}{\left.F'(r)\right|_{r=r_+}} = \frac{4\pi R^2 r_+}{4r_+^2 + 2R^2},
\end{equation}
where $F(r)$ is the prefactor of $dt^2$ in (\ref{dsadsSS}).
The $AdS_5$ black hole thus has a temperature  
\begin{equation}
T = \frac{1}{\beta} = \frac{4r_+^2 + 2R^2}{4\pi R^2 r_+}.
\end{equation}
The temperature as a function of the mass of the black hole has a minimum at $\frac{\partial T}{\partial M} = \frac{\partial T}{\partial r_+}  \frac{\partial r_+}{\partial M} = 0$. 
 Since
\begin{equation}
\frac{\partial M}{\partial r_+} \propto \frac{\partial \mu}{\partial r_+} = \frac{4r_+^3}{R^2} + 2 r_+ > 0 \quad (\text{voor }r_+ >0),
\end{equation}
 $T(M)$ reaches  its minimal value at  
\begin{equation}
\frac{\partial T}{\partial r_+} = \frac{1}{\pi R^2} - \frac{1}{2\pi r_+^2} = 0,
\end{equation}
where 
\begin{equation}
T_{min} = \frac{\sqrt 2}{\pi R}.
\end{equation}
The $AdS_5$ black hole is thermodynamically stable ($C = \partial M / \partial T > 0$) for $T>T_{min}$.

By adding a black hole to the $AdS_5$ background, the dual $\mathcal N =4$ SYM field theory living on the boundary of AdS space has acquired a finite temperature (as a consequence of the periodicity of Euclidean time). 
Adding the black hole is the so-called `Witten-prescription' for AdS/CFT at finite temperature, proposed by Witten in \cite{Witten:1998zw}.
 
The Euclidean partition function  
\begin{equation}
Z = \sum_{configurations} e^{-\beta F},
\end{equation}
with $F$ the free energy related to the Euclidean action through $S = \beta F$, contains a sum over all possible geometries 
(\ref{metriek1}) and (\ref{dsadsSS}).
The dominant geometry is the one with the smallest Euclidean action. An explicit calculation was performed by Witten in \cite{Witten:1998zw} 
with the result that
\begin{equation}
F_{AdS_5-Schwarzschild} - F_{AdS_5} = T \left( S_{AdS_5-Schwarzschild} - S_{AdS_5}\right) < 0
\end{equation}
for 
\begin{equation}
T > T_{critical} > T_{min},
\end{equation}
i.e.\ the $AdS_5$ background  dominates at low temperature, but the role of most dominant configuration is taken over by the thermodynamically stable  $AdS_5$ Schwarzschild-space at a critical value $T_{critical}$ of the temperature.

\section{Thermodynamics of the Sakai-Sugimoto model} \label{finiteTSSM}

QCD displays two strongly coupled effects at low energy: confinement and spontaneously broken (approximate) chiral symmetry. 
Both effects disappear at high temperature where the effective coupling becomes smaller.  
Because of the presence of massive dynamical quarks, no sharp order parameters can be defined in QCD for deconfinement 
and chiral symmetry restoration, 
i.e.\ there is no sharp phase transition between the low and high temperature phases 
but rather a crossover. 
Once a hot debate whether the deconfinement temperature $T_c$ and chiral symmetry restoration temperature $T_\chi$ coincided or not, see \cite{Aoki:2006we,Bazavov:2009zn} for 2 views on the $N_f=2+1$ case, it is by now accepted they are close in the cross-over region (see e.g.\ \cite{Heller:2006ub,Szabo:2014iqa}).

The Sakai-Sugimoto model 
offers a dual desciption (at low energy) of QCD in the large $N_c$ limit with massless bare quarks. 
We discuss why the absence of massive dynamical quarks in QCD corresponds to sharp phase transitions, that we hence expect to encounter in the Sakai-Sugimoto model as well. 
As the temperature rises in a model with exact massless quarks, we expect a sharp phase transition associated with chiral symmetry restoration, which was spontaneously broken at $T=0$. The corresponding order parameter is the chiral condensate. 
In pure QCD, 
dual to supergravity in the D4-brane background of the Sakai-Sugimoto model at strong coupling, we expect to see a sharp deconfinement phase transition. 
In the presence of dynamical quarks, the flux tube between a quark and antiquark will break when the energy stored in the string becomes larger than the amount needed to create a new $q \bar q$ pair. The confinement criterion of a linear confining quark potential hence can only be consistently defined for infinitely heavy, non-dynamical quarks, or more precisely,  in the absence of matter fields that can give rise to string breaking. These matter fields are the ones that in combination with a(n) (anti)quark can form a colour singlet, and are group theoretically identified with fields that transform non-trivially under the center $Z_{N_c}$ (the center $Z_{N_c}$ is the subgroup of $SU(N_c)$ consisting of elements $z_n = \exp \left( 2\pi i n/N_c \right) \mathbf 1_{N_c}$ (with $n=0,1,...,N_c-1$), which commute with the whole group).  
In pure QCD (with gluons in the adjoint representation automatically invariant under the center), it is thus possible to define a sharp order parameter associated with the deconfinement transition:
pure QCD at finite temperature has a global $Z_{N_c}$ symmetry 
which is unbroken in case of confinement and broken in the deconfined phase.

It is 
not a priori clear that chiral symmetry restoration and deconfinement will show up in the Sakai-Sugimoto model, which is not able to describe the asymptotic freedom of the dual QCD-like theory. 
The investigation of what happens when temperature is turned on in the D4/D8/$\overline{\text{D8}}$-configuration of the Sakai-Sugimoto model, was carried out in \cite{Aharony:2006da}. 
The effect of the $N_f$ D8-branes on the geometry is subleading  
in the limit of large $N_c$; the difference in free energies between competing bulk backgrounds in the partition function is of order $N_c^2$ while the contribution of the D8-branes is of order $N_c \times N_f$. 
We therefore discuss the thermodynamics of the D4-brane background (independent of the D8-branes), and 
only later add the D8-branes as probes to the dominating background at temperature $T$. 
(We will comment on corrections to this large-$N_c$ description at leading order, obtained in \cite{Ballon-Bayona:2013cta}, 
at the end of section \ref{resultstempartikel}.)

\begin{figure}[h!]
  \centering
  \scalebox{0.35}{
  \includegraphics{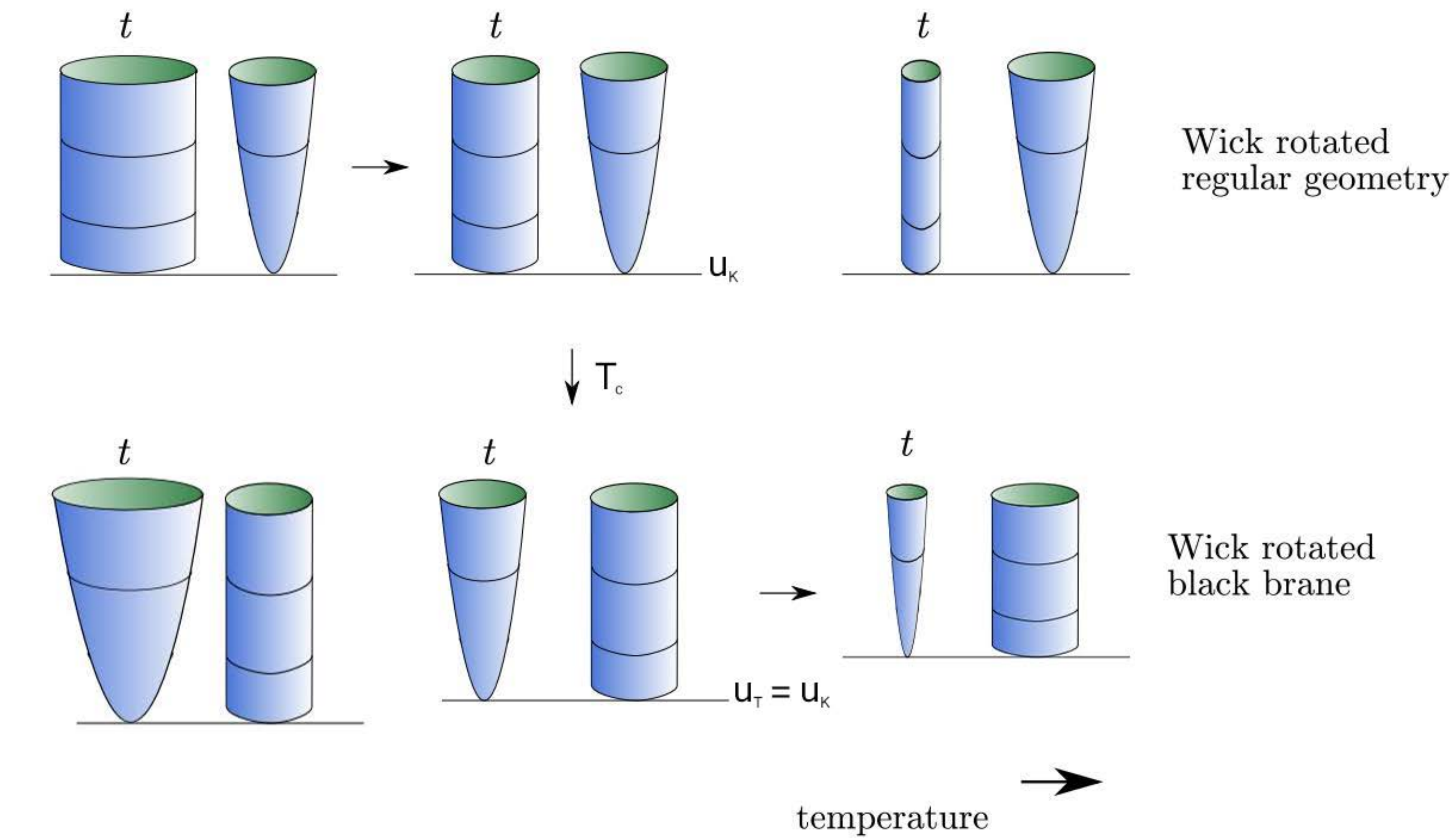}}
  \caption{The Wick-rotated D4-brane background (\ref{SSlowT}) with $\tau$-cigar (and $t$-cylinder) and the Wick-rotated black brane solution (\ref{SShighT}) with the same asymptotic geometry, but $\tau$-cylinder (and $t$-cigar), are the two regular Euclidean backgrounds competing with each other in the partition function  
    \cite{Peeters:2007ab}.}\label{peetersfig14}
\end{figure}

After a Wick rotation of the D4-brane background (\ref{D4SS}), we find the Euclidean signature 
metric 
\begin{equation} \label{SSlowT}
ds^2_{\tau\text{-cigar}} = \left(\frac{u}{R}\right)^{3/2} (dt^2 + \delta_{ij}dx^i dx^j + f(u)d\tau^2) + \left(\frac{R}{u}\right)^{3/2} 
\left( \frac{du^2}{f(u)} + u^2 d\Omega_4^2 \right), \quad f(u) = 1 - \left(\frac{u_K}{u}\right)^3 
\end{equation} 
which is regular 
for arbitrary periodicity $\beta = 1/T$ of Euclidean time $t \sim t+ \beta$, but shows a conical singularity at $u=u_K$, unless the $\tau$-circle of the cigar-shaped subspace $(\tau, u)$ is periodic  
$\tau \sim \tau + \delta \tau$ 
with period 
\begin{equation} \label{deltatauvooractie}
 \delta \tau 
= \frac{4\pi}{3} \sqrt{\frac{R^{3}}{u_K}}.
\end{equation}  

There is a second Euclidean background with the same asymptotic geometry as (\ref{SSlowT}), namely (\ref{SSlowT}) but with the roles of $t$ and $\tau$ interchanged:
\begin{equation} \label{SShighT}
ds^2_{t\text{-cigar}} = \left(\frac{u}{R}\right)^{3/2} (\hat f(u)dt^2 + \delta_{ij}dx^i dx^j + d\tau^2) + \left(\frac{R}{u}\right)^{3/2} 
\left( \frac{du^2}{\hat f(u)} + u^2 d\Omega_4^2 \right),  \quad \hat f(u) = 1 - \left(\frac{u_T}{u}\right)^3. 
\end{equation}
In this geometry it is the $t$-circle, with mandatory periodicity 
\begin{equation} \label{deltatvooractie}
\delta t = \beta = \frac{1}{T}  = \frac{4\pi}{3} \sqrt{\frac{R^{3}}{u_T}},
\end{equation}
that shrinks until it disappears at the temperature dependent minimal value of $u$,  given by $u_T = 16\pi^2 R^3 T^2/9$.
This is the black brane geometry with horizon at $u_T$. 
The $\tau$-circle forms a cylindric $(\tau, u)$-subspace of the geometry, with fixed radius $\delta \tau = 2\pi/M_K$. 

These two backgrounds, drawn in figure \ref{peetersfig14}, are the only known regular Euclidean solutions with the same asymptotic geometry. 
In the Euclidean partition function there is a competition between both, and the one with the smallest action dominates at a given temperature $T$. 
Both geometries are identical, modulo a redefinition of coordinates $t$ and $\tau$, when $\delta \tau$ equals $\beta$,  
which happens at the deconfinement temperature 
\begin{equation} 
T_c = (\delta \tau)^{-1} = \frac{M_K}{2\pi}. \label{TcSSM}
\end{equation} 
At this temperature a first order phase transition occurs (first order because the solutions do not come together continuously, but keep existing as separate solutions at temperatures lower and higher than the transition temperature: the free energy as a function of temperature is discontinuous at $T_c$). This is consistent with the large-$N_c$ nature of the transition in QCD (see the discussion of figure \ref{QCDphasediagramfig} in section \ref{statmech}).   
The background where the $\tau$-circle shrinks dominates at low temperatures $T < T_c$, 
while the background with shrinking $t$-circle dominates at high temperatures $T > T_c$. This is shown in \cite{Aharony:2006da} 
by determining the difference in free energy densities. The result can be remembered intuitively: the smallest circle shrinks.

The dominating background at low temperature, (\ref{SSlowT}), satisfies the confinement criterion (\ref{Teff}):
\begin{equation}
\left.\sqrt{G_{tt}(u_K) G_{xx}(u_K)}\right|_{T <T_c} = \left(\frac{u_K}{R}\right)^{3/2} \neq 0,
\end{equation}
whereas the dominating background at high temperatures, (\ref{SShighT}), does not:
\begin{equation}
\left.\sqrt{G_{tt}(u_T) G_{xx}(u_T)}\right|_{T > T_c} = \left(\frac{u_T}{R}\right)^{3/2} \sqrt{\hat f(u_T)} = 0.
\end{equation}
We therefore interpret the phase transition at temperature $T_c$ in (\ref{TcSSM}) as the deconfinement transition in the dual, strongly coupled ($g^2_{YM}N_c \gg 1$) QCD theory.

We should remark here that in \cite{Mandal:2011uq,Mandal:2011ws}, 
some problems 
are raised concerning the above identification of the deconfinement transition. 
The loop order parameter $W_0$ associated with 
the center symmetry is mapped to the action of a Nambu-Goto string wrapping the Euclidean time direction, which is finite in case of a $t$-cigar, and zero for a $t$-cylinder. 
While the transition from the D4-brane background to the black D4-brane background thus captures the correct behaviour of the confinement order parameter ($W_0=0$ in the confining phase and $W_0 \neq 0$ in the deconfining phase), it does not for 
a second $Z_{N_c}$ symmetry associated with the $\tau$-cigar or cylinder: the corresponding order parameter $W_4$ becomes zero at the proposed transition in supergravity, but remains finite in both phases in the 4-dimensional Yang-Mills theory, viewed as the Kaluza-Klein reduction of a 5-dimensional SYM theory. 
The authors of \cite{Mandal:2011ws} propose an alternative dual background for the deconfining phase (where $W_0 \neq 0$ and $W_4 \neq 0$), namely a localized D3-soliton geometry. This is the gravity solution corresponding to periodic boundary conditions on the adjoint fermions along the thermal circle, 
where they are antiperiodic for the black D4-brane background. 
This choice does not affect the pure 4-dimensional gauge theory. It does however affect the fundamental fermions, so there are some subtleties in this approach when adding 
flavour branes to the background.  
Moreover, it is necessary to consider a high-temperature approximation of the D3-soliton background to make calculations of transition temperatures feasible. 
Because of these technical complications, we will continue to treat the transition between the D4-brane backgrounds  in figure \ref{peetersfig14} as the dual of the deconfinement transition in the 4-dimensional pure QCD theory.

\begin{figure}[h]
\centering
 \subfigure[]{
\includegraphics[scale=0.7]{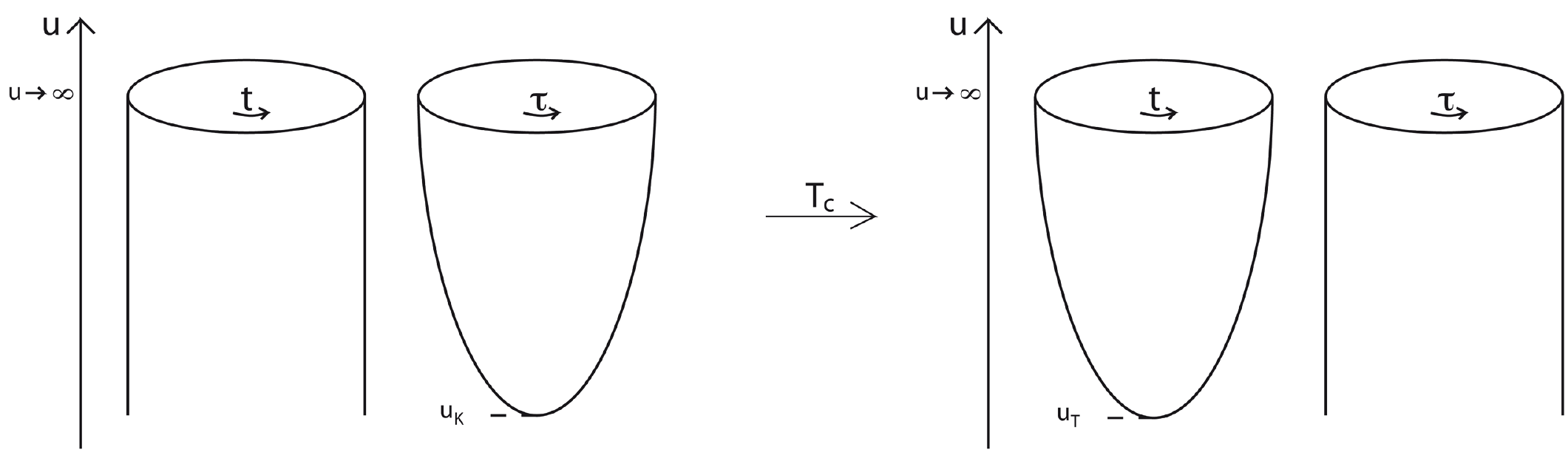}
}
\subfigure[]{
\includegraphics[scale=0.7]{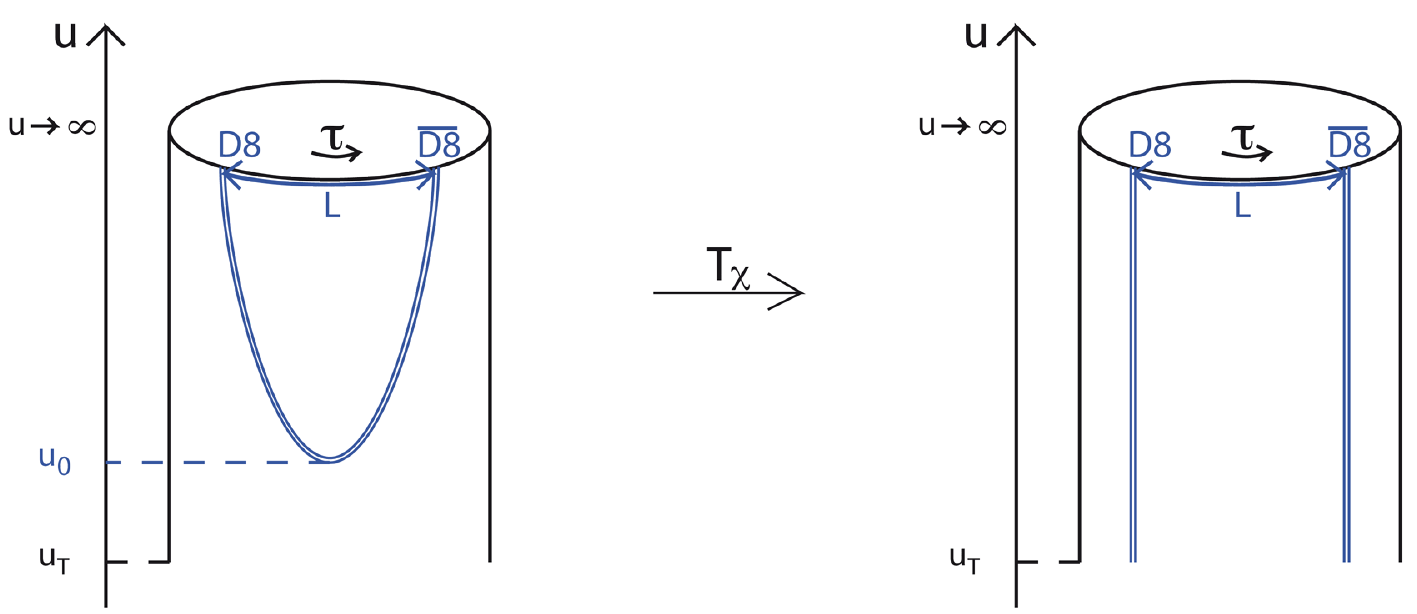}
}
\caption[]{(a) Deconfinement transition at $T_c$ and (b) chiral symmetry restoration at $T_\chi (\geq T_c)$ in the Sakai-Sugimoto model.} \label{figure2}
\end{figure}

\begin{figure}[h!]
  \centering
  \scalebox{0.7}{
  \includegraphics{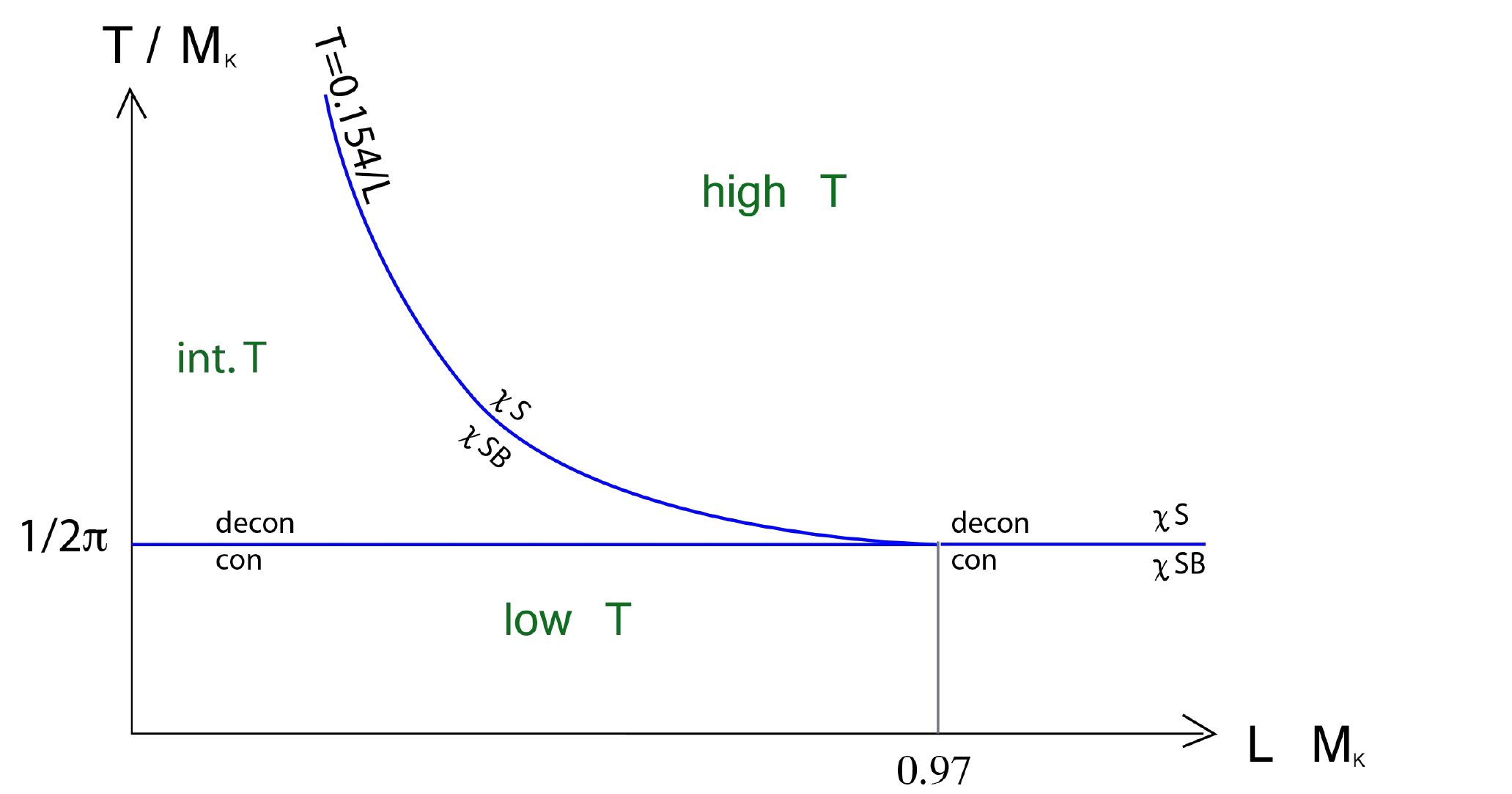}} 
  \caption{The $(T,L)$ phase diagram of the Sakai-Sugimoto model at finite temperature, in terms of the dimensionless parameters $T/M_K$ and $L M_K$ 
  \cite{Aharony:2006da, Peeters:2006iu}.}\label{finiteTfig7}
\end{figure}

We now reintroduce the probe D8-branes in the high-temperature background (\ref{SShighT}). 
In the deconfining phase the embedding of the flavour branes is no longer forced to be $\cup$-shaped, as the $(u,\tau)$-space is no longer cigar-shaped. At a certain value of the temperature, $T_\chi \geq T_c$, it will become energetically favourable for the flavour branes to fall straight down instead of merging in a $\cup$-shape, indicating chiral symmetry restoration (see figure \ref{figure2}). 
The (first order) transition point was shown in \cite{Aharony:2006da} 
to be located at 
\begin{equation} 
T_\chi \approx \frac{0.154}{L}, \label{TchiSSM}
\end{equation} 
by numerical evaluation of the DBI-actions for the possible embeddings (with $L$ the asymptotic separation between flavour branes and anti-branes). 
From the expressions (\ref{TcSSM}) and (\ref{TchiSSM}), it follows that the value of $L$ in units of $1/M_K$ determines whether or not the Sakai-Sugimoto model displays a separation between deconfinement and chiral restoration scales: for $L > 0.97/M_K$, deconfinement and chiral restoration will occur simultaneously, $T_c > T_\chi$, while for $L < 0.97/M_K$, an intermediate phase exists, $T_c < T < T_\chi$, where chiral symmetry remains broken in the deconfined phase. The $(T,L)$ phase diagram is displayed in figure \ref{finiteTfig7}. 

In the localized D3-soliton background proposed in \cite{Mandal:2011ws}, the obtained end result for the $(T,L)$ phase diagram, see Fig.\ 10 of \cite{Mandal:2011ws}, actually gives a qualitatively similar result as the above 
finite temperature Sakai-Sugimoto model.  
Based on this observation, we can expect that qualitative features of the $eB$-dependence of the chiral transition temperatures, which will be discussed in the next chapter, are not unlikely to be similar in both backgrounds.

\chapter{Magnetic splitting of chiral transition temperatures in the Sakai-Sugimoto model}
\chaptermark{Chiral transition temperatures}  \label{SStempchapter}

During heavy ion collisions, high temperatures and strong magnetic fields are generated. We use the finite temperature non-antipodal Sakai-Sugimoto model to study the $N_f=2$ QCD phase diagram under these extreme conditions in the quenched approximation and the chiral limit. 
We take the different coupling of up and down flavours to the magnetic field into account geometrically, resulting in a split of the chiral phase transition according to flavour. We discuss the influence of the magnetic field on the chiral temperatures -- in physical GeV units -- in terms of the choice of the confinement scale in the model, extending hereby our discussion of fixing the non-antipodal SSM parameters to the deconfinement phase. 
The flavour-dependent $(T,L,eB)$ phase diagram, with variable asymptotic brane-antibrane separation $L$, is  presented, as a direct generalization of the known $(T,L)$ phase diagram of the non-antipodal SSM at zero magnetic field. In particular, for sufficiently small $L$ we are probing an NJL-like boundary field theory in which case we do find results very reminiscent of the predictions in NJL models.

\section{Motivation} \label{motivationtemp}

\begin{figure}[h!]
  \centering
  \scalebox{0.55}{
  \includegraphics{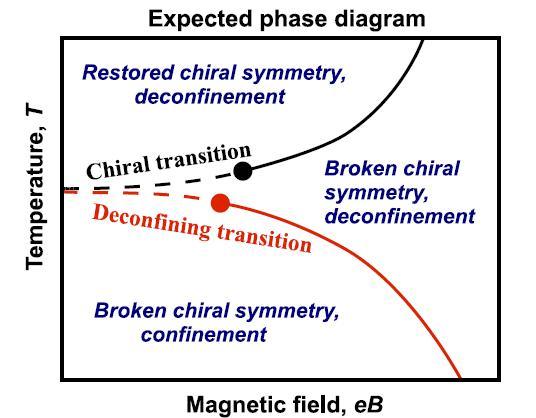}} 
  \caption{Expected $(T,B)$ phase diagram of QCD in 2008. The cross-overs (dashed lines) are expected to separate and  become stronger, transforming into first-order phase transitions (full lines) \cite{Fraga:2008um}. 
Figure from \cite{Mizher:2010zb}. } 
\label{TBexpected}
\end{figure}

In 2010 (approximately), a 
discussion evolved around the possibility that the deconfinement temperature $T_c$ and chiral transition temperature $T_\chi$ might separate under the influence of a constant magnetic background field, $\mathbf{B}=B\mathbf{e}_z$, and this for $N_f=2$ QCD, see e.g.~\cite{Mizher:2010zb,Gatto:2010pt,D'Elia:2010nq}. In figure \ref{TBexpected} the expected $(T,B)$ phase diagram of QCD is shown \cite{Mizher:2010zb} 
as it was conjectured to look like at the time, 
based on two ``observations''. The first one was the rising of $T_\chi$ with $B$ due to the chiral magnetic catalysis effect \cite{Miransky:2002rp} (which we discussed 
in the section on the non-antipodal embedding on page \pageref{testchiralmagncat}, more precisely in the paragraph on figure \ref{u0}). The second  was the decrease of $T_c$ with $B$ as argued in \cite{Agasian:2008tb} 
from a thermodynamic point of view. 
The reasoning  in \cite{Agasian:2008tb} is that the paramagnetic plasma of quarks and gluons at $T>T_c$ lowers its free energy in the presence of a magnetic field, and hence becomes thermodynamically favoured 
compared to the hadronic diamagnetic phase of mainly scalar pions at $T<T_c$. In \cite{Fraga:2012fs} 
this behaviour of $T_c(B)$ is recovered using the MIT bag model, 
but with the difference compared to \cite{Agasian:2008tb} 
 that $T_c$ saturates at large $B$ at a non-zero value. The chiral magnetic catalysis, in the sense of a boost of the chiral condensate, is observed in chirally driven low-energy effective models (i.e.\ models which are constructed to obey the global chiral symmetry of QCD) such as chiral perturbation theory \cite{Shushpanov:1997sf}, 
Linear Sigma model 
and NJL 
(see section \ref{sectioneffQCD}).

We briefly review the status of different analyses with regard to figure \ref{TBexpected}. 
The $N_f=2$ lattice results of \cite{D'Elia:2010nq} indicated a weak rise in the transition temperatures $T_\chi$ \'and $T_c$, while both remained compatible with each other (a split of $\sim 2\%$). 
Different variants and extensions (to include quarks and a deconfinement order parameter) of Linear Sigma- and NJL-models, found that both $T_\chi$ and $T_c$ rise with $B$ \cite{Mizher:2010zb,Gatto:2010pt,Gatto:2010qs,Fukushima:2010fe,Andersen:2012zc,Agasian:1999sx}
(with or without a ``noticeable' split between them), with the exception of the scenario in \cite{Mizher:2010zb} 
where vacuum corrections are not taken into account, which results in the opposite behaviour: $T_\chi$ and $T_c$ decreasing with $B$ (and no split). 

\begin{figure}[h!]
  \hfill
  \begin{minipage}[t]{.45\textwidth}
    \begin{center}
      \scalebox{0.44}{
  \includegraphics{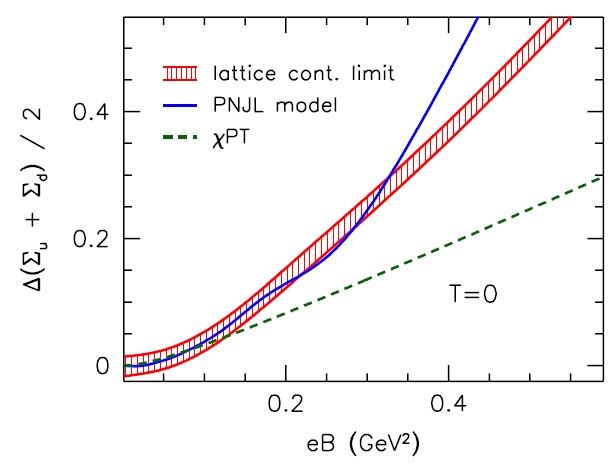}}
    \end{center}
  \end{minipage}
  \hfill
  \begin{minipage}[t]{.45\textwidth}
    \begin{center}
      \scalebox{0.44}{
  \includegraphics{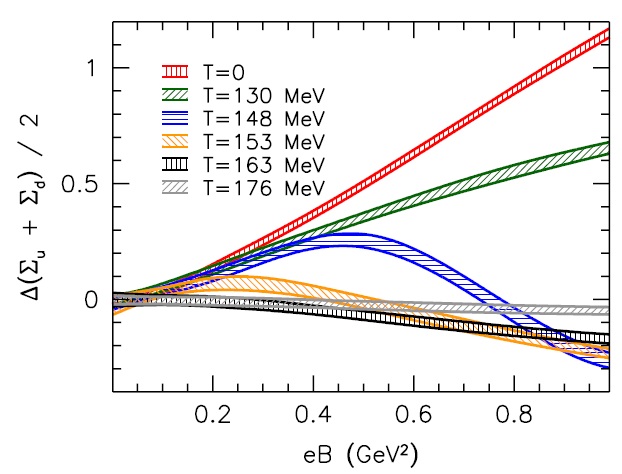}}
    \end{center}
  \end{minipage}
      \caption{Latest (continuum extrapolated) lattice results for the change of the chiral condensate as a function of $B$, at zero temperature on the left (and compared to model predictions) and at six different temperatures on the right, where $\Delta \Sigma_l(B,T) = \frac{2 m_u}{m_\pi^2 f_\pi^2} \left( \langle \bar \psi \psi \rangle_l(B,T) - \langle \bar \psi \psi \rangle_l (0,T) \right)$ with $m_u=m_d  < m_s$ \cite{Bali:2012zg}. 
}
	\label{figBali1}
  \hfill
\end{figure}

\begin{figure}[h!]
  \hfill
  \begin{minipage}[t]{.45\textwidth}
    \begin{center}
      \scalebox{0.46}{
  \includegraphics{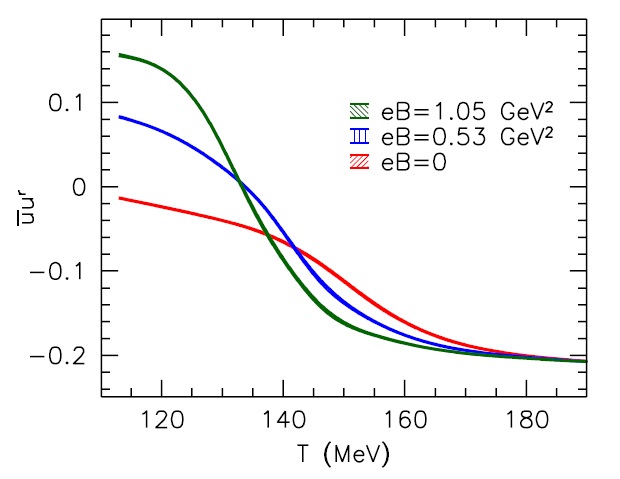}}
    \end{center}
  \end{minipage}
  \hfill
  \begin{minipage}[t]{.45\textwidth}
    \begin{center}
      \scalebox{0.46}{
  \includegraphics{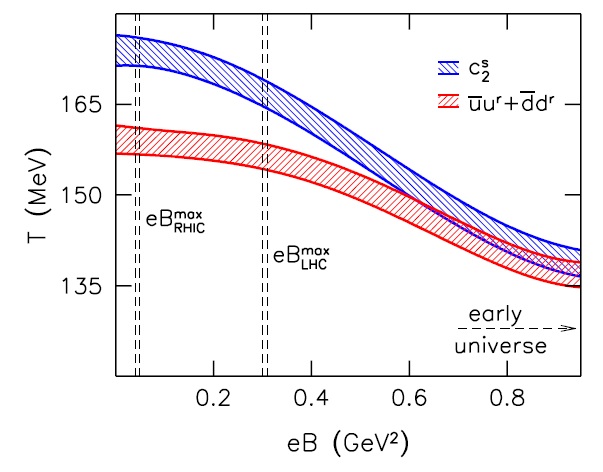}}
    \end{center}
  \end{minipage}
      \caption{Inverse magnetic catalysis observed in the most recent lattice simulations, with the  (pseudo-)critical temperature $T_\chi$ (on the right)  defined from inflection points of the renormalized chiral condensate (on the left) $\langle \bar u u \rangle^r + \langle \bar d d \rangle^r$ (red) and the strange quark number susceptibility $c_2^s$ (blue) \cite{Bali:2011qj}. 
 }  
\label{figBali}
\end{figure}

Somewhat later, a more thorough lattice study appeared using $N_f=2+1$ flavours with physical masses, leading to a much more complicated behaviour in the chiral/deconfinement (pseudo-)order parameters and ensuing critical temperatures \cite{Bali:2011qj}. It was motivated that the reported behaviour -- where contrasting with the results of \cite{D'Elia:2010nq} -- should be traced back to the lighter dynamical flavours and partially also to the present strange flavour, as the up ($u$) and down ($d$) quark of \cite{D'Elia:2010nq} were considerably heavier. Soon after, the first analytical papers appeared trying to explain the state-of-the art lattice data using backreacting pion dynamics \cite{Fraga:2012fs}. 
The results of \cite{Bali:2011qj} showed a more subtle picture regarding the chiral magnetic catalysis: 
the magnetic catalysis was confirmed for temperatures (sufficiently) below $T_c$, but for larger temperatures the (averaged over up and down) chiral condensate displayed a non-monotonous shape (shown in figure \ref{figBali1}), a feature translated into a similar behaviour in the transition temperature (see figure \ref{figBali}). This observation of an ``inverse magnetic catalysis'' seems to depend crucially on taking into account quark backreaction effects, see also \cite{Bruckmann:2013oba}, so we do not expect it to appear in the unquenched Sakai-Sugimoto setting we will use.

We compare our findings with the lattice results of \cite{Ilgenfritz:2012fw}. Although those results are referring to two-colour QCD, it contains an extrapolation to the chiral limit, which is the closest the available lattice results come to the $N_f=2$ SSM. 
In \cite{Ilgenfritz:2012fw}, no manifest split between $T_c$ and $T_\chi$ was reported, while the chiral condensate increases monotonically with the applied magnetic field for all temperatures in the confinement phase. We do present similar results here using prefixed values for the string theory parameters of the SSM (that is, a few physical QCD input values at zero magnetic field are chosen to match the corresponding SSM predictions). These results are a generalization to the 2 flavour case of the single flavour analysis of \cite{Johnson:2008vna}. To extend the scope of our analysis, we will also allow that the asymptotic D8-$\overline{\text{D8}}$ separation $L$ 
can vary and as such we construct the magnetic generalization of the $(T,L)$ phase diagram of figure  7 in the original work \cite{Aharony:2006da} concerning the SSM phase diagram. All results are presented in GeV units to make comparison with other QCD approaches more direct. 

Since a magnetic field couples to the up and down flavours with another strength, as they carry different electric charges, it seems natural that the up and down chiral restoration temperatures can be different, as well as the magnetic catalysis itself.  We recall that the classical chiral structure of QCD with and without magnetic field is different, since coupling a magnetic field to the quarks reduces $U(2)_L\times U(2)_R$ to $\left[U(1)_L\times U(1)_R\right]^u\times \left[U(1)_L\times U(1)_R\right]^d$, so that the eventually broken chiral invariances $U(1)_A^u$ and $U(1)_A^d$ can experience a different restoration temperature. Lattice simulations indeed confirm a larger value for the $\langle \bar u u \rangle$ 
than for the $\langle \bar d d \rangle$ chiral condensate at $T=0$ \cite{D'Elia:2011zu}, as does the $N_f=2$ NJL  model \cite{Boomsma:2009yk}. It would appear natural that $T_\chi^{u}$ should consequently be larger than $T_\chi^{d}$, and this is indeed what we will find. The splitting of degenerate order parameters, like $\langle \bar u u \rangle$ and $\langle \bar d d \rangle$ at $eB=0$, when an external field is switched on, is not that unfamiliar. In certain exotic superconductors, e.g.~$\text{Sr}_2\text{RuO}_4$, a similar phenomenon occurs \cite{Mackenzie:2003zz}.

\section{Set-up}

We employ the finite-temperature\footnote{
As remarked in section \ref{finiteTSSM}, there is also another proposal for the high-temperature background of the Sakai-Sugimoto model \cite{Mandal:2011ws}. It would be interesting to check whether the discussion in this chapter can be repeated in that model, but for the reasons mentioned in section \ref{finiteTSSM} (i.e.\ increased complexity but qualitatively similar end result for the $(T,L)$ phase diagram), we will consider the simpler black D4-brane background here. We thank Takeshi Morita for discussion on this point.}  
 SSM discussed in section \ref{finiteTSSM}, with $N_c=3$ and $N_f=2$, where in principle the limit $N_c\to\infty$ is always understood at the holographic level and the same remarks concerning these choices apply as discussed in the low-temperature phase.

\paragraph{Fixing holographic parameters at $eB=0$ up to $M$ } 

Instead of fixing all holographic parameters at $eB=0$, as in section \ref{B}, here we only fix them up to the value of the Kaluza-Klein mass $M_K$, which we will denote $M$ in this chapter. 

\begin{figure}[h!]
\centering
 \subfigure[]{
\includegraphics[scale=0.85]{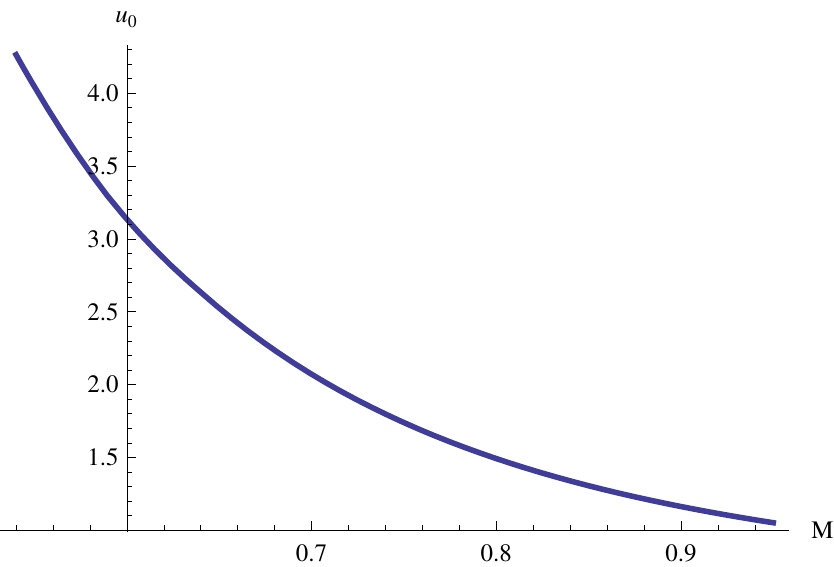} \label{a} 
}
\subfigure[]{
\includegraphics[scale=0.85]{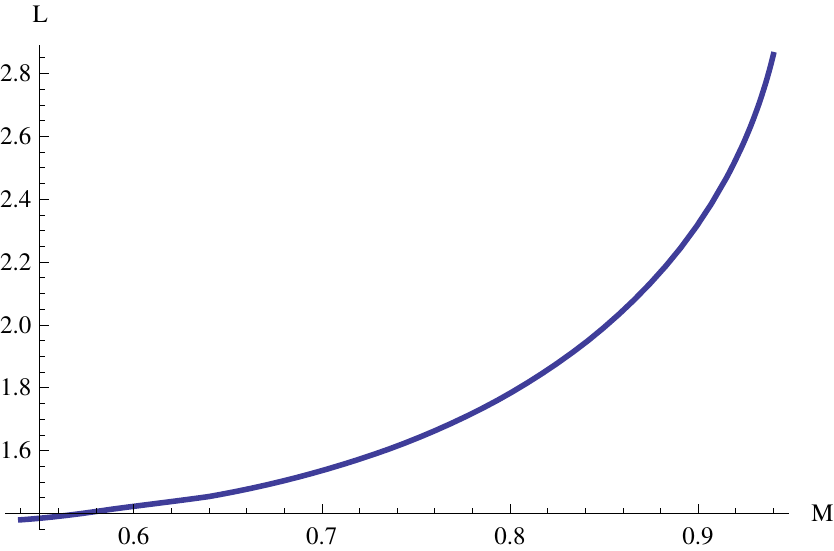} \label{b}
}
 \subfigure[]{
\includegraphics[scale=0.9]{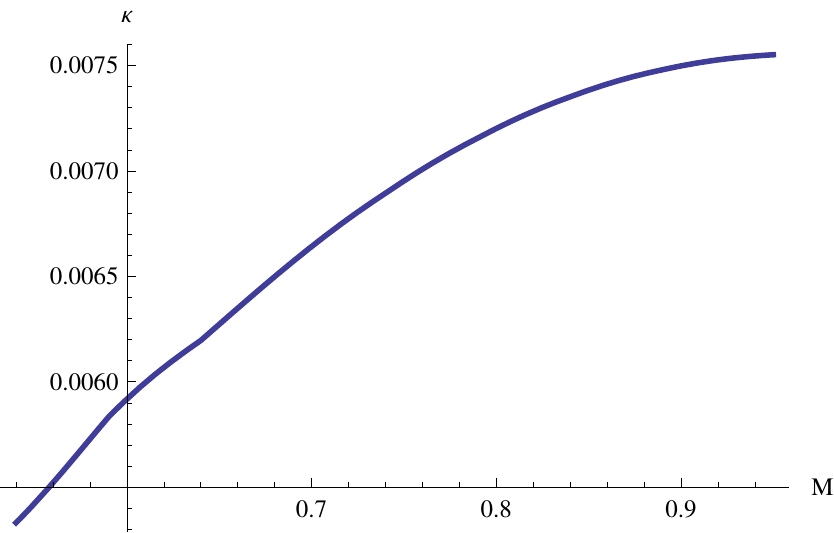} \label{c}
}
\caption[]{(a) Values of $u_0$ ($\text{GeV}^{-1}$) and (b) corresponding values of $L$ ($\text{GeV}^{-1}$)  compatible with $m_\rho=0.776$ GeV. (c) Values of $\kappa$ compatible with $m_\rho=0.776$ GeV and $f_\pi = 0.093$ GeV.}  \label{parameters}
\end{figure}

From the eigenvalue equation which determines $m_\rho$ holographically,
\begin{equation} \label{}
\partial_z \left( \frac{3}{u_0} u_z^{1/2} \gamma'^{-1} \partial_z \psi_\rho \right) = - \frac{4}{3}u_0 u_z^{3/2} \gamma' R^3 m_\rho^2 \psi_\rho, \quad \psi_\rho'(0)=0, \quad \psi_\rho(\pm \infty) = 0,
\end{equation} 
(this is (\ref{finiteTeigwvgl}) with $u_z^3$ defined in (\ref{uzdef}) and $\gamma'$ in (\ref{gammaaccentdef})), 
we extract the values of $u_0$ that, for a given $M$, lead to $m_\rho = 0.776 \mbox{ GeV}$. The resulting function $u_0(M)$ is plotted in figure \ref{parameters} for a range of $M$  -- the maximum value of $M$ corresponding to the limiting case $u_0 \rightarrow u_K = 1/M$ -- alongside with the function $L(M)$ for the corresponding asymptotic separation between branes and anti-branes, determined from (\ref{Lconf}).  
Next, demanding that the SSM-prediction for the pion decay constant $f_\pi$ in (\ref{fpiSSM}) equals $0.093$ GeV, leads to the function $\kappa(M)$ of allowed values for $\kappa$, as plotted in figure \ref{c}.
The string tension $(2\pi\alpha')^{-1} = 8\pi^2 M^2 \kappa(M)$ is then also known as a function of $M$.

\begin{figure}[h!]
\centering
\includegraphics[scale=1]{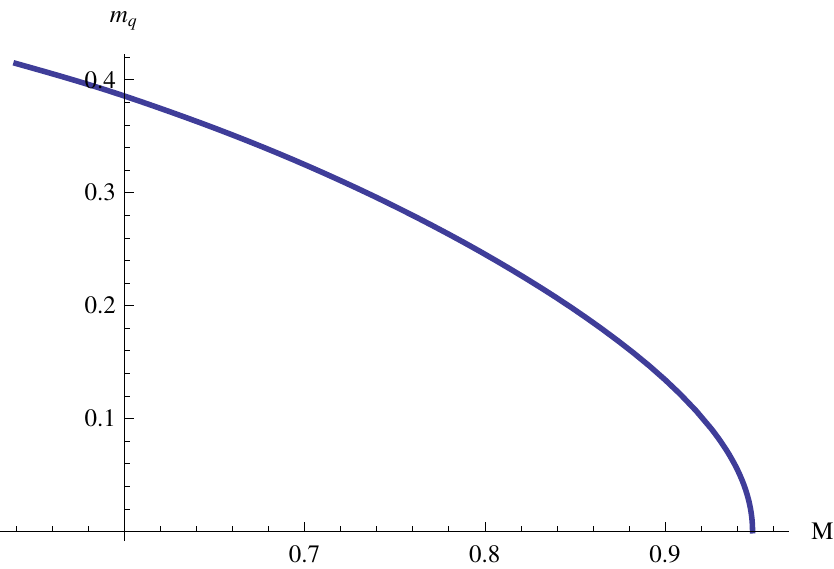}
\caption[]{SSM-prediction for $m_q$ ($\text{GeV}$) as a function of the confinement scale $M$ ($\text{GeV}$),  compatible with $m_\rho=0.776$ GeV and $f_\pi = 0.093$ GeV.} \label{mqMK}
\end{figure}

The remaining freedom of choosing the mass scale $M$ can be fixed for example by matching the SSM-prediction for the constituent quark mass $m_q(M,u_0,\kappa) $ in (\ref{mq}) 
to a phenomenologically reasonable value, as can be read off from figure \ref{mqMK}. 
In section \ref{B} we opted to reproduce $m_q = 0.310$ GeV, leading to the set (\ref{values}) of fixed holographic parameters.  
Here, we will however 
leave $M$ variable, or equivalently $L$ via figure \ref{b}, with the eye on drawing the $(T,L,eB)$ phase diagram later, and, more importantly, with the idea that the choice of $M$ or $L$ should be left free, as it determines the choice of holographic theory: $L$ very small ($\sim$ $\delta \tau=\frac{2\pi}{M}$ large $\sim M$ small) corresponding to an NJL-type boundary field theory  \cite{Antonyan:2006vw,Preis:2010cq,Aharony:2006da} versus $L=\delta \tau/2$ maximal ($\sim M$ maximal) corresponding to a maximal probing of the gluon background, i.e.~the original antipodal SSM.

\paragraph{Applying a magnetic field}

An electromagnetic background field $A_\mu^{em}$ is switched on through (\ref{Aachtergrond}) with corresponding field strength tensor (\ref{Fbardef}).

\FloatBarrier

\section{Results} \label{resultstempartikel}

\subsubsection{$eB$-dependent embedding in deconfinement phase}

The background at $T<T_c$ is identical to the zero temperature background up to the period $\beta = T^{-1}$ of Euclidean time, so the $eB$-dependent embedding  of the flavour branes is unchanged compared to the $T=0$ case, discussed in section \ref{D}. Things get more interesting once we enter the deconfinement region. We again have
an induced metric on each flavour brane,
\begin{eqnarray}
ds^2_{\text{D}8} &=& \left(\frac{u}{R}\right)^{3/2} (\hat f dt^2 + \delta_{ij}dx^i dx^j)
+ \left(\frac{R}{u}\right)^{3/2} u^2 d\Omega_4^2+\left(\frac{u}{R}\right)^{3/2} \left[ \frac{1}{\hat f} \left(\frac{u}{R}\right)^{-3} + \frac{1}{u'^2}\right] du^2\, 
\end{eqnarray}
with periodicity of the $t$-circle given by (\ref{deltatvooractie}),  
from which one 
can determine the action in the deconfined phase, completely analogous to the derivation of the action (\ref{SDBIconf}) in the confined phase.
For temperatures $T< T_{\chi,l}$, the $l$-brane's embedding remains $\cup$-shaped, with action
\begin{eqnarray}
{S}_l^{T<T_{\chi,l}}
&=& c_0 u_{0,l}^{7/2} \int_{1}^\infty dy \hspace{1mm} y \sqrt{y^3 A_l} \sqrt{\frac{1}{1 - \frac{\hat f_{0,l} A_{0,l}}{\hat{f}_l(y) y^3 A_l}y^{-5}}},
\end{eqnarray}
where $c_0= - 2 T_8 \mathcal{V}_4 V_4 g_s^{-1} R^{3/2}$,
$y=u/u_{0,l}$, $y_{T,l}=u_T/u_{0,l}$, $\hat{f}_l(y) = 1-(y_{T,l}/y)^3$ and $\hat{f}_{0,l} = 1-y_{T,l}^3$. If $T>T_{\chi,l}$, the $l$-branes are falling straight down, $u' = \infty$, with action
\begin{eqnarray}
{S}_l^{T>T_{\chi,l}}
&=& c_0 u_{0,l}^{7/2} \int_{y_{T,l}}^\infty dy \hspace{1mm} y \sqrt{y^3 A_l}.
\end{eqnarray}
The chiral transition temperature $T_{\chi,l}$ is the temperature for which $\Delta S_{l}$ becomes zero \cite{Aharony:2006da},
with
\begin{eqnarray}\label{kritiek}
\Delta S(u_0,eB,y_T) &=& \text{action}_{\cup\text{-shape}} -  \text{action}_\text{straight}\,.
\end{eqnarray}
The correspondence between $u_{0}$ and $L$ in the deconfined phase is modified into 
 (again suppressing the flavour index here)
\begin{eqnarray}
L^{dec}(u_0,eB,y_T)  &=& \frac{2}{3} \frac{R^{3/2}}{\sqrt {u_0}} \sqrt{\hat f_0 A_0} \int_0^1 d\zeta \frac{\hat f^{-1/2} \zeta^{1/2}}{\sqrt{\hat fA - \hat f_0 A_0 \zeta^{8/3}}}. \label{LeindigeT}
\end{eqnarray}

\begin{figure}[h]
\centering
\subfigure[]{\includegraphics[scale=1]{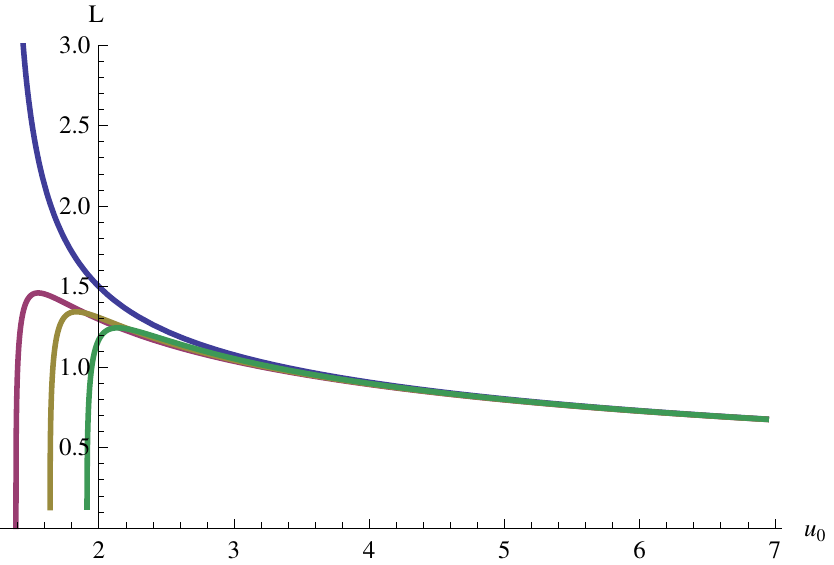} \label{Lu0fig}}
\hspace{1cm}
\subfigure[]{\includegraphics[scale=1]{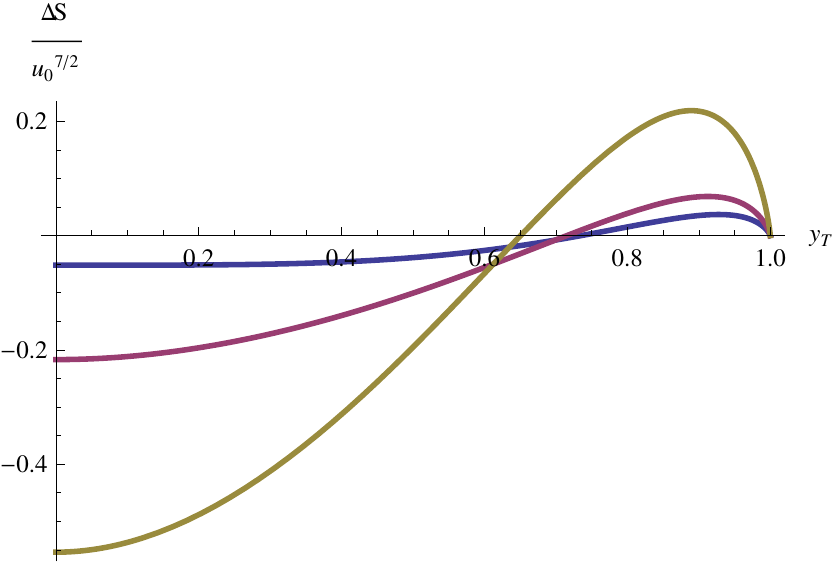}  \label{DeltaSyT} }
\caption[]{(a) $L^{conf}$ (blue) and $L^{dec}$ ($\text{GeV}^{-1}$) for $T=T_c$ (red)  and increasing values of $T(>T_c)$ at $eB=0$, (b) $\Delta S/u_0^{7/2}$ as a function of $y_T$ for $eB=0$ (blue), 0.5 (purple) and 1.6 GeV$^2$ (yellow). Both figures for $M=0.7209$ GeV.  }\label{}
\end{figure}

As before, we will hold the asymptotic separation fixed at its starting value $L$ at $eB=0$ and $T=0$.
This allows us to determine the $eB$- and $T$-dependence of $u_0$ (from here on also explicitly writing the dependence on $M$, through $R$ and $2\pi\alpha'$):
\begin{equation} \label{u0BTL}
L^{dec}(u_0,eB,y_T,M) = L \quad \Longrightarrow \quad u_0(eB,y_T,L,M).
\end{equation}
From figure \ref{Lu0fig} it can be seen that the one-to-one correspondence between $u_0$ and $L^{conf}$ is not preserved in the deconfinement phase, where each value of $L^{dec}$ corresponds to two possible values of $u_0$, as long as it does not exceed its maximum possible value (i.e.~as long as $T<T_\chi$). We numerically verified that the energetically favoured solution for $u_0$ is the largest one, consistent with the intuition that the lower-$u_0$ solution contains more energy as it probes a larger portion of the background. Keeping $L$ fixed during the deconfinement transition causes a jump in $u_0$, as well as in the constituent quark and meson masses \cite{Peeters:2006iu}.

With the expression found for $u_0$, the expression for $\Delta S$ at fixed $L$ is also known,
\begin{equation}
\Delta S(u_0(eB,y_T,L,M),eB,y_T,M) \equiv \Delta S(eB,y_T,L,M),
\end{equation}
so the chiral temperature can be determined from the point where the $\cup$-shaped embedding breaks into separated branes, i.e. when $\Delta S = 0$ (see figure \ref{DeltaSyT}):
\begin{equation}
\Delta S(eB,y_T,L,M) = 0 \quad \Longrightarrow \quad y_T^\chi(eB,L,M).
\end{equation}
The corresponding value of $u_0$ at the chiral transition is then given by
\begin{equation}
u_0^\chi = u_0(eB,y_T^\chi(eB,L,M),L,M) \equiv u_0^\chi(eB,L,M).
\end{equation}
Plugging the obtained $y_T^\chi(eB,L,M)$ and $u_0^\chi(eB,L,M)$ into the definition for the chiral temperature
\begin{align}
T_\chi  &= \frac{3}{4\pi} \sqrt{\frac{u_T^\chi}{R^3}} = \frac{3}{4\pi} \sqrt{y_T^\chi} \sqrt{\frac{u_0^\chi}{R^3}} = \frac{3}{4\pi} \sqrt{y_T^\chi(eB,L,M)} \sqrt{\frac{u_0^\chi(eB,L,M)}{R^3}} \equiv T_\chi(eB,L,M),
\end{align}
we obtain $T_\chi(eB,L,M)$.

\begin{figure}[h]
  \centering
  \scalebox{0.7}{
  \includegraphics{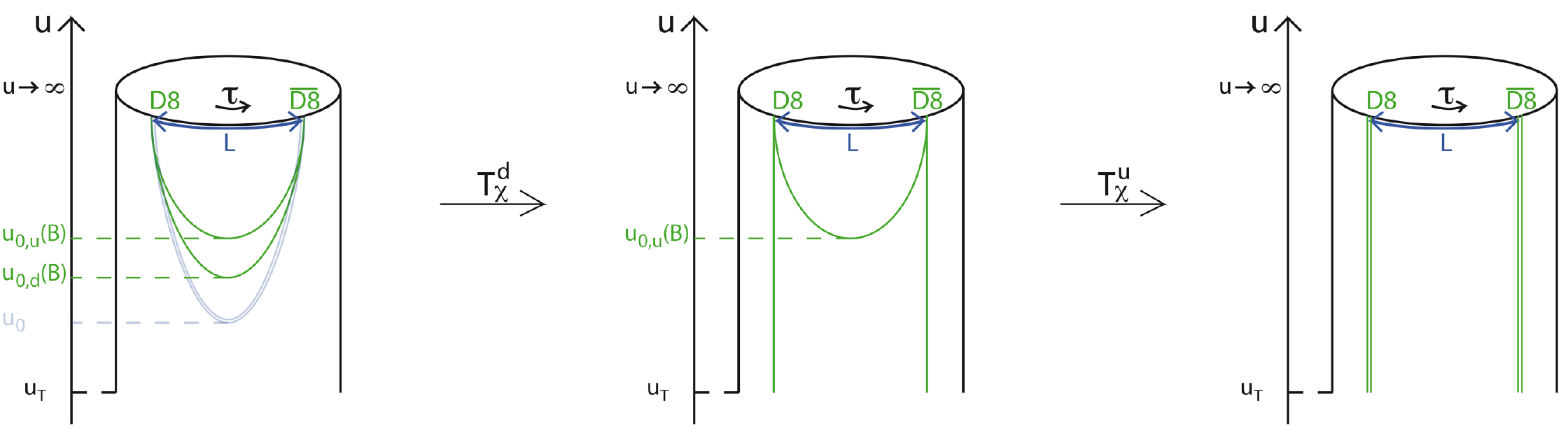}}
  \caption{Embeddings in the deconfined phase with magnetic field. }\label{TchiralSS}
\end{figure}

From the parameter discussion at $eB=0$ we know the value of the fixed asymptotic separation given a value for $M$ such that $m_\rho=0.776$ GeV (figure \ref{b}), hence we obtain $T_\chi(eB,M)$ or $T_\chi(eB,L)$, to be compared with the deconfinement temperature $T_c = M/(2\pi)$.
The deconfinement temperature $T_c$ will not change as it is determined from the background D4-brane metric, which could only become $eB$-dependent when the backreaction of the D8-branes would be taken into account\footnote{
This is only true to leading order in the large-$N_c$ limit, we will comment on subleading $B$-dependent corrections to $T_c$ \cite{Ballon-Bayona:2013cta} at the end of the chapter. 
}.  Or, field theoretically: $T_c$ is $eB$-independent in a quenched set-up, because the magnetic field can only couple to the neutral gluons indirectly via the quark interactions.
For every choice of $M$ (or $L$), $T_\chi(eB)$ rises with $eB$ (``chiral magnetic catalysis'').
But depending on the choice, there will or will not arise a split between $T_\chi$ and $T_c$.
Doing this for each flavour, we find $T_\chi^u(eB,L)$ and $T_\chi^d(eB,L)$, with $T_\chi^u$ consistently higher than $T_\chi^d$ for a given value of $L$, as expected. This leads to an intermediate phase where the chiral symmetry for up quarks is still broken while the chiral symmetry for down quarks is already restored:
\begin{equation}
U(1)_V^u \times U(1)_V^d \stackrel{T_\chi^d}{\rightarrow}  U(1)_V^u \times  (U(1)_L \times U(1)_R)^d \stackrel{T_\chi^u}{\rightarrow}(U(1)_L \times U(1)_R)^u \times (U(1)_L \times U(1)_R)^d,
\end{equation}
as sketched in figure \ref{TchiralSS}.

\begin{figure}[h]
\centering
\subfigure[]{\includegraphics[scale=0.8]{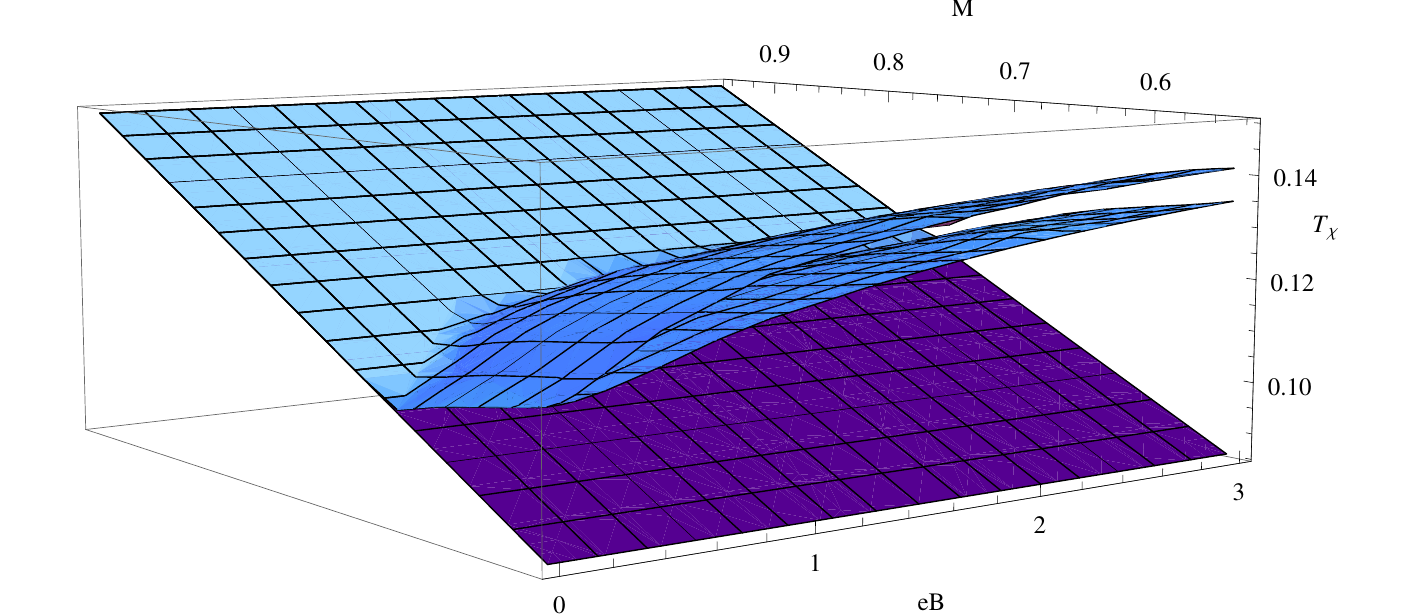} \label{TchiBM}}
\subfigure[]{\includegraphics[scale=1.1]{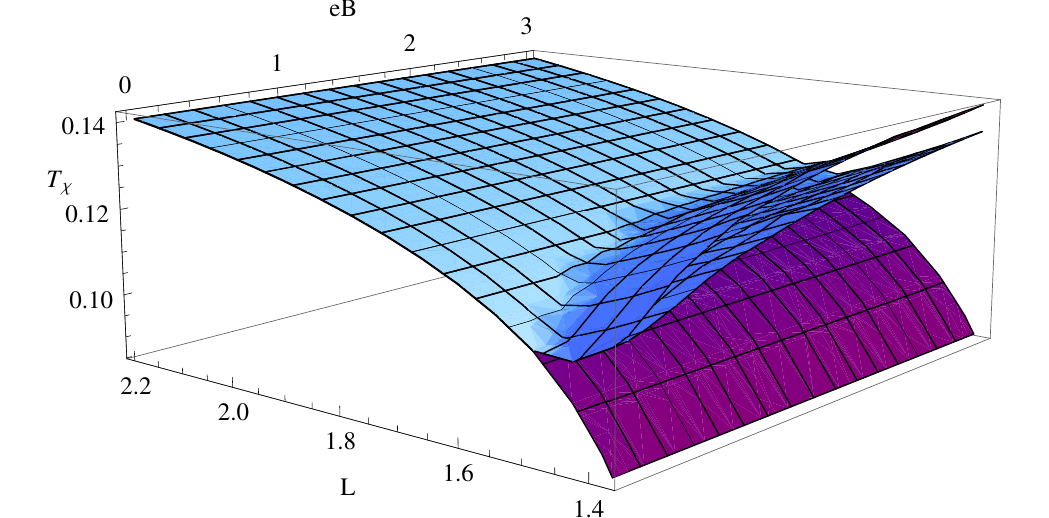} \label{TchiBL}}
\caption[]{(a) $T_\chi^u$ (GeV) (upper blue surface) and $T_\chi^d$ (GeV) (lower blue surface) as functions of $eB$ ($\text{GeV}^2$) and $M$ (GeV) compared to $T_c(M)$ (GeV) (purple), (b) same with $M$-dependence replaced by $L$-dependence compatible with $m_\rho=0.776$ GeV. }\label{Tchifig}
\end{figure}

\begin{figure}[h!!]
\centering
\subfigure[]{\includegraphics[scale=0.8]{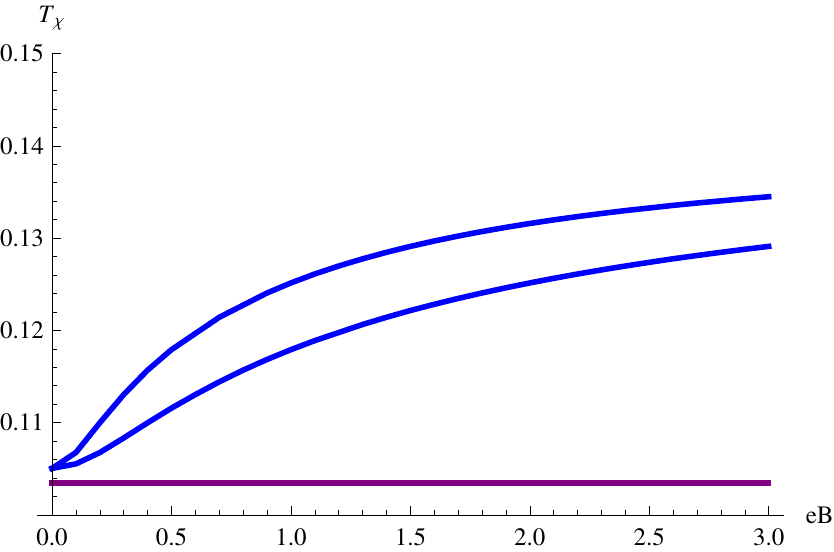} \label{T65}}
\subfigure[]{\includegraphics[scale=0.8]{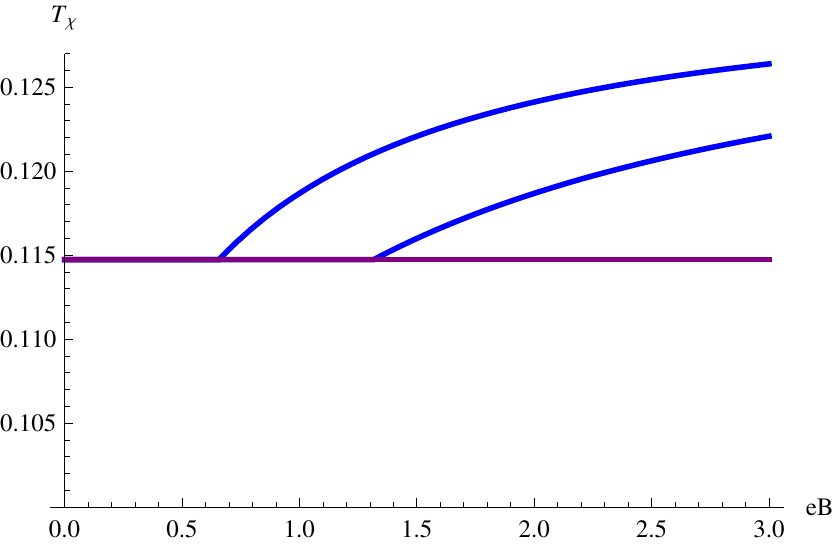} \label{T7209}}
\subfigure[]{\includegraphics[scale=0.8]{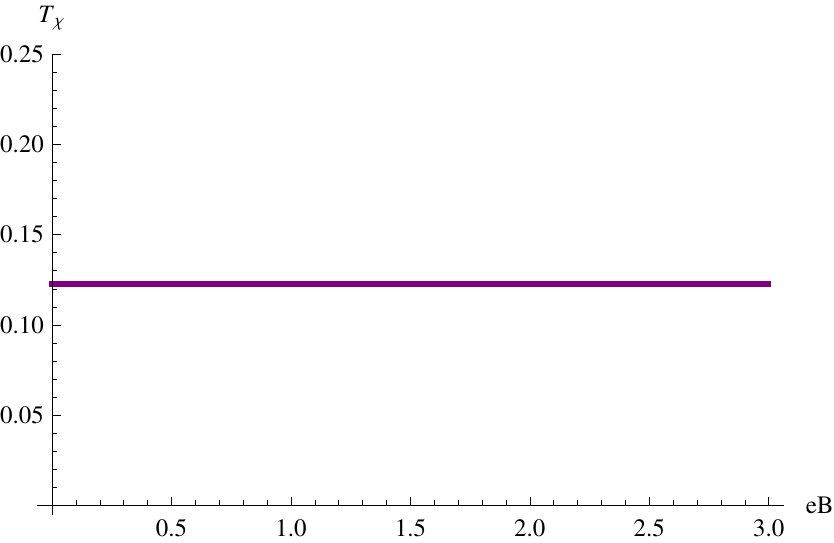}  \label{T77}}
\caption[]{Cross sections of figure \ref{Tchifig} for (a) $M=0.65$ GeV, (b) $M=0.7209$ GeV and (c) $M=0.77$ GeV, respectively corresponding to $m_q(eB=0) = 0.357, 0.310$ and $0.272$ GeV and $T_c=0.103, 0.115$ and 0.123 GeV. The appearance of a split between $T_\chi$ (GeV) (blue) and $T_c$ (GeV) (purple) depends on the choice of $M$, or equivalently $L$. }\label{Tchicrosssections}
\end{figure}

\begin{figure}[h!]
  \centering
  \scalebox{1 }{
  \includegraphics{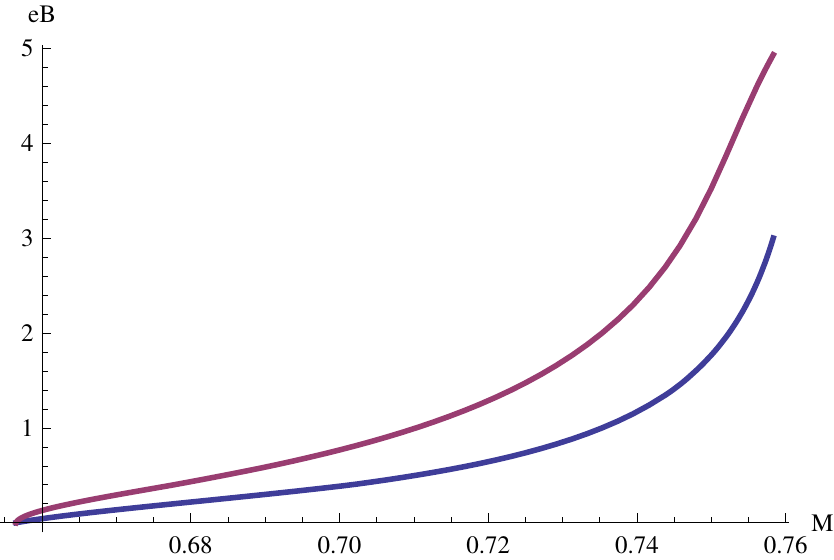}}
  \caption{Value of $eB$ where $T_\chi^u$ (GeV) (blue) resp.~$T_\chi^d$ (GeV) (purple) becomes larger than $T_c$ (GeV) for confinement scale values 0.657 GeV $< M < 0.767$ GeV.} \label{BcM}
\end{figure}

In figure \ref{Tchifig} the $(T,M,eB)$ and $(T,L,eB)$ phase diagrams of the two-flavour non-antipodal SSM are plotted. This generalizes the $N_f=1$ SSM phase diagram  in figure  7 of \cite{Aharony:2006da} to the $N_f=2$ magnetic case. For set-ups with large values of $M$, namely $M>0.767$ GeV corresponding to $m_q(eB=0) < 0.274$ GeV or $L>1.681$ GeV$^{-1}$,
there is no split between $T_\chi(eB,M)$ and $T_c=M/(2\pi)$, no matter how large the applied magnetic field is. This is a consequence of the saturation of the rising of $T_\chi$ with $eB$.
In a SSM with $M<0.657$ GeV, corresponding to $m_q(eB=0) > 0.353$ GeV or $L<1.473$ GeV$^{-1}$, there is already a split between chiral and deconfinement transition before the magnetic field is turned on: $T_\chi(eB=0,M) > T_c$, which becomes larger as $eB$ increases. This regime is probably the least physically relevant, as the values for constituent quark masses are too large and the values for the deconfinement temperature smaller than 0.105 GeV, which is rather small compared to the chiral limit value we can extrapolate ``by hand'' from \cite{Bornyakov:2009qh}, giving $T_c\sim 0.150~\text{GeV}$. The third possible case is that the value of $M$ is such, 0.657 GeV $< M < 0.767$ GeV ($\sim 0.274$ GeV $< m_q(eB=0) < 0.353$ GeV or 1.473 GeV$^{-1} < L < 1.681$ GeV$^{-1}$), that $T_\chi(eB=0,M) = T_c$ but a split between $T_\chi$ and $T_c$ arises at some value of $eB$, plotted in figure \ref{BcM}.
For each of the above possible cases, an exemplary cross section of the $(T,M,eB)$ phase diagram is shown in figure \ref{Tchicrosssections}, the middle one corresponding to the best matching parameters for reproducing a reasonable $m_q(eB=0)\approx 0.310$ GeV, although the corresponding value for $T_c \approx 0.115$ GeV is still on the small side.
A deconfinement temperature $T_c \approx 0.150$ GeV would correspond to $M=0.942$ GeV, very close to the antipodal value so in the regime where no split arises between $T_\chi$ and $T_c$, leading however to an unphysically\footnote{This might be related to the shortcoming of the SSM (in the used form, not considering possible modifications as in \cite{Dhar:2008um}) that the bare quark masses always remain zero.} small $m_q(eB=0) \approx 0.046$ GeV. 

The papers \cite{Johnson:2008vna,Preis:2010cq} also pointed out the splitting of the critical temperatures for the one flavour version of the SSM, but leaving the parameters of the SSM undetermined, making an explicit comparison with other approaches less straightforward. The explicit breaking of the global flavour symmetry by the different electromagnetic coupling of the up and down flavour is also  taken into account now for the first time, leading to a split between the two chiral transitions themselves.

The fact that a split between $T_\chi$ and $T_c$ can emerge only for sufficiently small values of the asymptotic brane separation $L$, i.e. sufficiently close to an NJL effective description of QCD, seems to be supported by NJL model calculations \cite{Gatto:2010pt}, see also the discussions in \cite{Mizher:2010zb,Boomsma:2009yk,Klevansky:1992qe}, that seemingly contrast with lattice data.
Selecting the holographic parameters in a way that brings the SSM as close as possible to (the chiral limit of) QCD, rather leads to a picture of the form of 
figure \ref{T7209} or \ref{T77}, namely no split at all or a small split that only emerges at rather large values of $eB$. Our findings are in this perspective consistent with lattice data of \cite{Ilgenfritz:2012fw}, where a split was neither reported. However, we must also repeat here that our results are obtained in a quenched framework, hence important QCD effects at the level of transitions can be missing (e.g.~pion loop effects). In particular, magnetic effects on the deconfinement temperature cannot be taken into account in the SSM without including backreaction of the probe branes on the background D4-brane metric. 

A remark is in order here regarding the last statement on $T_c(B)$. In \cite{Ballon-Bayona:2013cta} 
a perturbative correction $\delta T$ of order $\lambda^2 N_f/N_c$ to the deconfinement temperature $T_c$ is calculated, including its dependence on a background magnetic field. The result is obtained without backreacting the geometry 
(but it is argued that the deformed D4-brane background upon including backreaction effects would not alter that result, for it would only contribute to second order in the perturbative parameter $\lambda^2 N_f/N_c$; more precisely: the linear corrections to the metric and dilaton would not modify the on-shell supergravity action at linear order).  
The result comes from including the contribution to the pressure from the quarks when probe branes are added to the D4-brane background. In section \ref{finiteTSSM} we discussed the thermodynamics of the background, where $T_c$ is determined from equating the free energies of the confining and deconfining metric, or equivalently the pressures $P = -T \partial S/\partial V_3$ (from the supergravity actions): 
\begin{equation}
P_{conf}^{gluon} = P_{deconf}^{gluon}(T) \quad \Rightarrow \quad T_c, 
\end{equation}
with $P_{conf}^{gluon}$ and $P_{deconf}^{gluon}$ the gluonic pressures associated with the respective backgrounds, both of order $\lambda N_c^2$. 
The probe branes themselves add quark contributions to the pressure (from the DBI-actions), which are of order $\lambda^3 N_f N_c$ and for that reason neglected in \cite{Aharony:2006da}. When included they give rise to a corrected deconfinement temperature 
\begin{equation}
P_{conf}^{gluon} + P_{conf}^{quark}(B) = P_{deconf}^{gluon}(T) + P_{deconf}^{quark}(T,B) \quad \Rightarrow \quad T_c \rightarrow T_c(B) =  T_c + \delta T(B), 
\end{equation}
with $T_c(B)$ a decreasing function (but not saturating), consistent with the thermodynamic argument for such behaviour in figure \ref{TBexpected}. 
This correction is already present for zero magnetic field ($\delta T(0) \neq 0$). 
In light of this work, we stress that our results are obtained at leading order in the large-$N_c$ probe approximation, neglecting $\lambda^2 N_f/N_c$ effects.

\FloatBarrier

\subsubsection{Remark on the antipodal SSM}
In the original antipodal Sakai-Sugimoto model, with $u_0=u_K$ and the asymptotic separation $L$ taking
its maximum possible value, the embedding of the flavour
branes is unaffected by the presence of the magnetic field.
From this we can conclude that the antipodal Sakai-Sugimoto
model is unable to capture the magnetically induced explicit
breaking of chiral symmetry, as well as the chiral magnetic
catalysis. Chiral symmetry restoration and deconfinement coincide
for all values of the magnetic field.

\section{Summary and outlook}

In conclusion, we have investigated the phase diagram of the two flavour version of the non-antipodal Sakai-Sugimoto model in the presence of a temperature $T$ and external magnetic field $eB$. In particular we payed attention to fixing the holographic parameters, presenting a discussion of how they can be fixed by matching to carefully chosen QCD input parameters, in order to be able to present the phase diagram and related results in physical GeV units. This makes comparison to other approaches more direct. We indeed could compare our results with lattice and NJL results, the main conclusion being that the SSM results are consistent with other quenched settings that are able to model chiral magnetic catalysis. 

The main results are presented in the $(T,L,eB)$ phase diagram in figure \ref{Tchifig} and cross sections of that plot for different values of $L$ in figure \ref{Tchicrosssections}. Here, $L$ is the asymptotic separation between the flavour probe branes.  Keeping $L$ fixed serves as a boundary condition for the bulk dynamics, the effective boundary theory ranging from the NJL-type for small $L$ to a chiral QCD-like theory where gluon dynamics are fully taken into account for maximal $L$. The value of $L$, i.e.\ the choice of the type of boundary model in a sense, determines if a split between the chiral and deconfinement temperature may or may not arise, as summarized in figure \ref{Tchicrosssections}. Due to the different coupling to the magnetic background of the $u$ and $d$ flavour brane, we also find a split between the separate chiral transition temperatures.

It remains a challenge to construct a holographic dual of realistic QCD that could also describe the complicated finite temperature (above and below $T_c$) behaviour of the chiral condensate in figure \ref{figBali1}, as found in the latest lattice results  \cite{Bali:2011qj}. This would require taking backreaction of the flavour branes on the background metric into account. 
To our knowledge, this backreaction is studied in the Sakai-Sugimoto model only in \cite{Burrington:2007qd}, 
where the leading order backreaction in $N_f/N_c$ is discussed for antipodal embedded flavour branes in  the D4-brane background at zero temperature.  
The backreacted black D4-brane background at $T>T_c$ would also be needed 
to extract $T_c$ from the difference in free energies of the two possible geometries. 

Another holographic model which might prove interesting for this problem, 
is the Kuperstein-Son\-nen\-schein model \cite{Kuperstein:2008cq}, in which chiral magnetic catalysis was observed \cite{Alam:2012fw} in the non-backreacted case, and which was extended to include backreaction (be it perturbatively and for vanishing magnetic field) in \cite{Ihl:2012bm}, using a smearing technique. A downside of this model is however that it does not incorporate confinement due to the choice of the background in which the flavour branes are placed. 
The complexity of the problem suggests a bottom-up approach may be of more use.

\chapter{Time dependent spectral functions in AdS-Vaidya models} \label{vaidyachapter}

In this chapter we will be concerned with dynamics of a quantum system out of equilibrium. 
The thermalization process of forming the thermal equilibrium state of QGP (section \ref{statmech}) starting 
from the initial highly excited state right after the heavy ion collision, is for example poorly understood. 
Techniques such as linear response theory and hydrodynamics are only applicable for the study of equilibration of near-equilibrium initial states. The main AdS/CFT result in the near-equilibrium regime is that late-time hydrodynamic evolution with a minimal shear viscosity emerges naturally in the study of perturbations of the AdS$_5$-black hole metric \cite{Policastro:2001yc,Janik:2005zt}. 
We will be interested in 
far from equilibrium systems, thereby resorting to holography. 
The approach we will take involves the use of a bottom-up model, which also allows interpretation of the results in a condensed matter theory context. However, as the results discussed in this chapter involve work that was only being completed at the stage of writing this thesis, we will not go into much detail concerning possible interpretations (apart from the discussion in section \ref{latestpaper}). Instead we will focus on the used numerical techniques in obtaining time-dependent spectral functions of scalar and spinor bulk fields in thin-shell AdS$_3$-Vaidya.

\section{AdS-Vaidya background} \label{vaidyasection} 

Holographically, the process of thermalization in a boundary CFT has been argued to be dual to black hole formation in the bulk AdS space. The authors of \cite{Bhattacharyya:2009uu} 
turn on a source $\phi_0$ that is coupled to a marginal operator for a brief amount of time, dual to a rapid injection of energy on the boundary that forces the strongly coupled conformal field theory out of its vacuum. They then observe that it sets up a 
wave that propagates inwards into the bulk and collapses into a black hole. At leading order in the 
 source amplitude the resulting bulk spacetime takes the form of a Vaidya metric, which is an exact solution for the propagation of a null dust -- a fluid whose stress tensor is proportional to $\rho k_\mu k_\nu$ for a lightlike vector $k_\mu$ specifying the direction in which massless radiation (with intensity $\rho$) is moving.

We will use the Vaidya metric of black hole formation, which 
interpolates between an AdS and an AdS-Schwarzschild 
metric, as a toy model for thermalization. It is a bottom-up model, and the action to 
which it is a solution will be specified. 
A general form of the Vaidya metric is given by \cite{Galante:2012pv} 
\begin{equation}
 ds^2 = \frac{1}{z^2} \left( -f(v,z) dv^2 - 2 dz dv + d\vec x^2 \right). 
\label{Vaidyametric}
\end{equation}
It is asymptotically AdS$_{d+1}$ near the boundary at $z \rightarrow 0$: the function $f(v,z)$ there approaches 1. 
The coordinates $\vec x = (x_1, \cdots, x_{d-1})$ are the $d-1$ spatial coordinates at the boundary. 
A function $f(v,z) = 1- m(v) z^d$ can model 
a neutral in-falling shell collapsing into an AdS-Schwarzschild black hole, 
and $f_{RN}(v,z) = 1- m(v) z^d + q(v)^2 z^{2d-2}$ ($d \geq 3$) a charged in-falling shell collapsing into a Reiss\-ner-Nord\-str\"om black hole, with $m(v)$ and $q(v)$ modeling the change in mass and charge of the black hole as a function of $v$. 
The metric, together with an electromagnetic potential $A_\mu = - \sqrt{\frac{d-1}{2(d-2)}} q(v) z^{d-2}\delta_{\mu v}$, form a solution of the equations of motion obtained from the Einstein-Hilbert-Maxwell action accompanied with an external term,  
\begin{equation}
S = -\frac{1}{16 \pi G_{d+1}} \int d^{d+1} x \sqrt{|\det g_{\mu\nu}|} \left( \mathcal R  + d(d-1)\right)  - \frac{1}{4} \int d^{d+1} x \sqrt{|\det g_{\mu\nu}|} F_{\mu\nu}F^{\mu\nu} +  S_{ext},  
\end{equation}
with negative cosmological constant $\Lambda = -d(d-1)/2$ and $S_{ext}$ providing an external energy-momentum tensor $T_{\mu\nu}^{ext}$  
dependent on $m'(v)$, $q(v)$ and $q'(v)$, and external current $J_\mu^{ext}$ 
dependent on $q'(v)$. The external energy-momentum tensor models the effect of the scalar bulk field sourced by $\phi_0$ in the set-up of  \cite{Bhattacharyya:2009uu}.  Its $m(v)$-dependence is given by $T_{\mu\nu}^{ext} \sim m'(v) \delta_{\mu v} \delta_{\nu v}$. 

\begin{figure}
	\parbox[b]{.8\linewidth}{
		\includegraphics[width=\linewidth]{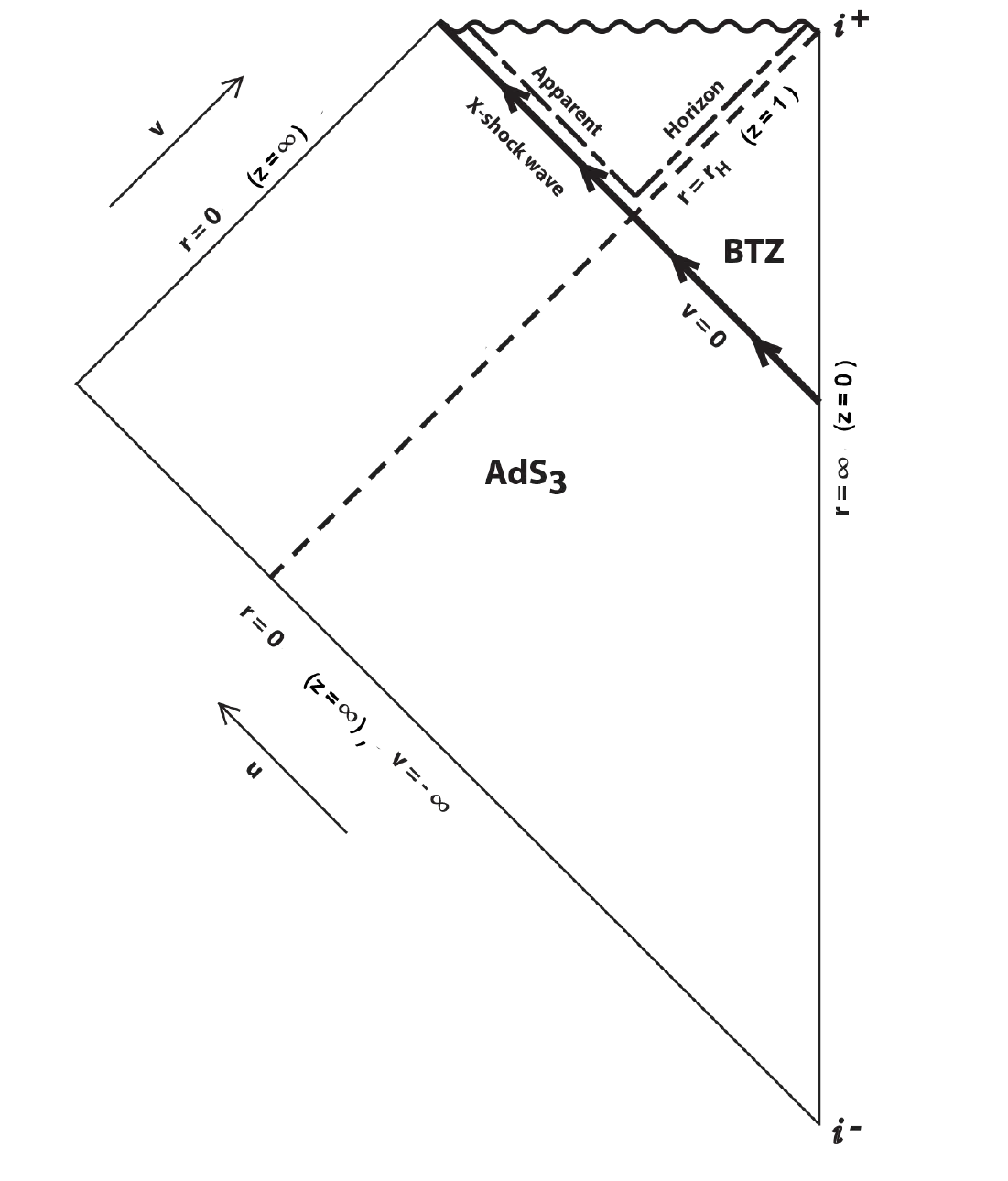}}\hfill
	\parbox[b]{.48\linewidth}{
		\caption{Penrose diagram of the $d=2$ Vaidya me\-tric (\ref{Vaidyametricfig}) interpolating between AdS$_3$ at $v<0$ and BTZ at $v>0$. The coordinate $r$ in the figure is related to the radial coordinate $z$ in (\ref{Vaidyametric}) by $z=R^2/r$ with the AdS radius $R$ set to 1. The AdS boundary is at $r=\infty$ or $z=0$, the dashed line represents the horizon of the black hole, the wavy line the singularity itself and the arrow the shock wave of collapsing null matter or ``shell'' at $v=0$. $i_+$ and $i_-$ mark future and past timelike infinity 
\cite{Ebrahim:2010ra}.
}
		\label{Penrose}
	}
\end{figure}

In \emph{thin-shell} Vaidya models the background changes instantaneously at the shell $v=0$, with $m(v)$ and $q(v)$ step functions. With the choice $m(v)=M \theta(v)$ (constant $M$) and $q(v)=0$, 
and for a moment changing radial coordinate to $r=R^2/z$ with AdS radius $R$ set to 1 throughout the chapter, 
the thin-shell AdS$_{d+1}$-Vaidya spacetime is given by 
\begin{equation}
ds^2 = -(r^2 - \frac{\theta(v) M}{r^{d-2}}) dv^2 + 2 dv dr + r^2 d\vec x^2 \label{Vaidyametricfig}
\end{equation}
and we will in particular focus on the $d=2$ case. 
The coordinate $v$ ($-\infty < v < \infty$) can now be recognized as the Eddington-Finkelstein coordinate, defined as the non-singular combination 
\begin{equation}
 dv = dt + \sqrt{\left|  \frac{g_{rr}}{g_{tt}} \right|} dr  
\end{equation}
of coordinates $r$ ($0<r<\infty$) and 
$t$ of an AdS-Schwarzschild metric at $v>0$ and an AdS metric at $v<0$.  
In Eddington-Finkelstein coordinates 
the horizon is a regular place for free in-falling observers.   
For the AdS$_3$-Schwarzschild metric, also known as the Ba\~{n}ados-Teitelboim-Zanelli (BTZ) solution 
\begin{equation}
ds^2 = -(r^2 - M) dt^2 + \frac{dr^2}{r^2-M} + r^2 dx^2,   
\end{equation}
the horizon is located at $r=r_H=\sqrt{M} R$, while for the Poincar\'e patch of AdS$_3$ (with a fake boundary at $r=0$)  
\begin{equation}
ds^2 = -r^2 dt^2 + \frac{dr^2}{r^2} + r^2 dx^2,  
\end{equation}
it is at $r=0$. 
The thin-shell AdS$_3$-Vaidya background is shown in figure \ref{Penrose} as a Penrose diagram. This type of diagram is a tool for depicting the causal structure of an infinite spacetime in a finite diagram by shrinking the manifold through conformal transformations. Because conformal transformations leave the null cones invariant, 
Penrose diagrams are characterized by 
null rays at 45 degrees. 

\section{Boundary 2-point functions and spectral function} \label{bdytwopf}

In a Lorentzian signature spacetime such as the Minkowski boundary of AdS(-Vaidya), different types of 2-point functions of field operators 
can be defined. For free field operators, vacuum expectation values of various products of them can be identified (see (\ref{Green})) with various `Green functions' of the wave equation, in the mathematical sense of the term: a solution of a differential equation with a delta function as source term which 
can be used to find a general solution of the inhomogeneous differential equation through a convolution with  
the general source 
(that is, when the Green function is time-independent and translation invariant).  
In that sense they are referred to as `response functions' as they determine response of a system to a source (response $=$ convolution of Green function and source).    
  
We give here the definitions of the Feynman, retarded and advanced 2-point functions for 
a bosonic operator $\mathcal O$: 
\begin{align}
 i G_F(x,x') &= \langle \mathcal T \mathcal O(x) \mathcal O(x') \rangle  \label{GFdef} \\ 
 i G_R(x,x') &= \theta(t-t') \langle [\mathcal O(x), \mathcal O(x')] \rangle \\ 
 i G_A(x,x') &= -\theta(t'-t) \langle [\mathcal O(x), \mathcal O(x')] \rangle  \label{GAdef} 
\end{align}
(with $x^0=t$). 
For a system at zero temperature, the $\langle \cdots \rangle$ denote expectation values in the pure vacuum state 
\begin{equation}
 \langle A \rangle = \langle 0|A|0 \rangle,  
\end{equation}
while for an equilibrium system at temperature $T$ and chemical potential $\mu$, described by a grand canonical ensemble of states, it denotes an ensemble average 
\begin{equation}
\langle A \rangle = \langle A \rangle_\beta = \sum_i \rho_i \langle\psi_i | A | \psi_i \rangle = \text{tr} (\rho A), 
\end{equation}
with $\rho_i = e^{-\beta(E_i - \mu n_i)}/Z = \langle \psi_i | \rho | \psi_i \rangle$ the probability for the system to be in 
a pure state $|\psi_i \rangle$  with 
ener\-gy eigenvalue $E_i$ and number eigenvalue $n_i$, with $\beta = 1/k_B T$, 
the grand partition function $Z = \sum_j e^{-\beta(E_j-\mu n_j)}$, 
and the quantum density operator $\rho = e^{-\beta(H - \mu N)}/Z$  satisfying tr$(\rho) = 1$. 

\begin{figure}[h!]
  \centering
  \scalebox{0.4}{
  \includegraphics{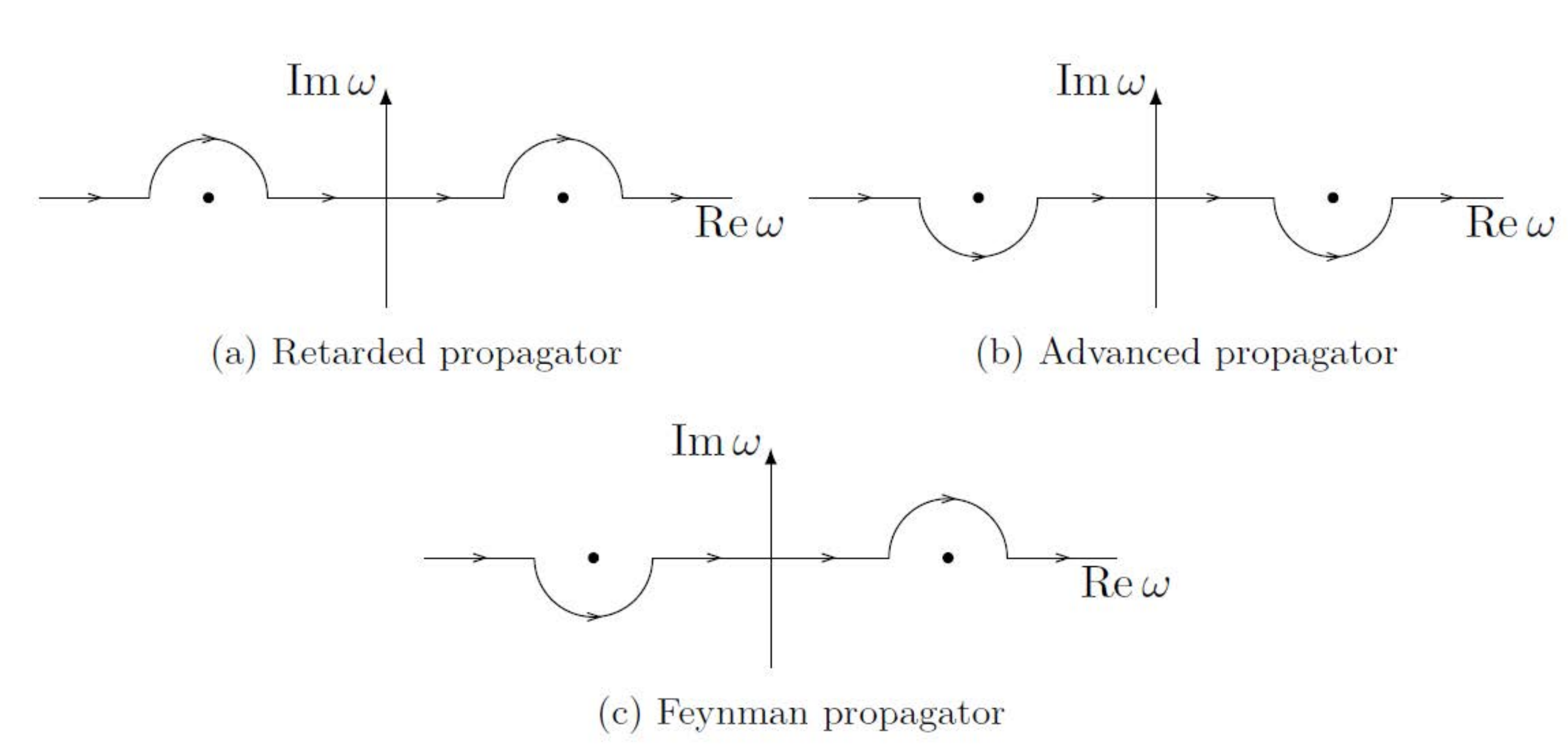}}
  \caption{ Different contours (assumed closed by an infinite semicircle) correspond to different causal propagators $G_{F,R,A}$. }
\label{contours}
\end{figure}

From $\partial_{t} \theta(t-t') = \delta(t-t')$ and if $\mathcal O(x)$ satisfies the scalar field equation $(\Box + m^2) \mathcal O = 0$ with d'Alembertian $\Box = \eta^{\mu\nu} \partial_\mu \partial_\nu$, we can indeed identify the above 2-point functions as Green functions, 
\begin{equation}
 (\Box + m^2) G_{F,R,A}(x,y) = - \delta^{d}(x-x'),  \label{Green} 
\end{equation}
describing the propagation of field disturbances subject to certain boundary conditions. Different boun\-da\-ry conditions correspond to different choices of contour around the poles at $\omega = k^0 = \pm (\vec k^2 + m^2)^{1/2}$ present in the integral representation of $\mathcal G = G_{F,R,A}$, 
\begin{equation}
 \mathcal G(x,x') = \int \frac{d^d k}{(2\pi)^d} \frac{e^{i k \cdot (x-x')}}{k^2-m^2}. \label{freeprop}
\end{equation}
The possible choices of contour in the complex $\omega$ plane, associated with different $i \epsilon$-prescriptions ($\omega^2 + i \epsilon$ for $G_F$, $\omega + i \epsilon$ for $G_R$ and $\omega - i \epsilon$ for $G_A$) 
are shown in figure \ref{contours}. 
From the contour for $G_F$ in the last picture, it can be seen that by rotating it  
$90^\circ$  counterclockwise the topological relation between the contour and the poles remains unchanged, but the integral runs along the imaginary axis from $-i \infty$ to $i \infty$ such that by changing the integration variable $\omega$ to $\omega_E = i \omega$ and the variables $t$ and $t'$ to $-i \tau$ and $-i \tau'$, one finds the relation 
\begin{equation}
G_F(t,\vec x; t',\vec x') = -i G_E(i t, \vec x; i t',\vec x') \label{usefulrel1}
\end{equation}
with 
\begin{equation}
 G_E(\tau,x;\tau',x') =  \int \frac{d^d k}{(2\pi)^d} \frac{e^{i \omega_E (\tau-\tau') + i \vec k \cdot (\vec x- \vec x') 
}}{\omega_E^2+\vec k^2  + m^2},  
\end{equation}
\begin{equation}
G_E(\tau,x;\tau',x') = \langle \mathcal T_E \mathcal O(\tau,x) \mathcal O(\tau',x') \rangle 
\end{equation}
the (unique) Euclidean Green function, satisfying (\ref{Green}) with $\Box$ the Laplacian. 
Another useful relation between the 2-point functions is 
\begin{equation}
G_R(x,x') = \theta(t-t') \left( G_F(x,x') + G_F^*(x,x') \right) \label{usefulrel2} 
\end{equation}
which can be easily obtained from writing out $G_F$ in (\ref{GFdef}) as $G_F(x,x') = \theta(t-t') \langle \mathcal O(x) \mathcal O(x') \rangle + \theta(t'-t) \langle \mathcal O(x') \mathcal O(x) \rangle$ from which it follows that $G_F + G_F^* = G_R + G_A$, and thus (\ref{usefulrel2}). 

In a translation invariant system (in time and space), the \emph{spectral function} -- describing the fluctuation spectrum -- can be defined in terms of the Fourier transformed $\tilde G_R(\vec k, \omega)$ as 
\begin{equation}
 \rho(\vec k, \omega) = -2 \text{ Im } \tilde G_R(\vec k, \omega). \label{eqspectral}
\end{equation}
From $G_R$ in (\ref{freeprop}) with $\omega \rightarrow \omega + i \epsilon$, the spectral function for free scalars is given by delta functions $\delta(\omega \pm  (\vec k^2 + m^2)^{1/2})$ (using $\frac{1}{x+i \epsilon} = \mathcal P_v \frac{1}{x}  - i \pi \delta(x)$ with principal value $\mathcal P_v\frac{1}{x} = \lim_{\epsilon \rightarrow 0+} \frac{x}{x^2 + \epsilon^2}$). The effect of interactions is typically a broadened peak at renormalized single-particle energies and, 
when the area under the peak has decreased, an additional term representing a continuum. In the case of fermions (for which the above discussion can be straightforwardly reformulated), there is a large class of interacting fermions termed `Fermi liquids' that displays this type of spectral function behaviour and hence can be qualitatively understood in terms of a picture of non-interacting fermions or quasi-particles, associated with the peaks. They are described by Landau's Fermi liquid theory (examples are the normal states  of most metals at sufficiently low temperatures). 
Understanding non-Fermi liquids (encountered for example in the normal state 
of high-$T_c$ superconductors and metals close to a quantum critical point) is a long-standing problem in condensed matter physics. In the modern renormalization group picture, the Landau Fermi liquids versus non-Fermi liquids form low-energy effective theories for interacting fermions corresponding to a trivial fixed point 
versus a non-trivial, critical fixed point \cite{Polchinski:1992ed,Shankar:1993pf}. 
In both cases, the presence of the
Fermi surface in momentum space controls the low-energy physics, effectively allowing only 
freedom in the normal direction to the surface 
due to the Pauli principle. This quasi one-dimensional behaviour in the  low-energy effective theory is reflected in macroscopic properties of the system such as the temperature dependence of the conductivity. 
Holographically, bottom-up models consisting of spinor bulk fields in a Reissner-Nordstr\"om-AdS black hole background (in the most basic set-up) have been constructed and are interpreted as (non-)Fermi liquids, which cannot be classified as Landau Fermi liquids \cite{Liu:2009dm,Cubrovic:2009ye,Faulkner:2009wj}.

\paragraph{Time-dependent spectral function} 

In \cite{Balasubramanian:2012tu}, 
with the goal of providing a measure for the degree of thermalization, the notion of a generalized time-dependent spectral function in a time-dependent system was introduced, via a Wigner instead of a Fourier transform. In such a system, 
the correlators depend explicitly on two instants of time: $G_R(\vec x_2,t_2;\vec x_1,t_1)$, or equivalently, on an average time $T$ and relative time $\Delta t$ defined by 
\begin{equation}
t_1 = T- \frac{\Delta t}{2}, \quad t_2 = T + \frac{\Delta t}{2}.
\end{equation}
 If there is still translation invariance in space, the space coordinates can be Fourier transformed and it is convenient to work in a mixed representation with $G_R(\vec k,t_2;t_1)$. The Wigner transform then consists of Fourier transforming relative time $\Delta t$, 
\begin{equation}
\tilde G_R(\vec k,T,\omega) = \int_{-\infty}^\infty d(\Delta t) \, e^{i \omega \Delta t} G_R(k,T+ \frac{\Delta t}{2}; T - \frac{\Delta t}{2}),  \end{equation}
and the $T$-dependent spectral function can be defined as 
\begin{equation}
 \rho(\vec k, T, \omega) = -2 \text{ Im } \tilde G_R(\vec k, T, \omega). \label{spectralf}
\end{equation}
It is this object that we will  calculate holographically for certain systems. 
In AdS$_3$-Vaidya, the correlator $G_R(\vec k,t_2;t_1)$ 
is thermal for $t_2 > t_1 > 0$ (determined entirely in a black hole background), thermalizing for $t_2 > 0 > t_1$ (background interpolating between AdS and BTZ), and reduces to the vacuum expression for $t_1 < t_2 < 0$ (within AdS).

\subsection{How to obtain $G_R$ holographically}

According to the Euclidean AdS/CFT dictionary discussed in section \ref{adsEucdictionary}, studying a scalar operator $\mathcal O(x)$ (with scaling dimension $\Delta_+$) in the boundary field theory corresponds to studying a field $\phi(x,z)$ (with scaling dimension $\Delta_-$) in the bulk that approaches $\phi_0(x)$ at the boundary $z \rightarrow 0$. In the boundary theory $\phi_0(x)$ serves as a source for the operator, and a term $\int d^d x \, \phi_0 \mathcal O$ is added to the action. 

Consider a massive scalar field in a Euclidean AdS$_{d+1}$ background  
\begin{equation}
ds^2 = \frac{R^2}{z^2} \left( \sum_{i=1}^d dx_i^2 + dz^2 \right) \label{EuclAdS}
\end{equation} 
with quadratic bulk action 
\begin{equation}
 S(\phi) = \frac{1}{2} \int d^{d+1} x \sqrt{|\det g_{ij}|} \left( (\partial_i \phi)^2  + m^2 \phi^2 \right).   
\end{equation} 
The form of $\phi$ satisfying the asymptotic equations of motion is 
\begin{equation} 
\phi(z,x) = z^{\Delta_-} \phi_0(x) \left( 1 + \cdots \right)  + z^{\Delta_+} \tilde \phi(x) \left(1 + \cdots \right)  
\qquad (z \rightarrow 0)   \label{1.2.6}
\end{equation} 
with $\phi_0$ and $\tilde \phi$ the two independent solutions of the second order (in $z$) differential equation (before imposing extra e.g. regularity conditions on the solution in the bulk), and  the ellipsis  
representing higher order terms in $z$ per independent solution. 
There is a unique solution to the equations of motion that is regular in the bulk and approaches $\phi_0(x)$ as $\phi(z,x) = z^{\Delta_-} \phi_0(x)$ near the boundary \cite{Balasubramanian:1998de,Witten:1998qj}: 
\begin{align}
\phi(z,x) &= \int d^d x' K(z,x;x') \phi_0(x')  
 \\ 
&= c \int d^d x' \frac{z^{\Delta_+}}{(z^2 + |x-x'|^2)^{\Delta_+}} \phi_0(x') \label{1.2.5} 
\end{align}
with $K(z,x;x')$ the `bulk-to-boundary propagator' 
satisfying 
\begin{equation}
\lim_{z\rightarrow 0} z^{\Delta_+ - d} K(z,x;x') = \delta^{(d)}(x-x')
\end{equation}
 and 
\begin{equation}
 \Delta_{\pm} = \frac{d}{2} \pm \sqrt{\left(\frac{d}{2}\right)^2 + m^2} = \frac{d}{2} \pm  \nu, \quad \nu = \sqrt{\left(\frac{d}{2}\right)^2 + m^2}, \quad \Delta_+ + \Delta_- = d. 
\end{equation}  
The uniqueness of the solution corresponds to the uniqueness of $G_E$ in the boundary field theory.  
In this simple Euclidean set-up, we can sketch the connection between the expectation value $\langle \mathcal O \rangle_{\phi_0}$ acquired by $\mathcal O$ in the presence of the source term $\int \phi_0 \mathcal O$, the vacuum 2-point function $\langle \mathcal O \mathcal O \rangle$ and the subleading component $\tilde \phi$ of the bulk field:  
\begin{align}
 \langle \mathcal O(x) \rangle_{\phi_0} &= \langle \mathcal O(x) e^{\int \phi_0 \mathcal O} \rangle \approx \int d^d x' \phi_0(x') \langle \mathcal O(x) \mathcal O(x') \rangle \label{firstline} \\ 
& \sim \int d^d x' \frac{\phi_0(x')}{|x-x'|^{2 \Delta_+}} \label{secline}\\  
&\sim \tilde \phi(x),  \label{phitilde}
\end{align} 
suggesting a relation between the 2-point function and the subleading component $\tilde \phi$ (which will be more clear in Fourier space). 
The first two lines reside in the CFT, where in the first line the assumption was made that $\langle \mathcal O(x) \rangle = 0$ and the approximation is to linear order in the source $\phi_0$ (again, the 2-point function is a response function in linear response theory). Scale invariance imposes the form of the 2-point function 
in the second line. Finally, the proportionality with $\tilde \phi$ follows from the identification of $\tilde \phi$ in (\ref{1.2.6})  
after the expansion in $z$ of (\ref{1.2.5}). 
The expression in (\ref{secline}) is also the one obtained by using the AdS/CFT prescription\footnote{
We refer to \cite{Balasubramanian:1998de,Witten:1998qj,Freedman:1998tz}, and \cite{Bianchi:2001kw} for the  derivation including careful holographic renormalization performed later.   
}
 $\langle \mathcal O(x) \rangle_{\phi_0} = \delta S^{on-shell}/\delta \phi_0(x)$. 
Alternatively this can be written as \cite{Iqbal:2008by} 
\begin{equation}
 \langle \mathcal O(x) \rangle_{\phi_0} = \lim_{z\rightarrow 0} z^{\Delta_-} \Pi(z,x) \label{canonic}
\end{equation}
with $\Pi(z,x) = \partial \mathcal L/\partial \partial_z \phi$ the canonical momentum conjugate to the field $\phi$ with respect to a foliation in the $z$-direction, and 
$z^{\Delta_-} 
\Pi(z,x)|_{bdy} =\left. z^{\Delta_-} (\partial \mathcal L/\partial \partial_z \phi 
\right)|_{bdy} = \delta S^{on-shell}/\delta \phi_0(x)$ 
as is clear from the point-particle analogy $p|_{bdy} = \left. 
(\partial \mathcal L/\partial \dot q) \right|_{bdy} = 
\delta S^{on-shell}/\delta q_{bdy}$ with $\phi$ as $q$ 
and $\langle \mathcal O \rangle$ as $p$.

In a time- and space translation invariant system, the linear response relation (\ref{firstline}) in Fourier space for a \textit{Lorentzian} field theory turns out to involve the \textit{retarded} 2-point function, if the derivation is done carefully \cite{Iqbal:2008by,Kapusta:2006pm}: 
\begin{equation}
 \langle \mathcal O(k) \rangle_{\phi_0} = - \tilde G_R(k) \phi_0(k). \label{fourierO}
 \end{equation} 
The non-uniqueness of correlator choice in the boundary field theory corresponds to non-uniqueness of the bulk solution: in the Lorentzian case, the requirement of regularity of the bulk field is not enough to obtain a unique solution, there are extra normalizable modes $\phi_n$ added to the solution (\ref{1.2.5}), and extra boundary conditions need to be imposed at the horizon of the background. A Lorentzian AdS/CFT prescription is required. 
From comparison with (the Fourier transform of) (\ref{canonic}), now evaluated in Lorentzian signature, the following holographic prescription for the calculation of $\tilde G_R(k)$ was suggested in \cite{Iqbal:2008by}: 
\begin{equation}
 \tilde G_R(k) = - \lim_{z\rightarrow 0} z^{\Delta_-} \frac{\Pi(z,k)}{\phi(z,k)} 
\end{equation}
 with $\phi(z, k)$ satisfying in-falling boundary conditions at the black hole horizon. This prescription gives the correct CFT result and is equivalent (at linear level) to the prescription for retarded 2-point functions in \cite{Son:2002sd}. 

In this subsection 
we have used $x$ and $k$ as 4-vector notation for $x^\mu$ and $k^\mu$, where from the context it should have been clear whether to interpret it in Euclidean ($x^0 = \tau$) or Lorentzian ($x^0 = t$) signature. From here on, we will switch to a notation where the time-component is written out, to make transitions between different coordinate systems more clear.

\paragraph{Holographic $G_R$ in time-dependent system}

Being interested in time-dependent systems, we cannot directly use the Lorentzian prescription in Fourier space. 
However, a similar relation as (\ref{fourierO}) can be obtained in coordinate space if  
$\phi_0$ is a delta function source: 
\begin{equation}
 \langle \mathcal O(\vec x,t) \rangle_{\delta} = -G_R(\vec x,t;\vec  x_0,t_0)   
\end{equation}
gives the response of an operator at $(\vec x,t)$ in the presence of a delta function source $\phi_0(\vec x') = \delta(\vec x' -\vec x_0) \delta(t'-t_0)$ 
at $(\vec x_0,t_0)$ (see for example appendix F of \cite{Balasubramanian:2012tu} 
for a derivation).  
On the other hand the dictionary gives 
\begin{equation}
 \langle \mathcal O(\vec x,t) \rangle_{\delta} =  \frac{\delta S^{on-shell}}{\delta \phi_0(\vec x,t)} = - 2 \nu \tilde \phi(\vec x,t) 
\end{equation}
with $\phi_0$ and $\tilde \phi$ respectively the leading and subleading asymptotic behaviour of the bulk scalar field $\phi$ satisfying in-falling boundary conditions at the background horizon 
\begin{equation}
 \phi(z,\vec x,t) = z^{\Delta_-} \phi_0(\vec x,t) + z^{\Delta_+} \tilde \phi(\vec x,t) + \cdots \qquad (z \rightarrow 0), 
\end{equation}
and  where the factor $2\nu$ arises from holographic renormalization, which involves adding an appropriate counterterm $S_{ct}$ to the scalar bulk action. 
   We obtain 
\begin{equation}
 G_R(\vec x,t; \vec x_0,t_0) = 2 \nu \tilde \phi(\vec x,t) = 2 \nu \lim_{z \rightarrow 0} z^{-\Delta_+} \phi(z,\vec x,t). \label{GRfromphi}
\end{equation}
Instead of specifying the boundary condition at the Poincar\'e or event horizon and integrating towards the boundary, the strategy used in \cite{Balasubramanian:2012tu} 
is to construct the bulk field  using the retarded boundary-to-bulk propagator (evolving initial data from the boundary into the bulk -- opposite to the bulk-to-boundary propagator):   
\begin{equation}
 \phi(z,\vec x,t) = \int d \vec x' dt' \,  K_R(z,\vec x,t;\vec x',t') \phi_0(\vec x',t') = K_R(z,\vec x,t;\vec x_0,t_0).  \label{bulkKR}
\end{equation}

\section{Spectral functions in AdS$_3$-Vaidya: numerics and results} \label{vaidyaresults}

The goal is to obtain time-dependent spectral functions (\ref{spectralf}) for field theory operators coupled to scalar and spinor bulk fields in AdS$_3$-Vaidya. This requires solving the equations of motion for the bulk field and extracting the subleading component in the asymptotic expansion as a first step, in order to obtain $G_R$ via (\ref{GRfromphi}), which subsequently needs to be Wigner  transformed. 
The bulk solution (\ref{bulkKR}) for scalar and spinor fields is known analytically in the AdS region $t_0<t<0$ of the background (for all $d$) and in the black hole region ($t>t_0>0$) for $d=2$, but remains to be determined numerically in the `thermalizing region' $t>0>t_0$. 

\subsection{Scalar bulk field}

\begin{figure}[h!]
  \hfill
  \begin{minipage}[t]{.45\textwidth}
    \begin{center}
      \scalebox{0.75}{
  \includegraphics{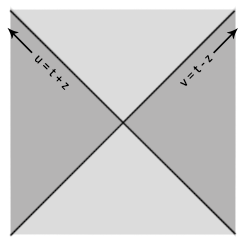}}
    \end{center}
  \end{minipage}
  \hfill
  \begin{minipage}[t]{.45\textwidth}
    \begin{center}
      \scalebox{0.75}{
  \includegraphics{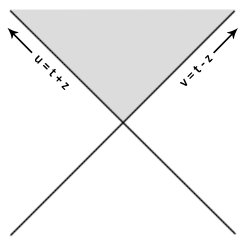}}
    \end{center}
  \end{minipage}
      \caption{Causal behaviour of the bulk-to-boundary propagators in a Penrose diagram, 
with lightcone coordinates $u=t+z$ and $v=t-z$ at an angle of 45 degrees. The Feynman propagator $K_F \sim \theta\left(\Delta t^{2}-z^{2}\right) (\cdots) + \theta\left(z^2-\Delta t^{2}\right) (\cdots)$ is real inside the lightcone (light grey) and imaginary outside the lightcone (grey).  The retarded propagator $K_R \sim \theta(\Delta t) \theta\left(\Delta t^{2}-z^{2}\right)$ is non-zero in the forward lightcone (light grey). }
	\label{causalsketch}
  \hfill
\end{figure}

The scalar equation of motion $(\Box-m^2)\phi = 0$ in the AdS$_{d+1}$-Vaidya background (\ref{Vaidyametric}) with $f(v,z) = 1- \theta(v) R^d z^d$ (and $R=1$) is given in Eddington-Finkelstein coordinates by 
\begin{equation}
\sum_{i=1}^{d-1} \partial^2_{x_i} \phi + f \partial_z^2 \phi - 2 \partial_v \partial_z \phi + \left( \partial_z f - (d-1) \frac{f}{z} \right) \partial_z \phi + (d-1) \frac{\partial_v \phi}{z} - \frac{m^2}{z^2} \phi = 0. \label{eomgend} 
\end{equation}
For negative $v$ the Vaidya background is just the metric of AdS$_{d+1}$. 
The 
Euclidean boundary-to-bulk propagator $K_E(z,\vec x,\tau;\vec x_0,\tau_0)= K_E(z,\vec x-\vec x_0,\tau-\tau_0) = K_E(z,\Delta \vec x,\Delta \tau)$ (in Euclidean Poincar\'e coordinates (\ref{EuclAdS}))  
associated with a boundary delta source $\phi_0 = \delta(\vec x'-\vec x_0) \delta(\tau'-\tau_0)$ is well-known: 
\begin{equation}
K_E(z,\Delta \vec{x},\Delta \tau)=C_{\Delta}\left(\frac{z}{z^{2}+\Delta\tau^{2}+\Delta\vec{x}^{2}}\right)^{\Delta_+} \label{KEscalar}
\end{equation}
with constant $C_{\Delta}$ in terms of Gamma functions 
$C_{\Delta}=
\Gamma\left(\Delta_+\right)/\{\pi^{d/2}\Gamma\left(\Delta_+-\frac{d}{2}\right)\}$.  
A Fourier transform $\Delta \vec x \rightarrow \vec k$ to a mixed representation expression $K_E(z,\Delta \vec k, \Delta \tau)$ is carried out first (but not in Euclidean time as we will want to continue the solution across the shell). 
We can subsequently use the relations (\ref{usefulrel1}) and (\ref{usefulrel2}) to Wick rotate to the Feynman propagator 
\begin{equation}
 K_F(z,\vec k, \Delta t) = -i K_E(z, \Delta \vec k, i \Delta t)  \label{wickrot}
\end{equation}
and obtain the requested 
retarded one 
\begin{equation}
 K_R(z, \vec k, \Delta t) = \theta(\Delta t) \left( K_F(z, \vec k, \Delta t) + K_F^*(z, -\vec k, \Delta t) \right). \label{KRspace}
\end{equation} 
In figure \ref{causalsketch} a sketch of the causal behaviour of these propagators is shown. 
The result for $K_R(z,\vec k, \Delta t)$ for non-integer $\Delta_+$ is\footnote{
To avoid any problems related to extra contact terms we will only consider non-integer values of $\Delta_+$. 
At the end of the calculations however, the spectral function result for integer $\Delta_+$ can be obtained from the result for a non-integer value $(\Delta_+ + \delta)$ by taking the limit $\delta \rightarrow 0$.  
} 
\begin{align}
K_{R}(z,\vec{k},\Delta t) 
&=-2\pi^{\frac{d+1}{2}}\frac{C_{\Delta}}{\Gamma(\Delta_+)}z^{\Delta_+}\theta(\Delta t)\theta\left(\Delta t^{2}-z^{2}\right)\left(\frac{2\sqrt{\Delta t^{2}-z^{2}}}{|\vec{k}|}\right)^{\frac{d-1}{2}-\Delta_+}J_{\frac{d-1}{2}-\Delta_+}\left(|\vec{k}|\sqrt{\Delta t^{2}-z^{2}}\right)  \label{KRscal}
\end{align} 
with spherical Bessel function $J$. 
This expression as well as the expression for the boundary correlator $G_R(\vec k, t; t_0)$ obtained from it, do not form well-defined distributions. They are to be supplemented with contact terms with diverging coefficients to regulate their singular behaviour at $\Delta t^2 ( - z^2) \rightarrow 0$.  In coordinate space these contact terms correspond to (derivatives of) delta functions of $\Delta t^2 ( - z^2)$. 
They can therefore be omitted from the propagator expressions, which only have a physical meaning at $\Delta t^2-z^2 > 0$ (in the bulk) or  $t>t_0$ (at the boundary). 
However, under the integral over $\Delta t$ to obtain the Fourier transformed $\tilde K_R(z,\vec k,\omega)$ or $\tilde G_R(\vec k,\omega)$, they are vital to render the integrals well-defined. 
It can be shown that the necessary contact terms are real and hence will not influence the physical spectral function (\ref{eqspectral}) (see appendix C of \cite{Balasubramanian:2012tu}).   
An alternative for calculating the contact terms explicitly in terms of a regulator is to perform the Fourier transform to $\omega$ by making use of analytic continuation in $\Delta_+$ of the result 
obtained through identities that are valid for limited ranges of $\Delta_+$. 
In this way, an analytic expression for the vacuum spectral function is derived, and similarly for the thermal one (from the analogous analysis in BTZ instead of AdS). (The expression for $\tilde G_R(\vec k,\omega)$ obtained through analytic continuation still has poles at integer $\Delta_+$, but we exclude these values from our analysis.)    

\begin{figure}[h!]
  \centering
  \scalebox{0.3}{
  \includegraphics{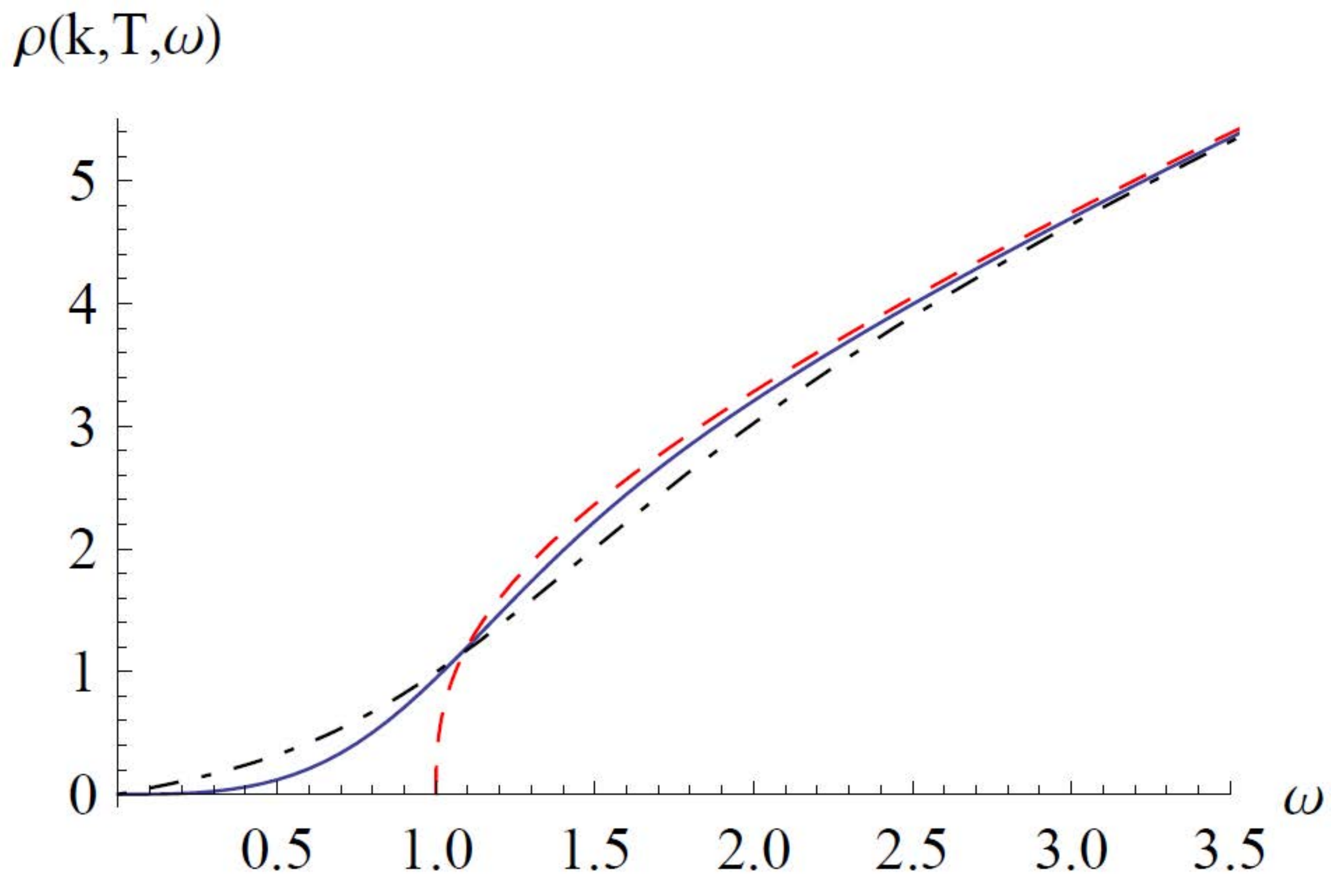}}
  \caption{Spectral function $\rho(\omega)$ at average time $T=0$ for a scalar operator of scaling dimension $\Delta_+ = 11/8$ and momentum $k=1$, dual to a scalar bulk field in AdS$_3$-Vaidya. It is compared to  
 the analytic vacuum (red dashed) and thermal (blue dashed) spectral functions.}
\label{figartikel}
\end{figure}

Going back to Eddington-Finkelstein coordinates through $t = v + z$, the retarded bulk solution 
\begin{equation}
K_R^{\mathcal E \mathcal F}(z,\vec k,v;t_0) =  K_R(z,\vec k, v + z - t_0) 
\end{equation}
is known in the whole negative $v$ region of the metric. 
The value $K_R^{\mathcal E \mathcal F}(z,\vec k,v=0;t_0)$ at the shell is used as initial condition for the numeric integration of the equation of motion (\ref{eomgend}) in the positive $v$ region (where the metric is BTZ for $d=2$), imposing continuity of the field $\phi$ at $v=0$. The scalar bulk field 
is also required to vanish at the boundary $z=0$ for all $v>0$ (since $\phi_0 \sim \delta(t-t_0)$ with $t_0 < 0$). 
From the numeric solution for all $v$, the boundary 2-point function is extracted as 
\begin{equation}
G_R(\vec k,t;t_0) = 2 \nu \, \lim_{z \rightarrow 0} z^{-\Delta_+} K_R^{\mathcal E \mathcal F}(z,\vec k,v \rightarrow t;t_0)
\end{equation}
and the time-dependent spectral function as 
\begin{equation}
 \rho(\vec k, T, \omega) = -2 \text{ Im }  \int_{-\infty}^\infty d(\Delta t) \, e^{i \omega \Delta t} G_R(\vec k,T+ \frac{\Delta t}{2}; T - \frac{\Delta t}{2})
\end{equation}
with average time $T=\frac{t_0 + t}{2}$ and relative time $\Delta t = t- t_0$.   
Following the outlined procedure we were able to reproduce the numeric result for $\rho(k, T, \omega)$ for a scalar field in AdS$_3$-Vaidya, first derived in \cite{Balasubramanian:2012tu}. In figure \ref{figartikel} the thermalizing spectral function  $\rho(k, 0, \omega)$ is shown to interpolate between the analytic thermal $\rho(k,\infty,\omega)$ and vacuum one $\rho(k,-\infty,\omega)$, as expected. 
Details on the numerical method will be explained in the following section for the case of fermion bulk fields. In the simpler case of scalars, the built-in Mathematica function NDSolve is in fact sufficient for solving the differential equation.  

\FloatBarrier

\subsection{Spinor bulk field} \label{subs932}

We consider a spinor bulk field $\psi$ instead of a scalar $\phi$, 
references on the relevant AdS/CFT dictionary including \cite{Iqbal:2009fd,Henningson:1998cd,Mueck:1998iz,Henneaux:1998ch
}. 
The addition $\int \phi_0 \mathcal O$ to the boundary action is to be   
replaced by $\int (\bar \chi_0  \mathcal O + \bar{\mathcal O} \chi_0)$ where $\mathcal O$ is now a fermionic operator.

There is an important consequence to the fact that spinor source $\chi_0$ and spinor bulk field $\psi$ exist in different spacetime dimensions, as the dimension determines the number of Dirac spinor components:  
we will see that the number of components of $\chi_0$ (and of $\mathcal O$) is half that of $\psi$ (in all dimensions $d$) \cite{Iqbal:2009fd}. Accordingly,  boundary conditions should only be applied to one half of the bulk spinor. 

The bulk action for a spinor field $\psi$ with mass $m$ in a general curved background $ds^2 = g_{MN} dx^M dx^N$  is 
\begin{equation}
S(\psi)  = \int d^{d+1}x \sqrt{|\det{g_{MN}}|} \left( \bar \psi \Gamma^M D_M \psi - m \bar \psi \psi \right)  
\end{equation}
with 
\begin{equation}
 \slashed D = \Gamma^M D_M = \Gamma^M \left( \partial_M + \frac{1}{2} \omega_{M ab} S^{ab} \right) 
\end{equation}
and $\bar \psi = \psi^\dagger \Gamma^t$. 
Latin indices $a,b$ refer to Minkowski indices of the local tangent frame defined through the vielbein $e_M^a$, sometimes referred to as the ``square root of the metric" because it obeys $g_{MN}(x) = e_M^a(x) e_N^b(x) \eta_{ab}$. It relates the spacetime-dependent bulk gamma matrices $\Gamma^M(x)$ with $\{ \Gamma^M,\Gamma^N\} = 2 g^{MN} \mathbf 1$ to the constant bulk gamma matrices $\Gamma^a$ with $\{ \Gamma^a,\Gamma^b\} = 2 \eta^{ab}\mathbf 1$ through $\Gamma^M = e^M_a \Gamma^a$. When using gamma matrices with specific indices (such as $\Gamma^t$, $\Gamma^z$, $\Gamma^\mu$, ...) they will always refer to the $\Gamma^a$ ones in the tangent frame.  
Boundary gamma matrices are denoted by $\gamma^\mu$. 
For the Dirac equation to be invariant under a Lorentz transformation for the spinor field, with antisymmetric generators  $S^{ab} \sim [\Gamma^a,\Gamma^b]$, the derivative has to be covariant with `spin connection' $\omega_{M ab} = e_a^N \nabla_M e_{bN}$, where $\nabla_M$ itself is a covariant derivative containing the affine connection to parallel transport four-vectors in curved spacetime.  

The spinor field can 
be decomposed in canonically conjugate components $\psi_\pm$ by introducing a projector  $\mathcal P_\pm$ in terms of $\Gamma^z$ ($z$ the radial direction): 
\begin{equation}
 \psi = \psi_+ + \psi_-, \quad \psi_\pm = \mathcal P_\pm \psi, \quad \mathcal P_\pm = \frac{1}{2}( \mathbf 1 \pm \Gamma^z), 
\end{equation} 
with corresponding Euclidean canonical momenta 
\begin{equation}
\Pi_\pm = \partial \mathcal L/\partial(\partial_z \psi_\pm) \sim \bar \psi_\mp. \label{canonicmompsi}
\end{equation} 
For even boundary theory dimension $d$, a convenient choice of bulk gamma matrices is $\Gamma^\mu = \gamma^\mu$ and $\Gamma^z = \gamma^{d+1}$, with $\gamma^{d+1}$ the analogue of $\gamma_5$ in $d=4$. The $\psi_\pm$ are then recognized as Weyl spinors of  opposite  
chirality from the $d$-dimensional point of view:  a Dirac spinor field $\psi$ in the bulk is mapped to a chiral spinor operator $\mathcal O$ on the boundary. 
Similarly it can be shown that   
for odd $d$ the boundary operator $\mathcal O$ is a Dirac spinor with half the number of components of its dual bulk  Dirac spinor $\psi$ (for more details see \cite{Iqbal:2009fd}).  
To determine on which of the components to impose the boundary value $\chi_0$, we have to examine the asymptotic behaviour of the solutions.  Hereby we assume that $\Gamma^z$ is chosen such that $\psi_+$ corresponds to the `top half' of $\psi$. Solving the Dirac equation of motion $(\slashed D - m)\psi = 0$ in an asymptotically (Euclidean) AdS background near the boundary leads to the following form for the asymptotic spinor bulk fields (in analogy with (\ref{1.2.6})): 
\begin{equation}
  \psi(z,x) =  \left( \begin{array}{cc} \psi_+(z,x) \\ \psi_-(z,x) \end{array} \right) = \left( \begin{array}{cccc}  z^{\Delta_- + 1} C(x) & + & z^{\Delta_+} \tilde \psi(x) &+ \cdots  \\ z^{\Delta_-} \chi_0(x) & + & z^{\Delta_+ + 1} B(x) &+ \cdots \end{array} \right) \quad (z \rightarrow 0), \label{psiassol}
\end{equation}
with 
\begin{equation}
 \Delta_\pm = \frac{d}{2} \pm m, \quad \Delta_+ + \Delta_- = d. 
\end{equation}
We discuss the various $x$-dependent fields defined in (\ref{psiassol}). 
As we take $m \rightarrow - m$, the roles of $\psi_\pm$ are simply interchanged, so we can restrict the discussion to non-negative values of $m$.   
The term in $z^{\Delta_-}$ is the overall leading term so the boundary condition is imposed on $\psi_-$ as 
\begin{equation} \label{eq9315}
\lim_{z \rightarrow 0} z^{-\Delta_-} \psi_-= \chi_0  
\end{equation}
with $\chi_0$ the spinor source coupling to an operator $\bar{\mathcal O}$ of dimension $\Delta_+$.   
The prescription for the expectation value of the corresponding conjugate operator $\mathcal O$, in analogy with (\ref{canonic}), is given in  \cite{Iqbal:2009fd} by (see in particular appendix C of \cite{Iqbal:2009fd} or \cite{Henneaux:1998ch}): 
\begin{equation}
\langle {\mathcal O}(x) \rangle_{\chi_0} = \lim_{z\rightarrow 0} z^{\Delta_-} \bar \Pi_-(z,x).  
\end{equation} 
From (\ref{canonicmompsi}) and (\ref{psiassol}), it follows that $\tilde \psi$ is related to the expectation value (in analogy with $\tilde \phi$ in the scalar case (\ref{phitilde})):  
\begin{equation}
\langle {\mathcal O}(x) \rangle_{\chi_0} \sim \tilde \psi. 
\end{equation} 
It is remarked in  \cite{Iqbal:2009fd} that it appears reasonable not to include the term proportional to $C(x)$ in the extraction $\lim_{z\rightarrow 0} z^{\Delta_-} \bar \Pi_-(z,x)$.  The reason is that $C$ is locally related to the source $\chi_0$ by the Dirac equation. This means the term in $C(x)$ could be considered analogous to the terms in the first ellipsis of (\ref{1.2.6}). Similarly, the field $B(x)$ in (\ref{psiassol}) is locally related to $\tilde \psi(x)$, and would belong to the analogue of the second ellipsis in (\ref{1.2.6}).

If we denote $\psi_0 = {0 \choose \chi_0}$ containing the source, we can write down the spinor equivalent of  (\ref{GRfromphi}). For the time-dependent system (all $x$ in (\ref{psiassol}) become ($\vec x,t)$) with a delta function source $\chi_0 = \delta(\vec x-\vec x_0)\delta(t-t_0)$, 
the 
Lorentzian prescription\footnote{
The Lorentzian prescription in Fourier space is given in \cite{Iqbal:2009fd}. 
} directly gives the retarded 2-point function of $\mathcal O$: 
\begin{equation}
 G_R(\vec x,t; \vec x_0,t_0) = 
\tilde \psi(\vec x,t) = 
\lim_{z \rightarrow 0} z^{-\Delta_+} \psi_+(z,\vec x,t) \label{GRfrompsi} 
\end{equation}
with $\psi$ the spinor bulk solution with in-falling boundary conditions at the horizon 
\begin{equation}
\psi(z,\vec x,t) =  \int d \vec x' dt' \,  K_R(z,\vec x,t;\vec x',t') \cdot \psi_0(\vec x',t') = K_R(z,\vec x,t;\vec x_0,t_0) \cdot {0\choose 1}, \quad \psi_+(z,\vec x,t) = \mathcal P_+ \psi(z,\vec x,t),  \label{bulkKRspinor} 
\end{equation}
or, in mixed representation: 
\begin{align} 
G_R(\vec k,t; t_0) &=
\tilde \psi(\vec k,t) = 
\lim_{z \rightarrow 0} z^{-\Delta_+} \psi_+(z,\vec k,t) \label{GRfrompsik} \\ 
 \psi(z,\vec k,t) &=  \int  dt' \,  K_R(z,\vec k,t;t') \cdot \psi_0(t') = K_R(z,\vec k,t;t_0)\cdot {0\choose 1}, \quad \psi_+(z,\vec k,t) = \mathcal P_+ \psi(z,\vec k,t).  \label{bulkKRspinorbis} 
\end{align}
Note that in (\ref{GRfrompsi}) we did not include the analogue of the factor $2 \nu$ in (\ref{GRfromphi}). This stems from the absence of subtleties regarding the holographic renormalization in the spinor case that were present in the scalar case \cite{Henneaux:1998ch}. 
The fermionic versions of the boundary correlator definitions are given by (\ref{GFdef})-(\ref{GAdef}), 
but with the second $\mathcal O$ in the expressions replaced by its complex conjugate $\bar{\mathcal O} = \mathcal O^\dagger \gamma^0$, and all commutators replaced by anti-commutators. 

$K_R$ is determined starting from the analogue of (\ref{KEscalar}) for the Euclidean spinor bulk-to-boundary propagator $K_E(z,\vec x,\tau;\vec x_0,\tau_0)= K_E(z,\vec x-\vec x_0,\tau-\tau_0) = K_E(z,\Delta \vec x,\Delta \tau)$ in Poincar\'e EAdS$_{d+1}$: 
\begin{equation}
K_E(z,\Delta \vec{x},\Delta \tau)=-K_{\Delta}\left(\frac{z-\vec x \cdot \vec \Gamma}{\sqrt{z^{2}+\Delta\tau^{2}+\Delta\vec{x}^{2}}}\right) \left( \frac{z}{z^{2}+\Delta\tau^{2}+\Delta\vec{x}^{2}}\right)^{\Delta_+} \mathcal P_-.   \label{KEspinor}
\end{equation}
with constant 
$K_{\Delta}=
C_{\Delta + 1/2 }$. 
This is a solution of the Dirac equation $(\slashed D - m)\psi = 0$ in the vielbein basis (with $x^0 = \tau$ Euclidean time) 
\begin{equation}
 e^i = \frac{dx^i}{z}, \quad e^z = \frac{dz}{z}
\end{equation} 
($e^a = e^a_M dx^M$ with frame field choice $e_M^a = \frac{1}{z} \delta_M^a$ to fix local Lorentz freedom),   
and corresponding spin connections $\omega_M^{ab} = \omega_i^{0j}=-\omega_i^{j0} = \frac{1}{z} \delta_j^i$ (all others zero). 
It is subsequently Fourier transformed in space to $K_E(z,\vec k, \Delta \tau)$, Wick rotated to $K_F(z,\vec k, \Delta t) = - i K_E(z,\vec k, i \Delta t)$ (in the process of which the gamma matrix $\Gamma^t = - i \Gamma^\tau$ is introduced), 
and recombined into the retarded solution through 
\begin{equation}
 K_R(z,\vec k, \Delta t) = \theta(\Delta t) \left( K_F(z,\vec k, \Delta t) +  C K_F^*(z, -\vec k, \Delta t)  C^\dagger  \right),  \label{KRrulespinor} 
\end{equation}
where  $C $ is the charge conjugation operator in the spinor representation. It satisfies $C^\dagger C = \mathbf 1$ 
and $(\Gamma^\mu)^* = \epsilon C^\dagger \Gamma^\mu C$ for any representation of the gamma matrices $\Gamma^\mu$ (in Minkowski signature), with 
$\epsilon=1$ in dimension 
2,3,4, mod 8 
such that the charged conjugated spinor $\psi^c = C \psi^*$ is also a solution of  the Dirac equation if $\psi$ is. Assuming furthermore that the source $\psi_0$ is a Majorana spinor, $\psi_0^c = \psi_0$, which can be done in dimension 2,3,4, mod 8, the conjugated spinor can be written as $\psi^c  = C \int K_F^* \cdot \psi_0^* = \int C K_F^* C^\dagger \cdot \psi_0$ and the linear combination $\theta(\Delta t) (\psi + \psi^c)$  is identified as the retarded solution (\ref{bulkKRspinorbis}), with $K_R$ from (\ref{KRrulespinor}) showing the correct 
causal behaviour within the lightcone $\sim \theta(\Delta t)\theta\left(\Delta t^{2}-z^{2}\right)$ (as in figure \ref{causalsketch}). 
The result for the AdS$_{d+1}$ retarded bulk spinor for non-integer $\Delta_+$ is given by 
\begin{align} 
K_{R}(z,\vec{k},\Delta t) 
&=-C_{\Delta+\frac{1}{2}}\frac{2\pi^{\frac{d+1}{2}}}{\Gamma(\Delta_+ +\frac{1}{2})}z^{\Delta_+}\theta(\Delta t)\theta\left(\Delta t^{2}-z^{2}\right)\sqrt{\Delta t^{2}-z^{2}}\left(\frac{2\sqrt{\Delta t^{2}-z^{2}}}{|\vec{k}|}\right)^{\frac{d}{2}-\Delta_+-1} \nonumber \\
&\left(\frac{(z\mathbf{1}+t\Gamma^{t})}{\sqrt{\Delta t^{2}-z^{2}}}J_{\frac{d}{2}-\Delta_+-1}\left(|\vec{k}|\sqrt{\Delta t^{2}-z^{2}}\right)+\frac{(i\vec{k}\cdot\vec{\Gamma})}{|\vec{k}|}J_{\frac{d}{2}-\Delta_+}\left(|\vec{k}|\sqrt{\Delta t^{2}-z^{2}}\right)\right)\mathcal{P}_{-}. 
\end{align} 
Finally, the retarded solution in Eddington-Finkelstein coordinates $K_R^{\mathcal E \mathcal F}(z,\vec k,v;t_0) =  K_R(z,\vec k, v + z - t_0)$ (in the frame $e^{\vec x} = \frac{d \vec x}{z}$, $ e^z = \frac{dz}{z}$, $e^v = \frac{dv + dz}{z}$)  
can be used as input for the numeric integration of the Dirac equation in the BTZ region ($v>0$) of the AdS$_3$-Vaidya metric.

\subsubsection{Pseudospectral method}

The pseudospectral method is a numerical technique for solving 
differential equations \cite{Boydbook}. Roughly\footnote{We follow here \cite{Boydbook}, who warns for different uses of the terminology in the literature.}, the word `spectral' refers to the idea of expanding the unknown in a basis of \emph{global} trial functions, extending over the whole spatial domain of interest (as opposed to local trial functions in finite difference and finite element methods). 
The prefix `pseudo' is added when space is discretized and the differential equation is only required to be satisfied at the grid points. 
In between grid points, the solution is approximated by an interpolating function. 

For partial differential equations, it is generally most efficient to 
apply the spectral method only to the spatial coordinate(s), 
and use a time-stepping method for the computation of the time-dependence, from one time level to another. Computing the solution simultaneously over all spacetime is much more expensive than marching in time. 
With a space-only discretization, the pseudospectal method reduces a partial differential equation to a set of ordinary differential equations in time, which can be subsequently solved with a method of choice. 

\begin{figure}
	\parbox[b]{.4\linewidth}{
		\includegraphics[width=\linewidth]{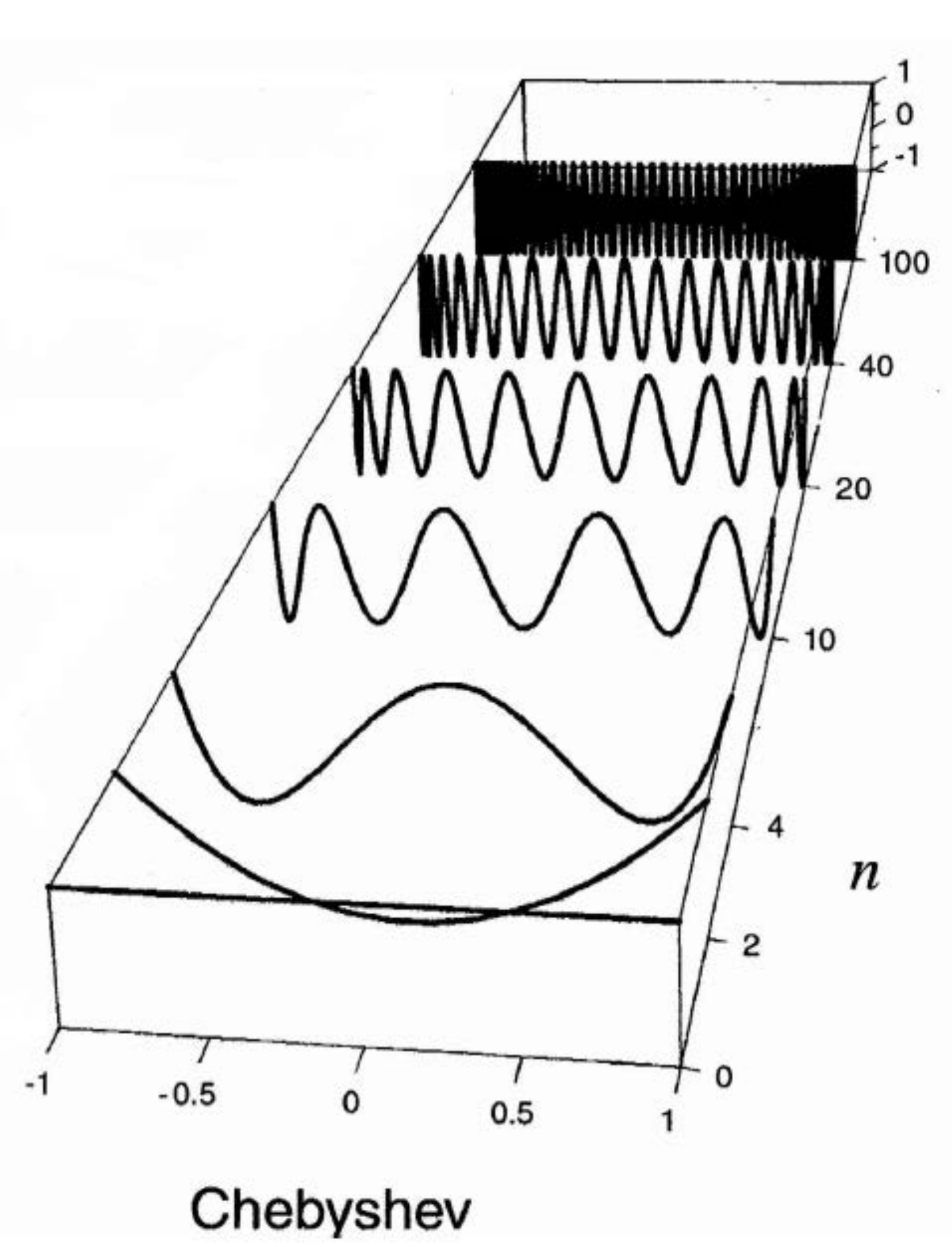}}\hfill
	\parbox[b]{.5\linewidth}{
		\caption{Chebyshev polynomials $T_n(x)$ of degree $n$, with $n$ roots or `nodes' and $n+1$ extrema (Max $T_n(x) = +1$, min $T_n(x) = -1$ for all maxima and minima). The first Chebyshev polynomials are $T_0(x) = 1$, $T_1 = x$ (not shown), $T_2(x) = 2 x^2 - 1$, $T_3 = 4 x^3 - 3 x$ (not shown), $\cdots$ \cite{Fornbergbook}. 
}
		\label{chebys}
	}
\end{figure}

A set of trial functions with favorable properties are the Chebyshev polynomials. They can be defined as the unique polynomials of degree $n$ satisfying 
\begin{equation}
T_n(\cos \theta) = \cos (n\theta) \quad (n=0,1,2,\cdots). 
\end{equation}
It follows that a Chebyshev polynomial expansion $f(z) = \sum_{n=0}^\infty a_n T_n(z)$ is a Fourier cosine series in disguise, $f(\cos \theta) = \sum_{n=0}^\infty a_n \cos(n \theta)$. The Chebyshev polynomials are then able to combine useful features of the Fourier series and orthogonal polynomials. 
The representation of a function by a series of Chebyshev polynomials converges much more rapid than the regular power series expansion (the basic theorem was proved by Chebyshev) \cite{Arfkenbook}. 
(For a detailed analysis of the convergence properties we refer to e.g. \cite{Boydbook}.) 
The `Chebyshev trunctation theorem' moreover says that the error in approximating $f(z)$ by the sum of its first $N$ terms is bounded by the sum of the absolute values of all the neglected coefficients, the proof of which is based on $|T_n(z)| \leq 1$.  
One can then show that numeric error of the pseudospectral method with Chebyshev polynomial trial functions $T_n(x)$ is minimized when the grid points are placed either at the $n$ Chebyshev nodes or the $n+1$ Chebyshev extrema. 
The most efficient grid points are thus unequally spaced, and more closely packed near the endpoints of the domain $-1<x<1$ over which the polynomials $T_n(x)$ are defined (see figure \ref{chebys}).

We discuss in detail our application of the 
 pseudospectral method 
to the problem of obtaining the retarded bulk profile of a 2-component spinor field $\psi(z,k,v) = {\psi_+ \choose \psi_-}$ in the thin-shell AdS$_3$-Vaidya metric ($d=2$) 
\begin{equation} \label{referlater}
 ds^2 = \frac{1}{z^2} \left( -f(v,z) dv^2 - 2 dz dv + dx^2 \right) 
\end{equation}
in Eddington-Finkelstein coordinates and with $f(v,z) = 1- \theta(v) z^2$.  
First, a suitable vielbein basis is chosen: 
\begin{equation} \label{vielbein}
 e^x = \frac{dx}{z}, \quad e^z = \frac{dz}{z} + \frac{f-1}{2z} dv, \quad e^v = \frac{dz}{z} + \frac{f+1}{2z} dv,
\end{equation}
which reduces to the one used in the AdS-part of the background at $v<0$ where $f=1$. 
The bulk field is required to approach  
a delta function source $\psi_0 = {0 \choose \chi_0}$, $\chi_0 = \delta(x-x_0) 
\delta(t-t_0)$ at the AdS boundary, 
with time of injection $t_0 < 0$. 
The Dirac equation of motion $(\slashed D - m)\psi = 0$ in this background is 
\begin{equation} \label{DiraceomVaidya}
 z \Gamma^z \partial_z \psi + (1-f)\frac{z}{2} (\Gamma^v - \Gamma^z) \partial_z \psi + z (\Gamma^v - \Gamma^z) \partial_v \psi - (\Gamma^z - i k z \Gamma^x) \psi - m \mathbf 1 \psi = 0.  
\end{equation}
Next, we choose the following representation of gamma matrices $\Gamma^a$ satisfying $\{ \Gamma^a,\Gamma^b\} = 2 \eta^{ab}\mathbf 1$,   
\begin{align}
\Gamma^{v}&=\Gamma^{t}=-i \Gamma^{\tau} = -i \sigma_2 = \left( \begin{array}{cc}  0& -1 \\ 1 & 0  \end{array}\right) \ , \quad \\
\Gamma^{z}&=\sigma_3 = \left( \begin{array}{cc}  1 & 0 \\ 0 & -1 \end{array}\right) \ , \quad
\Gamma^{x}= \sigma_1 = \left( \begin{array}{cc}  0&  1 \\  1 & 0  \end{array}\right) 
\end{align}
in terms of the Pauli matrices $\sigma_a$ which satisfy $\{ \sigma_a, \sigma_b \} = 2 \delta_{ab} \mathbf 1$. 
The equation of motion then reduces to a coupled system of first-order differential equations for $\psi_+$ and $\psi_-$. 
Motivated by the near-boundary behaviour (\ref{psiassol}) and with an eye on extracting $\tilde \psi$ to obtain $G_R$ via (\ref{GRfrompsi}), we will rescale the fields by a factor $z^{-\Delta_+}$. This way, the retarded correlator value will be directly extractable from the value of the numeric bulk field at the boundary $z=0$.   
It is moreover convenient to work with suitable combinations of the spinor components $\psi_+$ and $\psi_-$. We  define the new field combinations (with $\Delta_+ = 1+m$) 
\begin{equation}
 w = \frac{ \psi_+  - \psi_-}{z^{1+m}}, \quad y = \frac{ \psi_+  + \psi_-}{z^{1+m}}.  \label{fieldcomb}
\end{equation}
In terms of $w$ and $y$, the equations of motion are: 
\begin{align} 
& \text{eq1: } \quad \frac{m}{z} (w-y)  + i k y - \partial_z w  = 0    \label{eq1} \\
& \text{eq2: } \quad  i k (y-w) - z (1+m) y - 2 \partial_v y + \partial_z w + f \partial_z y = 0.  \label{eq2} 
\end{align}
In this form, only one equation needs to be evolved in time\footnote{In this section ``time '' will always refer to Eddington-Finkelstein time $v$.} and the other serves as a `constraint equation' which has to be obeyed at each value of $v$. 
The system is first order in $v$ and $z$ and needs to be solved in the range $v \geq 0$ and $0 \leq z < 1$ (with $z=1$ the event horizon in our choices of units). We impose as initial condition at $v=0$ the known analytic solution 
\begin{align}
 {\psi_+^0 \choose \psi_-^0} &= \psi(z,k,v=0) = K_R^{\mathcal E \mathcal F}(z,k,v=0;t_0) \cdot {0\choose 1} \label{ic} \\ 
 w^0 &= \frac{ \psi_+^0  - \psi_-^0}{z^{1+m}}, \quad y^0 = \frac{ \psi_+^0  + \psi_-^0}{z^{1+m}}, 
\end{align}
and as boundary condition at the boundary $z=0$ that 
\begin{align}
\psi_-(z=0,k,v\geq 0) &= 0, \quad \\ 
y(z=0)-w(z=0) &= 0,  \label{bcspinor} 
\end{align}
which follows directly from the nature of the source $\psi_0$, having delta function support only at $t=t_0 < 0$. 
The initial condition ensures that the obtained numeric solution will be the retarded bulk profile $\psi(z,k,v) = K_R^{\mathcal E \mathcal F}(z,k,v;t_0) \cdot {0\choose 1}$ obeying ``infalling boundary conditions at the horizon". Here, this demand means that no information is allowed to come out of the past horizon, which is the $v=-\infty$ part of the Poincar\'e horizon ($r=0$ or $z=\infty$) of the AdS part of the background (see figure \ref{Penrose}). This is automatically satisfied by imposing the initial condition (\ref{ic}) because it expresses that the fields are zero for times earlier than the injection of the delta function source, $v<t_0$ (see the causal behaviour of $K_R$ in figure \ref{causalsketch}).   
From the sketch in figure \ref{causalsketch} and the Penrose diagram of the background in figure \ref{Penrose} it should also be clear that there is no causal connection between the area behind the future event horizon $z>1$ and the region $0 \leq z < 1$. It is thus not necessary to impose an extra boundary condition on the retarded bulk field at $z=1$. 

The strategy for the  
integration of the differential equations is as follows: given $y$ and $w$ at $v=0$ for all values of $z$, eq2 in (\ref{eq2}) can be used to obtain $y$ at the next timestep $v=\delta v$ (for all $z$). This serves as input for eq1 in  (\ref{eq1}) at the fixed time-slice $v=\delta v$, which needs a boundary condition on $w$ at $z=0$. However, as $y(z=0$) is already known, the boundary condition (\ref{bcspinor})  on $(y-w)$ is sufficient to obtain $w(v=\delta v, z)$.  With $y$ and $w$ known at $v=\delta v$ and all values of $z$, we are back to step one, i.e. integrating eq2 
again, and so on. For the integration in the $z$-direction we will use a pseudospectral method, for the time-integration the NDSolve method of Mathematica (alternatively a fourth-order Runge-Kutta time-stepping or any other method of choice can be used).

The solution ansatz 
consists of an expansion in Chebyshev polynomials,  
\begin{align} \label{solansatz}
w_{ansatz}(z,k,v) &= w^0(z,k) + \sum_{i=0}^{N_z} w_i(k,v) ChT_i(z)  \\ 
y_{ansatz}(z,k,v) &=y^0(z,k)  + \sum_{i=0}^{N_z} y_i(k,v) ChT_i(z), 
\end{align}
with $ChT_i(z)$ the $i$-th order Chebyshev polynomials mapped to the relevant domain in $z$, namely 
$z$ going from 0 to an infinitesimal distance 
away from the horizon, $z_{c}=1 - \epsilon$. That is, 
\begin{equation}
 ChT_i(z) = T_i \left(2\frac{z}{z_{c}}-1 \right) \quad  (i=0,...,N_z).   \label{chebyT}
\end{equation}
The range in $z$ 
is divided in $N_z+1$ grid points 
\begin{equation}
z_j = \frac{z_c}{2} (1 -  \cos(\pi j /N_z)) \quad  (j=0,...,N_z), 
\end{equation}
located at the extrema  $-\cos(\pi j /N_z) = 2\frac{z_j}{z_{c}} - 1$ of the polynomial (\ref{chebyT}).  
The solution ansatz is subsituted into (\ref{eq1}) and the equation is evaluated at the grid points. 
Then we rewrite it as a matrix equation 
\begin{align}
 A \cdot w + B \cdot y = 0 
\end{align}
or $A_{ji} w_i(k,v) + B_{ji} y_i(k,v) = 0$ with  
$A_{ji}= \frac{m}{z_j} ChT_i(z_j) + ChT_i'(z_j)$ and $B_{ji} = (i k - \frac{m}{z_j}) ChT_i(z_j)$ for the gridpoints in the bulk $u_j \neq 0$ ($j=1,...,N_z$), and $A_{0i} = -ChT_i(0)$ and $B_{0i}=ChT_i(0)$ from (\ref{bcspinor}) at the AdS boundary $u_j=0$ (i.e. the first grid point $j=0$).  
There is no  contribution from the exact AdS solution part of the ansatz, because the $f$-independent eq1 (\ref{eq1}) is in fact equal to the AdS ($f=1$) equation. 
When inverted, 
\begin{align}
w = -A^{-1} \cdot B \cdot y,  \label{eq1replace}
\end{align}
eq1 gives us $N_z+1$ equations (from evaluating the equation at $N_z+1$ grid points $u_j$) for $N_z+1$ undetermined coefficients $w_i(k,v)$. 

The same routine of substituting the ansatz and evaluating the equation is repeated for eq2 (\ref{eq2}), in matrix form then given by 
\begin{align}
D\cdot w + E \cdot y + F \cdot \dot y + G = 0 
\end{align}
with notation $\dot y = \partial_v y(k,v)$ for the $v$-derivative at fixed $k$. 
Written out in components we have $D_{ji}  w_i(k,v) + E_{ji} y_i(k,v) + F_{ji} \partial_v y_i(k,v) + G_{j} = 0$ with, for completeness, $D_{ji} = -i k ChT_i(z_j) + ChT'_i(z_j)$, 
$E_{ji} = i k ChT_i(z_j) - z_j (1+m) ChT_i(z_j) + f ChT'_i(z_j)$, $F_{ji} = -2 ChT_i(z_j)$ for $i,j=0,...,N_z$, 
$G_{j} =  i k (y^0(z_j,k)-w^0(z_j,k)) - z_j (1+m) y^0(z_j,k) 
+ \partial_z w^0(z_j,k) + f \partial_z y^0(z_j,k)$ for $j=1,...,N_z$ and $G_0 = 2 \partial_v y_{exact}(z\rightarrow 0)$ with $y_{exact} = \frac{1}{z^{1+m}} \{ \mathcal P_+ K_R^{\mathcal E \mathcal F}(z,k,v;t_0) \cdot {0\choose 1} + \mathcal P_- K_R^{\mathcal E \mathcal F}(z,k,v;t_0) \cdot {0\choose 1} \}$ the exact AdS solution for $y$, since at the AdS boundary ($f\rightarrow 1$), the $v$-independent $y^0$ solves the asymptotic eq2 exactly up to that term. 

After inverting eq2 and  
substituting (\ref{eq1replace}),  
\begin{align}
\dot y = F^{-1} \cdot D \cdot A^{-1} \cdot B \cdot y  -F^{-1} \cdot E \cdot y - F^{-1} \cdot G, 
\end{align}
we are left with a system of $N_z+1$ coupled differential equations for the $N_z+1$ undetermined coefficients $y_i(k,v)$, which we can solve using NDSolve, with the initial condition at the shell 
\begin{equation}
y_i(k,v=0) = 0  \quad (\Rightarrow \text{and through (\ref{eq1replace}) also $w_i(k,v=0) = 0$ as required})  
\end{equation}
ensuring through the solution ansatz (\ref{solansatz}) that at the shell the solution is given by the exact known AdS solution. The solution of NDSolve for the coefficients $y_i(k,v)$ is then substituted back into eq1 (\ref{eq1replace}) to find the coefficients $w_i(k,v)$. 
This concludes the numeric routine for obtaining the spinor bulk solution $ \psi(z,k,v)$. 

The code is tested to reproduce the analytic bulk solutions in AdS (replacing the after-shell region by AdS in the procedure above) and BTZ (replacing the pre-shell region by BTZ) with an absolute accuracy of $10^{-8}$ for $N_z=25$. The magnitude of the exact solutions depends on the values of the parameters but is on average of the order $\approx 0.5$ to 1. The  
residual (i.e.\ the result of plugging the numeric solutions back into the differential equations) is of the order $\approx 10^{-7}$. The AdS$_3$-Vaidya result is trusted based on a stability check with respect to an increase in the amount of grid points $N_z$: the solution quickly reaches a stable value, with the $N_z=25$ result differing only $\approx 10^{-9}$ compared to the $N_z=20$ result. The corresponding residual is of order $\approx 10^{-6}$ for $N_z=25$. These numbers were checked for fixed values of $m$, $k$ and $T$ (namely $m=0.1$, $k=1$, $T=0$) and the range in $t$ from 0 to 5. For much higher values of $t$ an increase in $N_z$ is necessary to maintain the same level of accuracy.

The boundary 2-point function is 
extracted from the numeric bulk solution as 
\begin{align}
G_R(k,t;t_0) = \tilde \psi(k,t) &=  \lim_{z \rightarrow 0} z^{-\Delta_+} \mathcal P_+ K_R^{\mathcal E \mathcal F}(z, k,v \rightarrow t;t_0)\cdot {0 \choose 1} \\
&= \frac{w_{ansatz}(0,k,v\rightarrow t) + y_{ansatz}(0,k,v\rightarrow t)}{2} 
\end{align}
and the time-dependent spectral function as 
\begin{equation}
 \rho(k, T, \omega) = -2 \text{ Re }  \int_{0}^\infty d(\Delta t) \, e^{i \omega \Delta t} G_R( k,T+ \frac{\Delta t}{2}; T - \frac{\Delta t}{2})  \label{rhospinor} 
\end{equation}
(with real instead of imaginary part in the used conventions, compared to the scalar definition, and lower integration limit raised to zero because of the retarded nature $t \geq t_0$ of $G_R$).  

\begin{figure}[h!]
  \hfill
  \begin{minipage}[t]{\textwidth}
    \begin{center}
      \scalebox{1.1}{
  \includegraphics{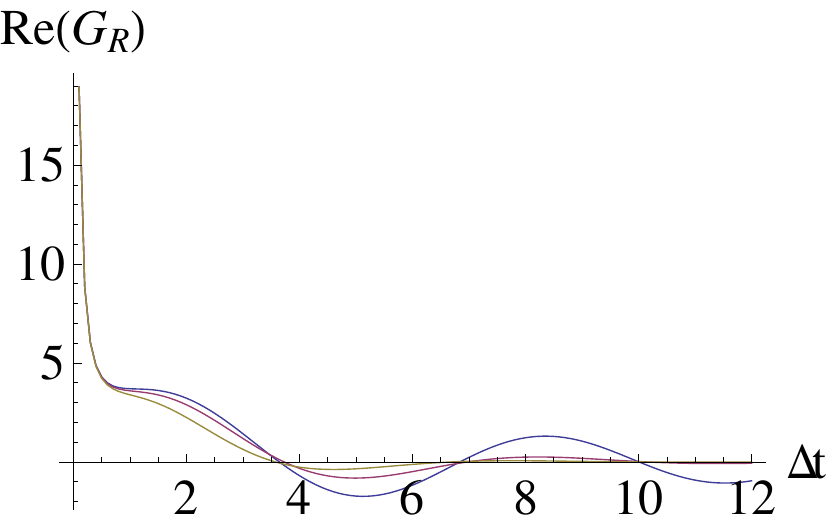}}
    \end{center}
  \end{minipage}
  \hfill
  \begin{minipage}[t]{\textwidth}
    \begin{center}
      \scalebox{1.1}{
  \includegraphics{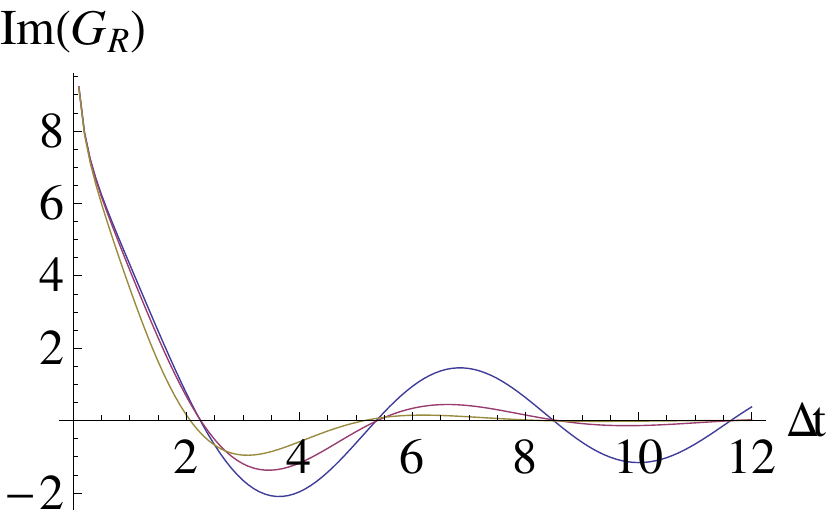}}
    \end{center}
  \end{minipage}
      \caption{The retarded boundary correlator $ G_R(k,T+ \frac{\Delta t}{2}; T - \frac{\Delta t}{2})$ as a function of relative time $\Delta t$ for fixed average time $T=0$, bulk spinor mass $m=0.1$ and momentum $k=1$, real part in the left and imaginary part in the right figure. The numerically obtained thermalizing $G_R$ (in red) is compared to the analytical $G_R^{vacuum}$ (in blue) and $G_R^{thermal}$ (in yellow). As expected, it interpolates nicely between those extremes. }
	\label{GRfigs}
  \hfill
\end{figure}

\begin{figure}[h!]
  \centering
  \scalebox{1.1}{
  \includegraphics{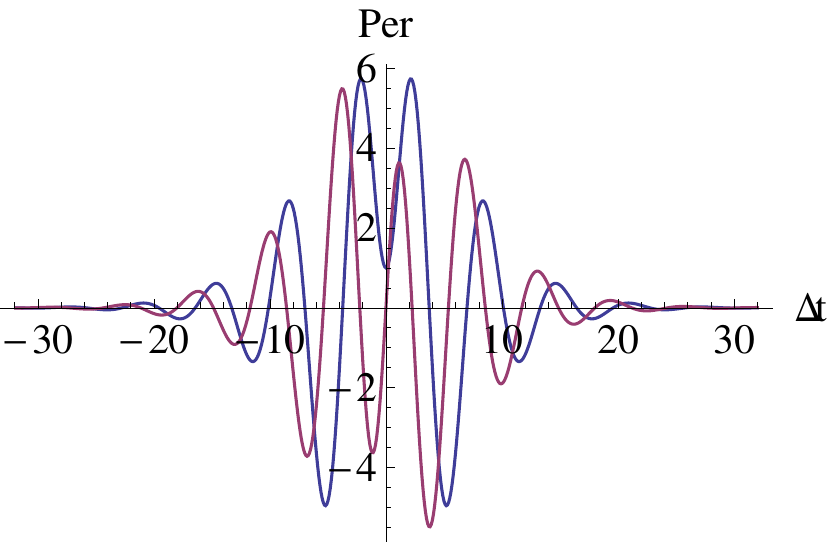}}
  \caption{Real (blue) and imaginary part (red) of the perturbation (\ref{Perdef}), doubled in range (for $m=0.1$, $k=1$ and $T=0$).}
\label{Perdoubledfig}
\end{figure}

\begin{figure}[h!]
  \hfill
  \begin{minipage}[t]{.45\textwidth}
    \begin{center}
      \scalebox{0.8}{
  \includegraphics{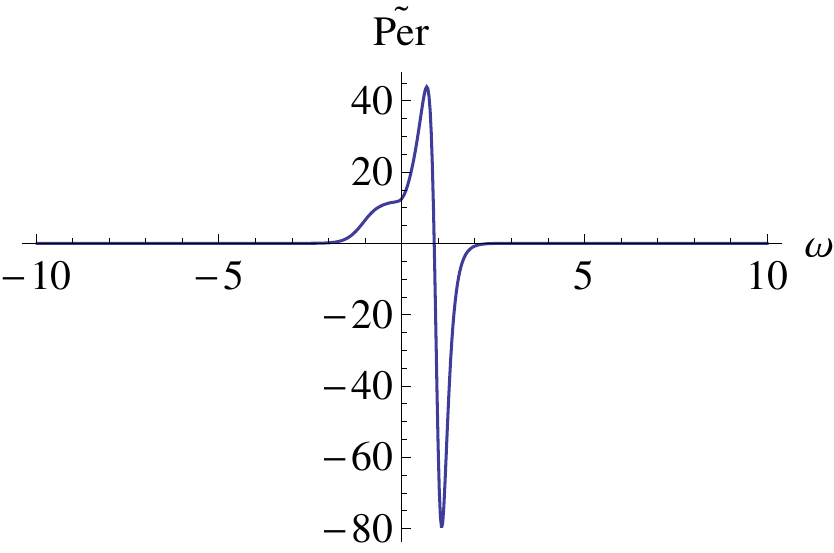}}
    \end{center}
  \end{minipage}
  \hfill
  \begin{minipage}[t]{.45\textwidth}
    \begin{center}
      \scalebox{0.8}{
  \includegraphics{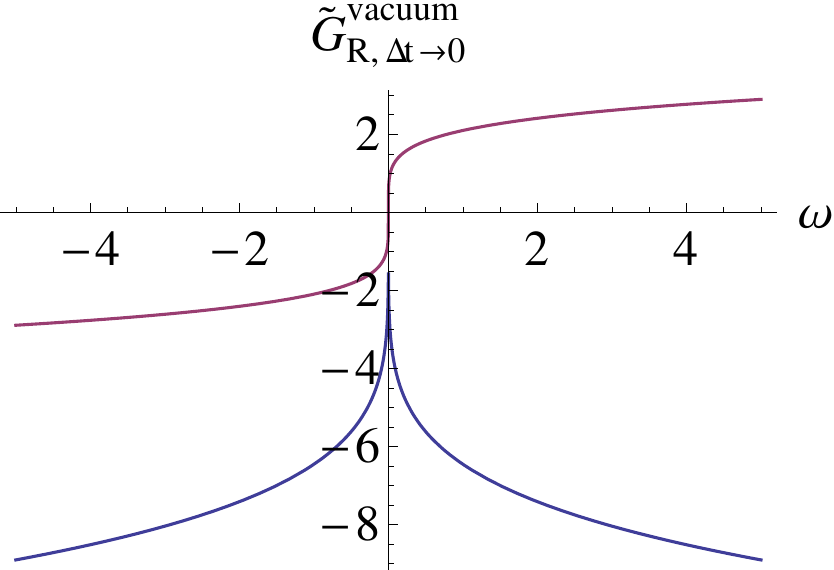}}
    \end{center}
  \end{minipage}
  \hfill
  \begin{minipage}[t]{.9\textwidth}
    \begin{center}
      \scalebox{1.4}{
  \includegraphics{figures/Untitled-6-result-rho-for-zeroT}}
    \end{center}
  \end{minipage}
      \caption{Functions $\tilde Per(\omega)$ (from numerics) and $\tilde G_{R,\Delta t\rightarrow 0}^{vacuum}$ (from analytics) in the convolution (\ref{convdef}) with resulting spectral function $\rho(\omega) = -2 \text{ Re } \tilde G_R(\omega)$ for a spinor operator of scaling dimension $\Delta_+=1.1$ 
dual to 
a spinor bulk field in AdS$_3$-Vaidya, with $k=1$ 
and at average time $T=0$. The real part of $\tilde G_{R,\Delta t\rightarrow 0}^{vacuum}$ is in blue and the imaginary part in red. 
The spectral function (in blue) is compared to its vacuum (red dashed) and thermal (blue dashed) analytic counterparts. }
	\label{convfigs}
  \hfill
\end{figure}

For positive average times $T>0$, the correlator $G_R( k,t=T+ \frac{\Delta t}{2}; t_0=T - \frac{\Delta t}{2})$ is thermal for $t > t_0 > 0$ 
or $0 < \Delta t < T$ (with relative time $\Delta t=t-t_0$ necessarily non-negative) and thermalizing ($t > 0 > t_0$) for $\Delta t > 2T$. 
 Similarly, for  negative average times $T>0$, it  
takes its vacuum value ($t_0 < t < 0$) for $0 < \Delta t < -2T$ and becomes thermalizing for $\Delta t > -2T$. 
In figure \ref{GRfigs} a numerically obtained $G_R$ as a function of relative time is shown and compared to its thermal ($T\rightarrow \infty$) and vacuum ($T\rightarrow -\infty$) equivalents for fixed values of the parameters (in particular non-integer $m$). 
They all show the same divergent behaviour for $\Delta t \rightarrow 0$. This divergence can be dealt with under the integral in (\ref{rhospinor}), by making use of the fact that the analytically known small-$\Delta t$ behaviour of $G_R^{vacuum}(\Delta t \rightarrow 0) \sim (\Delta t)^{-2 \Delta_+ + 1}$ 
has a well-defined Fourier-transform (for non-integer $\Delta_+$) 
\begin{equation}
\tilde G_{R,\Delta t\rightarrow 0}^{vacuum} \sim  \int_0^\infty d(\Delta t) \, e^{i \omega \Delta t} (\Delta t)^{-\alpha} = |\omega|^{\alpha-1} \Gamma(1- \alpha) \left( i \cos \left(\frac{\pi \alpha}{2} \right) \text{Sign}(\omega) + \sin \left(\frac{\pi \alpha}{2} \right) \right)  			\label{Gtildevac}
\end{equation}
with $\alpha = 2 \Delta_+ - 1$, obtained by analytic continuation in $\Delta_+$ (the integral identity used in (\ref{Gtildevac}) is only valid for $0 < \alpha < 1$).   
We define the perturbation 
\begin{equation}
 Per(\Delta t) = \frac{G_R(\Delta t)}{G_{R,\Delta t\rightarrow 0}^{vacuum}(\Delta t)} \label{Perdef}
\end{equation}
as the retarded propagator divided by the divergent small-$\Delta t$ behaviour of the corresponding vacuum retarded propagator. 
Because both the numerator and denominator contain a $\theta(\Delta t)$, we can without any problem (i.e.\ consequence to the spectral function)  double the range of $Per$ in $\Delta t$. This is illustrated in figure \ref{Perdoubledfig} with the real part of $Per$ (which by construction goes through 1 at the origin) extended to an even function, and the imaginary part to an odd one (the imaginary part goes through zero because its leading divergence is overcompensated by the definition (\ref{Perdef}), designed to compensate the \textit{overall} leading divergence).     
This doubling is convenient to subsequently perform the discrete Fourier transform to $\tilde Per(\omega)$. By construction (Re($Per$) even and Im($Per$) odd), $\tilde Per(\omega)$ is real. 
The requested Fourier transformed retarded propagator is then attainable through the convolution (suppressing for a moment all  
dependences but the frequency for notational clarity) 
\begin{equation}
 \tilde G_R(\omega) = \left( \tilde Per  * \tilde G_{R,\Delta t\rightarrow 0}^{vacuum} \right) (\omega) = \int_{-\infty}^\infty \tilde Per(\Omega) \tilde G_{R,\Delta t\rightarrow 0}^{vacuum}(\omega - \Omega) \, d\Omega.  \label{convdef}
\end{equation}
With $\tilde Per$ going to zero sufficiently fast, it is possible to restrict the integral to a finite range and perform a discrete numeric convolution. An example of the participating functions in the convolution and the result for $\rho(k,T,\omega) = -2 \text{ Re }\tilde G_R(k,T,\omega)$ is given in figure \ref{convfigs}. 
$G_R(\Delta t)$ is calculated up to the value of $\Delta t$ where its amplitude approximately reaches $\approx 10^{-6}$, and further values of $G_R(\Delta t)$ are neglected. Depending on the value of $T$, it is necessary to go further in $\Delta t$ to reach this required accuracy.  

\begin{figure}[h!]
  \hfill
  \begin{minipage}[t]{.6\textwidth}
    \begin{center}
  \hspace{-7cm}    \scalebox{1.7}{
  \includegraphics{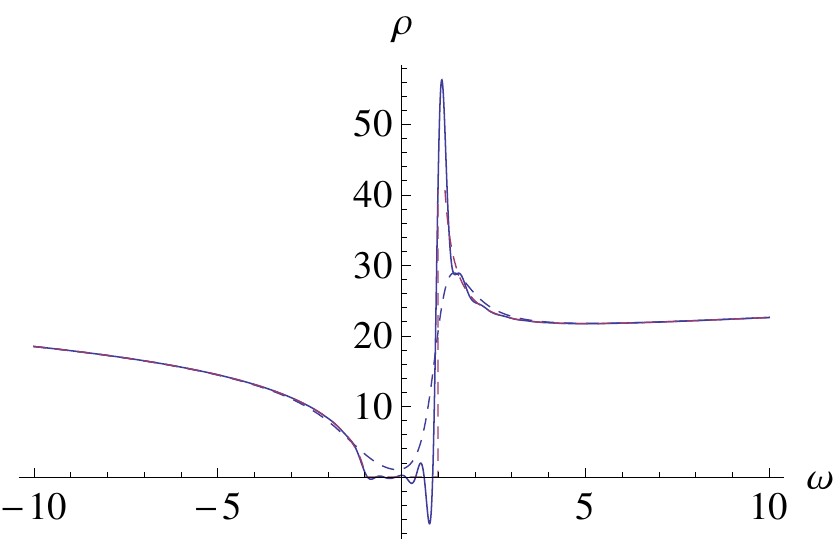}}
    \end{center}
  \end{minipage} 
  \hfill
  \begin{minipage}[t]{.47\textwidth}
    \begin{center}
      \scalebox{0.9}{
  \includegraphics{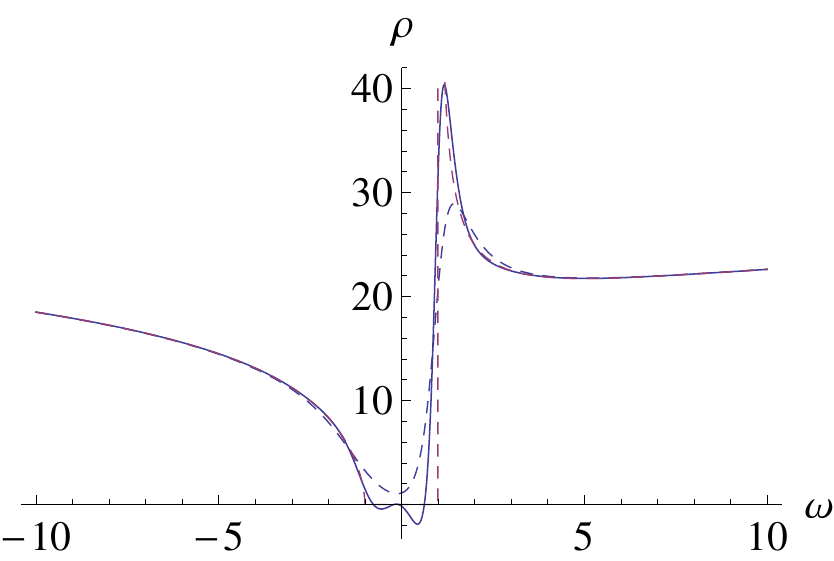}}
    \end{center}
  \end{minipage}
  \hfill
  \begin{minipage}[t]{.47\textwidth}
    \begin{center}
      \scalebox{0.9}{
  \includegraphics{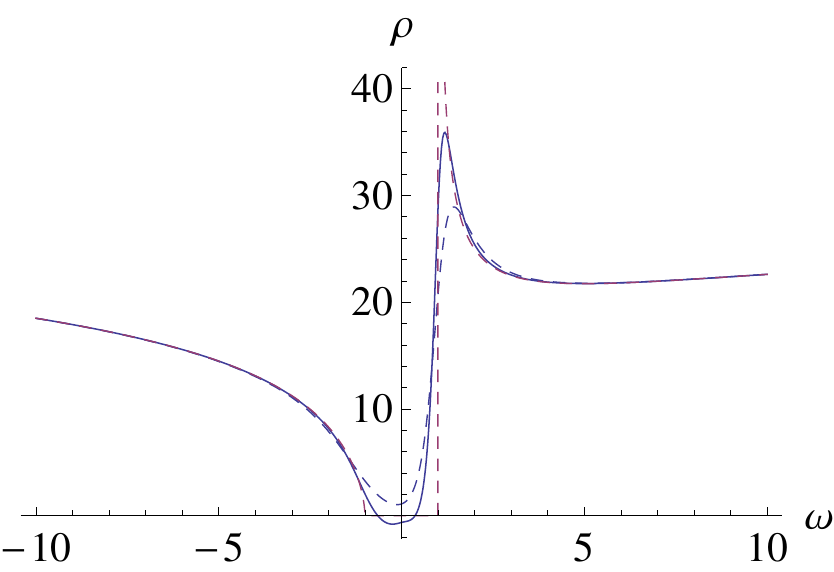}} 
    \end{center}
  \end{minipage}
  \hfill
  \begin{minipage}[t]{.45\textwidth}
    \begin{center}
      \scalebox{0.9}{
  \includegraphics{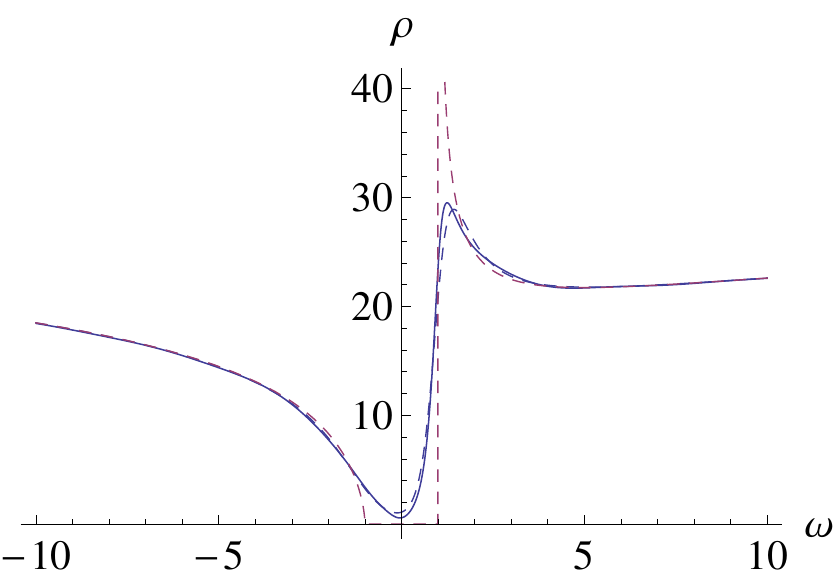}}
    \end{center}
  \end{minipage}
  \hfill
  \begin{minipage}[t]{0.45\textwidth}
    \begin{center}
      \scalebox{0.9}{
  \includegraphics{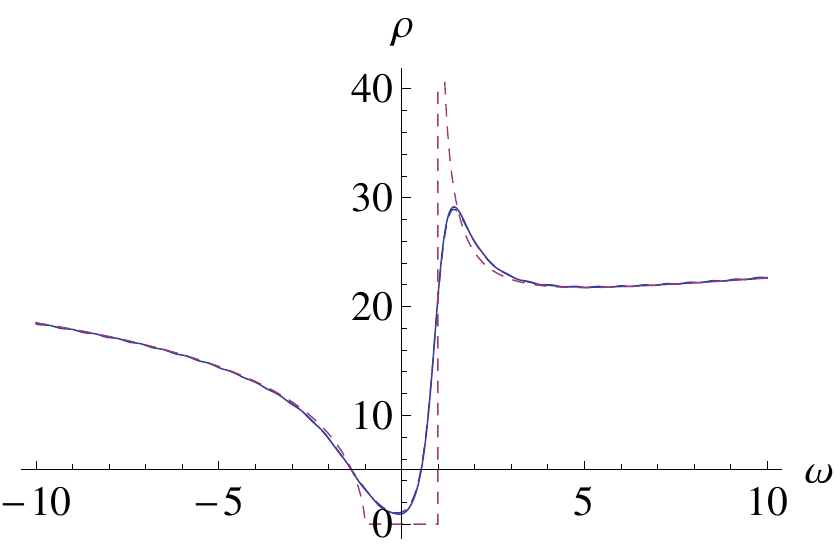}} 
    \end{center}
  \end{minipage}
      \caption{Spectral function $\rho(\omega)$ for a spinor operator of scaling dimension $\Delta_+=1.1$  
dual to 
a spinor bulk field in AdS$_3$-Vaidya, with $k=1$, at increasing values of average time $T=-5, -1, 0, 1, 5$, interpolating between the analytic vacuum (red dashed) and thermal (blue dashed) ones.}
	\label{finalrhofigs}
  \hfill
\end{figure}

A sequence of $\rho(k,T,\omega)$ for increasing values of average time $T$ is shown in figure \ref{finalrhofigs}. 
A pronounced feature are the `negative peaks' which point to strong non-equilibrium effects (as previously observed in the example of a quenched harmonic oscillator in \cite{Balasubramanian:2012tu}). 
A straightforward generalization of the code should enable to construct spectral functions for spinors in higher-dimensional Reissner-Nordstr\"om AdS$_{d+1}$-Vaidya backgrounds, more relevant for interpretations in a condensed matter context.

\section{Spectral functions in Reissner-Nordstr\"om-AdS$_4$-Vaidya} \label{latestpaper} 
 \label{VUBartikel}

The necessary extensions  
of the code for calculating fermionic spectral functions in Reiss\-ner-Nord\-str\"om-AdS$_4$-Vaidya were obtained in the group while this thesis was being written, resulting in Ref.\ \cite{preliminary}. For completeness, the strategy and physical interpretations presented in \cite{preliminary} will be summarized in this section. 

As mentioned earlier in this chapter, more precisely in the paragraph discussing the definition of the equilibrium spectral function (\ref{eqspectral}),  
 Reissner-Nordstr\"om-AdS backgrounds are studied as holographic models of non-Fermi liquids. 
Finite temperature is modeled by the addition of the black hole and finite charge density by imposing a chemical potential $\mu$ as boundary condition on the bulk gauge field, which backreacts on the metric, giving rise  
to a charged or `Reissner-Nordstr\"om' (RN) black hole in the bulk. In 
experiments on strongly correlated electron systems, the electronic structure can be directly examined by the technique of angle-resolved photoemission spectroscopy (ARPES), which provides an approximately 
direct measure of the fermionic spectral function weighted with the occupation number (i.e.\ multiplied with the occupation number which in equilibrium is given by a Fermi-Dirac distribution). 
It is therefore interesting to holographically calculate fermionic spectral functions by considering spinor bulk fields in the background. One finds \cite{Liu:2009dm,Cubrovic:2009ye,Faulkner:2009wj}  
a single fermion retarded Green function of the form 
\begin{equation}
G_R(\omega,k) \sim \frac{1}{\omega-v_F(k-k_F)+\Sigma(\omega,k)}, 
\end{equation} 
with self-energy $\Sigma$. 
This would signal a Landau Fermi liquid, characterized by a Lorentzian quasi-particle peak around $\omega = 0$ at the Fermi surface $k=k_F$, if the imaginary part of the self-energy (inversely proportional to the lifetime of the quasi-particle) scaled quadratically with the frequency at $k=k_F$. Instead, in this holographic set-up, Im$(\Sigma) \sim \omega^{2 \nu_{k_F}}$ with non-analytic self-energy scaling $\nu_{k_F}(m,q)$ dependent on the mass and charge of the spinor bulk field. For $2 \nu_{k_F} > 1$ the quasi-particle is stable but has irregular peak width, apart from the case where $2 \nu_{k_F} = 2$, whereas for $2 \nu_{k_F} < 1$ it is unstable and the quasi-particle picture breaks down.
When backreaction of the spinor bulk field on the metric is taken into account, the charged black hole is shown to be unstable towards formation of an ``electron star'' 
(see \cite{Hartnoll:2011fn} for a review). 

In time-resolved angle-resolved photoemission spectroscopy (tr-ARPES), the studied material (e.g.\ a high-$T_c$ superconductor) is first prepared in a highly  excited non-equilibrium state by a ``pump'' pulse of radiation. 
A second ``probe'' pulse of higher energy photons ejects photoelectrons which are detected with energy (and angle) resolution, allowing to probe out-of-equilibrium behaviour of the (strongly-coupled) system \cite{Freericks}. In this context, it would be interesting to holographically calculate time-dependent notions of spectral functions, using Vaidya type of metrics.   

The problem discussed on page \pageref{referlater} can be generalized to the problem of obtaining the retarded bulk profile of a 4-component spinor field $\psi(z,k,v)$ (with mass $m$ and charge $q$) 
in the thin-shell RN-AdS$_4$-Vaidya metric ($d=3$), which can be used to study non-Fermi liquids 
in out-of-equilibrium situations. 
The RN-AdS$_4$-Vaidya metric and  
gauge field are given in Eddington-Finkelstein coordinates by 
\begin{equation}
\begin{aligned}
&ds^2 = \frac{1}{z^2} \left(-f(v,z) dv^2 -  2 dv dz  + d\vec{x}^2 \right), \\
&A = g(v,z) dv, 
\end{aligned}
\end{equation}
with $f$ and $g$ given by 
\begin{equation}
\begin{aligned}
f(v,z) &=1+  \theta(v)\left( - ( 1 + Q^{2}) z^3 +  Q^2 z^4 \right), \label{fRN} \\
\quad g(v,z) &= \theta(v) \mu (1- z)    
\end{aligned}
\end{equation}
when working with dimensionless quantities such that the rescaled horizon is at $z=1$ (as in \cite{Liu:2009dm}). 
The black hole in the bulk with charge $Q$ and mass $M = 1+ Q^2$ corresponds to a chemical potential $\mu = Q$ (when U(1) gauge coupling is set to one) and temperature $\mathcal T = \frac{1}{4\pi} (3-Q^2)$ in the boundary field theory. An extremal black hole, at $\mathcal T=0$, has charge $Q = \sqrt 3 \approx 1.732$. 
The background interpolates between pure AdS$_4$ at $v<0$ and RN-AdS$_4$ at $v>0$, and the shell at $v=0$ represents a quench of the system from zero to non-zero $\mu$. More precisely, we will quench the system to a near-extremal final state, from $\mu=0$ to $\mu=1.7$, keeping the temperature close to zero. A small non-zero temperature has the advantage of stabilizing the numerics, while being closer to what one would obtain in a real experimental situation.     

This set-up does not literally mimic the experimental context, because there typically 
energy instead of charge is injected by the pump pulse. However, one can consider the system just after the charged shell is sent in as a concrete example of an out-of-equilibrium initial state, whose subsequent evolution can then be followed. The pre-shell pure AdS$_4$ phase can be interpreted as modeling the $(2+1)$-dimensional system of graphene right at the quantum phase transition to the Mott insulator.  
We deliberately choose the parameters $m$ and $q$ such that $2 \nu_{k_F} > 1$, i.e.\ sharp quasi-particles are formed in the finite density system in the post-shell phase. In practice, we used $m=0$ (or conformal dimension $\Delta_+ = \frac{3}{2}+m = \frac{3}{2}$) and $q=1$ in the numerics. We moreover treat 
the fermion bulk field as a probe that does not backreact on the metric.  Its equation of motion is 
\begin{equation}
(\slashed D - m)\psi = 0 
\end{equation}
with the covariant derivative containing the gauge field,  
\begin{equation}
\slashed D 
= \Gamma^M \left( \partial_M + \frac{1}{2} \omega_{M ab} S^{ab} \right) - i q \Gamma^M A_M \psi.   
\end{equation} 
Again, the Latin indices $a,b$ refer to the local tangent frame, using the same vielbein basis as before, (\ref{vielbein}), but with $f$ replaced by its Reissner-Nordstr\"om equivalent (\ref{fRN}).   A suitable choice of gamma matrices $\Gamma^a$ is 
\begin{equation}
\Gamma^{v}=\left( \begin{array}{cc}  0& i \sigma_{2} \\ i\sigma_{2} & 0  \end{array}\right) \ , \quad
\Gamma^{z}=\left( \begin{array}{cc}  \mathbf{1}_{2} & 0 \\ 0 & -\mathbf{1}_{2} \end{array}\right) \ , \quad
\Gamma^{1}=\left( \begin{array}{cc}  0&  \sigma_{1} \\   \sigma_{1} & 0  \end{array}\right) \ , \quad
\Gamma^{2}=\left( \begin{array}{cc}  0&  \sigma_{3} \\   \sigma_{3} & 0  \end{array}\right). 
\end{equation}
Writing the spinor as 
\begin{equation} \label{eq:spinorresc}
\psi = z^{m+ \frac{1}{2}} e^{i k x^{1}  }   \left(  \begin{array}{c} y_{+} (v,z) \\  z_{+} (v,z) \\ y_{-} (v,z) \\ z_{-} (v,z)\end{array} \right),  
\end{equation}
where we chose to align the spatial momentum with the $x^1$ direction and introduced a convenient resca\-ling,  
the equations of motion for $y_+$ and $z_-$ can be seen to decouple from those of $y_-$ and $z_+$. In terms of the new field combinations 
\begin{equation}
\alpha = y_+ + z_-, \quad \beta = y_+ - z_-, 
\end{equation}
the equations of motion are: 
\begin{align} 
&(-1 + m) \alpha  - (m + i k z) \beta + z \partial_{z}\alpha  = 0     \\
&-2 (m - i k z) \alpha + \left( z \partial_{z}f +  2 f (m-1 )+4 i q z g \right) \beta  + 2 z f \partial_{z}\beta - 4 z \partial_{v}\beta = 0.  \label{eq:eqms}
\end{align}
Similar equations for $y_{-}$ and $z_{+}$ can be obtained by substituting $k \leftrightarrow - k$ in these.    

The retarded 2-point function $G_R$ is now a $2\times 2$ matrix, but is diagonalized in the chosen basis of gamma matrices. It is extracted 
completely analogous as in section \ref{subs932} from the response of the spinor bulk field to a delta function source defined at the AdS boundary. 
For the rescaled fields, the asymptotic expansion (\ref{psiassol}) with $d=3$ becomes 
\begin{align}
y_{+} &\approx z^{2-2m} (C + O(z)) + z ~(D + O(z)),  \\
z_{-} &\approx z^{1-2m} (A + O(z)) + z^{2} (B + O(z)),  
\end{align}
where $A$ is identified as the source and $D$ is identified as the expectation value. $C$ and $B$ are local functions of $A$ and $D$ (and of their derivatives), respectively, as determined by the asymptotic Dirac equation\footnote{See for instance \cite{Liu:2009dm}. Notice however the different rescaling and chirality conventions.}.
To invoke a delta function source at a time $t_1$ on the boundary one thus enforces, as in (\ref{eq9315}),  
\begin{equation}
\lim_{z\to0} z^{2m -1} z_{-} (v,z)=\lim_{z\to 0} z^{2m-1}\frac{\alpha(v,z) -\beta(v,z)}{2} = \delta(v-t_1),
\end{equation}
and to ensure that the propagator is causal (i.e.\ the retarded propagator) we impose that 
\begin{equation}
\beta(v<t_1, z ) = 0. 
\end{equation}
By virtue of the Dirac equation, this is sufficient to ensure that also $\alpha(v<t_1,z)= 0$. With these conditions one can then determine the appropriate solution to \eqref{eq:eqms}. 
For $t_2 > t_1$ the source has no support and the expectation value of the boundary operator can be easily extracted. The upper component of $G_R$ can  be calculated as  (see (\ref{GRfrompsi}))
\begin{equation} 
G_{11}(t_2, t_1)  = -i \lim_{z\to0} z^{-1} y_{+}(v_2,z) = -i \lim_{z\to0} z^{-1} \frac{\alpha(v_2,z) +\beta(v_2,z)}{2},  
\end{equation} 
with a similar  definition for the lower component $G_{22}$ coming from the analogous equations for  $y_{-}$ and $z_{+}$. 

The numerical procedure for solving the equations of motion is identical to the one described in section \ref{subs932}, except for the case where the time interval $t=t_2-t_1$ occurs entirely after the quench. Whereas in the AdS$_3$-Vaidya metric analytic expressions for $G_R$ were available in the BTZ region ($v>0$), they are not for the RN-black hole region ($v>0$) of RN-AdS$_4$-Vaidya. The approach taken in this case is to first calculate $G_R(\omega,k)$ in Fourier space  as in  \cite{Liu:2009dm} and to then perform an inverse Fourier transform to find the desired Green function in the mixed representation.

\begin{figure}[ht]
 \begin{center}
\includegraphics[width=0.32 \textwidth]{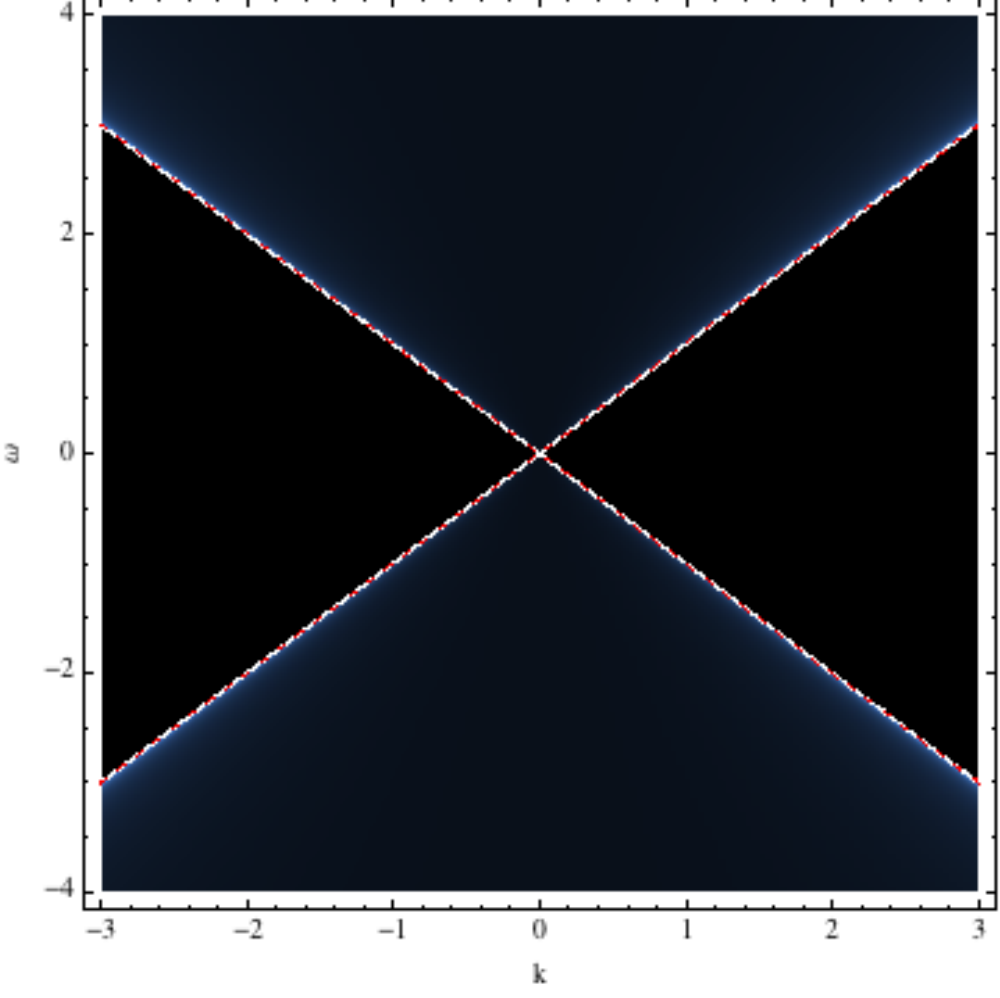} 
 \includegraphics[width= 0.32 \textwidth]{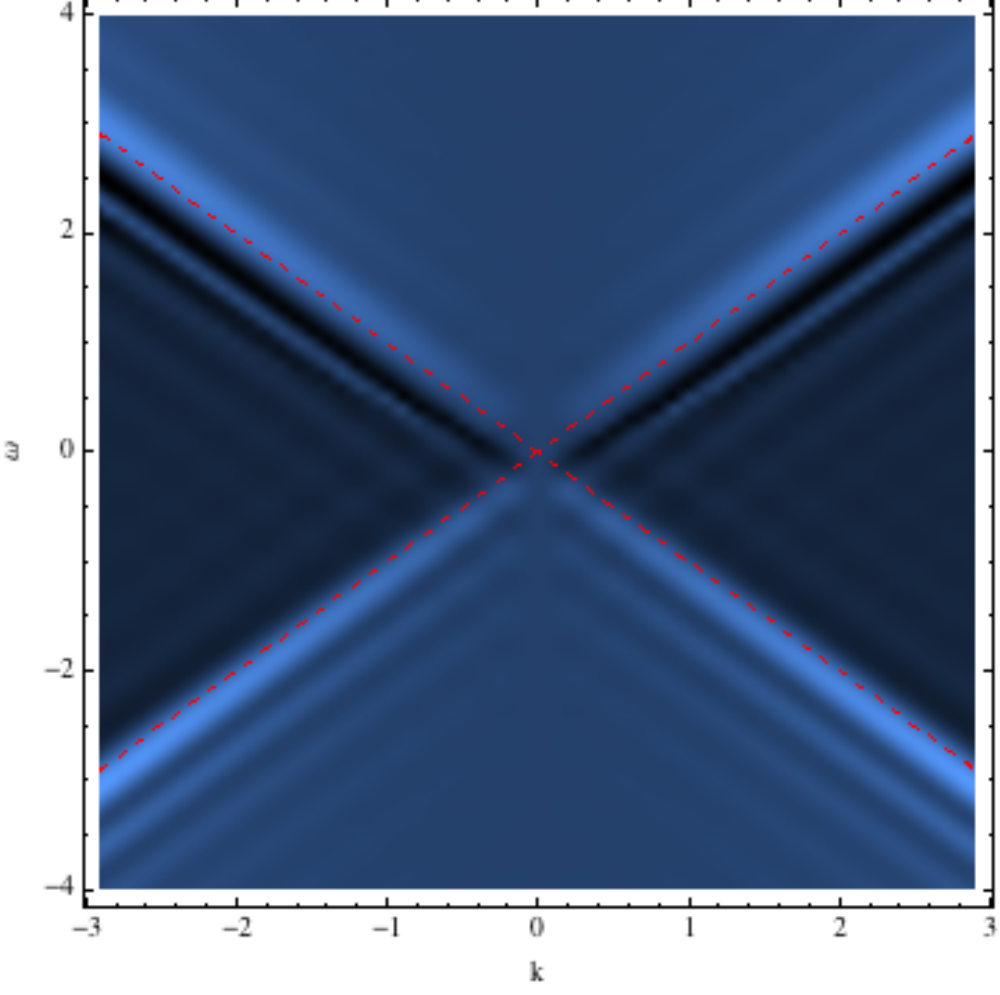}  
\includegraphics[width=0.32 \textwidth]{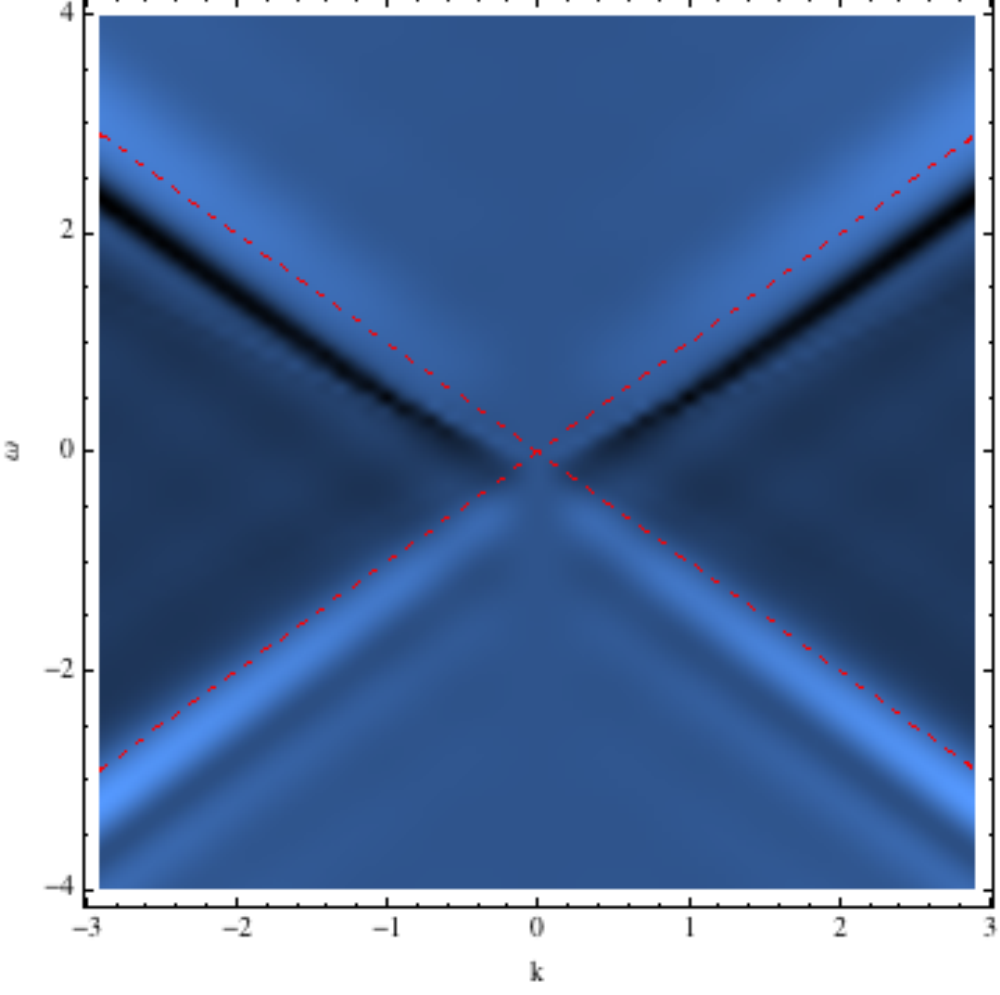}  \\
 \includegraphics[width=0.32 \textwidth]{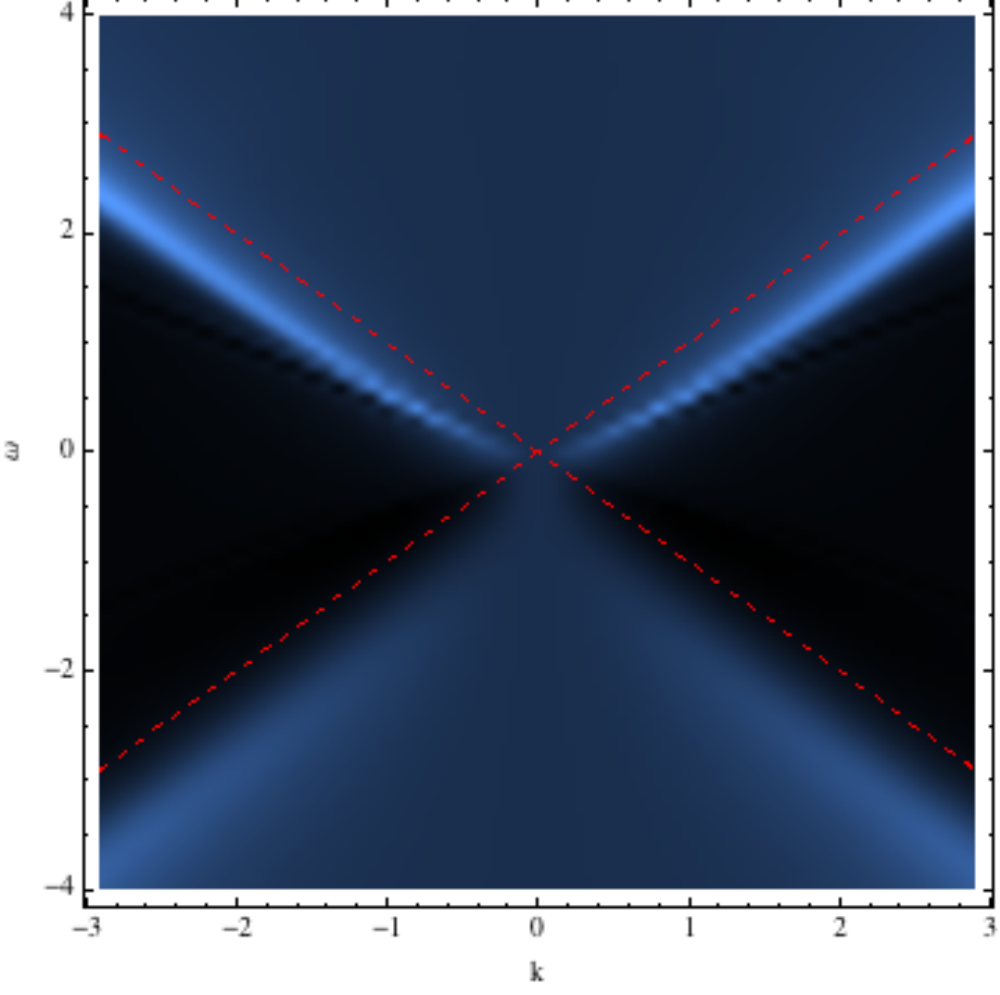} 
 \includegraphics[width=0.32\textwidth]{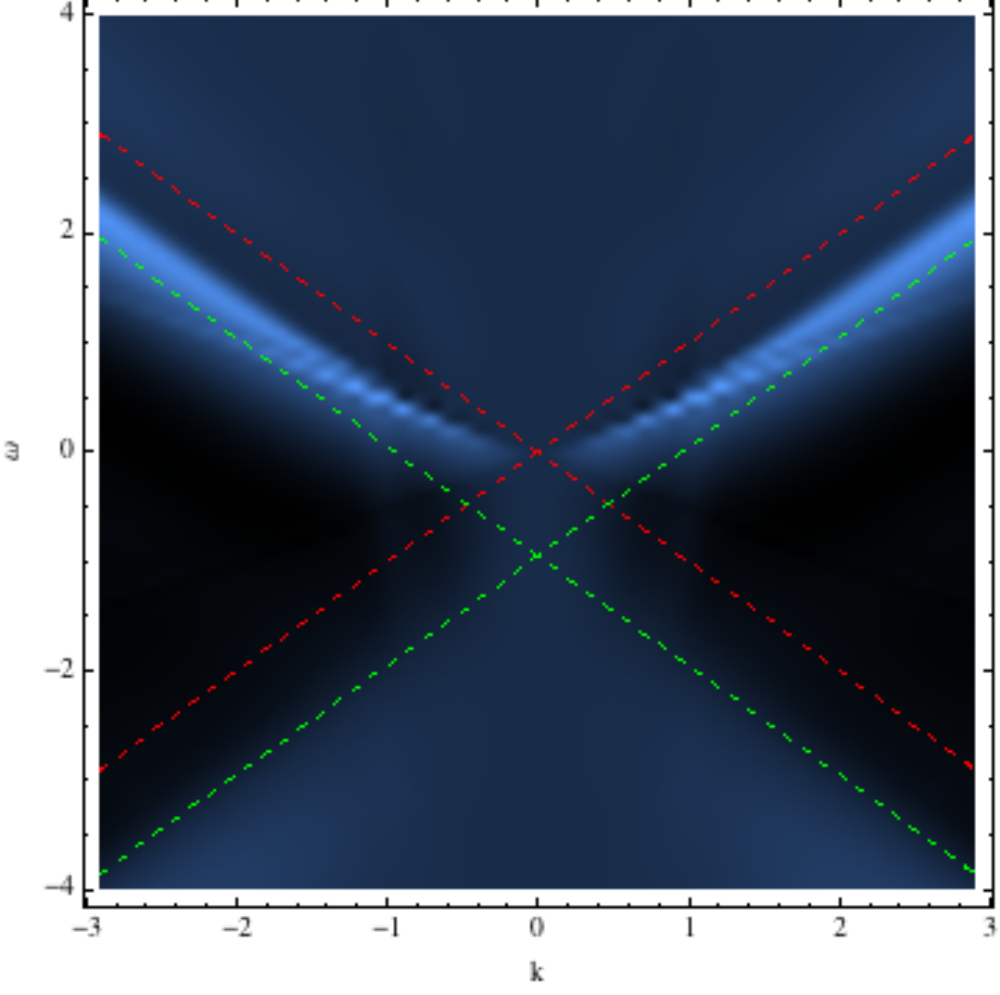}
\includegraphics[width=0.32\textwidth]{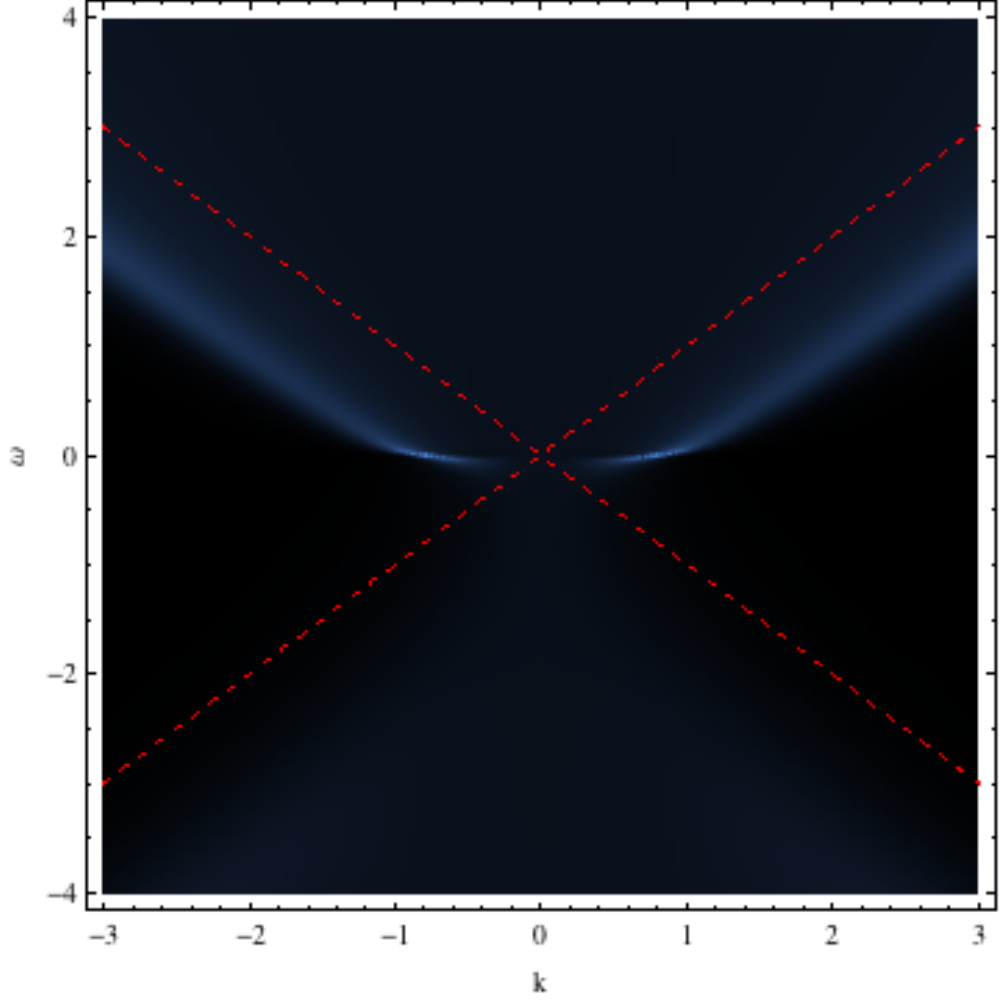}
  \end{center}
\caption{False colour density plots  for spectral sum $A= -2 \left[ \text{Im}(G_{11})+ \text{Im}(G_{22})\right]$  for bulk field of mass $m=0$ when quenched to $Q=1.7$,   at   (from  top to bottom) $T=\{-\infty,
-5,-2.5,  0,   1.5,+\infty
\}$. For reference we plot, in red dashed lines, lightcones centered around $\omega =0$.} 
  \label{fig:DensityPlots}
\end{figure}

Figure \ref{fig:DensityPlots} shows the result for the spectral sum (the trace of the spectral function matrix),
\begin{equation}
A(T, \omega, k)=-2\left[ \text{Im}(G_{11}(T, \omega, k))+ \text{Im}(G_{22}(T, \omega, k))\right]  
\end{equation}
in the $\{\omega, k\}$ plane for a fermionic operator of conformal dimension $\Delta = 3/2$. 
The top central panel of figure \ref{fig:DensityPlots} shows the result at an early average time $T=-5$ and we see that whilst the density  is roughly symmetric around the light cone centered at $\omega=0$ (indicated in the figure by the dashed red lines), a number of oscillations have already begun to influence the profile. In the top right panel, at $T=-2.5$, the period of the oscillations has roughly doubled. 
The oscillations and negative regions 
we believe to be present due to simple interference between the pre-quench and post-quench periods, as was the case for the quenched harmonic oscillator discussed in \cite{Balasubramanian:2012tu}. 
In the  bottom left panel, at $T=0$, the oscillations have essentially disappeared, and the spectral function has developed a clear asymmetry between positive and negative frequency branches.  In the bottom central panel, at $T=1.5$, it is now clear that the asymptotic behaviour of the peak of the spectral function has shifted down still further, characteristic of a system at finite density.
Taking figure \ref{fig:DensityPlots} at face value, after a transient regime at short times, characterized by oscillations  of the ``critical Dirac cones", it appears that at longer times a Fermi energy and Fermi surface develop. Observing that the asymptotic ``Dirac cones'' gradually move down, it is tempting to define a notion of time-dependent effective Fermi energy, or equivalently time-dependent effective chemical potential, $\mu_{eff}(T)$, which measures how much the asymptotic cones have moved down. (This is indicated by the green dashed lightcone in the bottom central panel of figure \ref{fig:DensityPlots}.)\footnote{At a practical level we numerically calculate the location of the maximum value $\omega_{\star}$ of the spectral sum at a fixed large $k$ and use this to define  $\mu_{eff}(T)= k-\omega_{\star}$. However, since the peaks of the spectral functions have a finite width and a profile that depends on both $Q$ and $\Delta$, in what follows to make true comparisons we always take the ratio of this quantity with the same quantity calculated in the final equilibrium state.}

\begin{figure}[h]
  \begin{center}
    \includegraphics[width=0.5\textwidth]{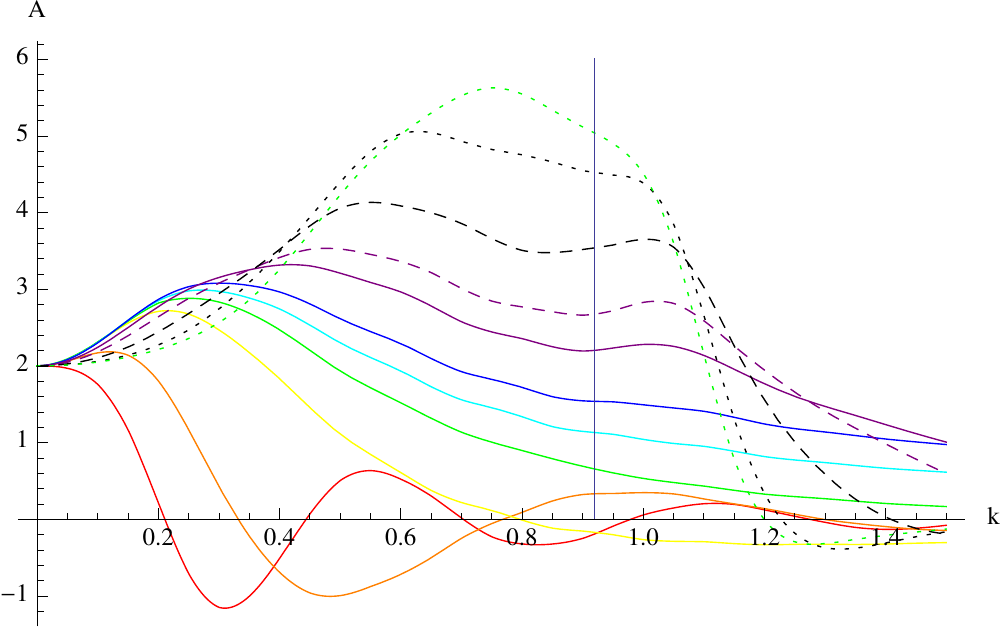}
       \end{center}
\caption{Plots of spectral sum at $\omega = -10^{-3}$ for field of mass $m=0$ when quenched to $Q=1.7$ for $T= \{-5,-2.5,-0.5,0,0.25, 0.5, 1.5,2.5,5, 10,15 \}$. The higher $T$, the higher the maximal value.  Also shown is a line at $k_F \approx 0.92$, the value of the Fermi momentum established in \cite{Liu:2009dm}. }   \label{fig:omega0Plots}
\end{figure}

Figure \ref{fig:omega0Plots} 
shows the accumulation of a peak at zero frequency for a number of average times.  We see for very early times (the curves with lowest peaks in figure \ref{fig:omega0Plots}) that  the spectral function attains its maximum very close to zero momenta, as would be expected at zero density,  and exhibits clear oscillations including negative spectral weight.  As the average time evolves one finds that these oscillations die out, the location of the maximum at zero frequency migrates to larger values of $k$ and the value at the maximum increases.  The edge at larger $k$ sharpens and a clear peak develops with the location of the peak moving to where ultimately a sharp spike at $k=k_F \approx 0.92$ \cite{Liu:2009dm} will occur. 
At late times, the system seems to settle in an equilibrium (quasi) Fermi liquid 
with a Fermi surface that obeys the Luttinger Volume theorem (which states that  the volume enclosed by the Fermi surface is proportional to the fermion charge density) at a finite temperature, chosen to be small compared to the chemical potential. 
The effective chemical potential can be viewed as a new non-local probe of thermalization. Some features it displays are that the effective chemical potential is built up on a time scale of the order of the inverse final chemical potential, and that chemical potential is acquired quicker for smaller $\Delta$ \cite{preliminary}.

This toy holographic system cannot be directly related to tr-ARPES yet. What is measured in ARPES experiments is the product of the spectral function and the occupation number (given by the Fermi factor), which technically equals the ``lesser Green function'' or Wightman function in Fourier space (see for instance \cite{Bellac}). Because in equilibrium the Fermi factor is known, the spectral function can be extracted from ARPES data.   
Away from equilibrium, the lesser Green function still carries the relevant information, as shown in \cite{Freericks}, but there is no straightforward relation between the lesser and retarded Green functions. To come closer to measurable quantities in tr-ARPES experiments, it would therefore be interesting to calculate the lesser Green function in our set-up, using techniques from  \cite{Herzog:2002pc,Skenderis:2008dg}.

\chapter{Conclusion} \label{conclchapter} 
 
In this thesis we have examined two background field effects in QCD using the Sakai-Sugimoto model, and one time-dependence effect in condensed matter physics using the AdS-Vaidya model. The overall strategy has been to use holographic models, which involve a duality to study strongly coupled physics via a higher-dimensional gravitational theory.

The mentioned background field refers to a magnetic field. The LHC and RHIC provide an experimental setting to study QCD, and have  in fact been able to produce a quark-gluon plasma state of matter as the resulting state after a collision of heavy ions. It is in this phase that a remnant magnetic field, induced in the collision, might influence the physical  
processes and hence teach us about QCD itself. The produced magnetic field is larger than any previously known magnetic field in nature, but the effects of its short lifetime and localization in space are hard to take into account. Many studies therefore approximate it by a static and constant background magnetic field, as do we. Moreover, the quark-gluon plasma is a complicated phase of finite density and temperature, and is in our work used as motivation for the study of magnetically induced QCD effects but not as the actual setting of the problems we address. Namely, the rho meson condensation effect is studied in the simplified set-up of zero density and temperature, and the effect of the splitting of chiral transition temperatures at zero density.  

The rho meson condensation effect is a magnetically induced conjectured instability of the QCD vacuum (as opposed to the quark-gluon plasma)  
to a superconducting phase where charged rho mesons with their spins aligned with the magnetic field are condensed, the new vacuum taking the form of an Abrikosov lattice of rho meson vortices. It was first studied using phenomenological effective QCD models (such as the DSGS- and the NJL-model), and also  
in lattice QCD by the time we published our result on rho meson condensation in the Sakai-Sugimoto model. Soon after, bottom-up holographic studies (as opposed to our top-down holographic approach) on rho meson condensation appeared as well. We were thus the first to show that the effect of rho meson condensation can be described in a holographic QCD-model. 

We considered increasingly complex, i.e.\ less simplified, set-ups within the Sakai-Sugimoto model to investigate if the top-down approach could add something to the established picture in phenomenological models, rather than just rebuilding it from a holographic viewpoint. After all, the fact that it can be rebuilt is not so surprising seeing that the holographic models lead to effective QCD models that range from similar to identical to, amongst others, the DSGS-model. This in itself is remarkable, and the holographic approach offers a deep meaning to the phenomenological models in this sense, but it made us wonder if possible modifications to the DSGS-model, and hence the description of rho meson condensation, could be obtained from the Sakai-Sugimoto model.  

We showed that the so-called antipodal Sakai-Sugimoto model with two flavours ($N_f=2$, necessary to describe electrically charged mesons) reproduces exactly the DSGS-model and hence Landau levels for the rho mesons, regarded as structureless point particles. The Landau levels immediately describe the tachyonic instability of the condensing rho mesons and lead to the prediction $B_c = m_\rho^2 \approx 0.6$ GeV$^2$ for the critical magnetic field that is needed for the phase transition to occur. We fixed the number of colours $N_c=3$ and all other holographic parameters by matching to QCD input parameters, to be able to give a result for $B_c$ in GeV units (notice that we absorb the electromagnetic coupling constant $e$ into the notation of $B$). This set-up ($\sim \text{Tr } F^2$) comes closest to the bottom-up ones that are used to describe 
rho meson condensation. 

In the non-antipodal Sakai-Sugimoto model, the embedding of the $N_f=2$ flavour branes is more general, allowing the description of a constituent quark mass and the possibility to take into account effects of the magnetic field $B$ on the quark constituents of the rho mesons. It greatly complicates the involved mathematics, as a result of the flavour branes being pulled apart by the presence of $B$ in this set-up. There are two new mechanisms associated with the now $B$-dependent embedding, that contribute to the mass of the rho mesons and hence a rise in $B_c$: a holographic Higgs mechanism and the chiral magnetic catalysis effect on the constituents. In this model the Landau levels are modified and we find $B_c = \mathit{m_{\rho,eff}^2}(B_c) \approx 0.78$ GeV$^2$. 

The results described in the two last paragraphs are obtained in a much used simplification of the model, approximating the flavour brane DBI-action to an action which is second order in the flavour gauge field strength $F$. Using the full non-Abelian DBI-action leads to the expression (\ref{omegasquared}) (combined with (\ref{defomega})) for the most general modified Landau levels, 
and further increased values for the critical magnetic field, $B_c \approx 1.07$ and 0.85 GeV$^2$, for respectively the antipodal and non-antipodal case. These predictions come closer to $B_c$ of the order 1 GeV$^2$ as obtained in the NJL-model and on the lattice.  

Our main result for the effective 4-dimensional action for a vector field $\rho_\mu^a$ in an external electromagnetic field, or our ``modified DSGS-model" as obtained from the Sakai-Sugimoto model, is 
\begin{align*}
S_{4D}
= \int d^4 x 
& \left\{- \frac{1}{4}(\mathcal F_{\mu\nu}^{a})^2 -\frac{1}{2} m_\rho^2{\color{Mediumblue}(B)} (\rho_\mu^a)^2    -\frac{1}{2}{\color{Mediumblue} m_{+}^2(B)}  (\rho_i^a)^2  
-\frac{1}{2} k{\color{Mediumblue}(B)} \overline F_{ij}^3 \epsilon_{3ab}\rho_i^a \rho_j^b  
\right.  \nonumber\\  & \left. \qquad \qquad
 {\color{Darkgreen} - \frac{1}{2} b(B) (\mathcal F_{12}^a)^2} {\color{Darkgreen} -\frac{1}{2} a(B) ((\mathcal F_{i 3}^a)^2 + (\mathcal F_{i0}^a)^2)} \right\} \hspace{2.5cm}
\end{align*}    
and the resulting most general expression for the rho meson mass' $B$-dependence 
\begin{align*}
\mathit{m_{\rho,eff}^2}(B) = \frac{m_\rho^2{\color{Mediumblue}(B)}+{\color{Mediumblue}m_{+}^2(B)}}{1+{\color{Darkgreen}a(B)}} - \frac{k{\color{Mediumblue}(B)}}{1+{\color{Darkgreen}a(B)}} B, 
\end{align*}   
where modifications in blue are present for the non-antipodal embedding and the ones in green when the full DBI-action is considered. The final plots for $\mathit{m_{\rho,eff}^2}(B)$, which form our main result for this part of the thesis, are presented in figure \ref{summaryfig} on page \pageref{summaryfig}. 

We performed a stability analysis in the scalar sector as well, showing stability of the $B$-dependent embedding of the flavour branes in the geometry. Furthermore, we demonstrated that there is no influence of the pions on the rho meson condensation (at a second order in the fluctuations analysis) from the Chern-Simons part of the action. 
   
The second magnetically induced QCD effect that we studied in the two-flavour Sakai-Sugimoto model is the behaviour of the chiral symmetry restoration temperatures (one for each flavour). We showed that they rise as a function of the magnetic field (known as the chiral magnetic catalysis effect), 
inducing a split between them and the deconfinement temperature or not, depending on the value of the asymptotic separation $L$ between flavour and anti-flavour branes. Our results are compatible with NJL-results for small $L$ and  consistent with all other approaches in the quenched approximation.  
We remark that the latest, unquenched lattice data however show the appearance of inverse magnetic catalysis, i.e.\ chiral transition temperatures decreasing with $B$, in contrast to all previous phenomenological model and lattice studies. A direct verification of these lattice results in the setting of the Sakai-Sugimoto model would require taking backreaction into account, which does not seem to be a straightforward task.     
In our analysis we took into account the different electric charges of the flavour branes, resulting in the chiral transition temperatures  
splitting per flavour. This marks an intermediate phase where chiral symmetry is only partially restored, i.e.\ for one of the two flavours.   

In the second part of the thesis we considered bottom-up holographic models for thermalization processes, called Vaidya models. They allow to study far-from-equilibrium behaviour of strongly coupled electron systems. To a great extent still mysterious materials of this type, such as high-$T_c$ superconductors, are currently being studied  
in (time-resolved) photoemission spectroscopy (ARPES) experiments. We presented the calculation of time-dependent spectral functions in Reissner-Nordstr\"om-AdS$_4$-Vaidya  as a first step towards extracting in principle measurable quantities  
in time-resolved ARPES. The focus was put on explaining the used numerical method (pseudospectral method) in the context of the most simple AdS-Vaidya model, interpolating between a pure AdS and a BTZ metric.

The AdS/CFT correspondence, and by extension gauge-gravity dualities, form an intriguing new way of studying strongly coupled field theories. 
Attempts at generalizations to holographic QCD models are however not without problems (both in the top-down and the bottom-up approach), and before doing any calculations you could conclude it would be difficult to make them work. 
It is therefore somewhat surprising but reassuring that for example the Sakai-Sugimoto model is in fact able to present a nice record of QCD-effects and properties that can be modeled, suggesting the influence of the redundant modes 
is not necessarily substantial. We were able to add two conjectured effects to that list, namely the rho meson condensation effect and the splitting of transition temperatures in the presence of strong magnetic fields. 
In the last chapter, we were able to study a far from equilibrium strongly coupled system by using 
the AdS-Vaidya model. 

In conclusion, holographic models do provide a useful framework for studying background field and time dependence effects in strongly coupled field theories. With the growing availability of data on quark-gluon plasma from LHC and RHIC experiments, as well as experiments in condensed matter theory, 
many more effects remain to be examined and might eventually lead us to new insights in QCD or condensed matter physics.

\appendix

\chapter{Anti de Sitter space} \label{adsspaceapp}

The (Anti) de Sitter space is defined as the space with Lorentzian signature and constant (negative) positive curvature, i.e.\ while de Sitter space is the Lorentzian version of a sphere, Anti de Sitter space is a Lorentzian version of a hyperboloid.  
The $d$-dimensional Anti de Sitter space is defined by the hyperboloid-like embedding in $d+1$ dimensions: 
\begin{equation} \label{AdSdmetriek}
ds^2_{AdS_d} = -dx_0^2 + \sum_{i=1}^{d-1} dx_i^2 - dx^2_{d+1}, \qquad 
\end{equation}
\begin{equation} \label{hyperboloide}
-x_0^2 + \sum_{i=1}^{d-1} x_i^2 - x^2_{d+1} = -R^2.
\end{equation}
Different parameterizations of the `hyperboloid' (\ref{hyperboloide}) lead to different forms of the $AdS_d$ metric  (\ref{AdSdmetriek}) 
 In the so-called  Poincar\'e coordinates, $-\infty < t, x_i < +\infty$ and $0<x_0<+\infty$: 
\begin{equation} \label{adspoincare}
ds^2_{AdS_d} = \frac{R^2}{x_0^2} \left(-dt^2 +  \sum_{i=1}^{d-2} dx_i^2 + dx_0^2\right).
\end{equation}
This coordinate system does not describe the whole AdS space. Global coordinates do and give the metric  
\begin{equation} \label{adsglobal}
ds^2_{AdS_d} = R^2 \left(-\cosh^2 \rho \hspace{1mm} d\tau^2 + d\rho^2 + \sinh^2 \rho \hspace{1mm} d\Omega^2_{d-2}\right)
\end{equation} 
with $\tau \in [0,2\pi]$ and $\rho \in \mathbb R^+$. 
Other commonly used metrics for the Poincar\'e patch are 
\begin{align}
ds^2_{AdS_d} &= R^2 \left( \frac{du^2}{u^2} + u^2 dx_\mu dx^\mu \right) \label{adspetersen} \\
 &= R^2 \left( dr^2 + e^{2 r} dx_\mu dx^\mu \right)
\end{align} 
with the different radial coordinates related to each other by $u = 1/x_0 = e^r$ and $dx_\mu dx^\mu$ short for $-dt^2 +  \sum_{i=1}^{d-2} dx_i^2$.

\chapter{Solving the mass eigenvalue equation} \label{appendixnum}

We numerically solve the eigenvalue equation (\ref{finiteTeigwvgl}) with normalization condition \eqref{orthonormconditiepsi}. Since
\begin{equation}\label{extra3}
u^{-1/2}\gamma^{1/2}\sim u^{-1/2} \qquad\text{for } u\sim \infty,
\end{equation}
normalizability requires that
\begin{equation}\label{extra4}
    \psi_n(u=\infty)=0.
\end{equation}
For numerical purposes, it is easier to work on a compact interval. The following transformation
\begin{equation}\label{extra5}
    \frac{u^3}{u_0^3}=\frac{1}{\cos^2x}
\end{equation}
maps  the interval under investigation onto $x\in[-\pi/2;\pi/2]$, with the boundary conditions \eqref{extra4} now reading
\begin{equation}\label{mass11}
    \psi_n(\pm \pi/2)=0.
\end{equation}
The differential equation \eqref{finiteTeigwvgl} transforms into
\begin{equation}\label{extra6}
- \frac{9}{4} u_0 \frac{\cos^4x}{\sin x}    \sqrt{\frac{1-\frac{u_K^3}{u_0^3}\cos^2x}{\cos^{16/3}x}-f_0}~\frac{\partial}{\partial x}\left(\frac{\cos^{8/3}x}{\sin x}    \sqrt{\frac{1-\frac{u_K^3}{u_0^3}\cos^2x}{\cos^{16/3}x}-f_0}~\frac{\partial}{\partial x}\right)\psi_n(x)=R^3\lambda_n\psi_n(x),
\end{equation}
where we denoted the mass eigenvalue $m_n^2$ with $\lambda_n$. Due to the reflection symmetry $x\rightarrow -x$ we can split up the eigenfunction set in even/odd $\psi_n(x)$'s and focus on the interval $[0,\pi/2]$. Analogously as explained in \cite{Sakai:2004cn,Peeters:2006iu}, the even/odd eigenfunctions correspond to odd/even parity mesons. We can thus demand that
\begin{equation}\label{mass10}
    \psi_n(0)=0\qquad\text{or}\qquad\partial_x\psi_n(0)=0.
\end{equation}
For the $\rho$ meson, we must look at the odd parity sector, in particular the even eigenfunction with the lowest eigenvalue.  We can temporarily replace \eqref{finiteTeigwvgl} and associated boundary conditions with the initial value problem
\begin{equation}\label{mass16}
- u_0 \frac{\cos^4x}{\sin x}    \sqrt{\frac{1-\frac{u_K^3}{u_0^3}\cos^2x}{\cos^{16/3}x}-f_0}~\frac{\partial}{\partial x}\left(\frac{\cos^{8/3}x}{\sin x}    \sqrt{\frac{1-\frac{u_K^3}{u_0^3}\cos^2x}{\cos^{16/3}x}-f_0}~\frac{\partial}{\partial x}\right)\psi=\frac{1}{M^3}\Lambda\psi\,,\psi(0)=1\,,\psi'(0)=0
\end{equation}
where $\Lambda$ is treated as a ``shooting'' parameter. We numerically solved the previous differential equation for each value of $\Lambda$ to give a unique $\psi_\Lambda(x)$, consistent with the initial conditions. Since the coefficient functions appearing in \eqref{mass16} display a delicate behaviour $\sim 0/0$ around $x=0$, some care is needed when using a numerical package, as this is precisely where we are imposing our initial conditions. The solution around $x=0$ can however be easily obtained using a Taylor expansion, and fed into the numerical procedure. Once the $\psi_\Lambda(x)$ is known, we can solve the equation
\begin{equation}\label{mass17}
    F^{\text{even}}(\Lambda)\equiv\psi_\Lambda(\pi/2)=0
\end{equation}
for $\Lambda$, which is then precisely the mass eigenvalue of the original eigenvalue problem.\\

For the $eB$-dependent eigenvalue problem (\ref{Beigvproblem}) with normalization condition (\ref{Bnormalization}), the corresponding initial value problem can be written as (\ref{mass16}) with every $f_0 \rightarrow f_0 A_0/A$,  where in the case of the approximation of coincident branes $A$ is replaced by $A_{average} = \frac{(\sqrt{A_u} + \sqrt{A_d})^2}{2}$, and where it is understood that each $u_0$ appearing in the eigenvalue equation is now $eB$-dependent, see $u_0(eB)$ from (\ref{u0Baverage}). This can accordingly be solved numerically, using the explained shooting method, for the $eB$-dependent mass eigenvalue $\Lambda(eB) = m_\rho^2(eB)$.

\chapter{STr-prescription} \label{appendix}

\paragraph{Prescription}
We write down the prescription for the evaluation of the symmetrized trace STr to second order in fluctuations in the presence of a constant Abelian background, as derived in \cite{Hashimoto:1997gm} and \cite{Denef:2000rj}.

For an even function $\mathcal H(F)$ of a diagonal background field $F = F^0 \sigma^0 +F^3 \sigma^3$ and fluctuation $\tilde X = \tilde X^a t^a$ (generator $t^a = -\frac{i}{2} \sigma^a$), one finds that
\begin{equation} \label{prescr}
\text{STr}\left(\mathcal H(F) \tilde X^2 \right) = -\frac{1}{2} \sum_{a=1}^2  (\tilde X^a)^2 \hspace{2mm} I(\mathcal H) -\frac{1}{2}\sum_{l=u,d} (\tilde X^l)^2 \hspace{2mm} I_l(\mathcal H)
\end{equation}
with
\begin{equation}  \label{Ifctions}
I(\mathcal H) = \frac{\int_0^1 d\alpha \mathcal H(F^0 + \alpha F^3) + \int_0^1 d\alpha \mathcal H(F^0 - \alpha F^3)}{2},
\end{equation}
\begin{equation}  \label{Ilfctions}
I_u(\mathcal H) = \mathcal H(F^0+F^3), \quad I_d(\mathcal H) = \mathcal H(F^0-F^3),
\end{equation}
\begin{equation}
\tilde X^u = \frac{\tilde X^0 + \tilde X^3}{\sqrt{2}}, \quad \tilde X^d = \frac{\tilde X^0 - \tilde X^3}{\sqrt{2}};
\end{equation}
and
\begin{equation} \label{typeI}
\text{STr}\left(\mathcal H(F) \tilde X \right) = \text{Tr}\left(\mathcal H(F) \tilde X \right).
\end{equation}

\paragraph{Generalized prescription} A straightforward generalization of the prescription when dealing with two Abelian background fields can be written down.

For even functions $\mathcal H(\partial \overline \tau)$ and $\mathcal G(F)$ of diagonal background fields $\partial \overline \tau = \partial \overline \tau^0 \sigma^0 + \partial \overline \tau^3 \sigma^3$  and $F = F^0 \sigma^0 + F^3 \sigma^3$, and fluctuation $\tilde X = \tilde X^a t^a$ (generator $t^a = -\frac{i}{2} \sigma^a$), it reads
\begin{equation}
\text{STr}\left(\mathcal H(\partial \overline \tau) \mathcal G(F) \tilde X^2 \right) = -\frac{1}{2} \sum_{a=1}^2 (\tilde X^a)^2 \hspace{2mm} I(\mathcal H \mathcal G) -\frac{1}{2}\sum_{l=u,d} (\tilde X^l)^2 \hspace{2mm} I_l(\mathcal H \mathcal G)
\end{equation}
with
\begin{equation} \label{IfctionsGeneral}
I(\mathcal H\mathcal G) = \frac{\int_0^1 d\alpha \mathcal H(\partial \overline \tau^0 + \alpha \partial \overline \tau^3) \mathcal G(F^0 + \alpha F^3) + \int_0^1 d\alpha \mathcal H(\partial \overline \tau^0 - \alpha \partial \overline \tau^3)\mathcal G(F^0 -\alpha F^3)}{2},
\end{equation}
\begin{equation} \label{IlfctionsGeneral}
I_u(\mathcal H\mathcal G) = \mathcal H(\partial \overline \tau^0+\partial  \overline \tau^3)\mathcal G(F^0 + F^3), \quad I_d(\mathcal H\mathcal G) = \mathcal H(\partial \overline \tau^0-\partial \overline \tau^3)\mathcal G(F^0 - F^3),
\end{equation}
\begin{equation}
\tilde X^u = \frac{\tilde X^0 + \tilde X^3}{\sqrt{2}}, \quad \tilde X^d = \frac{\tilde X^0 - \tilde X^3}{\sqrt{2}};
\end{equation}
and
\begin{equation} \label{typeIgeneral}
\text{STr}\left(\mathcal H(\partial \overline \tau) \mathcal G(F) \tilde X \right) = \text{Tr}\left(\mathcal H(\partial \overline \tau) \mathcal G(F) \tilde X \right).
\end{equation}

\section{Derivation of the prescription}
For completeness, let us schematically recapitulate
how the above prescription was obtained. In this derivation we will temporarily write $U(2)$-indices as lower instead of upper indices, to avoid notational clutter.

\begin{itemize}
\item Properties of the Pauli matrices ($a=1,2,3$):
\[ \text{Tr} (\sigma_a) = 0, \quad \text{Tr} (\sigma_a \sigma_b) = 2 \delta_{ab}, \quad \sigma_a \sigma_b = \delta_{ab} \mathbf{1} + i \epsilon_{abc} \sigma_c  \]
\[ \{\sigma_a,\sigma_b\} = 2 \delta_{ab} \mathbf{1}, \quad [\sigma_a,\sigma_b] = 2 i \epsilon_{abc} \sigma_c \]

\item $\text{STr}(\sigma_3^m \sigma_a \sigma_b)$:
\begin{align}
\text{STr}(\sigma_3^m \sigma_a \sigma_b) &= \frac{1}{(m+2)!} \sum_{\text{all permutations}} \text{Tr}(\sigma_3^m \sigma_a \sigma_b) \nonumber\\
&= \frac{1}{m+1} \sum_{k=0}^m \text{Tr}(\sigma_3^k \sigma_a \sigma_3^{m-k} \sigma_b) \nonumber\\
&= \left\{ \begin{array}{ll} 2 [\delta_{0a} \delta_{0b} + \delta_{3a} \delta_{3b} + \frac{\delta_{ab}}{m+1}|_{a,b=1,2}] \qquad \text{for $m$ even} \\
 2 [\delta_{0a} \delta_{3b} + \delta_{3a} \delta_{0b}]\qquad \text{for $m$ odd} \end{array} \right.
\end{align}
where now $a,b=0,1,2,3$ with $\sigma_0 = \mathbf{1}$,
and where we used
\begin{equation}
\sum_{k=0}^m \text{Tr}(\sigma_3^k \sigma_a \sigma_3^{m-k} \sigma_b) = \sum_{k=0}^m \text{Tr}((-1)^k \sigma_3^{m} \sigma_b \sigma_a).
\end{equation}
\item $\text{STr}(F^m \tilde X^2)$ with $m$ even, $F=F_0 \sigma_0 + F_3 \sigma_3$ and $\tilde X = \tilde X_a t_a$ with $t_a = -i \left(\frac{\mathbf{1}}{2},\frac{\sigma_a}{2}\right)$:
\begin{align}
\text{STr} (F^m \tilde X^2) &= F_3^m \text{STr}(\sigma_3^m \tilde X^2) + F_3^{m-1} F_0 \binom{m}{1} \text{STr}(\sigma_3^{m-1} \tilde X^2)  + F_3^{m-2} F_0^2 \binom{m}{2} \text{STr}(\sigma_3^{m-2} \tilde X^2) \nonumber \\& + \cdots + F_0^m \text{STr}(\tilde X^2) \nonumber\\
&= -\frac{1}{2}F_3^m [\tilde X_0^2 + \tilde X_3^2 + \sum_{a=1}^2 \frac{\tilde X_a^2}{m+1}] -\frac{1}{2} F_3^{m-1}F_0 \binom{m}{1} [\tilde X_0 \tilde X_3 + \tilde X_3 \tilde X_0] \nonumber\\ &\quad -\frac{1}{2} F_3^{m-2}F_0^2 \binom{m}{2} [\tilde X_0^2 + \tilde X_3^2 + \sum_{a=1}^2 \frac{\tilde X_a^2}{m-1}]  + \cdots -\frac{1}{2} F_0^m \sum_{a=0}^3 \tilde X_a^2 \nonumber\\
&= -\frac{1}{2} \sum_{a=1}^2 \tilde X_a^2 \left\{ \frac{F_3^m}{m+1} + \frac{F_3^{m-2}F_0^2}{m-1} \binom{m}{2} + \cdots + \frac{F_3^{2}F_0^{m-2}}{3} \binom{m}{2} + F_0^m \right\}  \nonumber\\
&\quad -\frac{1}{2} (\tilde X_0^2 + \tilde X_3^2) \left\{ F_3^m + F_3^{m-2}F_0^2 \binom{m}{2} + \cdots + F_0^m \right\} \nonumber\\
&\quad -\frac{1}{2} (2\tilde X_0 \tilde X_3) \left\{ F_3^{m-1} F_0 \binom{m}{1} + F_3^{m-3} F_0^3\binom{m}{3} + \cdots+ F_3 F_0^{m-1}\binom{m}{1}\right\}
\end{align}

\item $\text{STr}(\mathcal H(F) X^2)$ with $\mathcal H(F) = a_0  + a_1 F^2 + a_2 F^4 + \cdots + a_m F^{2m} + \cdots$ an even function of the background field $F$:

\begin{align}
\text{STr}(\mathcal H(F) \tilde X^2)
&= -\frac{1}{2} \sum_{a=1}^2 \tilde X_a^2 \left\{ a_0 + a_1 \left[ \frac{F_3^2}{3}+F_0^2\right] + a_2\left[ \frac{F_3^4}{5} + \binom{4}{2} \frac{F_3^2 F_0^2}{3} +  F_0^4\right] + \cdots \right\}  \nonumber\\
&\quad -\frac{1}{2} (\tilde X_0^2 + \tilde X_3^2) \left\{ a_0 + a_1 \left[ F_3^2+F_0^2\right] + a_2 \left[ F_3^4 + \binom{4}{2} F_3^2 F_0^2 + F_0^4\right] + \cdots \right\} \nonumber\\
&\quad -\frac{1}{2} (2\tilde X_0 \tilde X_3) \left\{ a_1 \left[ \binom{2}{1}F_0 F_3\right] + a_2 \left[ \binom{4}{1} F_0^3 F_3 + \binom{4}{1} F_0 F_3^3 \right]+ \cdots \right\} \nonumber\\
&= -\frac{1}{2} \sum_{a=1}^2 \tilde X_a^2 \left\{ \frac{\int_0^1 d\alpha \mathcal H(F_0 + \alpha F_3) + \int_0^1 d\alpha \mathcal H(F_0 - \alpha F_3)}{2}\right\} \nonumber\\
&\quad -\frac{1}{2} (\tilde X_0^2 + \tilde X_3^2) \left\{\frac{\mathcal H(F_0+F_3)+\mathcal H(F_0-F_3)}{2}\right\} \nonumber\\
&\quad -\frac{1}{2} (2\tilde X_0 \tilde X_3) \left\{\frac{\mathcal H(F_0+F_3)-\mathcal H(F_0-F_3)}{2}\right\}
\end{align}
which is the prescription (\ref{prescr}).

\end{itemize}

\bibliography{bibfile,bibfile2}
\bibliographystyle{utphys}   
\end{document}